\documentclass[12pt,twoside,a4paper,openany]{cernrep} 
\usepackage{rep_common}
\usepackage{hyperref}
\hypersetup{
    colorlinks=true,
    citecolor=blue,
    linkcolor=blue,
    filecolor=magenta,      
    urlcolor=blue,
}
\usepackage[pagewise,displaymath]{lineno}
\usepackage{lscape}
\usepackage{enumitem}
\usepackage{rep_common}
\usepackage{makecell,multirow}
\usepackage{pdflscape}
\usepackage{rotating}
\usepackage{tabularx}
\usepackage{hhline}
\usepackage{mathtools}
\usepackage{mathptmx}

\def\bibfiles{\main/bib/chapter,\main/ewksection/bib/section,\main/Strong-Interaction/bib/section,\main/Flavour/bib/section,\main/Neutrino/bib/section,\main/Cosmic/bib/section,\main/BSM/bib/combined,\main/Darkmatter/bib/maindm,\main/Accelerator/bib/section,\main/Instrum/bib/section,\main/section1/bib/section}
\def\invfb   {\ensuremath{\mbox{\,fb}^{-1}}\xspace}
\def\upgradetwo {\mbox{Upgrade~II}\xspace}

\newcommand{\FCChh}{FCC-hh\xspace}
\newcommand{\FCCee}{FCC-ee\xspace}

\newcommand{\FCCeh}{FCC-eh\xspace}
\newcommand{\CLIC}{CLIC\xspace}
\newcommand{\CEPC}{CEPC\xspace}
\newcommand{\ILC}{ILC\xspace}
\newcommand{\CLICThreeHundredEighty}{CLIC$_{380}$\xspace}
\newcommand{\CLICThreeThousand}{CLIC$_{3000}$\xspace}
\newcommand{\ILCTwoHundredFifty}{ILC$_{250}$\xspace}
\newcommand{\ILCFiveHundred}{ILC$_{500}$\xspace}
\newcommand{\ILCGigaZ}{ILC$_{91}$\xspace}
\newcommand{\HLLHC}{HL-LHC\xspace}
\newcommand{\HELHC}{HE-LHC\xspace}
\newcommand{\LHeC}{LHeC\xspace}
\newcommand{\CLICfifteen}{CLIC$_{1500}$\xspace}
\newcommand{\ILCOneThousand}{ILC$_{1000}$\xspace}
\newcommand{\CLICGigaZ}{CLIC$_{91}$\xspace}
\newcommand{\FCC}{FCC\xspace}

\newcolumntype{Y}{>{\centering\arraybackslash}X}
\newcolumntype{S}{>{\hsize=.5\hsize}Y}
\newcolumntype{M}{>{\hsize=.55\hsize}Y}

\begin{document}
\newcommand{\main}{.}
\def\biblio{}
\pagenumbering{roman}
\begin{titlepage}

\begin{flushright}
\vspace*{-12mm}
CERN-ESU-004\\ 
10 January 2020\\
\vspace*{5mm}
\end{flushright}
\centerline{\huge\bf Physics Briefing Book}
\rm 
\vspace*{6mm}
\centerline{\it Input for the European Strategy for Particle Physics Update 2020}
\vspace*{4mm}
\begin{center}
\begin{small}
{\vspace*{-3mm}\bf Electroweak Physics:}\quad \rm Richard Keith Ellis$^1$, Beate Heinemann$^{2,3}$ {\it (Conveners)} \\
Jorge de Blas$^{4,5}$, Maria Cepeda$^6$, Christophe Grojean$^{2,7}$, Fabio Maltoni$^{8,9}$, Aleandro Nisati$^{10}$, Elisabeth~Petit$^{11}$, Riccardo Rattazzi$^{12}$, Wouter Verkerke$^{13}$ {\it (Contributors)}\\
{\vspace*{3mm}\bf Strong Interactions:}\quad \rm Jorgen D'Hondt$^{14}$, Krzysztof Redlich$^{15}$ {\it (Conveners)}\\
Anton Andronic$^{16}$
% ...
and the Conveners
\end{small}
\end{center}

\footnotesize\twocolumn\noindent
$^1$ IPPP, University of Durham, UK\\
$^2$ DESY, Hamburg, Germany\\
$^3$ Albert-Ludwigs-Universit\"at Freiburg, Germany\\
$^4$ University of Padova, Italy\\
$^5$ INFN Sezione di Padova, Italy \\
$^6$ CIEMAT, Madrid, Spain\\
$^7$ Humboldt-Universit\"at, Berlin, Germany\\
$^8$ Universit\'{e} catholique de Louvain, Belgium\\
$^9$ Universit\`a di Bologna and INFN, Bologna, Italy\\
$^{10}$ INFN Roma, Rome, Italy\\
$^{11}$ Aix Marseille University, CNRS/IN2P3, CPPM, \\
\mbox{\quad}Marseille, France\\
$^{12}$ EPFL, Lausanne, Switzerland\\
$^{13}$ NIKHEF and University of Amsterdam, Netherlands\\
$^{14}$ IIHE, Vrije Universiteit Brussel, Belgium\\
$^{15}$ University of Wroc\l aw, Poland \\
$^{16}$ Westf\"alische Wilhelms-Universit\"at M\"unster, \\
\mbox{\quad}Germany \\
% ...
\onecolumn
\end{titlepage}
\begin{small}
\setcounter{tocdepth}{1}
\tableofcontents
\end{small}
\newpage
\pagenumbering{arabic}
\pagestyle{headings}

\chapter{Introduction}
\label{chap:summ}
\addtocontents{toc}{\protect\setcounter{tocdepth}{0}}
The European Particle Physics Strategy Update (EPPSU) process takes a bottom-up approach, whereby the community is first invited to submit proposals (also called inputs) for projects that it would like to see realised in the near-term, mid-term and longer-term future. National inputs as well as inputs from National Laboratories are also an important element of the process. All these inputs are then reviewed by the Physics Preparatory Group (PPG), whose role is to organize a Symposium around the submitted ideas and to prepare a community discussion on the importance and merits of the various proposals. 
The  results of these discussions are then concisely summarised in this Briefing Book, prepared by the Conveners, assisted by Scientific Secretaries, and with further contributions provided by the Contributors listed on the title page. This constitutes the basis for the considerations of the European Strategy Group (ESG), consisting of scientific delegates from CERN Member States, Associate Member States, directors of major European laboratories, representatives of various European organizations as well as invitees from outside the European Community. The ESG has the mission to formulate the European Strategy Update for the consideration and approval of the CERN Council.

For the 2020 EPPSU, the call for inputs was issued at the end of February 2018 with the deadline for submission set to 18 December 2019. In total 160 submissions were received. The list is to be found in Appendix C. All the submitted inputs were considered in 11 different categories. Two categories were singled out for review by the ESG. These were the ``National Road Maps'' and ``Other'' submissions related to communication, outreach, strategy process, technology transfer or individual contributions.
The remaining nine categories dedicated to large experiments and projects, to physics, instrumentation and computing, and accelerator science were handled by the PPG, with an evident overlap between the various categories. The Open Symposium to review these inputs was hosted by the Spanish Community in Granada on 13-16 May 2019.

For the purpose of the Symposium, the PPG members (excluding the chair) took charge of organizing the parallel discussion sessions according to the following eight themes\footnote{Note that astrophysics and non-accelerator neutrino and dark matter experiments are under the purview of APPEC; those topics were included in the parallel session discussions at the Open Symposium to assess complementarities and enhance synergies, but will not be the subject of recommendations in the EPPSU.}
\begin{itemize}
\vspace*{-1mm}\item{B1} Electroweak Physics (physics of the $W$, $Z$, $H$ bosons, of the top quark, and QED)
\vspace*{-2mm}\item{B2} Flavour Physics and CP violation (quarks, charged leptons and rare processes)
\vspace*{-2mm}\item{B3} Dark matter and Dark Sector  (accelerator and non-accelerator dark matter, dark photons, hidden sector, axions)
\vspace*{-2mm}\item{B4} Accelerator Science and Technology 
\vspace*{-2mm}\item{B5} Beyond the Standard Model at colliders (present and future)
\vspace*{-2mm}\item{B6} Strong Interactions (perturbative and non-perturbative QCD, DIS, heavy ions)\footnote{A glossary of acronyms is available in Appendix~A.}
\vspace*{-2mm}\item{B7} Neutrino Physics (accelerator and non-accelerator)
\vspace*{-2mm}\item{B8} Instrumentation and Computing
\end{itemize}
Two half-days separated by half-a-day were dedicated to each of the discussion sessions, with four running in parallel (B1 to B4 and B5 to B8). Each session was convened by two PPG members, one theorist and one experimentalist where appropriate. The summary of these discussions was then presented in the plenary session, thus enabling all the Symposium participants to contribute to the discussions.\footnote{The presentations that made up the scientific programme of the Open Symposium are listed in Appendix~B.}

The main purpose of the Symposium was to reach an understanding of the potential merits and challenges of the proposed research programmes. For that purpose, the conveners enlisted many experts to summarise concisely the state-of-the-art and the potential progress expected in the future. The results are presented in the chapters of this Briefing Book: electroweak physics (Chapter~\ref{chap:ew}), strong interactions (Chapter~\ref{chap:si}), flavour physics (Chapter~\ref{chap:flav}), neutrino physics and cosmic messengers (Chapters~\ref{chap:neut} and \ref{chap:cosm}) and the high- and low-energy Beyond the Standard Model physics (Chapters~\ref{chap:bsm} and~\ref{chap:dm}). They are preceded by Chapter~\ref{chap:th} in which the state of the theoretical thinking which led the experimental efforts in the last decades is briefly outlined. The latter 
also helps to motivate the need for a vigorous experimental programme, which is required to make progress towards a deeper understanding of the physical laws that govern the Universe.
The advances in tools necessary to reach new horizons, in accelerator science (Chapter~\ref{chap:acc}) and instrumentation and computing (Chapter~\ref{chap:inst}) are presented at the end. 

In this introduction, the emerging physics landscape and its potential future is summarised, broadly following the structure presented in the last Strategy update, while keeping track of the advances that have been achieved since 2013. This should determine the set of priorities for the current Strategy update, which may well be different from the previous one.


\section{High Luminosity LHC}
Within the high priority large-scale scientific projects, the exploitation of the full potential of the Large Hadron Collider, including the high-luminosity (HL-LHC) upgrade of the machine and detectors to collect ten times more data than in the initial design, was top of the list of the 2013 Strategy update. The HL-LHC upgrade of the accelerator and of the ATLAS and CMS detectors was approved by the CERN Council in June 2016.

By the end of Run\,2 in December 2018, the ATLAS and CMS experiments collected about $160 \units{fb^{-1}}$ and the LHCb experiment about $10 \units{fb^{-1}}$ of $pp$ interactions at centre-of-mass energy  $\sqrt{s}=13 \units{TeV}$, exceeding the expectations. In the heavy-ion mode, the collected integrated luminosity also exceeded the projections. 
The very successful Run\,2 data collection allowed ATLAS and CMS to develop new methodologies to study the properties of the Higgs boson ($H$), substantially improving in many channels the projected precision on the $H$ couplings by the end of HL-LHC. Typically, the experimental uncertainty matches the statistical one and the total uncertainty is dominated by that on the theory input, which enters in the interpretation of $pp$ scattering data (see Fig.~\ref{fig:higgsnow} in Chapter~\ref{chap:ew}). The couplings of $H$ to the SM bosons and to the third generation fermions can be measured to the percent level, provided an improvement in the theory input by a factor of at least two is achieved. 

The LHCb Upgrade\,II, combined with the enhanced $B$-physics capabilities of ATLAS and CMS Phase\,II upgrades, will enable a wide range of flavour observables to be determined at HL-LHC with unprecedented precision, complementing and extending the reach of Belle II, and of the high transverse-momentum physics programme.

It is thus clear that the next two decades will see a very dynamic HL-LHC programme occupying a large fraction of the community.  Its success will rely not only on the experimentalists involved in the LHC experiments but also on a strong support of the theory community and, last but not least, new advances in the computing software and infrastructure.

\section{Design studies for pushing the energy frontier}
The recommendation of the previous Strategy update was for CERN to undertake design studies for accelerator projects in a global context, with emphasis on $pp$  and $e^+e^-$ high-energy frontier machines. These design studies were to be coupled to a vigorous accelerator R\&D programme, including high-field magnets and high-gradient accelerating structures, in collaboration with national institutes, laboratories and universities worldwide.

Two relevant inputs were submitted to the present strategy deliberations: the project implementation plan for the compact linear $e^+e^-$ collider (CLIC)~[ID146]\footnote{This notation is used to refer to the submitted documents, in this case ID $=146$, accessible via Appendix~C.} and the Conceptual Design Report for a Future Circular Collider (FCC)~[ID132, ID133, ID135] in two operational modes, as a $e^+e^-$ collider and a $pp$ collider, staggered in time in that order.

The CLIC linear collider would start as a Higgs, $WW$ and $t\bar{t}$ factory at $\sqrt{s}=380 \units{GeV}$, while the tunnel could be extended with time to achieve $\sqrt{s}$ of $1.5 \units{TeV}$ in the second stage and to $3 \units{TeV}$ in the final stage. The whole programme would last 34 years from the start of the construction.

The FCC design is such that it could start as an $e^+e^-$ collider (FCC-ee) evolving in time from a $Z$, $H$, $WW$ and $t\bar{t}$  factory by increasing $\sqrt{s}$ from about $90 \units{GeV}$ to $365 \units{GeV}$. In the second stage, the FCC would be turned into a $\sqrt{s}=100 \units{TeV}$ $pp$ machine (FCC-hh) with high-field magnets of up to $16 \units{T}$, also suitable for heavy-ion collisions. With the addition of an energy recovery electron linac (ERL) of $60 \units{GeV}$, also $ep$ interactions could be explored providing additional input to achieve the ultimate precision of the FCC-hh. This integrated FCC programme would last 70 years from the start of the project implementation. 

As part of the FCC project studies, also a high-energy version of the LHC (HE-LHC) with FCC-hh magnets and conversely a low-energy FCC-hh with the LHC-type magnets were considered. While these options would push the energy frontier, they were deemed less attractive than the FCC integrated programme.

The readiness of these projects was subject to intense scrutiny during the Granada Symposium and the conclusions are summarised in Chapter~\ref{chap:acc}. No show-stoppers were found on the technical side, however there are still challenges ahead with time scales for addressing them quite uncertain, more so in the case of FCC-hh than for CLIC.
In the global context, CLIC and FCC-ee are ``competing'' with the International Linear Collider (ILC) project proposed to be built in Japan [ID77], and with the circular CEPC of China [ID29]. In the latter case, the CEPC could be turned at a later stage into a $pp$ collider similarly to the FCC project. 
As Higgs factories, all the four contenders have a similar reach, as established during the Open Symposium (see Chapter~\ref{chap:ew}).  There are no major technical obstacles for their realisation, however more effort is required before construction of any of them could start.

The accelerator community, led in Europe by CERN with partners in the US and Japan, is investing efforts in the design of high-field magnets based on the Nb$_3$Sn superconductor and first successful tests of dipole magnets with $11 \units{T}$ field have recently been reported\footnote{Recently a field of 14.1\,T was achieved in a demonstrator dipole magnet at Fermilab.}. This is motivated by the needs of the HL-LHC upgrade programme. Substantial progress has been achieved in the development of superconducting and normal conducting high-gradient accelerating structures, needed for the  $e^+e^-$ colliders, which is also driven by light source facilities all over the world. CERN has also invested in developing novel accelerator technologies such as the dual-beam acceleration for CLIC or proton-driven plasma wake-field acceleration (AWAKE project) [ID35, ID58]. Lately, the idea of a $\mu^+\mu^-$ collider [ID120] is gaining traction in Europe as it represents a unique opportunity to achieve a multi-TeV energy domain beyond the reach of $e^+e^-$ colliders, and within a much shorter circular tunnel than for a $pp$ collider. The biggest challenge remains to produce an intense beam of cooled muons, but novel ideas are being explored. 

The details and the time-lines needed to develop some of these technologies are discussed in Chapter~\ref{chap:acc}. 
An interesting observation is that the estimated time quoted for development of 16\,T magnets for the FCC-hh is comparable to the one projected, albeit with lesser confidence level, for the development of the novel acceleration technologies from proof-of-principle towards an accelerator conceptual design.

\section{An $e^+e^-$ collider complementary to the LHC}
Already the previous Strategy update expressed interest in the initiative of the Japanese particle physics community to host the ILC and welcomed this initiative. The negotiations in Japan are ongoing but no clear statement has been made at this time. As described above, three additional potential Higgs factory projects have been submitted for consideration in this European Strategy Update process.

From the national inputs submitted to the present Strategy update process,  a clear support is evident for an $e^+e^-$ Higgs factory as the next large-scale facility after the LHC. In the absence of clear signs of physics beyond the Standard Model, the hierarchy problem between the mass of the Higgs boson and the Planck scale still remains a strong argument to look for new physics at the energy frontier, and Higgs coupling measurements provide a powerful probe of the EW symmetry-breaking mechanism. The new physics, it is argued, would influence the values of the Higgs couplings to the fundamental constituents of matter and interactions, and could be detected provided they are measured with sufficient precision to be sensitive to the relevant energy scales (for detailed discussion see Chapters~\ref{chap:ew} and~\ref{chap:bsm}). Thus a Higgs factory could already provide the first hints of new physics.

The comparison of the performance of the various proposed Higgs factories was thus very much in the focus of the Open Symposium. The most precise determination of the Higgs couplings would be achieved by an $e^+e^-$ factory in combination with improvements of the knowledge of the Standard Model couplings from a Tera-$Z$ facility (as proposed for the FCC-ee), followed by the high yield production of Higgs at a $100 \units{TeV}$ $pp$ collider. Beyond Higgs physics, FCC-ee would also offer an interesting flavour physics program as well as searches for the dark sector.

On a time scale of 70 years, the integrated FCC programme would allow to determine the Higgs self-coupling to explore the nature of the electroweak phase transition with a precision of 5\%.
A similar sensitivity for this particular aspect could emerge from the CLIC integrated programme on a shorter time scale.


\section{Beyond the Standard Model at Colliders}

The aim of future colliders is the exploration of the unknown at very short distances, in the search for an understanding of the fundamental physical laws and an explanation of the many mysteries that still surround the world of particle physics. In this context, the study of physics beyond the Standard Model  (BSM) is a primary element of any future collider programme. 
The discovery of the Higgs boson has triggered the need to start a new physics programme of precise determinations of the Higgs properties and, correspondingly, of electroweak measurements with improved precision. These tests provide powerful probes of any kind of new physics that affects directly the electroweak symmetry breaking sector. A typical example is the case of Composite Higgs, for which future facilities can probe the degree of naturalness (fine-tuning) well below the percent level. There are several proposed projects around the world to carry out the Higgs precision programme (see Chapter~\ref{chap:ew}) and their respective physics reach for BSM physics  is documented in Chapter~\ref{chap:bsm}.

The exploration of short distances can proceed through direct or indirect searches. Proposed future colliders can explore new physics extensively, up to multi-TeV scales, through direct searches. Just to take some quantitative examples, FCC-hh can probe gluino masses\footnote{Natural units are adopted throughout this document, i.e.\ taking $c=1$.} up to $17 \units{TeV}$, stop masses up to $10 \units{TeV}$ and masses of scalar particles from a second Higgs doublet up to the range of $5-20 \units{TeV}$. CLIC at $\sqrt{s}=3 \units{TeV}$ can perform general searches for any new particle with electroweak interactions essentially up to the kinematical limit, which corresponds to masses of 1.5 TeV for pair production. Direct searches provide the only way to have hands-on access to new phenomena.

The indirect searches consist of looking for deviations from the Standard Model expectations, like for example in modification of Higgs couplings or of kinematic distributions with sensitivities to virtual effects of the theory content.  Indirect searches can probe in a model dependent way masses well beyond the collider kinematical limit, but typically cannot identify the specific source of new physics.  For a selection of physics scenarios, addressed in Chapter~\ref{chap:bsm}, lepton and hadron colliders are complementary. Hadron colliders have a better reach for direct searches of new states because at the lepton colliders there is a natural limit imposed by the available centre-of-mass energy. Lepton colliders tend to perform better in indirect searches in spite of the substantially lower centre-of-mass energy.  Weakly coupled theories where high luminosity is an important factor are better explored at hadron colliders. 

Colliders operating at very high energy also contribute in a complementary way to the Higgs, the electroweak precision and the flavour programmes. This includes rare Higgs decays
(e.g.\ $H \to \mu \mu,\, \nu \nu , \, Z\gamma$) which benefit from the large luminosity of hadron colliders and of effective operators whose contribution to scattering processes grows with the collision energy.

Diversity in research is a key element for a future strategy in particle physics, especially in view of the rapidly evolving status of  theoretical understanding. Feebly-interacting and long-lived particles are good examples of motivated paradigms, as discussed in Chapter~\ref{chap:bsm}. The investigation of these and other alternative paradigms requires a variety of experimental facilities, not limited to colliders, but complemented by beam-dump, fixed-target and other experiments.

\section{Neutrino Physics}

The discovery of neutrino oscillations is a ``laboratory'' proof of physics beyond the Standard Model, because new particle states or new interactions are required to generate the relevant mass term in the theory. The neutrino sector looks very different from the charged fermion one. The neutrinos are known to be orders of magnitude lighter than the charged leptons.  The explanations for this lightness span many orders of magnitude in the scale of new physics.  There could be light (sterile neutrinos) or heavier neutral leptons. Neutrinos could be their own antiparticles  in which case  lepton number conservation would be violated. This property could potentially be linked to the matter-antimatter asymmetry observed in the Universe. Their mixing pattern is also very different from the one observed for the charged fermions, with some terms still not fully known. Neutrino physics is an integral part of the flavour quest. It is thus essential to pursue the exploration of the neutrino sector with accelerator, reactor, solar and atmospheric neutrino experiments.  The rich programme already approved and the outlook for future  progress are discussed in detail in Chapter~\ref{chap:neut}. 

In 2013 CERN was mandated with developing a neutrino programme to pave the way for a substantial European role in future leading long-baseline experiments in the US and in Japan. This recommendation led to the establishment of the CERN Neutrino Platform (NP) in 2014 in which about 90 European institutions are involved. The main goal of the CERN NP is to support and participate in detector R\&D and construction for projects with European  interest and expertise (e.g.\ prototypes for the DUNE experiment at LBNF in the US and the near detector ND280 for T2K in Japan). In addition a neutrino group was set up at the CERN EP department in 2016 to help enhance coherence of efforts in the European neutrino community.

The CERN NP has been successful in fostering the European effort in advancing the study of neutrino oscillations.
This is reflected in
a very successful town meeting organised as part of the preparations for the European Strategy Update and its conclusions [ID45], which provide a broad view of the European present and future activities in this domain. From the European perspective,  the world-wide,  broad and challenging experimental programme in unravelling the neutrino sector, with synergies in particle, nuclear and astroparticle physics, requires a continuous balanced support. Continuation of the CERN NP may be an appropriate way to facilitate this. 

\section{Theory}
As already acknowledged in the previous Strategy update, theory is a strong driver of particle physics and provides essential input to experiments.
In addition to the important work connected closely to experiments, such as precision calculations and event generators, there are innumerable examples that show how apparently abstract
theoretical investigations have led to central developments
influencing experimental physics. By speculating on the extravagant
concept of local hidden variables while working at CERN, John Bell
laid the foundations of the most insightful experimental tests of
quantum mechanics and quantum entanglement. It was abstract research
on extensions of space-time symmetries in quantum field theory that
prompted experimentalists to design hermetic detectors with optimal
rapidity coverage and to improve triggering techniques. Peter Higgs'
celebrated 1964 paper had the purely theoretical aim to show that
Gilbert's theorem is invalid for gauge theories. At the time,
applications to electroweak interactions were well beyond the horizon.

Theoretical research in fundamental physics needs to be broad in scope
and not only limited to the goals of ongoing experimental
projects. A free and diverse theoretical activity, although prone to a number of unsuccessful attempts, is much more likely to lead to scientific breakthroughs than if limited to a targeted research programme.

\section{Flavour Physics and CP violation}
The observed pattern of masses and mixings of the fundamental constituents of matter, quarks and leptons, remains a puzzle (often called the flavour puzzle) in spite of a plethora of new experimental results obtained since the last Strategy update. It is hard to imagine that the new physics necessary to stabilise the Higgs mass would have no impact on the flavour sector. Conversely, solving the flavour puzzle may indicate the way to the new physics. 

The field of flavour and CP violation, with its many parameters entering the predictions of the Standard Model only through measurements, is traditionally explored through a wide spectrum of experiments all over the world.  These include measurements of electric dipole moments of charged and neutral particles and molecules, rare muon decays with high intensity muon beams at PSI, FNAL and KEK, rare kaon decays at CERN and KEK, and a variety of charm and/or beauty particle decays at the LHC with, in particular, the LHCb experiment. New results are expected in the near future from the Belle II experiment at KEK in Japan and from LHCb (currently undergoing an upgrade). A detailed discussion of the short-, mid- and long-term programme is presented in Chapter~\ref{chap:flav}. All these experiments are very challenging as they require large statistics and excellent control of experimental uncertainties commensurate with the expectations of the Standard Model, which in many cases are very precise. The reward is sensitivity to large scales for new physics, often by orders of magnitude higher than from direct detection experiments or precision electroweak measurements, as illustrated in Fig.~\ref{fig:NPscales}.  In the mid-term planning in Europe, much can be gained from the Upgrade II of the LHCb experiment for the HL-LHC, that is still pending approval, in addition to the hope that the pending question of lepton number universality will be fully resolved.  On the longer term, the Tera-$Z$ option of the FCC-ee also offers an attractive program of exploring flavour physics with high precision.

 From both the experimental and the theory side, a novel synergy  between the searches for flavour violating decays and for feebly interacting and dark particles is emerging. High energy colliders will explore  the high-mass range (above $10 \units{GeV}$).  Nevertheless fixed target experiments, the proposed LHC projects dedicated to long-lived particles, and beam-dump facilities, may provide complementary information to explore a lower mass range  ($1 \units{MeV}$ to $10\units{GeV}$, and even beyond in some cases) and open interesting new research lines. 

The search for flavour and CP violation  in the quark and lepton sectors at different energy frontiers has a great potential to lead to new physics at moderate cost and therefore flavour physics should remain at the forefront of the European Strategy. 
\vfill

\section{Dark Matter and the Dark Sector}

Within the context of General Relativity there is ample evidence from galactic and cosmological observations that dark matter (DM) is the dominant form of matter in the Universe, and detecting it in the laboratory remains one of the great challenges of particle physics. The existence of DM is another compelling evidence of physics beyond the Standard Model. It is highly plausible that DM is part of a richer 
hidden sector (HS), whose constituents may include multiple species of massive particles, one or more of which may mix with Standard Model particles such as the Higgs boson, the photon or neutrinos, via the so-called HS-SM portals. The current understanding of the basic properties of dark matter and its interactions is poor. 

Historically direct-detection DM experiments have been dominated by WIMP searches, motivated by the so-called ``WIMP miracle'': the qualitative observation that particles with masses of the order of~$100 \units{GeV}$, and weak interactions with SM particles, will end up with roughly the observed thermal relic density after freeze out in the standard Big Bang cosmology. However given the present limits from multiple overlapping direct detection experiments, the paradigm is changing and in principle the mass of dark matter particles could be anything from as light as $10^{-22} \units{eV}$ to as heavy as primordial black holes of tens of solar masses. A comprehensive suite of  experiments and techniques are required in order to cover the many possibilities. 
 
 Accelerator-based beam-dump and fixed-target experiments  can perform sensitive and comprehensive searches of sub-GeV DM and its associated dark sector mediators. They will broadly test models of thermal  light DM that are as yet underexplored.  Future colliders (ILC/ CLIC, FCC-ee/hh/eh and HL/HE-LHC) all have an excellent potential to explore models of thermal DM in the GeV to $10 \units{TeV}$ mass range. New search strategies at the LHC, for long-lived feebly-interacting particles, with detectors located far away from the interaction point, offer a complementary reach. The search for ultralight DM particles like the axion has gained significant momentum. They would arise as a consequence of one solution to the strong CP problem: why QCD appears to preserve CP symmetry. The axions or  axion-like particles could be detected directly in dedicated experiments, or produced in the laboratory in prospective light-shining-through-wall experiments. A detailed account of the various scenarios and relevant experimental programme is presented in Chapter~\ref{chap:dm}.

Europe has the opportunity to play a leading role in the searches for DM by fully exploiting the opportunities offered by the CERN facilities, such as the SPS, the potential Beam Dump Facility (BDF), and the LHC itself, and by supporting the programme of searches for axions to be hosted at other European institutions. The preparatory study of the BDF facility is now mature and a decision should be taken on its implementation following this Strategy update; its potential sensitivity to new physics should be compared to that of competing proposals, such as long-lived particle searches at the LHC, to inform this decision. 
There is a strong complementarity and synergy between direct DM detection experiments, under the auspices of APPEC, and the programme for its production and discovery in accelerator-based experiments. CERN  support for direct dark matter searches based on technologies for which CERN has expertise could deliver a decisive boost to their sensitivity.

\section{Strong Interactions}
Quantum Chromodynamics (QCD) is firmly established as the theory of strong interactions. It encodes the dynamics of quarks and gluons (partons). The dependence of the QCD coupling $\alpha_s(Q)$ on the energy scale $Q$ is predicted in QCD to evolve from a strong coupling at low energy scales to a weak coupling at high energy scales. As a consequence, quarks and gluons are confined into hadronic bound states at low energies, while they behave as asymptotically free at high energies. It is worth noting that the binding energy of QCD dynamics generates $\sim 95 \%$ of the proton and neutron mass, and thus ordinary matter, and only $\sim 5 \%$ originates from the coupling of quarks to the Higgs field.

The concept of asymptotic freedom allows for precise quantitative predictions for QCD processes at high energy colliders to be obtained through systematic perturbation expansion. For a full exploitation of precision collider data, these calculations need to attain high accuracy, which comes with many conceptual and technical challenges. Moreover, the perturabtive calculations are performed at the level of partons and have to be translated into experimental observables which involve hadronic states. This step introduces the low-energy, non-perturbative, QCD dynamics into the predictions for which the quantitative understanding  is less fully developed. However, the QCD factorisation theorem allows to separate the low-energy and high-energy dynamics, enabling predictions for collider processes by parametrising the strong-coupling dynamics into empirical quantities such as decay form-factors, parton distributions or hadronisation models. These have often dual relevance, as fundamental objects of investigation and as input to predictions.

To turn present and future hadron colliders into precision machines, without compromising their sensitivity to a wide spectrum of novel physics effects, and to take full advantage of the investment into theoretical calculations, an independent determination of the proton structure is very important. A programme based on fixed target experiments and on dedicated $ep$ machines has been proposed in Europe, in the US and in China. It is discussed in detail in Chapter~\ref{chap:si}. The high-energy end of the proposed facilities at CERN such as the LHeC and/or FCC-eh have in addition the potential to complement the programme of BSM physics discussed above. 

The early Universe has undergone a series of phase transitions of fundamental quantum fields, in particular the transition of matter from a quark-gluon plasma (QGP) in which partons are deconfined. This high temperature phase is experimentally accessible in the heavy-ion collision experiments. The challenge  is in understanding how collective phenomena and macroscopic properties, involving many degrees of freedom, emerge under extreme conditions from the microscopic laws of QCD. Though the creation of QGP as an almost perfect liquid has been experimentally established, studies of heavy-ion collisions at the LHC (by the dedicated experiment ALICE, as well as the other experiments)  and at RHIC (Brookhaven) have been a constant source of surprises, driving the theory developments. The observation of collective effects in $pp$ collisions came as another surprise and opened a new area of studies for the heavy-ion community. A high-energy $AA/pA/pp$ research programme  at present and future colliders would be unique to Europe and would lead to a profound understanding  of hot and dense QCD matter. The lower-energy research programme of QCD matter at the SPS at CERN, is complementary to other emerging facilities worldwide in the US (BES at BNL), in Germany (FAIR), in Russia (NICA at JINR)  or in Japan (J-PARC), and brings valuable contributions in the exploration of the QCD phase diagram.   

The mathematical framework for quantitative predictions of QCD for low scales and/or high density processes is numerical lattice QCD (LQCD). Over the past years, an increased computing power, together with the development of new algorithms and  analytical frontier techniques, have enabled precise  determination of a wide range of hadronic observables. Continued efforts  and support in developing new theoretical methods and better algorithms are needed to reach a fully predictive power of LQCD. Future progress in fundamental understanding and precision phenomenology of QCD will rely on a diverse research programme with close interplay between theoretical advances and experimental measurements.

\section{A diverse experimental physics programme}

There is a variety of tools to progress in addressing the fundamental puzzles of Nature, with no solution in the Standard Model, the list of which (on the particle-physics side: the origin of electroweak symmetry breaking, the nature of the Higgs boson, the pattern of quark and lepton masses, the neutrino nature and mass; on the cosmology side: dark matter, dark energy, inflation, the matter-antimatter asymmetry) did not change in the post-Run\,2 LHC era. To decipher the fundamental laws of nature, a judicious combination of abstract methods in theoretical physics and precise experimental scrutiny is needed. These two elements are  complementary and are both essential for progress in particle physics.
Experimentally, one way ahead in the exploration is to increase the reach of direct searches by increasing the energy scale at which the puzzles can be explored. The alternative way is to perform precision measurements of rare processes fuelled by quantum-mechanical effects of the theory at short distances. These indirect searches can in principle be performed at much lower energies. While the selection of the next large-scale collider project is currently focused upon, there is thus a strong case for also maintaining a diverse physics programme of smaller-scale experiments.

In preparation for the 2020 European Strategy Update, a study group for ``Physics Beyond Colliders'' was initiated by CERN, originally to explore the full scientific potential of CERN's accelerator complex and its scientific infrastructure in the next two decades through projects with unique physics reach, complementary to the LHC, HL-LHC and other possible future colliders [ID42]. 
It became a forum for discussing projects that target fundamental questions, some of which could in fact be realised outside CERN. Furthermore, given the long time-scales involved in planning and realising large-scale projects, it is essential to propose a parallel research programme to attract and educate next generations of scientists capable of carrying out the ever more challenging experiments. This was also recognised in the previous Strategy update which recommended that such experiments be supported in Europe, as well as the European participation in experiments in other regions of the world. The proposals for new experiments of this smaller-scale type are discussed in Chapters~\ref{chap:flav}, \ref{chap:bsm} and \ref{chap:dm}. 

\section{Essential tools for the future of particle physics}

Physicists and engineers have designed and constructed generations of accelerators with increasing centre-of-mass energy and beam intensity, and complex particle detectors, confronting also the associated computing challenges.
The discovery at the LHC of the Higgs boson was the result of different scientific and technical expertise coming together with a common goal.  This amazing achievement could not have been possible without the success of each of these contributions, and any future particle-physics programme must rely on the synergy between all of these different components. Supporting and developing them is therefore an essential prerequisite for continuing our exploration of the fundamental laws of Nature.

There is no lack of original ideas for how to exploit existing infrastructures and how to explore energy scales higher than those presently available. The way forward involves challenges that cannot be addressed without constant progress in advancing accelerator science, designing better detectors, and developing proper computer infrastructures. These issues are discussed in Chapters~\ref{chap:acc} and~\ref{chap:inst}, and summarised below. Moreover, the support of the theory community is essential in these endeavours, not only through generation of new ideas but also in the development of expert computational tools as discussed below.

\subsection{Accelerator Science}
New accelerator facilities are considered for the future scientific advances in particle physics. Their development is driving progress in accelerator science which will have a tremendous impact on the size, performance and cost of future facilities for societal applications such as in medical or industrial applications, not to mention for advances in other accelerator-driven fields of science which need light and neutron sources. 

Future $pp$ colliders drive the development of high field magnets, with the goal of achieving fields of $16 \units{T}$ with the Nb$_3$Sn superconductor. The estimates of time-scales necessary to develop new approaches and technologies range from as little as five years for optimising the existing technologies to as much as 20 years for the $16 \units{T}$ option. Higher fields will require advances in the development of high temperature superconductors (HTS) and the accelerator science in Europe could motivate these developments [ID105].

For the proposed $e^+e^-$ options the main challenges for both linear and circular colliders are the RF cavities (energy) and nanobeam (luminosity) performances. In that respect there are strong synergies with modern synchrotron and FEL light-source requirements, which should therefore be fully exploited through close collaboration between European and overseas partners. The designs for the first phases of CLIC and ILC are mature and complete. Their power and cost budget are estimated to be on a scale similar to the LHC making them well suited for implementation.

The design of a $\mu^+\mu^-$ collider, previously studied in the US, is gaining traction in Europe thanks to new ideas in muon cooling. The manipulation of muon beams is similar to that of the proton ones and the LHC tunnel could be used to achieve $\sqrt{s}=14 \units{TeV}$.  A strong R\&D programme would be needed to develop this facility as a possible candidate for a high-energy physics project; for that to happen, the formation of a global collaboration will be essential to carry out the work coherently and efficiently. 
 
Plasma-based particle accelerators, where accelerating fields are created by the collective motion of plasma electrons driven by lasers or particle beams, have shown capability of reaching an order of magnitude higher gradients than presently achieved, although the possibility to reach the beam quality needed for HEP applications remains to be demonstrated. In the past decade significant progress has been made in plasma wake-field acceleration. In view of the great promise of these novel acceleration techniques and the substantial effort worldwide to develop them, the advanced linear collider study group, ALEGRO [ID7], aims to foster studies on accelerators for applications to high-energy physics, with the ambition of proposing a machine that would address the future goals of particle physics.

There is a rich R\&D programme for improving the existing facilities, building and developing new facilities, as described in detail in Chapter~\ref{chap:acc}. An important issue, which has been brought into focus with the ambitions to push the energy frontier, is energy management. It is the HEP community's responsibility to develop sustainable models and optimised technologies in terms of energy consumption, aiming also at exporting improved technologies for other applications in society. It is essential for the future of particle physics that accelerator science be supported with high priority and that the already existing expertise be preserved. A strong cooperation between national institutes and CERN is vital for the progress of the field. 

\subsection{Instrumentation and Computing}

 The landscape of proposed next generation experiments is broad in terms of detector technologies designed to fulfill the physics programmes of the future.  These various technologies are being developed to address well-defined technological challenges such as as micron-scale spatial resolution and low mass, picosecond time resolution, high-performance photodetectors (also operating at cryogenic temperature and low dark current), radiation tolerance, large number of channels, high readout speed, and large sensitive area at low cost. The need for combined features (adding time and/or energy measurement in 4D tracking and 5D imaging) becomes more and more pressing.  
Also, for all types of particle detectors, the integration of advanced electronics and data transmission functionalities plays an increasingly important role.  Beyond R\&D activities driven by the needs to fulfill specific experimental requirements, it is essential that the community maintains the ability to carry out generic detector R\&D work that has the potential to bring about tool-driven revolutions, and this, concurrently through the design, planning and execution of large experimental projects.  Generic technology innovation often emerges from synergies within the field of particle physics, with other fields of science, or with industry.  
Therefore it is important to ensure that European programmes such as AIDA2020 or ATTRACT be appropriately supported in order to preserve and stimulate the community's potential for innovation. 

The development of novel particle physics instruments requires specialized infrastructures, tools and access to test facilities. National labs and large institutions play a central role in support for the community by providing access to these types of specialized infrastructures, tools and facilities. 
One example is the European network of test beam and specialized irradiation facilities that currently exists, and for which the continued and coordinated support has been identified as of utmost importance for the community.  In addition, technical personnel are required to efficiently exploit the specialized infrastructures, tools and facilities needed for detector R\&D.  The support of these personnel often remains a challenge, that must be addressed. 

In addition to detector development activities, the scientific outcomes of an experiment are made possible by the development of an efficient computing and software infrastructure. In the coming years, however, the science programmes at the HL-LHC Run\,4 and beyond, Belle-II at SuperKEKB, future circular and linear colliders, and large neutrino experiments, will together require about an order of magnitude more computing resources than presently available, while increase in funding for computing is not expected.  To meet the challenges the particle physics community must carry out carefully planned and coordinated R\&D programmes to improve the efficiency of HEP software and algorithms, adopt new hardware, and take advantage of industrial trends and emerging technologies.  To carry out these activities, a significant investment in skilled developers is of the highest importance.
Furthermore, exploiting synergies among experiments, other disciplines and with industry will be vital to provide a sustainable future for software and computing in the field.  There are many vehicles for these synergies to be  exploited: for example, both the WLCG and HEP Software Foundation (HSF) will have an important role promoting coherence in various development activities. 

More generally, the requirement of efficiency to extract the maximum physics potential from an experimental research programme requires an increasingly holistic approach to the design of experiments and their associated computing and software systems. For example, the evaluation of various detector designs must include the computing burden as a metric.  Yet the detector and the computing/software communities have been drifting apart, and individuals that can bridge the growing gap are rare. This is a challenge to the community.

Another challenge faced by the particle physics community is the limited amount of success in attracting, developing and retaining instrumentation and computing experts, which poses a growing risk to the field.  It is of utmost importance that activities carried out by these experts be recognized correctly as fundamental research activities bearing a large impact on the final physics results.  

\subsection{Theoretical and Phenomenological tools}
While, as discussed above, speculative theoretical research is a powerful driver of progress in particle physics, theoretical physics has another essential role for collider projects. The interpretation of LHC data would be impossible
without theoretical input from higher-order perturbative calculations at the parton level, in combination with parton distribution functions
(PDF) and the modelling of parton showers. The recent advances in theoretical calculations, together with progress in data analysis and
detector performance, have allowed previously unimaginable precision in measurements at the LHC. This is a critical element for future
collider programmes, since their ability to discover new phenomena heavily relies on accurate background determinations. Only with
significant advances in theoretical calculations can one hope to perform a valuable programme of precision Higgs and electroweak
measurements at future high-energy colliders (see Sect.~\ref{sec:ewktheory}).

The need for more refined theoretical calculation will only grow in the future, both for HL-LHC and for colliders at higher energy (see
Sect.~\ref{precision_sec}). Fully automated NLO tools are now available and the next challenge is to upgrade them to the
NNLO level. To describe the full hadron collision perturbative calculations are matched with parton showers using automated Monte
Carlo generators. Efficiency and accuracy improvements of these tools are needed to fully exploit the higher precision of perturbative
calculations. The high collision energy of future facilities is opening up new challenges (together with new theoretical
opportunities) related to the all-order resummation of large logarithmic terms, due to the presence of different energy scales in a
single scattering process. A better characterisation of the theoretical uncertainties in the PDFs is also needed. 

Monte Carlo event generators [ID114] are indispensable workhorses of particle physics, bridging the gap between theoretical ideas and first-principle calculations on the one hand, and the complex detector signatures and data of the experiments. They add to the theoretical input the low-scale transition of partons to hadrons and the multiple partonic interactions which contribute significantly to the overall particle yield.  All  experiments are dependent on event generators to design and tune the detectors and analysis strategies. As the precision of data increases, the imperfections in the event generators become visible and may lead to extra uncertainties. To mitigate these effects constant improvements in the event generators is mandatory. The development of these tools  is overwhelmingly driven by a vibrant community of academics at European universities and at CERN. 
All of this requires very particular skills, theoretical and computational, that have to be preserved and fostered for the future. 

Another theoretical research area with direct impact on future experimental programmes in particle physics is lattice gauge theory,
which is the only known method to compute consistently and systematically QCD observables in the non-perturbative regime. Lattice
calculations can provide reliable results for hadronic decay constants, form factors, and matrix elements that enter many
observables relevant for low-energy and flavour physics, for the determination of $\alpha_s$, for the extraction of the PDFs, and for
properties of the quark-gluon phase transition (see Sect.~\ref{si-LQCD}).

Advances on all these fronts require not only more resources (funding, manpower, computing time for numerical and algebraic calculations, organised collaborative networks), but also truly conceptual breakthroughs in theoretical techniques. New ground-breaking ideas are rapidly
developing, although severe challenges remain.  An adequately supported programme of theoretical activities is indispensable for the success of any experimental project and should be an integral part of the planning of future strategies for particle physics. 

\section{Synergies}

There are obvious synergies between the various areas of particle physics research and they are highlighted in the following chapters.
There are also clear synergies between particle physics and nuclear physics, through the ambition to achieve first-principle  understanding of strong dynamics based on QCD, but also because of similar experimental tools, as evident in the nuclear physics programme conducted at CERN. The latter includes not only the low-energy (by particle physics standards) programmes of the ISOLDE [ID39] and n\_TOF facilities, but also the heavy-ion programme at the SPS and the LHC. There are also synergies with atomic physics as exemplified by the $\overline{\rm H}$ experimental programme at the AD antiproton decelarotor. 

There are strong synergies with astroparticle physics, which addresses some of the same fundamental questions as particle physics. These connections are through neutrino physics, dark matter searches, cosmic ray physics and, potentially in the future, gravitational waves.
The precision measurements of the neutrino properties rely on solar and atmospheric neutrinos for the determination of several mass and mixing parameters (see Chapter~\ref{chap:neut}). Large underground neutrino detectors are used in long-baseline accelerator experiments and in astroparticle physics, with strong synergies between the two domains (see Sect.~\ref{sec:CM-synHEP}). Searches for dark matter from the halo are performed by dedicated underground experiments but also by large astroparticle detectors like H.E.S.S., Antares or IceCube, and in the near future the CTA observatory expected to start operations in 2022. The complementarity is not only through technological advances but also in the parameter space probed by astroparticle and accelerator-based experiments (see Sect.~\ref{sec:DM-Colliders}). 

The future Einstein Telescope for gravitational wave detection [ID64] will use infrastructure and techniques that are very similar to those deployed for large underground accelerator complex. For cosmic ray physics, precise simulation of the properties of air showers is needed to properly determine the mass of the primary cosmic rays. The shower development is driven mostly by hadron-nucleus interactions from the highest (1000~TeV c.m.) to the lowest (10~GeV lab) energies whose theoretical description relies heavily on collider and fixed-target data. 

There are thus multiple synergies between particle and astroparticle physics, at the level of infrastructure, 
detectors,
interaction models and physics goals. These synergies and the need to foster them has been clearly identified in the national inputs. This can be facilitated by the newly established EuCAPT Astroparticle Theory Centre as a joint venture of APPEC and CERN, as well as by further discussions among the many experts in the field. How to enhance the cooperation between the particle, astroparticle and nuclear physics communities, fully benefiting from the close collaboration between ECFA, APPEC and NuPECC, must be part of the discussions around the European Strategy Update.

The development of novel accelerator technologies has always been driven by the needs of high-energy physics. Today, diverse fields of research and applications  benefit from these developments but also contribute to advances in the field. Examples include fusion energy, high temperature superconductors, medical applications, photonics and neutronics.  And, last but not least,  plasma acceleration promises developments of compact facilities with a wide variety of applications compatible with university capacities and small and medium sized laboratories. A detailed discussion is presented in Sect.~\ref{sec:ACC-synergies}.


An important aspect of the European Strategy Update is to recognize the potential impact of the development of accelerator and associated technologies on the progress in other branches of science, such as astroparticle physics, cosmology and nuclear physics. Moreover, joint developments with applied fields in academia and industry have brought about benefits to fundamental research and may become indispensable for progress in the field.

Similar considerations apply to the field of instrumentation and computing. Synergies with other fields of science with similar challenges and with industry are essential to meet the needs of the next generation of experiments. The way forward has been extensively discussed during the Open Symposium in Granada and is summarised in Chapter~\ref{chap:inst}.
Strengthening the synergies in research and technology with adjacent fields is an important element of the progress in particle physics. Global platforms, networks and national institutes have the potential to enhance the research exchange among experts worldwide, to provide training opportunities, and more generally stimulate the community's potential for innovation.

\addtocontents{toc}{\protect\setcounter{tocdepth}{1}}


\baselineskip=14pt
\chapter{Theoretical overview}
\label{chap:th}
\section*{The role of exploration in particle physics}

Exploration of the unknown is the main driver of fundamental science. 
  The goal of particle physics is to push the frontier of knowledge deep into the smallest fragments of spacetime and unravel the natural phenomena that occur at the most minute distance scales. This line of research has delivered some of the most extraordinary discoveries in science, which not only have revealed the inner workings of particle interactions but have truly revolutionised our understanding of the physical laws that govern the Universe. Those laws have allowed us to decipher the properties of the Universe at the largest distance scales and reconstruct its time evolution back to the earliest stages. This path of discoveries and knowledge has continued with the latest generation of experimental projects in particle physics, among which the LHC is the most prominent example. Although this broad research programme is still ongoing, particle physics is already planning the next stage of exploration. While research that culminated with the LHC has established the Standard Model (SM) as the successful description of particle interactions, we are still confronted with many unresolved puzzles and open problems that can be tackled only with a bold experimental programme and with substantial technological advances.

\section*{Can we predict new discoveries?}

When the goal is exploration of the unknown, by its very nature it is difficult (or impossible) to foretell the discoveries that an experimental project may encounter. The value of an exploratory project should not be measured by the number of promised new discoveries, but by the importance of the questions addressed and by the amount of fundamental knowledge that can be extracted from its results.
There is a remarkable exception to the general rule concerning our inability to predict the unknown.
 It follows from a peculiar property of quantum field theories (QFT). A QFT can ``predict its own destruction,''\footnote{This expression was used by Pilar Hernandez in her talk at the Open Symposium in Granada.} in the sense that from low-energy measurements alone one can infer the existence of new phenomena that must {\it necessarily} occur below a calculable high-energy scale $\Lambda$, even if the theory is unable to predict what these new phenomena are. It is of course a very privileged situation for experimental searches.  
 This was indeed the case of the SM without the addition of the Higgs, which predicted its own destruction at energies below a TeV, as subsequently confirmed by the LHC with the discovery of the Higgs. 
 
 Does the SM today, after the inclusion of the Higgs boson, predict its own destruction?  The answer is negative: unlike the circumstances at the start of the LHC, today particle physics is not in the (rare and special) situation to predict new discoveries  with mathematical certainty at or under an energy scale within reach.  
 Remarkably, the SM properties are just right to keep all its coupling constants under control, up to meaningful high energies. The Higgs quartic coupling evolves towards an instability, but this is reached at sufficiently high energy to ensure that the lifetime of the electroweak vacuum is much longer than the age of the Universe. This remarkable self-consistency of the SM is very sensitive to the values of the coupling constants and it would not hold if the couplings were only slightly different from what we observe.  In a different realm, as soon as gravity is included the SM does predict its own destruction at an energy scale of $10^{19}$~GeV or below. In addition, neutrino masses suggest the self-destruction of the SM in the neutrino sector somewhere below (and possibly much below) $10^{15}$~GeV. However, these upper bounds on the cutoff energy $\Lambda$ are too weak to guarantee discoveries with absolute certainty at foreseeable future colliders. The situation could suddenly change if the LHC, or any other current experimental project, found evidence for new 
 interactions with low scale $\Lambda$.   

\section*{Open questions in particle physics}

 What drives the field towards the next generation of experiments is the awareness that we are facing fundamental questions that can be addressed by the scientific method,  and whose answers will significantly enrich human knowledge. 
 In the following we present some of the open questions in particle physics today. 
  Their breadth clearly requires 
  a diversified research programme with different experimental objectives and techniques, with bold projects pushing the energy and precision frontiers, and with substantial theoretical involvement. The open questions 
  are not independent, but deeply interconnected. This reflects the maturity of our global understanding of the particle world and the strong links between all aspects of particle physics, reaching out to neighbouring fields like cosmology and astrophysics. 
  Any new discovery is thus likely to affect our understanding of particle physics in multiple directions.

\medskip\noindent 
{\it 1~~~Electroweak Symmetry Breaking}

\smallskip \noindent
 With a ground-breaking result, the LHC has established the existence of the Higgs boson as the main agent of the spontaneous breaking of electroweak symmetry. In the context of the SM, all the parameters associated with the Higgs (scalar potential and couplings to gauge bosons and fermions) are 
 related to measured quantities. And yet, our understanding of the electroweak symmetry breaking dynamics is far from being satisfactory. The Higgs sector remains a conceptual mystery. 

The problem is related to the nature of the Higgs boson, which is an object different from any other particle we have encountered so far because, according to the SM, it is a fundamental particle with no spin. Contrary to particles that carry spin, for which the massless and massive cases are distinct as they correspond to different numbers of physical degrees of freedom, a massless spinless particle can be turned into a massive one without adding any new physical excitation. This property becomes lethal in the quantum world, since the mass of the Higgs boson becomes wildly sensitive to quantum fluctuations. Its spinless nature leads to another distinguishing feature: the existence of new types of interactions that are different from the gauge interactions that characterise the familiar four fundamental forces of nature. While the structure of gauge forces is restricted by the mathematical properties of symmetry, the new forces introduced by the Higgs boson are less constrained, and this leads to a large number of undetermined parameters. The Higgs alone requires the introduction of 15 new free parameters, as opposed to the strong and electroweak forces which are described by only 3 parameters. When compared with the structural simplicity of the gauge sector, the Higgs sector looks suspiciously provisional. Essentially all problems or unsatisfactory aspects of the SM are ultimately related to the structure of Higgs interactions. Our poor understanding of the Higgs sector at a deeper level, and its novelty in terms of physical properties, make Higgs precision measurements one of the most pressing issues of any future programme in particle physics.

The discovery of the Higgs boson has opened a new research programme, which is a clear priority for the future of particle physics. Precision measurements of Higgs properties enable us to study in depth the most puzzling sector of the SM, opening the door towards a deeper understanding of the mechanism for electroweak symmetry breaking. Future colliders promise an unprecedented scrutiny of the Higgs properties (see Chapter~\ref{chap:ew}). They can explore extensively the nature of the Higgs boson and the question of whether the Higgs is accompanied by other related spinless particles or not. Moreover, if the Higgs were a composite state rather than a fundamental particle as predicted by the SM, its size could be probed at future Higgs factories down to distances of $10^{-20}$~metres, about five orders of magnitude below the size of the proton (see Chapter~\ref{chap:bsm}). 

 The Higgs programme goes hand-in-hand with the programmes of electroweak precision measurements and flavour physics, which can probe the existence of new physics in a way complementary to direct searches (see Chapters~\ref{chap:ew} and \ref{chap:flav}). Moreover, very high-energy collisions offer the opportunity of studying the interplay between short and long-distance effects. 
 This is a new environment in which one can test the infrared properties of non-Abelian gauge theories in a regime which, unlike the case of QCD, is fully perturbative.

\medskip
\noindent {\it 2~~~Higgs Naturalness}

\smallskip
\noindent
A related puzzle in particle physics is the question of Higgs naturalness. 
 The problem arises because of the quantum sensitivity of the Higgs mass to possible new physics scales, while  various experimental measurements point
 towards a large separation of the latter and the Higgs mass ($\Lambda \gg m_h$).  The discovery of the Higgs boson has made the problem more concrete, and the lack of evidence for new physics has widened the gap between $\Lambda$ and $m_h$, making the tension more severe.
 In the language of  Effective Field Theories (EFTs), naturalness arises by viewing EFT parameters as functions of more fundamental ones: any specific structure in the EFT, like the presence of a very small parameter, should be accounted for by symmetries and selection rules rather than by accidents. When the criterion is applied to the Higgs mass, one famously finds that $\Lambda \gg m_h$ is inconsistent with the predicate of naturalness.  Overall, the lack of novel signals suggests that the purely accidental symmetries of the SM appear to be mysteriously maintained up to the high energies tested at present, directly or indirectly. 

An interesting point of view 
 formulates the problem as arising from the clash between two concepts:
 Infrared (IR) simplicity and Naturalness.\footnote{This formulation of the problem is due to Riccardo Rattazzi and the discussion here follows closely his talk at the Open Symposium in Granada.} The SM enjoys IR simplicity in the sense that certain crucial experimental facts,  not mandated by the SM gauge structure, are nevertheless an expected consequence of global symmetries that emerge by pure accident because of the specific matter content found in Nature, e.g.\ approximate baryon and lepton number conservation, lightness of neutrinos, custodial symmetry, and suppression of flavour-changing neutral currents. The problem appears when the SM is embedded in a broader underlying framework. When the latter is probed at energies much below its fundamental scale $\Lambda$,  there is no generic reason for it to abide by the same accidental symmetries as the SM does. This would mean that IR simplicity in the SM can be obtained only at the price of a loss of naturalness in the formulation of the high-energy theory. Indeed, 
  present models aiming to realise naturalness, such as supersymmetry or composite Higgs, invariably sacrifice simplicity. Those 
 extensions of the SM have concrete structural difficulties in reproducing the observed simplicity in flavour, CP violating and electroweak precision observables. In order to become phenomenologically viable, 
 they must rely on artificial constructions mostly associated with ad-hoc symmetries, which in the SM are either not needed or automatic. In this perspective, the tension between simplicity and naturalness is what defines the problem.
 
  Alternatively, the concept of naturalness may appear in a different guise when applied to the electroweak sector. 
  The increasing tension could well herald an exciting change of paradigm. 
 In fact, the LHC results have already prompted theorists to broaden their perspective on the problem and pursue alternative solutions that may lead to unconventional experimental signatures. While much of this research is still ongoing, it is already clear that searches for new physics must take a broad approach. 
 
  In conclusion, Higgs naturalness remains a crucial open question. Whatever the underlying rationale,  more experimental investigation is mandatory to delimit and 
  clarify the issue. Understanding its role in the SM by probing its consequences at even higher energies will give us knowledge about the governing principles of Nature, and critical information for the future course of research in particle physics (see Chapter~\ref{chap:bsm}).

\medskip
\noindent {\it 3~~~Strong Interactions}

\smallskip
\noindent
Strong interactions play a central role in particle physics today, but the relevant questions are not about the validity of the theory or the search for its  possible extensions, because QCD gives a successful and satisfactory explanation of strong interactions. The questions are about how to relate QCD to long-distance phenomena (e.g.\ confinement), how to characterise the collective behaviours that emerge under extreme conditions (e.g.\ high temperature or high density), how to obtain reliable predictions in the non-perturbative regime (e.g.\ hadronic matrix elements) and precise predictions in the perturbative regime (e.g.\ higher-order calculations). Particularly challenging is the question of deriving from the first principles of QCD a description of phenomena at the interface between low and high energies. An example is understanding how fast-moving quarks and gluons cluster into colour-singlet hadrons. As well as being conceptually challenging, these questions are relevant in practice since they lead to a real limitation in the theoretical prediction of observables in hadronic collisions and flavour physics. 

QCD is the necessary tool for describing particle interactions with applications that range from heavy ions to proton collisions, from neutron stars to early-Universe cosmology. One of the most striking successes of the LHC has been to show that, by combining advanced data analysis and detector performance with refined theoretical QCD calculations, hadron colliders can perform measurements with previously unimaginable precision. This is an important legacy for future collider projects, which establishes proton colliders as precision machines. A number of the challenges listed above are of crucial importance for the full exploitation of the physics potential of present and future colliders,  both in searches for new phenomena and in performing precision measurements (see Chapter~\ref{chap:si}).

\medskip
\noindent {\it 4~~~Strong CP}

\smallskip
\noindent
There is only one parameter, among those that specify the SM renormalisable interactions, which has not yet been measured: the strong $\theta$ angle. This parameter characterises the vacuum structure of the theory and contributes to the neutron EDM. Current experiments set an upper bound 
$|\theta |< 10^{-9}$ and no accidental symmetry in the SM can justify such a small value. Finding an experimental confirmation of the reason for the surprising smallness of $\theta$ is still an open question.

Out of the several possible proposed solutions, the axion remains the all-time favourite by theory. Based on a spontaneously-broken global symmetry, the axion solution turns $\theta$ into a dynamical variable, which relaxes to zero in the presence of a potential generated by non-perturbative QCD effects. The search for the axion is a central task in particle physics, offering a window into new physics which may take place at very high energies (see Chapter~\ref{chap:dm}). The axion can also have important effects in stellar evolution, in early-Universe dynamics, and its coherent oscillations could explain the dark matter we observe today.

\medskip
\noindent {\it 5~~~Flavour Physics}

\smallskip
\noindent
The pattern of quark and lepton masses and mixings is one of the most puzzling open questions in particle physics, directly connected with the Higgs since its couplings to fermions are at the heart of the problem.
Generations of dedicated experiments have provided us with precision measurements of the corresponding parameters, revealing a pattern which has a highly non-generic structure and suggests an underlying organising principle. However, the origin of this structure and the nature of the organising principle remain mysterious. 

There is another reason that makes experimental exploration of the flavour sector particularly important. Because of special accidental symmetries and structural aspects (like the GIM mechanism), the SM predicts strong suppressions of certain  flavour-changing transitions. These suppressions have been confirmed experimentally. However, the suppression mechanisms in the SM are fragile and any small deformation of the theory can drastically change the predictions for flavour-changing processes. This property makes the study of rare flavour processes one of the most powerful probes of new physics, in some cases testing scales up to $10^5$ or even $10^6$ GeV.  Experimental hints for deviations from SM predictions in flavour processes are one of our best hopes to direct research towards the right energy scale where new physics may lurk. Dark matter itself may have flavour-violating interactions and an understanding of its structure would require advances in interdisciplinary explorations. Furthermore, flavour experiments are often sensitive to new light particles, possibly related to dark matter.

To the class of flavour processes belong $B$, $K$, $D$ meson and $\tau$ lepton decays, rare muon transitions, anomalous magnetic moments and electric dipole moments (EDM). Future experimental projects will be able to push further the exploration on all these fronts (see Chapter~\ref{chap:flav}), providing us with new fundamental knowledge about the particle world. In particular, testing new sources of CP violation in EDM is a powerful probe of theories beyond the SM, and may turn out to be a decisive tool to test hypothetical mechanisms for generating the observed cosmic baryon asymmetry.
Finally, a true understanding of the flavour puzzle must encompass both the quark and the lepton sector including neutrinos, to which we turn next. 

\medskip
\noindent {\it 6~~~Neutrino Physics}

\smallskip
\noindent
Neutrinos are unique exploratory tools in particle physics. The special nature of their mass makes them sensitive to new physics at very short-distances. Their special propagation properties make them a penetrating probe into the far structure of the Universe and a precious instrument to peek into the dark sectors of the cosmos. Many intriguing open questions in particle physics are linked to the properties of neutrinos.

An active ongoing experimental programme aims at establishing the nature and mass ordering of neutrinos and 
at measuring, with increasing precision, their overall mass scale and mixing parameters (see Chapter~\ref{chap:neut}). An elegant explanation for the lightness of neutrinos in terms of IR Simplicity and separation of scales is encoded in a dimension-five operator with a new-physics scale $\Lambda_\nu$ in the range of $10^{15}$~GeV. This result gives a conclusive proof for the existence of physics beyond the SM. The scale $\Lambda_\nu$ has the dimension of mass divided by coupling squared, so its value could be explained by either a large mass or a small coupling (or a combination of the two). Since neutrino masses require the breaking of both chiral symmetry and lepton number, they are sensitive to fundamental ingredients of the symmetries of the particle world. Furthermore, the neutrino mixing angles show a pattern distinctively different than that observed in the quark sector, exposing another puzzling aspect of the flavour problem. Understanding this structure is a central question in particle physics today. Firmly establishing CP violation in the lepton sector would be a milestone in neutrino physics. A possible consequence of CP violation and neutrino physics is leptogenesis, which is the simplest and most robust known mechanism for generating the cosmic baryon asymmetry in the early Universe. 
Another interesting aspect of the neutrino experimental programme is the search for new light particles that could hide behind the origin of neutrino masses, contributing to the great physics potential offered by the exploration of neutrino physics.

\medskip
\noindent {\it 7~~~Dark Matter}

\smallskip
\noindent
There is overwhelming observational evidence for the existence of Dark Matter (DM), whose contribution to the mass density of the Universe is 5.3 times larger than for ordinary baryonic matter. While its gravitational imprint is well established at galactic and cosmic scales, the microscopic nature of DM is still a mysterious and outstanding open question. If DM is made of particles or compact objects, its constituents could have masses that vary by some 90 orders of magnitude, ranging from Fuzzy DM of $10^{-22}$~eV to primordial black holes of tens of solar masses. During the last decades, a significant experimental effort has focused on DM masses around 100~GeV:  these are motivated by the  observation that particles in this mass range predicted by supersymmetry or other weak-scale theories automatically lead to a particle density in excellent agreement with the observed DM density. This is usually referred to as the `WIMP miracle'. Recently, there has been growing interest in widening the scope of these searches. On one side, the lack of discoveries of weak-scale particles that could act as mediators in primordial annihilation processes has changed the emphasis, since the choice of masses around 100~GeV was completely driven by specific model considerations, especially related to supersymmetry. A more generic WIMP, which annihilates into gauge bosons via ordinary weak interactions, prefers larger masses. The WIMP miracle occurs when the mass is 1.1~TeV for a weak doublet, 2.9~TeV for a triplet, and even larger masses for larger SU(2) representations. 

Renewed interest has flared for DM masses well below the weak scale. These cases are theoretically motivated by axions or axion-like particles, asymmetric DM, light mediators, or non-thermal relics. Experimentally, the search for light DM has stimulated new remarkable ideas using unconventional techniques (see Chapter~\ref{chap:dm}).

Besides the exciting prospect of discovering a new form of matter so common in the Universe, the search for DM is fascinating  because it brings together different fields (particle physics, cosmology, astrophysics) and different experimental techniques (accelerators, underground detection, cosmic rays). Future accelerator-based projects (from high-energy colliders to fixed-target and beam-dump experiments) can contribute to the search for DM in a distinctive and unique way (see Chapter~\ref{chap:bsm}).

\medskip
\noindent {\it 8~~~Dark Sectors and Feebly Interacting Particles}

\smallskip
\noindent
 In the exploration of the unknown, the high-energy frontier remains the best motivated direction to concentrate research efforts in particle physics. Nevertheless, the puzzling questions that confront us require a broad approach, both in terms of experimental strategies and theoretical hypotheses. One alternative research direction that has recently gained momentum is the search for new families of particles which are either very light, but only rarely produced in collisions among ordinary particles, or have very long lifetimes, thus travelling macroscopic distances. These particles are usually referred to as Feebly Interacting Particles (FIP). The hypothetical existence of FIPs is a valid open question in particle physics today.

Superficially FIPs may appear as very unconventional and exotic objects, but we can take the SM for comparison. We know about the existence of very weakly-interacting and penetrating light particles (neutrinos), of particles with relatively long lifetimes due to tiny mass differences (neutrons) or mass hierarchies (weakly-decaying hadrons). Thus, in the presence of hidden sectors with a structure as rich as the SM, it is not unrealistic to expect new particles behaving as FIPs. The existence of FIPs is also motivated by theoretical models for DM, approximate Goldstone bosons, mechanisms for neutrino masses, Higgs naturalness and various hidden sectors. The search for FIPs (see Chapters~\ref{chap:bsm} and \ref{chap:dm}) is an interesting complement to high-energy explorations, with the additional feature of bringing together different experimental strategies ranging from particle physics (collider, beam dump, fixed target, rare decays) to neighbouring fields (DM detection, astrophysics, multi-messenger astronomy, and even atomic or condensed-matter physics).

\medskip
\noindent {\it 9~~~The Cosmos}

\smallskip
\noindent
One of the greatest successes of particle physics was to show that knowledge derived from very short distances is crucial to understand our Universe at large scales. This path of knowledge started from nuclear physics explaining why stars shine and how chemical elements are created, and led to the present understanding of galaxy distribution in terms of quantum fluctuations of a primordial field active during inflation. The connection between the Universe at the smallest and largest scales is a monumental conceptual achievement, which has not only produced some of the most mind-boggling results in physics, but also revealed new fundamental open questions. 

One question that future colliders will be able to address is the nature of the electroweak phase transition in the early Universe. While the phase transition is a high-temperature phenomenon that cannot be recreated experimentally, precision measurements of Higgs properties---in particular of the triple-Higgs self-coupling---will give us decisive elements to reconstruct the dynamics that occurred when the Universe changed its vacuum state. According to the SM, the Higgs mechanism took place as a smooth crossover when the Universe cooled down to temperatures below 160~GeV, but the transition could be very different in the presence of new physics. A particularly interesting possibility is that the Universe underwent a first-order phase transition, which would open the door to the exciting prospect of explaining the cosmic baryon asymmetry with weak-scale physics or of observing gravitational waves produced by the abrupt transition at that epoch. Independently of these speculations, testing the nature of the electroweak phase transition is an important task for future colliders that will considerably expand our knowledge about the early history of the Universe.

Inflation and dark energy are two other crucial ingredients of cosmology that require input from particle physics. They are recurrent themes of theoretical physics studies and targets for projects in observational cosmology. Any progress addressing these two fundamental problems would have revolutionary impact on our understanding of the particle world and on our future research priorities. 

The Universe also provides a unique laboratory for particle physics, offering opportunities to test new ideas in environments that cannot be reproduced on Earth. This is the case of ultra-high-energy cosmic rays, containing particles produced at energies beyond those attainable at human-made accelerators, and whose origin and acceleration mechanisms are still largely not understood. It is the case of black-hole observations in gravitational-wave detectors, of particle production in the core of dense stellar bodies and, more in general, of multi-messenger particle astrophysics.

\medskip
\noindent {\it 10~~~Gravity}

\smallskip
\noindent
Gravity is the most familiar of all forces in nature and yet it hides some of the most perplexing open questions in particle physics today. At the classical level, it is elegantly understood as a gauge theory in which the gauge symmetry acts on spacetime coordinates, according to General Relativity. At the quantum level, the theory ``predicts its own destruction" somewhere below the Planck mass, at the extraordinary energies of $10^{19}$~GeV. In spite of the great developments in string theory, the ultimate theory bringing together quantum gravity and the SM has not been identified yet.

Early cosmology and black-hole physics provide the two known training grounds where ideas about gravity in the quantum regime can be tested. The thermodynamical properties of black holes and the information paradox have stimulated new ideas that are revolutionising the approach towards the quantum properties of gravity. Although this research is revealing surprising connections that range from quantum information to condensed-matter physics, this is still a highly speculative and theoretical activity. However, the observation of black hole collisions through gravitational waves has opened a new field that holds promise for experimental tests of modifications of gravity (see Chapter~\ref{chap:cosm}).

Another open question related to gravity is the value of the cosmological constant. From the particle-physics point of view, the amount of dark energy measured by astronomers is ridiculously small. The vacuum energy typically predicted by particle theories contributes to the cosmological constant by some 120 orders of magnitude more than what is observed. The problem is particularly interesting from a particle-physics perspective because it is conceptually identical to the naturalness problem encountered with the Higgs mass, suggesting that there could be hidden connections between the two puzzles.

\subsection*{Advancing particle physics}

Particle physics is a central node of the interconnected network of scientific disciplines that defines human knowledge. It has been able to distill the essence out of natural phenomena and translate it into universal laws expressed in terms of a few mathematical equations. Those laws follow from a handful of fundamental principles and show a remarkable structural unity. Their power lies in allowing us to make certain and precise predictions that have been empirically verified in the domain of simplicity -- the Universe at very small and very large distance scales. 

  The success of particle physics relies on the ability to build experiments where controlled phenomena can be studied, leading to certain and precise measurements. The combination of precise predictions and measurements is the hallmark of particle physics among the sciences, which allows us to ask fundamental questions about the Universe and obtain definite answers. 

 The present success of the Standard Model is the very reason why we can embark on future missions. This success 
  gives us a reliable and solid starting point to formulate consistently our questions and to advance cogently and systematically into the exploration of the unknown.   In particular, the pressing  open questions have been formulated precisely and rigorously.  Their crucial nature is tantalising: they may herald a change of paradigm awaiting discovery. 
   Addressing those questions 
  requires a diversified scientific strategy. Within this broad research programme, high-energy colliders are an indispensable and irreplaceable tool to pursue our exploration of the fundamental laws of nature.

\documentclass[../report.tex]{subfiles}
\providecommand{\main}{..}
\IfEq{\jobname}{\currfilebase}{\AtEndDocument{\biblio}}{}
\newcommand{\ew}{electroweak\xspace}
\newcommand{\met}{E_\textrm{T}^\textrm{miss}}
\newcommand{\epem}{e^{+}e^{-}}
\newcommand{\commentsinout}[1]{#1}
\newcommand{\KE}[1]{\commentsinout{\textbf{\color{brown} [KE: #1]}}} 
\newcommand{\FM}[1]{\commentsinout{\textbf{\color{orange} [FM: #1]}}} 
\newcommand{\AN}[1]{\commentsinout{\textbf{\color{teal} [AN: #1]}}} 
\newcommand{\BH}[1]{\commentsinout{\textbf{\color{blue} [BH: #1]}}} 

\newcommand{\ifb}{fb$^{-1}$}
\newcommand{\iab}{ab$^{-1}$}
\newcommand{\sintheta}{\sin^2\theta_W}
\newcommand{\etmiss}{E_\textrm{T}^\textrm{miss}}
\newcommand{\BR}{BR}
\newcommand{\BRinv}{BR$_{\textrm{inv}}$}
\newcommand{\BRunt}{BR$_{\textrm{unt}}$}

\def\permille{\ensuremath{{}^\text{o}\mkern-5mu/\mkern-3mu_\text{oo}}}

\begin{document}
\chapter{Electroweak Physics}
\label{chap:ew}
In this chapter the status of the \ew physics programme and its future prospects are discussed. Particular emphasis is given to the exploration of the Higgs boson at the future colliders discussed in Chapter~\ref{chap:acc}.

\section{Introduction}
\label{sec:ewkintro}
The \ew sector of the Standard Model (SM) of particle physics is extraordinarily rich, and its theoretical elucidation and experimental exploration over the past 80 years is among the most outstanding scientific achievements of humankind. Components of that achievement are the invention of quantum electrodynamics (QED)~\cite{Dirac:1927dy}, the discovery of the weak interaction\cite{Fermi:1934hr,Sudarshan:1958vf,Feynman:1958ty}, the unification of the weak and electromagnetic interactions in the late 1960s~\cite{Weinberg:1967tq,Salam:1968rm}, the discoveries of the $W$ and $Z$ bosons in the 1980s~\cite{Rubbia:1985pv}, the precision tests of the electroweak theory on the $Z$ pole~\cite{ALEPH:2005ab}, and last but not least the discovery of the Higgs boson in 2012~\cite{Aad:2012tfa,Chatrchyan:2012xdj}. In the perturbative regime, QED has been tested with a precision of one part in $10^{12}$, an amazing achievement of both experiment and theory. For the \ew theory, many tests have been made at the per mille level at high-energy colliders and low-energy experiments, all confirming the SM predictions.

Despite this huge success, the \ew sector of the SM is puzzling. In particular, if new physics occurs at a higher mass scale, there is generally no explanation for why the Higgs boson mass should be at $\sim 125$~GeV, rather than at the much higher scale. 
Indeed, quantum corrections to the Higgs boson mass, $\Delta m_H$, due to e.g.\ the top quark, are much larger than the Higgs boson mass itself. The natural expectation is that $(\Delta m_H)^2 \sim \Lambda^2$ where $\Lambda$ is the energy scale of new physics. 
This issue is called the naturalness problem (see also Chapter~\ref{chap:th}). The naturalness problem can be quantified by the ratio of the experimentally measured Higgs mass to the quantum corrections to the Higgs mass, i.e.
\begin{equation}
    \epsilon \equiv \frac{m_H^2}{(\Delta m_H)^2}\,,
\end{equation}
where $\Delta m_H$ is the sum of all quantum corrections to the Higgs boson mass, and can be calculated in any model. In the SM, where there is no new physics below the Planck scale, the value is $\Delta m_H\sim 10^{19}$~GeV, corresponding to an extreme fine-tuning $\epsilon\sim 10^{-34}$. Values of the parameter $\epsilon \sim 1$ correspond to no fine-tuning. Note that in the literature, it is common to express the fine-tuning in terms of $\Delta$ where $\Delta=1/\epsilon$.

Depending on how the new physics couples to SM particles, new physics models can be classified as \textit{soft}, \textit{super-soft} and \textit{hyper-soft}~\cite{Contino:2017moj}. An example for a soft model is the Minimal Supersymmetric Model (MSSM) with high-scale mediation of the soft terms, for super-soft examples are Composite Higgs (CH) models and SUSY models with low-scale mediation
(see Chapter~\ref{chap:bsm}).
For hyper-soft models Neutral Naturalness is a prime example (see Chapter~\ref{chap:bsm}). Generally, the fine tuning can be related to the mass of a putative top quark partner, $m_T$, as shown in Table~\ref{tab:finetuning}. Measurements of the Higgs boson couplings can also be related to the fine tuning parameter $\epsilon$ as shown in Table~\ref{tab:finetuning}.
Furthermore, $\epsilon$ can be related to the oblique parameters~\cite{Peskin:1990zt,Peskin:1991sw,Altarelli:1991fk,Golden:1990ig, Barbieri:2004qk}, $O$, which are introduced to quantify possible modifications of the \ew precision observables due to new physics. Here, parameters $S$ and $T$ are the focus; $T$ measures the difference between the new physics contributions of neutral and charged current processes at low energies and $S$ describes new physics contributions to neutral current processes at different energy scales. $S$ and $T$ quantify universal loop corrections to the photon, $W$ and $Z$ propagators, i.e. do not depend on the lepton and quark flavours.

\begin{table}[htbp]
    \centering
     \caption{Constraints on the fine tuning parameter, $\epsilon$, as determined via direct searches and via precision measurements of Higgs boson couplings and oblique parameters. For direct searches $m_T$ is the mass of the top quark partner, and $y_t$ and $\lambda_h$ are coupling parameters which are both assumed to be $\sim 1$ in natural theories. 
     For the direct searches and the oblique parameters, the mass value used for interpretation as $\epsilon$ is $m_T=1$~TeV, motivated by current limits on top partners\cite{atlasstop,cmsstop,atlasvtop,cmsvtop}. CH stands for ``composite Higgs'' and SUSY for ``supersymmetry''.
    \label{tab:finetuning}}
    \begin{tabular}{l|l|r}
    \hline
    Method & Dependence & Current Constraint \\\hline\hline
         Direct searches: soft models  & $\Delta m_H^2 \sim m_T^2$ & $\epsilon\lesssim 1\%$ \\
        Direct searches: super-soft models & $\Delta m_H^2 \sim 3y_t^2/(4\pi^2)
    m_T^2$  & $\epsilon\lesssim 10\%$\\
        Direct searches: hyper-soft models & $\Delta m_H^2 \sim 3\lambda_h/(16\pi^2) m_T^2$  & $\epsilon \lesssim 100\%$\\\hline
         Higgs couplings & $m_H^2/\Delta m_H^2\sim \delta g_h/g_h$ & $\epsilon \lesssim 10\%$\\
         Oblique parameters (CH models)& $
         m_H^2/\Delta m_H^2 \sim 
         \delta O\times 3$ 
          & $\epsilon \lesssim 30\%$ \\
          Oblique parameters (SUSY models)& $
         m_H^2/\Delta m_H^2 \sim 
         \delta O\times 10^{3}$ 
          & n.a. \\
          \hline
    \end{tabular}
\end{table}
Based on these arguments, it is clear that measurements of Higgs boson couplings at the \% level or better test new physics models at or beyond the current constraints from direct searches. The same is true when the oblique parameters are measured to better than $\sim 3\times 10^{-3}$. Future direct searches at the FCC-hh or a muon collider are expected to improve the sensitivity to the top quark partner to 10~TeV, and thus be sensitive to values of $10^{-3}-10^{-5}$ (see Chapter~\ref{chap:bsm}). 

\subsection{Higgs studies at hadron colliders}
\begin{figure}[t]
\centering
\includegraphics[width=.8\linewidth]{\main/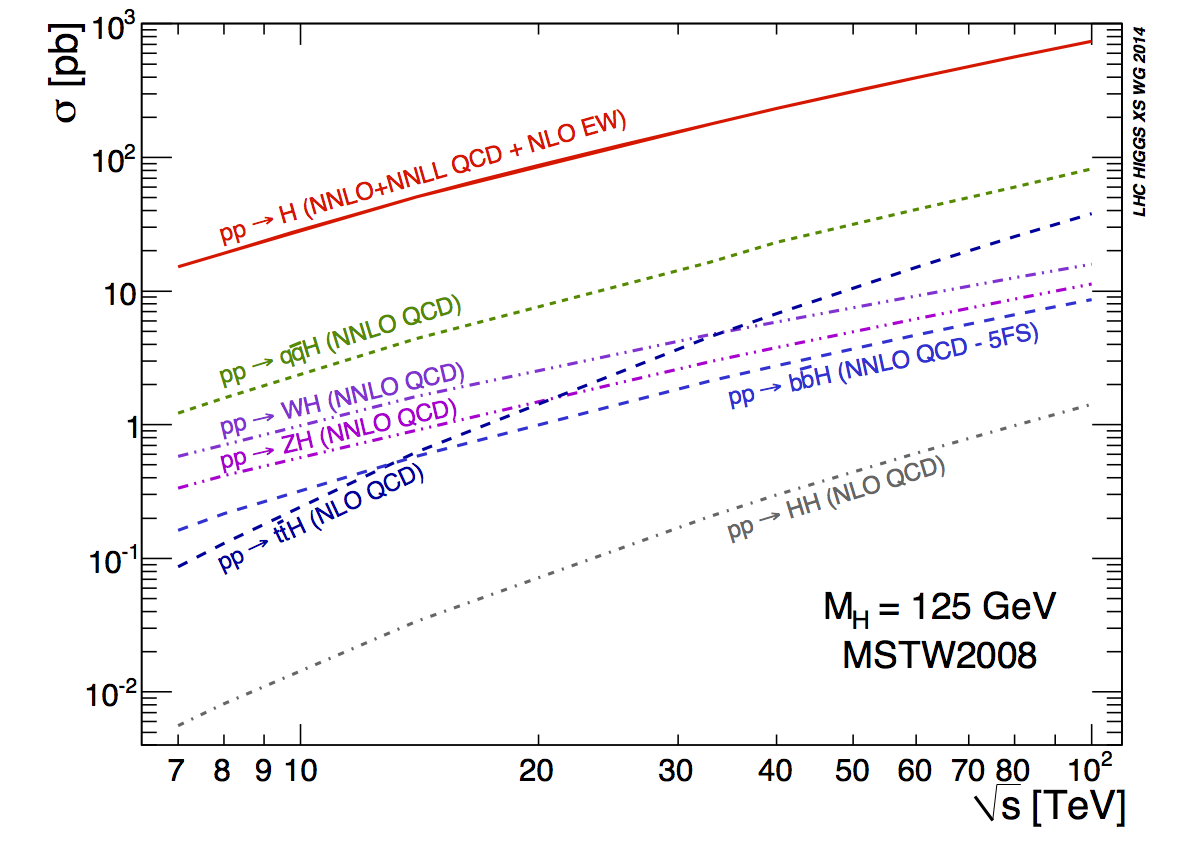}
\caption{\label{fig:Higgs_XS_7-100TeV_HH}
Higgs production cross sections in hadronic collisions.}
\end{figure}

Figure~\ref{fig:Higgs_XS_7-100TeV_HH} shows the cross
sections for the production of Higgs bosons from LHC energies to the energy of \FCChh. 
So far at the LHC, about 8 million Higgs bosons have been produced and the data sample will further increase by a factor of 20 at the conclusion of \HLLHC.
The coupling parameters themselves cannot be extracted at hadron colliders without further assumptions which depend on the new physics model. 
In particular, to resolve a multiplicative ambiguity in inferring couplings from measured cross section times decay branching ratio ($\sigma(H) \times {\rm BR}$) values, it is usually assumed either that there are no new light states that the Higgs boson can decay to (i.e.\ the Higgs width is fully determined by the couplings to the SM particles), or that the coupling to the gauge bosons can not be larger than the SM value.  This latter assumption is valid for the vast majority of BSM models, and is made for results and projections presented here. 

\begin{figure}[htbp]
    \centering
    \includegraphics[width=.44\linewidth]{\main/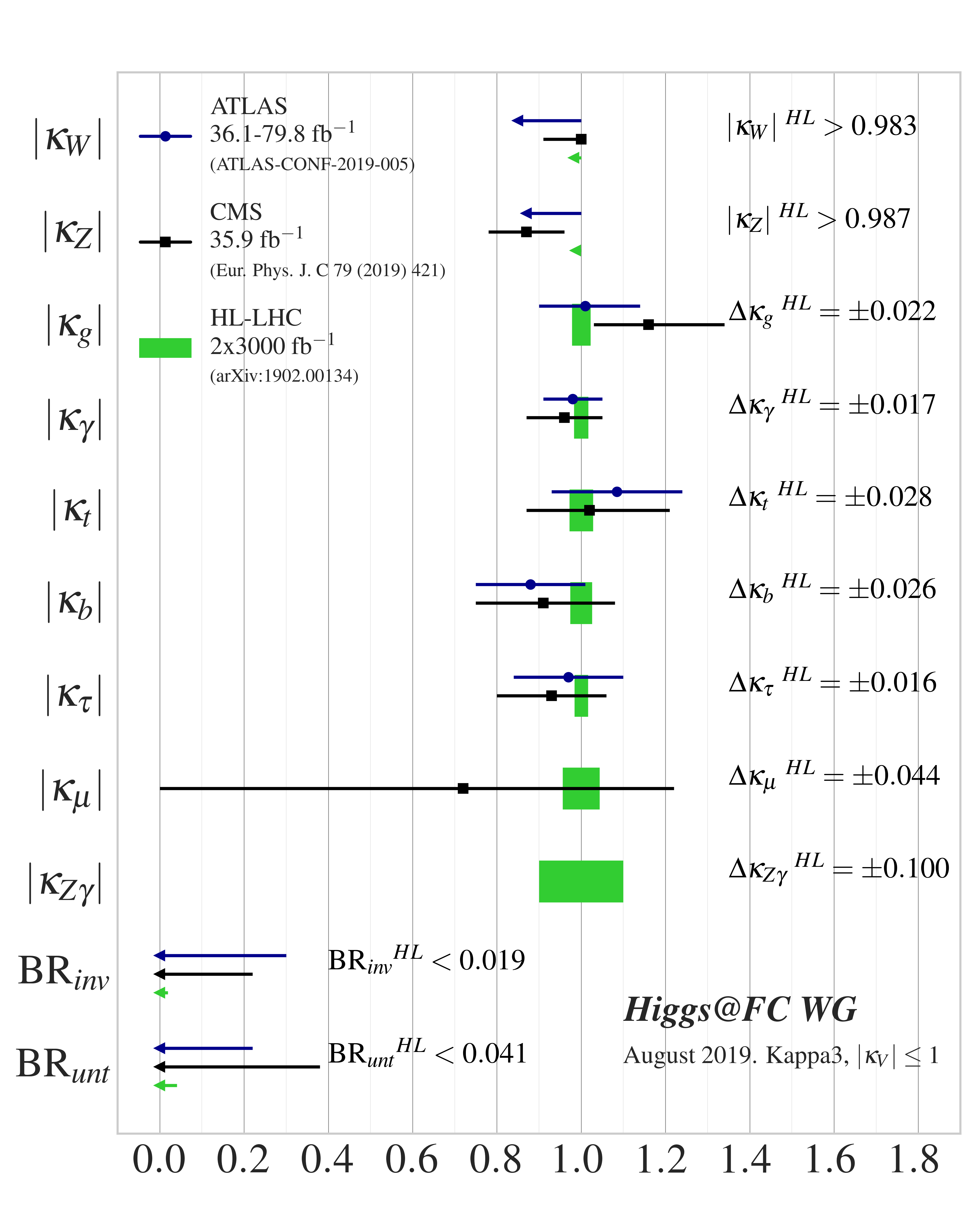}
    \includegraphics[width=.45\linewidth]{\main/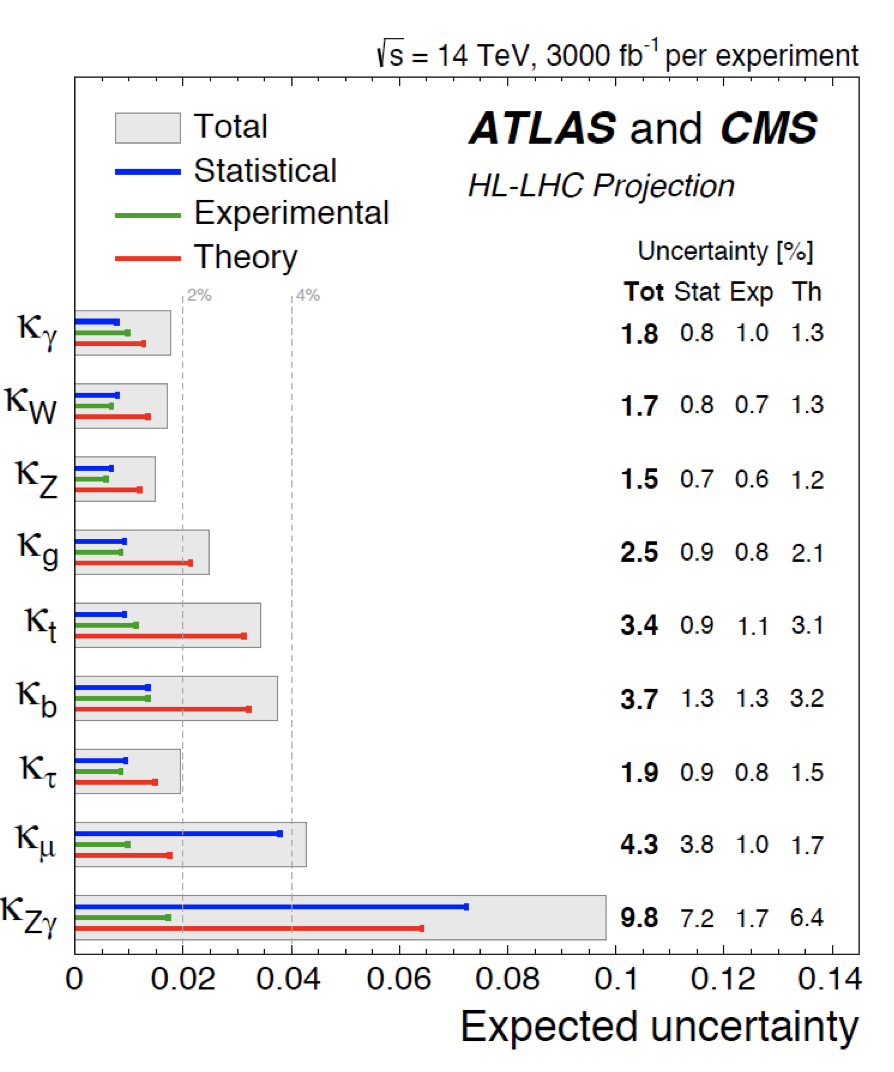}
    \caption{Left: Relative precision on Higgs coupling modifiers, $\kappa$, determined by ATLAS and CMS with the LHC data at present, and as expected for \HLLHC with the constraint $\kappa_V\leq 1$. Also shown are the constraints on invisible and undetected decay branching ratios, \BRinv and \BRunt. Right:  Expected uncertainty on Higgs coupling parameters at \HLLHC, showing separately the statistical, experimental and theoretical uncertainties. Here, it was assumed that the branching ratios (BR's) to untagged and invisible decays are zero.
    \label{fig:higgsnow}}
\end{figure}

Figure~\ref{fig:higgsnow} presents a selection of current Higgs coupling measurements~\cite{atlashcomb2,cmshcomb2} and the precision projected for HL-LHC~\cite{Cepeda:2019klc}, with the constraint $|\kappa_V|\leq 1$. Here, $\kappa_i$ is a parameter which specifies by how much the coupling of the Higgs boson to a given particle $i$ deviates from the SM expectation, see Sec.~\ref{sec:higgsfuture} for more details. The current uncertainties are typically 10-20\% for the bosons and 3$^{\rm rd}$ generation fermions. For the muon coupling modifier the uncertainty is about 100\%, and the upper limits on new invisible or undetected particles are 20-30\%. With the \HLLHC, the precision will be improved by about a factor of 5-10 on all observables. Figure~\ref{fig:higgsnow} also shows the composition of the expected \HLLHC uncertainties in a $\kappa$-fit where the width is assumed to be fully determined by the couplings to the SM particles. The only channels which are expected to be limited by data statistics are the rare decays to muons and $Z\gamma$. In all other cases, the experimental systematic uncertainties are similar to the statistical uncertainties, but the dominant source of uncertainty arises from theory. Here, it is already assumed that the theory uncertainties can be reduced by a factor of two compared to the current uncertainties which is challenging to achieve. For both hadron and lepton colliders, a further reduction of theory uncertainties is pivotal to fully capitalise on the experimental data. Section~\ref{sec:ewktheory} discusses the status and prospects for theory uncertainties.

\subsection{Higgs studies at $e^+ e^-$ colliders}
\begin{figure}[!ht]
\centering
\includegraphics[width=.6\linewidth]{\main/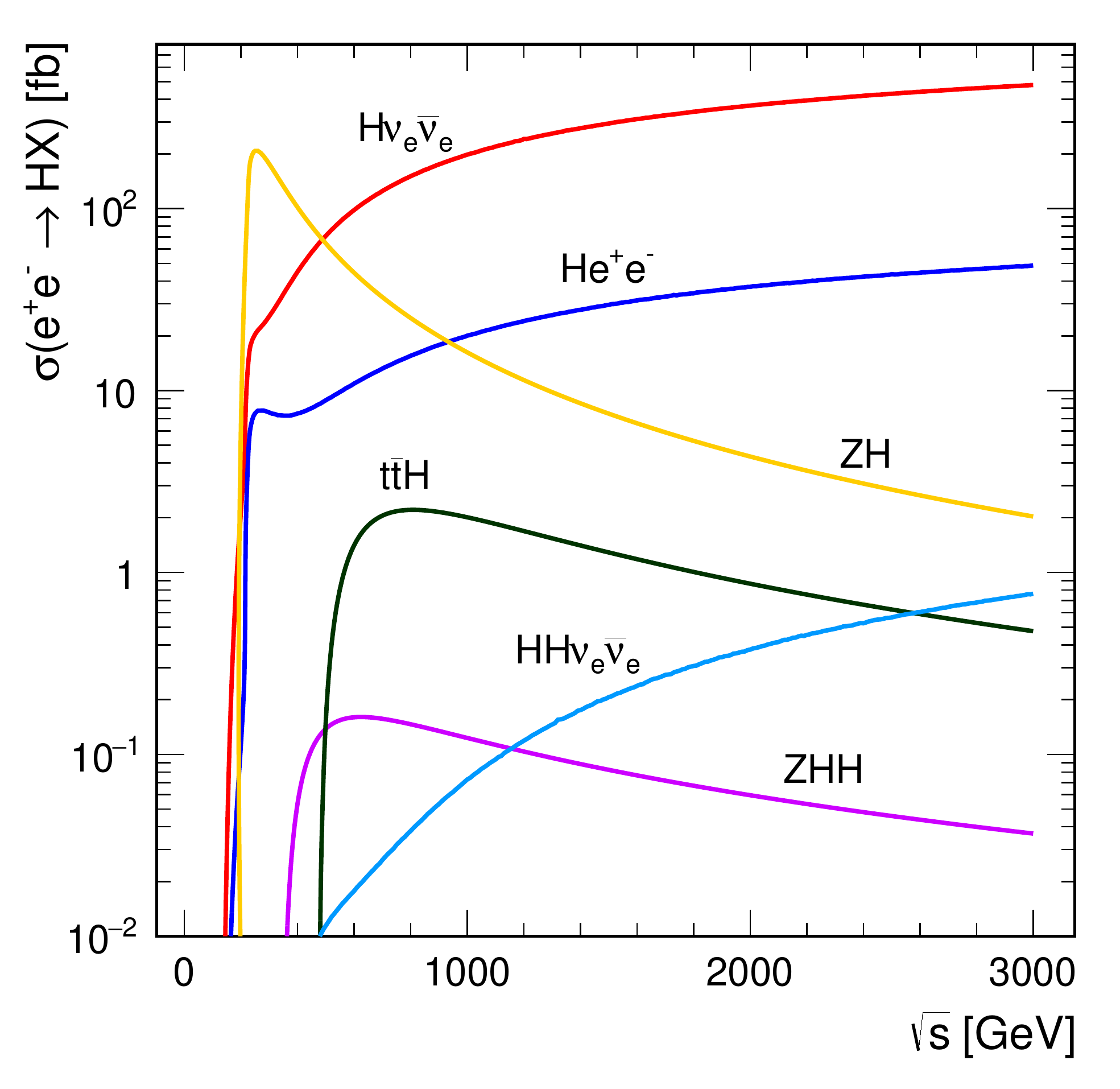}
\caption{\label{fig:xsec_vs_cme}
Higgs production cross sections in $e^+e^-$ collisions\cite{Abramowicz:2016zbo}. The cross section of different production processes for single and double Higgs production are shown as function of $\sqrt{s}$.}
\end{figure}

The Higgs production processes in unpolarised $e^+e^-$ collisions are shown in Fig.~\ref{fig:xsec_vs_cme}.
Importantly the total $ZH$ cross section can be measured  independently of the Higgs boson decay, using a missing mass technique. From this measurement the coupling $g_{ZZH}$ can be derived. Consequently, at an $e^+ e^-$ collider, the Higgs total width ($\Gamma_H$) can be determined
from $\Gamma(H\to ZZ^*)/BR(H \to ZZ^*)$	thus removing the ambiguity on the Higgs width that afflicts all measurements at hadronic machines.
Longitudinal polarisation is expected at the linear machines $e^+e^-$ machines, e.g.
$|P(e^{-})|=0.8,|P(e^{+})|=0.3$ is projected to be achievable for the \ILC.
As shown in Table~\ref{tab:higgs:polarisation}, with the appropriate polarisation this can enhance the Higgs boson production cross section.
In addition, because the importance of different subprocesses can be tuned by changing the polarisation, it plays an important role in effective operator fits.
Thus, the presence of polarisation can sharpen these analyses, and help to compensate for the lower luminosities at linear machines.
\begin{table}[tb]
\caption{The dependence of the event rates for the $s$-channel
$\epem\to Z H$ process and the pure $t$-channel
$\epem\to H \nu_e \nu_e$ and $\epem\to H\epem$ processes for several
example beam polarisations~\cite{Abramowicz:2016zbo}.
\label{tab:higgs:polarisation}}
\begin{center}
  \begin{tabular}{cccc}
    Polarisation                              & Scaling factor                 &&       \\
\hline
    $P(e^{-}):P(e^{+})$                  &
    $\epem \!\to ZH$ &
    $\epem \to H\nu_e \bar{\nu}_e$ &
    $\epem \to H\epem$ \\
    unpolarised                                  & 1.00                & 1.00        &   1.00        \\
    $-80\,\%\,:\phantom{+3}\,0\,\%$   & 1.12                & 1.80        &    1.12             \\
    $-80\,\%\,:\,+30\,\%$                    & 1.40                & 2.34        &    1.17           \\
    $-80\,\%\,:\,-30\,\%$                    & 0.83                & 1.26        &    1.07           \\
    $+80\,\%\,:\phantom{+3}\,0\,\%$  & 0.88                & 0.20        &    0.88             \\
    $+80\,\%\,:\,+30\,\%$                    & 0.69                & 0.26        &    0.92           \\

    $+80\,\%\,:\,-30\,\%$                    & 1.08                & 0.14        &    0.84           \\
\hline
\end{tabular}
\end{center}
\end{table}


\subsection{Electroweak Precision Observables}
Loop corrections to \ew precision observables (EWPO) provide a powerful test of the consistency of the SM. The relation between e.g.\ the  Fermi constant ($G_F$), Weinberg angle ($\sin^2\theta_W$), and the masses of the $Z$, $W$ and $H$ bosons ($m_Z$, $m_W$, $m_H$) and the top quark ($m_\textrm{top}$) is precisely predicted in the SM. Inconsistencies between these would indicate contributions from new physics. In the following we concentrate on oblique observables, discussed in Section~\ref{sec:ewkintro}.

\begin{figure}[h]
    \centering
    \includegraphics[width=.54\linewidth]{\main/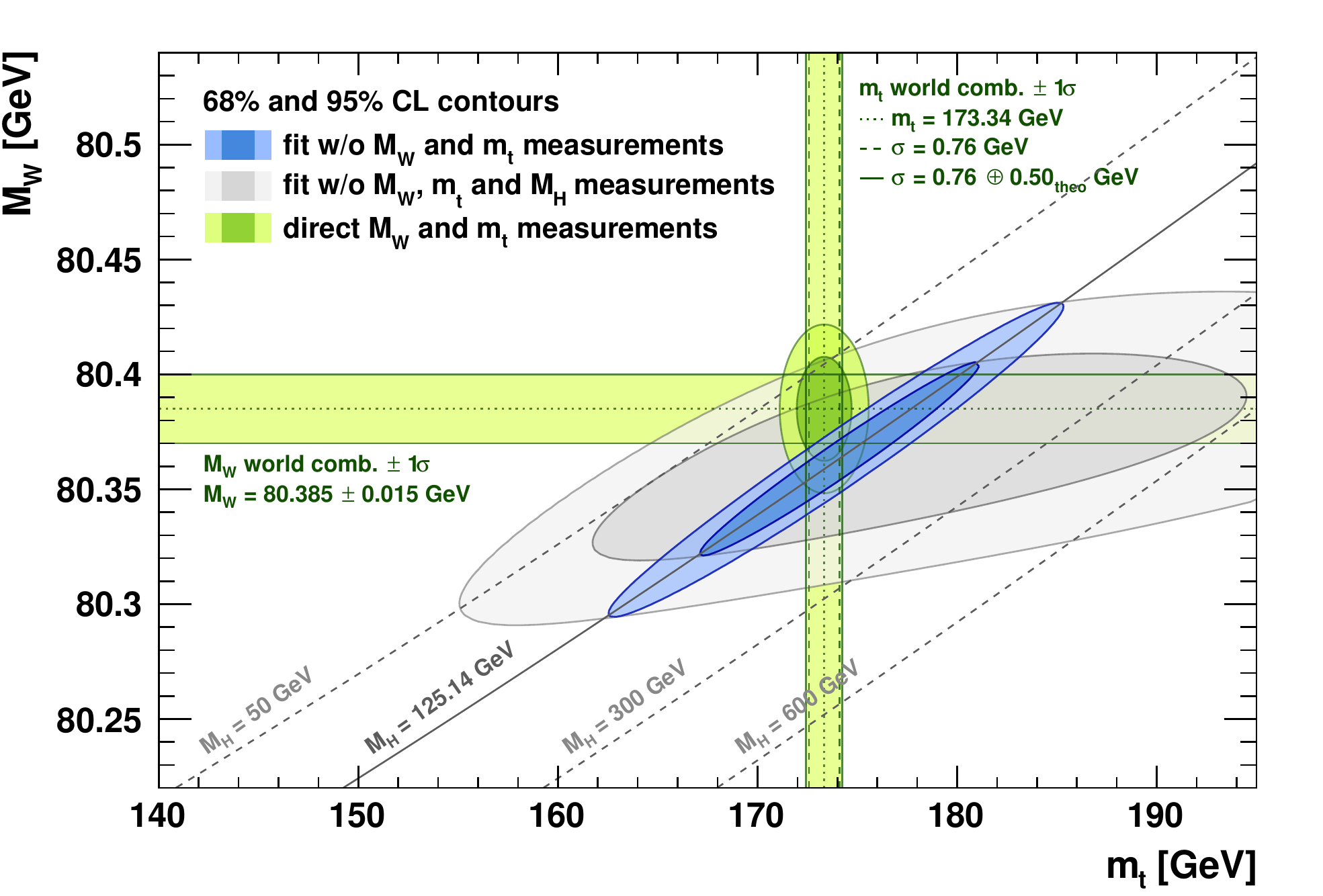}
    \includegraphics[width=.44\linewidth]{\main/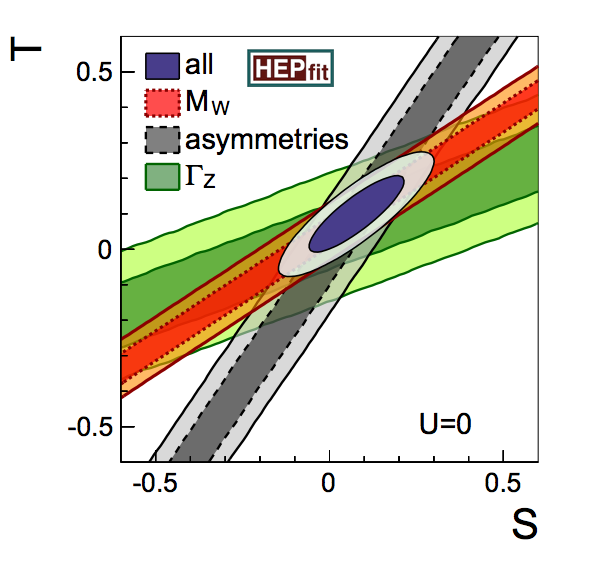}
\caption{
Left: Constraints on the $W$-boson and top-quark mass from direct measurements and indirect constraints~\cite{Baak:2014ora}. 
Right: Constraints on the oblique parameters $S$ and $T$ (setting all other oblique parameters to zero), 
together with the individual constraints from $M_W$, the asymmetry parameters $\sin^2\theta^{\rm lept}_{\rm eff}$, $P_{\rm pol}^{\tau}$
and the forward-backward asymmetries $A_{\rm FB}^{f}$ with $f = \ell, c, b$, and $\Gamma_Z$. The dark (light) region corresponds to 68\% (95\%) probability~\cite{deBlas:2016ojx}.
    \label{fig:ewknow}}
\end{figure}

These contributions are currently constrained primarily by the $Z$ pole measurements made at the LEP experiments and SLD~\cite{ALEPH:2010aa}, measurements of $WW$ production at LEP-2~\cite{LEP-2}, measurements of $W$-boson and top quark masses at the Tevatron~\cite{tevmw,tevmtop} and LHC~\cite{Aaboud:2017svj,Tokar:2019wny} experiments, and $m_H$ measurements at the LHC~\cite{Aaboud:2018wps, Sirunyan:2017exp}. The current constraints on the EWPO are shown in Fig.~\ref{fig:ewknow}. All measurements agree within the current precision.

Based on the \ew precision measurements, the 95\% CL upper limits on the oblique parameters~\cite{Peskin:1990zt} are $S<0.18$ and $T<0.26$~\cite{Tanabashi:2018oca}. Fig.~\ref{fig:ewknow} shows $T$ vs $S$ and illustrates how the various precision measurements on $\Gamma_Z$, $M_W$ and asymmetries contribute. 
 
Measurements of diboson production are also sensitive to the \ew symmetry breaking mechanism as they depend on the trilinear gauge couplings of the bosons to each other. In addition, some processes, such as  $pp\to W^\pm W^\pm +2$~jets,  are  also sensitive to the quartic coupling of the $W$ bosons and to the coupling of the Higgs boson to $W$ bosons. 

In addition, at future $e^+e^-$ colliders, a campaign of \ew  measurements at the $Z$-pole and at the $WW$ threshold is also foreseen.

\subsection{The Higgs potential}
A very important aspect of the \ew physics programme is the measurement of the trilinear and quartic couplings of the Higgs boson to itself. These couplings are directly related to the shape of the Higgs potential, 
\begin{equation}\label{eq:SMpotential}
V(h)= \frac{1}{2} m_H^2 h^2 + \lambda_3 v h^3 + \frac{1}{4}\lambda_4 h^4,\qquad \textrm{with}\quad \lambda_3^{\rm SM}= \lambda_4^{\rm SM}= \frac{m_H^2}{2 v^2},
\end{equation}
where $v=1/\sqrt{\sqrt{2}G_F}\approx 246$\,GeV is the vacuum expectation value of the Higgs field, and $m_H\approx 125$~GeV.

The shape of the Higgs potential can have important consequences for our Universe as it determines how the early universe went through a phase transition. In the early universe, the \ew phase transition is determined by the scalar potential at finite temperature, whereas collider measurements probe the potential at
zero temperature.
At a temperature of about 100~GeV it went from a symmetric state into a state with a broken \ew symmetry. For the SM Higgs potential, this phase transition is a crossover, but alterations to the potential could result in this phase transition having been first order. 
If the phase transition is strongly first order ($\lambda_3$ is modified by $O(1)$), and there is a new mechanism for CP violation, two of the Sakharov conditions necessary (but not sufficient) for an explanation of matter and anti-matter asymmetry of the Universe are fulfilled.
Gravitational waves stemming from that phase transition could be discovered by the Laser Interferometer Space Antenna (LISA)~\cite{Caprini:2015zlo}.  

The parameters $\lambda_3$ and $\lambda_4$ can be measured in processes where two or three Higgs bosons are produced, and via loop-contributions in single Higgs production processes. At present, the LHC data are not sensitive, but with \HLLHC it is expected that the trilinear Higgs self-coupling can be determined with a precision of about 50\%~\cite{Cepeda:2019klc}. 

\subsection{Higgs boson decays to new particles}
The Higgs boson is also potentially a window of discovery for new particles, in particular of those having a mass $m<m_H/2$, into which the Higgs boson can decay. Thus even particles that do not interact with any SM particles, except the Higgs boson, can be discovered using this new channel. These might be invisible, such as e.g.\ dark matter candidates, and result in the experimental signature of $\met$, or may be in principle detectable. These can be searched for via global coupling analyses as well as through targeted analyses, see Chapter~\ref{chap:bsm}. The present constraints and the anticipated \HLLHC constraints for global analyses are presented in Fig.~\ref{fig:higgsnow}. With \HLLHC (using the constraint $|\kappa_V| \leq 1$) such invisible and untagged decays will be probed at the few-\% level. However, without the $|\kappa_V|\leq 1$ constraint, the BR to untagged decays is essentially unconstrained at the LHC through the inclusive coupling analysis, and only direct searches for anomalous decays provide sensitivity, as discussed in Chapter~\ref{chap:bsm}. 

\section{Future prospects}
\subsection{Electroweak precision measurements}
\begin{figure}[htbp]
    \centering
    \includegraphics[width=.75\linewidth]{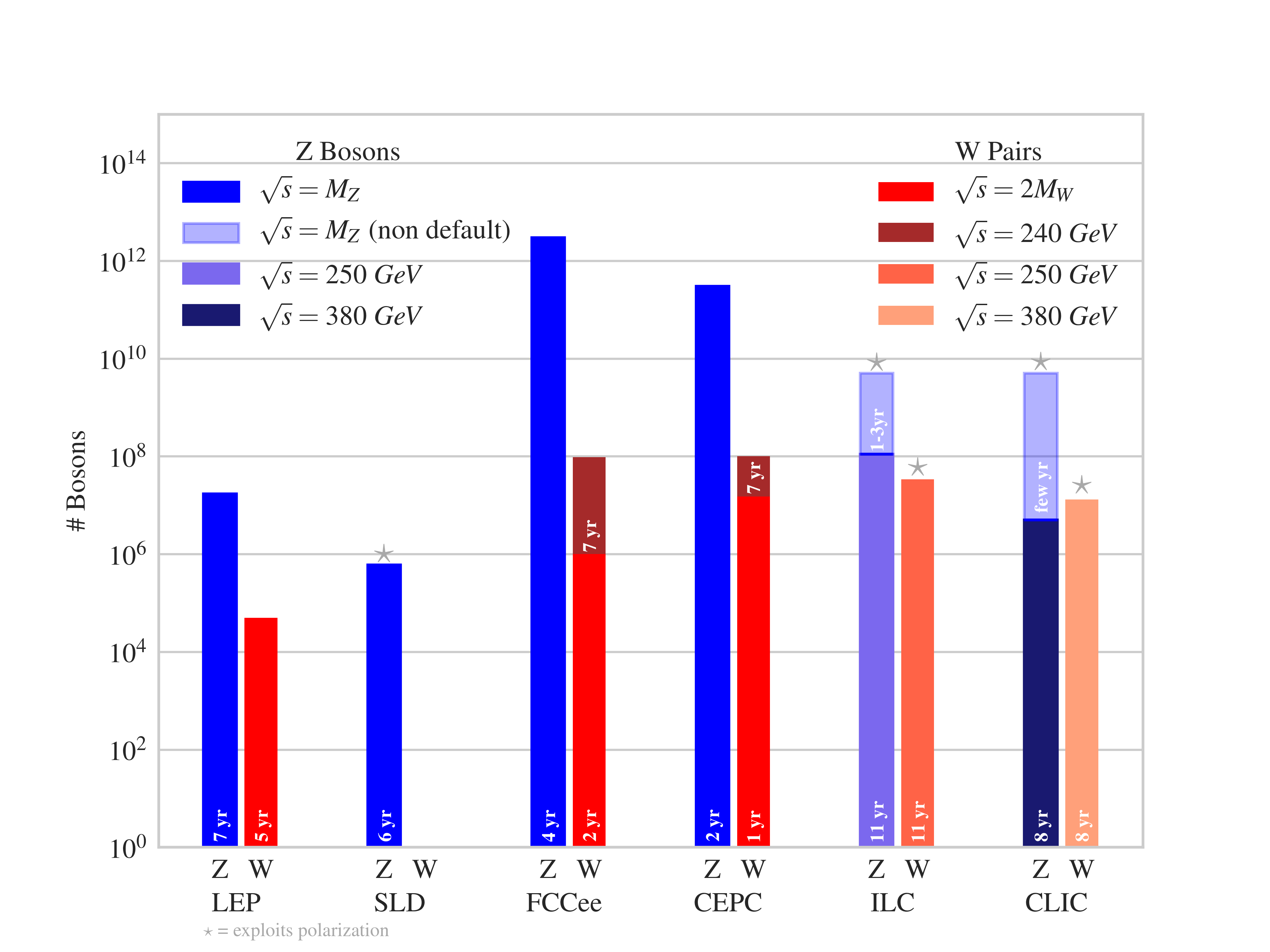}
    \caption{Number of $Z$ bosons and $W^+W^-$ boson pairs at past and future $e^+e^-$ colliders. The numbers are summed over experiments (four for LEP, two for \FCCee and \CEPC and one for the other colliders). For LEP the number of $W$ pairs shown includes all energies $\sqrt{s}\gtrsim 2M_W$.
    \label{fig:ewkprogramme}}
\end{figure}

The precision of many observables related to the \ew bosons can be improved at future experiments. In particular, the proposed $e^+e^-$ colliders will be able to advance the precision measurements of the $W$- and $Z$-boson properties significantly. Figure~\ref{fig:ewkprogramme} shows the number of $Z$ and $W$ bosons that will be recorded at the various lepton colliders. For the circular colliders there are dedicated runs planned on the $Z$ pole and at the $WW$ threshold to make precise measurements of $Z$ boson properties and the $W$-boson mass, respectively. Since for circular colliders the luminosity increases with decreasing $\sqrt{s}$ about $5\times 10^{12}$ $Z$ bosons will be recorded for \FCCee within four years. For the linear colliders, \ILC and \CLIC, within a few years a sample of a few $10^9$ $Z$ bosons could be recorded~\cite{Fujii:2019zll,gigazclic}. In addition, a significant improvement for some of the $Z$ boson properties can also be achieved using $Z$ bosons during the default running at higher energies; those numbers are also shown in Fig.~\ref{fig:ewkprogramme} for \ILC and \CLIC.  

\begin{figure}[htbp]
    \centering
    \includegraphics[width=.9\linewidth]{\main/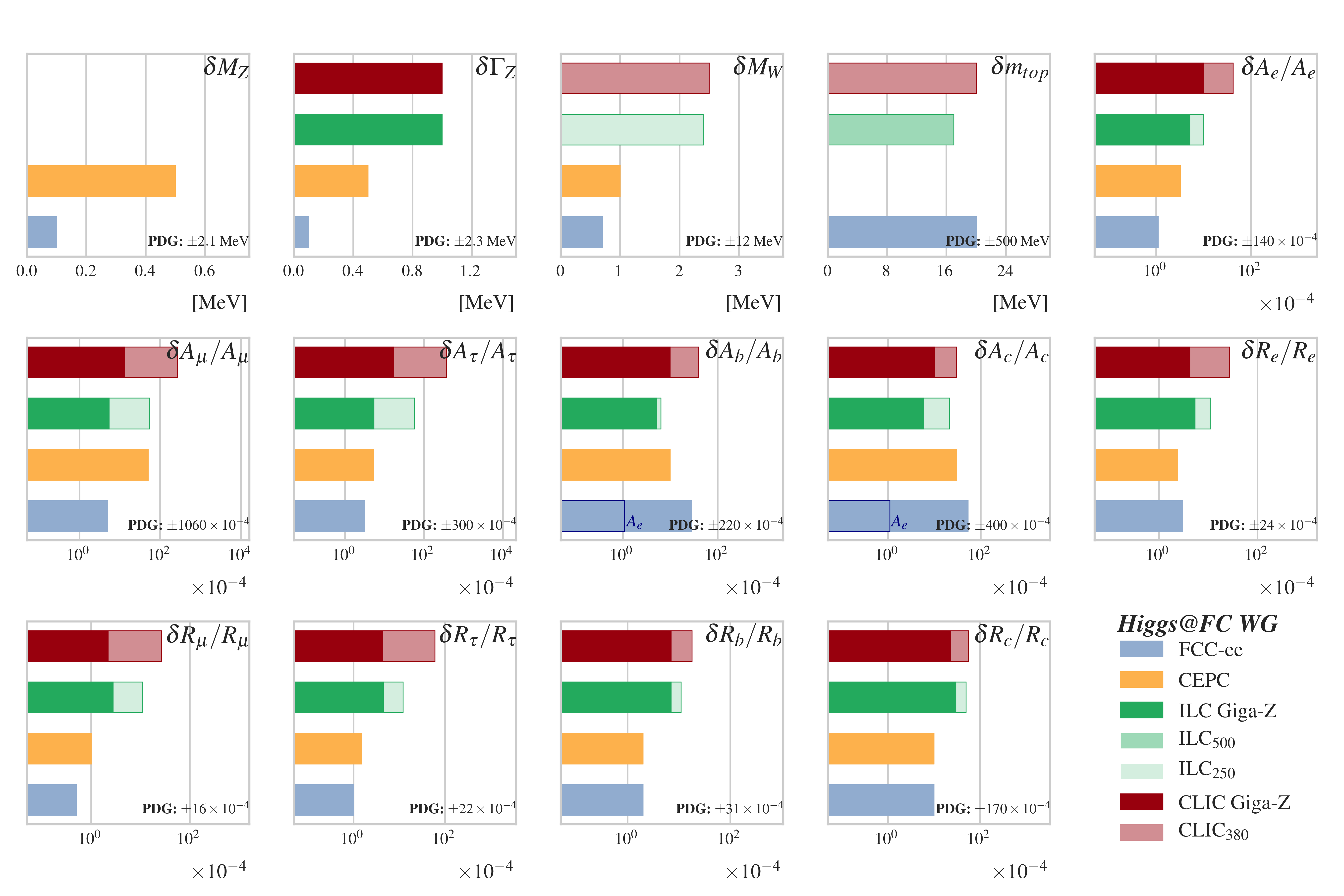}
    \caption{Uncertainty on several observables related to the properties of the \ew bosons: the masses of the $Z$ and $W$ boson and the top quark, the $Z$ boson width, and for fermion $f$ the polarisation asymmetries ($A_f$) and ratios of decay rates relative to the total hadronic decay rate ($R_f$). For the asymmetries and decay rate ratios relative uncertainties are shown. The fermions considered are leptons and $b$- and $c$-quarks. For $A_b$ and $A_c$, \FCCee considers uncertainties due to modelling of heavy quarks not considered by the other colliders. If these are neglected the uncertainty is similar to that on $A_e$ (indicated by a vertical line). The uncertainty on $m_\textrm{top}$ is only the experimental uncertainty, currently there is also a theoretical uncertainty of 40~MeV which is not shown.
    \label{fig:ewkpar}}
\end{figure}

Figure~\ref{fig:ewkpar} shows a selection of important EWPO, comparing the current precision to the future prospects at various future colliders. It is seen that all colliders will result in a significant improvement with respect to the current precision. For instance, at circular colliders, where the beams are transversely polarises, the mass and width of the $Z$ boson will be improved by about a factor of 20 by the \FCCee and 4 by the \CEPC. The decay rates and asymmetries (which are important for constraining the left- and right-handed couplings of the fermions) are improved by factors between 5 and 50. For the linear colliders, even with the running at $\sqrt{s}=250-380$~GeV a significant improvement compared to the current precision is achieved on most observables but dedicated running can add an additional large factor, close to the precision achieved by the circular machines, in many observables\footnote{The reason for this similarity is that all measurements are expected to be dominated by systematic uncertainties; if the circular colliders can use the higher statistics to constrain these effectively the situation could change.}. 

In Ref.~\cite{deBlas:2019rxi} a fit of the \ew precision data was performed to assess the impact on the oblique parameters mentioned earlier. For this fit, only parametric uncertainties and no intrinsic uncertainties are considered.

\begin{figure}[htbp]
    \centering
    \includegraphics[width=.7\linewidth]{\main/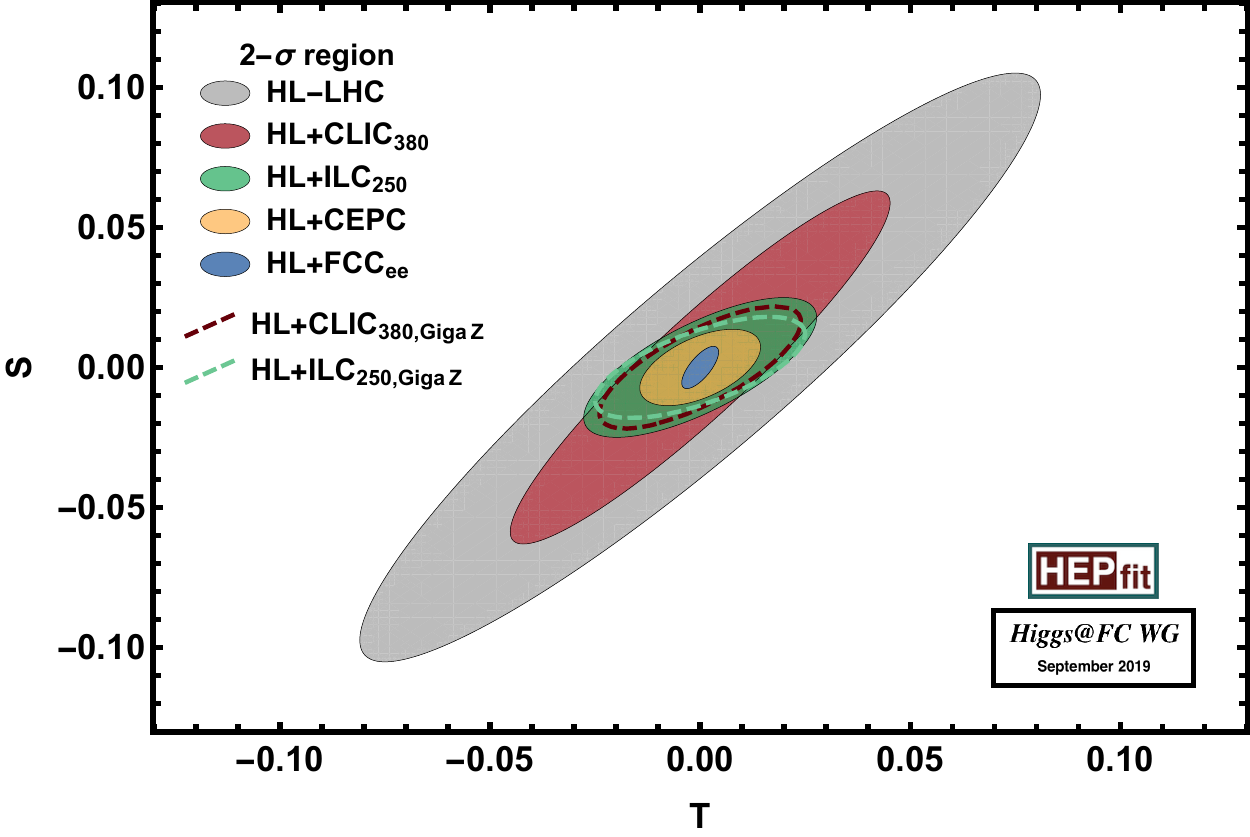}
    \caption{Expected uncertainty contour for the $S$ and $T$ parameters for various colliders in their first energy stage. For \ILC and \CLIC the projections are shown with and without dedicated running at the $Z$-pole. All other oblique parameters are set to zero.
    \label{fig:oblique}}
\end{figure}

Values for $S$ and $T$ are listed in Table~\ref{tab:oblique2}. It is seen that the sensitivity is ${\cal O}(10^{-2})$, making it sensitive to fine-tuning values of 3\% for composite Higgs models but not competitive with direct searches for SUSY models (see Table~\ref{tab:finetuning}).  
Figure~\ref{fig:oblique} shows the correlation between the $S$ and $T$ parameters for the different colliders.

\begin{table}[]
    \centering
    \caption{Values for 1$\sigma$ sensitivity on the $S$ and $T$ parameters. In all cases the value shown is after combination with \HLLHC. For \ILC and \CLIC the projections are shown with and without dedicated running at the $Z$-pole. All other oblique parameters are set to zero. The intrinsic theory uncertainty is also set to zero.
    \label{tab:oblique2}}
    \begin{tabular}{c|c|c|cc|c|c|cc}
    & Current & \HLLHC & \multicolumn{2}{c|}{\ILCTwoHundredFifty} & \CEPC & \FCCee & \multicolumn{2}{c}{\CLICThreeHundredEighty} \\
    & & &  & (\& \ILCGigaZ) & & & & (\& \CLICGigaZ) \\\hline\hline
        $S$ & 0.13 & 0.053 & 0.012 & 0.009 & 0.0068 & 0.0038 & 0.032 & 0.011 \\
        $T$ & 0.08 & 0.041 & 0.014 & 0.013 & 0.0072 & 0.0022 & 0.023 & 0.012 \\
        \hline
    \end{tabular}
\end{table}

In addition to measurements that probe the \ew sector of the SM, there are also several approaches at low-energy which provide interesting and complementary information. 
The forward-backward asymmetry $A_{FB}^{b}$ for the production of $b$ quarks measured at zero polarisation disagrees with the SM prediction 
by $2.3\,\sigma$~\cite{Tanabashi:2018oca}.
There is also a long-standing discrepancy of about $3\,\sigma$ between the value for the weak mixing angle, $\sintheta$ measured at LEP/SLC,
and that measured in neutrino deep-inelastic scattering by the NuTeV experiment~\cite{Zeller:2001hh}. The discrepancy may well be due to nuclear effects in the latter measurement~\cite{Beneke:2007zg}. The DUNE~\cite{Abi:2018dnh} experiment, primarily designed to measure the neutrino oscillations, plans to measure $\sintheta$ with a precision of about 1\% using its near detector. This should clarify the discrepancy further and serve as a complementary probe for the $Z$-boson to neutrinos at low energies $\sqrt{s} \ll M_Z$.  The electron-ion collider (EIC~\cite{Accardi:2012qut}), planned in the US, also plans to measure the dependence of $\sintheta$ on $Q^2$ in the range $Q^2\sim 10-70$~GeV$^2$ using polarised electrons scattered off unpolarised deuterons with a precision better than 1\%.

QED is the world's most precisely tested theory. The most impressive comparison between data and theory is the anomalous magnetic moment of the electron, $(g_e-2)$~\cite{Odom:2006zz,Gabrielse:2006gg}, which has been measured with a precision of one part in $10^{12}$ and is found to agree with theoretical calculations performed up to order $\alpha^{5}$ (\!\cite{Volkov:2017xaq} and references therein). For the muon $(g_\mu-2)$, however, there is a $3-4\,\sigma$ discrepancy~\cite{Blum:2013xva,Jegerlehner:2018zrj} between theory and experiment which could hint at new physics breaking lepton universality. A new experiment is now running at FNAL to clarify the situation~\cite{Grange:2015fou}, aiming at a precision of $1.6\times 10^{-10}$ ($4\times$ better than the current precision). The uncertainty on the theoretical calculation is $\sim 5\times 10^{-10}$ and the largest source comes from hadronic contributions. The MUonE experiment~\cite{Abbiendi:2016xup} at CERN plans to make measurements of high-energy muons ($E=150$~GeV) scattering on atomic electrons ($\mu e\to \mu e$) to constrain the hadronic contributions to the theoretical value for $g_\mu$.

The fine structure constant at $M_Z$ is currently determined as $1/\alpha=128.952 \pm 0.014$, see Ref.~\cite{Tanabashi:2018oca}. 
Based on future measurements at BES III, Belle II and VHEP-2000, it should be possible to reduce the uncertainty to $\pm 0.006$~\cite{Blondel:2019vdq}. 
It has been estimated that it can be reduced to $\pm 0.004$ by measuring the forward-backward asymmetries for 
$e^+e^- \to \mu^+\mu^-$ production versus $\sqrt{s}$ near $M_Z$ using 40~\iab of data~\cite{Janot:2015gjr} with \FCCee. The current uncertainty on $\alpha=\pm 0.014$ limits the precision of the \ew precision tests when the experimental precision on $M_W$ is reduced to below 8~MeV, as expected possibly with \HLLHC but definitely at future $\epem$ colliders.

Tests of QED have so far been restricted mostly to the perturbative regime but already in the first half of the 20$^{\rm th}$ century it was pointed out that there is also a strong-field-limit in QED, where QED becomes non-perturbative~\cite{Sauter:1931zz,Heisenberg:1935qt}. This becomes relevant when the electrical field, seen by an electron, attains a value close to the Schwinger field~\cite{Schwinger:1951nm}. Several proposals exist to probe this regime with a high energy electron beam and a high-power laser using the AWAKE plasma wakefield accelerator at CERN~\cite{Caldwell:2018atq}, the European XFEL in Germany~\cite{Altarelli:2006zza}, or the FACET facility at SLAC~\cite{SLAC:2016van}. Previous experiments at SLAC and CERN did not quite reach the critical field value~\cite{Burke:1997ew,Andersen:2012ea}. The proposed new experiments will probe QED in the critical field regime, which is of relevance for instance for astrophysical phenomena (for instance magnetars~\cite{Kouveliotou:1998ze}), atoms with $Z>137$~\cite{pomeranchuk} and for high energy $e^+e^-$ colliders~\cite{Bell:1987rw, Blankenbecler:1988te}. 

\subsection{Higgs boson physics}
\label{sec:higgsfuture}

\subsubsection*{The Higgs boson couplings}
One of the most important open points after the discovery of the 125 GeV scalar at the LHC is how this particle couples to the known fermions and bosons, compared to the uniquely determined predictions from the Standard Model. Deviations in data from theory expectations would definitively indicate New Physics (NP), going Beyond the Standard Model (BSM), and as argued earlier in this chapter, they are a direct measure of the fine-tuning. 

Higgs boson couplings can be determined from the measurement of rates of events with given final states, which, using the fact that the Higgs width is very small, can be expressed in terms of production cross sections times the decay branching fractions.

A simple yet powerful method to parameterise possible deviations from SM couplings is the so-called $\kappa$-framework ~\cite{LHCHiggsCrossSectionWorkingGroup:2012nn,LHCHXSWG3}.
In this framework, the deviations of SM Higgs boson couplings are parameterised through rescaling factors, the $\kappa_{i}$, which are defined as the ratios of the extracted couplings of the Higgs bosons to particles $i$ ($i=W, Z, \gamma, b, \tau,...$) to their corresponding values as predicted by the Standard Model.
Hence, the SM case is recovered by taking $\kappa_{i}$ = 1, for all particles $i$. This interpretation framework has the significant advantage of being extremely simple in its experimental implementation as well as of not needing any additional prediction from theory beyond those in the SM. This simplicity comes at a price; away from $\kappa=1$ the $\kappa$-framework violates gauge invariance. It is, therefore, a tool to indicate deviations from the SM, but not to diagnose their cause.
The $\kappa$-framework can also be extended to probe new Higgs boson interactions, such as for instance, those of the Higgs boson with lighter BSM states. In this case, the total Higgs boson width, $\Gamma_H$ increases, and hence the branching fractions to SM final states are altered with respect to the SM predictions.  As discussed in Sec.~\ref{sec:ewkintro} Higgs boson decays to BSM particles can be separated in two classes: decays into invisible particles (with branching ratio \BRinv), and decays into all other untagged particles  (with branching ratio \BRunt).

The $\kappa$-framework, however,  by construction, does not parametrise possible effects coming from different Lorentz structures and/or the energy dependence in the Higgs couplings. Such effects could generically arise from the existence of NP at higher scales and could lead not only to changes in the predicted rates, but also in distributions. An efficient, robust and predictive parametrisation of these effects can be obtained  by extending the SM to an effective field theory, i.e.\ by including higher dimensional interactions, which respect the SM symmetries and are suppressed by a scale $\Lambda$. A plausible class of such effective Lagrangians is the so-called SMEFT, which can be expressed as a polynomial of gauge invariant operators 
 organised as an expansion in inverse powers of $\Lambda, {\cal L}_{d}= \sum_{i}  c_{i}^{(d)} {\cal O}_{i}^{(d)}/\Lambda^{d-4}\, $.
The {\em Wilson  coefficients} $c_i^{(d)}$ encode the virtual effects of the heavy new physics in low-energy observables. Their precise form in terms of masses and couplings of the new particles can be obtained via {\it matching} with an ultraviolet (UV) completion of the SM, or inferred using {\it power-counting} rules. New physics effects start at dimension $d=6$ for which  a complete basis of operators is known. When considering the EW observables,  one can reasonably focus on a relatively small subset of operators, involving Higgs-boson, gauge-boson and fermion fields.
In the study presented here, {\it Neutral Diagonality} (ND) is assumed for flavour, which implies no flavour-changing couplings to the Higgs and Z boson.
In addition, in the fermion sector, only the 2$^{\rm nd}$ and 3$^{\rm rd}$ generation fermions are considered here, and their operators are treated independently. It is also assumed that there are no non-SM particles the Higgs boson can decay to. These assumptions give rise to modifications to the Higgs self-interaction, Higgs coupling to vector bosons, trilinear gauge couplings, Yukawa couplings to fermions, and vector couplings to fermions, for a total of 30 independent parameters.  

This scenario can be used to study the sensitivity at future colliders to general departures from the SM in the global fit to EWPO, Higgs boson rates and diboson production.  

Studies of the impact of future collider projects on the Higgs boson coupling precision have been made in both $\kappa$ and SMEFT frameworks by the \textit{Higgs@FutureCollider} working group~\cite{deBlas:2019rxi}, and are the basis of the results presented in the following.

\begin{figure}[t]
\centering
\includegraphics[width=0.98\linewidth]{\main/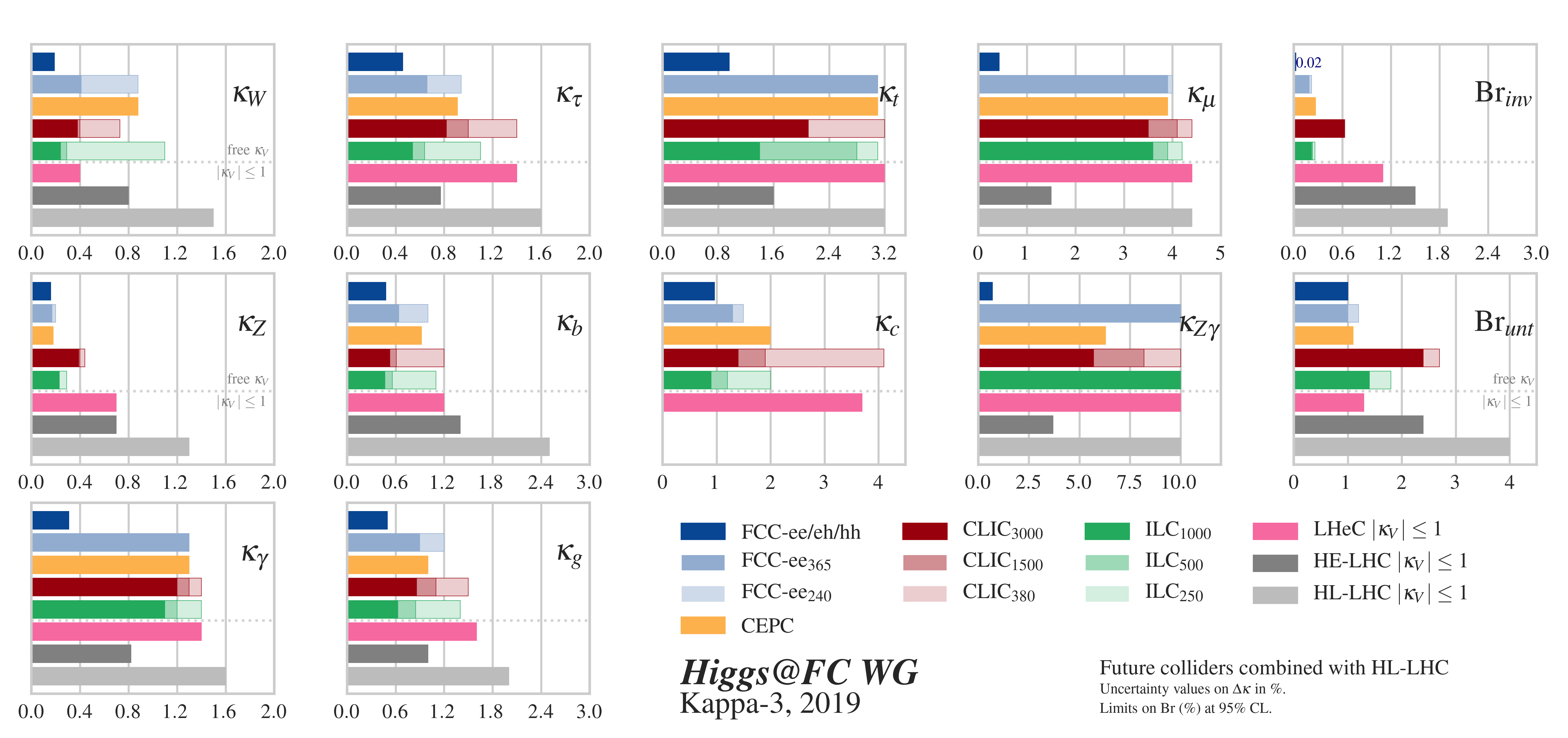}
\caption{\label{fig:Kappa3Summary}
Expected relative precision of the $\kappa$ parameters and 95\% CL upper limits on the branching ratios to invisible and untagged particles for the various colliders. All values are given in \%. For the hadron colliders, a constraint $|\kappa_V|\leq 1$ is applied, and all future colliders are combined with \HLLHC. For colliders with several proposed energy stages it is also assumed that data taken in later years are combined with data taken earlier. Figure is from Ref.~\cite{deBlas:2019rxi}.
}
\end{figure}

Figure \ref{fig:Kappa3Summary} shows the expected precision of the $\kappa_i$ parameters for various future colliders~\cite{deBlas:2019rxi}, all combined with the expected \HLLHC results \cite{Cepeda:2019klc}. 
The Higgs width, $\Gamma_H$, is left free in the fit, and for hadron colliders a constraint $|\kappa_V|\leq 1$ is applied as the width is otherwise not constrained.
In all cases, experimental statistical and systematic uncertainties are included. For hadron colliders uncertainties on the Higgs production cross section are included. For decay branching ratios only the parametric uncertainties are included while the intrinsic uncertainties are neglected, see discussion in Ref.~\cite{deBlas:2019rxi} and Sect.~\ref{sec:ewktheory}.

At the \HLLHC the Higgs boson couplings can be determined with an accuracy of ${\cal O}(1-3\%)$ in most cases, under the assumption $|\kappa_V| \leq 1$. Ratios of couplings are (mostly) model independent, and an accuracy of ${\cal O}(1-3\%)$ is expected in many cases~\cite{Cepeda:2019klc}.
Based on analyses of final states with large $\etmiss$, produced in Higgs VBF and $VH$ ($V=W$ and $Z$) processes, BR$_\textrm{inv}$ values of 1.9\% will be probed at 95\% CL. The constraint from the $\kappa$-fit on the BR to untagged final states is 4.0\% at 95\% CL. The \HELHC improves the precision typically by a factor of two, although much of the improvement comes from the assumption 
of a further reduction by a factor of two in the theoretical uncertainty, scheme $S2^{\prime}$~\cite{Cepeda:2019klc}.

Lepton colliders allow a measurement of the $ZH$ total production cross section, independently of its decay making use of the collision energy constraint. This measurement, together with measurements where the decay products of the Higgs boson are identified, can be interpreted as a nearly model-independent measurement of the total decay width. Therefore the constraint $|\kappa_V| \leq 1$, used for hadron colliders, is not needed for lepton colliders. 

Future $e^+e^-$ colliders improve the accuracy on Higgs coupling determination typically by factors between 2 and 10, except for $\kappa_t$, $\kappa_\gamma$, $\kappa_\mu$ and $\kappa_{Z\gamma}$ where no substantial improvement compared to \HLLHC is seen. \LHeC achieves a significant improvement for $\kappa_W$, $\kappa_Z$ and $\kappa_b$.
At $e^+e^-$  colliders, the couplings to vector bosons will be probed with a few 0.1\% accuracy. Higgs boson couplings to $b$-quarks can be measured with an accuracy between 0.5\% and 1.0\%, a factor of $2-4$ better than at the \HLLHC.  The coupling to the charm quark, not easily accessible at \HLLHC, is expected to be measured with an accuracy of ${\cal O}(1\%)$. The various $e^+e^-$ colliders do not differ significantly in their initial energy stages. 

The rate of rare Higgs boson decays such as $H\to \mu^+\mu^-$ that allows the study of the second generation lepton couplings, will be best measured by \HLLHC with an accuracy of about 4\%. 

It is difficult to access the couplings for the first generation. The current limit $\kappa_e<611$ \cite{Khachatryan:2014aep} is based on the direct search for $H \to e^{+} e^{-}$. A study at FCC-ee~\cite{Abada:2019zxq} has assessed the reach of a dedicated run at $\sqrt{s}=m_H$. In one year, an upper limit of $2.5$ times the SM value can be reached, while the SM sensitivity would be reached in a five-year run. For the light quark couplings, please see Ref.~\cite{deBlas:2019rxi} for further discussion.

When \FCCee is combined with \FCCeh and \FCChh a further significant improvement is seen, particularly for couplings to top quark, muons, photons and $Z\gamma$ where \FCChh will benefit from very large event samples. The improvement in $\kappa_W$ comes primarily from \FCCeh. A study of various other combination of aspects of the \FCC programme is documented in Ref.~\cite{deBlas:2019rxi}. 
 
The sensitivity of the Higgs branching ratio to BSM invisible final states is predicted to be improved by a factor 3 (CLIC) to 10 (\FCCee, \ILC) with respect to \HLLHC. For \FCChh a sensitivity to branching ratios as small as $0.025\%$ is expected to be achieved. Branching ratios to untagged decays are typically probed with a precision of $(1-2)\%$.

\begin{figure}[t]
\centering
\includegraphics[width=0.85\linewidth]{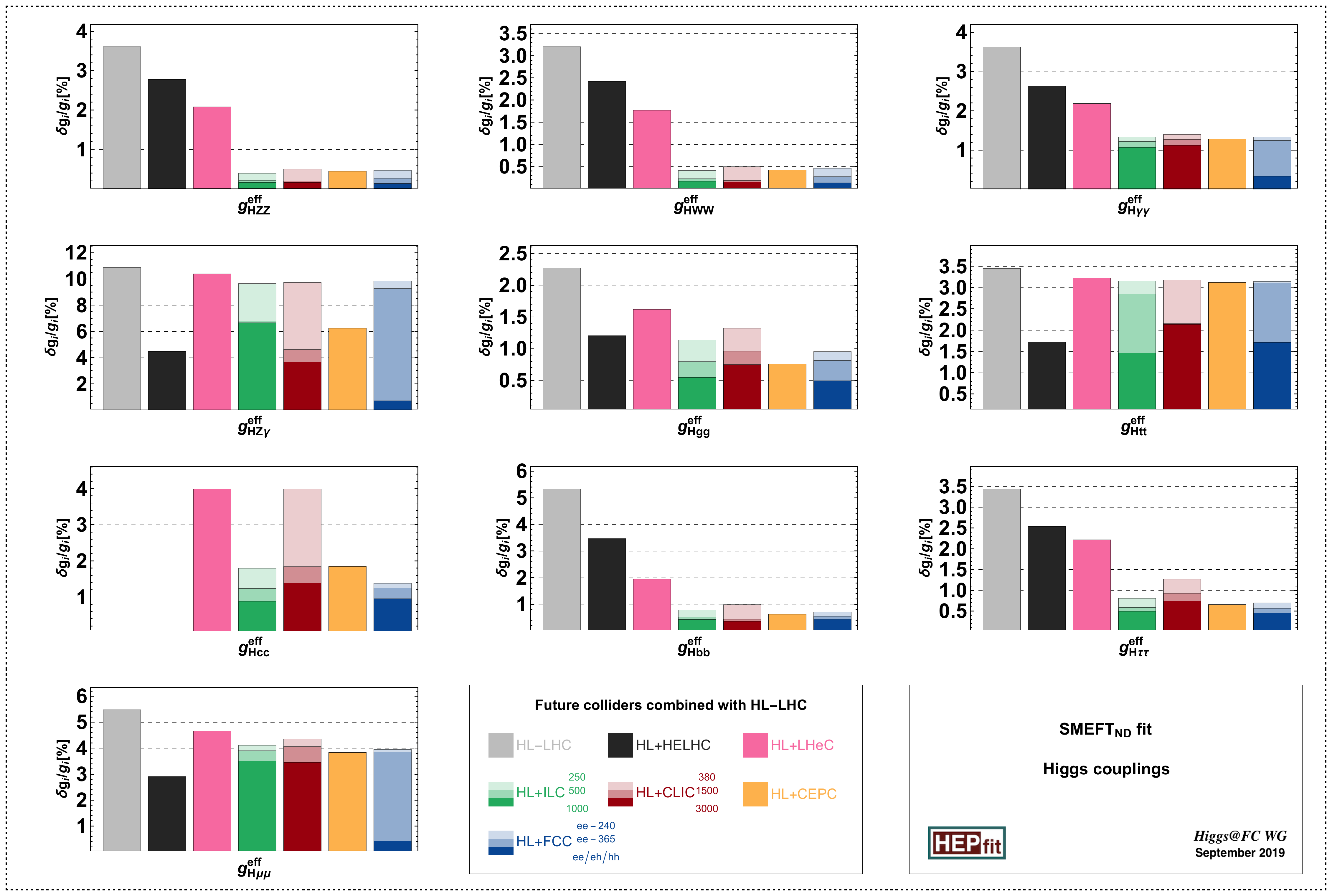}
\caption{\label{fig:EFT-results}
68$\%$ probability reach on Higgs couplings at the different future colliders from the Global fit SMEFT$_{\rm ND}$. For details, see Ref.~\cite{deBlas:2019rxi}.
}\vspace*{-5mm}
\end{figure}

In Fig.~\ref{fig:EFT-results}, the results of the fit corresponding on the EFT benchmark, expressed in terms of effective couplings, are shown. 
Again, it is seen that compared to the \HLLHC the $\epem$ colliders improve most parameters by about factors of 5-10. The exceptions are the coupling parameters related to top, $Z\gamma$ and $\mu$ couplings. The sensitivity of the different types of $\epem$ colliders is similar in their first stages. The improvements seen for \HELHC and LHeC are more modest.
For the $Z$ and $W$ a sensitivity below 0.3\% can be achieved by \ILC, \CLIC and \FCC. At this precision, the uncertainty is potentially limited by the intrinsic theory uncertainties which is not considered here (see discussion in Sect.~\ref{sec:ewktheory}). For fermions, the best sensitivity is reached for $b$-quarks and $\tau$-leptons, and it is about 0.5\%. 

\subsubsection*{The Higgs boson self-coupling}
The Higgs field is responsible for the spontaneous breaking of the \ew symmetry, and for the generation of all the SM particle masses.
Within the SM, the associated Higgs potential is characterised by one parameter, $\lambda$, that can be inferred from the experimental measurements of the Fermi constant $G_F$ and of the Higgs mass $m_H$,  (see Eq.~\ref{eq:SMpotential}). Beyond the SM, the parameter $\lambda$ is unconstrained experimentally, and could show particularly sizeable departures from the SM prediction. The determination of parameters related to the Higgs potential is therefore a high priority goal of the physics programme of all future colliders. 

The most direct way to assess the Higgs boson self-interaction and  in particular $\lambda_3$, is through the measurement of processes that feature two Higgs bosons in the final state.  At hadron colliders,  double Higgs boson production cross section is dominated by gluon fusion, $gg\to HH$, while at lepton colliders it proceeds via double Higgs-strahlung,  $e^+e^-\to ZHH$, particularly relevant at low energies, or via vector boson fusion (VBF), $e^+e^-\to HH\nu_e\bar{\nu}_e$, more important at centre-of-mass energies of 1\,TeV and above. 

For the \HLLHC, the cross section is predicted to be about three orders of magnitude smaller than the single Higgs production, which makes the double Higgs boson final state a challenging process to observe. The analysis relies on the combination of the $b\bar{b}\gamma\gamma$ and $b\bar{b}\tau\tau$ decay channels to reach about four standard deviation evidence for double Higgs production at \HLLHC (see Table~55 and Fig.~65 of Ref.~\cite{Cepeda:2019klc}). This corresponds to an accuracy on $\kappa_3 = \lambda_3$/$\lambda_3^{\rm SM}$ of about 50\%.

Higgs self-interactions also affect, at higher orders, the single Higgs processes~\cite{McCullough:2013rea,Degrassi:2016wml,Bizon:2016wgr}  and even the \ew precision observables~\cite{vanderBij:1985ww, Degrassi:2017ucl, Kribs:2017znd}. Therefore, single-Higgs boson production measurements can also be used to extract the Higgs self-coupling strength.

\begin{figure}[!ht]
\centering
\includegraphics[width=.8\linewidth]{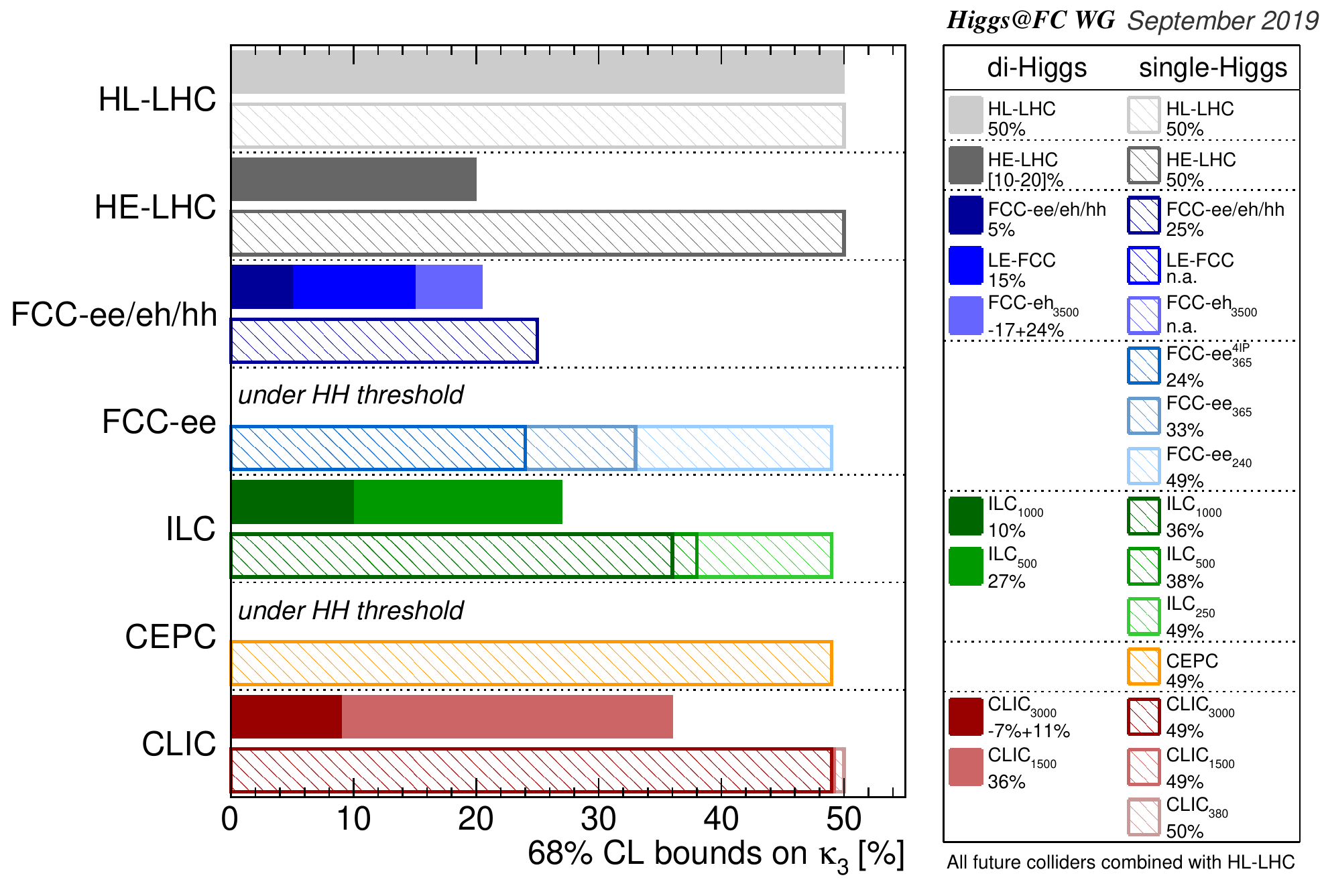}
\caption{\label{fig:h3-summary}
Sensitivity at 68\% probability on the Higgs self-coupling parameter $\kappa_3$ at the various future colliders. All the numbers reported correspond to a simplified combination of the considered collider with \HLLHC, which is approximated by a 50\% constraint on $\kappa_3$.
For each future collider, the result from the single-$H$ from a global fit, and double-$H$ are shown separately.
For \FCCee and \CEPC, double-$H$ production is not available due to the too low $\sqrt{s}$ value.
\FCCee is also shown with 4 experiments (IPs) as discussed in Ref.~\cite{Blondel:2019yqr} although this option is not part of the baseline proposal. LE-FCC corresponds to a $pp$ collider at $\sqrt{s}=37.5$~TeV.
}\vspace*{-5mm}
\end{figure}

Figure \ref{fig:h3-summary} shows the uncertainty expected on the measurement of $\kappa_3$ at the various proposed future colliders, in combination with the expectation from \HLLHC. The results have been obtained by studying the determination of $\kappa_3$ that can be obtained from single and double Higgs boson production processes using the EFT framework (see above). It is found that, when accessible, the $HH$ channel plays a pivotal role in constraining the trilinear self-coupling, as its addition to the fit allows to fully exploit single Higgs boson measurements to put severe constraints on the possible deviations of the other Higgs couplings, e.g.,  the top-quark Yukawa.  Details of this study are presented and discussed in Ref. \cite{deBlas:2019rxi}.

CLIC at $\sqrt{s} =3$~TeV and ILC at $\sqrt{s} =1$~TeV can extract $\kappa_3$ with a precision of about 10\%, and \FCChh expect to reach 5\%, respectively. With \HELHC or a $pp$ collider at $\sqrt{s}=37.5$~TeV (LE-FCC) a precision of about 15\% is expected~\cite{babyfcc}.
The accuracy on $\kappa_3$ is to about 30\% for \ILCFiveHundred and \CLICfifteen, while \FCCeh expects to reach about 20\%. Circular $e^+e^-$ colliders have insufficient energy to produce two Higgs bosons, and rely on the global fit of single-$H$ analyses, dominated by the high precision estimated for the $ZH$ cross section measurement. In this case, an accuracy of about 35\% can be achieved by \FCCee and \ILCFiveHundred, when combined with the \HLLHC. 

At \FCChh, a $2\sigma$ sensitivity to the quartic coupling, $\lambda_4$, is also expected. 

\subsection{Theoretical developments}
\label{sec:ewktheory}
The expected increase of precision in the measurements of EW observables at future colliders will demand a substantial improvement in the accuracy of theoretical predictions, (see also \cite{Blondel:2019qlh}). In this subsection, the needs are motivated and estimates are provided on what could be achieved based on the developments in the field in the last years, for both $e^+ e^-$ and $pp$ colliders. Figure~\ref{fig:higgsnow} has already shown that the dominant uncertainties in most Higgs couplings at the \HLLHC 
are theoretical, even after assuming a factor of two improvement 
with respect to the current state of the art. Higgs couplings will be approaching the percent level at \HLLHC.
At the $e^+e^-$ Higgs factories detailed measurements of the \ew Higgs production cross sections and (independently) of the decay branching ratios will be performed. Higgs couplings will be probed at approaching the per mille level.
At $e^+e^-$ colliders, a campaign of \ew  measurements at the $Z$-pole and at the $WW$ threshold is foreseen.  
The increase  in the number of $Z$ and $WW$ events with respect to LEP/SLD, as shown in Fig.~\ref{fig:ewkprogramme}, indicates that statistical errors will decrease by as much as two orders of magnitude at the future machines. As a consequence of this increased statistical precision, the requirements on the theoretical errors for EWPO~\cite{Blondel:2018mad} are even more stringent than for precision Higgs physics.

To interpret these precise results significant theoretical improvements  in several directions are required.  The first  is the increase of the accuracy of fixed order computations of inclusive quantities, e.g.\ from next-to-leading-order (NLO) to next-to-next-to-leading order (NNLO) and beyond. This reduces the so-called intrinsic uncertainties, i.e.\ those corresponding to the left-over unknown higher order terms in the perturbative expansion. Another important element is the accuracy in the logarithmic resummations that are needed to account for effects of multiple gluon or photon radiation in a large class of observables. In this case, different techniques and results are available, some numerical and some analytic, of different accuracy (from next-to-leading log (NLL) to next-to-next-to-leading log (NNLL) and beyond) and applicability. Improvements in the resummed predictions as well as their matching to fixed-order calculations of the highest accuracy will be needed also in the form of fully exclusive Monte Carlo generators that can be directly employed by the experimental collaborations. Finally, a reduction of the so-called parametric uncertainties, i.e.\ those coming from the imperfect knowledge of the input parameters, will also call for improvements in the theoretical predictions related to their extraction from data. Current values and future projections of the parametric uncertainties are given in Table~\ref{PUncertainties}. For example, for Higgs decay processes, the  principal parametric uncertainties are the Higgs mass and the quark masses ($m_t,m_b,\ldots$) and the value of $\alpha_s(M_Z)$~\cite{Lepage:2014fla}. 
For general \ew processes there are parametric uncertainties associated with $\alpha(M_Z),M_Z,M_W$.  For hadronic production, uncertainties from parton distribution functions also contribute.
\begin{table}
\begin{center}
\caption{Current and projected errors on input parameters. Where a lower bound 
is given on the projected uncertainties, the value is future machine dependent, see 
Fig.~\ref{fig:ewkpar}.}
\label{PUncertainties}
\begin{tabular}{|l|l|l|l|}
\hline
Error on&  Current value & Projected value & Uncertainty on  quantity\\
parameter &  Ref.\cite{deFlorian:2016spz} & Ref.\cite{Freitas:2019bre} & with projected value \\
\hline
$\delta M_H$  & 240~MeV & $>10$~MeV & $\Gamma_H(WW^*,ZZ^*),(0.1\%)$\\
$\delta m_t(m_t)$  & 1000~MeV & 50~MeV& \\
$\delta m_b(m_b)$  & 30~MeV & 13~MeV & $\Gamma_H(b \bar{b}),(0.6\%)$ \\
$\delta m_c$(3 GeV)  & 26~MeV & 10~MeV & $\Gamma_H(c \bar{c}),(1\%)$\\
$\delta M_Z$  & 2.1~MeV & $>0.1$~MeV & \\
$\delta M_W$  & 12~MeV & $>0.7$~MeV & \\
$\delta \alpha_S(M_Z)$  & $1.5\times 10^{-3}$ & $2\times 10^{-4}$& $\Gamma_H(gg),(0.5\%)$\\
$\delta \alpha(M_Z)$  & $10^{-4}$ &  $>5\times 10^{-5}$& \\
\hline
\end{tabular}
\end{center}\vspace*{-8mm}
\end{table}

Observables in the SM are calculable in terms of a double expansion in 
$\alpha_W =g_W^2/(4 \pi) \sim 0.034$ and in $\alpha_S$, which at high scales becomes small, for example,  $\alpha_S(M_Z)\sim 0.118$. Given the relative size of the interactions,  QCD corrections where applicable are more important than the \ew ones. Yet, QCD corrections at NNLO $O(\alpha_S^2)$ are expected to be of similar size as NLO $O(\alpha_W)$ \ew corrections. In addition, because of the special nature of QCD renormalization group improved perturbation theory, 
the naive expectation that $O(\alpha_s)$ NLO predictions should have an intrinsic uncertainty of $O(\alpha_s^2) \simeq 1\%$ does not hold. In fact, to achieve $1\%$ precision in a QCD computation, a combination of a higher fixed--order and resummed results has to be computed. 
The interpretation of the \ew production measurements of $e^+ e^- \to ZH, e^+ e^-\to \nu \bar{\nu} H$ requires predictions including EW corrections starting at one loop and QCD corrections starting at two loops. Decay branching ratios require both QCD and \ew corrections starting at one loop. 

At hadron colliders, QCD corrections to the production mechanisms are the dominant effect and their imperfect knowledge provides the main source of uncertainty. This is already the case 
at HL-LHC where, for all decay modes which are not statistically limited, the dominant uncertainty 
is theoretical as shown in Fig.~\ref{fig:higgsnow}. In addition, in order to be directly usable in experimental analyses, results need to be accurate, fully differential and include matching with parton showers, in order to be available in the form a Monte Carlo generator.

In order to judge the plausibility of success in reaching the desired theoretical improvements it may be useful to look backwards and note what has been achieved in the last twenty years in the context of $pp$ colliders. 
At the end of the last millennium, systematic algorithms existed for calculating differential NLO results, and a handful of NNLO results for QCD corrections, i.e.\ inclusive single boson ($\gamma^*,W,Z,H$) production, and relatively small set of results at NLO EW. The general purpose Monte Carlo programs, Herwig\cite{Corcella:1999qn} and Pythia\cite{Sjostrand:2000wi},
evaluated all cross sections at leading order. And fits to parton distribution functions were provided without any attempt to quantify uncertainties. 

Since then a number of achievements have been made; the complete automation of calculations for multi-leg processes, the matching to parton showers up to NLO in both QCD and in EW (see e.g.\ Refs.~\cite{Frederix:2018nkq,Alwall:2014hca,Ablinger:2017tan,Behring:2019tus}), 
intensive development of systematic NNLO subtraction and slicing algorithms
allowing fully differential predictions to be made at NNLO
for a number of ($2 \to 2$) processes, 
N$^{3}$LO calculations for $2 \to 1$ processes\cite{Mistlberger:2018etf}, now also including some differential information.
The calculation of DGLAP evolution kernels at N$^{2}$LO\cite{Vogt:2004mw,Moch:2004pa} has been further completed with partial results at higher loop order, and reached a large consensus on parton distributions with accompanying errors\cite{Butterworth:2015oua}. 

These technical advances have allowed predictions for the different mechanisms of Higgs boson production at $pp$ colliders. The results for the various mechanisms, gluon-gluon fusion (through a heavy quark loop), vector boson fusion, associated production with an \ew vector boson, associated production with a heavy quark pair, (top or bottom), and associated production with a top quark have already been shown in Fig.~\ref{fig:Higgs_XS_7-100TeV_HH}. The accuracy achieved has been reached as a result of the calculations, detailed in Table \ref{tab:StateofArt}.
\begin{table}
\begin{center}
\caption{State-of-the-art QCD corrections at fixed order for Higgs boson production in $pp$ collisions.  NLO EW corrections are known for all processes.  }
\label{tab:StateofArt}
\begin{tabular}{|l|l|l|l|}
\hline
Gluon fusion (effective theory)  & N$^{3}$LO & Total cross section & \cite{Mistlberger:2018etf,Dulat:2018rbf} \\
                                 & N$^{3}$LO & Rapidity distribution  & \cite{Dulat:2018bfe,Cieri:2018oms} \\
\hline 
Higgs + 1 jet (effective theory)   & NNLO & differential & \cite{Chen:2016zka}  \\
Higgs + 1 jet (full  theory)       & NLO  & differential & \cite{Jones:2018hbb} \\
\hline
Vector boson fusion   & NNLO (N$^{3}$LO) & differential (total) & \cite{Cruz-Martinez:2018dvl,Cacciari:2015jma,Figy:2007kv} (\!\cite{Dreyer:2016oyx})\\
\hline
$VH$   & NNLO & differential & \cite{Ferrera:2017zex,Caola:2017xuq} \\
\hline
$ttH$ production & NLO & differential  & \cite{Reina:2001sf,Beenakker:2002nc,Yu:2014cka} \\
$tHj$  production & NLO & differential  & \cite{Farina:2012xp,Demartin:2015uha} \\
\hline
\end{tabular}
\end{center}\vspace*{-5mm}
\end{table}
Predictions for Higgs decay widths to various possible final state are also known at rather high precision, with uncertainties ranging from a few per mille ($bb,\tau\tau,\mu\mu,WW,ZZ$) to a few percent level ($\gamma\gamma,gg,\gamma Z$). 

A comprehensive list of the specific calculations which will be needed, together with the development of new techniques, in the context of EW physics at the future colliders cannot be presented here. However, the main challenges can be illustrated by discussing a few key examples. 

At $e^+e^-$ colliders running at the $Z$ pole, one expects that the fully inclusive $Z$ decay rates EW and EW-QCD three loop computations as well as the EW 2-loop calculation for off-shell $e^+e^- \to f\bar{f}$ will be needed.  These calculations are expected to be challenging but in line with expected progress in the field. In addition, conceptual work towards a sound extension of the definitions of pseudo-observables at the targeted accuracy will be necessary. 

For the $WW$ cross section scan at $e^+e^-$ colliders, the full NNLO EFT calculation for $e^+e^- \to WW$ is foreseen together with the determination of the leading 3-loop Coulomb-enhanced EFT corrections, achieving the targeted accuracy of $(1-4) \times 10^{-4}$ for $\sigma(WW)$ at threshold. 
Finally, for Higgs production at $e^+e^-$ colliders, the full EW 2-loop calculation for on-shell $e^+e^- \to ZH$ and the leading contributions to $e^+ e^-\to \nu \bar{\nu} H$ will be needed. Obtaining these results is in line with the expected developments in the field of multi-scale two-loop computations. In addition, Higgs decay widths into massless partons in 4-/5-loop QCD calculations are expected. 

At $pp$ colliders a concerted effort in multiple directions will be needed for better determination of the signal as well as backgrounds. The inclusion of higher-order corrections in QCD and mixed QCD/EW, the improvement on the accuracy of the parton shower algorithms in line with the accuracy achievable with analytic resummations, the definition of new more perturbatively stable observables, are being pursued. The relative importance of each of such improvements for a given observable will depend on the specific process under consideration. 

For $pp$ colliders, the uncertainties on the parton distribution functions also play an important role when comparing measured cross sections to theoretical predictions. For the HL-LHC they are expected to be reduced by a factor of two compared to the current precision, and an additional reduction by a factor of two is assumed for HE-LHC. These reductions are assumed to be possible with precise measurements of various $pp$ cross sections~\cite{Khalek:2018mdn}. For \FCChh the projections are mostly based on ratios of cross sections where the uncertainties should largely cancel. If an $ep$ collider is available such as \LHeC or \FCCeh, the uncertainties on e.g.\ Higgs coupling measurements at these hadron colliders can be further improved.

\section{Summary and conclusions}
There is a very rich programme of precision measurements to be made of the properties of the Higgs boson, and of the other particles and parameters that are relevant to the \ew force. Both \ew precision observables and Higgs couplings can be related directly to the naturalness problem, the problem of why the Higgs mass is unstable in the SM and whether there is new physics that stabilises it at the \ew scale. 

Precise measurements of the $Z$ boson could be made at future $\epem$ colliders, in particular at circular colliders where $10^{12}$ $Z$-bosons would be available. At linear colliders a significant advance compared to the current precision is also expected, thanks in part to the polarisation that is available. The oblique parameters will be measured with a precision of $\sim 1\%$, about a factor 10 better than the current precision. Furthermore, several low-energy experiments are expected to also contribute to the precise determination of important parameters of the \ew aspects of the SM.

\begin{figure}[t]
\centering
\includegraphics[width=.8\linewidth]{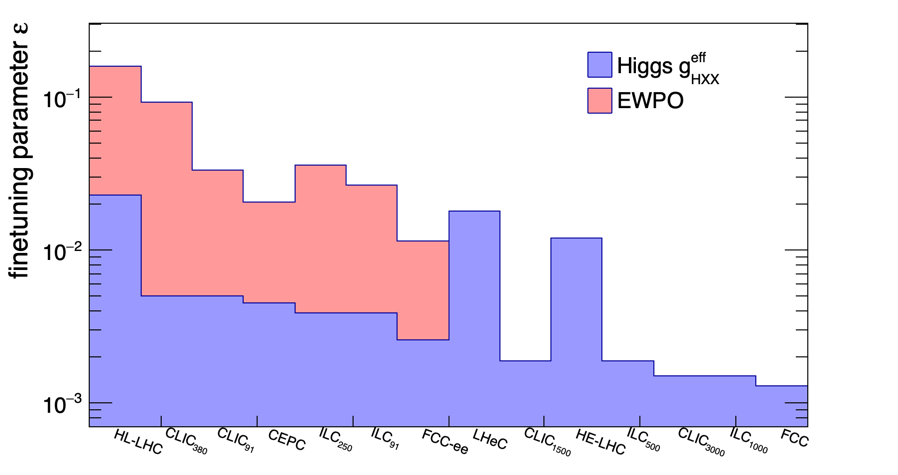}
\caption{\label{fig:h3-finetune}
Fine-tuning sensitivity as defined in Sect.~\ref{sec:ewkintro} based on the Higgs coupling and EWPO precision projections. In each case the highest precision Higgs measurement is shown based on the EFT analysis: for HL-LHC, HE-LHC and LHeC this is the $ggH$ coupling, and for all others it is the $VVH$ coupling. For the EWPO the value of $S$ is chosen, multiplied by three to be a measure of $\epsilon$, and only the low-energy stages of the lepton colliders are shown. The colliders are roughly ordered by the time it takes to take the data after a project start time $t_0$. For projects with multiple stages, $t_0$ is defined as start of data taking for the first stage.
}\vspace*{-4mm}
\end{figure}

The precision measurements of Higgs boson couplings are sensitive to BSM physics as it can alter them from their SM values. The \HLLHC will vastly improve the precision on the Higgs boson coupling parameters from typically 15\% today to a few percent with the full dataset, assuming that the present theory uncertainties will be reduced by a factor of two.  A further large improvement can be achieved with future $e^+e^-$ colliders, which also have the novel and powerful ability to measure Higgs production without any assumptions on its decay. The sensitivities in their initial stages are rather comparable. The most precise coupling measurements (to $Z$ and $W$ bosons), are measured to 0.2-0.3\% depending in part on the precision that can be achieved for the theoretical calculations. Additionally, Higgs decays to invisible particles (e.g.\ dark matter candidates) can be constrained to values much better than 1\%. The measurement of the total width to within a few percent, possible at $e^+e^-$ colliders, will provide an important constraint on many new physics scenarios.

The Higgs self-coupling is currently unconstrained by data, and a deviation of $\sim 1$ from the SM value has potentially dramatic consequences for cosmology, in particular the nature of the \ew phase transition that occurred in the early Universe. It can be measured to $\sim 50\%$ precision at \HLLHC, and ultimately a precision of 5-10\% could be reached at \FCChh at 100 TeV, \CLIC at 3 TeV or \ILC at 1 TeV. 

In summary, the \ew precision programme provides a complementary way to search for new fundamental particles, and its constraints can be related directly to the naturalness problem, which has traditionally been the main motivation for why there should be BSM physics at the weak scale. With the Higgs programme at future $\epem$ colliders, fine tuning can be probed in to a few $10^{-3}$, about 30 times better than today.
This is illustrated in Fig.~\ref{fig:h3-finetune} where the smallest uncertainty on the Higgs couplings (based on the EFT fit), and the $S$ parameter, at the various colliders are compared. These can be interpreted as measures of  fine-tuning as discussed in the introduction. It is seen that the HL-LHC will be able to probe fine-tuning at the 2\% level. Major improvements are expected by the first generation $e^+e^-$ colliders, pushing it to as low as $(0.3-0.4)\%$. With the higher energy colliders it can be even be pushed to $(0.15-0.2)\%$, an order of magnitude smaller than the HL-LHC. The sensitivites of the oblique parameters to fine-tuning are generally inferior in this measure. It is also worth noting that the mass scales probed are comparable to those probed via direct searches at the \HLLHC in many cases, see Chapter~\ref{chap:bsm}. 

In conclusion, the \ew physics programme is pivotal to the understanding of our Universe, relating to important open questions such as the naturalness of our theory, dark matter or the \ew phase transition in the early Universe and the matter-antimatter asymmetry. The proposed accelerators would hugely advance this field of research.


\end{document}


\chapter{Strong Interactions}
\label{chap:si}






\section{State-of-the-art}
Quantum Chromodynamics (QCD) is firmly established as the theory of strong interactions. The dynamics of quarks and gluons is encoded in 
a locally SU(3) invariant Lagrangian density, which predicts and explains a plethora of experimentally observed phenomena. The dependence of the QCD coupling $\alpha_s(Q)$ on the energy scale $Q$ is predicted in QCD to evolve from a strong coupling at low scales to a weak coupling at high scales. Quarks and gluons are confined into hadronic bound states at low energies, while behaving asymptotically free at high energies. 

The concept of asymptotic freedom enables precise quantitative predictions for QCD processes at high-energy colliders, obtained in a systematic manner using perturbation theory. To allow the full exploitation of precision collider data, these calculations need to attain high accuracy and precision, which goes along with conceptual and technical challenges in perturbative calculations and event simulation. Alterations of QCD may still become manifest, such as the embedding of QCD in a higher gauge theory possibly unifying electroweak and strong interactions, the discovery of unbound colour or of a new level of substructure. At low scales and/or high density, the quantitative understanding of QCD is less fully developed, with the first-principles understanding of confinement as an outstanding open question. The  evaluation of  non-perturbative contributions arising
in QCD  is, however,   possible     within numerical lattice QCD (LQCD). Furthermore,  synergies of QCD with adjacent fields are identified, e.g., with string theory through the use of the AdS/CFT correspondence for a strongly coupled gauge theory,
and the strong CP problem, connected to the neutron EDM, potentially related to
axion-like cold dark matter.

The separation of low-energy and high-energy dynamics through QCD factorisation is a cornerstone of particle physics phenomenology, enabling precision predictions for collider processes by parametrising the strong-coupling dynamics into empirical quantities such as decay form factors, parton distributions, and hadronisation models. These have often dual relevance, as fundamental objects of investigation and as input to predictions, depending on the research question under consideration. Future progress in fundamental understanding and precision phenomenology of QCD will rely on a diverse research programme with close interplay between theoretical advances and experimental measurements. 
\vfill
\label{section_low}

\subsection{The low-$\alpha_s$ or vacuum regime}
The partonic structure of hadrons is described in terms of Parton Distribution
Functions (PDFs) where QCD provides with the DGLAP equations only a prediction of the evolution of their structure with the energy scale. Especially the experiments at the HERA facility at DESY provided Deep Inelastic electron- and positron-proton Scattering (DIS) data to determine the PDFs covering a significant part of the $Q^2$ versus $x$ plane.
The use of these PDFs in certain kinematic ranges probed in $pp$ collisions at the
LHC (and even more at the FCC or SPPC) propagates into large systematic uncertainties
for key theoretical predictions.
Proposals are submitted to address this by extending the kinematic domain with new collider and fixed-target experiments.
%

Electroweak and Higgs physics at the LHC are entering the precision era, demanding a new level of understanding QCD, including questions related to proton structure and the value of the strong coupling ($\alpha_s$), calculation of higher-order matrix elements, the understanding of multi-scale problems through parton showers and resummations, and effects related to hadronisation.
Measurements performed at the HL-LHC of important electroweak parameters like the $W$ boson mass~\cite{ATL-Wmass} and $\sin^2\theta_W$~\cite{ATL} can be significantly improved when more profound (from PDFs to GPDs) and extended (to higher $Q^2$, and both higher and lower $x$) information becomes available on the proton structure.
Similarly, the full exploitation of the $gg \to H$ cross section at N$^3$LO (see Sect.~\ref{precision_sec}) envisaged at the HL-LHC requires a consistent N$^3$LO set of PDFs, which could be obtained from studies at future $ep$ colliders~\cite{Bruening:2013bga}.
This underlines the importance of measuring the strong coupling at the $Z$ mass, $\alpha_s(m_\text{Z})$,
to 1--2 per mille precision which is considered to be possible at future $ep$ and $e^+e^-$ colliders~\cite{dEnterria:2019its} and challenges LQCD to reach equivalent precision. 
In the search for new physics, the HL-LHC will explore the largest possible masses  requiring a new understanding of partons, gluon, sea and valence quarks at large $x$ in the proton. Estimates illustrate that large $x$ PDFs obtained at the future LHeC would extend the search range for example for new heavier vector bosons in contact interactions at the HL-LHC by 10~TeV.
Very precise PDFs obtained in $ep$ collisions would establish as well a new base for testing QCD factorisation and evolving QCD theory by isolating novel dynamics.
Before that, fixed-target collisions at the LHC (in particular, in the
``backwards'' hemisphere) will bring a few tens of percent reduction in the
high-$x$ PDF uncertainties.
HL-LHC detectors upgrades will enable the explorations into the forward region probing the low Bjorken-$x$ range. Future $ep$,  $eA$  and $pp$ colliders like the LHeC,  FCC-ep, as well as  FCC-pp and  SPPS  have the potential to 
answer the question of where 
non-linear parton (especially gluon) evolution sets in. These would fundamentally change the pattern of parton evolution and enhance the numerical predictions of cross sections envisaged at the high-energy FCC-hh 
collider, e.g.\ the Higgs production at 
central rapidities corresponding to $x \simeq 10^{-3}$. For a profound exploration at the energy frontier of strong interaction, electroweak, Higgs and beyond the SM (BSM) physics the QCD program made possible with future  $pp$, $ep$ DIS and $e^+e^-$ colliders would be highly beneficial.

Beyond the collinear parton model extending our description of the hadron structure, and especially the proton structure, to three dimensions is an overall objective to test QCD and to significantly enhance the physics programme at hadron colliders. Additional experimental efforts are required to address parton Transverse Momentum Dependent (TMD) features and towards measurements of Generalised Parton Distributions (GPDs). Proposals are submitted to make significant progress on this nucleon structure front 
[ID99, ID111, ID143].

The vast research areas of phenomenology of multi-parton interactions, forward 
and diffractive physics address specific QCD phenomena in both non-perturbative 
and perturbative QCD regimes, and their interplay. In order to understand the proton 
structure and QCD dynamics at very high energies and small-$x$, it is highly relevant 
to further explore the fundamental aspects of forward and diffractive physics, both 
theoretically and by performing new precision measurements, including development 
of novel experimental tools and techniques~\cite{N.Cartiglia:2015gve}.



\subsection{Collective properties of QCD matter}
The Standard Model implies that our early Universe has undergone a series of phase transitions of fundamental quantum fields. Specifically, for QCD, lattice calculations predict the transition of matter to a quark-gluon plasma (QGP) in which partons are deconfined and chiral symmetry is restored~\cite{Bazavov:2019lgz}. QCD is the only phenomenologically realised non-Abelian Quantum field theory whose high-temperature phase is experimentally accessible in the laboratory~\cite{Citron:2018lsq}. Most generally, the focus of experimentation with nuclear beams is on learning how collective phenomena and macroscopic properties, involving many degrees of freedom, emerge under extreme conditions from the microscopic laws of Quantum Chromodynamics. This includes assessing thermal properties of QCD matter, characterising the QCD non-equilibrium dynamics that evolves nucleus-nucleus collisions towards equilibrium, quantifying the initial conditions of the collective dynamics, e.g.\ in terms of nuclear parton distribution functions, and establishing the system-size of collective phenomena from proton-proton ($pp$) via proton-nucleus ($pA$) to central nucleus-nucleus ($AA$) collisions as well as their dependence on centre-of-mass energy. 

In the soft (low-transverse momentum) sector, the occurrence of numerically large and abundant signatures of collectivity in $AA$ collisions is by now firmly established. In particular,  measured hadronic particle distributions show large flow-like momentum correlations that are in one-to-one correspondence with the initial spatial anisotropies in the collision system and that are thus unambiguous telltale signs of collective evolution. Also, soft particle distributions are found to approach hadrochemical equilibrium~\cite{Andronic:2017pug}. Model comparisons support fluid-like behaviour of the system with close to minimal dissipative properties and statistical hadronisation. Similarly, the occurrence of numerically large jet quenching signals in all hard (high-transverse-momentum) hadronic observables is a generic feature of nucleus-nucleus collisions. These data indicate that nucleus-nucleus collisions rapidly and efficiently evolve towards equilibrium and that the detailed analysis of how hard out-of-equilibrium probes soften and isotropise, i.e.\ quench, can provide insight into the microscopic mechanisms underlying QCD equilibration phenomena. The discoveries of the LHC Run\,2 have added now a qualitatively novel dimension to these findings by establishing that smaller but non-vanishing flow-like signatures and medium-modified hadrochemical abundances persist in the smaller $pp$ and $pA$ collision systems, and that these signals of collectivity increase smoothly from minimum bias $pp$ to central $AA$ collisions. These qualitative phenomena cannot be accounted for in terms of physics effects commonly invoked for multi-particle production in $pp$ collisions, and they thus constitute a challenge for the understanding of both $pp$ and $AA$ collisions. In contrast, unambiguous signatures of jet quenching have not yet been established in these smaller collision systems, though they may become accessible with refined measurements in the future. 

Capitalising on these previous discoveries, the experimental collaborations at the LHC and the world-wide theory community working on heavy ion phenomenology have identified  four major motivations for future experimentation with nuclear beams, namely~\cite{Citron:2018lsq} (see also Sects.~\ref{epcoll} and \ref{hot}): i) Characterising the long-wavelength properties of QGP matter with unprecedented precision, ii) probing the inner workings of the QGP, iii) developing a unified picture of particle production across system size, and iv) exploring nuclear parton densities and searching  for a possible onset of parton saturation over a wide range of $(x,Q^2)$. 
The generic denominator of these multiple studies is to arrive at a detailed, experimentally tested dynamical understanding of how out-of-equilibrium evolution occurs and equilibrium properties arise in a non-Abelian quantum field theory. 

\section{Hadronic structure}





\subsection{QCD and collider experiments (LHC, HL-LHC, HE-LHC, FCC, CEPC)}

Although QCD is not the main driving force behind future colliders, QCD is
crucial for many $pp$, $ee$ measurements, both as signals and backgrounds. The
precision needed to fully exploit all future $ee/pp/ep/eA/AA$ Standard Model and
BSM programs require exquisite control of perturbative and
non-perturbative QCD physics. There are unique QCD precision studies accessible
at FCC-ee, ILC,  CEPC, the Super Charm-Tau Factory (SCT@BINP), and at FCC-pp and SPPC. The main
themes are detailed in the following.

The QCD coupling parameter $\alpha_s$ is the least-known coupling of the
standard model. Its value impacts all QCD cross sections and decays, it is
a leading parametric uncertainty (in some cases together with the charm
and bottom masses) in calculations of key quantities in Higgs, top quark, and
electroweak physics~\cite{dEnterria:2019its}, and
its energy evolution affects the range of stability of the electroweak vacuum
approaching the Planck scale.
Future colliders enable a per
mille precision determination of $\alpha_s$ via hadronic $Z$, $W$, and
$\tau$-decays, and jet shapes (FCC-ee, FCC-pp, SCT).
Parton distribution functions have a wide impact on new physics at high-$x$, as
well as on new QCD evolution at low-$x$. 
Experimental study of partonic
processes will provide key constraints and result in high-precision PDFs
(HL-LHC, HE-LHC, FCC-hh). Worth to note is that FCC-ep is
required
to reach 1\% uncertainty for various cross section at FCC-pp.
FCC-ee provides multiple handles to study gluon radiation and gluon-jet
properties, as well as studies on jet substructure (N$^n$LO+N$^n$LL, see Sec. \ref{precision_sec}) and flavour
tagging through quark/gluon/heavy-quark discrimination, also giving access to
high-precision parton fragmentation functions~\cite{Anderle:2017qwx}. FCC-pp enables unique studies of
highly-boosted dijets and multijets, as well as,  heavy flavours,  pentaquarks and other exotic hadron structures.
In the field of non-perturbative QCD, FCC-ee allows for a percent-level
control of colour reconnection, while all machines give access to
high-precision measurements on hadronisation (baryon and strangeness
production, final-state correlations, bound states).


\section{Electron-proton collisions (LHeC, EIC, FCC)}
\label{epcoll}

\begin{figure}

 \centering
 \includegraphics[width=0.49\textwidth]{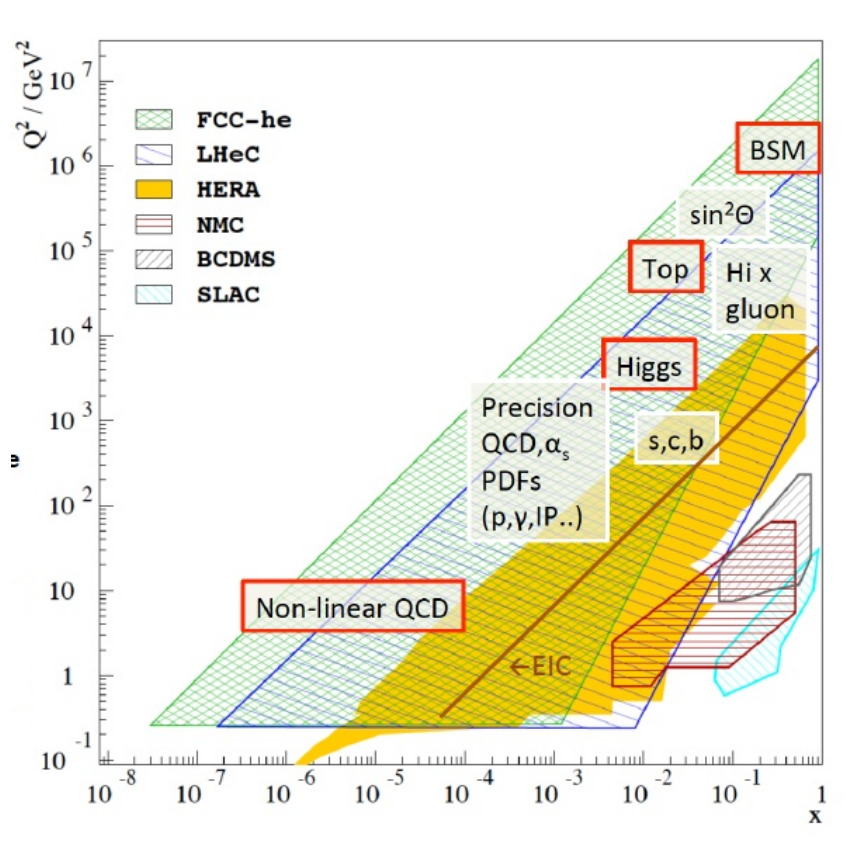}
 \includegraphics[width=0.49\textwidth]{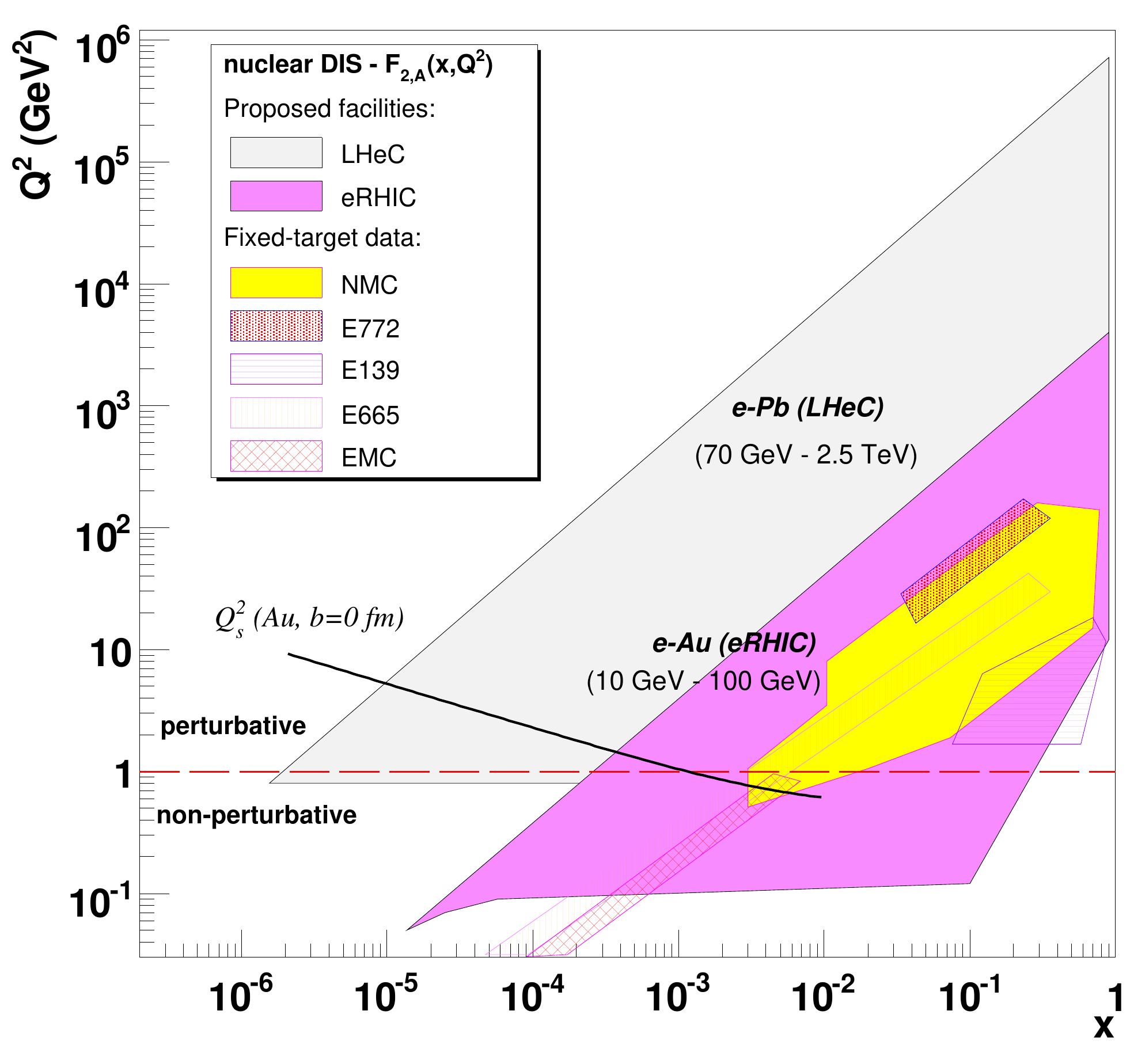}

 \caption{Kinematic $(x, Q^2)$ plane probed in $e-p$ (left) and $e-A$ processes
(right): existing data compared to proposed particle and nuclear DIS
facilities~\cite{Britzger:2017,dEnterria:2007jwb}.}

 \label{fig:xQ2}

\end{figure}

Deep inelastic lepton-nucleon scattering is a powerful and unique tool to study
nucleon structure (Fig.~\ref{fig:xQ2}) and unravel QCD with high-precision data and on firm
theoretical grounds. The future of DIS is bright with 
two proposed independent,
high-luminosity electron-hadron colliders in Europe [ID135, ID159]. DIS collider measurements extend substantially the
kinematic range of fixed target lepton-hadron scattering experiments. Excellent
precision is achieved through the redundant reconstruction of the leptonic and
hadronic final states while theoretically a next order of perturbation theory
in QCD and EW has to be controlled. Tagging of photons, electrons, protons, and
neutrons near the beam pipe will be a particular experimental challenge, as will the tagging of heavy flavour decays in a large rapidity range.

At medium energies, below that of HERA, electron-ion colliders can study the polarised
nucleon structure, contributing to the solution of the nucleon spin
puzzle and exploring the structure of the proton in three-dimensions at the medium
and high $x$ range relevant for the HL-LHC.
These include a US-based EIC\footnote{At the time of writing, DOE is moving forward to approve the
Mission Need soon, and has organised a panel to assess options for siting and
consideration of best value between the two proposed concepts~\cite{Hallman:2019}.}   (20 to
140 GeV c.m.\ energy) [ID74]  as well as one proposed in China (16 to 34 GeV). 

At the energy frontier, the CERN-based hadron-electron colliders (LHeC,
FCC-eh), with c.m.\ energies above that of HERA, will resolve the flavour
structure of unpolarised nucleons from $x$ about 10$^{-6}$ to near 1, measure
$\alpha_s(m_Z)$ 
to per mille accuracy, and
discover new parton dynamics (gluon saturation). The LHeC and FCC-eh are precision
Higgs- and EW-physics facilities with a remarkable BSM discovery potential. The $ep$ c.m.\ energies are 1.3~TeV for LHeC using
7~TeV $p$ from HL-LHC, and 3.5 (2.2)~TeV for FCC-eh using 50 (20)~TeV $p$ from
FCC-hh. The high-energy electron beams are produced using novel energy recovery
acceleration techniques (ERL), transforming the hadron colliders into an $eh$ and
$hh$ twin collider complex. Such a synergy will establish physics programmes
reaching much further than those of the HL-LHC and of future $hh$ colliders alone.

\subsection{Electron-ion collisions (LHeC, EIC, FCC)}

Several electron-ion ($eA$) colliders with per nucleon luminosities $\sim
10^{33}-10^{34}~\mathrm{cm^{-2}s^{-1}}$ are projected to start operating in the
2030s.
Colliding
electrons from an ERL with the HL-LHC or FCC nuclear beams, the LHeC is the
most powerful $eA$ facility that one can build in the next decades. It will
clarify the partonic substructure and dynamics in nuclei in an unprecedented
kinematic range. Also, it will unequivocally probe the new non-linear partonic
regime of QCD through density effects in $ep$ and $eA$, that increase both with
$1/x$ and mass number $A$. The LHeC will provide an accurate benchmark for perturbative probes,
the initial conditions for collective expansion, for the understanding of the
prior dynamics and the collective behaviour  in $pp$ and $pA$ collisions.

The EICs in the US~\cite{Accardi:2012qut} and China~\cite{Chen:2018wyz}, with
c.m.\ energies below 100~GeV/nucleon, are dedicated to a detailed mapping of
nuclear structure and its $A$ dependence in the medium $x$, lower $Q^2$ region,
extending the kinematic $(Q^2, 1/x)$ range as compared to existing DIS data by
up to a factor of 30. The flexible choice of lower energy but polarised beams, while limiting
access to small $x$, is optimal for pursuing a unique proton (and light ion)
spin programme.

The development of a broad QCD programme for the 2030s 
based on
synergies
and complementarities between different machines and collision systems,
$pp/pA/AA$ and $ep/eA$, should be encouraged.

\subsection{Fixed-target experiments (LHC, HL-LHC, SPS M2 beamline)}

The multi-TeV LHC proton- and ion-beams allow for the most energetic
fixed-target (LHC-FT) experiments ever performed opening the way for unique
studies of the nucleon and nuclear structure at high $x$, of the spin content
of the nucleon and of the nuclear-matter phases from a new rapidity viewpoint
at seldom explored energies~\cite{Brodsky:2012vg,Hadjidakis:2018ifr}.

On the high-$x$ frontier, the high-$x$ gluon, antiquark and heavy-quark content
(e.g.\ charm) of the nucleon and nucleus is poorly known (especially the gluon PDF for
$x \gtrsim 0.5$). In the case of nuclei, the gluon EMC effect should be measured to
understand that of the quarks. Such LHC-FT studies have strong connections to
high-energy neutrino and cosmic-ray physics.

The dynamics and spin of gluons and quarks inside (un)polarised nucleons is
also very poorly known; possible missing contributions are expected to come
from their orbital angular momentum. The LHC-FT mode enables to test the QCD
factorisation framework and to measure TMD
distributions, such as that of the linearly polarised gluons in unpolarised
protons or the correlation between the proton spin and the gluon transverse
momentum.

For heavy-ion studies, the proposed fixed-target experiments with LHCb and
ALICE enable the exploration of new energy regimes between SPS and RHIC
energies, across a wide rapidity domain to scan azimuthal asymmetries, and the
use of new physics probes (e.g.\ excited quarkonia, Drell-Yan pairs) to test the
factorisation of nuclear effects.
In addition, double crystal LHC-FT experiments give access to studies beyond
QCD, such as
MDM and EDM of heavy baryons.
 
There are two proposed ways towards LHC-FT
collisions [ID67]: a slow extraction with a
bent crystal, or internal gas target inspired by SMOG@LHCb, HERMES, H-Jet,
and others. The physics reach of the LHC complex can greatly be extended at a
very limited cost with the addition of an ambitious and long term LHC-FT
research program. The efforts of the existing LHC
experiments to implement such a programme, including specific R\&D actions on the
collider, deserve support.

The CERN M2 beamline [ID143] also allows for further dedicated nucleon-spin analyses
and innovative studies of the kaon structure (gluon content, spectroscopy,
polarisability, etc.). The CERN M2 beamline is key in the long-term multipurpose hadron
structure facility COMPASS++/AMBER proposal using the beamline beyond 2021 with a large community revolving around medium energy QCD, and for elastic muon scattering in the MUonE proposal.

\subsection{Opportunities of HL-LHC beams after the regular HL-LHC physics
programme}


The proposed LHeC
with a c.m.\ energy of 1.3~TeV will allow
high-precision measurements of the parton densities from high $x$ down to $x
\sim 10^{-6}$. It could be in operation at the earliest around 2030, but if not
realised during the regular HL-LHC programme, there could be an opportunity for
continued use of the HL-LHC beams afterwards. If in the future (beyond 2040) a
27~TeV c.m.\ energy HE-LHC
or an FCC-hh is realised, the corresponding
LHeC or FCC-eh
will have a c.m.\ energy of several TeV, enhancing
the kinematic reach further. Even higher energy electron-proton collisions may
be reached if in the future LHC-proton driven plasma wakefield accelerated 3
TeV electrons can be realised and collided with LHC protons. Such a 9~TeV
collider, called VHEeP (AWAKE++),
[ID58,ID103], 
would probe $x$ down to an
unprecedented $10^{-8}$. 
However, the target luminosity of
$10~\mathrm{pb^{-1}}$ over the entire lifetime is modest, so it should not be
viewed as a substitute for a high-precision machine like the LHeC. This also
applies to the SPS-driven Plasma Electron-Proton/Ion Collider (PEPIC) 
[ID58],
that will have a c.m.\ energy of 1.4~TeV, but a luminosity several orders of
magnitude lower than LHeC.

The region of large $x$, which is relevant for searches for new massive BSM
particles at the LHC and interesting for the nuclear dependence of the gluon
distribution, will not be covered by HL-LHC or EIC, but can be covered
by the LHeC and/or by fixed-target (FT) experiments,
either during
the regular HL-LHC programme or afterwards. FT experiments at an HE-LHC would
have a relatively modest increase in c.m.\ energy from 115 GeV to 163 GeV.

\subsection{Synergies between these programmes}

There are a number of striking synergy examples among future QCD physics
experiments:

\begin{itemize}

\item The determination of the strong coupling constant at per mille level
with LHeC/FCC-eh, FCC-ee, CEPC  and lattice gauge theory will lead to a new level of
understanding of QCD and to confidence in its predictive power. Agreement at
this high level of precision is required but will not likely result from the
first attempts in any of the above directions.

\item Precision measurements of flavour-separated parton distributions and the
correct theoretical description of the partonic content of the proton is a
necessity for precision physics and searches for new physics with the LHC and
subsequent higher energy hadron colliders. It can be provided to the required
accuracy and kinematic range only by a TeV $ep$ collider with a factor of 100
higher luminosity than that of HERA. Independent input on PDFs empowers e.g.\
precision interpretations within contact interaction and effective field theory
frameworks but also in the understanding of QCD background processes for e.g.\
novel SUSY searches.

\item It is essential to perform spin studies both at $pp$ and $ep$ machines to test
pQCD predictions with a sign change in some spin asymmetries.

\item The characterisation of the quark-gluon plasma suffers from  sizeable
uncertainties e.g.\ from our lack of knowledge of nuclear PDFs for quarkonium
suppression~\cite{Andronic:2015wma} and  charm production cross section in $AA$ collisions~\cite{Andronic:2017pug}  or of the initial conditions for collective
expansion for the extraction of the QGP transport
coefficients~\cite{Niemi:2015qia,Schenke:2012wb}, and of the role of parton fragmentation and hadronisation in cold nuclear
matter~\cite{Accardi:2009qv}. Therefore, it requires novel
input on nuclear parton structure, on nuclear multiparton correlations, on parton
fragmentation and hadronisation in-vacuum and in cold nuclear matter, which
should come from future electron-ion and
electron-positron colliders. The observation of long-range correlation effects
even in $pA$ collisions demands a detailed investigation of the lightest systems,
$ep$ and $eA$.

\item The development of accelerator technology (energy recovery) for a next
energy frontier $ep$ collider is supported by and also invites low energy ERL
facility developments. These have fundamental low-energy physics programmes, in
particle, astroparticle, and nuclear physics. Intense $\sim$1~GeV electron ERL facilities have
a wide range of important applications for material-, bio-, accelerator physics,
and other branches of science and technology. The high-quality requirements for
superconducting radio frequency are synergetic with developments of
electron-positron colliders. The first 802~MHz cavity, for example, has been
developed jointly for LHeC and FCC-ee, by CERN and JLab.

\end{itemize}

\section{Hot and dense QCD}
\label{hot}

\begin{figure*}[!ht]
\center
\includegraphics[width=3.355in]{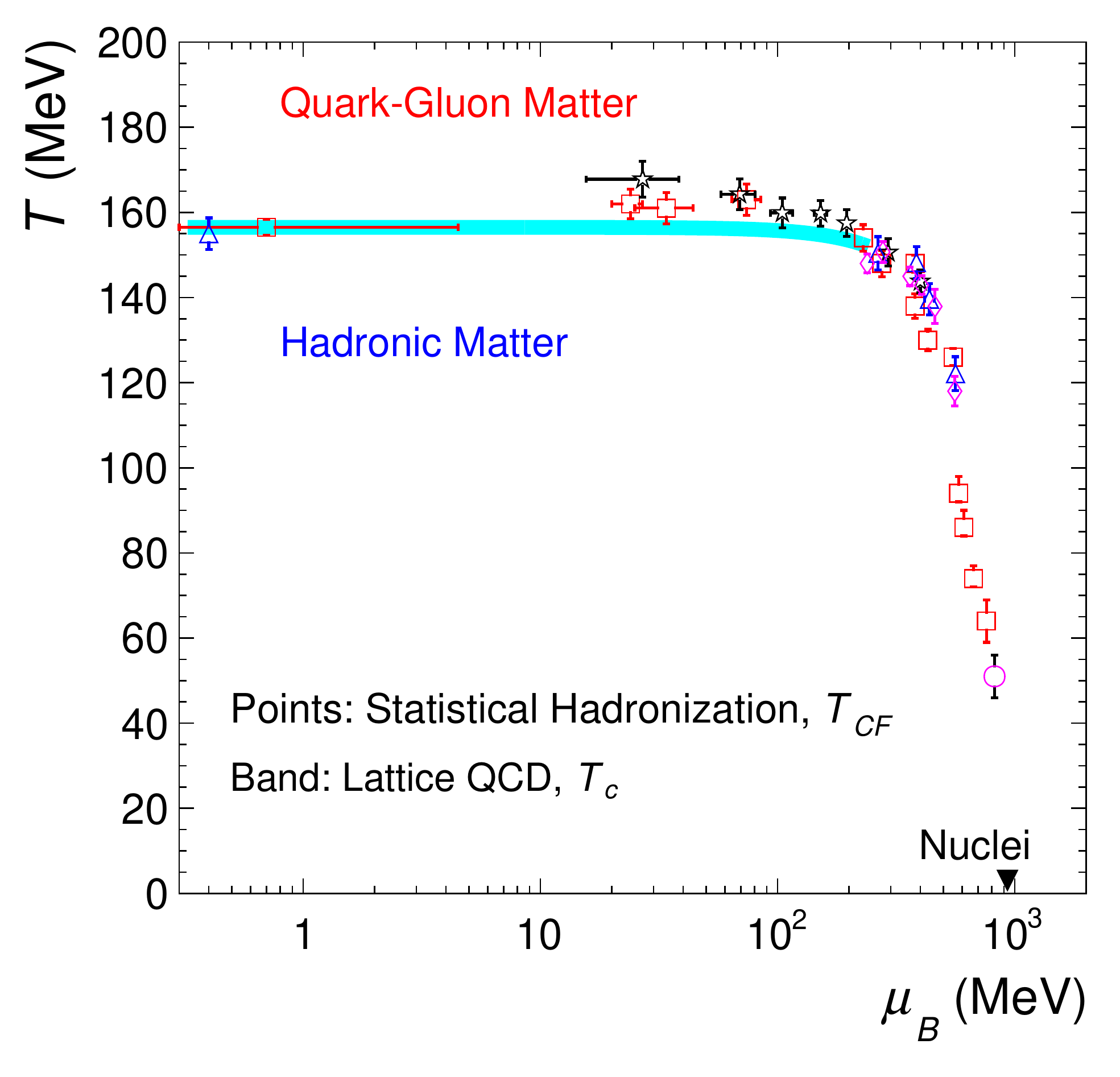}
 \caption{The QCD phase diagram in the temperature -- baryochemical potential plane. The band is from LQCD calculations~\cite{Bazavov:2018mes}, the points from statistical hadronisation model fits to hadron yields in heavy-ion collisions over a broad range of collision energy (updated from Ref.~\cite{Andronic:2017pug}).}
\label{fig:Tmub}
\end{figure*}

The study of hot and dense QCD matter is performed over a broad range of energies, both in fixed-target and collider experiments.
The connection to theory is achieved via the LQCD formalism, hydrodynamic description, microscopic transport models and statistical models.
An illustration is given in Fig.~\ref{fig:Tmub} of the results achieved to date concerning the phase diagram of QCD, where the points are extracted from fits of hadron yields with a statistical hadronisation (thermal) model, describing the chemical freeze-out---at higher energies, corresponding to $\mu_B\lesssim$ 300 MeV, they coincide with band showing the chiral crossover phase boundary as determined in LQCD.
Whether a critical point and a first order phase transition exist in this phase diagram is currently a subject of intense research.

A multitude of observables, like photon and dilepton yields, jets, heavy quarks, correlations and fluctuations, help to quantitatively describe the dynamics and the properties of the hot and dense matter.
Also  of particular interest 
is currently the question of the formation in high-multiplicity 
$p(D)A$ (or even $pp$) 
collisions of a droplet of deconfined matter that expands collectively, akin to the observations in $AA$ collisions.

\subsection{The ongoing experimental programme of relativistic heavy-ion collisions}

A very successful data taking for heavy-ion collisions in Run 2 at the LHC was completed, with Pb--Pb collisions at a nucleon-nucleon c.m. energy $\sqrt{s_{\mathrm{NN}}}$ = 5.02 TeV and $p$--Pb collisions at $\sqrt{s_{\mathrm{NN}}}$ = 5.02 and 8.16 TeV, recorded by all four experiments: ALICE,  ATLAS, CMS, and LHCb, with luminosities exceeding the initial goals.
The first physics results in a fixed-target mode were recently reported by the LHCb collaboration~\cite{Aaij:2018ogq,Aaij:2018svt}.

At BNL data taking is ongoing with the STAR experiment at RHIC in the beam energy scan (BES-II) program, spanning $\sqrt{s_{\mathrm{NN}}}$ = 7.7 -- 62 GeV in Au--Au collisions \cite{Adam:2019wnb,Adam:2019xmk}. At CERN,  the SPS programme is ongoing, with data taking in various collision systems at $\sqrt{s_{\mathrm{NN}}}$ = 7 -- 17 GeV with the NA61/SHINE fixed-target experiment \cite{Aduszkiewicz:2017sei}, which provides
important measurements for neutrino oscillation experiments \cite{Berns:2018tap}.
At the SIS18 at GSI, the HADES experiment continues to take data at $\sqrt{s_{\mathrm{NN}}}\simeq$ 2.5 GeV, while at JINR the BM@N detector is in operation at the Nuclotron in a similar energy range, currently with light ions \cite{Galatyuk:2019lcf}.

\subsection{Future opportunities for experiments at high-energy colliders}

The ongoing and foreseen upgrades of the four detectors at the LHC will meet the challenge of providing the required detector performance for the data taking with ion collisions with HL-LHC in Runs 3 and 4, a physics programme that is approved and will bring a significant advance in the field for the next decade~\cite{Citron:2018lsq}. 
The focus is on rare and challenging observables of the hot and dense phase, that could only be glimpsed with the existing data. Those include:
i) thermal radiation (photons and dileptons), to characterise the electromagnetic emissivity;
ii) heavy-flavour baryons  and (hyper)nuclei, to study in detail the hadronisation, on which the complex objects shall give unique insight; 
iii) quarkonium, both in the charm and beauty sector, to pin down quantitatively the basic mechanism of colour screening in QGP and the dynamics of (re)generation or quarkonia in QGP and at the QCD phase boundary;
iv) fluctuations of conserved charges, to establish experimentally signatures of the phase boundary that are predicted by solving QCD on the lattice;
v) highly energetic jets, to probe extreme regimes of parton energy loss (jet quenching) in the QGP.
Together, this suite of observables will provide new insights and a precise characterisation of the QGP (via transport coefficients) in the highest temperature and density regime. 

Beyond Run 4, a proposal for a next generation experiment at the LHC has been put forward~\cite{Adamova:2019vkf} to address hot QCD
physics, where the QGP initial temperature $T_{in}\sim 0.5$ GeV,   and soft  QCD   at very low $p_T< 10$  MeV, (while the particle physics experiments could continue data taking with heavy ions with focus on the hard regime). Advances in detector technology, for instance the possibility to curve thin silicon wafers, will open up the possibility to build a very light detector, enabling to reach very soft probes, both with hadrons and with dileptons, to probe in a direct way the quantum nature of the dense QGP state.
In addition, a complementary proposal has been
also presented to use in parallel the ATLAS, CMS and LHCb experiments in unique
BSM searches accessible in the high-luminosity ion-ion mode beyond
Run~4~\cite{Bruce:2018yzs}. In this time span, a complete LHC-FT program could also be realised with the ALICE and LHCb detectors in parallel to these measurements.

Advances in detector technology will be equally relevant for the FCC, where an attractive physics programme is envisaged~\cite{Dainese:2016gch}. The motivation for the FCC lies primarily in the unique availability of hard probes, enabled by the significant increase as a function of energy of their production cross section. Examples include beauty quarks, whose thermalisation in QGP can be determined, or top quarks, whose (boosted) decay to jets will provide a unique handle on the tomography of QGP. That goal could be realised, albeit in a less significant way, also at the HE-LHC.
Both at the FCC as in the HL(HE)-LHC a significant gain in luminosity can be achieved with lighter ions, which consequently may be more advantageous for rare observables. Collisions of protons with nuclei are part of the heavy-ion programme at HL(HE)-LHC and are envisaged too for the FCC~\cite{Abada:2019lih}.

It is important to notice that, at high-energy colliders, colliding heavy ions provide a significant photon flux, enabling unique studies of cold QCD matter in the regime of high gluon occupancy (low Bjorken $x$); for instance, exclusive photoproduction of J/$\psi$ at FCC can probe down to $x=10^{-7}$.
In addition, photon-photon collisions, already now providing interesting results at the LHC, have also the potential to bring sensitivity to BSM physics, providing discovery  potential (or exclusion limits) for magnetic monopoles or axions~\cite{Bruce:2018yzs}.
Ultra-strong transient (short-lived) magnetic fields are created in heavy-ion collisions and are probed via particle correlations which may provide insights relevant for astrophysics.
The high production rates of multi-strange hyperons at collider energies enables the study of their interaction properties, with results of relevance for the understanding of the neutron star structure, while abundantly-produced antimatter is of relevance for the search for antimatter in the Universe (currently with the AMS detector).

The long-term measurements in heavy-ion collisions will benefit from a sustained support from the theory community. The continuous improvement of existing theoretical tools and models, as lattice QCD or hydrodynamics, as well as the development of new techniques, in particular calculations in higher orders, will be crucial for the characterisation of QGP in its regime of highest temperature and densities.

This expected progress, both experimental and theoretical, will impact and benefit from the experiments in the next decade(s) at lower energies at RHIC (STAR and sPHENIX experiments) and at the NICA collider under construction at JINR, which will provide beams spanning $\sqrt{s_{\mathrm{NN}}}$ = 4--12 GeV, with the MPD experiment planned to start operation in stages in 2022~\cite{Golovatyuk:2016zps}.

\subsection{Future opportunities for fixed-target experiments} 

The RHIC fixed-target programme, planned to start in 2020, will cover $\sqrt{s_{\mathrm{NN}}}$ = 3.0 -- 7.7 GeV, corresponding to $\mu_B\simeq$ 400--700 MeV. 
The approved FAIR accelerator will deliver high-intensity beams ($\sqrt{s_{\mathrm{NN}}}$ up to 5 GeV) starting in 2025; the CBM detector aims at a collision rate of 10 MHz with continuous readout and online tracking and event selection. 
The  NA61/SHINE experiment at SPS, currently being upgraded with vertex capability (using pixel sensors developed for ALICE), will extend in the coming years its suite of observables into the charm sector.
An experiment at the SPS (NA60+) dedicated to thermal dimuon, open and hidden charm measurements is curently under design and aims at collision rates of 10 MHz~\cite{Dainese:2019xrz}.
The possibility of a heavy-ion programme with similar characteristics as that at FAIR is currently being considered for the J-PARC facility.
The physics motivation \cite{Friman:2011zz} is common for all these fixed-target experimental programs and it is shared as well by the BES programme at RHIC and by the NICA programme~\cite{Ablyazimov:2017guv,Bzdak:2019pkr}.
It is the investigation of hot and compressed baryon-rich matter, with special focus on the discovery of the critical point and (consequently) of a first order phase transition in the QCD phase diagram. Also prominent is the determination of the Equation of State of compressed baryonic matter, which is of relevance for neutron stars and for neutron star collisions.
This will be achieved with correlations and fluctuations observables and with rare probes like dileptons, multi-strange hyperons or hypernuclei, probes that will become for the first time available (with abundant statistics) for this energy regime.

The highest-energy fixed-target programme can be realised with LHC beams~\cite{Hadjidakis:2018ifr}, both in the ALICE and LHCb experiments. The hot QCD component of such a programme has as a special aspect the broad coverage in rapidity which is also of relevance as input for cosmic ray physics.
\section{Precision QCD}
\label{precision_sec}


As already introduced in Sect.~\ref{section_low}, the  interpretation  of LHC data and the searches for new physics 
require increased efforts to reach a higher level of precision and accuracy in key
theoretical predictions.
 This is the case in both the electroweak and the
   QCD  sectors of the Standard Model~\cite{Azzi:2019yne,Citron:2018lsq}.

The  QCD predictions for a scattering process,  owing to  the collinear factorisation theorem and the universality of PDFs and FFs,  are obtained by convolving perturbative 
scattering amplitudes        with the 
parton distribution functions,  (PDFs). Consequently, to reach sufficiently  high   accuracy in theoretical predictions of processes  involving quarks and gluons good control of     partonic   cross sections,  the value of the  strong coupling constant $\alpha_s$,  and  the    precise determination of PDFs is required.

Detailed modelling of strong interactions also
calls for a better 
understanding how the partonic final state in hard scatterings evolves to the hadrons observed in the experiment. The quantitative description of  fragmentation and  hadronisation  is part of the physics programmes at  fixed target and collider experiments.

\begin{figure}[hbt]
 \centering
\includegraphics[scale=0.78]{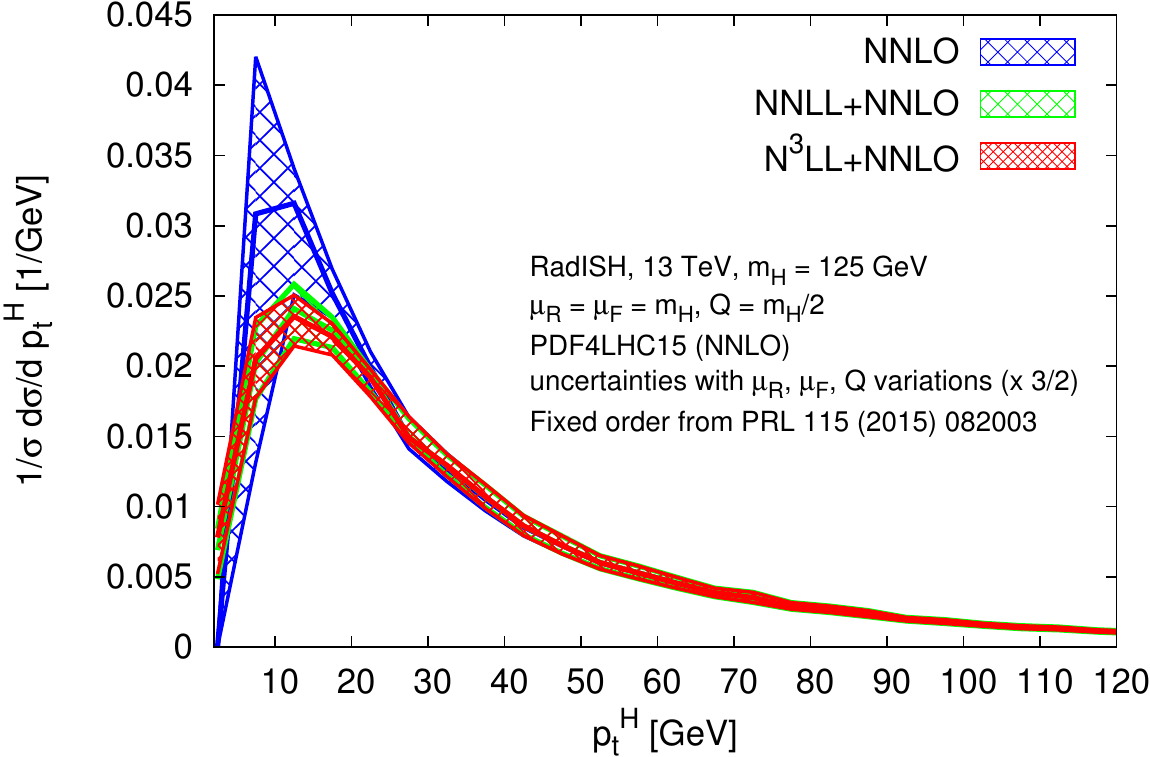}
 \caption{Comparison among the matched normalised distributions  at:  N$^3$LL+NNLO, NNLL+NNLO, and the pure NNLO of the Higgs-boson transverse-momentum spectrum at the LHC~\cite{Bizon:2017rah}.  }
 \label{NLL}
\end{figure}
Some non-perturbative aspects  of QCD are  directly  accessible in   first-principles
calculation by formulating QCD on the lattice and  solving it numerically. LQCD provides quantitative input on hadron structure and spectroscopy,  the properties  of matter  under extreme conditions and  hadronic  contributions  to the processes and matrix elements relevant for the  SM and BSM~\cite{Aoki,Cirigliano:2019jig,Bazavov:2019lgz,Lehner:2019wvv,DeGrand:2015zxa}.

\subsection{Methods and tools}

At  parton level, QCD cross sections at high momentum scales $Q$  are obtained through perturbative series expansion in the strong coupling $\alpha_s(Q)$.
This is the most  straightforward and successful approach that is also systematically improvable in accuracy  by calculating an increasing number of coefficients in the series. The current standard of such calculations is the next-to-next-to-leading order (NNLO) accuracy~\cite{Boughezal:2015dva,Ridder:2015dxa,Czakon:2015owf,Czakon:2016dgf,Currie:2013dwa,Currie:2016bfm}.  However, in view  of high-quality data at the  LHC  and the expected high-precision  measurements at HL-LHC,    there is a continuous theoretical  demand and efforts to go beyond NNLO and calculate QCD processes at the next-to-next-to-next-to-leading order, N$^3$LO.
   At present, the   hadron collider observables for which N$^3$LO QCD corrections have been calculated are e.g.\ the total cross section for Higgs boson production in gluon fusion~\cite{Anastasiou:2015ema,Mistlberger:2018etf} and in vector boson fusion~\cite{Dreyer:2016oyx}.
   Under QCD factorisation, the resulting predictions carry a residual uncertainty and dependence on the factorisation scheme due to the missing N$^3$LO (i.e., four-loop) splitting functions, recently motivating the computation of the QCD splitting functions at four loops \cite{Baikov:2006ai,Velizhanin:2011es,Baikov:2015tea,Davies:2016jie,Moch:2017uml}.  Furthermore, first steps have been taken towards more differential observables by computing  N$^3$LO  rapidity distribution of the Higgs boson at the LHC   in  gluon  fusion~\cite{Dulat:2017prg,Dulat:2018bfe,Cieri:2018oms}.  Moreover, fully differential distributions of jet production in deep inelastic scattering have been also derived    to N$^3$LO~\cite{Currie:2018fgr}.

Extending the perturbative  expansion  of QCD    to  the higher order is theoretically challenging as it    implies developing new methods and techniques to achieve the cancellation of infrared   divergences. At the NLO order there  are well established subtraction schemes and there are methods developed recently for NNLO calculations
\cite{Catani:2007vq,Bozzi:2005wk,Bonciani:2015sha,Boughezal:2015eha,Gaunt:2015pea,Czakon:2011ve,Boughezal:2011jf,Cacciari:2015jma,Ger}. 

In general,  N$^{\rm n}$LO  perturbation theory   is based on the expectation  that  calculating   a finite number of terms in the perturbative expansion is sufficient  since higher-order terms get progressively smaller and can be neglected once the desired accuracy is reached.  
In some processes, however,  logarithmically 
large  contributions appear at all orders  and spoil the convergence of the perturbative expansion.  In such cases,  it is necessary to rearrange   the perturbative expansion and perform resummation of these large logarithms to  all-orders in the perturbation theory.
Resummed calculations are typically needed in processes that depend on more than one scale.
\begin{figure}[t]
\begin{center}
  \includegraphics[scale=0.65]{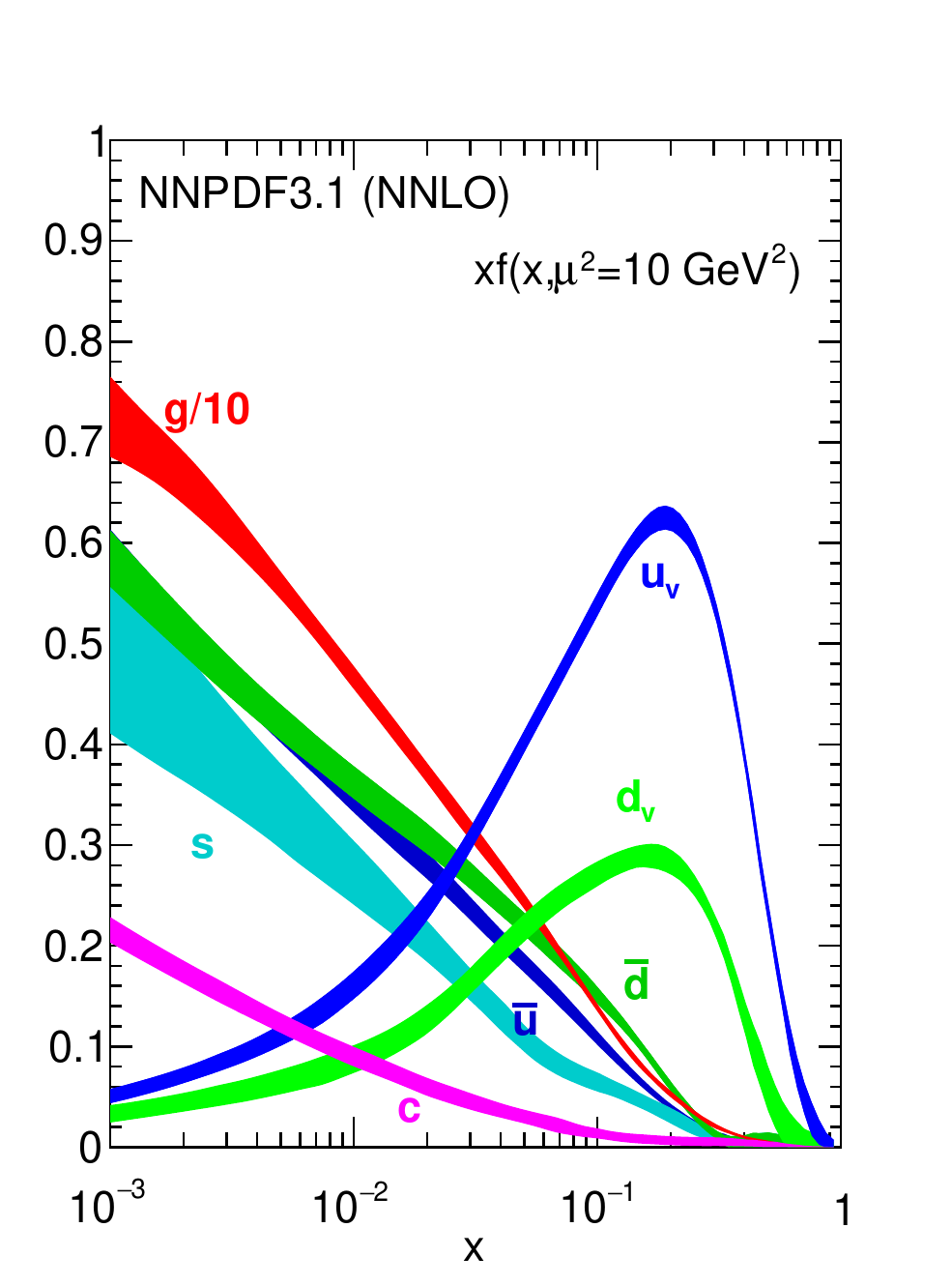}
  \includegraphics[scale=0.65]{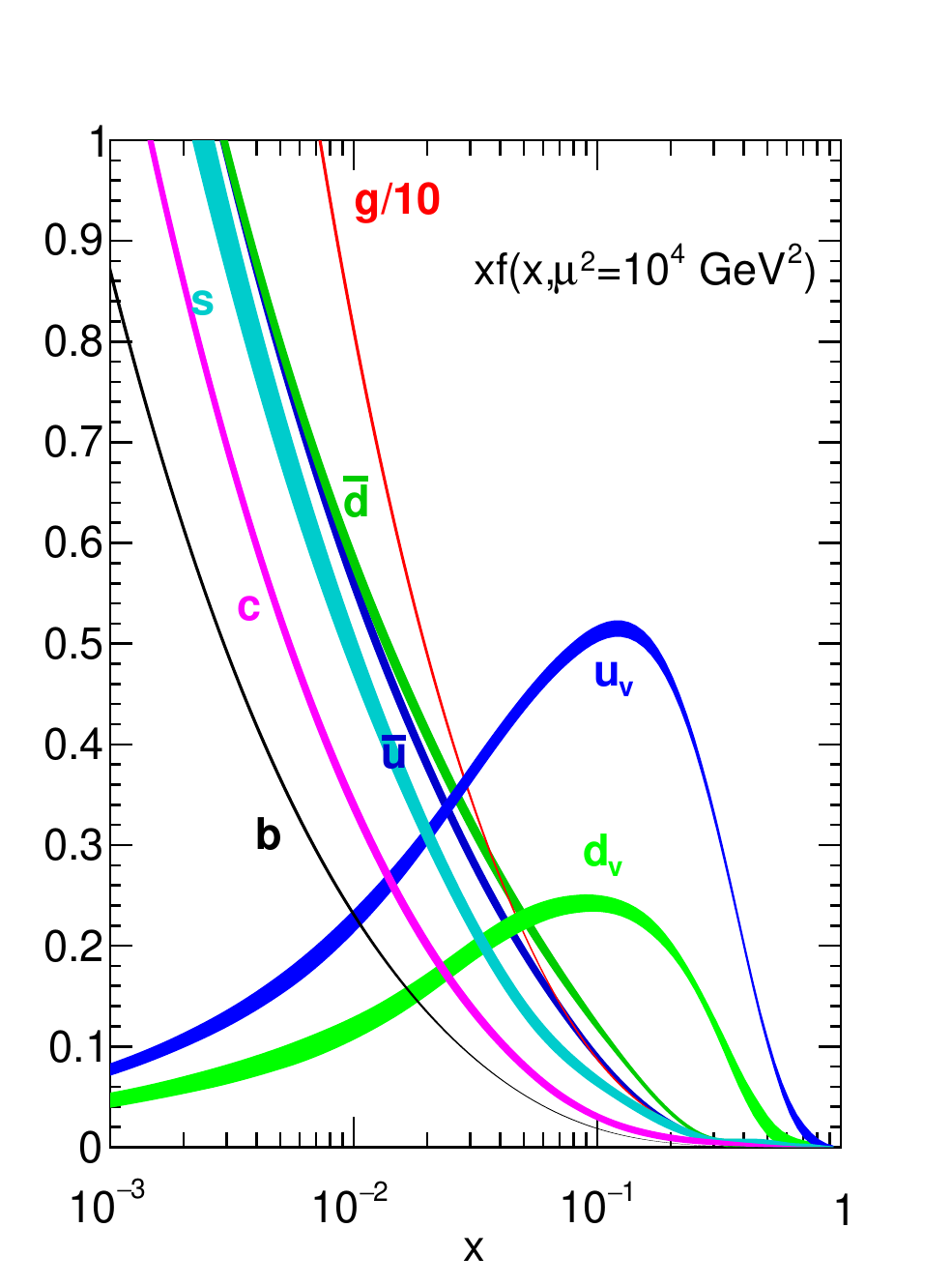}
  \caption{\small Different PDFs as obtained  at NNLO level (NNPDF3.1) from Ref.~\cite{Ball:2017nwa}, evaluated
    at $\mu^2=10~{\rm GeV}^2$ (left) and $\mu^2=10^4~{\rm GeV}^2$ (right).
    \label{fig.pdf}
  }
\end{center}
\end{figure}
Many of the observables that are
  studied at the LHC
  depend on more than one energy scale, and thus, all order resummation   become necessary to describe the kinematic regimes in which the logarithms of ratios of these scales become large.
There are different successful approaches to calculate the resummed expressions.
In the soft-collinear effective theory (SCET) of QCD~\cite{Bauer:2000ew,Bauer:2008jx}
the resummation follows the concept of Collins, Soper and Sterman~\cite{Collins:1984kg} where the starting point is  the derivation of a factorisation theorem for the specific cross section and then  calculations of all logarithmic terms from  the renormalisation group evolution equation, resulting in analytic expressions for the resummed cross section.  An alternative approach is based on the branching formalism\cite{Catani:1990rr,Catani:1991kz} where resummation is usually performed using a Monte Carlo algorithm.  Both approaches have already been applied to obtain higher order resummations in the hadronic collisions
~\cite{Bozzi:2005wk,Bauer:2002nz,Banfi:2004nk,Becher:2010tm,Stewart:2010pd,Banfi:2011dx,Berger:2010xi,Jouttenus:2011wh,Becher:2012qa,Zhu:2012ts,Banfi:2012jm,Becher:2013xia,Stewart:2013faa,Procura:2014cba,Li:2016ctv,Monni:2016ktx,Bizon:2017rah}.
The improvement of theoretical predictions   with increasing  accuracy of perturbative calculations    N$^k$LO  and  resummation  N$^k$LL is observed in different relevant  processes. In Fig.~\ref{NLL} we show 
 as an example   the  theoretical predictions for  the normalised  transverse-momentum  distribution  of the  Higgs  boson from  gluon fusion  at  13 TeV $pp$ collisions,
 calculated to  different orders in perturbative QCD~\cite{Bizon:2017rah}.
  In this figure,
   the Higgs-boson transverse-momentum spectrum at N$^3$LL   is matched to fixed  NNLO   and compared to NNLL   matched to NNLO, as well as  to the pure    NNLO results.  All curves in Fig.~\ref{NLL}  are normalised to the  total  N$^3$LO cross section.

 The  need of improvement of theoretical predictions with increasing order of perturbative calculations    is clearly identified  in  Fig.~\ref{NLL}.
Thus,  accurate theory calculations for collider processes are crucial to interpret the precise experimental data and to discern whether experimental measurements differ from  the SM predictions.  To match the precision of the data, theory uncertainties should be reduced to  the one  percent level for the core-  and to the few percent level for complex-processes. Furthermore, accurate theoretical predictions for BSM effects are highly desirable for  new physics searches. 
This requires a
relentless effort to improve the  understanding of QCD, by computing higher-order
corrections for a larger number of processes and by refining the  theoretical and numerical  methods.
Resummed calculations are instrumental to reach an accurate description of many observables at the LHC and beyond. The understanding of regions of validity of perturbation theory, non-perturbative  and collective effects, as well as     description of very high-multiplicity final states which 
  can  provide    insights  about  the  dynamics  of  multi-particle production arising from the saturated gluon fields inside the protons,
is theoretically challenging.

A further area of active research concerns general purpose Monte Carlo event generators. These are an essential part of the collider QCD toolkit, being crucial for the vast majority of collider measurements and studies. Over the coming years it will be important to sustain progress on a number of fronts: (1) perturbative improvements for matching and merging (e.g. generalisation of existing approaches for parton shower + NNLO merging); (2) understanding and exploiting the relation between parton-shower algorithms and resummation to obtain higher accuracy parton showers; and (3) the further development of phenomenological models (notably those for hadronisation and the underlying event).

%
\begin{figure}[t]
\begin{center}
  \includegraphics[scale=0.35]{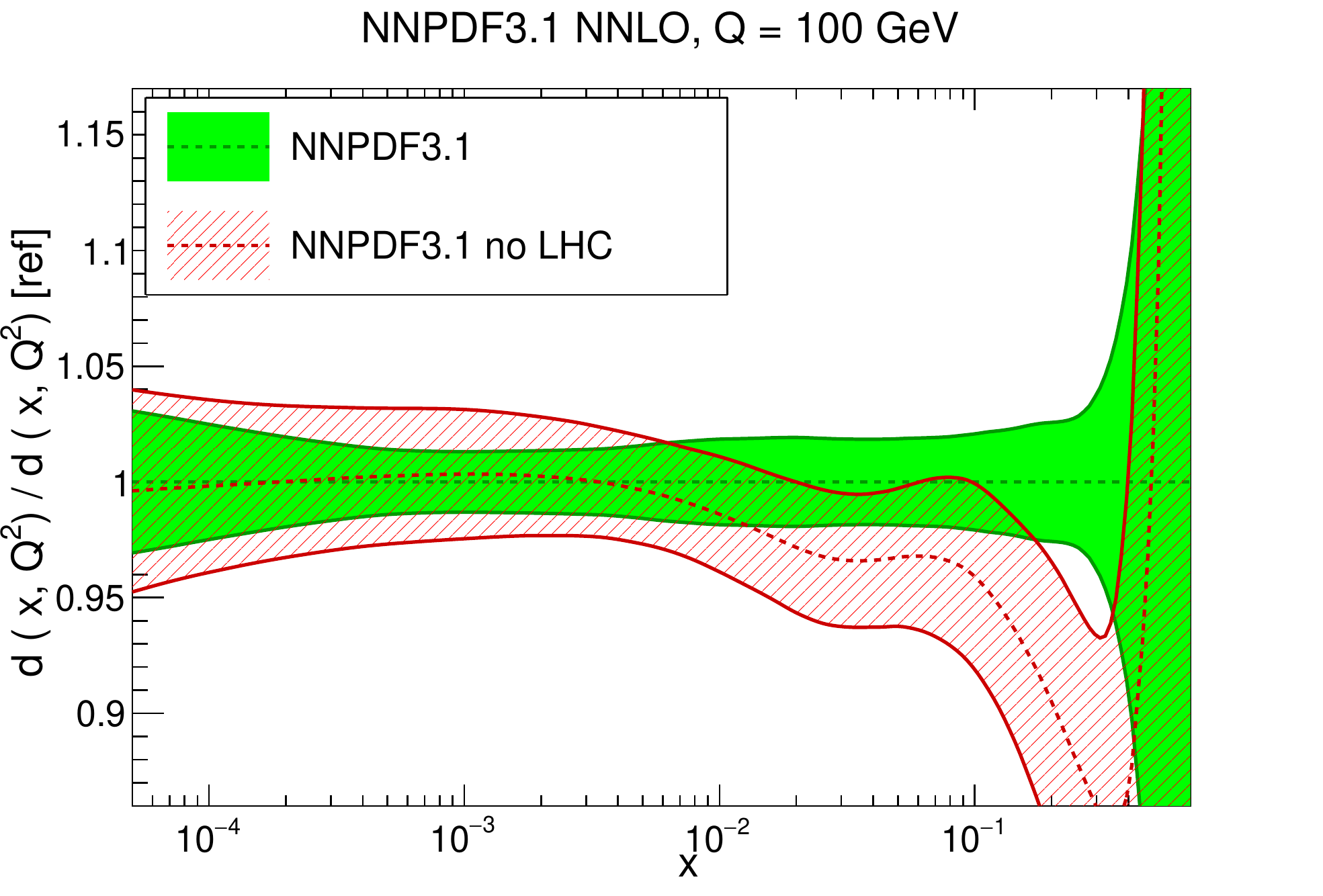}
  \includegraphics[scale=0.35]{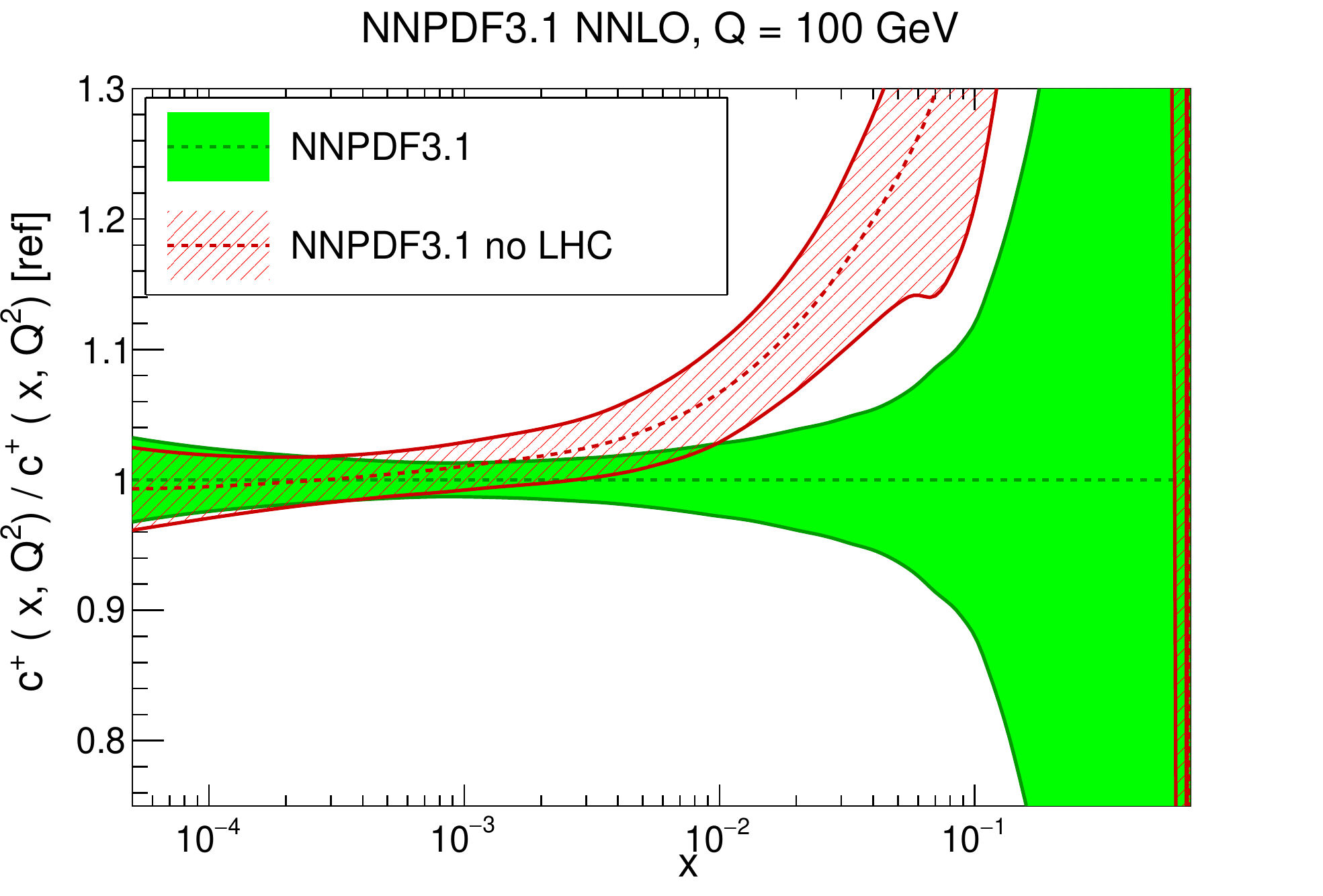}
  \caption{\small
The impact of the LHC data on the NNPDF3.1 NNLO PDFs  for the down and charm quarks~\cite{Ball:2017nwa}.\vspace*{-1cm}
\label{fig.impact}}
\end{center}
\end{figure}

 \subsection{ Parton Distribution Functions}

The quest for precision at the LHC  requires
a precise knowledge of the  Parton Distribution Functions  of the proton. An incomplete  knowledge of PDFs  is one of the main limitations in searches of new physics at the LHC.   At present,
PDFs cannot be  computed from first principles, although there are already
attempts to access the parton distribution functions based on LQCD simulations~\cite{Lin:2017snn}. The PDFs
have to be extracted from  the experiments, through  a careful comparison of theoretical predictions to experimental data.
Moreover, to obtain consistent predictions, it is crucial that the advancement in perturbative calculations goes together  with PDFs determination  with equal or higher  accuracy.

\begin{figure}[t]
\begin{center}
  \includegraphics[scale=0.5]{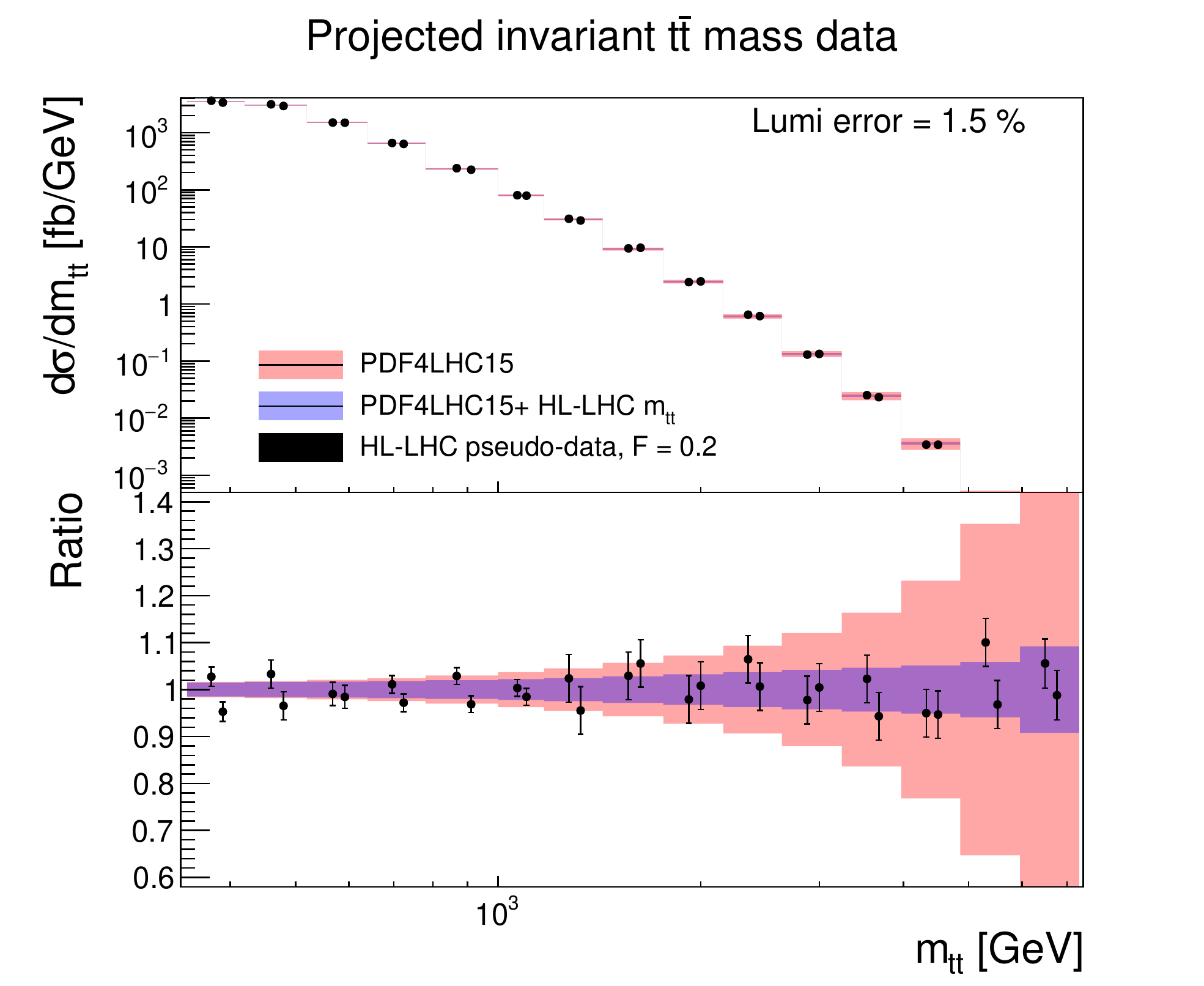}
  \caption{\small
Comparison between  HL$–-$LHC pseudodata and the theoretical predictions for
the $m_{t\bar t}$ distribution in top quark pair production   from Ref.~\cite{Khalek:2018mdn}. The theory calculations are shown both before (PDF4LHC15) and after  (PDF4LHC15-HL LHC) constraining PDFs  with  pseudodata  in the fit.}
\label{fig.hllhc}
\end{center}
\end{figure}

In recent years a new generation of PDF sets  have been developed for use at the LHC Run 2~\cite{Accardi:2016ndt, Alekhin:2017kpj}, and some of  these~\cite{Ball:2014uwa,Dulat:2015mca,Harland-Lang:2014zoa} have been combined  in the construction of the PDF4LHC15  sets which  broadly represent the present understanding of the proton structure.
Further  update  in global PDF
was  due to the progress  in  methodology  and  the available  set of new  data with
 the increasingly significant role played by LHC processes which provided stringent PDF constraints. The combination of high-precision LHC data of the ATLAS, CMS and LHCb experiments   with state-of-the art NNLO theory calculations for such hadronic processes as  the transverse-momentum spectrum of the $Z$ and $W$ bosons ~\cite{Boughezal:2015dva,Ridder:2015dxa}, the top-quark pair production~\cite{Czakon:2015owf,Czakon:2016dgf},  and inclusive jet production~\cite{Currie:2013dwa,Currie:2016bfm} have had  an important impact on precision PDF fits~\cite{Ball:2017nwa}. The  recent  PDF sets  NNPDF3.1 by the NNPDF Collaboration~\cite{Ball:2017nwa} are shown in   Fig.~\ref{fig.pdf}.
To illustrate the impact of the LHC  Run-1  data in the fits, Fig.~\ref{fig.impact}  compares the  NNPDF3.1 fit  with and without  LHC data at Q = 100 GeV   for the down and charm quarks~\cite{Ball:2017nwa}.  The  impact of the LHC data
can be observed in this figure,  both  for  central  values  and  for  the  PDF  uncertainties. 
   Thus, it is clear that the addition of data from Runs 2 and 3 and then    from the   HL-LHC,  for  which  the  precision  and  reach  will  be  greatly  increased,  should  lead  to further improvements in the determination of the proton structure. In  Fig.~\ref{fig.hllhc} we show
the comparison between the HL-LHC pseudodata and the theoretical predictions on  the   $m_{t\bar t}$ distribution in top-quark pair production~\cite{Khalek:2018mdn}.  There is  a clear reduction of the  PDF uncertainty    at large values of the invariant mass.

\begin{figure}[t]
\centering
\includegraphics[width=0.5\textwidth]{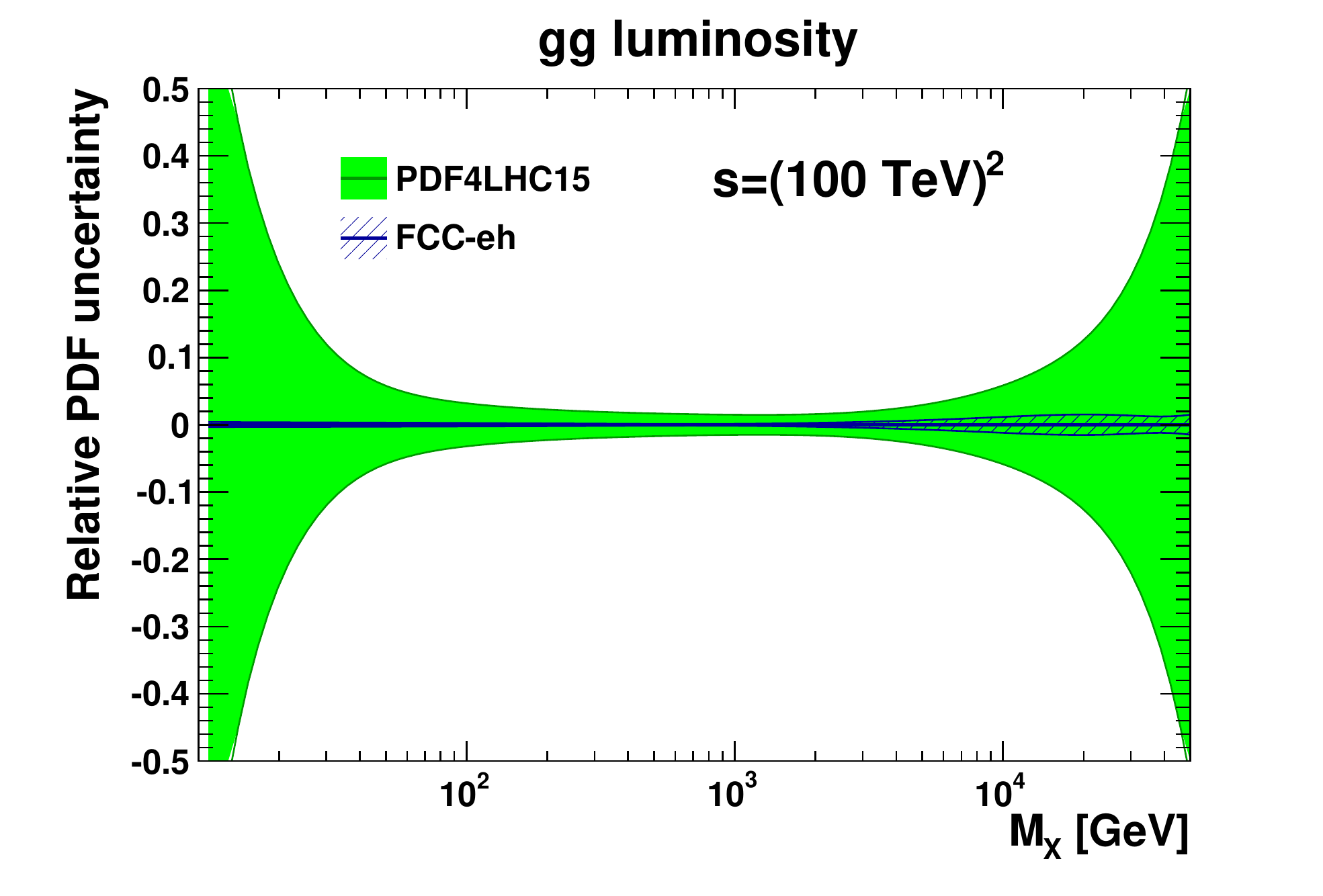}\includegraphics[width=0.5\textwidth]{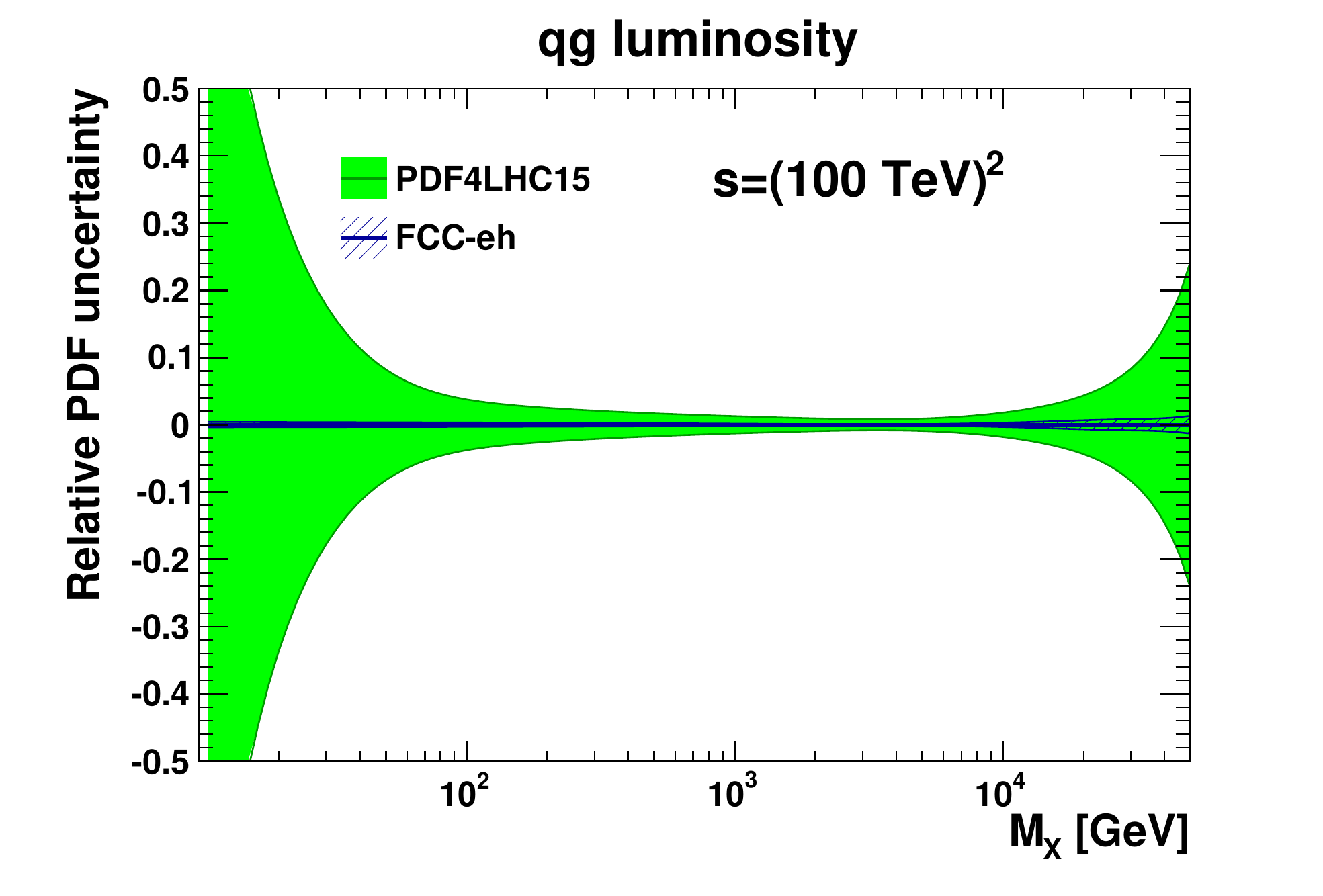}\\
\includegraphics[width=0.5\textwidth]{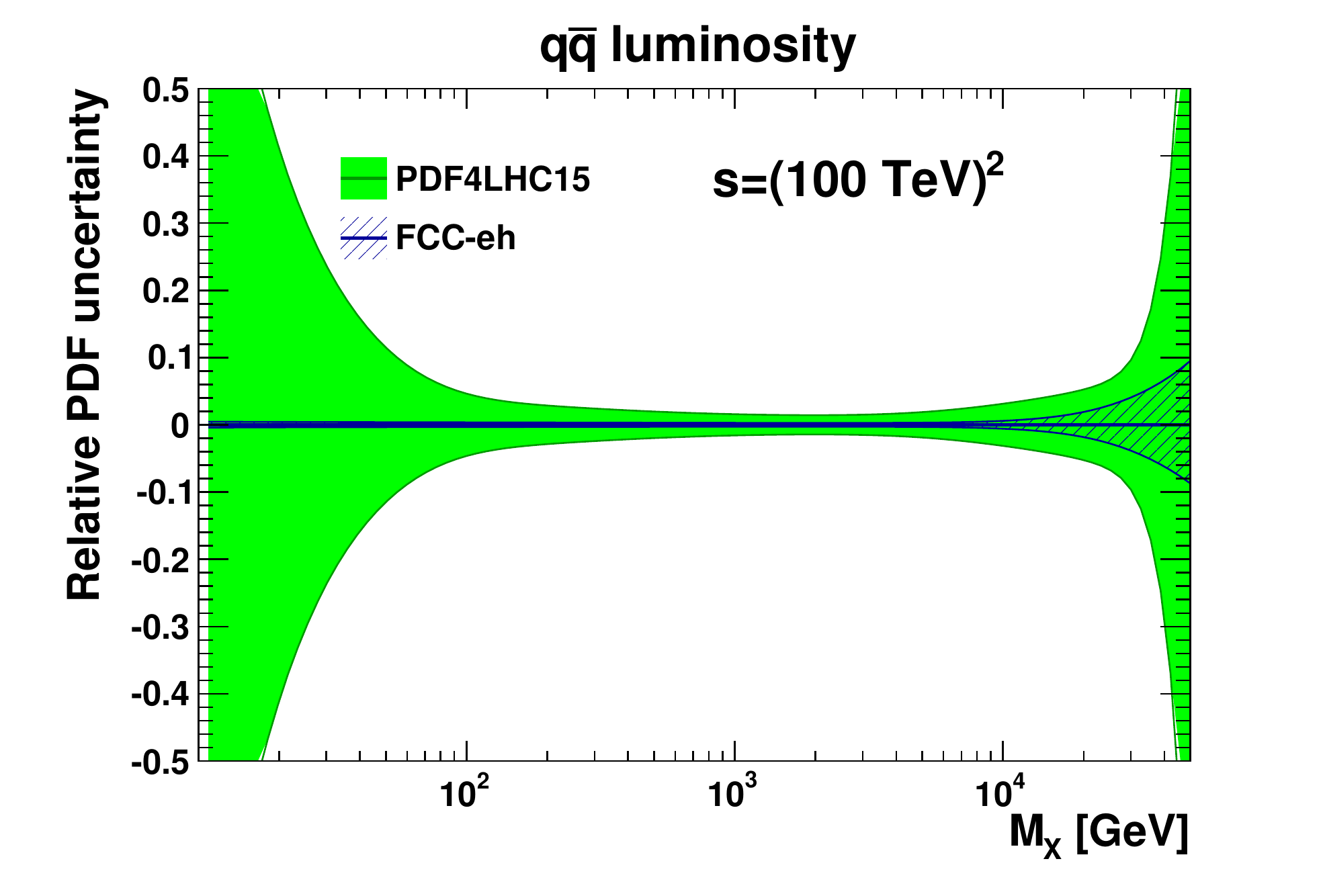}\includegraphics[width=0.5\textwidth]{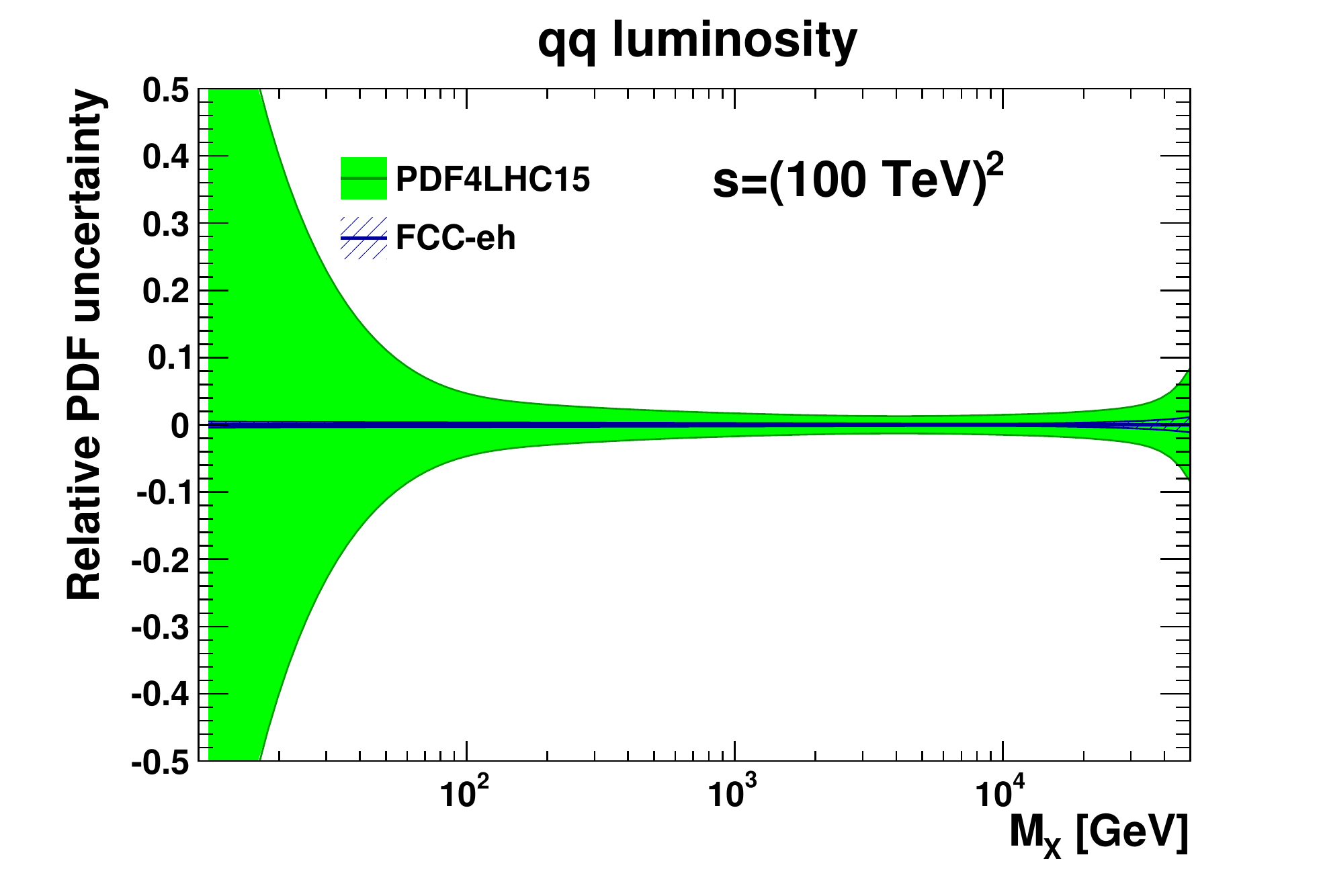}
\caption{{Relative PDF uncertainties on parton-parton luminosities from
the PDF4LHC15
and FCC-eh PDF sets, as a function of the mass of the produced heavy
object, $M_X$,
at $\sqrt{s} = 100$ TeV. Shown are the gluon-gluon (top left),
quark-gluon (top right), quark-antiquark (bottom left) and quark-quark
(bottom right) luminosities. The LHeC expectation is very similar but
misses one order of magnitude towards low $x$.}}
\label{figlumisep}
\end{figure}

 Currently, PDF uncertainties account mostly for the propagated statistical and systematic  errors  on  the  measurements  used  in  their  determination. Clearly, the missing higher-order uncertainties  from the truncation of the QCD perturbative expansion also affect predictions for the various processes that enter the PDF determination. The    PDF extraction that systematically accounts for these effects  in  QCD calculations   and assesses  their impact on the  uncertainties of the resulting PDFs, has been recently considered~\cite{AbdulKhalek:2019bux}.
%
The ultimate method to determine the parton structure of the proton is with high-precision, high-energy measurements in electron-proton
scattering. The LHeC and FCC-eh provide a complete resolution of PDFs, of both protons and ions,  as has been described recently in \cite{Ides159,Klein:2018rhq}. The high $x$ region, crucial for searches at $pp$ colliders,  would be clarified (the valence quark and small but important sea and gluon contributions) thanks to the high luminosity and large $Q^2$ lever arm of the measurements. At medium $x$ sub-percent precision for N$^3$LO PDFs is reached due to the unique measurement techniques of electrons, neutrinos and the hadronic final state. Flavour is completely resolved owing i) to the high energy enabling to access charged current cross sections for many orders of magnitude in $x$ and
ii) to the direct measurements of strange, charm and beauty densities. Finally, the long standing quest of a possible saturation of the gluon density at small $x$ and the existence of the  non-linear parton evolution 
will be
solved at the LHeC. All of this maybe illustrated, in a yet simplified manner, with the ultimate accuracy one would achieve with LHeC (and FCC-eh in an extended range) for the various parton-parton luminosities, see
Fig.~\ref{figlumisep}.

Of principal importance is to disentangle predictions and effects from QCD/PDFs from possible new physics. DIS and LHC would provide two independent configurations for achieving this while $pp$ alone could   run into the conceptual problem of possibly hiding new phenomena  behind
PDF or QCD uncertainties as
PDF or QCD effects. In nuclei one needs to disentangle nuclear environment (shadowing) from non-linear (gluon saturation) effects. All this requires very high precision, very high energy, large luminosity $ep$ and $eA$ collision data taken while LHC is operational (and later synchronously through FCC-hh and FCC-eh operation). 

Given the high precision expected 
at the HL-LHC,
 it will be crucial to include all sources of experimental, methodological, and theoretical uncertainties associated with PDFs in order to ensure robust predictions. In particular,
the impact of theoretical uncertainties due to  missing higher-order corrections in theoretical calculations,  the fits with high-energy and threshold logarithms resummed and    the contribution of  electroweak corrections  should    be analyzed and included in the determination of  global PDFs in the foreseeable future. Furthermore, as already pointed out in in Sect.~\ref{section_low},  a  precision physics programme at future hadron colliders requires    a more detailed  description  of  
the partonic substructure of hadrons   as encoded  e.g. in GPDs or  TMDs. Theoretically challenging is also the phenomenon  of  saturation of partonic densities  at small enough values of  the  fraction  of  momentum $x$, which has  developed into a complete and coherent formalism of the Colour Glass Condensate~\cite{Iancu:2002xk,Petreska:2018cbf,Altinoluk:2019fui}. Furthermore,  a reliable determination     of the parton distribution functions of nucleons bound within nuclei (nPDFs),  is particulary relevant for precision phenomenology and   fundamental understanding of the strong interactions in the nuclear environment~\cite{Kusina:2017gkz,AbdulKhalek:2019mzd}.

\subsection{ Numerical Lattice  QCD}
\label{si-LQCD}

Many of the SM  predictions require the knowledge  of  parameters and observables  which encode nonperturbative QCD effects.  They can only be calculated from first principles  by using  LQCD~\cite{Aoki,Bazavov:2019lgz,Cirigliano:2019jig}.

Over the past years, an increased computing power, together with   the development of better algorithms and  analytical frontiers  techniques have enabled realistic LQCD predictions   with controlled errors. LQCD allows for a  precise  determination of a wide range of hadronic observables, including the hadron masses and  decay constants, form-factors and mixing parameters characterising weak-decay amplitudes,  PDFs, as well as key SM parameters such as quark masses and the QCD coupling~\cite{Aoki}.
LQCD    provides   the most precise determination of $\alpha_s$ 
~\cite{Patrignani:2016xqp,Aoki:2016frl},  and  high-accuracy    results for  the  $D$, $D_s$, $B$, and $B_s$ heavy flavour meson decay constants  with sub percent precision~\cite{Bazavov:2017lyh}. Similar precision  has been reached in the determination of the  light ($u,d,s$) and heavy flavour ($c,b$) quark  masses~\cite{Bazavov:2018omf}.
 Further  LQCD results on  $B\to \pi$ and  $B\to K$ semileptonic form factors, neutral kaon mixing  and neutral $B_d$- and $B_s$-meson  mixing~\cite{Dowdall:2019bea}  led to significant improvements in the determination of the Cabibbo-Kobayashi-Maskawa   quark-mixing matrix elements and the global unitarity-triangle fit.
The  unprecedented  accuracy  of  lattice calculations  and  state-of-the-art  techniques    allowed one to extract the nucleons  electromagnetic and axial  form  factors and their electromagnetic radii and the magnetic  moments~\cite{Alexandrou:2018sjm,Alexandrou:2018lvq,Djukanovic:2019jtp}.
Inspired by the LHCb discovery of  a  new  narrow  charmonium  state,  the $X$(3842)~\cite{Aaij:2019evc}, LQCD presented  calculations   of   charmonium states  near the open-charm threshold by   means of  the  L\"uscher  formalism, and found  the charmonium resonance with $J^{PC}=3^{--}$ and the mass consistent with the $X$(3842)~\cite{Aaij:2019evc}.  The  recent  LQCD  calculations  of exotic states in  QCD,  also predicted    the existence of a  doubly-bottom $(\bar b\bar b ud)$ tetraquark bound state that is stable under the strong and electromagnetic interactions~\cite{Piemonte:2019cbi}.

In the  ongoing search for  BSM physics,
LQCD   provides results  for  the  nucleon  scalar and tensor charges~\cite{Aoki,Alexandrou:2017qyt}, as well as demonstrates   the feasibility of  computing the amplitude of the rare kaon decay $K^+\to \pi^+ \nu\bar\nu$~\cite{Bai:2018hqu}, a process   that is experimentally  studied  by   the NA62 experiment at the CERN SPS. LQCD has also addressed the issue of   CP violation in the QCD sector.   Interesting results are obtained  in recent LQCD calculations  for  the EDM  of the nucleon induced by the QCD $\theta$ term~\cite{Dragos:2019oxn}. The  results  indicate  that the EDM of the nucleon stays finite in  a continuum and the chiral limit.  This result together with the experimental bound on    the neutron EDM provide the  upper limit     for  the value of QCD $\theta$ term.   
Enormous progress has also been achieved in ab-initio  lattice  calculation of  the strong interaction contribution $a_\mu^{\rm had}$  to the anomalous magnetic moment of the muon, $g-2$~\cite{Meyer:2018til,Westin:2019tgc}.  LQCD
provides  results on   the   hadronic  vacuum polarisation (HVP) and hadronic light-by-light (HLbL) scattering contributions to $a_\mu^{\rm had}$. While the  error of current LQCD estimates of the HVP reached  the few-percent level, the 
 uncertainty in the  determination of HLbL is  at the $(10-15)\%$-level. The goal for the future LQCD calculations is  the sub-percent precision  of a HVP contribution  which is needed to reduce the SM uncertainties  on the value of  $g-2$ to possibly  identify  new physics  by   comparison to the  experimental data.

LQCD methods are also very successful  in describing  thermodynamic   properties
and structures of the strongly interacting matter under extreme conditions of high temperature $T$  and  density $\rho$~\cite{Bazavov:2019lgz,Andronic:2017pug,Soltz:2015ula,Ding:2015ona,Ratti:2018ksb}.  Such a QCD matter  is produced in ultra-relativistic $AA$ collisions, $pA$ and possibly  even   in  high-energy and high-multiplicity events in $pp$ collisions~\cite{Citron:2018lsq}.  LQCD   describes   the phase structure of  strongly interacting matter, the  nature of QCD  phase transition and the equation of state. Large-scale computing has allowed one to quantify  the value and the shifts of the chiral critical temperature from the  chiral limit  to the physical point. The shift of the pseudocritical temperature with finite chemical potential  $\mu_B$ has also been  recently established up to $(\mu/T)^4$ order~\cite{Bazavov:2018mes}  (shown in Fig.~\ref{fig:Tmub}). LQCD provides  calculation of  correlations, diffusion and transport properties  in  a QGP and  fluctuation observables~\cite{Soltz:2015ula,Friman:2011pf,Karsch:2010ck}  that are directly linked to experimental data. LQCD predictions are essential  inputs to hydrodynamic-   and transport-model description of experimental data   in $AA$ and  $pA$   collisions~\cite{Citron:2018lsq}.

Continued efforts  and support in developing new theoretical methods and better algorithms are needed in  LQCD  to reach the  anticipated progress and  precision   of  SM  predictions  and to  sharpen  the opportunity for new physics  discovery.  In high-density QCD, more detailed studies close to the chiral limit and on large lattices are needed to reach  definitive  conclusions  on  the  role  of the  $U_A(1)$ symmetry  breaking   and the order of the chiral phase transition. To  have  impact  on  the experiments in  future,  LQCD  calculations  of  electromagnetic  probes, fluctuation observables, spectral functions   and transport properties   of  QGP  need  to be carried out with high statistics on large lattices and with physical dynamical quarks. The problem of the finite chemical potential in LQCD calculations needs further attention, to develop an efficient  formalism that allows for lattice simulations
with a complex action.

\subsection{The strong coupling constant $\alpha_s$}


The strong coupling is the least known fundamental coupling in the Standard Model. This becomes a hindrance for precision measurements, such as of the Higgs production cross sections and couplings, and it impacts the uncertainty budget in predictions of electroweak vacuum stability and grand unification theories approaching the Planck scale.
Currently, the LQCD results are more precise than the values directly derived from experiment as summarised recently in Refs.~\cite{Bethke:2017uli,dEnterria:2019its} and shown in Table~\ref{tabl}. 

%
%
\begin{table}[h]
\caption{
World-average values of the strong coupling constant at the $Z$ mass,
$\alpha_s(m_Z)$, at NNLO accuracy for 5 active
flavours~\cite{Tanabashi:2018oca}.
}
\begin{center}
\begin{tabular}{|c|c|}
\hline
Method & $\alpha_s(M_Z^2)$ \\
\hline
LQCD & $ 0.1188 \pm 0.0011$ \\
$\tau$-decays & $ 0.1192 \pm 0.0018$ \\
DIS & $ 0.1156 \pm 0.0021$ \\
Hadron Collider & $ 0.1151 \pm 0.0028$ \\
Electroweak Fits & $ 0.1196 \pm 0.0030$ \\
$e^+e^-$  & $ 0.1169 \pm 0.0034$ \\
\hline
\end{tabular}
\label{tabl}
\end{center}
\end{table}

New measurements of $\alpha_s$ at the level of
per mille accuracy are mandatory and could be achieved with FCC-ee from the ratio of the hadronic-to-leptonic
width at the Z-pole and with LHeC/FCC-eh from the scaling violations of
the structure functions.
Both demand a new level of theoretical support to
develop theory one order of perturbation further. Both also require
extreme experimental care and additional cross
checks, e.g.\ from the $W$ in $e^+e^-$ and from jets in $e^+e^-$ and DIS,
respectively, to assure a maximum of confidence and enhanced precision.
The realisation of this complementary programme would be a major
milestone in the development of QCD and in the  reduction of important parametric uncertainties today in key SM and BSM theoretical calculations.
It would also be a
major triumph of experimental physics and its intimate collaboration with theory. The challenge of LQCD at this high level of accuracy will surely lead to new insight into the details~\cite{Aoki:2016frl} of
LQCD calculations as well.

\subsection{Low-energy precision QCD}

 
 Many aspects of QCD   are important for experiments at low energies and, vice versa, various experiments at low energies yield precision QCD benchmarks~\cite{Alemany:2019vsk,Dainese:2019xrz}.

The strong CP problem~\cite{Cheng:1987gp,Pospelov:2005pr}, i.e.\ the extreme smallness of the  {\it $\theta$-term} in the QCD Lagrangian, is evident from the non-observation of permanent hadronic electric dipole moments (EDM), in particular of the neutron and of $^{199}$Hg. Considerable international efforts aim at improving the
experimental sensitivity of the
 permanent EDM of the neutron, of heavy nuclei and recently also of the deuteron, the proton and heavier baryons ~\cite{Engel:2013lsa,Semertzidis:2003iq,Afach:2015sja,Lenisa:2017okq,Chupp:2017rkp}. The most sensitive and straightforward to interpret is the neutron EDM. Experimental sensitivities should improve by two orders of magnitude to a few $10^{-28}\,$e$\cdot$cm in the next decade. Further improvements will require R\&D into new experimental concepts to overcome statistical and systematics limitations. Similarly, the EDM of the proton can be searched for with a dedicated storage ring experiment, not statistically limited
till
$10^{-29}\,$e$\cdot$cm. R\&D, precursor experiments as, e.g.\ at COSY, and prototyping can pave the way to understand and tackle the corresponding systematic issues~\cite{Lenisa:2017okq}.

Besides EDMs, several finite electromagnetic form factors of the nucleons provide benchmarks for low-energy QCD studies:
\begin{itemize}
\item        Magnetic moments of proton, antiproton and neutron are measurable at much higher precision~\cite{Mooser:2014vla,Smorra:2018syp,Smorra:2016vxa,Afach:2014fha} than calculated by theory. Nevertheless, they are important for metrological reasons and strong tests of CPT and other fundamental symmetries. The AD and ELENA at CERN are essential facilities as are the neutron EDM experiments, in Europe at PSI and at ILL.

\item        The proton rms charge radius continues to puzzle, with discrepant results from muonic hydrogen spectroscopy at PSI, ordinary hydrogen spectroscopy and $ep$ scattering~\cite{Antognini:2015moa,Pohl:2010zza,Antognini:2005fe,Bernauer:2010wm,Antognini:1900ns}. Considerable theoretical and experimental efforts, notably also $\mu$-$p$ scattering with MUSE at PSI and potentially COMPASS$^{++}$ at CERN~\cite{Gilman:2013eiv,Denisov:2018unj}, have emerged aiming at resolving the puzzle.  Eventually LQCD calculations will contribute to corroborating the {\it true} value. The proton Zemach radius and the magnetic radius are targets of next generation muonic hydrogen experiments measuring the ground state hyperfine splitting. Experiments are planned for PSI, Riken-RAL, and J-PARC.

\item        The nucleon axial form factor $g_A$ at lowest momentum transfer is determined by measurements of neutron decay correlations, in particular at ILL~\cite{Markisch:2018ndu}. An order of magnitude improvements can be envisaged with a new generation of experiments, in particular at a new fundamental physics beamline at the ESS.

\item        The nucleon axial radius can be determined by muon capture on the proton~\cite{Hill:2017wgb} yielding complementary information to neutrino scattering and potentially crucial input to long baseline neutrino experiments.
\end{itemize}

QCD input is also essential in order to allow for comparisons between high-precision experiments and precision SM calculations. The anomalous magnetic moment $(g-2)$ of the muon is an example where substantial theoretical and experimental efforts are presently underway in order to improve the experimental sensitivity and the accuracy of the QCD corrections. The MUonE experiment  aims at determining the leading order hadronic contribution to the muon $g-2$ by measuring the hadronic part of the photon vacuum polarisation in the spacelike region. 

The  experimental studies of  strong force and  the structure and dynamics of atomic nuclei is a vital part of strong interaction physics.
The proposal revolving around the Di-meson Relativistic Atom Complex (DIRAC++)  aims to check
low-energy QCD predictions using double-exotic $\pi\pi$ and $\pi K$ atoms 
to  gain a deeper insight into the quantum theory of the strong force \cite{Adeva:2018fwr,DIRAC:2016rpv}.
The precision measurements of QCD are one of the research directions intensively
pursued at the CERN ISOLDE facility. The facility is unique worldwide using
energetic 1.4 GeV protons on thick targets, in combination with the Isotope
Separation On-Line method to produce pure low-energy (50 keV) radioactive ion
beams (RIB's) of more than 1000 different radioactive isotopes. Research is
performed in more than a dozen of permanent experimental set-ups for studies of
the structure of atomic nuclei, and for a variety of other studies in the fields
of astrophysics, medical and materials research and physics beyond Standard
Model. Phase 2 of its HIE-ISOLDE upgrade has reached completion, which permits most
of ISOLDE's radioactive isotopes to be re-accelerated to energies up to 9.2 MeV
per nucleon. Using 3 dedicated experimental set-ups these beams are used to
study a variety of nuclear reactions with radioactive isotopes, opening up new
possibilities for nuclear structure and astrophysics research. The intensity
upgrade and a final energy upgrade to reach 10 MeV/u for all isotopes is
foreseen for Phase 3. A further upgrade, including the construction of two
additional target stations and a compact low-energy storage ring are proposed as
part of the EPIC project. This upgrade will allow ISOLDE to remain at the
forefront of nuclear and astrophysics research for another few decades.

\section{QCD and other disciplines}



The LHeC and FCC-eh are the highest energy, highest electron current applications of energy recovery linac (ERL) technology. In itself, ERL entails major technology innovations, most obviously on high-quality, superconducting cavities such as those required for the FCC-ee. 
ERL reaches efficiencies in excess of 90\% reducing the required power, as for the  LHeC 60 GeV electrons, from a GW to below 100 MW. By decelerating the beam after the interaction, the power is taken to supply the accelerating part of the ERL while the beam energy is eventually reduced to the injection energy. 
The dump of ${\cal O}(500 {\rm MeV})$  is environment friendly, in contrast to proton or electron high-energy, radioactive beam dumps. 
In the context of the PERLE facility at Orsay (submission ID147), the first 802 MHz prototype was built for both LHeC and FCC-ee.
These facilities have major technical applications such as for lithography,
through their laser
application, or material tomography and photofission, through the intense ,
mono-energetic photon beams, generated by laser backscattering. Therefore, a new
facility (or facilities) may be established that also open new avenues to
photonuclear and particle physics, in complementarity with ELI-NP. This is a one
of the best examples of a direct synergy between energy frontier particle, low-energy nuclear and applied physics.

The low radiation and zero pile-up environment of $ep$ colliders, combined with the demand for very high quality vertex tagging makes detectors at these facilities a most suited application for an HV CMOS silicon vertex detector, of low material and high integration level.

Laser spectroscopy of light muonic atoms has been established as a game changer in the determination of nucleon/nuclear charge radii. For these, and for the determination of Zemach radii and other spectroscopy goals in those exotic systems, the development of high power, pulsed lasers, stochastically triggerable with low latency is key.

The development of higher intensity, higher quality, low-momentum muon beams of both polarities is an essential part to further improve the measurements of the muon anomalous magnetic moment, the proton radii, and the nucleon axial radius.

\subsection{QCD and astroparticle physics}
The production, propagation, and detection of the most energetic particles produced in the Universe, in yet unravelled sources, relies heavily on our understanding of QCD at high energies. For both charged cosmic rays (CRs) and neutrinos, cross sections linked to QCD processes in the non-perturbative and perturbative regimes are still the largest source of theoretical uncertainty. Dedicated measurements at FCC-pp, FCC-ep, and LHC-FT, in particular with heavy- and light-ions collisions, are needed to improve the QCD-based simulations of relevance for astroparticle physics:
\begin{itemize}
\item [$-$] Direct measurements of the production cross section of low-energy secondary particles, such as anti-nuclei, are of prime importance to determine the amount of anti-matter produced in standard astrophysical sources, which is the background for possible dark-matter signals in the cosmic-ray flux measured in satellite experiments.
\item [$-$] At CR energies above 10$^{15}$~eV, the flux of charged cosmic rays is too low for direct detection in satellites, and the air showers produced by their interaction with air molecules in the Earth's atmosphere are used instead~\cite{Kampert:2012mx}. A precise simulation of the properties of such air showers is needed to properly determine the mass of the primary CR. The shower development is driven mostly by hadron-nucleus interactions from the highest (1000~TeV c.m.\ energy) to the lowest (10~GeV lab energy) energies whose theoretical description relies heavily on collider and fixed-target data~\cite{dEnterria:2011twh}. Key detection techniques such as the measurement of the shower maximum position are very sensitive to the first hadronic interactions. The current theoretical uncertainty being even higher than the experimental one, direct particle production measurements at HL-LHC~\cite{Citron:2018lsq} (in particular with light ion beams) and FCC-pp~\cite{Mangano:2016jyj,dEnterria:2016oxo} would reduce significantly the uncertainties due to the extrapolations to high energies. If muons are used to measure the air shower properties, the data are even in contradiction with the simulations 
at the highest CR energies~\cite{Dembinski:2019uta}. The muon production in air showers is sensitive to all hadronic interactions (all energies) in the shower, and in particular to forward particle production, dominated by small-$x$ QCD phenomena that can be carefully studied at FCC-ep. In addition, final-state effects (such as collective parton hadronisation, leading e.g.\ to enhanced strangeness production) observed at the LHC in light systems may help solve the muon puzzle~\cite{Pierog:2019}, and new measurements are needed at higher c.m.\ energies.
\end{itemize}

\subsection{QCD and neutrino physics}
The understanding of the production of neutrinos in hadronic collisions at low and high energies will  highly benefit from dedicated QCD measurements at future collider facilities:
\begin{itemize}
\item [$-$] Very high-energy neutrinos detected by IceCube, and the future KM3NeT, are a key element of the multi-messenger detection of astrophysical objects such as black holes and neutron star mergers. A very precise knowledge of typical sources of energetic decay neutrinos, such as forward charm production, as well as of nuclear PDF at very small $x$, are required to understand the atmospheric neutrino background and the neutrino interaction in Earth allowing its detection.
\item [$-$] The new generation of low-energy neutrinos experiments such as DUNE or Hyper-\-Kamiokande, built to solve the mass hierarchy and CP violation in the neutrino sector, require very low systematic uncertainties in their theoretical production cross sections. Dedicated high-statistics studies of  hadronic interactions at accelerators and atmospheric observatories, are needed to provide a precise understanding of neutrino sources in the decay of primary and secondary particles.
\end{itemize}

\noindent\section{Overview and perspectives for  QCD}
\vskip 0.3cm

\noindent\textbf{Precision QCD program:} 
A globally concerted precision QCD research programme provides
 a unique avenue to support the search
for new physics
beyond the Standard Model.  A high-luminosity $e^+e^-$
collider at the EW scale and a high-energy ep
collider provide an excellent experimental  environment for high-precision QCD studies ($\alpha_s$, parton radiation, fragmentation and
hadronisation, higher-order  perturbative and  non-perturbative parton dynamics) which are essential in support of  our aspirations in particle physics.
\\

\noindent \textbf{Hadronic structure program:}
A hadronic structure research programme exploring the complementarity of $pp/ep/eA$ colliders and fixed-target facilities provides vital ingredients for the high-precision exploration in
searches for new physics and provides as well unique steps into unknown territories of QCD.
\\

\noindent\textbf{Hot and dense QCD program:} 
A high-energy $AA/pA/pp$ 
research programme  at the LHC, HL-LHC, HE-LHC and  FCC   
with newly designed detectors in the collider and fixed-target modes  is 
unique and provides essential
science at the frontline towards a profound understanding  of {\it hot and dense QCD matter.} 
A coherent  research programme of QCD matter at the SPS,
is complementary to other emerging facilities worldwide like BNL/BES, FAIR/CBM,  JINR/NICA  or J-PARC, and brings valuable and unique contributions in the exploration of the QCD phase diagram.   
\\

\noindent\textbf{QCD theory community:} It is
{\it essential} to support coherently the QCD theory community to succeed in the above programs and to link QCD
to
frontier research  in particle  and nuclear physics. This also 
 concerns  strong community  support of developments of general purpose Monte Carlo event-generators and numerical LQCD. 
\\ 

\noindent\textbf{Organization:} Strengthening the synergies in research and technology with adjacent fields has the potential to reinforce our efforts. Global platforms, networks and institutes have the potential  to enhance the research exchange among experts worldwide and to provide essential training opportunities.



\chapter{Flavour Physics}
\label{chap:flav}

This chapter discusses the present and future flavour quest, focusing first on the physics of the spectrum lighter than $\sim$\,GeV, and then on the heavier fermions, 
 the heavy SM bosons, and finally the flavour-dark sector connection. The terminology short-term, mid-term and long-term will denote, respectively, present experiments, updates of the latter (e.g.\ LHCb Upgrade~II) as well as approved projects (e.g.\ Belle II, HL-LHC, Mu3e, etc.), and future facilities under discussion.  

\section{Introduction/Theory of Flavour}
That fundamental forces arise as gauge interactions is experimentally an extremely successful prediction of the
Standard Model (SM) of particle physics. 
In contrast, within the SM 
a deep understanding of the rationale for the Higgs couplings and the flavour structure remains to be achieved.  There is a striking difference between the simplicity of the gauge sector, described by just three gauge couplings, and the complicated structure of the rest of the SM with over twenty Higgs related parameters describing the SM flavour structure. This  suggests that flavour physics
is  a unique portal to 
a more fundamental organizing principle.

Further fundamental questions emerge when attempts are made to 
couple the SM and gravity. The quantization of the latter remains a crucial open question. In addition, the data 
 point towards a Universe that cannot be understood solely in terms of the SM and gravity. 
There is strong evidence  for the existence of new particles and {\it new \underline{particle} physics} (NP): 
\begin{itemize}
    \item {\it Experimental evidence} which remains unexplained within the SM laws, to wit 
    \begin{enumerate}
        \item {\it Dark matter}, whose nature is unknown, and the mass range of possible dark matter candidates spanning across over eighty orders of magnitude. 
        \item {\it Neutrino masses}. The SM gauge group allows for Majorana neutrino masses, justifying their suppressed values, but the size of the putative Majorana
 scale is unknown.
        \item The observed {\it matter-antimatter asymmetry} of the Universe. 
    \end{enumerate}
%
    
    \item {\it Strong tensions and fine-tunings within the SM}, such as   
    \begin{enumerate}
        \item {\it The electroweak (EW) hierarchy problem}. ``Why is the Higgs so light?''  when its mass is {\it a priori} sensitive through quantum corrections to any putative NP scale higher than the EW one.  What stabilises  the Higgs vacuum expectation value? 
        \item {\it The strong CP problem}, that defines the QCD vaccuum.  Why is its $\theta$ parameter experimentally constrained to be extremely small? For {\it a priori} no good reason.
        \item {\it The flavour puzzle}. Why are there three generations of quarks and leptons? What accounts for the very different masses and mixings? What fixes the  
        size of CP-violation, largely insufficient to explain the observed dominance of matter over anti-matter?
    \end{enumerate}
           \end{itemize} 
The flavour puzzle, in particular, feeds into the first two tensions. For instance, within the SM the top loop gives the main contribution to the EW hierarchy problem, while the strong CP problem is an issue only in as much as all the quarks have non-zero masses. Furthermore, many NP models designed to solve the EW hierarchy problem tend to worsen the strong CP problem and generate unacceptably large contributions to electric dipole moments (EDMs), 
    as a consequence
    of the presence of CP-violation in  non-chiral flavour changing couplings.  
      All three tensions in their core amount 
 to the question of why certain parameters are very small. 
 In natural theories small numbers are explained by  symmetries or dynamical assumptions, suggesting that the SM needs to be extended in order to become a natural theory.

     The underlying nature of CP violation, which is at the heart of many open questions,
     deserves  
     special mention. On the one hand, the  combination of the discrete symmetries C, P and T is essential to the formulation of quantum field theory itself. On the other hand, CP violation  is at the backbone of the SM three-family flavour puzzle and of the strong CP problem. In addition,  it 
     is also an essential ingredient to generate the observed baryon asymmetry (assuming baryogenesis). From a 
     practical perspective, it is one of the main driving forces behind the present experimental efforts, especially in the neutrino sector. Finally,  dark matter itself may have flavour structure,   and 
     a true understanding of flavour would then require an interdisciplinary exploration. As a side benefit,  
      the present and planned flavour experiments  are  often, without special requirements, 
      sensitive to light dark matter candidates such as feebly interacting particles.
 
\begin{figure}
\begin{center}
\includegraphics[width=0.90\textwidth, angle=0]{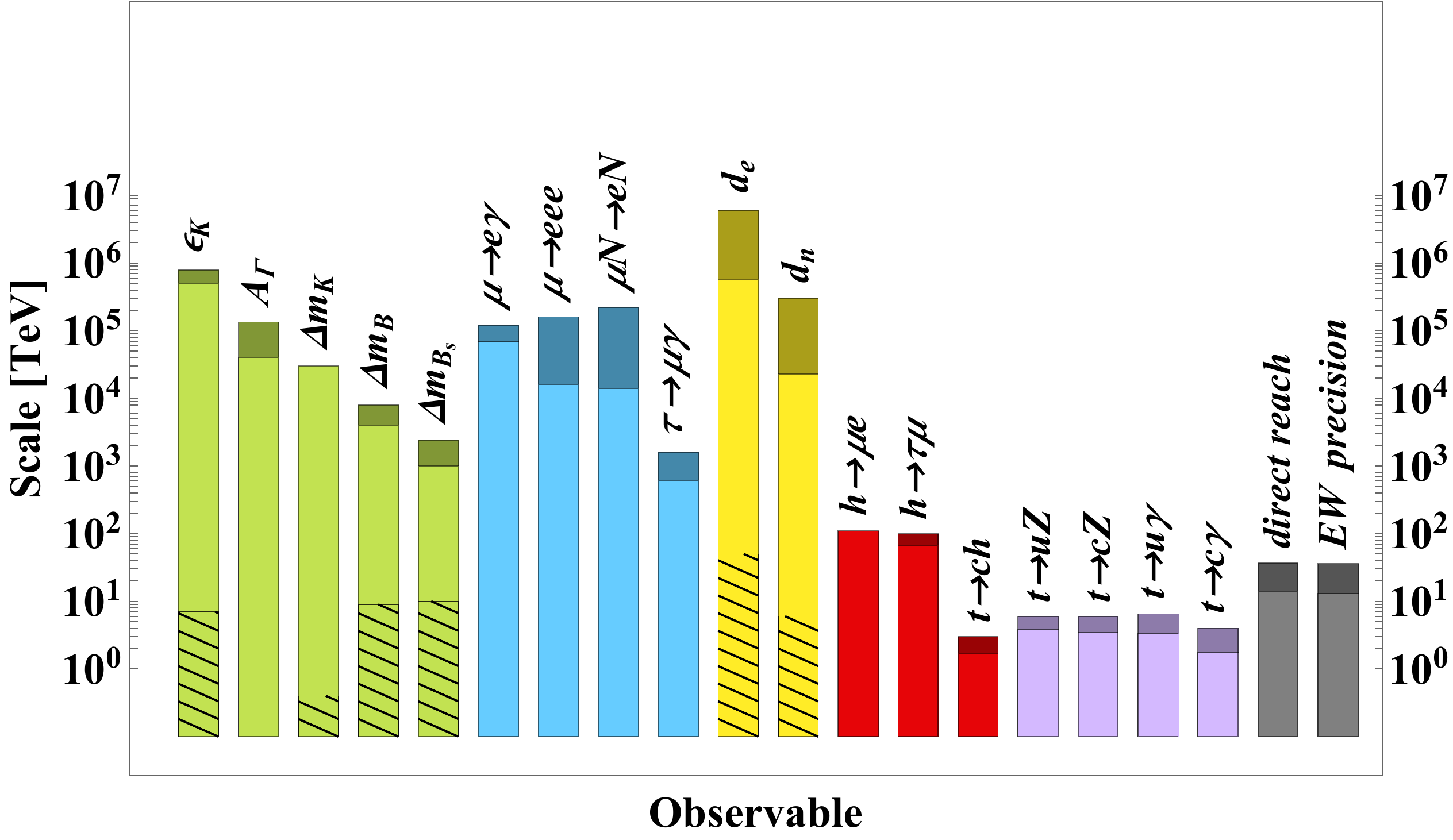}
\end{center}
\caption{Reach in new physics scale of present and future facilities, from generic dimension six operators. Colour coding of observables is: green for mesons, blue for leptons, yellow for EDMs, red for Higgs flavoured couplings and purple for the top quark. The grey columns illustrate the reach of direct flavour-blind searches and EW precision measurements.
The operator coefficients are taken to be either $\sim 1$ (plain coloured columns)
or suppressed by MFV factors (hatch filled surfaces). Light (dark) colours correspond to present data (mid-term prospects,  including HL-LHC, Belle II, MEG II, Mu3e, Mu2e, COMET, ACME, PIK and SNS).
}\label{fig:NPscales}
\end{figure}
  
The progress in understanding the above fundamental questions can be made through a variety of tools: directly by  increasing the energy at which  the world of fundamental particles and forces is explored, or indirectly by making precise measurements of rare or even SM forbidden processes, relying on quantum mechanical effects to probe shorter distances or effectively higher energies.  
The expected experimental progress, especially with regards to the indirect probes,  can be  neatly encoded in the model-independent tool of effective Lagrangians. As long as the NP particles are heavier than the energy released in a given experiment, their impact can be included via effective operators of increasing mass dimensions, constructed from the SM fields. The resulting effective field theory (SM-EFT) has the following form: 
\begin{equation}
    \mathcal{L}_{\text{eff}}= \mathcal{L}_{\text{SM}} + \frac{C_5}{\Lambda_M}\mathcal{O}^{(5)} +  \sum_a \frac{C_{6}^{a}}{\Lambda^2}\mathcal{O}^{(6)}_{a}+\cdots \,. 
    \label{effective}
\end{equation}
The dimension five ($d=5$) operator $\mathcal{O}^{(5)}$  
breaks lepton number and, if present, induces Majorana neutrino masses of order  $v^2/\Lambda_M$, where  $\Lambda_M$ is assumed to be 
much larger than the electroweak (EW) scale $v$. The $d=6$ operators ${O}^{(6)}_{a}$  
encode the effects of  
NP particles of generic mass $\Lambda$. 
Experiments probe the ratios $C^a / \Lambda^2$. 

For a qualitative appraisal, Fig.~\ref{fig:NPscales} illustrates the scales probed  by the present flavour  experiments (light colours) and mid-term prospects, assuming  $C_6^a\sim {\mathcal O}(1)$~\cite{Nir_Private}. This can be compared with the reach of direct high-energy searches and EW precision tests (in grey), illustrated  by using flavour-blind operators that have the optimal reach~\cite{Nir_Private}: the gluon-Higgs operator and the oblique parameters for EW precision tests, respectively.  The shown effective energy reach of flavour experiments do have several caveats. First of all, in many realistic theories either the coupling constants are  smaller than unity and/or the symmetries suppress the sizes of the coefficients.  This effect is illustrated by including in the quark sector the present bounds in tree level NP with Minimal Flavour Violation (MFV) pattern of couplings (hatch filled areas)~\cite{Chivukula:1987py,Buras:2000dm,DAmbrosio:2002vsn,Cirigliano:2005ck}. Furthermore, there could be cancellations among several higher-dimension operators. In addition, for theories in which the new physics contributes as an insertion inside a one-loop diagram mediated by SM particles,  
all the shown scales should be further reduced by extra
GIM-mass suppressions and/or
a factor $\alpha /4\pi \sim10^{-3}$ 
(where $\alpha$ denotes the generic gauge structure constants).

Finally and importantly, the  new physics scale behind the flavour paradigm may differ from the  electroweak new physics scale.  Despite these caveats, Fig.~\ref{fig:NPscales} does illustrate the unique power of flavour physics to probe NP.
 The next generation of precision particle physics experiments will probe significantly higher effective NP scales, as discussed in more detail below. 

\section {Light sector: spectrum below GeV (short-, mid- and long-term)}
\subsection{Electric dipole moments ($n$, $p$, charged leptons, atoms)}

Electric dipole moments  are the P- and CP-odd  
counterparts (from CPT conservation) to magnetic moments. They arise from a set of  ${O}^{(6)}_{a}$ operators in Eq.~(\ref{effective}) which flip the chirality of a fermion  and involve the EW field strengths and the Higgs doublet. After spontaneous symmetry breaking they result in a coupling 
of the fermion spin $\vec{\sigma}$ to the electric field $\vec{E}$ of the form $d\,\vec{\sigma} \cdot \vec{E}$, where $d$  denotes the  EDM strength.

\begin{figure}
\begin{center}
\includegraphics[width=0.43\textwidth]{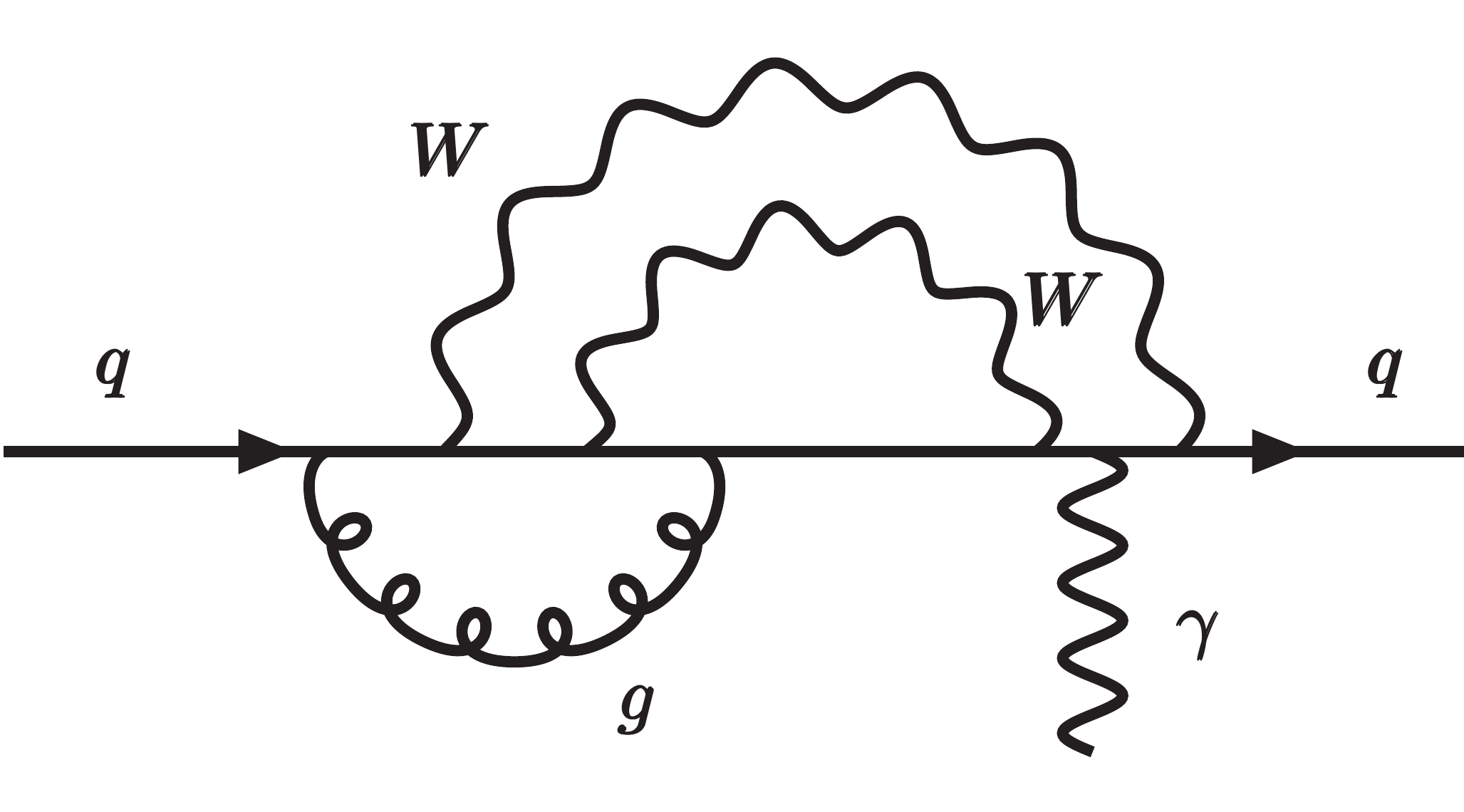} \hspace*{5mm}
\raisebox{10mm}{\includegraphics[width=0.37\textwidth]{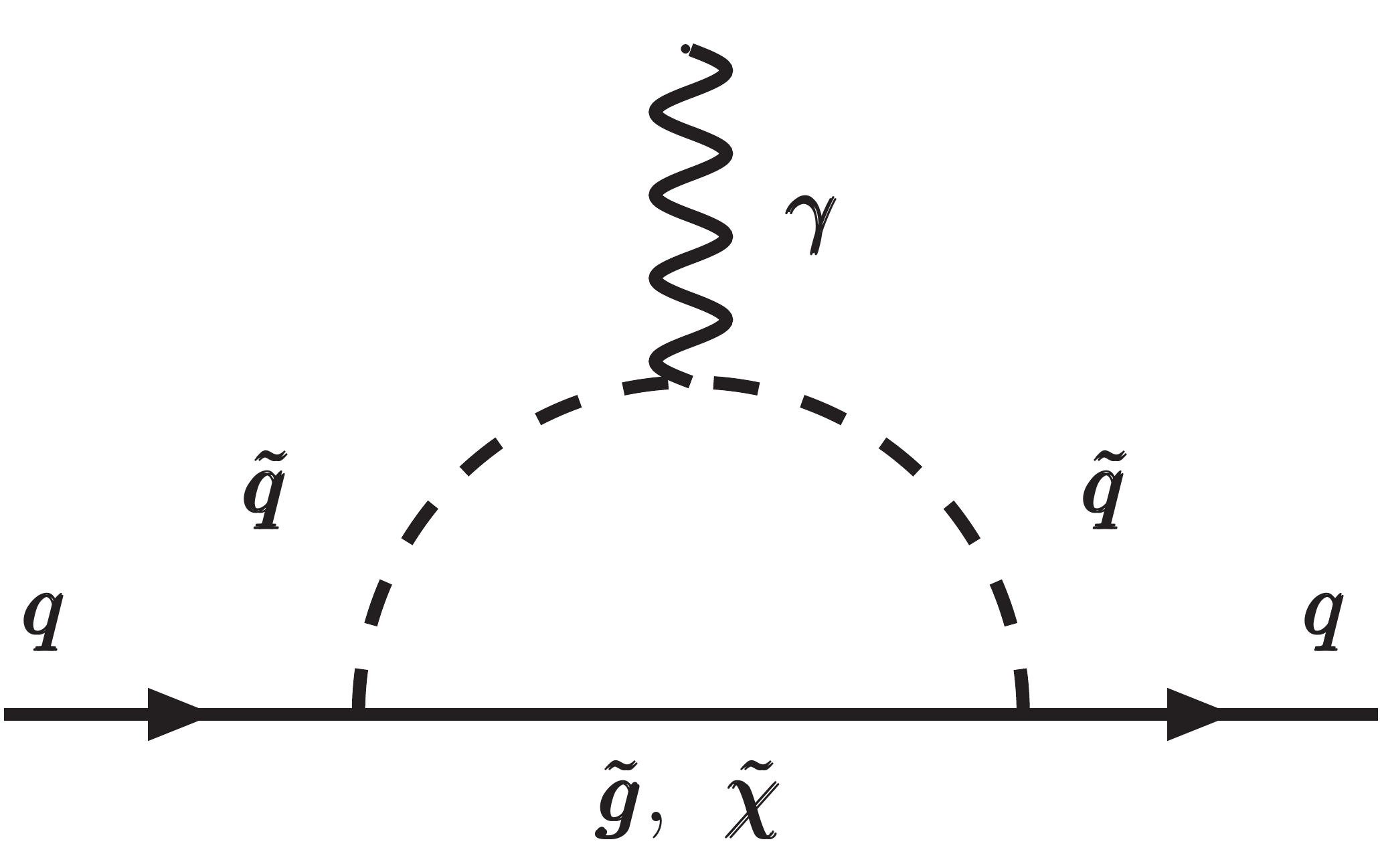} }
\end{center}
\caption{On the left, one example of SM contribution to the quark EDM; on the right, one-loop new physics contribution, exemplified by a generic supersymmetric  contribution mediated by squarks and gauginos. }\label{fig:FeynEDM} 
\end{figure}

The EDMs are clean and powerful probes of new physics (NP).  
In the SM, the
EW contributions to quark EDMs arise only at 3 loop-level and are
extremely suppressed due to the chiral nature of the SM flavour changing currents: the predictions lie well below present sensitivities. The same applies to leptons taking into account the known structure of neutrino masses and mixings. 
In contrast, most new physics models   
include new mediators and  new sources of CP violation that can generate  EDMs already at one-loop level, see Fig.~\ref{fig:FeynEDM}. The bounds on the EDMs thus severely constrain NP scenarios. For instance, 
in the absence of a suppression mechanism the new contributions to the EDMs can be easily in 
the range of $10^{-23}- 10^{-25}$~e$\cdot$cm, which is already in conflict with the present experimental bounds. Note that the EDMs are null searches; any nonzero signal at present or projected sensitivities would imply new physics.

The quark (or hadron) EDM and lepton EDM searches are complementary, as they may test different physics.  The leptonic  EDM is necessarily sourced by EW physics, while the hadronic EDMs are sensitive to both the EW and the strong interactions, for instance the $\theta$  parameter of QCD. 
The EDM searches are also
complementary to the high-energy CP probes, since they probe a different combination of NP parameters. 
 The combination of
quark and lepton searches for CP violation at different frontiers is
thus a formidable tool to test for NP.

The current experimental limit on the {\bf neutron EDM} is
$d_n < 3.6 \times 10^{-26}$~e$\cdot$cm at $95\%$ CL~\cite{Chupp:2017rkp},
and for the {\bf electron EDM} $d_e < 1.1 \times 10^{-29}$~e$\cdot$cm at $90\%$
CL~\cite{Andreev:2018ayy}.  
These bounds already test NP at mass scales above $10$ TeV and up to $\sim 10^6$ TeV, see  Fig.~\ref{fig:NPscales} (light yellow columns).  
A variety of other systems 
have also been explored as sensitive probes 
of the EDMs: atoms, protons,  deuterons, muons and different molecules. The current status of the measurements is reported in   Table~\ref{tab:EDM_all}.
The European collaborations lead current neutron EDM searches~\cite{Afach:2015sja}, while the best limits using molecules~\cite{Andreev:2018ayy}, 
diamagnetic atoms~\cite{Graner:2016ses} and 
muons~\cite{Bennett:2008dy} presently come from the experiments conducted in the US. 

{\setlength{\tabcolsep}{2pt} 
\begin{table}
\caption{Current EDM limits. In the table, $Q_m$ denotes the magnetic quadrupole moment, $C_S$ is the scalar form factor, and  $\mu_N$ and $R_{Cs}$ are the nuclear magneton and the nuclear radius of $^{133}C_s$, respectively. For 
further details and a complete review of the experimental 
scenarios see Ref.~\cite{Chupp:2017rkp}. } 
\label{tab:EDM_all} 
{\scriptsize
 \begin{center}
\begin{tabular}{cc}
 \begin{tabular}{lll}
\hline\hline
 & Result & 95\% u.l.\\ \hline 
 \\
& $\quad$ Paramagnetic  systems
 & \\ 
 \hline
 \\
Xe$^m$ & $d_A = (0.7\pm 1.4)\times 10^{-22}$ & $3.1 \times 10^{-22}$ 
e\, cm\\
Cs & $d_A =(-1.8 \pm 6.9) \times 10^{-24}$ &
$1.4 \times 10^{-23}$ 
e\, cm\\ 
& $d_e =(-1.5 \pm 5.7) \times 10^{-26}$& $1.2 \times 10^{-25}$ e\, cm\\ 
& $C_S =(2.5 \pm 9.8) \times 10^{-6}$& $2 \times 10^{-5}$\\
& $Q_m =(3 \pm 13) \times 10^{-8}$& $2.6 \times 10^{-7}$ $\mu_N R_{\rm{Cs}}$\\
Tl &$d_A =(-4.0\pm 4.3) \times 10^{-25}$ & $1.1 \times 10^{-24}$ e\, cm \\
& $d_e =(6.9 \pm 7.4) \times 10^{-28}$& $1.9 \times 10^{-27}$ e\, cm\\
YbF & $d_e =(-2.4 \pm 5.9) \times 10^{-28}$& $1.2 \times 10^{-27}$
e\, cm\\
$^*$ThO & $d_e =(4.3 \pm 3.1 \text{\scriptsize (stat.)}\pm$ \\
& $\pm 2.6 \text{\scriptsize (sys.)}) \times 10^{-30}$& 
$1.1 \times 10^{-29}$ e\, cm\\
{\scriptsize*90\% C.L.}& $C_S$ &$7.1 \times 10^{-10}$ \\
HfF+ & $d_e =(0.9 \pm 7.9) \times 10^{-29}$& 
$1.6 \times 10^{-28}$ 
e\, cm\\
\hline\hline
 \end{tabular}
\hspace*{2mm}&\hspace*{2mm}
\begin{tabular}{lll}
\hline\hline
 & Result & 95\% u.l.\\ \hline 
 \\
& $\quad $ Diamagnetic systems & \\ \hline
\\
$^{199}$Hg & $d_A =(2.2 \pm 3.1) \times 10^{-30}$& 
$7.4 \times 10^{-30}$ 
e\, cm\\
$^{129}$Xe & $d_A =(0.7 \pm 3.3) \times 10^{-27}$& 
$6.6 \times 10^{-27}$ e\, cm\\
$^{225}$Ra & $d_A =(4 \pm 6) \times 10^{-24}$& 
$1.4 \times 10^{-23}$ e\, cm\\
TlF & $d =(-1.7 \pm 2.9) \times 10^{-23}$& 
$6.5 \times 10^{-23}$ e\, cm\\
n& $d_n =(-0.21 \pm 1.82) \times 10^{-26}$&
$3.6 \times 10^{-26}$ e\, cm\\
\hline
\vspace*{1.5mm}\\
&  $\quad $ Particle systems& \\
\hline
\\
$\mu$ & $d_{\mu} =(0.0 \pm 0.9) \times 10^{-19}$&
$1.8 \times 10^{-19}$ e\, cm\\
$\tau$ & $Re(d_{\tau}) =(1.15 \pm 1.70) \times 10^{-17}$&
$3.9 \times 10^{-17}$
e\, cm\\
$\Lambda$ & $d_{\Lambda} =(-3.0 \pm 7.4) \times 10^{-17}$&
$1.6 \times 10^{-16}$
e\, cm\\
& \\
\hline\hline
 \end{tabular}
 \end{tabular}
\end{center}}
\end{table}
}

\setlength{\tabcolsep}{4pt}
\begin{table}
\caption{Summary of the neutron EDM facilities~\cite{PaulESPP19}.}
\label{tab: EDM_status}
\scriptsize
 \begin{center}
 \begin{tabular}{cccc}
\hline\hline
 Place & UCN source & sensitivity $\delta d_n$& start\\ \hline
 ILL &  $^4$He (SuperSANS) at reactor & $10^{-27}$ &2019\\
PIK (Gatchina) & sD$_2$ at PIK reactor  & $2 \times 10^{-28}$ & 2022\\
PSI& sD$_2$ at Spallation Source & $10^{-27}$ &2019\\
TRIUMF &$^4$He at spallation source & $10^{-27}$ &2020\\
SNS (Oak Ridge) &$^4$He at spallation source &
     $2 \times 10^{-28}$ & 2022\\
LANL & sD$_2$ at Spallation Source   &$1-3 \times 10^{-27}$ & 2019\\
RCNP & $^4$He at Spallation Source & $\text{few} \times 10^{-27}$ & under study\\
JPARC & Spallation Source&  $\text{few} \times 10^{-27}$& under study\\
TUM&sD$_2$ at FRMII reactor &$10^{-28}$& $>$ 2022\\
ILL& stack of 100 $^4$He source/EDM cells & $10^{-29}$ &  $>$ 2024\\
ESS& cold neutron beam & $10^{-25}-10^{-26}$ & $>$ 2024\\
    \hline\hline
       \end{tabular}
\end{center}
\end{table}
\setlength{\tabcolsep}{6pt}

In the {\bf short-} and  {\bf mid-term},  several  neutron EDM projects are in  various stages  of  development, as summarised in  Table~\ref{tab: EDM_status}. 
In Europe [ID123], they are grouped around reactor facilities (ILL Grenoble, FRM-2 Munich, PNPI Gatchina)  and
intense  proton-driven  spallation neutron sources  (PSI-Villigen,  ESS-Lund).  
Worldwide, there are three
important neutron EDM projects under development: 
SNS (Oak Ridge), LANL (Los Alamos) and TRIUMF (a  Japanese-Canadian collaboration).

The searches for the EDMs  of  charged  particles  such  as  {\bf protons} or  {\bf deuterons}  can  be  performed using  the  storage  rings \cite{Anastassopoulos:2015ura}. The {\bf short-} and  {\bf mid-term} strategies rely on a step-wise plan. 
The preparatory stages include the exploratory  measurements by the JEDI collaboration
\cite{Jedi_exp} (using COSY at FZJ) to demonstrate the technical feasibility. 
Since 2017, the CPEDM collaboration \cite{Rathmann:2019vtb} has investigated options for the design and construction of a storage ring for  
EDM measurements of charged states. 
The construction of the high-precision
electric-field storage ring could start in 2027, 
and achieve the {\bf proton} EDM sensitivity goal of $10^{-29}$~e$\cdot$cm, at the same level as the neutron EDM sensitivity prospects. 
The prototype ring and the CPEDM stages are host independent.

The ACME Collaboration \cite{Andreev:2018ayy} has recently obtained a new limit on the {\bf electron} EDM, $d_e < 1.1 \times 10^{-29}$~e$\cdot$cm at $90\%$ CL~\cite{Andreev:2018ayy}. In the mid-term, 
substantial improvements by a factor of $10-20$ in sensitivity are possible with further developments of the ACME technique, allowing to probe increasingly high scales of new physics. Fig.~\ref{fig:NPscales} illustrates in dark yellow the substantial improvements expected  in the mid-term for the electron and neutron EDMs. 

 So far, the {\bf muon EDM} searches  have been a byproduct of the muon $g-2$ experiments, as presently pursued at FNAL \cite{Chislett:2016jau} and projected at J-PARC \cite{Sato:2017sdn}. In contrast,  PSI intends to exploit their high-intensity muon source for a dedicated muon EDM measurement, using a compact storage ring \cite{Crivellin:2018qmi}.   Searches for {\bf molecular and atomic EDMs} are traditionally done in smaller groups at  university laboratories.   In  Europe  this  is,  for  instance,  the  case  for $^{129}$Xe and $^{199}$Hg  atomic EDM searches at TUM (Munich), PTB (Bonn) and FZJ, or the YbF and BaF molecular searches at Imperial College London and University of Groningen, respectively. 
 Some smaller projects  must be hosted at radioactive beam facilities such as ISOLDE. Several EDM experiments also rely on ideal magnetic environments which need advanced shielding  and  often  use  facilities  such as the  magnetically  shielded  room, BMSR-2, at the PTB Berlin or at TUM.

\begin{figure}[t]
\begin{center}
  \includegraphics[width=1.0\textwidth]{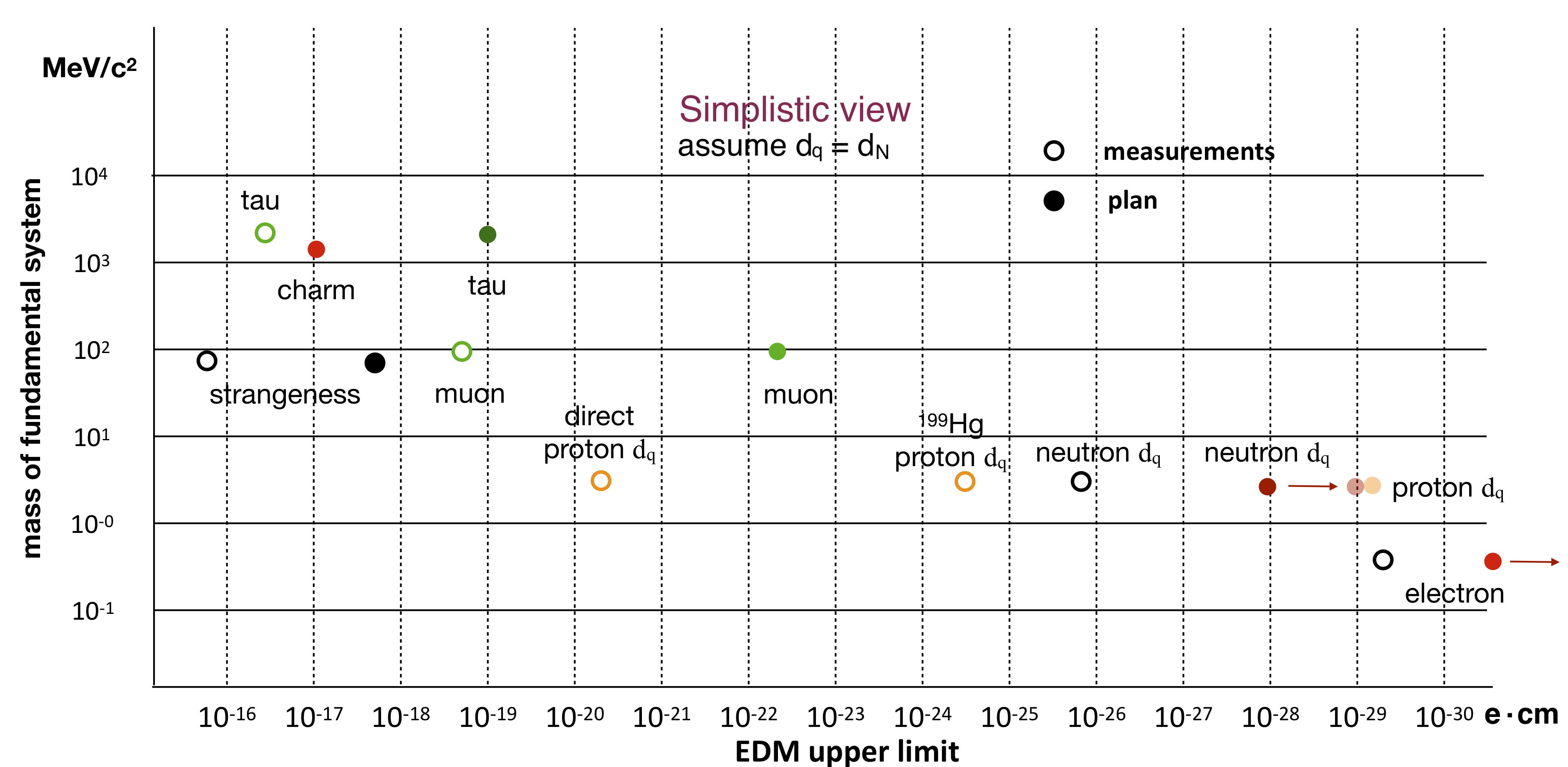}
  \caption{Summary of current EDM limits (empty circles) and short/mid-term  
  planned sensitivities (full circles) for light quarks, strange and charm quarks, electron, muon and tau~\cite{PaulESPP19}.} 
    \label{fig:EDM}
  \end{center}
\end{figure}

The current EDM upper limits for different fundamental systems and the expected sensitivities at present and planned facilities are summarised in Fig.~\ref{fig:EDM}.

Extending EDM searches to heavier probes, such as tau leptons and baryons, can provide qualitatively new BSM probes. These are especially interesting if the structure of the BSM contributions overcomes the inherently lower experimental sensitivities.

For the {\bf tau EDM}, searches could be performed in the {\bf mid-} and {\bf long-term} at Belle~II or, with even  higher sensitivity,  at the Super Charm-Tau (SCT) factories, such as SCT BINP at Novosibirsk~\cite{Bondar:2013cja} [ID49], and STCF/HIEPA in China~\cite{Luo:2018njj,Peng:2018}, 
exploiting the polarised beams and the large statistics of $10^{10} \tau \bar \tau$ pairs per year.

New ideas for {\bf heavy baryon electric and magnetic moment} searches  via spin precession of channeled particles in bent crystals are also being considered. 
These would require extraction of multi-TeV particle beams, which is being studied at
the LHC~\cite{Baryshevsky:2019vou, Mazzolari:2018hsu}. It could also open the door to measurements of magnetic moments of short-lived heavy-quark baryons.

\subsection{$\mu \to e$ transitions: short-, mid- and long-term}
\begin{figure}
\begin{center}
\includegraphics[width=1.0\textwidth]{\main/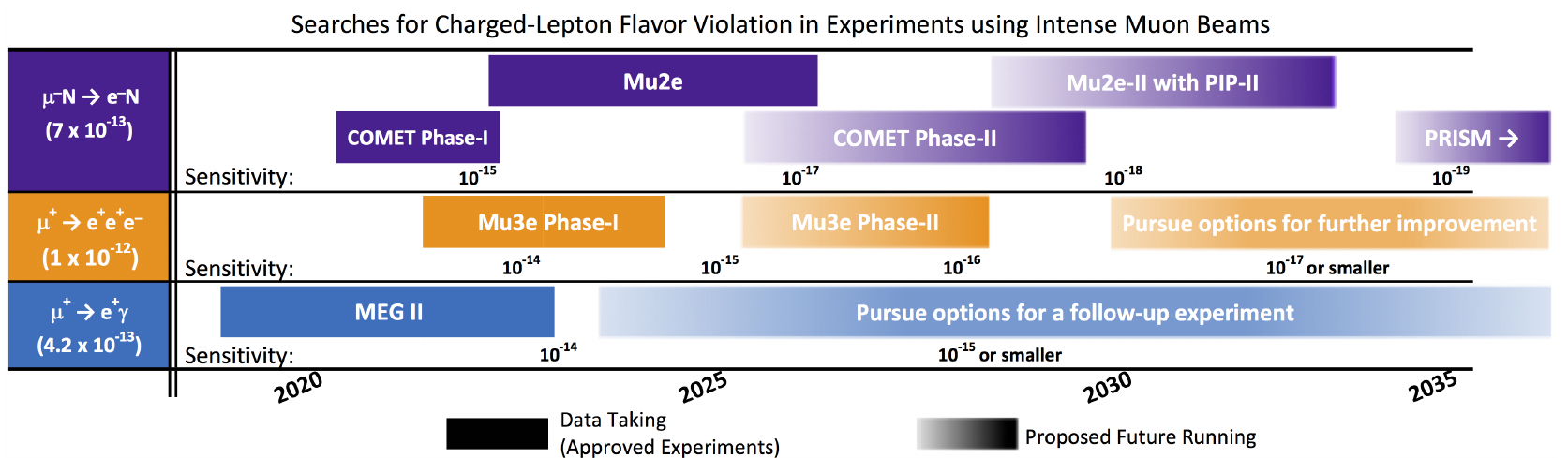} 
\end{center}
\caption{Planned data taking schedules for current experiments searching for LFV muon-electron transitions, as well as the schedules and expected sensitivities for future upgrades~[ID25].
  }\label{fig:cLFV:muon:chrono} 
\end{figure}

The searches for charged \textbf{lepton flavour violation (LFV)}   probe new physics in a manner that is complementary
to the collider, dark
matter, dark energy, and neutrino physics programmes.
They already test NP flavour scales up to  $10^4$~TeV (see Fig.~\ref{fig:NPscales}, light blue), while the sensitivity is expected to substantially increase in the mid-term, 
as shown in Figs.~\ref{fig:cLFV:muon:chrono} 
and~\ref{fig:NPscales} (dark blue). 
The highest sensitivity relies on the use of high-intensity muon beams to search for the ``golden'' $\mu \to e$ transitions~[ID25]. 

The {\bf short-term} prospects for the different muon channels include:  

\textbf{The ${\mu^+ \to e^+ \gamma}$ decay}: MEG II at PSI~\cite{Baldini:2018nnn}. The signature consists of a back-to-back
photon--positron pair coincident in time,
each with an energy of $m_\mu/2$.
To fully profit from the high intensity $\pi$E5
muon beam ($\sim 10^8$ stopped $\mu^+/$s), 
MEG II detector must succeed in mitigating 
the dominant 
accidental backgrounds. 
For the upgraded detector the 
improvements of resolutions  by roughly a factor 2 should allow 
a factor of ten improvement in the expected sensitivity at the end of the 2020-2023 physics run, giving the projected reach of
BR$(\mu \to e \gamma)< 6 \times 10^{-14}$~\cite{Baldini:2018nnn}. 

\textbf{The ${\mu^+ \to e^+ e^- e^+}$ decay}: Mu3e at PSI~\cite{Blondel:2013ia}.  
The experimental signature 
consists of three time-coincident charged particle tracks from a common
vertex, with the energy of the final state particles 
ranging from below 1~MeV up to $\sim m_\mu/2$.
Excellent timing and vertex resolutions can suppress the dominant
accidental backgrounds, while a very good
resolution of the final state energy successfully controls the intrinsic backgrounds.
In principle,
the sensitivity of Mu3e Phase-I is only limited by the maximal muon rate.
Commissioning is scheduled for 2021, and 
after 3 years of operation
Mu3e Phase-I aims at a sensitivity of $2\times 10^{-15}$ (90\% CL)
using the existing $\pi$E5 beamline at PSI~\cite{Blondel:2013ia}.

\textbf{The ${\mu^- - e^-}$ conversion in Al}: 
the neutrinoless conversion process $\mu^- \text{N} \to e^- \text{N}$ has a clean
experimental signature: an outgoing electron with an
energy near the muon mass. 
In the short term, two experiments, COMET at 
J-PARC~\cite{Adamov:2018vin} [ID38] and Mu2e at FNAL~\cite{Bartoszek:2014mya}, 
are under construction, both based on a system
of three solenoids with graded magnetic fields.
COMET Phase-I 
will first carry measurements of the muon yield and determine
rates for various background processes. 
With over $10^{9} \mu^-/$s from the 3.2~kW proton beam,
the projected sensitivity 
is $7 \times 10^{-15}$ (90\%CL). 
The Mu2e experiment aims at 
a sensitivity of $8 \times 10^{-17}$ (90\% CL)~\cite{Bartoszek:2014mya}, with the current
FNAL 8~kW proton beam giving $10^{10} \mu^-/$s. 

It is important to note that the programme of the LFV dedicated experiments is versatile and can be adapted to include other searches. Examples include: searches for exotic light particles $X$ via $\mu^{\pm} \to e^{\pm} X$ at MEG II, Mu3e and COMET; dark photons $A^\prime$ via $\mu^+ \to e^+ \nu \bar \nu A^\prime$ (with $A^\prime \to e^+e^-$) at Mu3e; lepton number violation (LNV) via $\mu^- + \text{N}(A,Z) \to e^+ + \text{N}^\prime(A,Z-2)$ and other LFV processes such as $\mu^- e^- \to  e^- e^-$, 
both at Mu2e and COMET.

{\bf Mid- and long-term prospects for LFV muon channels:} 
The next generation searches rely on the advent 
of very intense muon beams, which can increase
the stopped muon rate by a factor 10--100~[ID25]; these will be available at PSI, J-PARC and FNAL.

At PSI, the new High Intensity Muon Beamline (HiMB) project could deliver over $10^{10}$ stopped $\mu^+/$s.
However, going beyond a sensitivity of $10^{-15}$ in the \textbf{$\mu \to e \gamma$ decay}
will require a new experimental concept. For \textbf{${\mu \to 3e}$ searches}, the upgrade Mu3e Phase-II with larger acceptance and higher
rates can improve the sensitivity below $10^{-16}$, 
with the physics programme expected to span the years 2025-2030. 

Concerning \textbf{$\mu-e$ conversion}, COMET Phase II will reach a sensitivity for the conversion rate (CR) of 
CR($\mu-e$, Al)$\sim 2.6 \times 10^{-18}$ (90\% CL), by utilizing over
$10^{11}$ stopped $\mu^-/$s from the 56~kW proton beam at the J-PARC Main
Ring~[ID38, ID76]. 
COMET Phase II (whose original design~[ID38] aims at a sensitivity of $7 \times 10^{-17}$ (90\% CL)) is now being redesigned to improve to $\mathcal{O}(10^{-18})$ for the same beam power~\cite{KunoESPP19}. 
With further upgrades to the beam (1.3 MW) and to the detector systems, the 
high-intensity beam provided by PRISM will allow for the use of heavy targets, such as Pb, Au, and thus to achieve $\mathcal{O}(10^{-19})$ sensitivities~[ID38]. 
The Mu2e-II upgrade, with detector and solenoid improvements, will benefit from the
increased proton beam intensity of the PIP-II 
project at FNAL with 100~kW of protons and $10^{11} \mu^-/$s. For a  
Titanium target the Mu2e-II sensitivity can be improved by a factor ten
or more~\cite{Knoepfel:2013ouy,Abusalma:2018xem}.

In the {\bf long term}, the PIP-II beam at FNAL offers the possibility to study at the same facility distinct LFV observables, $\mu \to e\gamma$, $\mu \to 3e$, $\mu^- N\to e^\mp N^\prime$ and $\mu^- e^- \to e^- e^-$, relying on either pulsed or non-pulsed 
beams.  

\subsection{Kaons: short-, mid- and long-term}
Kaon decays are powerful probes of new physics in the quark sector. The sensitivity of kaon observables to new physics in general surpasses that of $B$-meson decays, because of the stronger flavour (CKM and GIM) suppression factors of the SM contributions.  
Current experimental sensitivity 
already probes new physics scales 
up to $\sim 10^5$~TeV, as can be seen from Fig.~\ref{fig:NPscales} (in light green).

\textbf{${K \to \pi \bar \nu \nu}$ decays}   are at present the driving goal of kaon physics. Their potential is due to the excellent theoretical precision, 
and to the possibility of disentangling different new physics models.
The SM predictions are BR($K^+ \to \pi^+ \bar \nu \nu$)$=(9.31 \pm 0.76)\times 10^{-11}$ and 
BR($K_L \to \pi^0 \bar \nu \nu$)$=(3.74 \pm 0.72)\times 10^{-11}$~\cite{Buras:2015qea,SozziESPP19}. 
An experiment aiming at the 10\% measurement for the $K^+$ decay rate is underway, and a discovery experiment for the $K_L$ decay is running. 
In addition, several $K_L$-dedicated experiments aiming at the $15-20\%$ sensitivity are under proposal, and ultimate experiments  aiming  at the precision of a few $\%$ for both modes are being discussed.  

\begin{figure}
\begin{center}
\includegraphics[width=0.45\textwidth]{\main/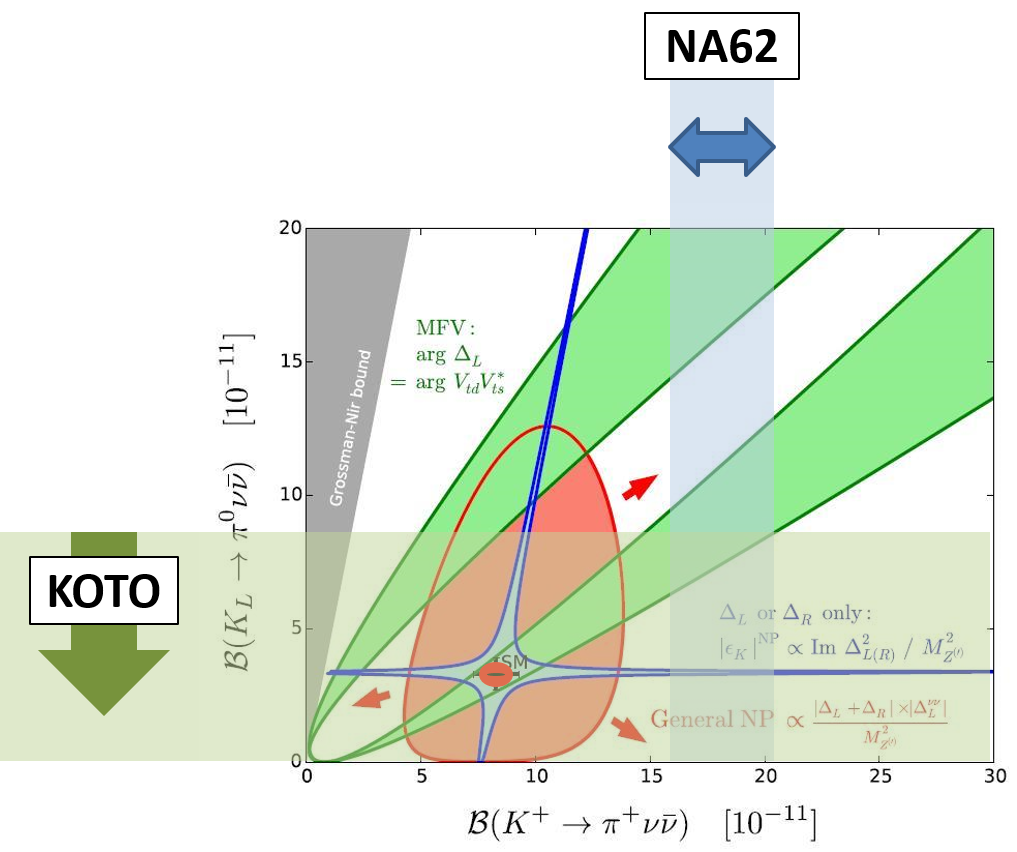} 
\includegraphics[width=0.49
\textwidth]{\main/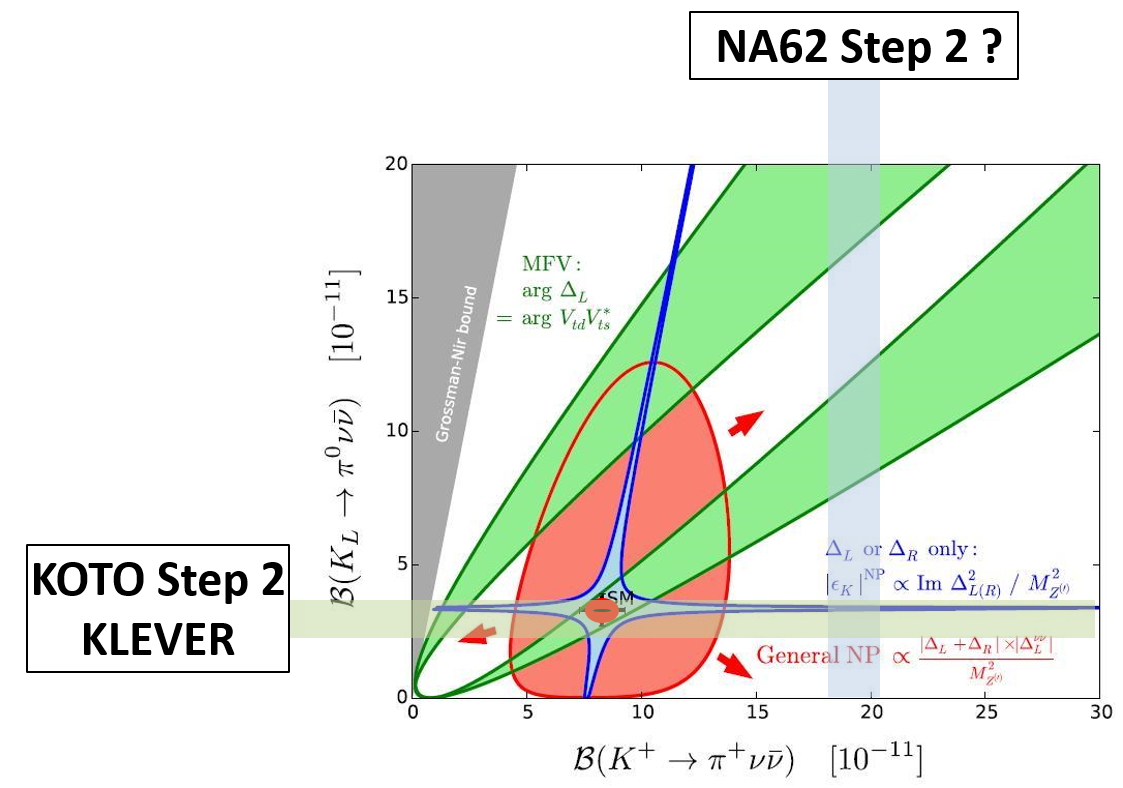} 
\end{center}
\caption{Educated guess for the future of the $K \to \pi \bar \nu \nu$
  decays (shaded vertical and horizontal regions): mid and long-term (respectively  
  left and right 
  panels)~\cite{SozziESPP19}. Original figure 
  from Ref.~\cite{Buras:2015yca}: the red region illustrates the lack of correlation for NP models with general left-handed and right-handed couplings; in green the correlation present in some MFV models; in blue the correlation induced by the constraint from $\epsilon_K$ if only left- or right-handed couplings are present.  
}
\label{fig:Kpinunu.KOTO.NA62}  
\end{figure}

The {\bf charged kaon decay mode} 
has been measured at E787/E949~\cite{Artamonov:2009sz},
BR($K^+ \to \pi^+ \bar \nu \nu$)$=17.3^{+11.5}_{-10.5} \times 10^{-11}$. 
In the {\bf short-term}, NA62 is committed to deliver a measurement 
with $\mathcal{O}(10\%)$ precision 
prior to the Long Shutdown 3 (LS3) (preliminary results from the analysis of 2017 data show two events in the signal region consistent with the SM expectation~\cite{NA62:2019}). A feasibility study is also underway
to improve the precision of measurements 
at the higher beam intensities available after LS3~\cite{SozziESPP19,NA62:Lazzeroni}. 
For the {\bf neutral decay} mode, $K_L \to \pi^0 \bar \nu \nu$, the dedicated KOTO experiment at J-PARC [ID76]
has recently obtained the limit BR$(K_L \to \pi^0 \bar \nu \nu) < 3 \times 10^{-9}$ (90\% CL)~\cite{Ahn:2018mvc}. In the {\bf mid-/long-term}, both KOTO as well as the KLEVER project at CERN [ID153] aim at significant progress. 
KOTO proposes to reach an $\mathcal{O}(100)$ SM event sensitivity 
by using the increase of the J-PARC Main Ring power to 100~kW and by upgrading the experiment~\cite{KOTO_prospects}. 
KLEVER~\cite{Ambrosino:2019qvz} [ID153], using the 400~GeV 
SPS proton beam, might start data taking during the LHC Run 4 (2026). The benchmark goal is to collect around 60 SM events in five years of data taking, assuming a
delivered intensity of $10^{19}$ pot/year. 
The joint prospects of KOTO and NA62 in the corresponding kaon decay
modes, as well as long term prospects including KLEVER and a second
phase KOTO-II (under discussion)~\cite{KOTO_II} [ID76],
are schematically depicted in Fig.~\ref{fig:Kpinunu.KOTO.NA62}. The  upcoming results expected  from NA62  and the evolution of the Japanese project will guide the future European steps in this research field.

Concerning the \textbf{$K_S$ decay modes}, the {\bf mid-term} prospects rely on the \HLLHC. 
The LHCb Upgrade II can approach the SM value 
BR($K_S \to \mu \mu$)$=5.2 \times 10^{-12}$~\cite{Cirigliano:2011ny}, and significantly improve the precision of the measurements done by NA48 (e.g.\  $K_S \to \pi \mu \mu, \pi \pi e e$). The ultimate reach of LHCb Upgrade II could be $\mathcal{O}(10^{-15})$ for some $K_S$  decay modes~\cite{SozziESPP19}.  

Kaon decays further offer a good laboratory for \textbf{LFV, LNV} and lepton flavour universality violation (\textbf{LFUV}) searches, given
the high statistics, clean signatures and controlable backgrounds.
The $K^+$ and $K_L$ fluxes allow ``parasitic'' \textbf{charged LFV} decay searches below the $10^{-12}$ branching ratio level~\cite{SozziESPP19}.  For \textbf{LNV} decays, 
NA62 has already obtained the
bounds ${\rm BR}(K^+ \to \pi^- e^+e^+ (\mu^+\mu^+))<2.2(0.42)\times 10^{-10}$ 
at $90\%$ CL,  
from the 2017 data set~\cite{CortinaGil:2019dnd}.
In the {\bf short-term}, the full data collected for the latter modes will be about three times larger.  
{\bf LFUV} in kaon decays can be probed via the comparison of the helicity-suppressed widths, 
through the ratio 
$R_K^\ell = \Gamma (K^\pm \to e^\pm \nu)/\Gamma (K^\pm \to \mu^\pm\nu)$. 
In the {\bf short-term}, 
both TREK at J-PARC~\cite{JPARC_TREK} and NA62 are expected to reduce the current errors, 
 to 2.5~per mille and sub-per mille level, respectively~\cite{SozziESPP19,NA62:Lazzeroni}.

New concepts to test {\bf discrete symmetries}
in kaon decays are being considered. 
As an example, TREK is  
exploring T violation in 
$K^+ \to \pi^0 \mu^+ \nu$. 
For the CP violation observables $\epsilon_K$ and $\epsilon^\prime/\epsilon$, no significant experimental developments are foreseen. 
On the theory side, however, progress in the computation of  the weak matrix elements and in the determination of the CKM angles will play a clear role. 
In particular, 
the first SM computation of $\epsilon^\prime/\epsilon$ with fully controlled uncertainties is expected in the short-term~\cite{Kelly:2019yxg}. 

The kaon sector also has a unique role in the determination of CKM elements, 
including new CKM unitarity tests through the emerging kaon unitarity triangle~\cite{Lehner:2015jga} (see Sect.~\ref{sec:CKM}). The sensitivity of related kaon observables to new physics  scales  is illustrated in Fig.~\ref{fig:NPscales} (present bounds from kaons and $B$ mesons in light green, LHCb Upgrade-II prospects in dark green).  

Ultra-rare $K_L$ and $K^+$ decays are clearly the most-promising goal of the field.  Existing machines are fully adequate for the next step in kaon physics. However, ultimate discoveries require very high hadron intensities which could be provided in the {\bf long-term} by intense, high-duty-cycle extracted proton beams, such as from the J-PARC main ring, or from
the proton drivers required for a future hadron collider, with a beam power in the MW range.

\section{Heavy sector (short-, mid- and long-term)}
 Due to their larger masses, the heavy fermions might offer a privileged  handle on the Higgs couplings and on the origin of the flavour puzzle. The flavour quest in the beauty, charm and tau sectors is discussed first, followed by that for the top and gauge bosons. 
\subsection{Charm and beauty: short-, mid- and long-term}
\begin{figure}[t]
  \centering  
  \includegraphics[width=0.45\textwidth]{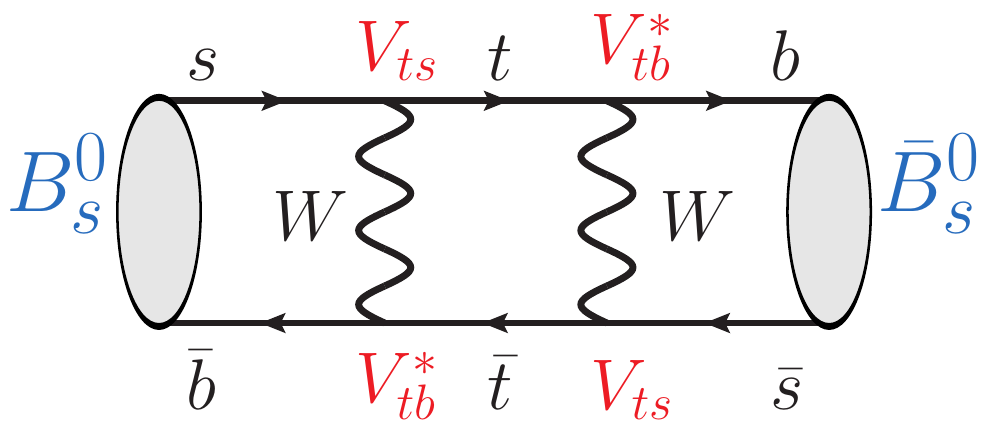}~~~~
  \raisebox{1.5mm}{\includegraphics[width=0.45\textwidth]{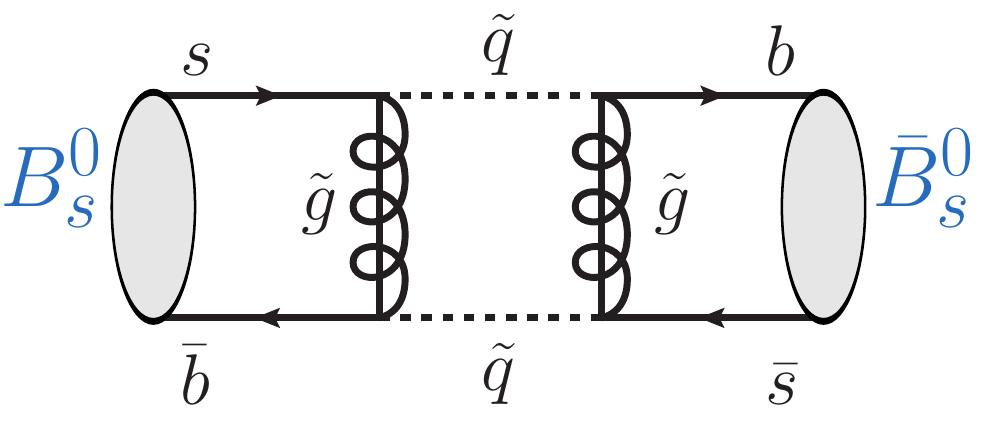}}
  \caption{Examples of diagrams contributing to $B_s-\bar B_s$ mixing in the SM (left) and example of SUSY new physics contributions, with a gluino-squark exchange shown as an example (right), adapted from \cite{Zupan:2019uoi}.}
  \label{fig:Bsmixing}
\end{figure}

The flavour physics results obtained at the LHC in the past decade have far exceeded expectations. 
A combination of measurements by LHCb and CMS resulted in the observation of the long sought-after $B_s\rightarrow \mu^+\mu^-$ decay mode in 2014~\cite{CMS:2014xfa}.
Remarkable progress has been made in CP violation studies in the beauty sector including  measurements of the CKM angle $\gamma$ by LHCb~\cite{Aaij:2018uns}, 
$B_s$ mixing phase $\phi_s$ by ATLAS and LHCb~\cite{ATLAS:2019akj,Govorkova:2019jlz}, and the discovery of CP violation in the $B_s$ system by LHCb~\cite{Aaij:2013iua}.
In charm physics, a landmark result was obtained by LHCb in 2019 with the observation of CP 
violation~\cite{Aaij:2019kcg}, opening a new field of study. Spectroscopic studies also yielded important results, with several new hadronic states found by the LHCb, including the observation of pentaquark states ~\cite{Aaij:2015tga,Aaij:2019vzc} in 2014 and 2019. The spectroscopic studies of heavy quark systems are expected to continue in the future at Belle II, LHCb, FAIR/PANDA...

\begin{figure}[t]
\centering
\includegraphics[width=0.45\textwidth]{\main/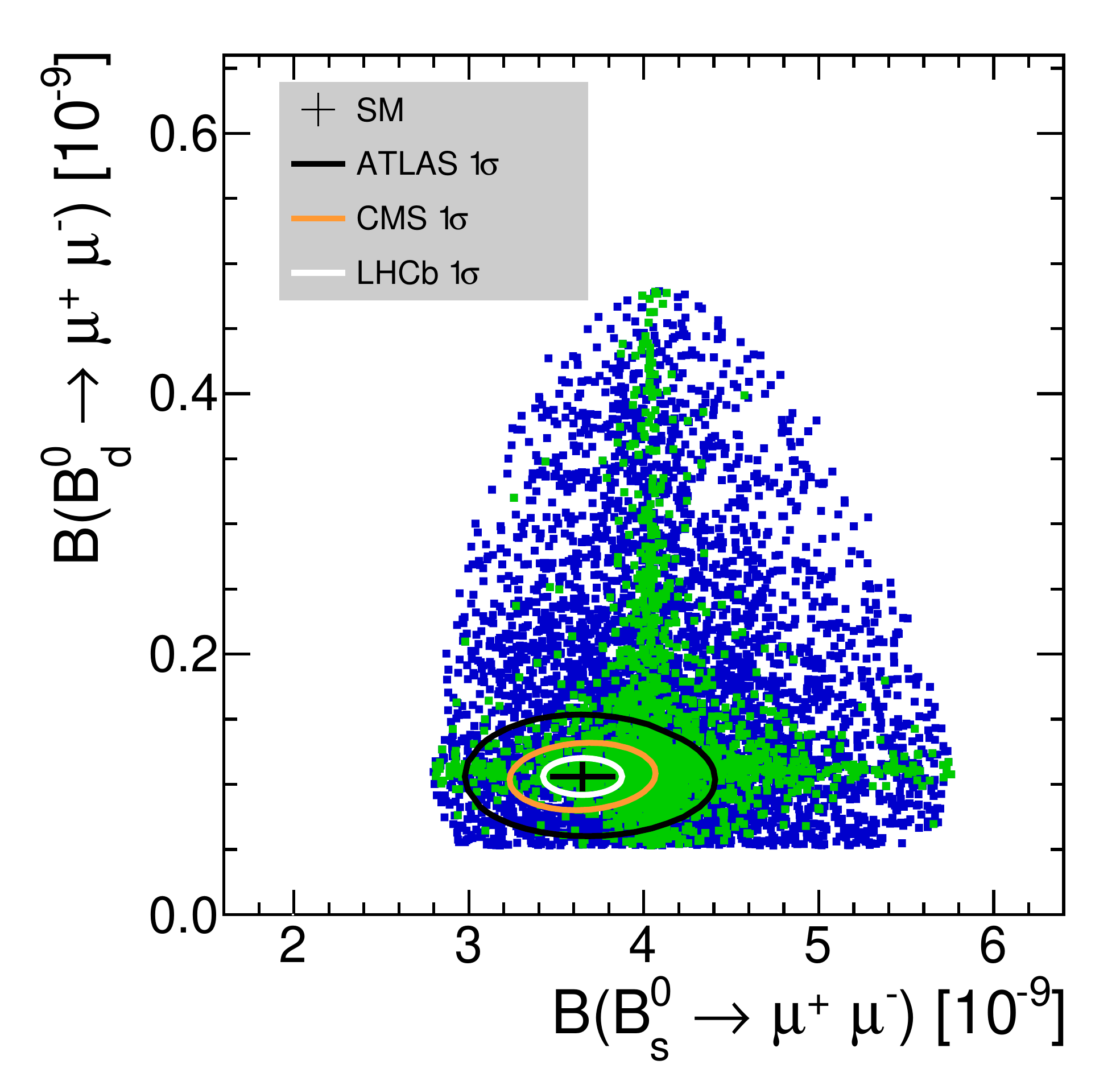}
\caption{ 
 BR$(B_s^0 \to \mu^+\mu^-)$ vs. BR$(B_d^0 \to \mu^+\mu^-)$ in the SM (black cross), and in a particular supersymmetric unified model (green points are consistent with other constraints). The coloured contours show the expected $1\sigma$ \HLLHC sensitivity of ATLAS, CMS, and LHCb Upgrade~II. From Ref.~\cite{Cerri:2018ypt}.}
\label{fig:Bmumu} 
\end{figure}

An important goal of the program is to search for potential new physics effects that could enter through virtual corrections, see Fig. \ref{fig:Bsmixing} for a $B_s-\bar B_s$ mixing example. It is important to stress that there are a number of measurable quantities that are {\it theoretically clean}, and where the knowledge will still be statistically limited after Belle II and LHCb Upgrade I. A list of such observables, which will actively drive the field, includes the CP-violating phase $\gamma$, the lepton-universality ratios $R_{K^{(*)}}$, $R_{D^{(*)}}$,  etc., the mixing phases in the $B_s$ and $D$ systems, as well as the ratio of branching ratios ${\rm{BR}} (B_d \rightarrow \mu^+ \mu^-)$/${\rm{BR}} (B_s \rightarrow \mu^+ \mu^-)$. 
A notable target of the physics programme at the \HLLHC is to probe at the $10\%$ level the ratio ${\rm{BR}}(B_d \to \mu^+\mu^-)/ {\rm{BR}}(B_s \to \mu^+\mu^-)$, which, in case that new physics deviations are found, is a powerful observable to test the MFV hypothesis.
The expected 1$\sigma$ sensitivities of ATLAS, CMS and LHCb in BR($B^0_s \to \mu^+ \mu^-$) versus ($B^0_d \to \mu^+ \mu^-$) are shown in  Fig.~\ref{fig:Bmumu}.

\begin{figure}[t]
  \centering  
  \includegraphics[width=0.7\textwidth]{\main/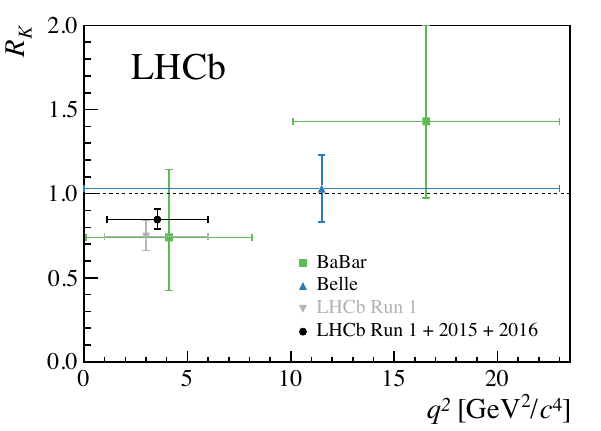}\hbox{\hspace{-0.2cm}}\\
  \includegraphics[width=0.715\textwidth]{\main/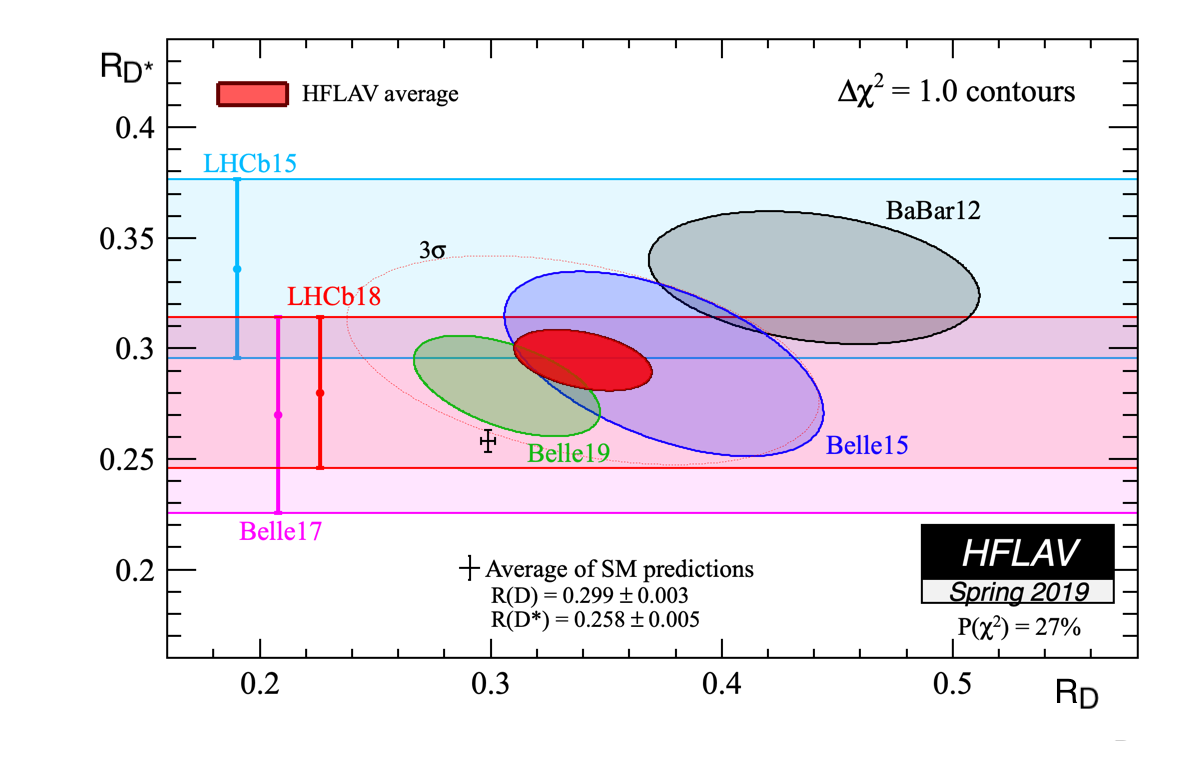}\hbox{\hspace{-0.2cm}} 
  \caption{Upper panel: experimental results for 
  $R_K$ as function of the di-lepton invariant mass squared, $q^2$. Lower panel: status of $R_{D^{(*)}}$
  measurements; the SM predictions are  
   in tension with the experimental world average at 
  the 3.08\,$\sigma$ level \cite{HFLAVSPRING19}.}
  \label{fig:RKRD}
\end{figure}

Intriguingly, some measurements, in both charged-current and neutral-current 
semileptonic \textbf{$B$ decays}, hint at a violation of one of the 
key predictions of the SM: the universality of interactions for leptons of different generations (\textbf{LFUV}). 
The statistical significance of the anomalies is not sufficiently high to 
claim a discovery, but the situation is very interesting. Regardless of whether or not their statistical significance increases with improved measurements, the anomalies do demonstrate the genuine discovery potential of the flavour programme at the LHC.
More precise measurements of some of these observables, in particular the  
 LFUV ratios
$R_{K^{(*)}} = \Gamma(B\to K^{(*)} \mu^+\mu^-) / \Gamma(B\to K^{(*)} e^+e^-)$ and 
$R_{D^{(*)}} = \Gamma(B\to D^{(*)} \tau\bar\nu) / \Gamma(B\to D^{(*)} \ell\bar\nu)$, where $\ell=e,\, \mu$,
could establish the presence of new physics, even 
with modest improvements in statistics.  
The current experimental situation for 
$R_{K}$~\cite{PhysRevLett.103.171801, PhysRevD.86.032012,PhysRevLett.113.151601,Aaij:2019wad,Abdesselam:2019wac} is shown in Fig.~\ref{fig:RKRD} (upper panel), and for $R_{D^{(*)}}$~\cite{Abdesselam:2019dgh,PhysRevD.97.072013,PhysRevLett.118.211801, PhysRevLett.115.111803,PhysRevD.94.072007,PhysRevD.92.072014,PhysRevD.88.072012,PhysRevLett.109.101802,Aaij:2017uff} in Fig.~\ref{fig:RKRD} (lower panel). While the discrepancies are still not statistically significant, it may be useful to address which type of new physics could explain them if they do become significant.  
To explain the $R_{D^{(*)}}$ discrepancy, new physics needs to enter at tree-level and be lighter than a few TeV, while for $R_{K^{(*)}}$ the tree-level new physics can be as heavy as several tens of TeVs, such as a $Z'$ with ${\mathcal O}(1)$ couplings. A combined explanation of both sets of anomalies is possible using leptoquarks. 
Interestingly, some tensions with the SM predictions are also seen in the $B \to K^{*} \mu^+\mu^-$ angular distributions \cite{Aaij:2015oid,Wehle:2016yoi,Sirunyan:2017dhj,Aaboud:2018krd}, possibly forming a coherent pattern with the LFUV measurements. However, compared to LFUV ratios the SM predictions in this case do have larger uncertainties due to the presence of charm-loops, making reliable theoretical description essential for future progress \cite{Jager:2012uw,Ciuchini:2017mik,Matias:2012xw,DescotesGenon:2012zf,Bobeth:2017vxj}.

In the {\bf short-term}, significant progress is expected in the precision of the measurements for core flavour physics observables, whose knowledge is still largely statistically limited. There is a concerted effort devoted to extensive studies of the $b\rightarrow s \ell^+\ell^-$, $b\rightarrow d \ell^+\ell^-$ and $b \rightarrow c \ell^- \bar\nu_\ell$ transitions, including analysis of $B$-hadron decay angular distributions. 
The two dedicated $B$-physics experiments, Belle~II, the $e^+e^-$ superflavour factory 
operating mostly at the $\Upsilon(4S)$ resonance, and the LHCb at the LHC, including its Upgrade~I,
are aiming at a rich programme of measurements to be performed, {\bf with a high level of complementarity}.  LHCb benefits from higher statistics in charged-track decay modes and the access to all $b$ hadrons, while Belle~II has a unique access to fully neutral final states and rare leptonic decays with final state neutrinos. Belle II is thus expected to notably contribute to the understanding of $B$-anomalies.
The ATLAS and CMS experiments 
will continue to contribute to flavour physics, notably in $B$ decays to final states containing muons.

\begin{figure}[t]
\centering
\includegraphics[width=0.4\textwidth]{\main/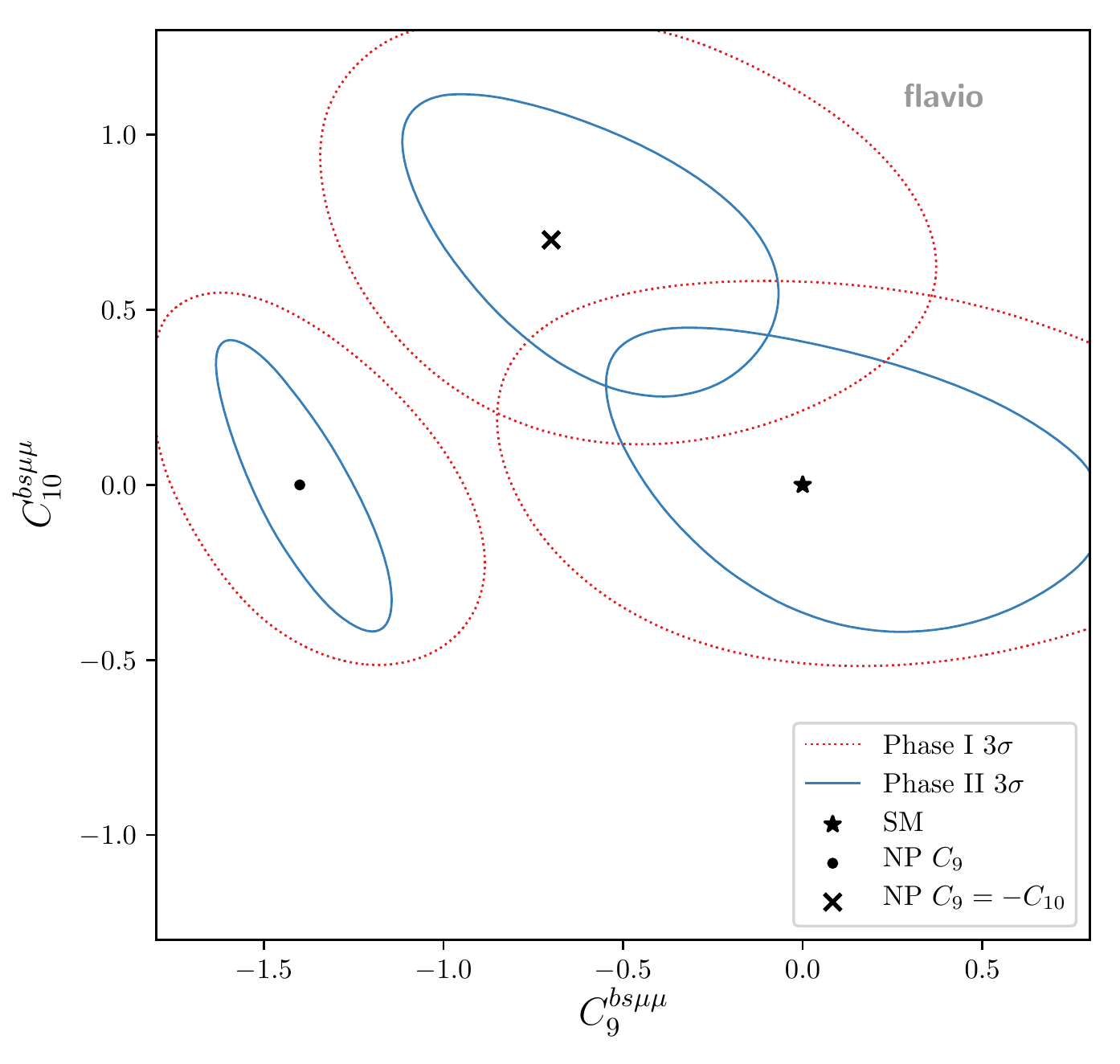}
\caption{ 
  Combined sensitivity of LHCb, ATLAS and CMS after the \HLLHC phase to potential new physics in $b\to s\mu\mu$ processes, motivated by recent anomalies (from Ref.~\cite{Cerri:2018ypt}). New physics benchmarks with leptonic vector current (new physics only in $C_9$) or pure left-handed current ($C_9=-C_{10}$), as well as the SM predictions are shown. The observables included are the branching ratio  of the $B_s\to\mu^+\mu^-$ decay and the angular observables of the decay $B^0 \to K^{\ast 0} \mu^+\mu^-$ in the low-$q^2$ region. 
See  Ref.~\cite{Cerri:2018ypt} for details.}
\label{fig:Bmumu2} 
\end{figure} 

In the {\bf mid-term}
the LHCb Upgrade~II, combined with the enhanced $B$-physics capabilities of ATLAS and CMS Phase~II upgrades, will enable a wide range of flavour observables to be determined at \HLLHC with unprecedented precision, complementing and extending the reach of Belle II, and of the high-$p_T$ physics programme.
Substantially improved tracking systems and the addition of timing layers in ATLAS and  CMS Phase-II detectors may significantly improve their capabilities for charged-hadron  particle identification.
The LHCb Upgrade~II will allow the experiment to run at instantaneous luminosities up to $2 \times 10^{34}{\rm cm}^{-2} {\rm s}^{-1}$ with a target integrated luminosity of 300 fb$^{-1}$, and thus exploit the full \HLLHC potential in flavour physics \cite{Bediaga:2018lhg, Cerri:2018ypt}.
Generically, and for fixed couplings, the new physics mass scale probed  will roughly double compared to the pre-\HLLHC era, see Fig.~\ref{fig:NPscales} (light vs.\ dark green for mid-term prospects with LHCb Upgrade~II).

Very recently, the suggestion that a factor of five increase in luminosity could be achieved at SuperKEKB was raised, aiming for an integrated luminosity of $250~{\rm ab}^{-1}$.
The clear complementarity of flavour physics at $e^+ e^-$ and ``$pp$'' colliders makes this possibility appealing. 
The major issues related to such Belle~III project  
are the feasibility from accelerator perspective and the detector challenges when running at 4 $\times 10^{36}$
cm$^{-2}$ s$^{-1}$.

The \HLLHC, combining ATLAS, CMS and LHCb Upgrade~II, has the potential to distinguish between some well-motivated  
new physics scenarios. 
The increasing precision of observables from measurements of statistically limited FCNC processes will provide significant improvements in terms of the reach to the energy scale of new-physics. 
As an example, the  plot in Fig.~\ref{fig:Bmumu2} shows the potential sensitivity to the Wilson coefficients $C_9$ (vector current) and $C_9=-C_{10}$ (pure left-handed current), for definitions see, e.g., \cite{Aebischer:2019mlg}.
These fit results take as inputs the measurements of the branching ratio of the  $B_s\to\mu^+\mu^-$ decay and the angular observables from the decay $B^0 \to K^{\ast 0} \mu^+\mu^-$ in the low-$q^2$ region. The reach for generic new physics at tree-level is found to exceed 100~TeV, and in terms of the constraints on new-physics contributions to the $C_9$ and $C_{10}$ Wilson coefficients the study shows an approximate gain of about a factor of two compared to the constraints prior to \HLLHC. 

In addition, a wide range of lepton-universality tests in $b \to c \ell \nu$ decays can be performed, exploiting the full range of
$B$ hadrons, to probe  models of new physics.
The copious yields of semileptonic decays allow high-precision searches to be made for CP violation in $B^0$ and $B_s$ mixing. 
In the {\bf mid-term}, the LHCb Upgrade~II dataset will allow the semileptonic asymmetries for both mesons to be measured at the level of a few $10^{-4}$, giving unprecedented new-physics sensitivity.
Precise measurements of \textbf{${\phi_s}$} and 
\textbf{${\sin 2 \beta}$} will be performed, with strategies to monitor and control possible  pollution from penguin contributions.
Advances in lattice-QCD calculations will also motivate better measurements of other critical observables, e.g., $|V_{ub}|/|V_{cb}|$, for details see Sect.~\ref{sec:CKM_prospects}. 

In the {\bf long-term},
at future colliders, the heavy-flavour physics is expected to be, as it is now, an integral part of the physics programme. 
 Comprehensive flavour-dedicated studies at colliders, which are not available for all future projects, would be necessary for a   thorough comparison. According to the studies available, the new $e^+e^-$  circular colliders 
are foreseen to collect 
 $\sim 10^{ 11}$ and 
 $ \sim  10^{12}$ 
 $Z \to b \bar b$ events at 
\CEPC~[ID29] and \FCCee~\cite{Blondel:2019yqr} respectively, which  will result in all species of heavy-flavoured hadrons.
 The boosted topologies of the decay particles at the $Z$ energy, in conjunction with the clean $e^+e^-$ environment, will be beneficial for a number of measurements. 
In this respect, the $B_{s,d} \to \tau^+\tau^-$ and  $ B \to K^{(*)} \tau^+ \tau^-$ decays are  natural candidates to study. For example about 1000  events with a reconstructed  $\bar {B}^0 \to K^{*0} \tau^+\tau^-$ are expected in the $5 \times 10^{12}$ $Z$ decays at \FCCee, which would 
allow for the first investigation of  the tau lepton polarisation in this mode 
\cite{Kamenik:2017ghi}. 
Recently, there has been renewed interest for the Giga-Z programme at a linear collider, i.e., for runs that would collect $\sim 10^9 Z$'s from collisions with polarised electrons~\cite{Irles:2019xny}.
In general, Tera-Z samples  of $\ge 10^{12} Z$'s are needed to further improve flavour-physics precision measurements and searches for rare decays after the LHCb Upgrade~II and   Belle~II (and possibly Belle~III).  
In particular, three different tests of lepton universality can be performed at a $Z$-factory. Charged current universality tests are best carried out with the $1.7 \times 10^{11}  \tau^+ \tau^-$ pairs (precision level $10^{-5}$ with Tera-Z samples).
Further tests can be also performed using rare decays of heavy-flavour hadrons from the $\sim 10^{12}$ $Z\to b\bar b$  and $Z\to c \bar c$ decays.
In this case there is no benefit from the longitudinal polarisation so that the polarised Giga-Z sample, with three orders of magnitude fewer events, is significantly more limited. 
Neutral current universality can be tested first from the comparison of the partial widths of the $Z$ into each of the three lepton pairs; a precision better than $10^{-5}$ is expected at the Tera-Z. Here again, the Giga-Z will  suffer from 3000 times less statistics. 
 The ratio of vector-to-axial-vector couplings can be accessed through measurements of initial- and final-state polarisation asymmetries, as well as forward-backward asymmetries. For such asymmetry measurements  the initial-state beam polarisation brings substantial improvements in the reach. However, these measurements can be well performed also without longitudinal beam polarisation 
 \cite{Blondel:2019yqr}.

In the \FCChh configuration, a dedicated experiment  \`{a} la LHCb could be conceived, for instance at the booster stage of the accelerator complex. Since a number of observables are expected to have negligible theoretical uncertainties even when compared with the experimental ones  after LHCb Upgrade~II, there could be a strong physics case for such an experiment.

The discussion so far focused on the physics with $b$ quarks. A related, yet different probe of new physics is offered by the {\bf charm quark decays}. In the SM, the FCNC processes involving charm hadrons are suppressed compared to those involving strange or beauty hadrons, in both the mixing and decay amplitudes. The reason is, on the one hand, that the charm quark, unlike the $b$ and $s$ quarks, can decay inside its own family. The characteristic time for the FCNC transitions is therefore much longer than the decay time, giving both $\Delta M/\Gamma\ll~1$ and $\Delta \Gamma/\Gamma\ll~1$. This fact is sustained, on the other hand, by the small breaking of the GIM mechanism controlled by the $b$ quark mass. 
The charm FCNCs can then be used as sensitive probes of new physics in the up-quark
sector, to the extent that theoretical uncertainties can be brought
under control, e.g., by constructing null tests, or circumvented by using the experimental data.

\begin{figure}
  \centering
  \includegraphics[width=0.85\textwidth]{\main/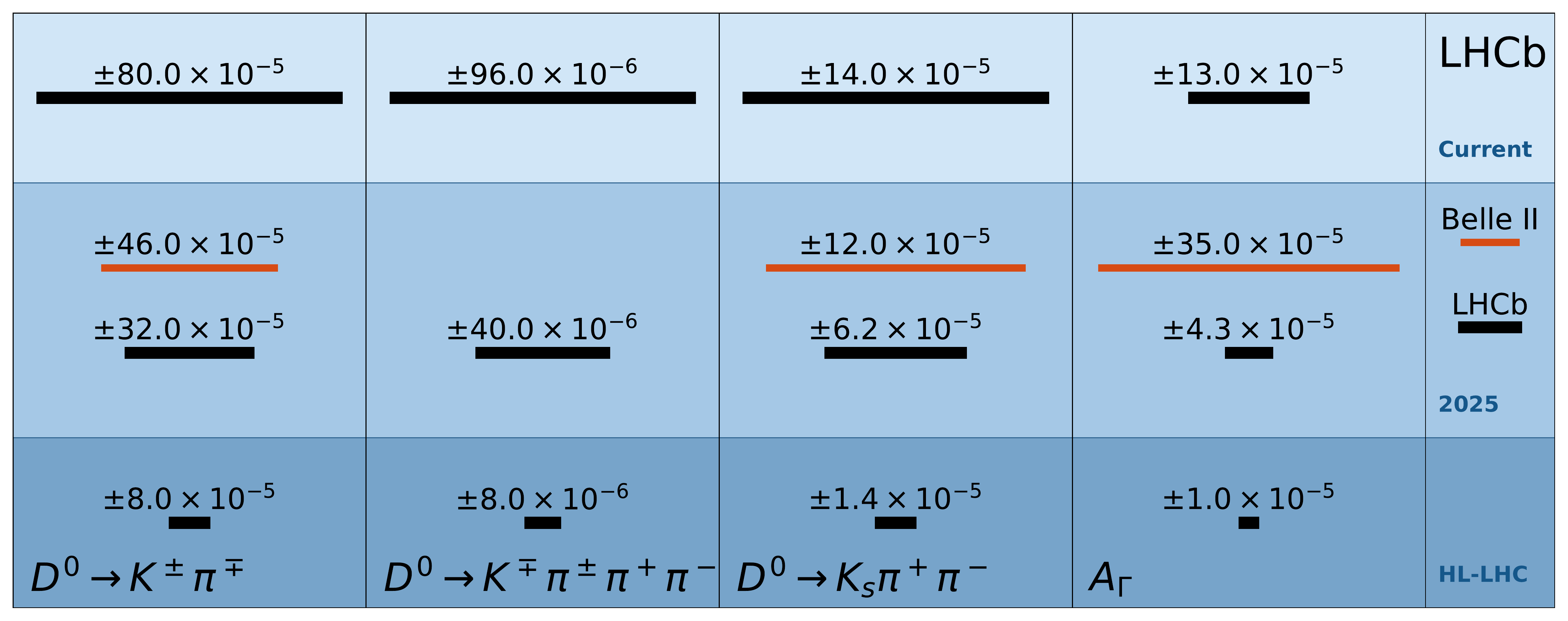}\vspace*{3mm}\\
 \includegraphics[width=0.65\textwidth]{\main/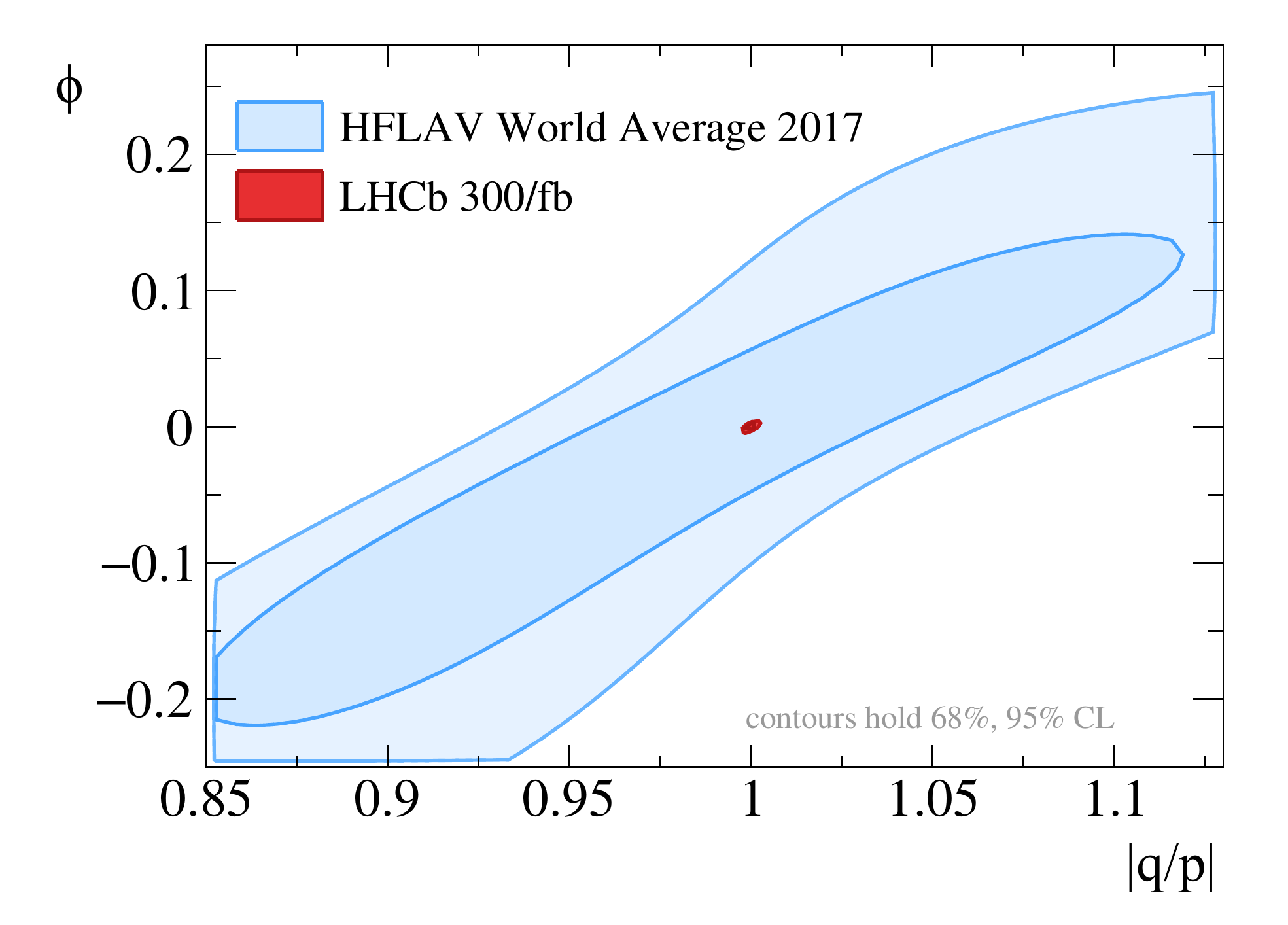} 
  \caption{Upper panel: Predicted constraints on the indirect CP violation
    asymmetry in
    charm from the decay channels indicated in the labels at the
    bottom of the columns. Predictions are shown in LS2
    (2020) from LHCb, LS3 (2025) from LHCb, at the end of Belle~II (2025), and at the end
    of the \HLLHC LHCb Upgrade~II\ program.
    Lower panel: Estimated constraints for LHCb \upgradetwo\ on $\phi$, $|q/p|$ from the combination of the analyses (red) compared to the current world-average precision (light blue). Both panels from Ref.~\cite{Bediaga:2018lhg}.} 
\label{fig:charm_indirect_channels}
\end{figure}

In the SM, the size of CP violation in charm decays is predicted  to be $\mathcal{O} (10^{-3} -10^{-4}$), and may be altered by  virtual new physics particles. 
The first observation of CP violation in the decay of charmed hadrons \cite{Aaij:2019kcg}
opens new opportunities across two-body, multi-body, direct and indirect CPV. The projected precisions of some analyses performed by LHCb are shown in   Fig.~\ref{fig:charm_indirect_channels} (upper panel) and are compared with the {\bf short-} and {\bf mid-term} precisions expected at Belle~II and \HLLHC~\cite{Bediaga:2018lhg}. 
The available experimental measurements are also combined to establish the sensitivity to the CP-violating parameters $q/p$ and $\phi$. 
The lower plot in Fig.~\ref{fig:charm_indirect_channels}  shows the projected sensitivity with the HFLAV world average as of 2017~\cite{Bediaga:2018lhg}.
At an integrated luminosity of $300\invfb$ the sensitivity to $|q/p|$ is expected to be $0.001$ and that to $\phi$ to be $0.1^\circ$.
The LHCb \upgradetwo will have the power to reach the SM estimates for $\phi$, and characterise possible new-physics contributions, if these are not too suppressed.
The charm sector investigation will be  exploited at BES\,III, LHC, Belle~II (Belle~III) and \HLLHC. The programme could be complemented by some specific initiatives: TauFV at the Beam Dump at CERN~[ID102] and $e^+e^-$ SCT factories. 
BESIII, and possibly future charm-tau factories, will exploit the $e^+e^- \to D^0 \bar {D^0}$ process to perform quantum-correlation measurements.

\subsection{$\tau$ lepton: short-, mid- and long-term}
In addition to probing BSM theories, taus offer a 
good laboratory for EW precision studies and for many observables (including $|V_{us}|$, $\alpha_s(m_\tau)$ and low-energy QCD quantities)~\cite{LusianiESPP19}.

\noindent
{\bf Lepton universality tests with taus:}
these are 
powerful probes to constrain new physics models (especially those 
designed to explain LFUV anomalies).
The most precise tests rely on measurements of the tau mass,
lifetime and leptonic branching ratios, respectively 
best measured by BES III, Belle and ALEPH 
with world-average uncertainties of 0.007\%, 0.172\% and 0.178\%~\cite{Tanabashi:2018oca}.
In the {\bf short- and mid-term}, these can be improved 
by exploiting high luminosity $e^+ e^-$ colliders
like Belle II~\cite{Abe:2010gxa,Kou:2018nap} and the SCT factories. In the {\bf long-term}, 
the proposed high-energy $e^+ e^-$ colliders with high-luminosity runs at the $Z$ peak (\FCCee~\cite{Blondel:2019yqr} [ID132], 
\CEPC~[ID29], possibly \ILC~\cite{Baer:2013cma,Fujii:2017vwa}) also offer opportunities for tests of tau lepton universality. 

{\bf Tau LFV searches:} In the {\bf short-term}, 
Belle II [ID11] has a reach of
order $10^{-9}$ for $\tau \to \mu \gamma$  and 
$10^{-10}$ for $\tau \to 3 \ell$~\cite{Kou:2018nap}, corresponding to a factor 10-50 improvement. This is illustrated in Fig.~\ref{fig:NPscales} (blue), for the reach in new physics scale. 
Belle II can also search for the rare decay $\tau \to \rho \gamma$ with a projected sensitivity of $2\times 10^{-10}$ (at 50~ab$^{-1}$)~\cite{Kou:2018nap}. 
In the {\bf mid-term},  
the two SCT projects~\cite{Bondar:2013cja} [ID49] and~\cite{Peng:2018} offer favourable conditions for a competitive reach on $\tau \to \mu \gamma$. 
TauFV (planned at the Beam-Dump Facility of the
SPS, upstream of SHiP)~[ID102]
is a fixed-target experiment designed to search for LFV in tau decays, 
benefiting from technological developments being
pursued for the \HLLHC experiments [ID152] and future hadron colliders.
TauFV's sensitivity to the $\tau \to 3 \mu$ decay
may improve that of Belle II.
In the {\bf longer term}, runs at the $Z$ pole of \FCCee (\CEPC)
can deliver 
$15\times 10^{10}$ ($3 \times 10^{10}$) $\tau$ pairs~\cite{Blondel:2019yqr} [ID29]; 
improvements of an
order of magnitude with respect to Belle II may thus be possible.

{\bf Precision tau measurements:}
In the {\bf short-term}, Belle II has the potential for improvements,
at the price of considerable work on the limiting systematics. 
In the {\bf mid-term}, the programmes of the STC factories include precision
measurements of low-multiplicity tau branching ratios, and
offer by far the best prospects regarding tau mass measurements~\cite{Bondar:2013cja} [ID49] \cite{Luo:2018njj,Peng:2018}: 
$\Delta m_\tau = \pm 0.012$~MeV, due to a reduction of systematic errors
(for Belle II, 
$\Delta m_\tau = 
\pm 0.10 - 0.15$~MeV~\cite{Kou:2018nap}). 
In the {\bf long term}, future $e^+ e^-$ colliders operating at the $Z$
pole offer the best conditions for significant advances. 
For BR($\tau \to \ell \bar \nu\nu$), 
 both \FCCee and \CEPC [ID29] can
reach a precision of 0.02\%~\cite{Blondel:2019yqr} [ID29]. For the tau lifetime, 
\FCCee aims at a precision  
$\Delta \tau_\tau \sim 0.01\%$ 
(to be compared with 0.026\% at Belle II). For tau mass measurements, \FCCee is less performant than the other possible projects: 
by calibrating on $m_{D^+}$, it aims at 
$\Delta m_\tau = 0.07$~MeV.

\begin{figure}
\centering
\hspace*{-3mm}
\includegraphics[width=0.5\textwidth]{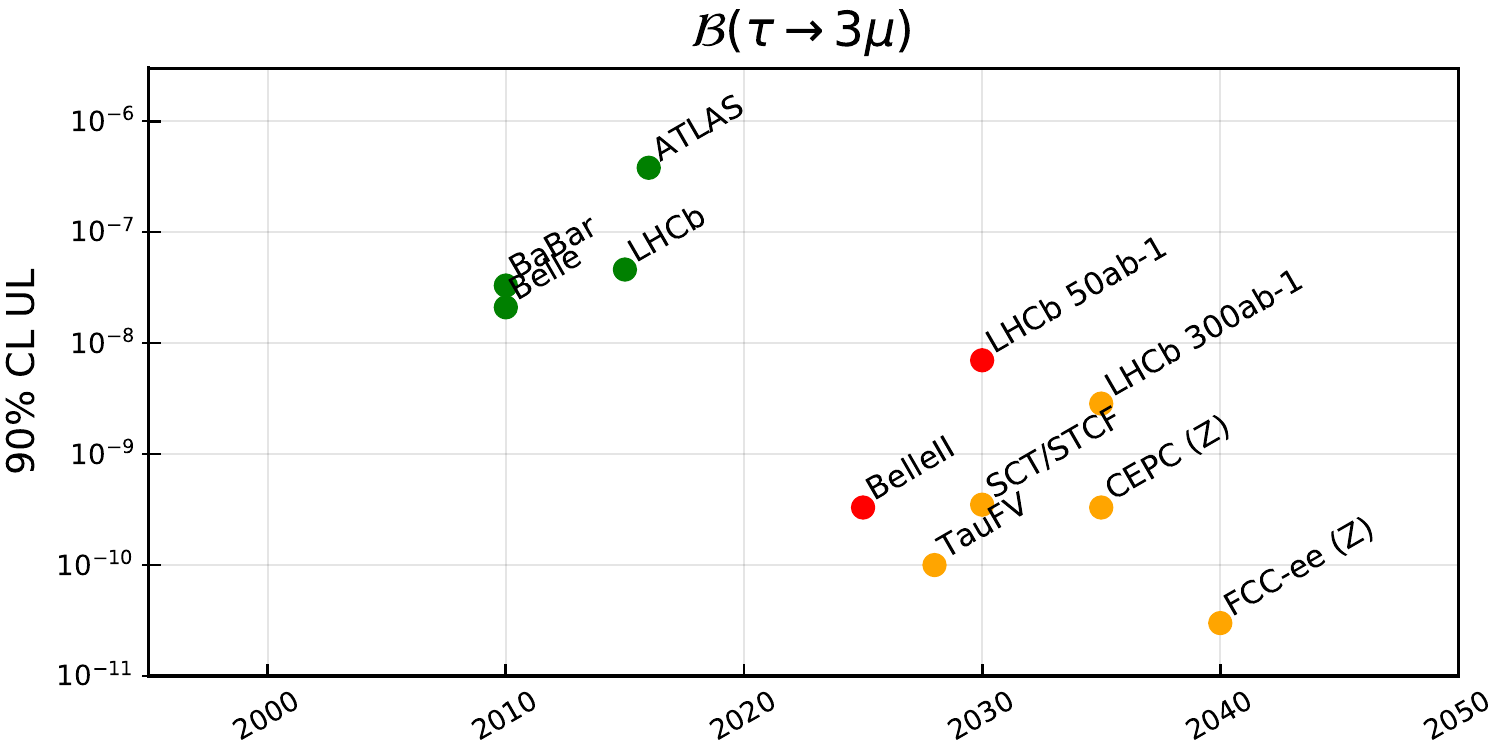}
\includegraphics[width=0.5\textwidth]{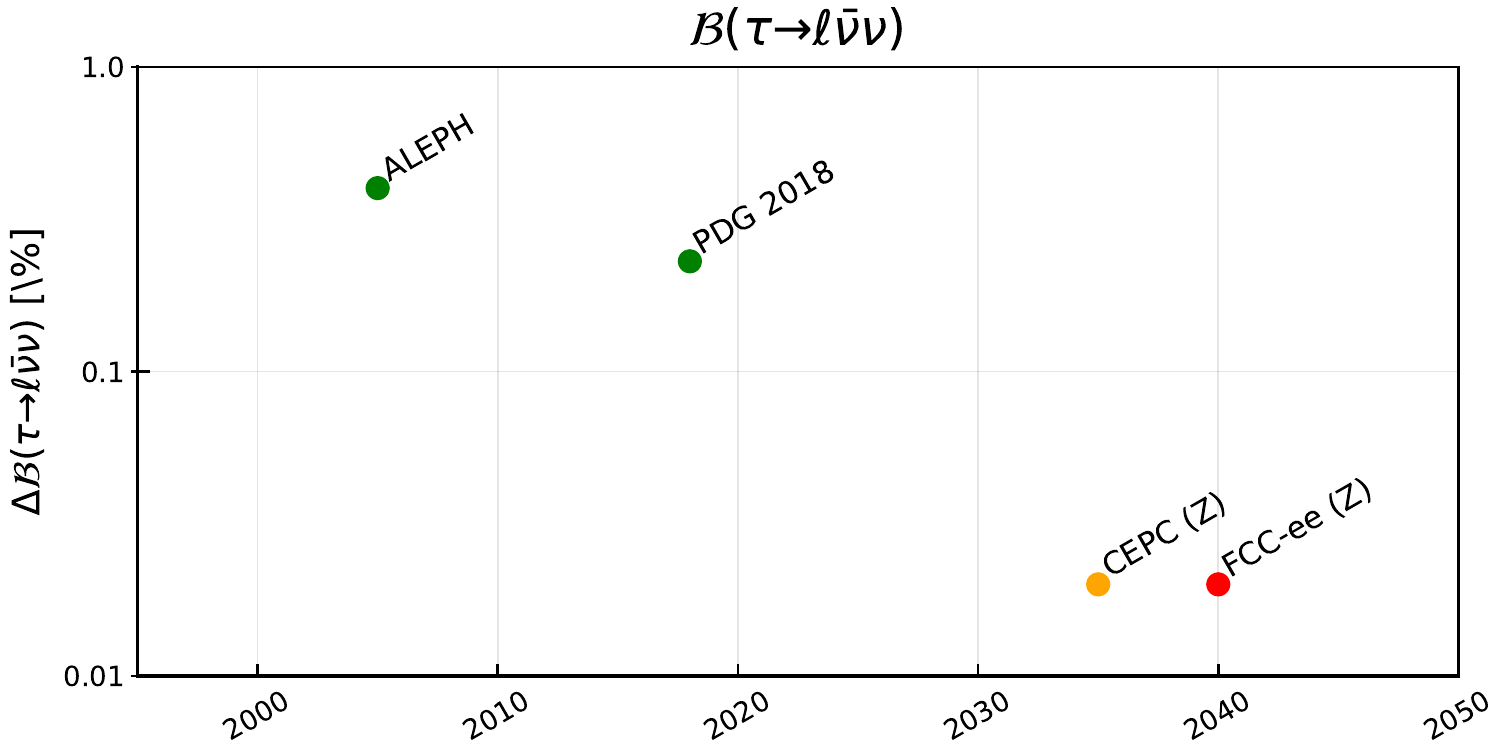} 
\vspace*{3mm}\\
\includegraphics[width=0.49\textwidth]{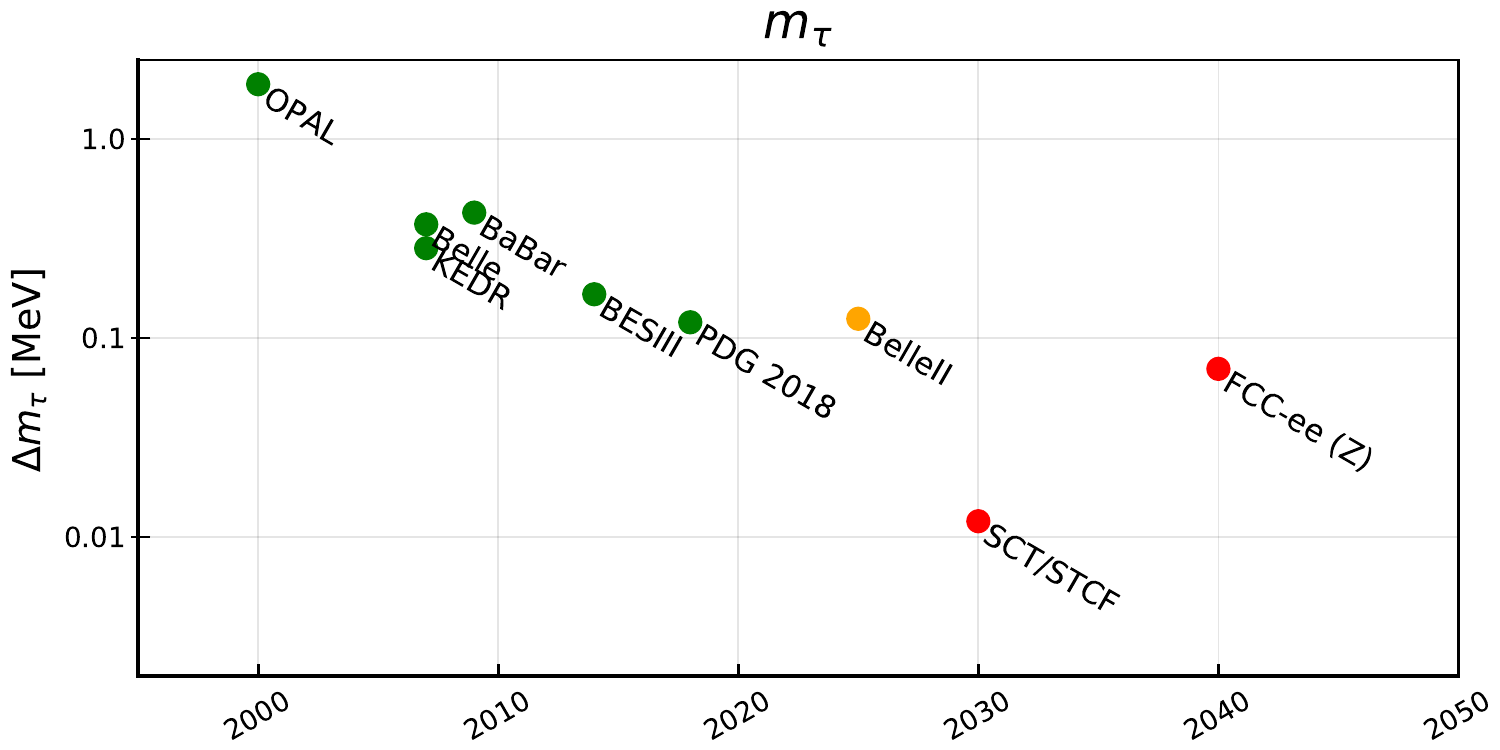}
\includegraphics[width=0.49\textwidth]{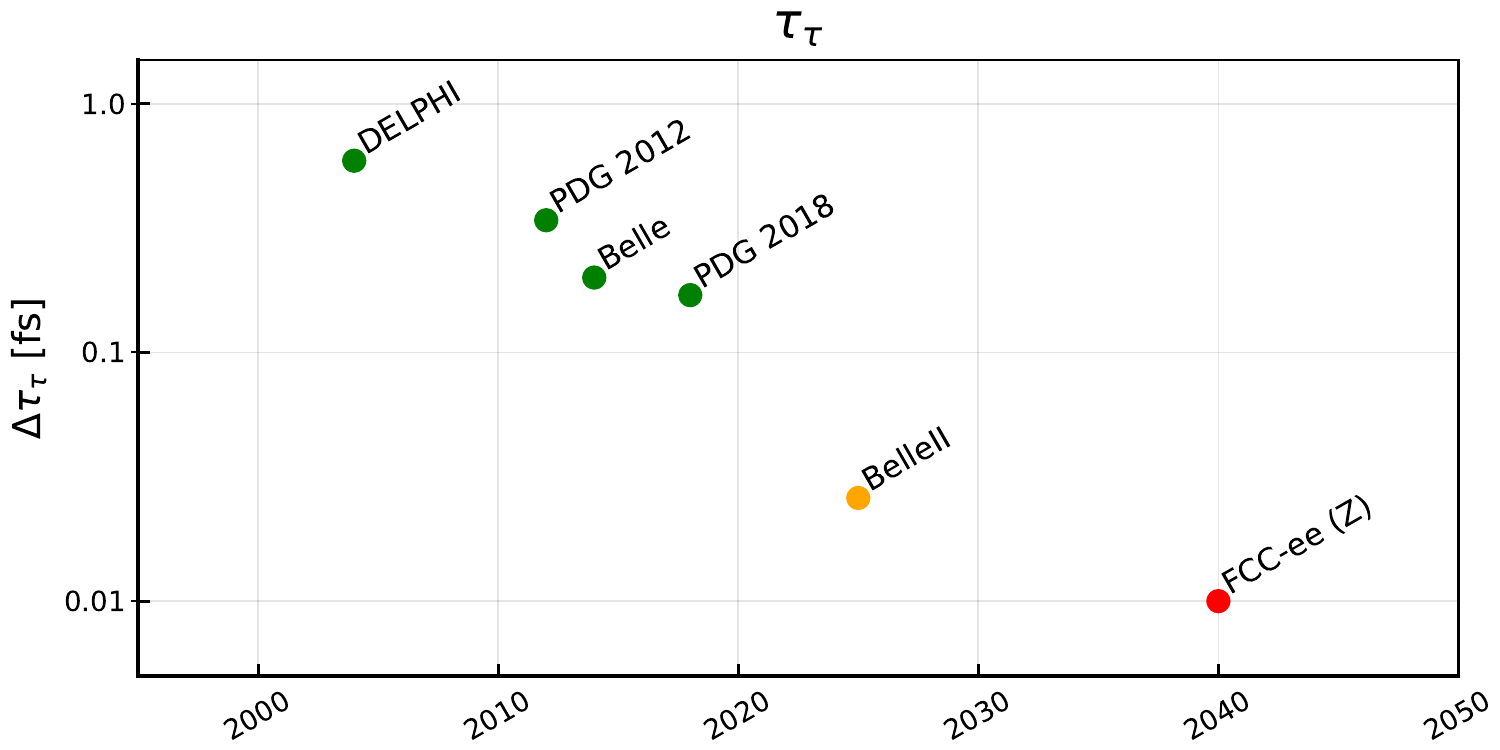} 
\caption{Expected sensitivity and projected precision of  present and future experiments for $\tau \to 3\mu$, $\tau \to \ell \bar \nu \nu$ decays, $m_\tau$ and $\tau_\tau$~\cite{LusianiESPP19}.
Green denotes current measurements, red  points correspond to estimates of future experimental sensitivities based on dedicated studies (relying on extrapolation from past established performances) while orange corresponds to estimates of experimental sensitivities including novel features (for which extrapolation from past experience is more difficult). }\label{fig:Lusiani_prospects} 
\end{figure}

A  summary of the above sensitivity goals and prospects concerning
$\tau \to 3\mu$, $\tau \to \ell \bar \nu \nu$ decays, $m_\tau$ and $\tau_\tau$
can be found in Fig.~\ref{fig:Lusiani_prospects}.

\subsection{The Higgs, top quark, gauge bosons (short-, mid- and long-term)}
 In the SM the flavour structure is encoded in 
 the Higgs couplings to the fermions. 
 
\textbf{ Fermion masses.} In the SM the  
diagonalized Yukawa couplings, $y_f$, are
proportional to the fermion masses, $m_f$,  with a common
factor,  $y_f = \sqrt{2} \kappa_f m_f / v$, where  $\kappa_f^{\rm SM}=1$;
moreover,
there are no tree-level flavour changing couplings of the Higgs. This may change in the presence of new physics, in which case the couplings of the Higgs to the fermions can in general take the form, $\mathcal{L}_{\rm eff} = -\kappa_{f_i} (m_{f_i}/{v}) h \bar f_i f_i + i \tilde \kappa_{f_i} (m_{f_i}/v) h \bar f_i \gamma_5 f_i  
	- \big[\big( \kappa_{f_i f_j} + i \tilde \kappa_{f_i f_j} \big) h \bar f_L^i f_R^j +{\rm h.c.}\big]_{i\neq j}. $
	Currently, only the third generation Yukawa couplings have been measured, having been found to be in agreement with the SM predictions, while for the Higgs couplings to the first two generations, only upper bounds exist. 
Experimentally, a number of SM predictions for the Higgs couplings to fermions 
must be tested as precisely as possible: {\em (i)} proportionality, $y_f\propto m_f$; {\em (ii)} the factor of proportionality, $\kappa_{f_i}=1$; {\em (iii)} diagonality (no off-diagonal flavour violating couplings at tree level, $\kappa_{f_i f_j}=\tilde \kappa_{f_i f_j}= 0$); {\em (iv)} reality (no CP violation at tree level, $\tilde \kappa_{f_i}=\tilde \kappa_{f_i f_j}=0$) \cite{Nir:2016zkd}.

\begin{figure}[t]
	\centering
	\includegraphics[width=0.75\textwidth]{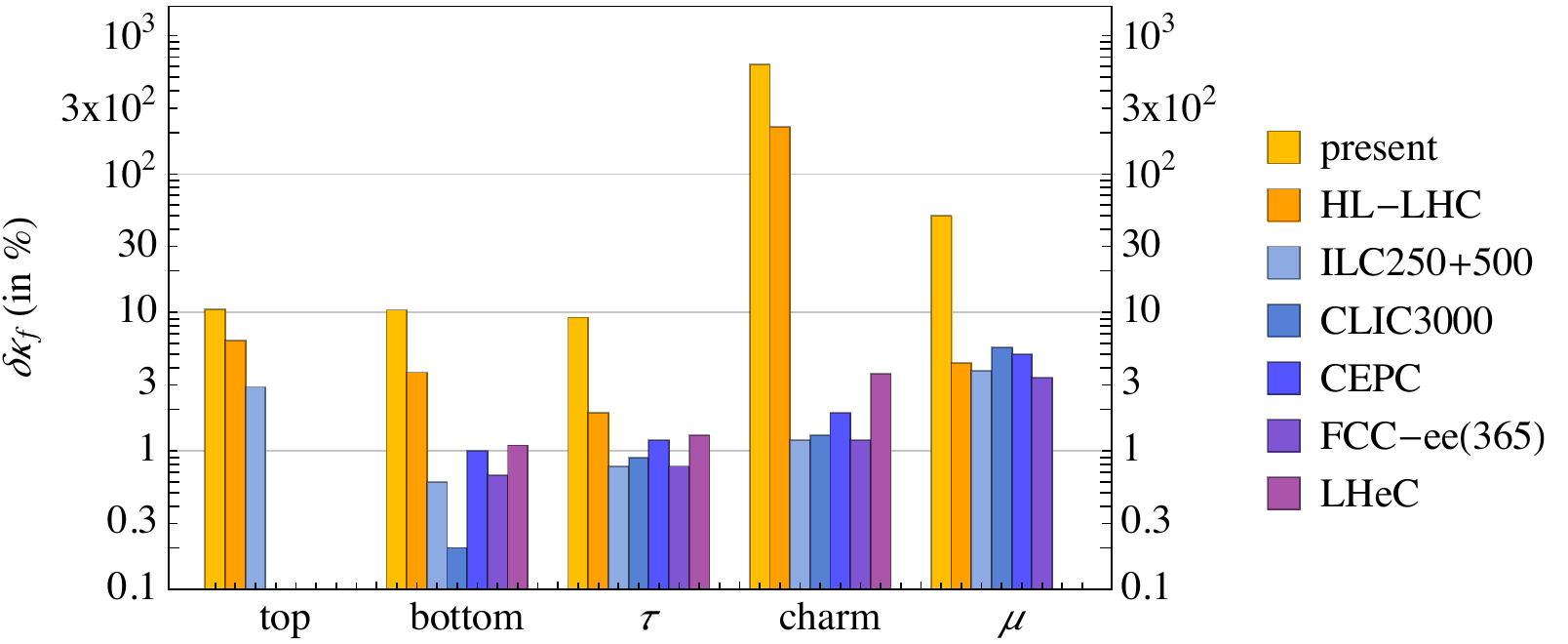}\vspace*{3mm}\\
       \includegraphics[width=0.75\textwidth]{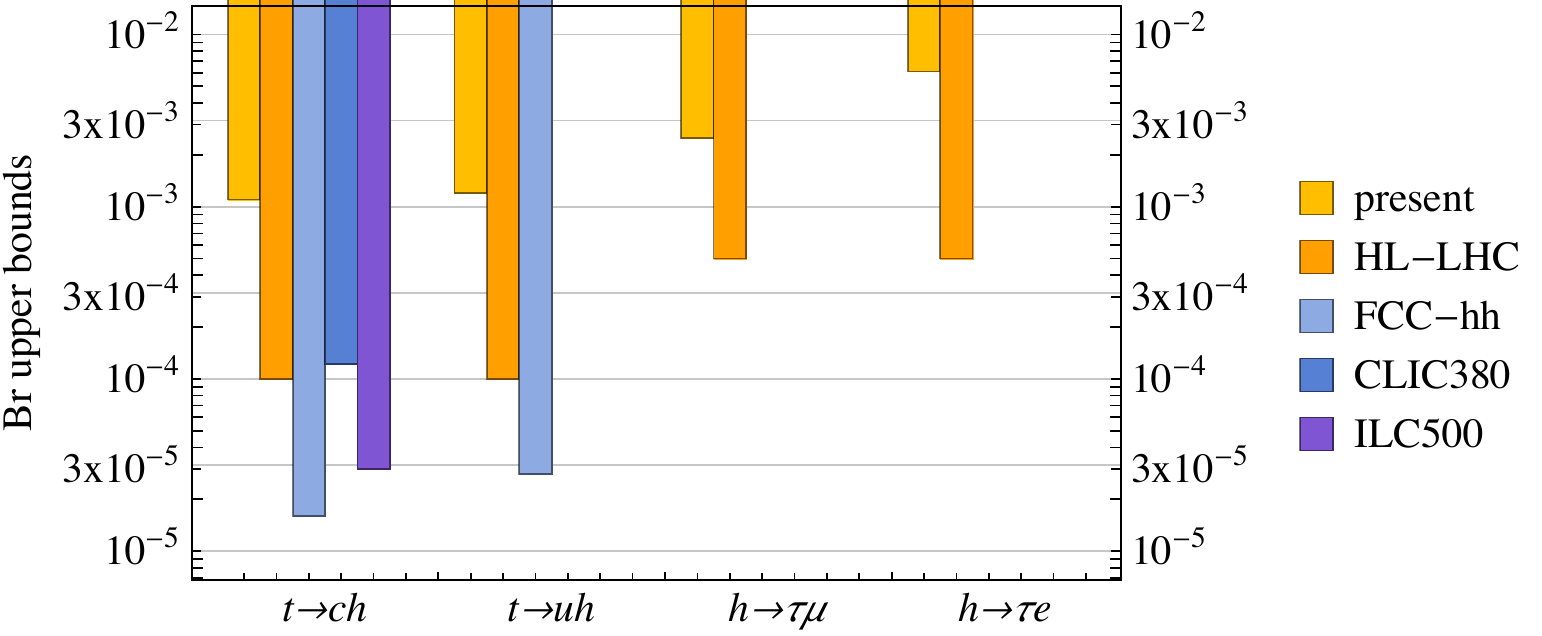}
	\caption{Summary of the available projections for measurements of Higgs Yukawa couplings to quarks and leptons (upper panel), and on select flavour violating decays (lower panel), adapted from Ref.~\cite{Heinemann:2019trx} with further input from Refs.~\cite{deBlas:2018mhx} and~\cite{Bambade:2019fyw}. 
	\label{fig:higgs:forecast}}
\end{figure}
The summary of the expected experimental sensitivies is shown in Fig.~\ref{fig:higgs:forecast} (upper panel) using the $\kappa_f$ framework as a toy approximation to show sensitivity in each channel (a global view of experimental constraints using SM-EFT can be found in Ref.~\cite{deBlas:2019rxi}). 
The sensitivity in the muon channel is now close to what is required to test the SM prediction for the muon Yukawa. This will be the first meaningful test of the 2$^{\rm nd}$ generation Yukawa couplings. A precision measurement does require larger datasets that will be provided in the {\bf short- and mid-term} by the LHC in Run 3 and by the \HLLHC.
In the {\bf mid-term}, the \HLLHC will bound the Yukawa couplings to the third generation fermions and to the muon to a few percent level. 
In the {\bf long-term}, the proposed large scale experiments, \ILC, \FCCee, \CEPC, and \CLIC can significantly improve this precision, to below percent level. They would also measure the charm Yukawa coupling for the first time, and probe it quite precisely, at the percent level. A slightly more modest improvement can be expected at the \HELHC \cite{deBlas:2019rxi}. 
In addition, at \FCCee the upper bound of $\delta y_e/y_e<1.6$ can be achieved after
one year of running at $\sqrt{s}=m_h$. 

\textbf{ Flavour violating Higgs decays}: for the branching ratios of $h\to \tau\mu, \tau e$ decays,  in the {\bf mid-term}  \HLLHC is expected to improve the reach by an order of magnitude, to the level of $5\times 10^{-4}$, see Fig.~\ref{fig:higgs:forecast} (lower panel).  
Figure~\ref{fig:NPscales} illustrates in light (dark) red the scale reach of present data (\HLLHC prospects). Taking $h\to \mu\mu$ and $h\to \tau\tau$ as guidance, one can expect {\bf in the long term} another one to two orders of magnitude improvements for $h\to \tau\mu$ at \FCChh \cite{deBlas:2019rxi}[ID133].

\textbf{The top quark} is unique among the SM fermions, with its ${\mathcal O}(1)$ coupling to the Higgs. Such a large Yukawa coupling for the top is also the origin of the weak scale hierarchy problem -- the quadratically divergent corrections to the Higgs mass are a problem precisely because of it. 
It is thus very common for new physics models that address the  hierarchy problem to also lead to modifications of the top quark properties.  
Studying precisely the latter may in addition give insight into the origin of the SM flavour puzzle, or at least as to why one and only one Yukawa coupling is large.  Experimentally, new physics is probed using FCNC top decays, $t\to c\gamma, cZ, cg, ch$. 
The present upper bounds on their branching ratios are in the range $10^{-3}-10^{-4}$, and will be improved by an order of magnitude or more in the {\bf mid-term} at \HLLHC, resulting in about a two-fold increase in the reach to the effective new physics scale, cf. Fig. \ref{fig:NPscales}.
Also in the mid-term, the reach for  $t\to hc, hu$ decays is expected to improve  by one order of magnitude from the present $10^{-3}$ level,   at \HLLHC; see Fig.~\ref{fig:NPscales} (dark red) for the corresponding sensitivity to new physics scales. This is expected to be further improved in the {\bf long-term}  at \ILCFiveHundred and \FCChh, see Fig.~\ref{fig:higgs:forecast} (lower panel; note that for certain channels the projections for some long-term projects are not available).

\textbf{The $Z$ boson} is also a sensitive probe of new physics: for instance, the observation of flavour violating $Z\to e\mu, \mu\tau, e\tau$ decays would be a clear evidence of new physics, for instance, the existence of sterile neutral fermions. The current limits on these decays are ${\mathcal O}(10^{-6}-10^{-5})$, and could be improved by several orders of magnitude, down to ${\mathcal O}(10^{-9})$ at \FCCee \cite{Abada:2019lih}[ID132].

Lepton flavour violating transitions such as $e^+e^-\to e^+\tau^-$ would also probe \textbf{contact interactions}. The LFV operators of the schematic form $(\bar e e)(\bar e \tau)$ could be well probed at future high-energy $e^+e^-$ colliders, and would increase the present bound of $\sim9$~TeV (on the scale of the contact interaction) to 35~TeV at a \CLIC running at 3~TeV~\cite{deBlas:2018mhx}[ID145].

\section{Flavour and dark sectors (short-, mid- and long-term)}

Flavour physics could be instrumental in searches for dark matter or dark sectors. Light dark sector particles can be produced in \textbf{flavour violating rare decays}, for instance in $B\to K X_{\rm dark}$, $B\to D X_{\rm dark}$, $D\to K X_{\rm dark}$, $K\to \pi X_{\rm dark}$, either through tree level or one loop mediated emissions of dark sector particles, see, e.g., Refs.~\cite{Bird:2004ts,Kamenik:2011vy}. Such decays are exciting and increasingly important drivers in searches for dark sector candidates, since they are often the dominant production channels. Particles that 
either carry nonzero flavour or have flavour violating couplings  may  thus be produced in the flavour violating transitions at tree level, as for instance, heavy neutral leptons \cite{Gorbunov:2007ak} and the  axiflavon~\cite{Calibbi:2016hwq,Ema:2016ops,Wilczek:1982rv}. 
Examples of particles with flavour diagonal couplings, produced in meson decays induced at one loop via $W^\pm$ exchange, include a light singlet scalar mixing with the Higgs~\cite{OConnell:2006rsp,Batell:2009jf,Winkler:2018qyg},  ALPs (axion-like particles)~\cite{Jaeckel:2010ni}, and some dark matter candidates. It is  interesting that even the experiments whose primary goals are in flavour physics, such as NA62 or the Upgrade II of LHCb, have significant discovery potential in dark sector searches. 

\begin{figure}[t]
	\centering 
	\includegraphics[width=0.8\textwidth]{\main/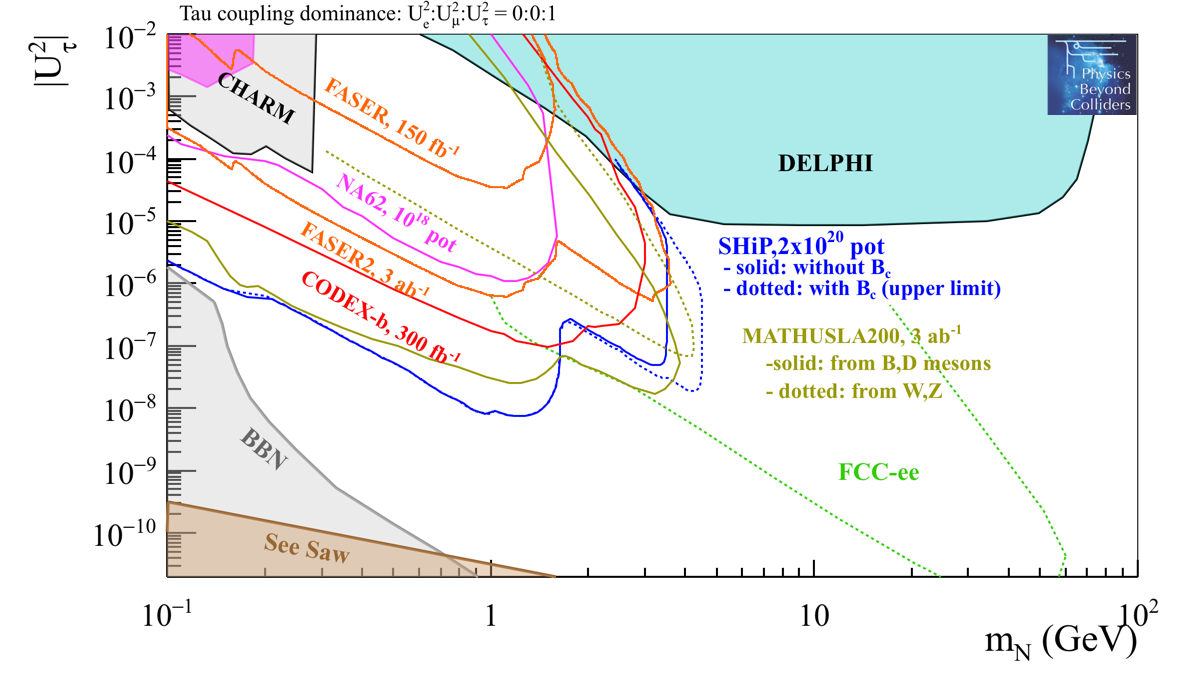}\\
	\caption{Reach of proposed experiments for heavy neutral leptons coupling predominantly to the tau flavour (from Ref.\cite{Beacham:2019nyx}), while for electron and muon flavours they are given in Figs. \ref{fig:FIPs-HNL} and \ref{fig:sec3_LLP}, respectively. For Higgs mixed scalar the reach is given in Fig. \ref{fig:FIPs-DarkHiggs}.
	\label{fig:dark:daulepton}}
\end{figure}

The dark sector particles in the final state, $X_{\rm dark}$, may decay back into visible particles and be detected or, alternatively, result in missing energy and momentum, if they are sufficiently long-lived
or decay to neutrinos. 
When visible, the vertex may be prompt or displaced, depending on the lifetime of $X_{\rm dark}$.   There are a number of proposed experiments, with a clear synergy between different search strategies, such as searches for displaced vertices, monojets, etc. In the {\bf short-term}  the existing and approved experiments, FASER~\cite{Feng:2017uoz}[ID94], NA62~\cite{Gonnella:2017hsz}, NA64~\cite{Banerjee:2016tad}[ID9], SeaQuest~\cite{Berlin:2018pwi}, LHCb, ATLAS and CMS could explore different dark sector models. In the {\bf mid-term} the proposed experiments, such as beam-dump mode of NA62~\cite{Beacham:2019nyx}, LHCb combined with CODEX-b~\cite{Gligorov:2017nwh}, Belle II combined with Gazelle, LDMX~\cite{Akesson:2018vlm} [ID36], MATHUSLA~\cite{Chou:2016lxi,Alpigiani:2018fgd} [ID75], and SHiP~\cite{Alekhin:2015byh} [ID12]   could explore large regions of presently unconstrained parameter space, see Fig.~\ref{fig:dark:daulepton} for two examples,
as well as the discussion in Sects.~\ref{hnl-sec-nu} and~\ref{sec:BSM-FIPs}. 
  Also in the mid-term, Upgrade II of LHCb will have improved sensitivity to dark photons from $D^*\to D\gamma_{\rm dark}$ decays or from bump hunting in the $\mu^+\mu^-$ spectra. In the {\bf long-term}  \FCCee  could significantly increase the reach for dark sector masses, e.g. to a further order of magnitude in mass. 

The  sensitivity of the proposed experiments  to {\bf heavy neutral leptons} mixing with active neutrino flavours is discussed in Sect.~\ref{hnl-sec-nu}.  The  reach for the case in which they mix predominantly with the electron, with the muon or with the tau flavour, as well as the present bounds (shaded regions), are  illustrated in Fig.~\ref{fig:FIPs-HNL}, Fig. \ref{fig:sec3_LLP}, and  Fig.~\ref{fig:dark:daulepton}, respectively.  There is clearly a very strong potential for a whole suite of experiments.  
The same holds for other sample models, for instance for a {\bf Higgs-mixed singlet scalar} $S$, shown in Fig.~\ref{fig:FIPs-DarkHiggs}. While smaller scale experiments such as NA62++, FASER and CODEX-b can cover substantial parts of the parameter space for light new dark sectors, the two
 experiments proposed in this regime
  with larger (and comparable) reach in many models  
 are MATHUSLA and SHiP. MATHUSLA and CODEX-b can also probe dark sector particles with masses above few GeV if they originate from decays of heavier states,  such as $h\to SS$ decays (for first estimates regarding FASER-2 see Ref.~\cite{Boiarska:2019vid}). Overall, the combination of LHCb Upgrade II and CODEX-b, and of ATLAS/CMS and MATHUSLA would cover a very diverse and wide range of new physics options. Heavier dark sectors can be probed in the long-term at \FCCee and/or \CEPC relying on their expected large numbers of $Z$ decays.

There is also sensitivity to dark sector mediators in \textbf{classical flavour observables} such as \textbf{meson mixings},  or \textbf{${(g-2)}$ of the muon}. They can contribute at tree-level, in which case the flavour experiments can probe very high scales for ${\mathcal O}(1)$ couplings, or at loop-level as is the case for some   DM candidates and the accompanying $Z_2$-odd mediators. 

An intriguing possibility is that flavour physics might be directly involved in the structure of the dark sector or in its cosmological consequences. For instance, the stability of dark matter could be due to flavour symmetries; striking examples of cosmological consequences of flavour structures are the option of low-scale ($1-100$ GeV) baryogenesis via heavy neutral leptons~\cite{Akhmedov:1998qx,Asaka:2005pn}, or 
having CP violation solely from the SM sector through meson mixing oscillations, whose signature would be seemingly baryon-violating flavour transitions with missing energy (such as $B\to \Lambda 
X_{\rm dark}$)~\cite{Nelson:2019fln,Elor:2018twp}. 
 Conversely, a better understanding of 
 the SM flavour puzzle could potentially come by studying the flavour structure of dark sector.

\section{The CKM matrix elements: prospects}\label{sec:CKM}
\label{sec:CKM_prospects}

Weak charged currents mix quarks of different generations. In the SM, the strengths of the corresponding transitions are encoded in the unitary Cabibbo-Kobayashi-Maskawa (CKM) matrix~\cite{Cabibbo:1963yz,Kobayashi:1973fv}.
Unitarity, in conjunction with invariance under field redefinitions, implies that all nine complex elements of the  $3\times 3$ CKM matrix are described by four physical parameters. 
In turn, this implies relations between different CKM elements, such as the closure of the standard CKM unitarity triangle (which may not hold  in the presence of new physics).  
Overconstraining the apex of the unitarity triangle from tree- and loop-level quark mixing processes is therefore a powerful way to probe for virtual new physics effects that may arise from mass scales above those which can be directly searched for at colliders. In many cases such indirect probes of new physics will not be limited by either experimental or theoretical systematics at least in the {\bf mid-term}, i.e., in the \HLLHC era, and potentially even for any {\bf long-term} programs (see Table \ref{table:Proj} for expected improvements in lattice QCD for a selection of observables).

\begin{table}
\caption{Current estimates and projections for experimental reach for a selection of observables at Belle~II and LHCb, including Upgrade~II, compared to lattice QCD determinations of hadronic inputs for the {\bf short-} and {\bf mid-term}, taken to be, respectively, the Phase I and II stages defined in Ref.~\cite{Cerri:2018ypt}. It is assumed that the QED corrections to lattice QCD results will be calculated. For a more complete listing of lattice QCD projections see Table 42 of Ref.~\cite{Cerri:2018ypt}.}
    \label{table:Proj}
    {\footnotesize
\begin{center}
\begin{tabular}{ccccc}
\hline\hline
Quantity & Ref. &present error & short-term & mid-term\\
\hline
$(\Delta m_s/\Delta m_d)_{\rm exp}$ & \cite{Tanabashi:2018oca} & $0.4\%$ & - & -
\\
$\xi$ for $(\Delta m_s/\Delta m_d)_{\rm theor}$ & \cite{Cerri:2018ypt} & $1.4\%$ & $0.3\%$ & $0.3\%$
\\
\hline
$B\to \pi$: $|V_{ub}|_{{\rm exp}}$ & \cite{Cerri:2018ypt,Bediaga:2018lhg,Kou:2018nap} &    $2.3\%$ & $1.6\%$ & $1.1\%$\\
$B\to \pi$: $|V_{ub}|_{{\rm theor}}$ & \cite{Cerri:2018ypt} &    $2.9\%$ & $1\%$ & $1\%$\\
\hline
$B\to D$: $|V_{cb}|_{{\rm exp}}$& \cite{Cerri:2018ypt,Kou:2018nap} &   $ 2.0\%$ & $1.4\%$ & -\\
$B\to D$: $|V_{cb}|_{{\rm theor}}$&\cite{Cerri:2018ypt} &    $1.4\%$ & $0.3\%$ & $0.3\%$\\
\hline
$B\to D^*$: $|V_{cb}|_{{\rm exp}}$ & \cite{Kou:2018nap}& $1.2\%$ &- & - \\ 
$B\to D^*$: $|V_{cb}|_{{\rm theor}}$ & \cite{Cerri:2018ypt}& $1.4\%$ & $0.4\%$ & $0.4\%$ \\  
\hline
$\Lambda_b\to p(\Lambda_c)$: $|V_{ub}/V_{cb}|_{{\rm exp}}$ & \cite{Bediaga:2018lhg} &   $ 6\%$ & $1\%$ & $1\%$\\
$\Lambda_b\to p(\Lambda_c)$: $|V_{ub}/V_{cb}|_{{\rm theor}}$  &\cite{Cerri:2018ypt} & $4.9\%$ & $1.2\%$ & $1.2\%$\\
\hline
\hline
\end{tabular}
\end{center}
}
\end{table}

\begin{table}
\caption{Relative uncertainties on the predictions of UT parameters and angles, using current and extrapolated input values for measurements and theoretical parameters (UTfit collaboration, from~\cite{Cerri:2018ypt}, with short-(mid-)term taken as Phase I(II) stages defined in~\cite{Cerri:2018ypt}).  }
\label{tab:utfit_global}
{\footnotesize
 \begin{center}
       \begin{tabular}{ccccccccc}
               \hline\hline
                & $\lambda$ & $\bar{\rho}$ & $\bar{\eta}$ & $A$ & $\sin2\beta$ & $\gamma$ & $\alpha$ & $\beta_s$ \\ \hline
                       Current &0.12\%& 9\% & 3\% & 1.5\% & 4.5\% & 3\% & 2.5\% & 3\% \\
                       short-term  &0.12\%& 2\% & 0.8\% & 0.6\% & 0.9\% &  0.9\% & 0.7\% & 0.8\%\\
                       mid-term  &0.12\%& 1\% & 0.6\% & 0.5\% & 0.6\% & 0.8\% & 0.4\% & 0.5\%\\ \hline
       \hline
       \end{tabular}
\end{center}
}
\end{table}

\begin{figure}[t]
\begin{center}
\includegraphics[width=0.65\textwidth]{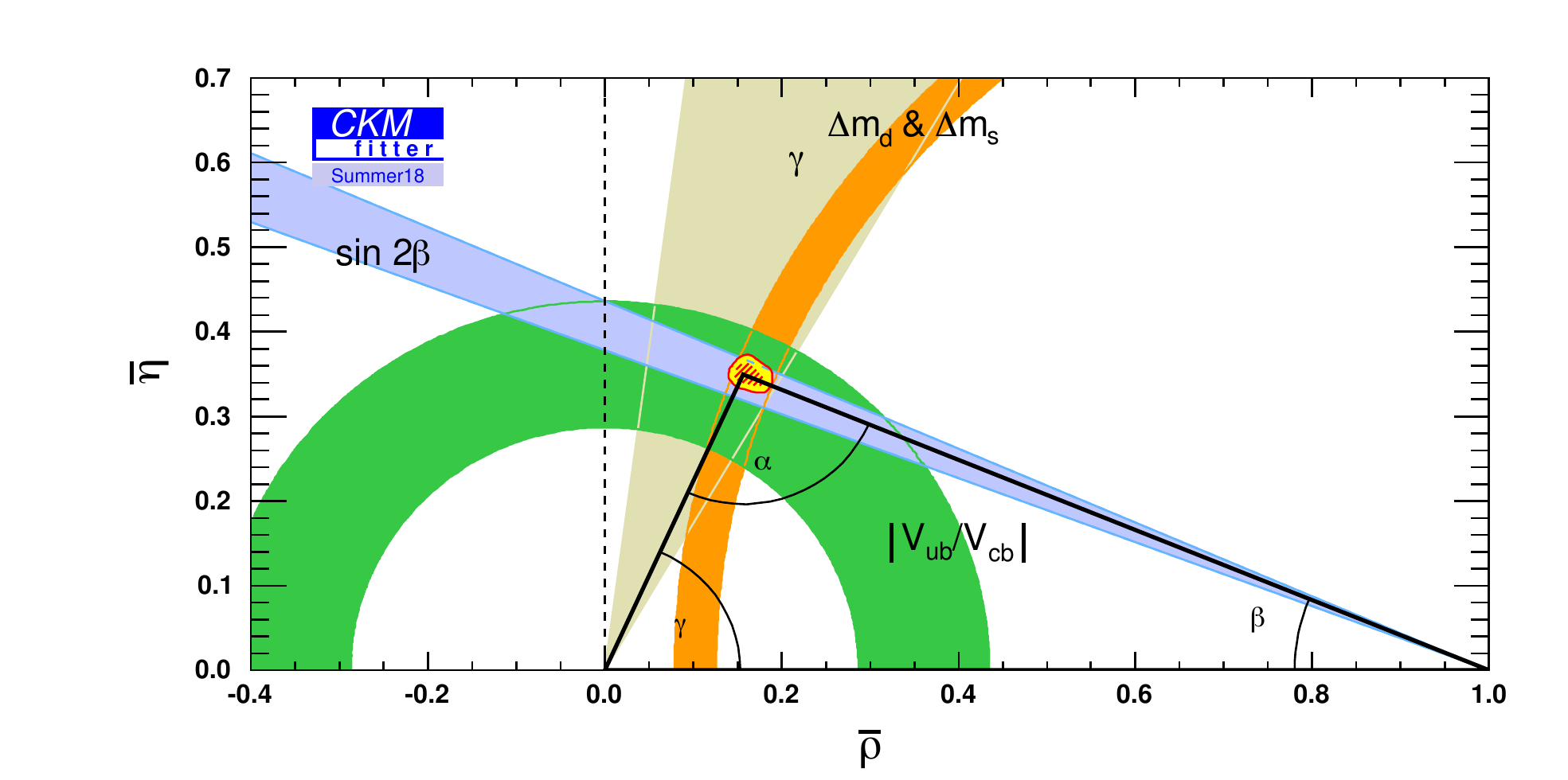}

\begin{minipage}{0.65\textwidth}
\includegraphics[width=\textwidth]{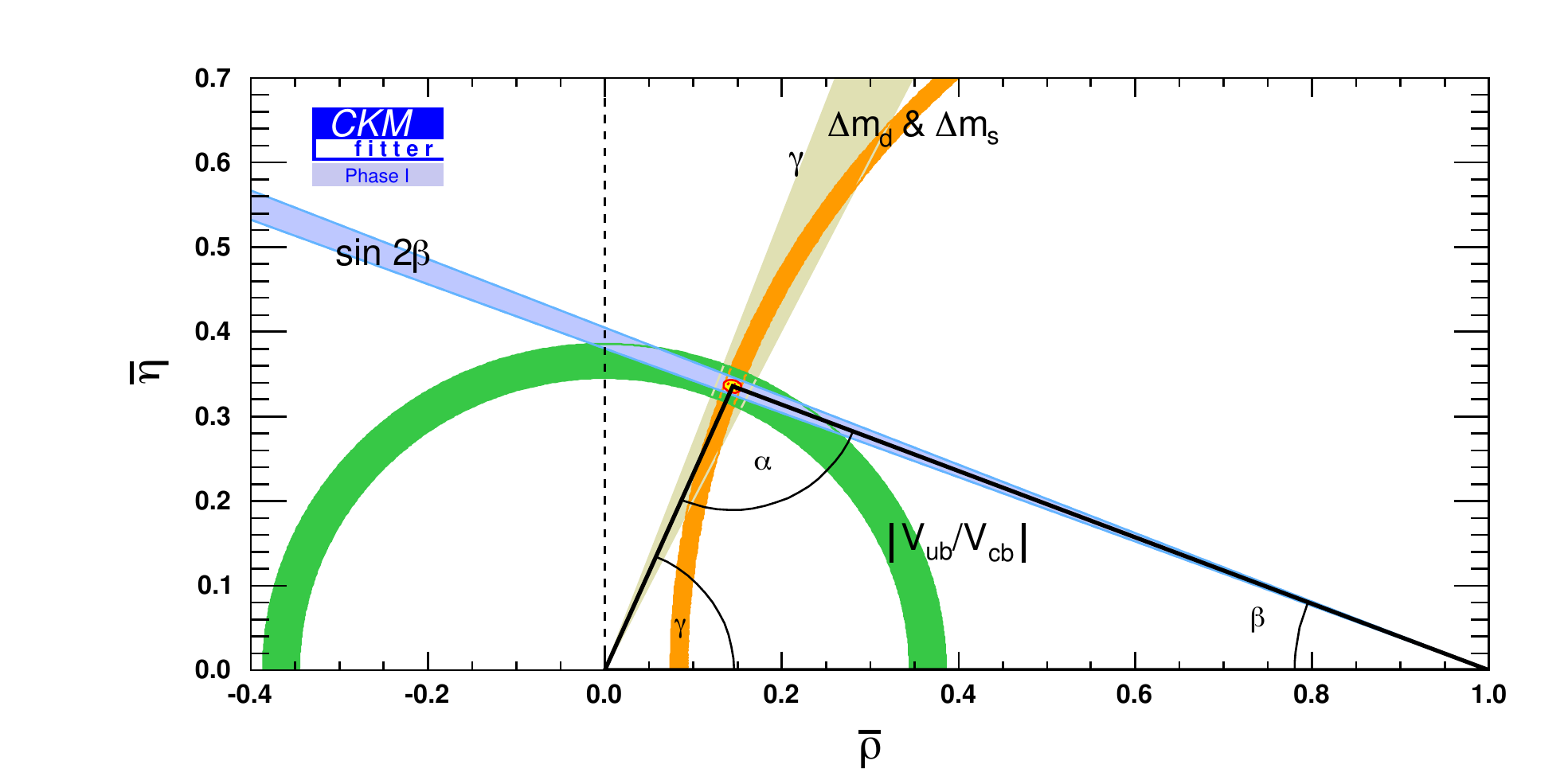}\vspace{-5.36cm}
\mbox{\hspace{7.9cm}\includegraphics[width=0.26\textwidth]{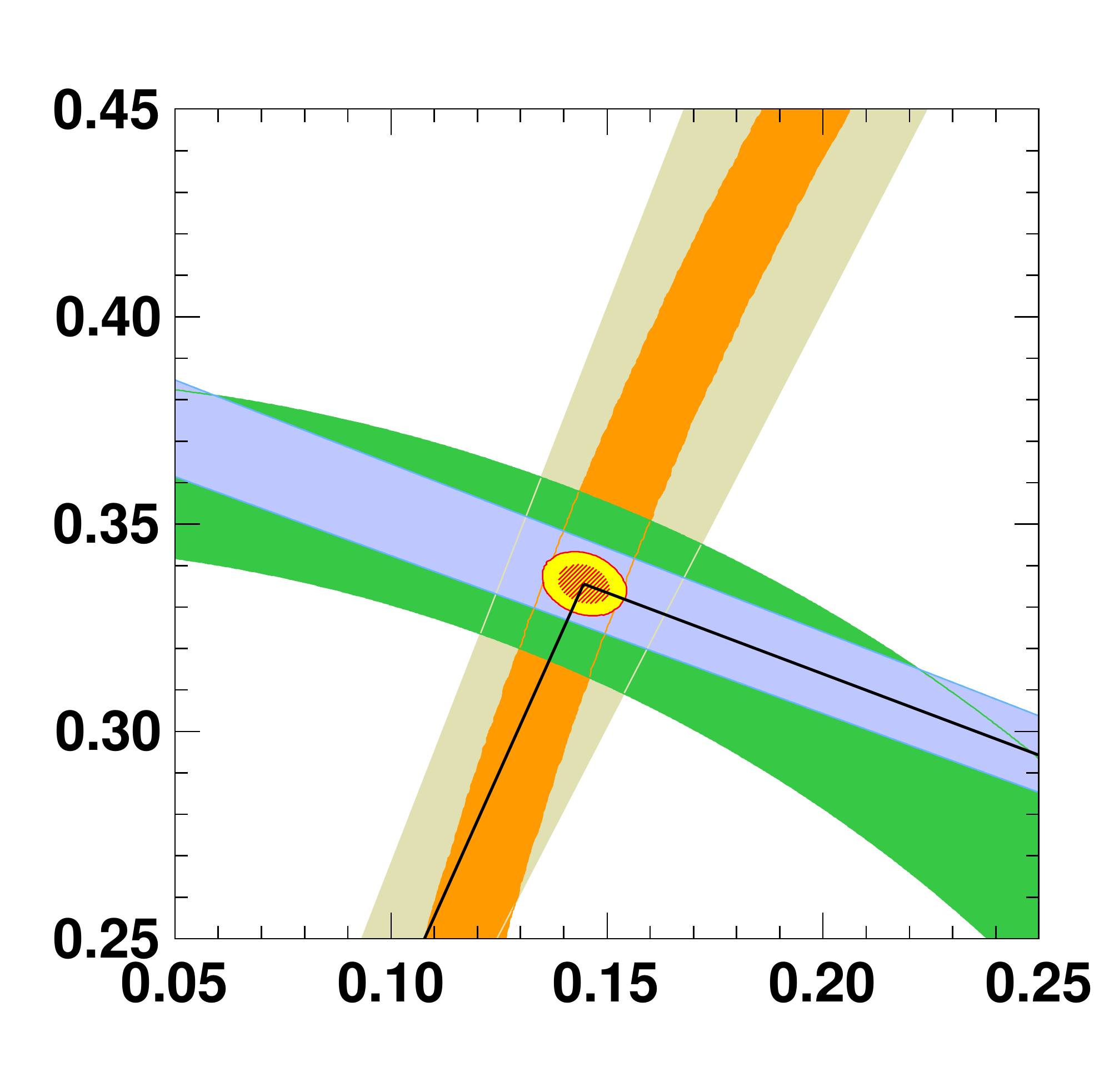}}
\vspace{2.25cm}
\end{minipage}

\begin{minipage}{0.65\textwidth}
\includegraphics[width=\textwidth]{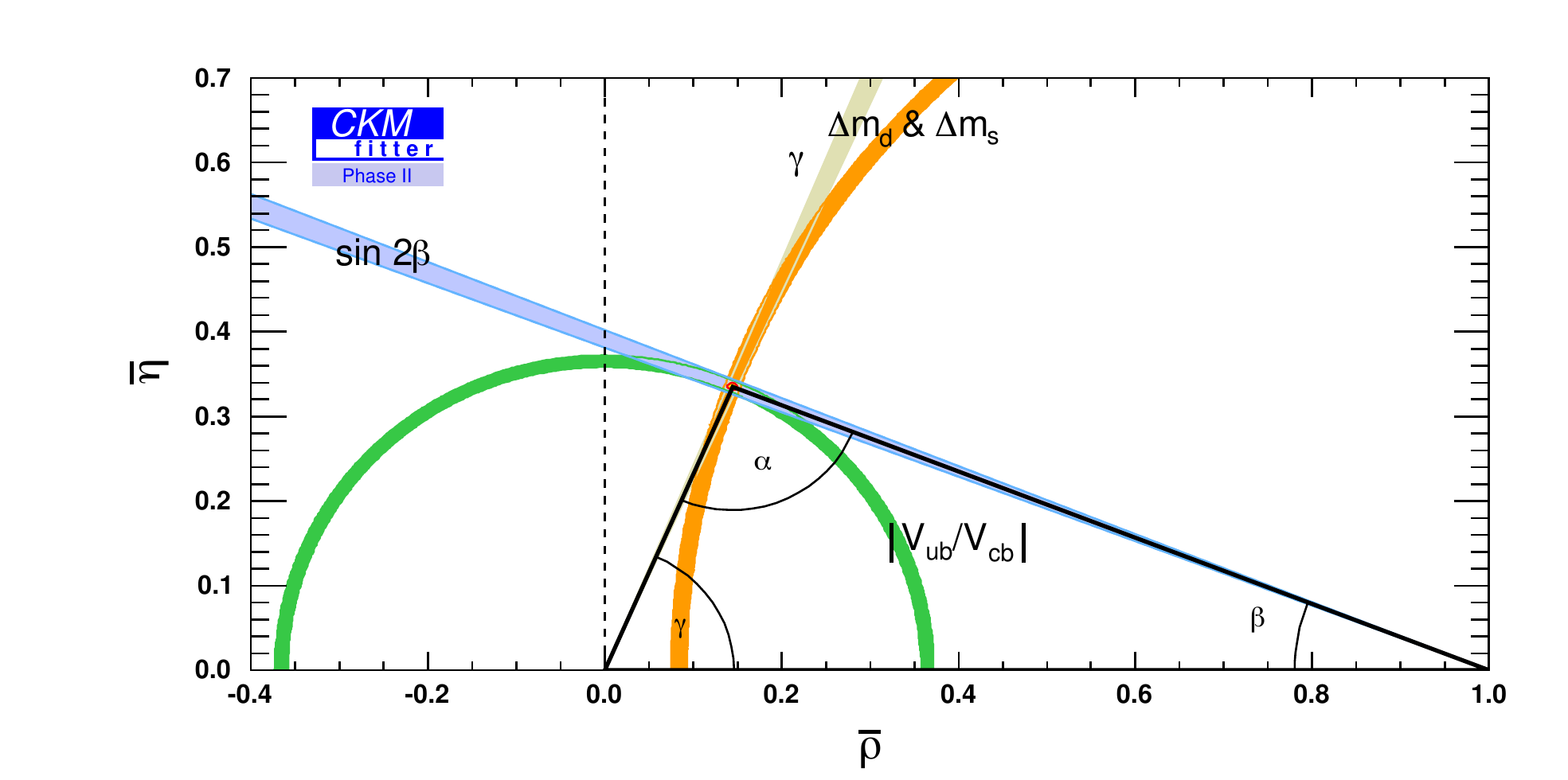}\vspace{-5.36cm}
\mbox{\hspace{7.9cm}\includegraphics[width=0.26\textwidth]{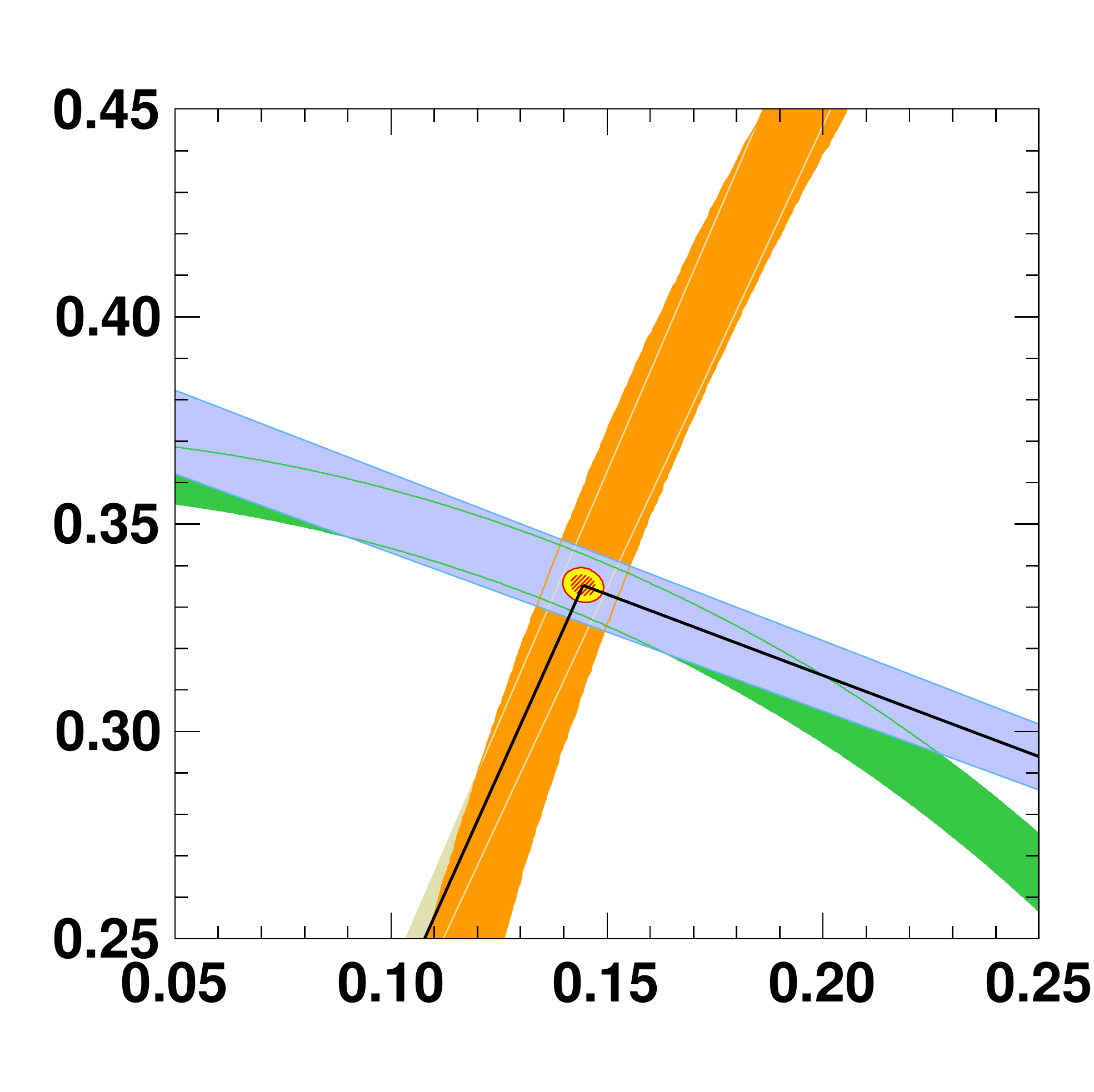}}
\vspace{2.25cm}
\end{minipage}

\caption{\small  Evolving constraints in the $\bar{\rho}-\bar{\eta}$ plane from LHCb measurements and lattice QCD calculations, alone, with current inputs (2018), and the anticipated improvements from the data accumulated by 2025  (23\,fb$^{-1}$) and 2035 (300\,fb$^{-1}$), from top to bottom, respectively.
Figures and underlying assumptions for future projections
from Ref.~\cite{Bediaga:2018lhg}.\vspace*{-1cm} }
\label{fig:UTprojection}
\end{center}
\end{figure}

Figure~\ref{fig:UTprojection} shows the projected {\bf short-} and {\bf mid-term} improvements on constraints in the plane of two unitarity triangle parameters, $\bar{\rho}$ and $\bar{\eta}$, using only expected improvements in LHCb inputs and lattice-QCD calculations, while Table \ref{tab:utfit_global} gives the expected improvements by using both LHCb and Belle-II results. The  precisions quoted in this table  combine statistical (experimental and/or computational) and theory errors, which are not generally expected to be Gaussian -- hence a careful appraisal of the error budget will become key wherever theoretical uncertainties contribute heavily. The increased sensitivity will allow for extremely precise tests of the CKM paradigm. In particular, it will permit the tree-level observables, which provide the SM benchmarks, to  be assessed against  those with loop contributions, which are more susceptible to new physics.  In practice, this already very powerful ensemble of constraints will be 
further strengthened by complementary measurements from Belle II, particularly in the case of $|V_{ub}|$ and $|V_{cb}|$, where $\sim $1\% precision is expected. Improvement on the determination of $|V_{cb}|$ will also greatly impact the constraints on the CKM matrix elements that follow from the measurement of $\varepsilon_K$. It is worth noting that the longstanding few-$\sigma$ tension between the exclusive (lattice-based) and inclusive determinations of $|V_{ub}|$ and $|V_{cb}|$ has not been resolved yet, and has to be subject to further scrutiny. In particular, recent progress in the analysis of $|V_{cb}|$ has not yielded clear conclusions (see Ref.\cite{Gambino:2019sif} for a recent summary).

The \textbf{angle} $\pmb{\gamma}$ is currently the least well known CKM parameter ($\pm 5^\circ$).
In the {\bf short-term} both LHCb Upgrade~I and Belle ~II will provide measurements at $\sim$ 1.5 $^\circ$. In the {\bf mid-term} LHCb Upgrade~II will allow for a  sensitivity at the sub-degree level.
The best sensitivity to $\gamma$ is from $B\to DK$ decays. The anticipated largest systematic uncertainty will be due to the external inputs on the strong phase differences in $D$ meson decays, for which 
BES~III and the SCT factories are expected to provide precise measurements.  Table~\ref{tab:utfit_global} shows the foreseen {\bf mid-term} impact of experimental and theoretical developments on global CKM fits, as estimated by the UTFit group~\cite{Cerri:2018ypt} (comparable CKMfitter estimates can also be found in Ref.~\cite{Cerri:2018ypt}).

The precision measurement of the $B_s$ weak mixing phase, $\phi_s$, will be another highlight of the {\bf mid-term} programme. The expected precision on $\phi_s^{c\bar{c}s}$ at the end of \HLLHC  will be $\sim 5$~mrad for ATLAS and CMS, and $\sim 3$~mrad for LHCb. 
This determination of $\phi_s$ from a loop induced process susceptible to new physics, will be at the same level of precision as the current one on the indirect determination (based on the CKM fit using tree-level measurements). This allows for a precise probe of new physics contributions.

\section{Conclusions}

 Since the last update of the European Strategy, a plethora of new experimental results  has been achieved in flavour physics.  No indisputable evidence of new physics has  emerged so far,  though. 
 The rationale for the observed pattern of masses and mixings of quarks and leptons thus still remains a fundamental open question, which calls for new physics laws.  Precision flavour physics is a fundamental tool to discover them.

  The probing power of flavour physics is manifest from the comparative effective analysis of Fig.~\ref{fig:NPscales}. In the near future, the sensitivities of several observables will reach very high NP scales, $10^2-10^5$~TeV---scales which are beyond the reach of high-energy colliders.  Note that this analysis does not claim that physics at---or below---the  scale depicted is guaranteed to provide a measurable deviation from the SM.  All we claim is that a new physics scale  might be first signalled via 
flavour measurements. In spite of its 
 caveats, the figure illustrates well the great physics reach of flavour observables versus that of  direct and electroweak precision searches. Overall, these three quests are complementary and essential.

The field of flavour has been traditionally explored through a wide spectrum of experiments, ranging from low- and intermediate-energies, such as  EDMs, dedicated muon, kaon, tau, charm and beauty experiments, to the high-energy frontier (the LHC experiments), all the way to feebly interacting particle searches. Measurements by these diverse experiments and facilities lead to valuable complementary information on the different pieces required to assemble the flavour physics mosaic. 

In the \textbf{short- and mid-term}, the expected progress of running experiments, the sensitivity goals of those already foreseen and the proposed upgrades of existing ones, will enable a wide range of flavour observables to be determined with unprecedented precision (or establish new impressive limits).

Searches for EDMs of various types of particles are currently being performed. 
Substantial improvements in the short and mid-term are expected: the neutron and proton EDMs down to $10^{-29} $ e$\cdot$cm, and  $10^{-30} $ e$\cdot$cm  for the electron. Dedicated searches for charged lepton flavour violation in the muon sector foresee nominal improvements by as much as a factor of $10\,000$,  owing to the  high-intensity muon beam programmes at PSI, FNAL and J-PARC.   Furthermore, current hints of lepton non-universality in both charged-current and neutral-current semileptonic $B$ decays are expected to be  unambiguously tested.

In the quark flavour arena, kaon physics proceeds with a steady pace of improvements: efforts towards a substantial improvement on sensitivity for the $K^+ \to \pi^+ \nu \bar \nu$ decay 
and a first measurement of the $K_L \to \pi^0 \nu \bar \nu$ decay are underway.  The upcoming results expected from NA62 and the evolution of the Japanese project will guide the future European steps in this research field. 
Heavy quarks, and in general heavy flavour physics will continue to play an important role also in the post-LHC era. Furthermore, the improvements in lattice-QCD calculations are expected to keep pace with advances in  experimental precision,  
 motivating better measurements of observables critical to test the CKM paradigm.  A flourishing new area is also opening in the mid-term for the exploration of NP in flavourful Higgs couplings, as well as in top and gauge boson interactions. 

The LHCb Upgrade II would allow 
 for the full exploitation of the flavour physics potential of \HLLHC  and increase the explored NP mass scale by close to a factor two with respect to the short-term LHCb Upgrade~I.
It would also provide a bridge towards larger scale collider facilities.
Analogous considerations are valid for a possible upgrade of Belle II (Belle III).
Essential to the success of the physics programme is the experiment capability of charge particle identification and systematic uncertainty control
in different  environments 
($pp$  and $e^+e^-$).

In the \textbf{long-term}, the heavy-flavour physics is expected to remain an integral part of the physics programme at future colliders. Experiments at a future high-luminosity $e^+e^-$ collider 
would perform unique heavy-flavour studies 
in specific channels. 
Circular colliders operating at the $Z$  
pole (in particular \FCCee) 
can strongly contribute to develop searches for  the charm Yukawa coupling and flavour-violating Higgs and $Z$ couplings, lepton flavour violation and precision tau physics, and dark sector searches.   
At a circular high-energy $pp$ collider, the top Yukawa coupling, flavour-violating top-Higgs couplings,  and other heavy flavour physics program related to $b$ and $c$ quark decays would be best studied with a dedicated experiment, along the lines of LHCb.

Furthermore,  from both the experimental and the theory side, a novel synergy  between the searches for flavour violating decays and that  for feebly interacting and dark particles is emerging. Searching for exotic signatures in flavour violating decays  
may have profound implications for our understanding of the Universe, and should be part of any broad program of searches for dark sectors. 
 High-energy colliders will explore a large number of signatures and cover a large fraction of the parameter space for the high-mass range (above 10~GeV). 
Nevertheless fixed-target smaller-scale experiments, LHC projects dedicated
to long-lived particles and beam-dump facilities may provide complementary information to explore a lower mass range  
(1~MeV - 10~GeV) and open new interesting research lines. 

The foreseen unprecedented improvements in sensitivity, together with the novel  channels to be explored, will make  the next decade particularly exciting for the flavour physics arena. 
In summary, the combination of
quark and lepton searches 
for flavour and CP violation  
 at different frontiers is 
a formidable tool to discover new physics. 
Flavour physics must be a crucial ingredient of the future strategy of particle physics.

\label{chap:flavour}
\chapter{Neutrino Physics} 
\label{chap:neut}

\section{Introduction}

The discovery of neutrino oscillation proves that neutrinos have non-zero masses. This is one of the few solid experimental proofs of physics beyond the Standard Model, as new interactions or new elementary particle states are needed to introduce this mass term in the Lagrangian.

Moreover, the extremely low mass of the neutrinos, well below the eV scale, sets them far apart from the other fermions. This extraordinary lightness might be related to new physics at a very high scale, as proposed by the see-saw models. There is the tantalizing possibility suggested by the leptogenesis hypothesis that the phenomena at these high scales could explain the baryon asymmetry in the Universe. Moreover neutrinos could be a completely new kind of particle, a Majorana fermion, identical to its antiparticle. If this is realised in nature, new processes violating the conservation of the lepton number are possible.
For all these reasons, neutrinos are therefore widely considered as a unique window to BSM physics.

These considerations have triggered a very vibrant experimental program world-wide that has made rapid progress in the last years. With the discovery of the third mixing angle $\theta_{13}$ the three neutrino mixing framework has been established.

A new exciting phase of experiments and discoveries opens up, that will be covered in this chapter. Moreover, as neutrinos are also a special probe of dense astrophysical systems, there is a strong synergy at many levels with astroparticle physics that will be covered in Chapter~\ref{chap:cosm}.

\subsection{The big questions related to the neutrino masses }



Thanks to the recent discoveries in the sector of neutrino oscillations
 we have now a clear zeroth-order picture of neutrino properties, which however, raises a number of theoretical and phenomenological questions: Why are neutrino masses many orders of magnitude smaller than any other fermion mass in the Standard Model? Are neutrinos their own antiparticles? What are the actual values of neutrino masses (absolute mass scale and mass ordering)? Is the CP symmetry violated in lepton mixing? What are the precise values of the mixing angles and why is lepton mixing so much different than quark mixing? Are there observable deviations from the standard three-neutrino picture (e.g., non-standard interactions or non-unitarity of the mixing matrix)? 
Answering these questions is the main focus of the present and future neutrino experimental program. It is of paramount importance as it offers a unique window on the physics beyond the Standard Model. Furthermore, the hypothesis of leptogenesis for the generation of the baryon asymmetry of the Universe links the origin of neutrino mass with the origin of matter.

The mechanism behind neutrino mass is unknown. 
To obtain finite neutrino masses, the Standard Model has to be extended in some way.
A minimal extension is to introduce gauge-singlet neutrinos (so-called right-handed or sterile neutrinos) which would allow to write down a Dirac mass term for neutrinos, in the same way as for all other fermions. This could indeed be the only source of neutrino masses, but in this case coupling constants need to be smaller than $10^{-11}$ and lepton-number conservation has to be postulated as a fundamental symmetry. However, the electric charge-neutrality of neutrinos offers also the possibility of a Majorana mass term, which would imply that neutrinos are their own antiparticles and break lepton-number by two units. Many possibilities for generating Majorana neutrino masses are known, including the see-saw mechanism with right-handed neutrinos and/or with an extended scalar sector, or radiative neutrino mass models. Therefore, the search for lepton-number violation, in particular via neutrinoless double-beta decay, addresses a fundamental property of the theory of neutrino masses. 

Neutrino masses by themselves do not provide guidance towards the energy scale of new physics responsible for generating them. There is a vast range for the scale of new physics extending from sub-eV up to the GUT scale of $10^{16}$~GeV. In order to make progress in view of this multitude of possibilities a wide range of complementary observables needs to be explored. These include (i) the search for sterile neutrinos at various different mass scales including oscillations at the eV scale and heavy neutral leptons at collider and beam dump experiments, (ii) lepton number violation in neutrinoless double-beta decay or at high-energy colliders, (iii)~charged lepton-flavour violation, (iv)~precision measurements in the neutrino sector, and (v)~search for non-standard neutrino properties such as exotic interactions or non-unitarity of the \mbox{3 $\times$ 3} mixing matrix.

Since new states or new BSM interactions are required to explain the neutrino masses, the experiments must be ready for unexpected phenomena, and if possible provide alternative complementary measurements to over-constrain the three-flavour parameters. An example of these phenomena is given by Non Standard Interactions (NSI) between neutrinos and the other fermions. NSI are today constrained only weakly and they might modify the propagation of neutrinos in matter.

\section{Present knowledge of neutrino mixing parameters}


The neutrino mixing is described by the Pontecorvo-Maki-Nakagawa-Sakata (PMNS) matrix, analogous to the CKM matrix in the quark sector, connecting the neutrino mass eigenstates to the flavour eigenstates. It is parameterized by three mixing angles, $\theta_{12}$, $\theta_{13}$ and  $\theta_{23}$, and by a complex phase $\delta_{\rm CP}$ (two supplementary phases are present in the case of Majorana neutrinos, but they do not affect the neutrino oscillations). The latter might produce CP violating effects if it is different from 0 and $\pi$. The oscillation experiments are also sensitive to the difference of the masses squared, $\Delta m^2_{21}= m^2_2 -m^2_1$ and $\Delta m^2_{31}= m^2_3 -m^2_1$
where $m_1$, $m_2$, and $m_3$ are the  mass eigenvalues. The neutrino mass ordering refers to the two present possibilities: $m_1 < m_2 < m_3$ (normal) or $m_3 < m_1 < m_2$ (inverted), where by convention $m_1 < m_2$ has been chosen, with $\Delta m^2_{21}$ being the parameter governing the oscillation of the solar neutrinos.

We summarize here the results of a global analysis of neutrino oscillation data in the three-flavour framework from the NuFit group \cite{Esteban:2018azc} comprising all available data from solar, atmospheric, reactor and accelerator neutrino experiments. Consistent results are obtained also by other groups \cite{Capozzi:2018ubv,lisi:2019EPSSU_Granada,deSalas:2017kay}.

To date the determination of the three mixing angles, the complex phase, and the two neutrino mass-squared differences is given in Table~\ref{tab:NuFit-results}. The best fit occurs for the normal mass ordering ($\Delta m^2_{31} > 0$), while the inverted ordering is disfavoured with a difference of $\Delta \chi^2 = 9.3$ between the two hypotheses. 
The mixing angles $\theta_{12}$ and $\theta_{13}$ are well measured with relative precision at $3\,\sigma$ of 14\% and 9\%, respectively. The uncertainty of the third angle $\theta_{23}$ is relatively large (24\% at $3\,\sigma$) and includes maximal mixing; the best fit occurs in the second octant ($\theta_{23} > \pi/4$), with the first octant being disfavoured with $\Delta\chi^2 = 6$. The preferred range for the complex phase $\delta_{\rm CP}$ is between $180^\circ$ and $360^\circ$ and CP conservation is allowed with $\Delta\chi^2 = 1.8$.

\begin{table}[htbp]
    \caption{Three-flavour oscillation parameters from the NuFit-4.0 global
    analysis \cite{Esteban:2018azc}.
    The numbers in the 1$^{\rm st}$ (2$^{\rm nd}$) column are obtained assuming normal (inverted) mass ordering,
    i.e.\ relative to the respective local $\chi^2$ minimum. The best fit point (bfp) is shown as the central value.
    Note that $\Delta m^2_{3\ell} \equiv \Delta m^2_{31} > 0$ for normal ordering and
    $\Delta m^2_{3\ell} \equiv \Delta m^2_{32} < 0$ for inverted ordering. Details on the parametrization and the used data samples can be found in Ref.\cite{Esteban:2018azc}.}
    \label{tab:NuFit-results}
    \vspace*{3mm}
    \centering
        \begin{tabular}{l|cc|cc}
      \hline\hline
      &
      \multicolumn{2}{c|}{Normal Ordering (best fit)} &
      \multicolumn{2}{c}{Inverted Ordering ($\Delta\chi^2=9.3$)}
      \\
      \hline
      & bfp $\pm 1\sigma$ & $3\sigma$ range
      & bfp $\pm 1\sigma$ & $3\sigma$ range
      \\ \hline
      \rule{0pt}{5mm}\ignorespaces
      $\theta_{12}/^\circ$
      & $33.82_{-0.76}^{+0.78}$ & $31.61 \to 36.27$
      & $33.82_{-0.75}^{+0.78}$ & $31.62 \to 36.27$
      \\[2mm]
      $\theta_{23}/^\circ$
      & $49.7_{-1.1}^{+0.9}$ & $40.9 \to 52.2$
      & $49.7_{-1.0}^{+0.9}$ & $41.2 \to 52.1$
      \\[2mm]
      $\theta_{13}/^\circ$
      & $8.61_{-0.13}^{+0.12}$ & $8.22 \to 8.98$
      & $8.65_{-0.13}^{+0.12}$ & $8.27 \to 9.03$
      \\[2mm]
      $\delta_{\rm CP}/^\circ$
      & $217_{-28}^{+40}$ & $135 \to 366$
      & $280_{-28}^{+25}$ & $196 \to 351$
      \\[2mm]
      $\dfrac{\Delta m^2_{21}}{10^{-5}\, {\rm eV}^2}$
      & $7.39_{-0.20}^{+0.21}$ & $6.79 \to 8.01$
      & $7.39_{-0.20}^{+0.21}$ & $6.79 \to 8.01$
      \\[2mm]
      $\dfrac{\Delta m^2_{3\ell}}{10^{-3} \,{\rm eV}^2}$
      & $+2.525_{-0.031}^{+0.033}$ & $+2.431 \to +2.622$
      & $-2.512_{-0.031}^{+0.034}$ & $-2.606 \to -2.413$
      \\[2mm]
      \hline\hline
    \end{tabular}
\end{table}

\section{Measurements of neutrino oscillation parameters}

The study of neutrino oscillations is today focused on the determination of the remaining unknowns in the three flavour framework,
namely the CP-violating phase $\delta_{\rm CP}$, the deviation of the $\theta_{23}$ angle from $\pi$/4 (also called the octant determination), and the mass ordering. This can be done by the long baseline experiments. Another important task is the high-precision determination of all the mass and mixing parameters. For a recent review on the subject refer to \cite{Giganti:2017fhf}. 

\begin{figure} [htbp!]
\begin{center}
\includegraphics[width=12cm]{\main/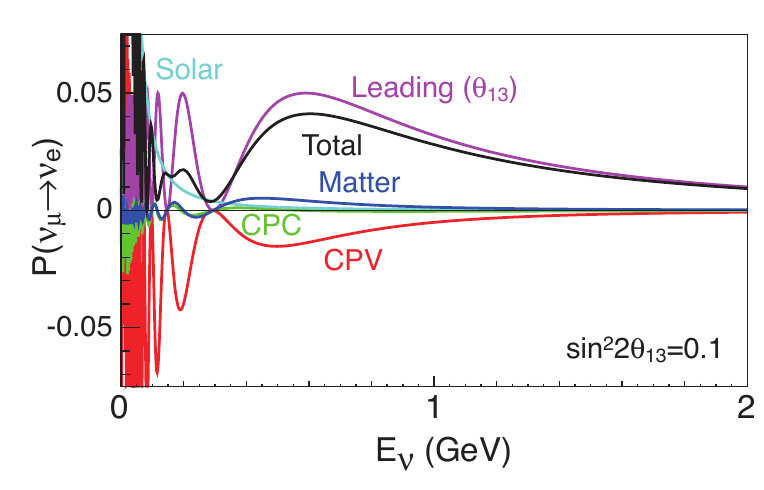}
\caption{\label{fig:t2kappprob} Oscillation probability $\nu_\mu \rightarrow \nu_e$ as a function of neutrino energy with $L = 295$ km, ${\sin}^2(2\theta_{13})$ = 0.1, $\delta_{CP}$= $\pi/2$ and normal mass ordering. The contribution of each term in the oscillation probability~Eq.~(\ref{eq:theta13app}) is shown separately. The first term in that equation is shown by the curve labeled "Leading", the second term is labeled "Solar" (so called because $\theta_{12}$ governs the oscillation of solar neutrinos), the third CP-violating term (CPV) proportional to $\sin \delta_{CP}$ is labeled CPV. The fourth CP-conserving (CPC) term proportional to $\cos \delta_{CP}$ is labeled CPC.}
\end{center}
\end{figure}

The long baseline oscillation experiments will precisely measure the neutrino oscillation probability beyond the leading order term, for a beam of muon neutrinos.
\newpage
The probability of electron neutrino appearance in a beam of muon neutrinos is given by:
\begin{eqnarray}
P(\nu_{\mu} \rightarrow \nu_e) & = & \sin^2 \theta_{23} \frac{\sin^2 2 \theta_{13}}{(A-1)^2} \sin^2 [(A-1) \Delta_{31}] \nonumber \\
&+& \alpha^2 \cos^2 \theta_{23} \frac{\sin^2 2 \theta_{12}}{A^2} \sin^2 (A\Delta_{31}) \nonumber \\
&-& \alpha \frac{\sin 2 \theta_{12}\sin 2 \theta_{13}\sin 2 \theta_{23}\cos  \theta_{13}\sin  \delta_{CP}}{A (1-A)} \sin \Delta_{31} \sin (A\Delta_{31})\sin[ (1-A) \Delta_{31}] \nonumber \\
&+& \alpha \frac{\sin 2 \theta_{12}\sin 2 \theta_{13}\sin 2 \theta_{23}\cos  \theta_{13}\cos  \delta_{CP}}{A (1-A)} \cos \Delta_{31} \sin (A\Delta_{31})\sin[ (1-A) \Delta_{31}] \nonumber\\
\label{eq:theta13app}
\end{eqnarray}
where $ \alpha = \Delta m^2_{21} / \Delta m^2_{31}$, $\Delta_{ji}= \Delta m^2_{ji} L / 4 E$ and $A= 2\sqrt 2 G_F n_e E/ \Delta m^2_{31}$.
The corresponding formula for $P(\bar \nu_{\mu} \rightarrow \bar \nu_e)$ can be obtained by reversing the signs of the terms proportional to $\sin \delta_{CP}$ and to $A$. The different terms contributing to $P(\nu_{\mu} \rightarrow \nu_e)$,  are plotted in Fig.~\ref{fig:t2kappprob} together with the total contribution from matter effects, assuming a baseline $L$ of 295 km.

The first term in Eq.~(\ref{eq:theta13app}) (labelled ``leading'' in Fig.~\ref{fig:t2kappprob}) is the leading order term and corresponds to the 1--3 sector oscillations driven by the squared-mass difference $\Delta m^2_{31}$. 
%
%
The third term (``CPV'') contains the CP-violating part. 
It modifies the leading order term by as much as $\pm$ 30 \% for $\delta_{CP} = \mp \pi/2$, respectively, in the case of neutrinos.
%
%
Matter effects enter through the $A$ term proportional to the ratio $E/E_{res}$, where $E_{res}$ is the energy for which the resonance condition is attained in the medium of constant density. Since the resonance condition can be satisfied either for neutrinos or for antineutrinos, depending on the sign of $\Delta m^2_{31}$ and therefore on the mass ordering, matter effects either enhance $ P (\nu_\mu \rightarrow \nu_e)$ or  $ P (\bar{\nu}_\mu \rightarrow \bar{\nu}_e)$.
The $A$ term is approximately 5\% for T2K and Hyper-Kamiokande, about 20\% for DUNE.

In principle the precise measurement of $P(\nu_{\mu} \rightarrow \nu_e)$ and $P(\bar{\nu}_{\mu} \rightarrow \bar{\nu}_e)$ allows by itself to determine all the remaining unknowns among the the neutrino mixing parameters: as long baseline experiments based on rather pure muon neutrino (or muon antineutrino) beams allow to conduct this measurement with  well-defined experimental conditions, they have received a lot of attention and are today the central focus of the experimental effort. Today they are a strategic scientific priority for the international community and will continue to be for the next two decades.

While multiple ambiguities could play a role in the extraction of physics parameters from the observables, in practice the impact of these ambiguities will be greatly mitigated by two facts. First, reactor neutrino experiments have provided a clean and high-precision measurements of $\theta_{13}$. Second, present and future experiments will provide measurements at different baselines, and therefore with different strength of the matter effects, going from the shorter baseline of T2K and Hyper-Kamiokande (295 km) to the longer baselines of NOVA (810 km) and DUNE (1300 km). This will help decorrelate the effect of CP violation from matter effects.


\subsection{The present experiments T2K and NOVA}

The presently running experiments T2K in Japan and NOVA in the US will continue to produce world-leading results in the coming years. As reported above, in conjunction with other measurements, the data from these experiments already today provides an indication favouring the normal ordering at the 3\,$\sigma$ level, and a weak hint of CP violation in neutrino oscillation. 

\begin{figure} [bhp!]
\begin{center}
\includegraphics[scale=0.35]{\main/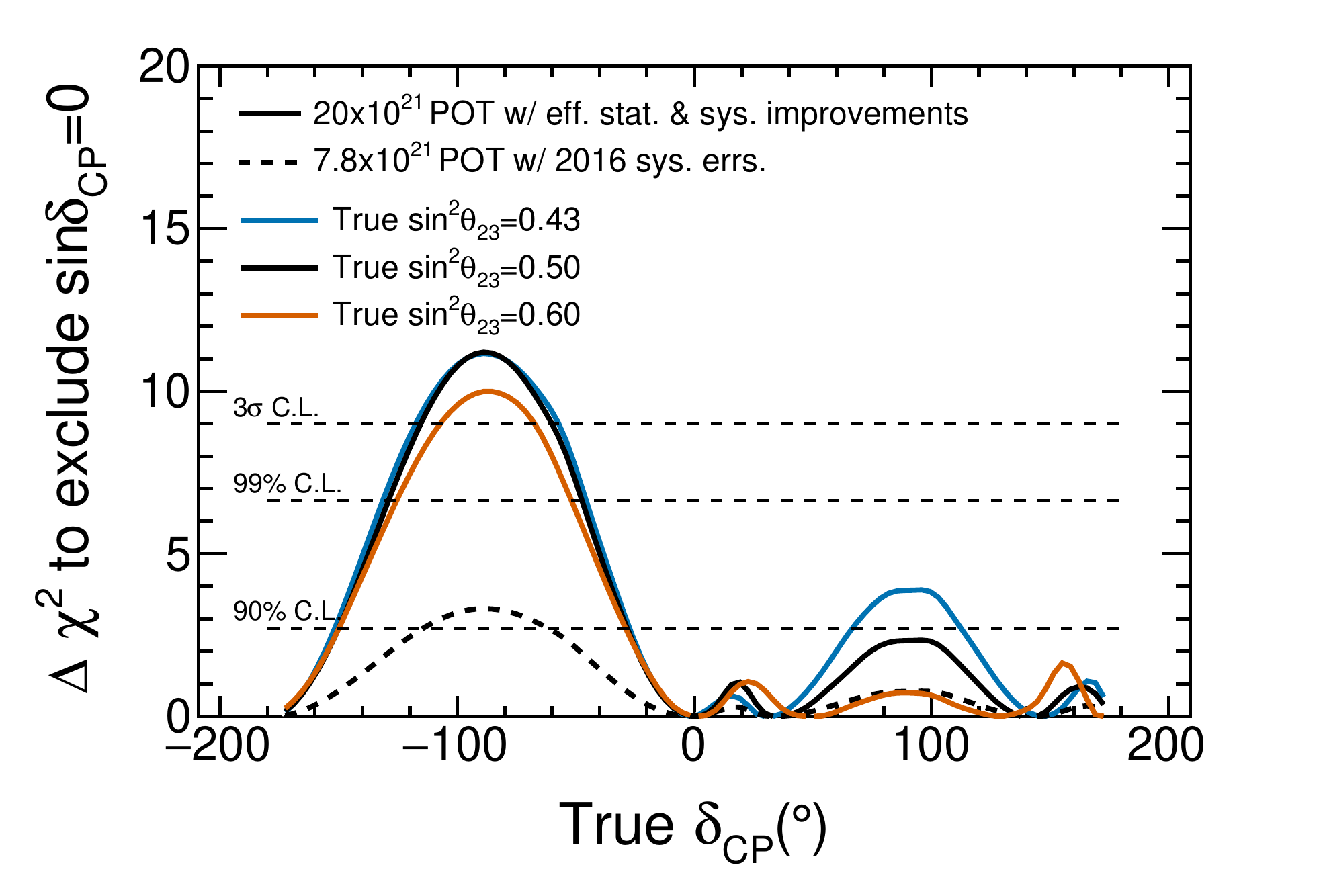}
\includegraphics[scale=0.35]{\main/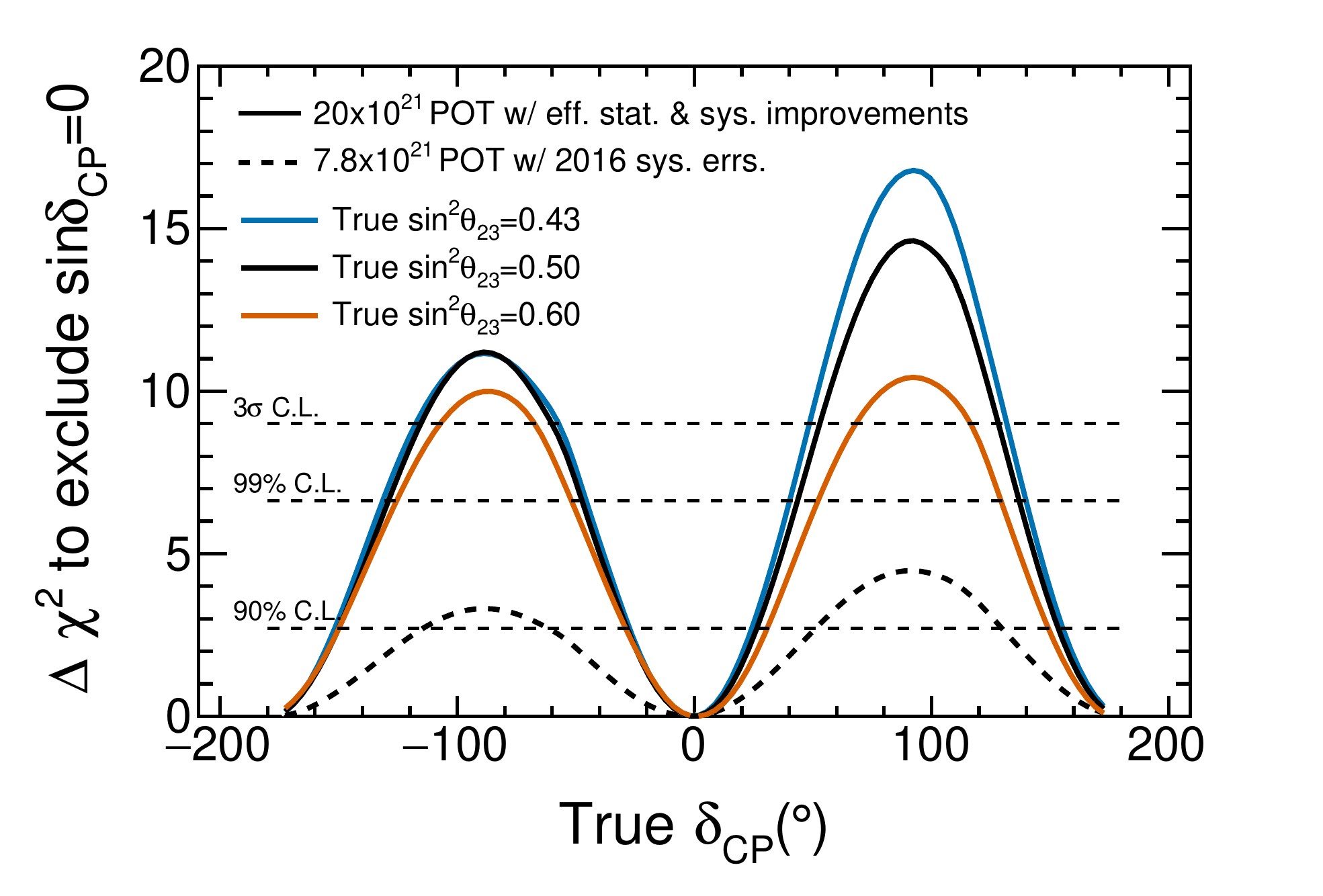}
\caption{\label{fig:t2k2sensi} Sensitivity to CP violation as a function of the true
$\delta_{CP}$ for three values of $\sin^2 (2 \theta_{23})$ (0.43, 0.50, 0.60) and normal ordering, for the full T2K-II exposure of $20\times 10^{21}$ POT and a reduction of the systematic error to 2/3 of the 2016 T2K uncertainties. On the left plot the mass ordering is considered unknown, while on the right plot it is considered known~\cite{Abe:2016tez}.}
\end{center}\vspace*{-5mm}
\end{figure}

The J-PARC main ring, presently operating at 485 kW beam power for the T2K neutrino beam line is currently undergoing a series of upgrades allowing it to reach 1 MW around 2022 and later 1.3 MW in the Hyper-Kamiokande era [ID158]. 
The T2K Collaboration has submitted a proposal for an extension of T2K running (T2K-II) accumulating 20$\times$10$^{21}$ protons-on-target (POT), that is 6 times the present exposure.
This aims at an initial observation of the $\delta_{CP}$ parameter at the 3\,$\sigma$ level for a significant range of the possible values (Fig.~\ref{fig:t2k2sensi}) by 2026. T2K has also launched an upgrade of its near detector complex~\cite{Abe:2019whr} in order to reduce the systematic uncertainties to match this increase of the data sample. 

NOVA [ID167] 
is based on the NuMI beamline operating at 700 kW, with opportunities to increase to 900 kW by 2021. The NOVA far detector is a 14 kt liquid scintillator calorimeter located in Ash River, at 810 km from the neutrino source. With its longer baseline NOVA has interesting sensitivity to the mass ordering. For the currently favoured set of oscillation parameters, NOVA has potential sensitivity to the mass ordering at 3\,$\sigma$ in 2020, and similar sensitivity to CP violation by 2024. 


\subsection{Future long-baseline experiments}

While the measurements with the existing facilities continue to provide a very important contribution to the field, only for favourable parameters values will the statistical significance of the measurements of the CP violation and mass ordering exceed the 3\,$\sigma$ level. An almost complete and reasonably precise coverage of the parameter space will be provided by the next generation of experiments, Deep Underground Neutrino Experiment (DUNE) in the US and Hyper-Kamiokande in Japan. 

The 2013 strategy recommended that ``CERN should develop a neutrino programme to pave the way for a substantial European role in future long-baseline
experiments.'' This recommendation took the form of the CERN Neutrino Platform (NP) that has rapidly become the hub for neutrino physicists to develop innovative detectors for the next generation long and short-baseline experiments in USA and Japan, and beyond. The 2013 strategy recommendation for neutrino physics was further discussed and articulated in a series of Town Meetings jointly organized by APPEC.

Among the most successful realizations of the CERN NP are the ProtoDUNE-SP (NP04) and ProtoDUNE-DP (NP02) that demonstrated the feasibility of very large liquid argon TPC. The NP is involved in the Japanese program with the Baby-MIND detector for J-PARC, the ongoing upgrade of the T2K Near Detector ND280 (NP07)~\cite{Abe:2019whr} and supports several other detector R\&D relevant for the next generation neutrino experiments.
A next phase of the Neutrino Platform has been recently approved, including most notably a second campaign of test beam for the ProtoDUNEs until 2021.

There is a wide support of the neutrino community at large, as shown by the recent Neutrino Town Meeting in 2018 and its conclusions~[ID45],    
to focus on the long baseline experiments in the US and Japan. Europe should
continue to provide a balanced support for DUNE and Hyper-Kamiokande, to secure the determination of the remaining unknowns in neutrino oscillations, aiming at the determination of CP violation and testing the three-neutrino framework. 

The DUNE experiment [ID126] 
is a top priority of the US High Energy Physics program. It is based on the Fermilab LBNF wide-band beam with an initial beam power of 1.2 MW of 120 GeV protons. The DUNE far detector will consist of four similar liquid argon TPCs, each with 10 kt fiducial mass, placed in the Sanford Laboratory in South Dakota, 1300 km away from Fermilab.
This detection technique will provide ``bubble-chamber-like'' quality data.
Large prototypes of these detectors, using the Single Phase (SP, wire chambers in the liquid argon) and Double Phase (DP, gas amplification in gaseous argon on top of the liquid argon) have been built at CERN: the ProtoDUNE-SP and ProtoDUNE-DP.
These projects have been carried out in the framework of the CERN Neutrino Platform which offered crucial support to these developments where European teams played a major role.
ProtoDUNE-SP has been exposed to a charged particle beam and has demonstrated performance exceeding the design values. 
The construction of ProtoDUNE-DP has been completed and it will be tested with cosmic rays in the second half of 2019.
In Homestake the excavation of the main cavern hosting the DUNE far detectors has started in 2017, and the next major milestones are the installation of the first Far Detector in 2022 and the beam readiness for operation in 2026. 

\begin{figure} [htbp!]
\begin{center}
\includegraphics[width=10cm]{\main/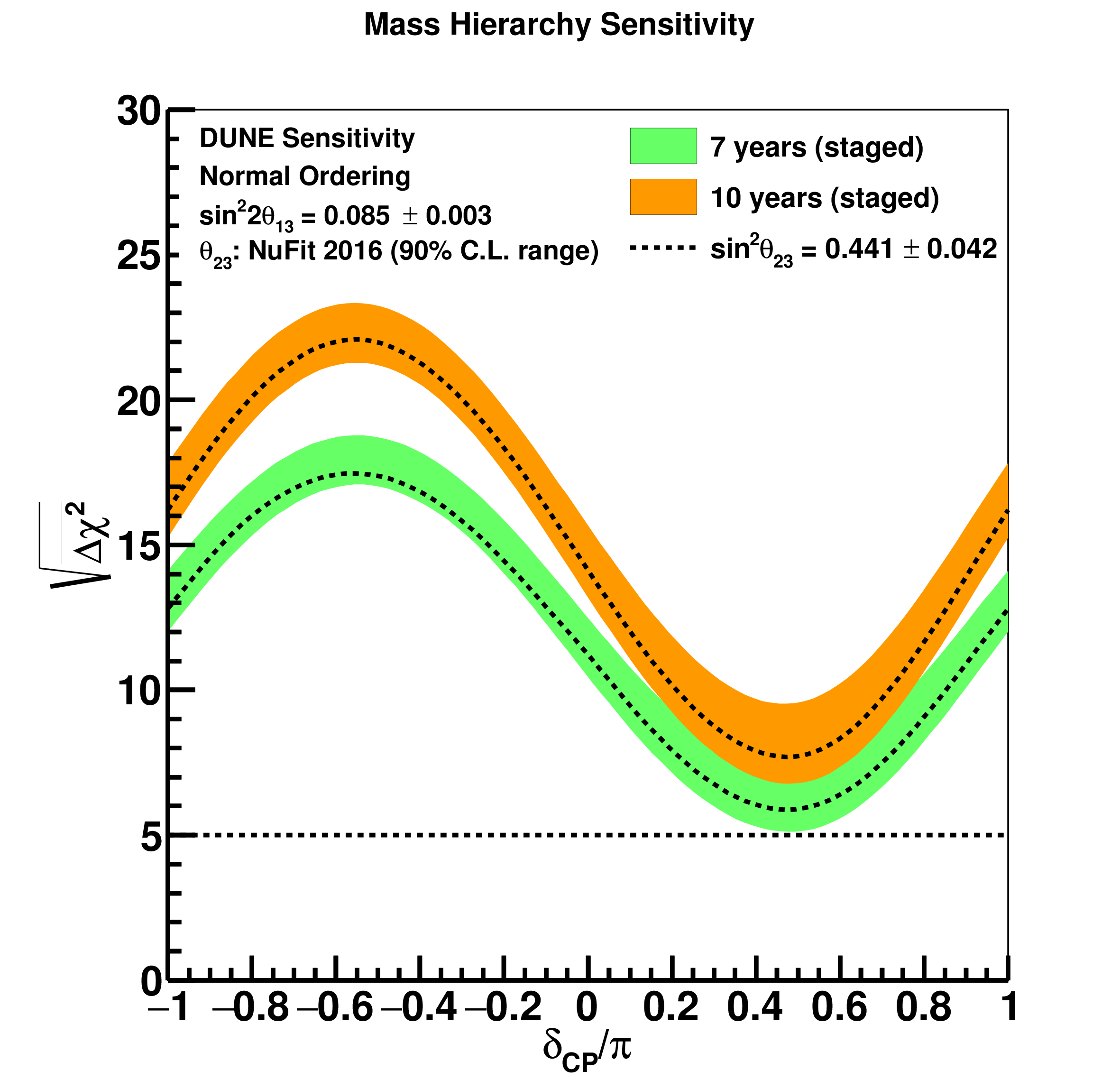}
\includegraphics[width=10cm]{\main/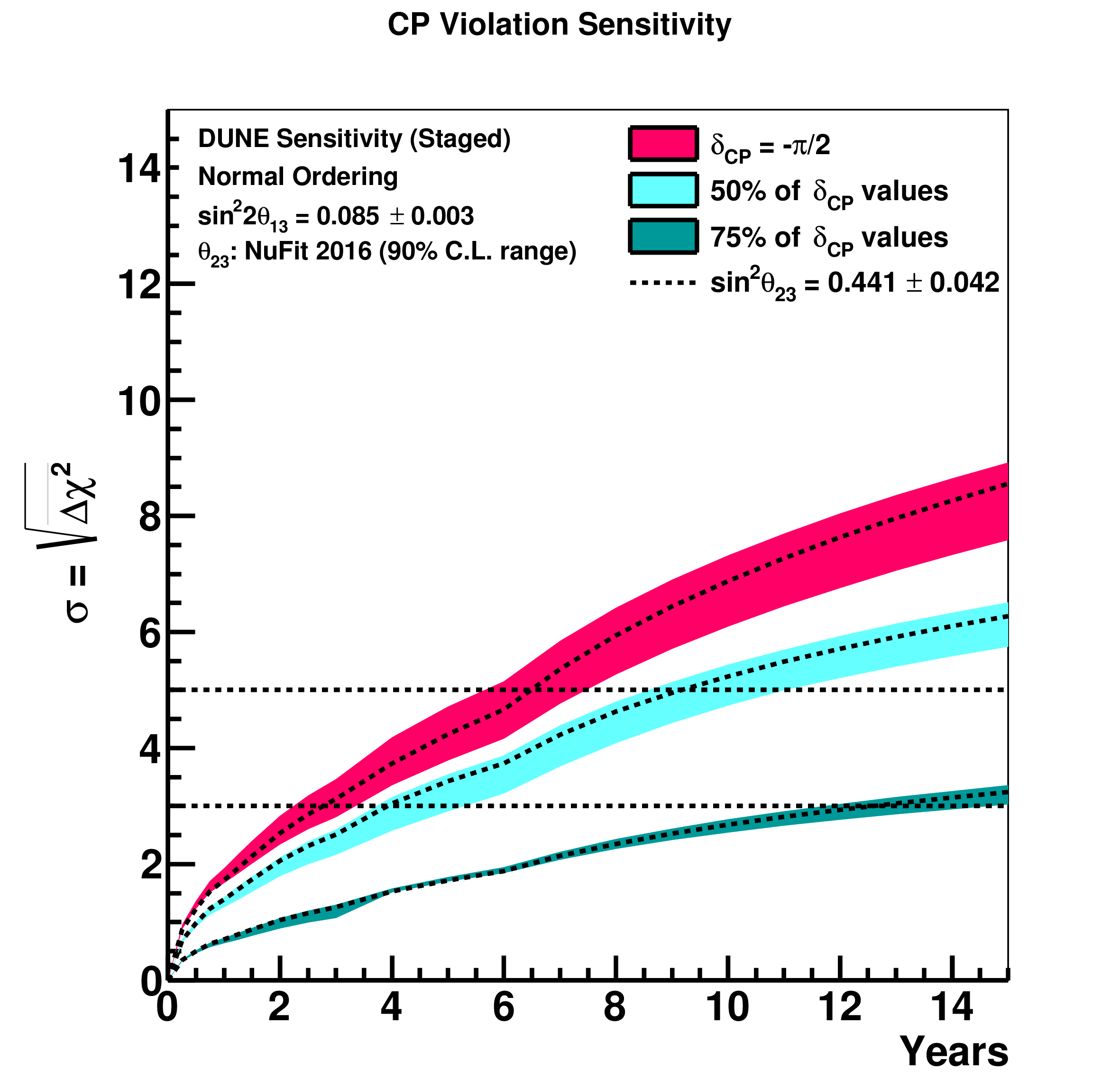}
\caption{\label{fig:dunesensi} Top: DUNE expected sensitivity~\cite{Abi:2018dnh} to the mass ordering (shown as the 
the square root of the mass ordering discrimination metric $\sqrt{\Delta \chi^2}$)
\ as a function of $\delta_{CP}$ for an exposure of 7 and 10 years. The exposure for 7 (10) years in a staged scenario is equivalent to 336 (624) kt $\times$ MW $\times$ years.
Bottom:  
the significance with which CP violation can be determined by DUNE for 75\% and 50\% of $\delta_{CP}$ values
and for $\delta_{CP}=-\pi/2 $.
}
\end{center}
\end{figure}

Thanks to the very long baseline in DUNE, the matter effects will have a strong impact on the $\nu_{\mu} \rightarrow \nu_e$ appearance probability in the far detector, providing a complete coverage of the mass ordering with $\sqrt{\Delta \chi^2} > 5 $ \footnote{For the case of the mass ordering determination, the usual association of this test statistic with a $\chi^2$ distribution for one degree
of freedom is not strictly correct and its interpretation 
in terms of number of $\sigma$ of a Gaussian probability density
is not exact. This applies also to the other experiments sensitive to the mass ordering.}, and DUNE will provide a 5\,$\sigma$ measurement of CP violation for 50\% of all $\delta_{CP}$ values (see Fig.~\ref{fig:dunesensi}). Revised figures for the physics sensitivity will appear in the DUNE TDR to be released in Autumn 2019.

\begin{figure} [htbp!]
\begin{center}
\includegraphics[width=11cm]{\main/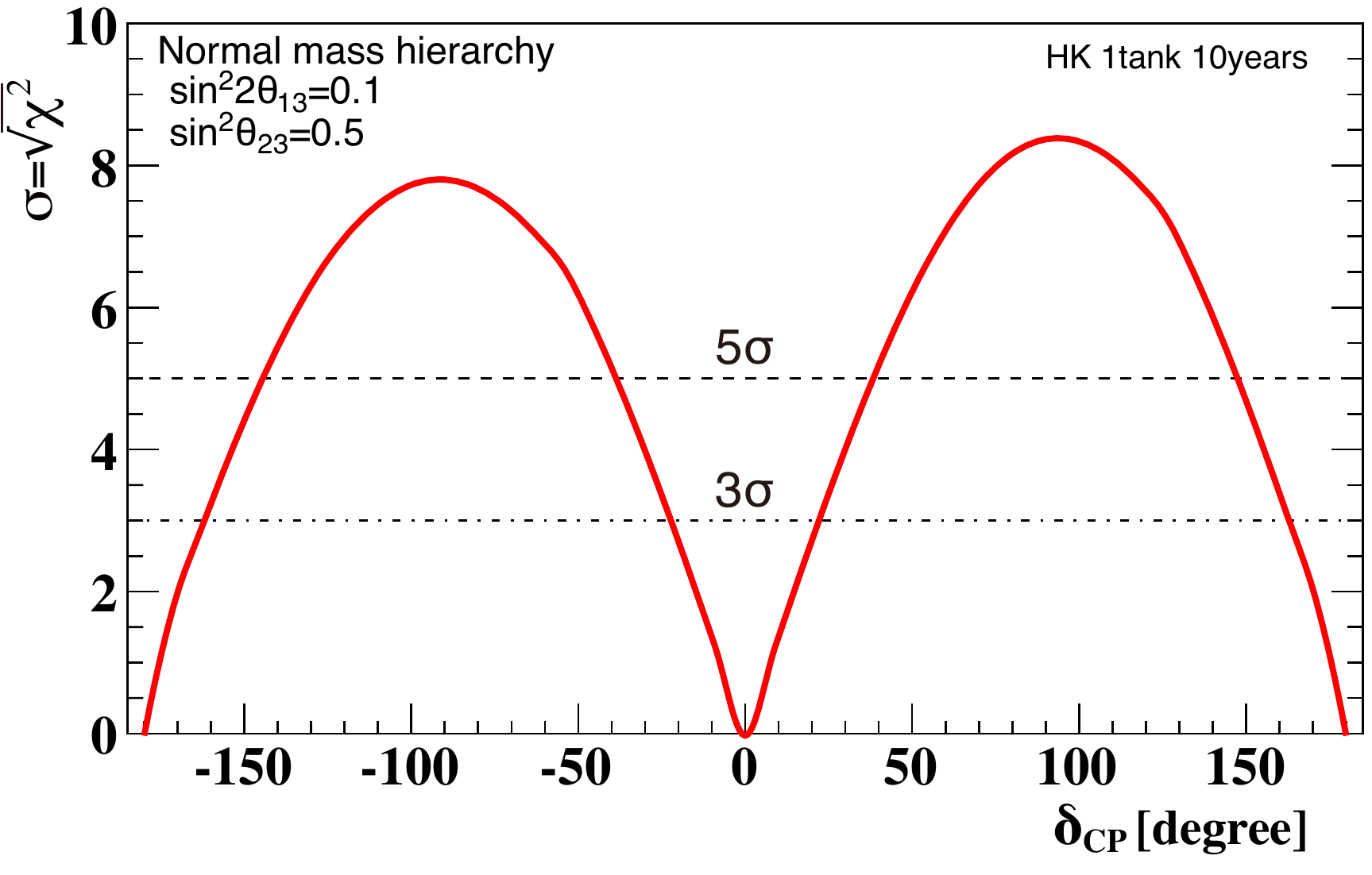}
\caption{\label{fig:hkcpv} 
Hyper-Kamiokande expected significance to exclude $\sin \delta_{CP} = 0 $  in case of normal hierarchy~\cite{Abe:2018uyc} for an exposure of 10 years. The mass ordering is
assumed to be known. }
\end{center}\vspace*{-1cm}
\end{figure}

The Hyper-Kamiokande collaboration [ID158]  
has been recently formally established and the project has been selected in 2017 by the Science Council of Japan for its Master Plan. The detector based on the well-established water Cherenkov technique will have a fiducial mass of 187 kt for one tank, located in the Gifu prefecture close to the Super-Kamiokande site (baseline 295 km). 
In August 2019 an important step toward the approval of the project was made: the Japan Ministry of Science (MEXT) officially added to its fiscal year 2020 budget request a part of the cost for constructing Hyper-Kamiokande.
The target date is to complete the detector construction by 2027.
There are plans for a second similar tank, possibly located in South Korea, at a different baseline of about 1000 km. 
Concerning the long baseline physics program the experiment sensitivity allows the establishment of CP violation in neutrino oscillations at 5\,$\sigma$ significance for 50\% of the allowed phase-space (Fig.~\ref{fig:hkcpv}) for an exposure of 10 years. The neutrino mass ordering can be investigated either with the addition of a second tank at a longer baseline or using the atmospheric neutrino sample.

The sensitivities of DUNE and Hyper-Kamiokande regarding the possible discovery of leptonic CP violation and other precision oscillation measurements are quite similar.
Nevertheless, given the pivotal importance of systematic uncertainties in these measurements, the
availability of two experiments with orthogonal choices regarding beam design, detector technology
and baseline will be essential for reaching authoritative conclusions~\cite{Cao:2015ita}.
The complementarity between DUNE and Hyper-Kamiokande becomes even more evident for tests of new
physics scenarios: because their relative sensitivities to the standard 3-family oscillation and to
non-standard scenarios, such as e.g.\ a new type of neutrino matter effect~\cite{Liao:2016orc}, are different, their
combination will allow to probe and characterise new physics effects more comprehensively.
Finally, DUNE and HyperK are highly complementary in their physics program beyond
accelerator-based neutrinos.

DUNE and Hyper-Kamiokande, as well as JUNO, that will be described below, will also be the largest and most sophisticated neutrino detectors underground, offering sensitivity to a range of exotic and astrophysical processes like:
the search for proton decay, the study of neutrinos produced by core-collapse supernovae (CCSN), and new studies of solar and atmospheric neutrinos. 
Proton decay is a generic prediction of Grand Unified Theories. The current best limit  at 10$^{34}$ years has been established by Super-Kamiokande~\cite{Miura:2016krn}. DUNE and Hyper-Kamiokande have the sensitivity to increase this limit by an order of magnitude. 

The observation of 25 neutrinos emitted by SN1987A has marked the birth of neutrino  and multimessenger astrophysics, as detailed in Chapter~\ref{chap:cosm}. DUNE, Hyper-Kamiokande and JUNO will be excellent detectors of the neutrinos produced by the next CCSN in our galaxy, providing complementary samples of several 10$^4$ events and  allowing to follow the time development of neutrino emission during the collapse and rebound phases to better understand the still largely unknown physics of these spectacular stellar collapses. 
Hyper-Kamiokande will provide the largest sample due to its large mass, with event-by-event determination of the energy down to 3 MeV. It will mainly detect anti-electron neutrinos using the inverse beta decay reaction. In addition it can provide a 
$1^{\circ}$
pointing accuracy for a supernova at 10 kpc, enabling multi-messenger studies. On the other hand DUNE will detect electron neutrinos via $\nu_e + {}^{40}{\rm Ar} \rightarrow e^- + {}^{40}{\rm K}^*$. This is particularly interesting because supernova neutrino emission begins with the neutronization burst, an initial sharp, bright flash of $\nu_e$  from $ p + e \rightarrow  n + \nu_e$.

The European Spallation Source in Lund (Sweden) is designed to provide a 5 MW proton beam. There is a proposal to modify this facility to double the proton beam power to 10 MW in order to provide a neutrino beam for long baseline studies in Europe [ID98] with a far detector located at the second oscillation maximum, where CP violation effects are enhanced. At the moment this project is in its design phase. 
This facility would start several years after the DUNE and Hyper-Kamiokande experiments, but some improvement in precision in the measurement of $\delta_{CP}$ with respect to DUNE and Hyper-Kamiokande can be expected, which may be relevant if $\delta_{CP}$ happens to be close to $\pm \pi/2$ or if $ \sin \delta_{CP}$ is close to 0. Independently of the outcome, ESS could be an important test-bed for R\&D on new intense neutrino beams along the road towards a Neutrino Factory. Indeed a Neutrino Factory~\cite{Bogomilov:2014koa}, where the neutrinos are produced in the decay of muons in a storage ring, would be the natural next-to-next step for long-baseline experiments beyond DUNE and Hyper-Kamiokande. It would be a unique facility to reach a precision of $6^{\circ}$ on the measurement of $\delta_{CP}$.

\subsection{Future experiments with reactor and atmospheric $\nu$ to determine mass ordering}

The study of the neutrino mass ordering has spurred many new ideas since the discovery of the last mixing angle $\theta_{13}$. As $\theta_{13}$ is relatively large, several non-leading effects are within experimental reach and could allow the mass ordering to be established with new methods.

The Jiangmen Underground Neutrino Observatory (JUNO)~[ID10] 
collaboration is building underground a large liquid scintillator detector, with a mass of 20 kt. It is located in China, at a distance of 53 km from two clusters of nuclear reactors. One of the main goals of JUNO is the detection of antineutrinos from these reactors. At these distances the electron neutrino survival probability will be  
modulated by $\Delta m^2_{21}$, with slow oscillations, on which a term governed by $ \Delta m^2_{31}$ will superimpose much faster oscillations with a much smaller amplitude. The latter carry information about the mass ordering. In order to do this, an unprecedented 3\%/$\sqrt{E}$ energy resolution has to be reached, which is an experimental challenge.
The measurement of the antineutrino spectrum with excellent energy resolution will also lead to the precise determination of the neutrino oscillation parameters $\sin^2 \theta_{12}$, $\Delta m^2_{12}$, and $|\Delta m^2_{ee} |$ ($ \Delta m^2_{ee}= \cos^2 \theta_{12}  \Delta m^2_{31}  + \sin^2 \theta_{12}  \Delta m^2_{32} $) to an accuracy of
better than 1\%. JUNO will start taking data in 2021.

Another approach for the study of neutrino oscillations is to use atmospheric neutrinos. These are produced by the decay of pions created in high-energy cosmic ray interaction in the atmosphere. The advantage of this neutrino source is its range in energy (typically from sub-GeV to hundreds of GeV) and in oscillation length, reaching 12,000 km.
The disadvantage is that this source contains both neutrinos and anti-neutrinos, $\nu_\mu$ and $\nu_e$. Since the large Cherenkov detectors used to detect these neutrinos cannot differentiate between neutrinos and antineutrinos, matter effects are convoluted with flux and cross-section effects. 
Moreover, the critical energy region to carry out this measurement is in the few GeV range, which is difficult to attain with a sparse detector array. Nevertheless, the IceCube experiment has shown a good sensitivity to the $\nu_\mu$ disappearance. The ORCA experiment~[ID84] 
 is part of the KM3NeT project and foresees to deploy an array of instrumented lines in the Mediterrenean, close to Toulon (France), with a distance between PMT of about 10\,m. Sensitivity studies show that this detector might provide a 3\,$\sigma$ measurement of the mass ordering after 3 years of data-taking. The detector construction has started and the target date for its completion is 2024. A new neutrino beam from Protvino (Moscow) to ORCA has been proposed [ID124] with a baseline of 2595 km. The detector configuration,  performance and physics case need to be established and studied in more detail.

\subsection{ Precision flux and cross-section measurements}

A crucial point for the overall neutrino program is the precision knowledge of neutrino fluxes and cross sections. Indeed, as the modulation of the $\nu_\mu \rightarrow \nu_e$ appearance probability induced by the PMNS $\delta_{CP}$ phase is about 30\% on the first oscillation maximum, the precision extraction of oscillation parameters requires good control of the interaction rates, at the percent level, to be compared to the current level of 5-7\% reached for instance by T2K.

This impressive reduction of the systematic uncertainties requires a fully-fledged programme to develop new techniques and detectors, innovative measurements and phenomenological models. Europe, and CERN in particular, could play a crucial role in implementing this program on the basis of the present expertise, and of existing or new facilities.
In the following we will briefly sketch some important developments along this road.

The neutrino beams used in long baseline experiments rely on the production of hadrons by impinging a primary proton beam on a target. The precise knowledge of the hadroproduction cross section is a crucial ingredient to improve the precision on the neutrino fluxes. Europe has a unique facility with the CERN NA61/SHINE experiment~[ID13] 
at the SPS, capable of large acceptance and high resolution, that has already produced several crucial measurements for T2K. It plans to upgrade its detector and to continue its program of measurements with replica targets for the DUNE and Hyper-Kamiokande experiments. 

The ENUBET (NP06)~[ID57] 
collaboration  proposes a dedicated facility to precisely measure the $\nu_\mu$ and $\nu_e$ cross section improving by an order of magnitude over the present knowledge. They propose to do so with a combination of narrow-band neutrino beams and monitored beams, instrumenting the decay tunnel with a segmented calorimeter.

The $\nu$STORM~[ID154] 
collaboration proposes a new facility capable of delivering a precisely (1\%) known neutrino beam. It relies on a new concept, where the neutrinos are produced in a storage ring by the decay of muons. Besides precise neutrino-nucleus cross sections useful for DUNE and Hyper-Kamiokande, the $\nu$STORM facility can also be used for searches for sterile neutrinos beyond the capabilities of the Fermilab short baseline experiments and can also serve as the test facility for the development of  a neutrino factory and muon accelerators
towards a multi-TeV lepton-antilepton collider. The operation of ENUBET and/or $\nu$STORM by 2027 would maximize the impact of these measurements for the world neutrino program. 
Both $\nu$STORM and ENUBET are to a large extent site-independent concepts, studies and R\&D; however both consider a possible implementation at CERN. For $\nu$STORM, under the auspices of the PBC program, 
an initial study of implementation at CERN was carried out, and no show-stoppers have been identified. 
For ENUBET the option of using SPS as the proton driver has been considered in greater detail with a possible site in the North Area and the ProtoDUNEs as neutrino detectors.   
A dedicated study should be set-up to evaluate the possible implementation, performance and impact
of a percent-level electron and muon neutrino cross-section measurement facility (based on e.g.\ ENUBET or $\nu$STORM) with conclusion in a few years time.

Together with these developments aimed at precisely known neutrino fluxes and cross sections, a corresponding development plan is unfolding for high precision near-detectors [ID106, ID131]~\cite{Abe:2019whr} 
and for the development of precise and reliable neutrino interactions models~\cite{Alvarez-Ruso:2017oui}. The latter point will require a close collaboration and much increased effort between neutrino physicists, nuclear physicists and phenomenologists, where Europe is today playing a leading role and can continue to do so in the future provided
the community receives the much needed support.






\section{Determination of neutrino mass and nature}

The absolute values of neutrino masses and the neutrino nature (Dirac versus Majorana) are among the most fundamental open questions in physics today. The determination of these neutrino properties will be an essential ingredient to understand the origin of neutrino mass, and the role of neutrinos in the Universe. Due to their abundance and their being initially relativistic, neutrinos play a crucial role in large-scale structure formation of the Universe. Their potential Majorana nature would be a key to solve the puzzle of the matter-antimatter asymmetry of the Universe. 

\begin{figure} [t]
\begin{center}
\includegraphics[width=12cm]{\main/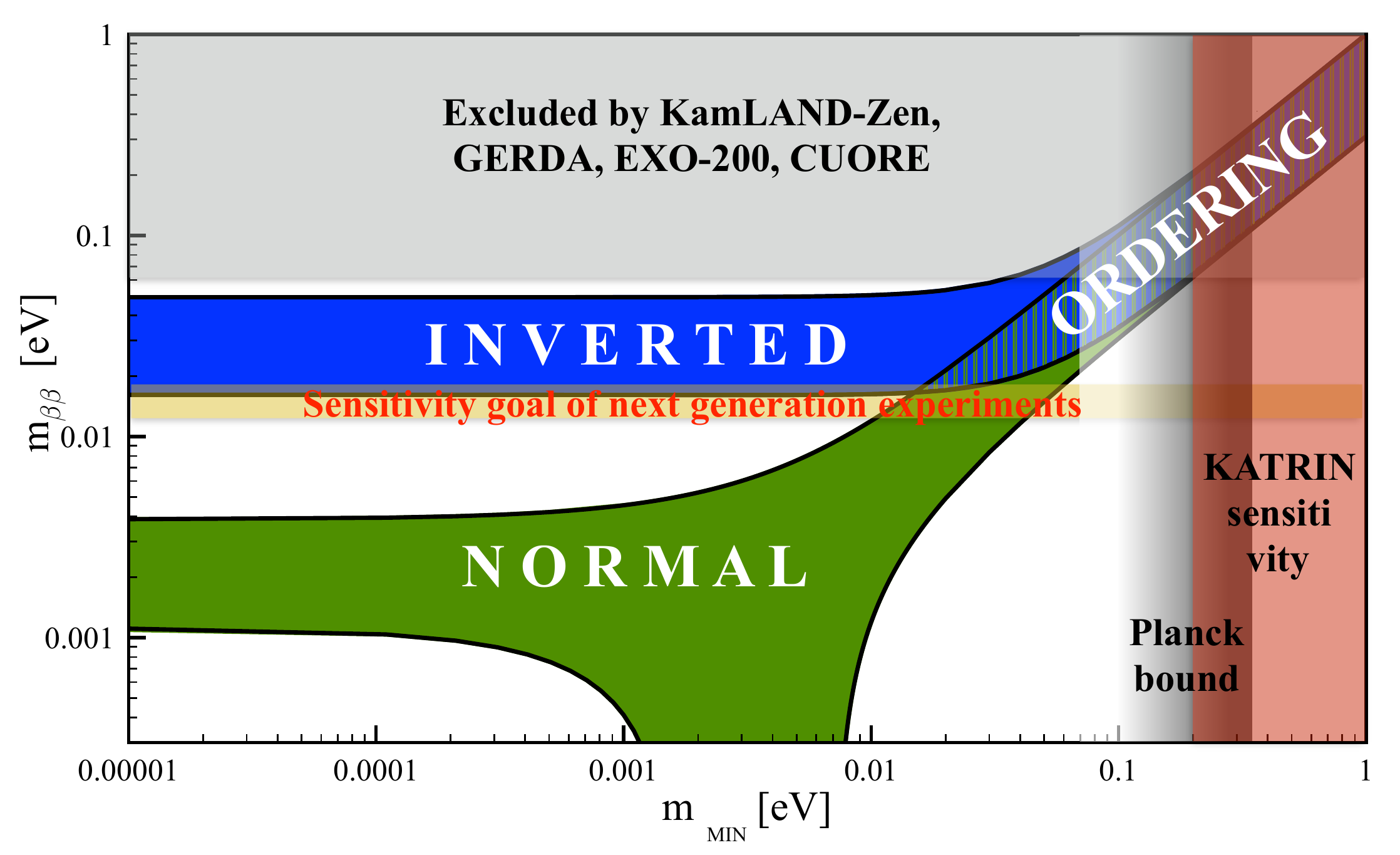}
\caption{\label{fig:doublebeta} 
The effective mass $ m_{\beta \beta}$ controlling the rate of the neutrino-less double beta decay versus the mass of the lightest eigenstate\cite{pascoli}. The limits from the current and future experiments are shown.}
\end{center}\vspace*{-1cm}
\end{figure}

\subsection{Cosmology}
Cosmological observations themselves provide powerful probes of the neutrinos mass. They are mainly based on the fact that neutrinos, due to their relativistic velocities and large free-streaming lengths, prevent the formation of small-scale structures in early epochs of the universe. Current limits based on the Planck satellite set limits of $m_{\nu} = \sum_i m_{\nu_i}<230 - 540$~meV (95\% CL)~\cite{Aghanim:2018eyx}. Future surveys, such as the imaging and spectroscopic space telescope EUCLID~\cite{Laureijs:2011gra} (to be launched in 2022 for a six year mission), aim for the first time at an actual measurement of the neutrino mass with a precision of $\sigma({m_{\nu} }) = 10$~meV. 
It is important to note that, despite their sensitivity, these results will depend on the underlying cosmological model, and could be affected by degeneracies with other cosmological parameters~\cite{Boyle:2017lzt, Boyle:2018rva}.  This stresses the importance of performing the measurement of the absolute neutrino masses both via cosmological observables and with the even more challenging direct methods. A good agreement between these two approaches would be a crowning confirmation of the cosmological model.

\subsection{Neutrinoless double beta decay}

The 
neutrinoless double $\beta$-decay ($0\nu\beta\beta$) is a unique probe of the nature of neutrinos. Discovering this process
would indeed prove that neutrinos are Majorana particles and hence that lepton number is violated in Nature. 
This discovery would in addition provide insight into the absolute neutrino mass scale.

The half-life for this process $T_{1/2}$ is inversely proportional to the square of the Majorana neutrino
mass $m_{\beta \beta}= | \sum_i  U^2_{ei} m_i |$. The predictions for $m_{\beta \beta}$ depend on the neutrino mass spectrum. For the inverted ordering there is a lower bound of 15 meV, while for the normal ordering $m_{\beta \beta}$ can go from the current bounds to zero, for specific values of the neutrino masses. Apart from light Majorana neutrinos, other lepton number violating mechanisms can mediate this process.

The current best neutrino mass limit of $m_{\beta\beta} = 50 - 160$~meV is provided by the KamLand Zen experiment~\cite{KamLAND-Zen:2016pfg} in Japan. The best half-life sensitivity of $T_{1/2}>11\times 10^{25}$ years (90\% CL) was achieved by the \textsc{Gerda} experiment~\cite{zsigmond} (LNGS, Italy). Thanks to their scalability, future $0\nu\beta\beta$ experiments plan to reach sensitivities down to $m_{\beta\beta} \approx 10$~meV, covering the allowed parameter space for inverted neutrino mass ordering (Fig.~\ref{fig:doublebeta}). 

The European roadmap for this experimental program is presently being developed under the aegis of APPEC.
The next-generation LEGEND experiment~\cite{DAndrea:2019umy}, successor of the \textsc{Gerda}~\cite{Agostini:2018tnm} and \textsc{Majorana}~\cite{Aalseth:2017btx, Alvis:2019sil} experiments, targets to operate 1000~kg of Ge-semiconductor detectors (enriched in ${}^{76}$Ge) to reach a 3$\sigma$ discovery sensitivity of $>10^{28}$~years, corresponding to a mass limit of $m_{\beta\beta} < 10 - 17 $~meV. The nEXO experiment~\cite{Albert:2017hjq}, successor of EXO~\cite{Albert:2017owj}, plans to instrument a 5-tonne liquid xenon (enriched in ${}^{136}$Xe) time projection chamber to reach a 3\,$\sigma$ discovery sensitivity of $5.6\times 10^{27}$~years, corresponding to a mass limit of $m_{\beta\beta} < 5.7 - 17.7$~meV. NEXT~\cite{Alvarez:2011my} explores an alternative approach based on gaseous xenon TPC, which has advantageous features both with respect to energy resolution and background suppression. It is interesting to note that also future xenon-based dark matter experiments, such as DARWIN~\cite{Aalbers:2016jon}, may have competitive sensitivity to $0\nu\beta\beta$. The next-generation \mbox{CUPID} experiment~\cite{Azzolini:2018tum}, the successor of CUORE~\cite{Alduino:2017ehq}, plans to employ cryogenic detectors, which allow for both heat and light signal readout. Promising results have been obtained with $^{130}$TeO$_{2}$~\cite{Artusa:2016mat}, Zn$^{82}$Se~\cite{Azzolini:2018dyb}, Li$_2^{100}$MoO$_4$~\cite{Bekker:2014tfa} crystals. 

\subsection{Direct neutrino mass experiments}
The least model-dependent technique to determine the absolute neutrino mass scale is based on the kinematics of single-$\beta$-decay~\cite{Drex13}. Here, the impact of the so-called effective electron (anti-)neutrino mass $m^2_{\nu_e}  = \sum_i |U_{ei}|^2 m_{\nu_i}^2$ is a reduction of the kinematic endpoint energy $E_0$ and a distortion of the beta-decay spectrum close to this endpoint. 

Single-$\beta$-decay experiments are also ideally suited to perform searches for eV- to keV-scale sterile neutrino searches. These hypothetical neutrinos would manifest themselves as spectral distortion at an energy $E = E_0 - m_s$, where $m_s$ is the mass of the sterile neutrino. These searches are highly complementary to oscillation-based searches~\cite{Boser:2019rta, Mer:2015a}. 

The Karlsruhe Tritium Neutrino (KATRIN) experiment, a large-scale tritium beta decay experiment~\cite{Angrik:2005ep}, started data taking in 2019. KATRIN combines an ultra-luminous gaseous tritium source ($10^{11}$ decays per second) with a high-resolution magnetic-adiabatic collimation and electrostatic filter~\cite{Otten:2008zz,Lobashev:1985mu, PICARD1992345}. The design sensitivity of 200~meV (90\% CL) will be reached after 5 calender years of data taking. 

A complementary approach is pursued by the ECHo~\cite{Gastaldo:2017edk} and Holmes~\cite{Giachero:2016xnn} experiments, which exploit the electron-capture decay of $^{163}$Ho. New ideas based on cyclotron emission spectroscopy (CRES) and the usage of atomic tritium are being explored by Project-8~\cite{Esfahani:2017dmu} to push the sensitivity beyond the degenerate mass regime to about 40~meV (90\% CL).

\section{Search for new neutrino states}

An important question in neutrino physics is whether the 3-neutrino paradigm is complete or whether new neutrino states exist in Nature. Their masses can range from the sub-eV scale up to the GUT scale. In the context of oscillation experiments eV to keV mass neutrinos are conventionally called sterile neutrinos, while heavier neutrino states with masses above MeV are often called Heavy Neutral Leptons (HNL).

\subsection{Searches for sterile neutrinos}

The hypothesis of sterile neutrinos is in part motivated by the LSND, MiniBooNE,
Reactor, and Gallium anomalies: 
these are results from short baseline accelerator-, reactor-, and source-based experiments that cannot be interpreted in terms of three neutrino mixing.  
However there is strong tension between electron neutrino appearance and muon neutrino disappearance results and even with several new neutrino states,
sterile neutrino oscillations
are unable to explain all the experimental data.
Each of these anomalies is at the 2--4\,$\sigma$ level: they motivated a program of  new experiments to understand the nature of the anomalies.  

A host of reactor-based experiments are now underway at both power and research reactors to look for anti-electron neutrino disappearance.  
The experiments NEOS (South Korea), DANSS (Russia), Neutrino-4 (Russia), SoLiD (Belgium), STEREO (France), and Prospect (US), have presented first results which provide hints and limits, and 
continue collecting data
for more definitive results.  Some of these experiments, for example the Prospect experiment, 
are also performing measurements which will inform and constrain the prediction for the un-oscillated reactor flux.   

A series of accelerator-based experiments~[ID137] 
at Fermilab in the US are addressing the accelerator based anomalies. The MicroBooNE experiment uses the same neutrino beam as MiniBooNE, but a different detector technology, liquid argon Time Projection Chambers, to address the low-energy excess observed by MiniBooNE.  While MicroBooNE is running and first results are anticipated soon, the SBND and ICARUS experiments, representing the second phase of this programme, are under construction and installation.  SBND serves as a near detector to measure the un-oscillated flux and ICARUS as a large mass far detector to fully address the LSND anomaly. ICARUS has benefited from CERN support (WA104, NP01) to move the existing detector from LNGS to CERN, overhaul it, and move it to Fermilab. 

In addition to these dedicated experiments, there are a number of other experiments which can help understand the anomalies.  These include the MINOS+ experiment at Fermilab, IceCube (South Pole), the KATRIN (Germany) and $^{163}$Ho neutrino mass experiments, and results from cosmology.  The BEST source experiment is in the planning stage and the JSNS2 experiment, a direct test of LSND, is about to start running in J-PARC (Japan).  Results from these experiments in the upcoming years should help to clarify the situation, be it in terms of new physics or of other explanations.

\subsection{Searches for Heavy Neutral Leptons} \label{hnl-sec-nu}


Heavy Neutral Leptons (HNLs) are the simplest extension envisaged to the SM:
there are several specific models that predict the existence of HNLs, such as the $\nu$MSM~\cite{Asaka:2005pn}, Left-Right symmetric model~\cite{Pati:1974yy,Mohapatra:1974gc,Senjanovic:1975rk}, some SUSY scenarios~\cite{LopezFogliani:2005yw} and several others (see for instance ~\cite{Akhmedov:1998qx,Kersten:2007vk}). Particularly interesting are models predicting the existence of these particles below the electroweak breaking scale where they can be searched for experimentally. 
If the mass of HNLs is in the GeV region they can be produced in meson decays, such as $D_s^+\to \mu^+ N$. HNLs  decay by mixing with active neutrinos, 
where $U_{\ell}^2$ gives the mixing probability with an active neutrino of a specific flavour. Notice that the full process might violate lepton flavour and lepton number
conservation. 
HNLs have been searched for at colliders (see~\cite{Sirunyan:2018mtv,ATLAS:2012ak,Aaij:2014aba}). In addition, sensitivity studies have been performed for the FCC
~\cite{Blondel:2014bra,Drewes:2016jae,Antusch:2017pkq} 
The FCC has the possibility to set strong constraints on HNLs for masses above a few tens of GeV and below the $Z$ mass. In the region below 10 GeV the lifetime of HNLs becomes in general very long such that experiments with large fiducial volume, like those designed for searches for hidden sector particles, are optimal. 
Proposals for the LHC-based experiments FASER~\cite{Feng:2017uoz}, MATHUSLA~\cite{Curtin:2018mvb} and CODEX-b~\cite{Gligorov:2017nwh}, include sensitivity studies for HNLs. Their experimental signature  consists of isolated displaced vertices. The critical parameters in these experiments are the size of their fiducial volume and the distance from the interaction point. These parameters have to be balanced with the level of background, and the possibility to measure momentum and identify the decay channels. While MATHUSLA has the best sensitivity among such experiments because of the large decay volume, it does not allow measuring of decay channels, therefore HNL-signatures are indistinguishable from other hidden sector particles. On the other hand FASER, which allows measuring final states and masses, has a small fiducial volume and therefore a limited sensitivity. 

Proton beam-dump experiments allow  
 both 
issues to be addressed simultaneously, i.e.\ they have a relatively large acceptance while allow measuring mass and final states of the signal. 
The NA62 experiment at the SPS of CERN~\cite{Anelli:2005ju} is collecting data and its primary goal is to measure rare decays of $K$-mesons. 
In addition, NA62 plans to run in beam-dump mode (NA62$^{++}$), collecting roughly $10^{18}$ protons on target (POT) during the LHC Run\,3 (2021-2023). This would allow to set limits in a presently unconstrained region. 

SHiP~\cite{Anelli:2015pba} is a beam-dump experiment designed and optimised to search for Hidden Particles. It will take advantage of the high-intensity proton beam of the Beam-Dump Facility~\cite{SHiP:2018yqc} (BDF) under study in the context of the Physics Beyond Collider  activity, that would allow starting data taking in the LHC Run\,4 of the CERN accelerator schedule. SHiP will accumulate the full statistics of NA62$^{++}$ every few days, thanks to the larger acceptance and beam intensity ($2\times 10^{20}$ POT in 5 years). 
Figures  \ref{fig:FIPs-HNL}, \ref{fig:sec3_LLP} and \ref{fig:dark:daulepton} in chapters \ref{chap:bsm}, \ref{chap:dm} and \ref{chap:flav}, respectively, show the sensitivity as a function of the mass of HNLs and the mixing angle with active electron, muon and tau neutrinos, for various planned and proposed experiments.
Note that
thanks to the high proton intensity of the primary beam,
neutrino experiments 
are competitive in searching for HNLs in the near detectors, as demonstrated in \cite{Abe:2019kgx, Ballett:2019bgd, Arguelles:2019xgp}.



\section{Conclusions} 

The study of the neutrino nature and properties, motivated by the unique window that this particle offers on BSM physics, is entering a new phase with a rich array of experiments using neutrinos from accelerators but also from other sources (atmospheric, solar, reactor, cosmic).

The first priority is today on the completion of the program of measurements of the oscillation parameters, most notably the CP-violating phase of the mixing matrix and the neutrino mass ordering. 
Two strong and complementary experimental programmes are in preparation towards this goal in the US and Japan with the DUNE and Hyper-Kamiokande experiments. Following the recommendations of the 2013 European Strategy, there is a strong participation of European physicists in both programmes, benefiting of CERN support, most notably through the CERN Neutrino Platform. This approach receives today the support of the neutrino community at large as shown by the recent Neutrino Town Meeting in 2018 and its conclusions~[ID45]. A balanced support to this world-wide effort from Europe will allow to secure the determination of the oscillation parameters, aim at the discovery of CP violation and test for possible deviations from the three-neutrino framework.

To extract the most physics out of DUNE and Hyper-Kamiokande, a complementary program of precision supporting measurements is needed. 
NA61 and its upgrade are an important component of this programme for the determination of the neutrino fluxes. A study should be set up to evaluate the possible implementation and impact of a facility (based on the ENUBET or $\nu$STORM concepts) to measure the neutrino cross sections at the percent level.

Other important complementary experiments are in preparation in China (JUNO) and in Europe with the KM3NeT/ORCA programme, using reactor and atmospheric neutrinos respectively. They have the potential to discover the mass ordering and to do other precision oscillation measurements.
The study of the neutrino absolute mass and nature (Dirac or Majorana) is the other priority in the field, covered by both laboratory and cosmology measurements.





\chapter{Cosmic Messengers} 
\label{chap:cosm}



From the early observation of a supernova by Tycho Brahe in 1572 to the recent observation in 2017 of a neutron star merger with both gravitational waves and electromagnetic waves ranging from $\gamma$ rays to infrared, the cosmos has always been a rich source of unexpected phenomena and wide-ranging discoveries. Some of these had a deep impact on our understanding of the physical world and its fundamental laws.

This is also true in the realm of particle physics, where many new particles were discovered with cosmic rays. Neutrino oscillations are a case in point, as their study was motivated for a long time by the puzzles posed by solar and atmospheric neutrinos. Cosmic rays are accelerated to energies far beyond the reach of present and future accelerators by processes that are still largely unknown. They offer therefore the possibility to test particle physics in an otherwise inaccessible domain, both in terms of energies and of cosmological propagation distances.

Numerous observatories have inaugurated a new way of observing the Universe through various messengers that appear tightly related one to the other: cosmic rays (Pierre Auger), gamma-rays (Fermi from space and the ground based MAGIC, H.E.S.S. and VERITAS) and cosmic neutrinos (IceCube and ANTARES), recently discovered by IceCube.  

\begin{figure} [t]
\begin{center}
\includegraphics[scale=0.5]{\main/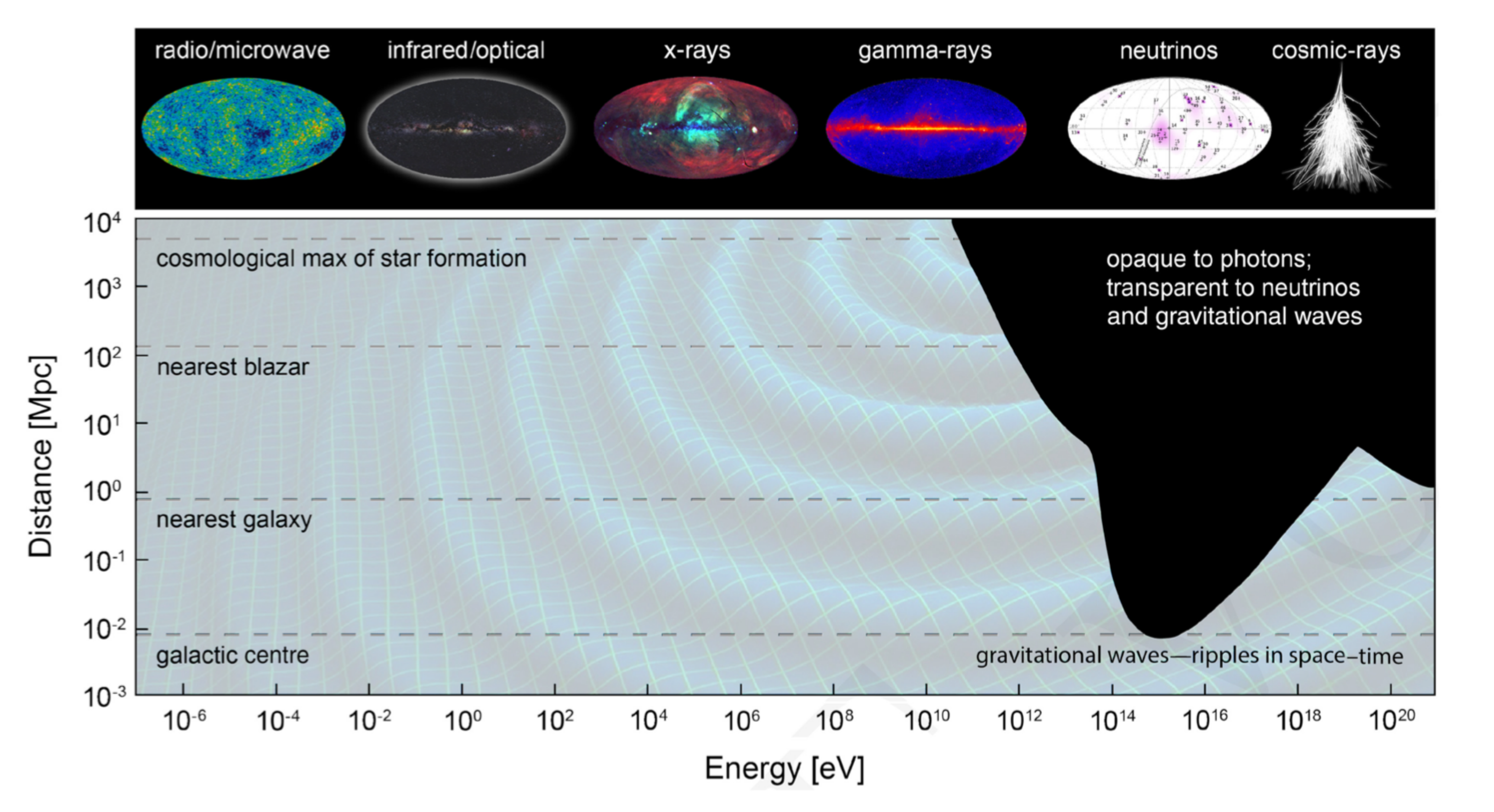}
\caption{\label{fig:multimessenger} Distance horizon at which the Universe becomes optically thick to electromagnetic radiation. The Universe is transparent to gravitational waves and neutrinos, making them suitable probes of the high-energy sky.
}
\end{center}\vspace*{-8mm}
\end{figure}

With the discovery of gravitational waves, LIGO and Virgo proved them to be powerful messengers as well, and additionally probes for fundamental properties of nature and for cosmology. 
On 17 August 2017 Adv.\ LIGO and Adv.\ Virgo observed the gravitational wave event GW170817 from a binary neutron star in-spiral at a distance of $\sim 40$ Mpc. Fermi-GBM (Gamma-ray Burst Monitor) detected a short gamma-ray burst GRB 170817A 1.7 s after the GW merger with a location compatible to the region of provenance established by LIGO/Virgo. This confirmed the model that short gamma-ray bursts are caused by neutron star mergers.  

Additionally, this event improved fundamental physics limits on the speed of gravity, equivalence principles, Lorentz invariance, alternatives to General Relativity, the Hubble constant and $r$-process nucleosynthesis.  Given the exciting science behind these observations, the ground-based observatories with large European components are upgrading their detectors  (Pierre Auger, IceCube, GVD, Adv.\ Virgo), others are starting construction (KM3NeT and CTA) and new infrastructures are being proposed (ET). The simultaneous observation of cosmic events with multiple messengers provides new insights in the astrophysical properties of compact objects and also stringent tests of physical laws and particle properties, and additionally contributes to the search for dark matter with the potential of probing its particle or non-particle  nature and its location in the Universe.

In this chapter the various fields and the  observatories
most relevant for this Strategy update are briefly reviewd, before concluding with some considerations on the synergy with particle physics. The European roadmap for this field has been discussed and prepared by APPEC~[ID84].

\section{Ultra-High Energy charged particles} 
Cosmic accelerators deliver the highest energy atomic nuclei, single protons, photons, and neutrinos for probing new physics. The physics scope of the research field ``Ultra-High Energy Cosmic Rays'' (UHECR) is focused on the detection of 
cosmic rays of energies above PeV and the related physics questions. The detection is performed with 
large-scale experimental infrastructures via the observation of extensive air showers generated by the impinging primary particles. Approximately 98\% of the flux corresponds to hadronic particles, and less than 1\% to antimatter, photons or neutrinos.

The main goals of the research are the characterization of the particles reaching the Earth's atmosphere and the unveiling of their sources, the understanding of the acceleration and propagation physics of these particles, 
the search for extreme high-energy neutrinos and photons, as well as the study of
particle physics at energies and in phase-space regions not accessible at man-made accelerators, i.e.\ interaction physics and cross section measurements with energies typically above 100 TeV. Furthermore, the observatories of UHECR are important instruments for the multimessenger
observations of the Universe due to their capabilities in extending the energy range on photon 
and neutrino diffuse fluxes compared to dedicated gamma-ray and neutrino facilities as well as the  study of possible ultra-high energy neutrino and photon signals in correlation with gravitational wave events. In addition, this allows  also an extension of the mass range in indirect searches of dark matter. 

The interpretation of air shower data heavily relies on our knowledge of particle interaction, production, and decay over a very wide range of energies and phase-space regions. Particle physics theory and measurements made at accelerators provide indispensable and complementary input for understanding extensive air showers. 
Here the role of fixed-target and collider experiments for providing important data is recognized and a coherent  measurement programme  included in the plans for accelerator-based experiments would significantly improve the accuracy of the UHECR measurements. In particular, the research field supports plans of an LHC
run with light ions, like proton-oxygen (and further O-O, N-N and Si-Si collisions), which would fill a very important gap in data needed
for air shower physics. Similarly, fixed-target and collider measurements with very good forward coverage for hadron production will be very valuable. Such an experimental programme and corresponding  joint theoretical studies of questions in the overlap region between particle and astroparticle physics is of fundamental importance for making scientific progress. One example is the development and tuning of hadronic event generators needed in high-energy physics and air shower simulations (like EPOS). To mention just one aspect, both theoretical and experimental work is needed to address the well-established muon discrepancy in air showers. While the measured muon excess relative to predictions is probably related to shortcomings in the simulated hadronic interactions it could also indicate new particle physics at energies beyond the reach of LHC.

At the same time, studying UHECR generated air showers provides a window to energies far beyond 
those accessible at existing accelerators and also emphasises phase-space regions of particle production
that cannot be covered in collider experiments. Important here are the measurements of proton-air cross 
sections above 20 TeV c.m.~energy. 
Moreover, cosmic-ray generated air showers provide a very good possibility for searching for physics beyond the 
Standard Model for various classes of models,  including production of micro-black holes, magnetic monopoles, other heavy states and violation of Lorentz invariance.
The combination of UHECR data with astrophysical information, in addition, gives  physics reach  to 
fundamental phenomena, like providing important constraints on theories of quantum gravity involving 
Lorentz invariance violation.

\section{High-Energy gamma rays}

The physics with high-energy gamma rays was not covered in the Open Symposium at Granada due to lack of time but a short section is included here due its relevance for particle physics. The physics questions answered by these observatories include the study  
of the sites and mechanisms of high-energy particle acceleration in the Universe, the physical processes at work close to neutron stars and Black Holes, the nature of dark matter, and the search for axion-like particles and for quantum gravitation effects on photon propagation~\cite{Acharya:2017ttl}. 

This messenger is studied both by satellites like Fermi in the lower energy range (100~MeV to 100 GeV) and by ground-based telescopes like MAGIC, VERITAS and H.E.S.S. (100~GeV to 100 TeV).
The latter detect Cherenkov radiation generated by the cascade of relativistic charged particles produced when a high-energy gamma ray strikes the atmosphere.
The 500 GeV to 100~TeV range is also covered by extensive air shower arrays (HAWC, Milagro) and future projects like LHAASO.
The present sensitivity will be greatly extended by the Cherenkov Telescope Array (CTA) observatory, both in energy (up to 300 TeV), flux and angular resolution. CTA is today under construction and will be based in two sites, one in the Northern Hemisphere and one in the Southern Hemisphere.

Among the most significant results are the limits on dark matter by searching for cosmic radiation emitted from annihilations of pairs of WIMPs in
surrounding regions of the Universe with a high dark matter density. These results are discussed in Chapter~\ref{chap:dm}. Another interesting result is the discovery in 2016 by H.E.S.S. of a PeVatron accelerator close to the Galactic Center.
As discussed below, gamma rays are an important component of multimessenger studies.

Apart from the obvious impact of these results on particle physics, similarly to the physics of cosmic rays, the detailed quantitative understanding of hadron interactions at high energy provided by the LHC experiments is of great relevance to these experiments, as it feeds into the discrimination methods between gammas and charged hadrons. 

\section{Ultra High Energy neutrinos} 

Cosmic accelerators also deliver the highest energy neutrinos for probing new physics. The operation of a generation of detectors such as ANTARES in the Mediterranean, GVD in Lake Baikal and IceCube at the South Pole has already yielded new results relevant to particle physics, as was once the case in the pioneering days of cosmic ray physics. 
Neutrino physics beyond the Standard Model tops their list of the big questions to be explored. In the next few years, two deep-sea sites will be deployed for KM3NeT-ORCA and KM3NeT-ARCA. Together with the IceCube Upgrade, IceCube-Gen2, HyperK and the future GRAND proposal~[ID119], these ambitious projects will lead the assault exploiting the atmospheric neutrino flux and cosmic events, as SuperK, ANTARES and IceCube are doing today. 

Specifically, these experiments have a unique capability to improve the precision of atmospheric tau-neutrino appearance measurements. These are complementary to the accelerator-based programme because tau neutrinos are scarce in the existing neutrino beams. Already, IceCube is performing competitive measurements of the oscillation parameters with neutrinos in the energy range of 5 to 55\,GeV, an order of magnitude above the energy of all present experiments, with the primary goal of detecting variations in the oscillation parameters, signaling new physics. KM3NeT-ORCA will be densely configured to determine fundamental properties of atmospheric neutrino oscillations. It has a window of opportunity to be the first experiment to determine the neutrino mass ordering, as discussed in Chapter~\ref{chap:neut}.

\begin{figure} [htbp!]
\begin{center}
\includegraphics[scale=0.9]{\main/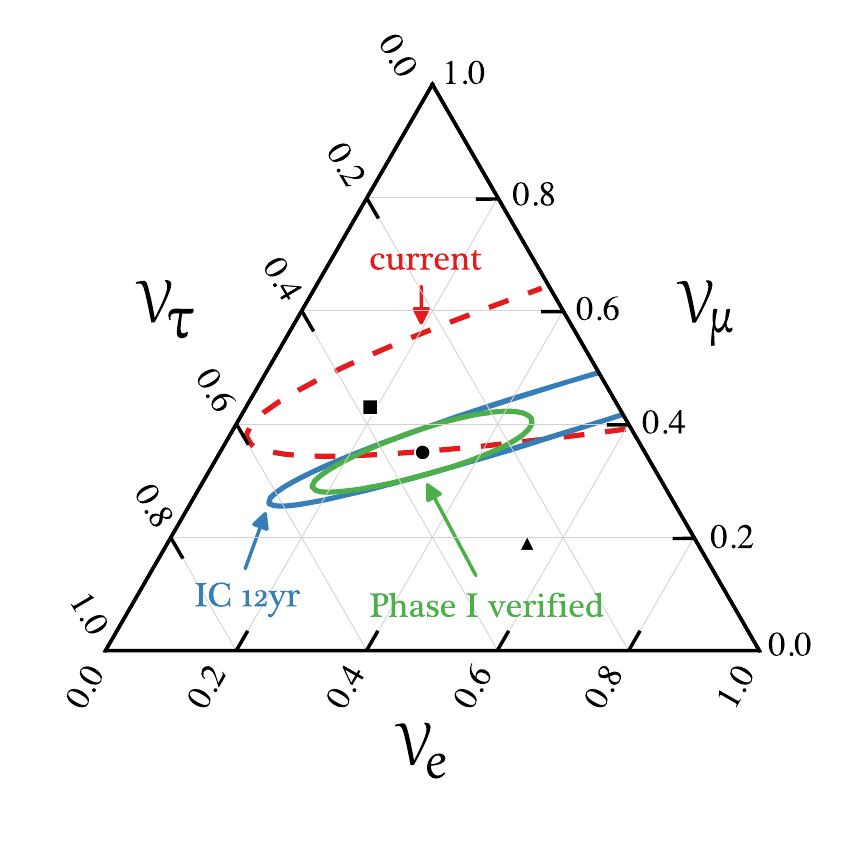}
\vspace*{-3mm}
\caption{\label{fig:nutriangle} 
Present and future sensitivity of IceCube to the flavour composition of cosmic neutrinos. 
Each point in the triangle corresponds to a ratio
$\nu_e$  : $\nu_\mu$  : $\nu_\tau$ as measured on Earth, the individual contributions are read off the three sides of the triangle.
The red dashed contour shows the current IceCube measurement, the blue contour (label ``IC 12 yr'') shows the result expected with 12 years of the denser array DeepCore data, and the green contour (labeled ``Phase I verified'') shows the result with one year of data after the deployment of 7 new strings inside DeepCore in 2022. The square, round and triangular markers show three possible results depending on the properties of the cosmic neutrino source. Any measurement outside the line connecting these three points would signal new physics in the neutrino propagation. A more precise measurement might therefore reveal the nature of the cosmic accelerator producing these neutrinos as well as non standard propagation of these particles.
}
\end{center}
\vspace*{-8mm}
\end{figure}

More than fifty years after pioneering experiments in deep underground mines in India and South Africa revealed atmospheric neutrinos, the IceCube neutrino telescope discovered a new flux for neutrino physics of extragalactic origin reaching energies over five orders of magnitude higher in energy than those of the highest energy neutrinos produced in the laboratory and ten million times those that reached us from Supernova 1987A~\cite{Ahlers:2018fkn}. The observations led to the first identification of a cosmic ray source~\cite{IceCube:2018dnn}, a rotating supermassive Black Hole at a distance of 4 billion lightyears. IceCube performed the first oscillation measurements over cosmic distances and found evidence for the appearance of tau neutrinos, including one event where a tau travels 17 metres through the ice before decaying. These measurements (Fig.~\ref{fig:nutriangle}) represent a powerful tool to reveal BSM neutrino physics that can be performed independently of the properties of this UHE neutrino source. IceCube identified a first Glashow-resonance event where an intermediate boson is produced in the interaction of a 6300-TeV antielectron neutrino with an atomic electron. KM3NeT-ARCA will consist of two building blocks, sparsely configured to make a very large volume detector for neutrinos of TeV to PeV energy of astrophysical origin. KM3NeT-ARCA will provide superior pointing resolution, which will enable the discovery of the neutrino sources, and a field of view including the Galactic plane.

These facilities provide unique opportunities including precision tests of fundamental symmetries, most prominently Lorentz invariance, and the search for new physics covering proton decay, dark matter trapped in the Sun, sterile neutrinos, magnetic monopoles and, in general, any hints of deviations from Standard Model physics using a neutrino beam in a new energy regime. The operating experiments have clearly demonstrated the potential of using neutrino ``telescopes'' for exploring the physics of neutrinos themselves. Finally, the next Galactic supernova explosion will not only be the astronomical event of the century, it will also provide an extraordinary opportunity to do neutrino physics, as was the case with 1987A.

\section{Gravitational waves}
The past four years have seen a new revolution in astronomy with the Nobel prize-winning detection of gravitational waves from binary Black Holes and neutron stars by the US LIGO and the European Virgo interferometers. The future generation of GW ground-based detectors, which in Europe is the Einstein Telescope project, has the unique potential to strengthen the synergy between cosmology, particle, nuclear, astroparticle physics and astrophysics science domains since: 1) It tests the nature of gravity, through extremely precise measurements of GW speed of propagation, behaviour at extreme curvatures and matter concentration (Black Holes and pulsars) and through the testing of the Black Hole hypotheses (no-hair theorem, horizon structure, echoes, etc.);  2) It explores the nature of dark matter in a complementary way to colliders and underground direct search experiments, since the existence or not of primordial Black Holes has a large impact on WIMP cosmological density.  It will furthermore explore other dark-matter candidates and/or new particles currently searched for at CERN and elsewhere, e.g.\ axions or ultra-light bosons; 3) It provides an independent measure of the Hubble constant either solving the current tension between its {\em far} and {\em near} determinations or alternatively opening a portal to new physics (e.g.\ sterile neutrinos, new particles etc.). It furthermore provides an alternative way to measure the equation of state of Dark Energy. 4)  It reveals phase transition from nucleons to free quarks giving insight into the QCD phase diagram, explores the state of ultra-dense nucleons and the origin of heavy elements and has the potential, through the studies of gravitational waves from supernova, to  determine the physics of core-collapse supernova associated to neutrino and electromagnetic radiation emission. 5) It studies the primordial Universe, through primordial stochastic backgrounds, early-Universe phase transitions, cosmic strings, etc.; 6) It  probes the nature of space-time at the interface with quantum mechanics,  through the study of alternative gravity  and quantum gravity theories.
 
 ET aims to reach a sensitivity for signals emitted by astrophysical and cosmological sources about a factor of ten better than the design sensitivity of the advanced detectors currently in operation. To reduce the effects of the residual seismic motion, ET will be located underground at a depth of about 100 m to 200 m and, in the complete configuration, it will consist of three nested detectors, each in turn composed of two interferometers. The topology of each interferometer will be the dual-recycled Michelson layout with Fabry-Perot arm cavities, with an arm-length of 10 km. The configuration of each detector devotes one interferometer to the detection of the low-frequency (LF) components of the signal ($2-40$ Hz) while the other one is dedicated to the high-frequency (HF) components. In the former (ET-LF), operating at cryogenic temperature, the thermal, seismic, gravity gradient and radiation pressure noise sources will be particularly suppressed; in the latter (ET-HF) the sensitivity at high frequencies will be  improved by high laser light power circulating in the Fabry-Perot cavities, and by the use of frequency-dependent squeezed light technologies. 

ET science not only  reflects CERN's own pursuit of fundamental physics, but strong synergies exist concerning the precision engineering that was required to build and operate the LHC. The core areas where CERN expertise could benefit ET are:
\begin{enumerate}
\item
The vacuum and cryogenics R$\&$D is an enabling factor for the new generation infrastructures both in accelerators and GW interferometers;
\item
Extreme photonics solutions are necessary both in particle physics and GW interferometers for stray light control and mitigation;
\item
The civil infrastructure cost is a large part of the next generation projects (particle physics including neutrino and GW interferometer arms on ground or underground), and therefore, smart and resilient solutions have to be found for their realisations;
\item
Innovative solutions will have to be found for the computing and data-analysis infrastructure, including the distribution of low latency alerts among continents as well as new data harvesting methods (big data analytics, machine learning, etc.);
\item
Last but not least the consortia building GW large infrastructures would profit enormously from the previous experience of the particle physics community of building and sustaining large communities and infrastructures.    
\end{enumerate}

\section{Multimessenger astroparticle physics}
The spectacular observations of Supernova SN1987A in both MeV neutrinos as well as electromagnetic radiation provided an early taste of the enormous scientific potential of multimessenger astronomy, not just for astronomy but also for fundamental physics. It then took 30 years to conquer the extra-Galactic distance scale with the result that  today we can observe events such as neutron star mergers in both gravitational waves and electromagnetic radiation, or coincident high-energy cosmic neutrinos from active galaxies that are flaring in high-energy gamma rays. New generations of observatories with significantly improved sensitivity are either in construction or being planned,  further accelerating the discovery rates and exploiting the scientific opportunities provided by the combination of the messengers. 

Gamma-ray observatories, as well as neutrino or cosmic ray detectors such as AMS II, allow for sensitive searches of dark matter via  indirect signatures. In fact, the European Cherenkov Telescope Array (CTA), which just entered the construction phase, will have sensitivity to probe the natural WIMP annihilation cross sections in the TeV mass range. 


The observations of high-energy neutrinos  in temporal coincidence with a flaring Blazar \cite{IceCube:2018dnn} has not
only launched a new era in the study of the origins of high-energy
cosmic rays, but also made possible a breakthrough in the exploration
of Lorentz symmetry using neutrinos. A new generation of neutrino detectors, such as GVD, IceCube Gen2, KM3NeT or GRAND aim at increasing the available event statistics, as well as the energy range, and  will significantly improve the sensitivity to new physics.
\vspace*{-2mm}
 


\section{Synergies with HEP}
\label{sec:CM-synHEP}
The field of astroparticle physics, and in particular the exciting field of multimessenger astrophysics, has deep connections at many levels with the field of particle physics,  
spanning from acceleration mechanisms of particles, neutrino physics, dark matter and cosmology.

The physics cases for gravitational waves, neutrinos and gamma-rays are particularly revealing. They offer a complementary approach to the search for indirect signatures of dark matter (see Chapter~\ref{chap:dm}), and in addition explore its location in the cosmos. Moreover, the cosmos provides bright sources of neutrinos with various energies and baselines and different matter densities, spanning from the Sun to supernovae and atmospheric neutrinos. This deep complementarity will contribute to the future determination of the neutrino mass ordering, the study of sterile neutrinos and maybe other surprises with UHE neutrinos and exotic interactions. Additionally, synergies with the particle physics communities and beam-dump experiments are extremely relevant for astroparticle physics to improve particle interaction models (for instance for neutrino and cosmic rays). 

These many fields of synergy between particle and astroparticle communities require a strong theoretical support, which will be fostered by the European Center for AstroParticle Theory (EuCAPT)
with its first hub at CERN.

The future generation of Gravitational Wave ground-based detectors, which in Europe is the Einstein Telescope, has the unique potential to explore the dark matter, its location in the cosmos and to understand the fraction of it that is not of particle origin. With its capacity of testing gravity at extreme curvatures and matter in extreme conditions, such as in Black Holes or pulsars, it explores new frontiers in cosmology and particle physics and specifically the unification of all forces, including gravity. ET can explore heavy element formation and exotic forms of matter, and has also the potential to strengthen synergy between the nuclear, particle physics and cosmology communities. 

Multimessenger  astrophysics has a striking potential for cosmic exploration, reaching out to the inspiration of every citizen. 
Such exploration concerns also technology and computing challenges, and open-access policies, which need to be developed in cooperation with CERN. There are multiple synergies between particle and astroparticle physics at the level of infrastructures (large underground excavations, vacuum, engineering and management of large projects), detectors (including photosensor and other detection techniques, electronics, computing),  etc. All of these synergies might clearly benefit from a closer collaboration between CERN and the astroparticle community, to be developed through focused discussion between CERN and APPEC. 

\chapter{Beyond the Standard Model} 
\label{chap:bsm}

\section{Introduction}

The search for physics Beyond the SM (BSM) is the main driver of the exploration programme in particle physics. The initial results from the LHC are already starting to mould the strategies and priorities of these searches and, as a result, the scope of the experimental programme is broadening. Growing emphasis is given to alternative scenarios and more unconventional experimental signatures where new physics could hide, having escaped traditional searches. This broader approach towards BSM physics also influences the projections for discoveries at future colliders. Rather than focusing only on a restricted number of theoretically motivated models, future prospects are studied with a signal-oriented strategy. In this chapter an attempt to reflect both viewpoints and to present a variety of possible searches is made. Since it is impossible (and probably not very useful) to give a comprehensive classification of all existing models for new physics, the choice is made to consider some representative cases which satisfy the following criteria: {\it (i)} they have valid theoretical motivations, {\it (ii)} their experimental signatures are characteristic of large classes of models, {\it (iii)} they allow for informative comparisons between the reach of different proposed experimental projects.

In considering the physics reach of any experimental programme, there are two key questions: what new physics can be discovered and, in the absence of discoveries, what information can be extracted from the measurements. For many of the current models, any discovery will require several observations in different channels in order to characterise the nature of the phenomenon. As an example, the well-known missing energy signature arises in numerous models and, while any large excess of such events would signify a departure from the SM, a real understanding of the underlying physics will be possible only with multiple experimental studies. For this reason, the results presented in this chapter do not attempt to characterise the potential for `discovery'; instead, all results are expressed in terms of the extent to which one can be sensitive to new physics. This is quantified by the exclusion reach on some key physical parameters, such as the distance down to which the Higgs boson still behaves like a point-like particle, or of the masses of new hypothetical particles. Unless otherwise stated, all limits correspond to 95\% Confidence Level (CL) limits. The results presented in the current summary originate from a large number of physics studies of varying degrees of sophistication, which vary from detailed simulations, to fast simulations, simple detector parametrisations, or direct extrapolations of results from data. In some cases, simple rescalings from similar results are used. We caution the reader to bear in mind this large variation in the precision of the estimates.

This chapter is organised as follows. In Sect.~\ref{sec:BSM-EWKNR} various signatures related to the dynamics of electroweak symmetry breaking, i.e.\ composite Higgs, vector resonances, contact interactions, are considered. The search for supersymmetry is the theme of Sect.~\ref{sec:BSM-SUSY}. Extensions of the Higgs sector with new scalar particles, neutral naturalness and high-energy dynamics associated with the flavour problem are presented in Sect.~\ref{sec:BSM-ExtendedScalars}. In Sect.~\ref{sec:BSM-DM} the prospect for exploring dark matter at colliders is presented. Section~\ref{sec:BSM-FIPs} is dedicated to a study of how different experimental facilities compare in the search for feebly-interacting or long-lived particles. Finally, in Sect.~\ref{sec:BSM-Conclusions} the results are summarised and put in a broader perspective.

\section{Electroweak symmetry breaking and new resonances}
\label{sec:BSM-EWKNR}

\subsection{Contact interactions}
\label{sec:contactint}

Higher-dimensional operators in an EFT are the most robust and model-independent way of describing BSM virtual effects whenever there is a large separation between the EW and new-physics mass scales. The only drawback of this approach is that, without any theoretical bias on the underlying BSM model, one is confronted with a large number of independent operators, all compatible with SM symmetries. Here, representative sets of operators, which are typically generated in BSM extensions of the EW symmetry breaking (EWSB) sector and which allow for an informative comparison between different collider facilities, are considered.

\begin{figure}[htb]
    \centering
    \includegraphics[width=1.0\textwidth]{\main/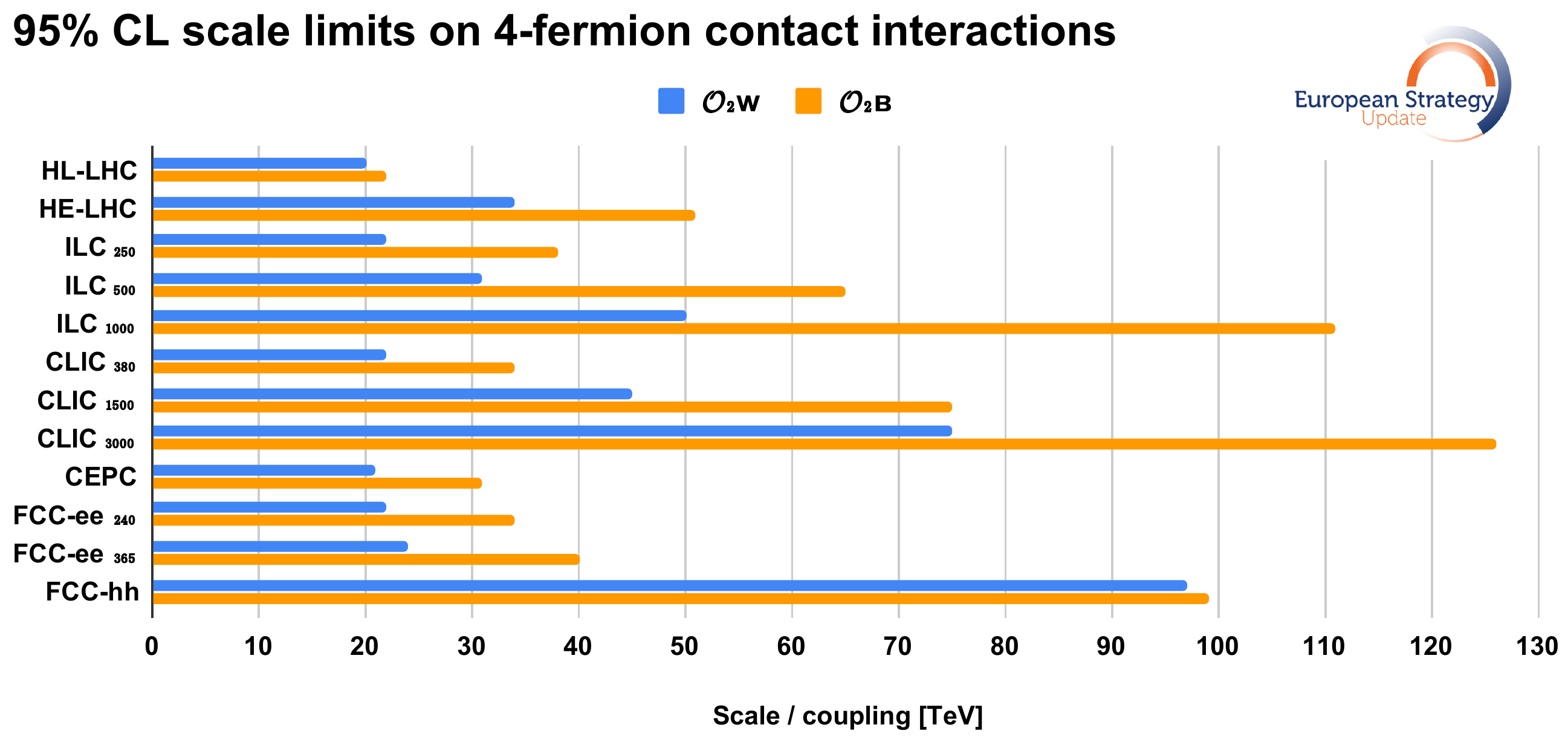}
    \caption{Exclusion reach of different colliders on four-fermion contact interactions from the operators $\mathcal{O}_{2W}$ and $\mathcal{O}_{2B}$. The blue bars give the reach on the effective scale $\Lambda /(g_2^2 \sqrt{c_{2W}})$ and the orange bars on $\Lambda /(g_1^2 \sqrt{c_{2B}})$, where $c_{2W,2B}$ are the Wilson coefficients of the corresponding operators and the gauge couplings come from the use of the equations of motion.}
    \label{fig:4f}
\end{figure}

As a first example, consider the operators ${\cal O}_{2W}$ and ${\cal O}_{2B}$ (defined in Ref.~\cite{Giudice:2007fh}), which correspond to the oblique parameters $W$ and $Y$~\cite{Barbieri:2004qk} and describe the leading higher-derivative corrections to the SM gauge boson propagators. Using the equations of motion, these operators lead to charged and neutral current {\it four-fermion contact interactions} of universal type.
The 95\% CL exclusion reach of different colliders on the effective scales of the operators ${\cal O}_{2W}$ and ${\cal O}_{2B}$ is shown in Fig.~\ref{fig:4f}. The precise definition of each collider option is given in Ref.\cite{deBlas:2019rxi}. In some cases (ILC, CLIC, FCC) different phases or running modes of the same collider program are shown. All collider sensitivities shown in Fig.~\ref{fig:4f} are derived after combination with the results expected from the HL-LHC, which will precede all future colliders. 
The projected limits come from a variety of di-fermion final states. 
Lepton colliders with suitable luminosity are particularly powerful in testing the neutral-current case thanks to the clean signatures, the small theoretical uncertainties, and the capability to perform detailed differential analyses for a large set of di-fermion final states. Linear colliders can also benefit from different longitudinal polarisations of the two beams.
Hadron colliders have the additional advantage of excellent sensitivity via Drell-Yan (DY) production for both neutral and charged currents, because of the accessible charged initial states (e.g.\  $u\bar{d}$). As a matter of fact, the reach of hadron colliders for charged and neutral currents is expected to be about the same. The apparent exception for the HE-LHC collider is simply due to the absence of dedicated charged Drell-Yan studies in the present inputs. As expected for contact interactions, the experimental sensitivity increases significantly with $\sqrt{s}$ in all cases. The results from $ep$ colliders are not shown because they are not competitive: LHeC does not improve the reach of HL-LHC, and FCC-eh does not affect the global fit of FCC-ee/hh.

\begin{figure}[htb]
    \centering
          \includegraphics[width=1.0\textwidth]{\main/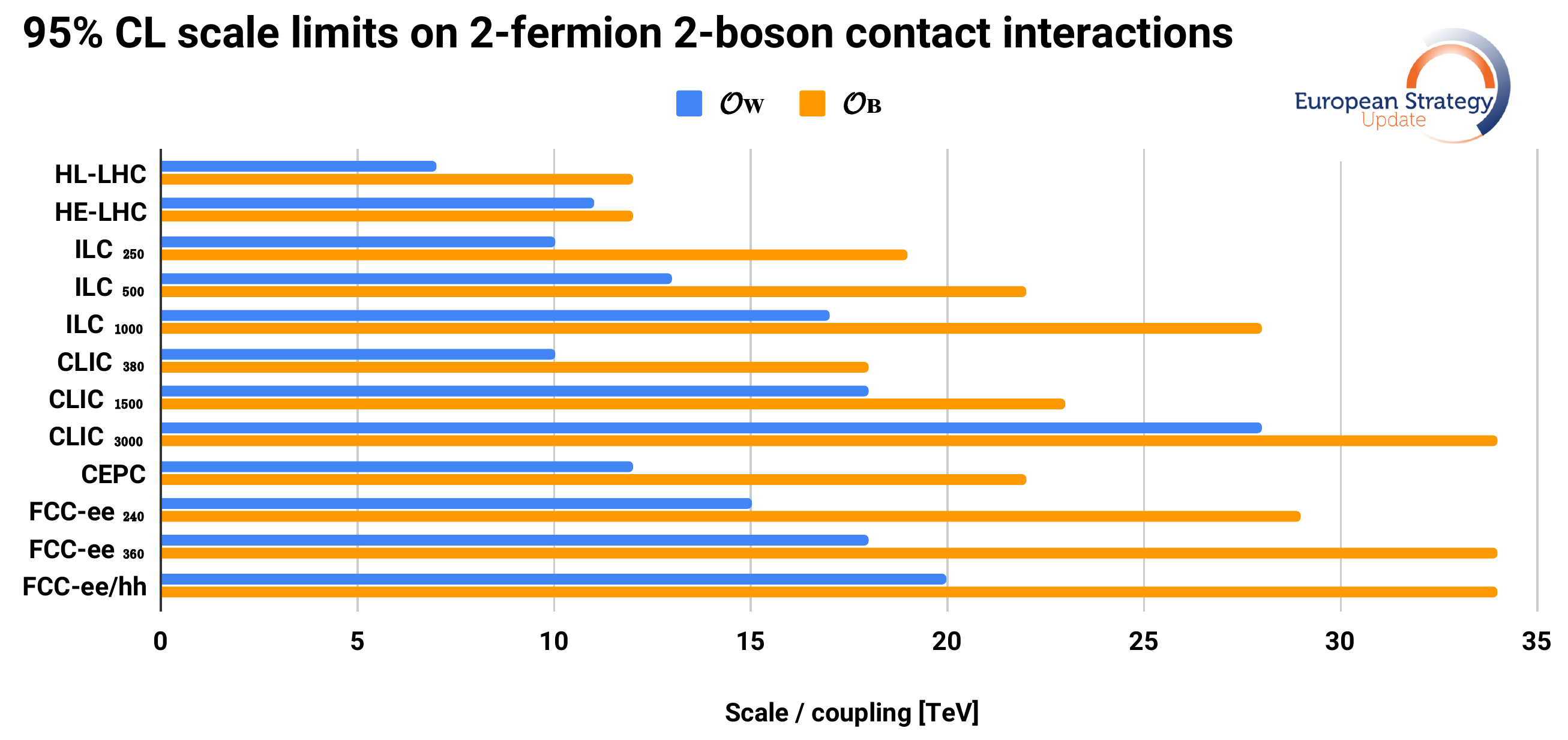}
    \caption{Exclusion reach of different colliders on the two-fermion/two-boson contact interactions from the operator $\mathcal{O}_{W}$ and $\mathcal{O}_{B}$. The blue bars give the reach on the effective scale $\Lambda /(g_2^2 \sqrt{c_{W}})$ and the orange bars on $\Lambda /(g_1^2 \sqrt{c_{B}})$, where $c_{W,B}$ are the Wilson coefficients of the corresponding operators and the gauge couplings come from the use of the equations of motion.}
    \label{fig:2f2b}
\end{figure}

As a second example, consider the operators ${\cal O}_{W}$ and ${\cal O}_{B}$ (defined in Ref.~\cite{Giudice:2007fh}), which are of special phenomenological relevance for BSM theories of EWSB because they describe new-physics effects in the interaction between the gauge and Higgs sectors. Using the equations of motion, they can be turned into {\it two-fermion/two-boson contact interactions}. The reach on the effective scales of the operators ${\cal O}_{W}$ and ${\cal O}_{B}$ is shown in Fig.~\ref{fig:2f2b}. The projected limits come from new-physics contributions that can interfere with SM di-boson production processes. For CLIC, the leading sensitivity on ${\cal O}_{W}$ comes from a detailed differential analysis of ${e^+e^-}\to {ZH}$~\cite{Beneke:2014sba}, whereas the power of FCC-hh comes through an analysis of the $p_T$ distribution of the $Z$ in ${pp}\to {WZ}$~\cite{Franceschini:2017xkh}. The largest sensitivity of lepton colliders at lower $\sqrt{s}$ and even on the ${\cal O}_{B}$ operator alone at large $\sqrt{s}$ comes from EW precision measurements of the oblique parameter $S$, which constrains directly the combination ${\mathcal{O}}_{W}+{\mathcal{O}}_{B}$~\cite{Giudice:2007fh}. 

\subsection{New vector bosons: the $Y$-Universal $Z^\prime$}

New vector bosons are common in many BSM theories, ranging from new models of EWSB to extensions of the SM gauge group. As a representative example of these classes of theories, the ``$Y$-Universal $Z^\prime$'' (see e.g. \cite{Appelquist:2002mw}) is considered. The model consists of a new neutral gauge boson $Z^\prime$ with mass $M$ and charges to SM particles equal to hypercharge, although the coupling constant $g_{Z^\prime}$ is taken to be a free parameter, in general different from the one of the SM \mbox{U$(1)_Y$}. The perturbative limit is taken to correspond to $g_{Z^\prime}<1.5$ since for larger values the width of the $Z^\prime$ exceeds $0.3\, M$.

The $Y$-Universal $Z^\prime$ is selected instead of one of the standard benchmarks (such as the Sequential or $B-L$ models) for several reasons. It has comparable couplings to quarks and leptons, allowing for a fair comparison between hadron and lepton colliders. Its couplings are flavour-diagonal, making the model safely compatible with flavour constraints. When integrated out at tree-level, it generates only the universal operator ${\mathcal{O}}_{2B}$ in the SM EFT, with coefficient ${c_{2B}}/{\Lambda^2}={g_{Z^\prime}^2}/{(g_1^4 M^2)}$.  Since the sensitivity to ${\mathcal{O}}_{2B}$ is available for all colliders~\cite{deBlas:2019rxi}, a straightforward and rigorous assessment of the indirect reach is possible for the $Y$-Universal $Z^\prime$ model, while additional input would be needed for the standard benchmarks.

\begin{figure}[t]
    \centering
    \includegraphics[width=0.6\textwidth]{\main/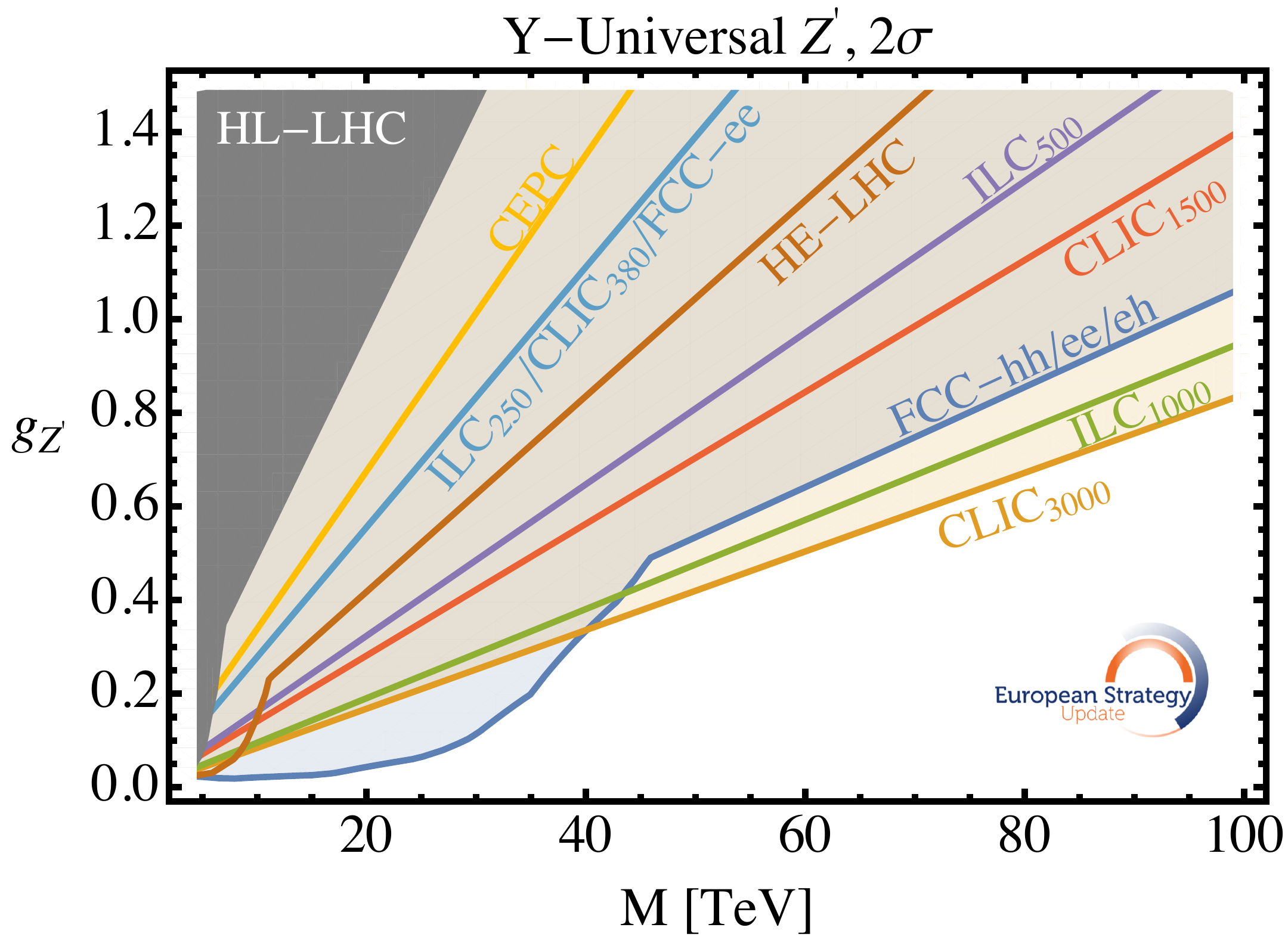}
    \caption{Exclusion reach of different colliders on the $Y$-Universal $Z^\prime$ model parameters.}
    \label{fig:Universal_Zp}
\end{figure}

Figure~\ref{fig:Universal_Zp} displays the $95\%$~CL exclusion reach on $g_{Z^\prime}$ and $M$, at various colliders. For hadron machines, the reach of direct searches (round curves at small $g_{Z^\prime}$) is obtained from recasting the results in Refs.~\cite{CidVidal:2018eel,Jamin:2019mqx}, overlaid with the indirect sensitivity (diagonal straight lines at large $g_{Z^\prime}$) discussed previously. It is seen that the direct mass reach is inferior to the indirect one for high $g_{Z^\prime}$, in agreement with the generic expectation that strongly-coupled new physics is better probed indirectly. Moreover, the indirect reach  benefits greatly from higher collider energies. These two observations explain both the competitiveness of lepton colliders in indirect searches and the good indirect performances of the FCC-hh and HE-LHC colliders.

\subsection{The Composite Higgs scenario}

Decades of theoretical investigation (for a review, see \cite{Panico:2015jxa}) provided us with a rather clear picture of how a viable Composite Higgs model should look like, and with effective parametrisations of the relevant phenomenology that can be employed in the present context. The central idea is that the Higgs emerges as a bound state of a new strongly-interacting confining Composite Sector, analogue to QCD but with a much higher confinement scale. The Higgs, similarly to the pions in QCD, emerges as a Goldstone boson associated with a spontaneously-broken global symmetry of the Composite Sector. This explains why it is lighter than other bound states (collectively called `resonances') that will unavoidably emerge from the Composite Sector dynamics. In analogy with the pion in QCD, the spin-one resonances will be called $\rho$.

The phenomenology is mainly controlled by two parameters: the mass scale $m_*$ and the coupling $g_*$. The mass $m_*$ is the Composite Sector confinement scale, analogue to $\Lambda_{\rm{QCD}}$. It controls the mass of the resonances and sets the scale of the EFT operators that describe at low energy the indirect effects of Higgs compositeness. Its inverse can be interpreted as the \emph{geometric size} of the Higgs, $\ell_H=1/m_*$. The sensitivity of future colliders to $\ell_H$ is the most important parameter to answer the question whether the Higgs is elementary ($\ell_H=0$) or composite ($\ell_H\neq0$). The coupling parameter $g_*$ represents the interaction strength among particles originating from the Composite Sector. It controls the strength of the Higgs couplings to the $\rho$ resonance and it sets the scale of couplings that appear in the EFT Lagrangian. The internal coherence of the construction requires $g_*$ to be larger than the EW coupling ($g_*\gtrsim 1$) but smaller than the perturbative unitarity limit ($g_*\lesssim 4\pi$). 

Among the operators in the Composite Higgs EFT,  $\mathcal{O}_{\phi}$ (defined as in \cite{deBlas:2019rxi}), $\mathcal{O}_{W}$ and $\mathcal{O}_{2W}$ are the most representative and offer the best sensitivity at all colliders. Parametrically, their Wilson coefficients are
\begin{equation}\label{eq:CHOp}
\frac{c_{\phi}}{\Lambda^2}\sim \frac{g_*^2}{m_*^2}\,,\;\;\;\;\;
\frac{c_{W}}{\Lambda^2}\sim \frac{1}{m_*^2}\,,\;\;\;\;\;
\frac{c_{2W}}{\Lambda^2}\sim \frac{1}{g_*^2m_*^2}\,.
\nonumber
\end{equation}
These relations are merely \emph{estimates} of the expected magnitude of the Wilson coefficients, which hold up to model-dependent order-one factors. In the current analysis, these relations are taken as exact equalities, so the results should not be interpreted as strictly quantitative, but only as a fair assessment of the sensitivity. 

\begin{figure}[t]
    \centering
    \includegraphics[width=0.48\textwidth]{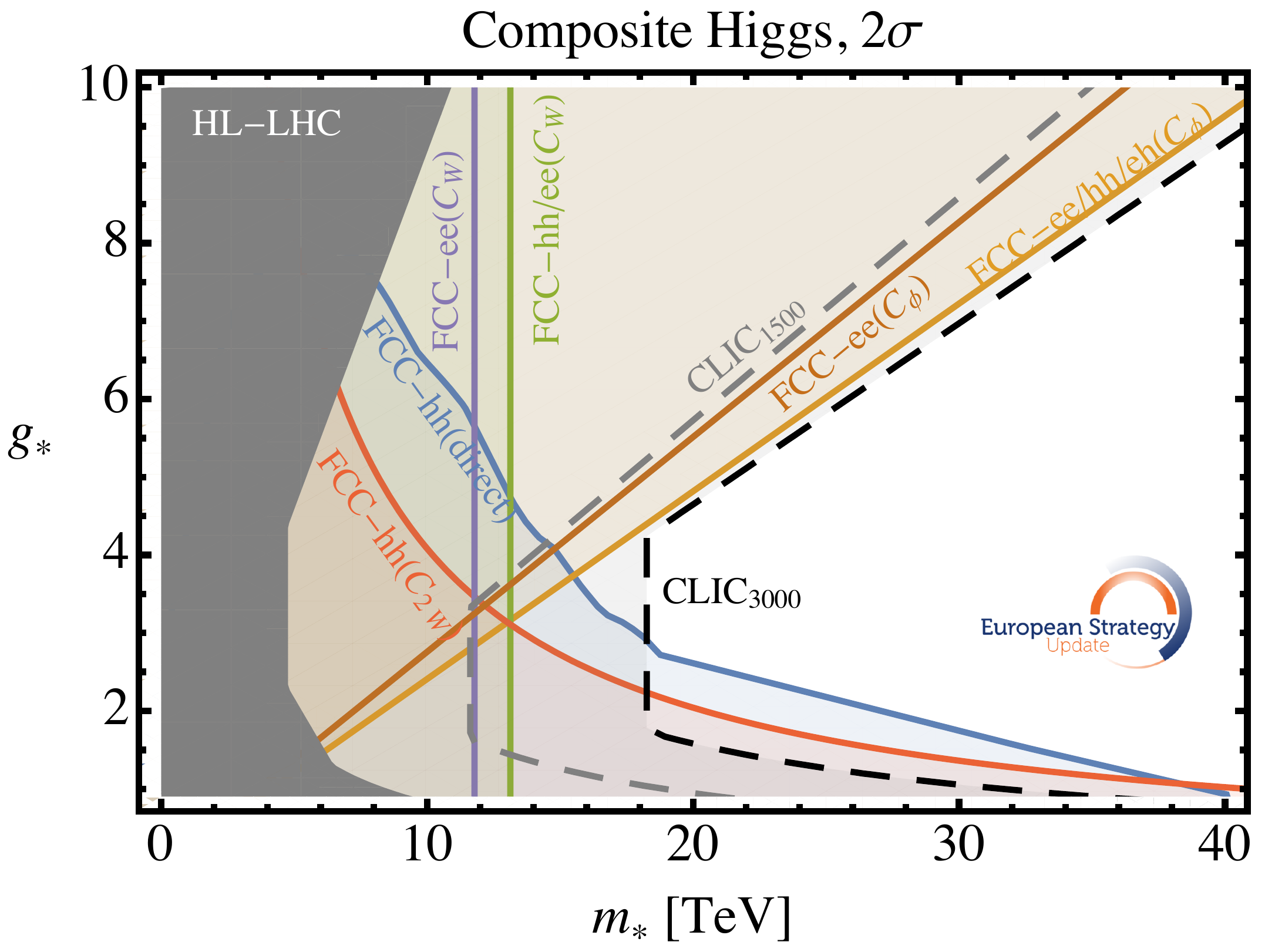} ~
        \includegraphics[width=0.48\textwidth]{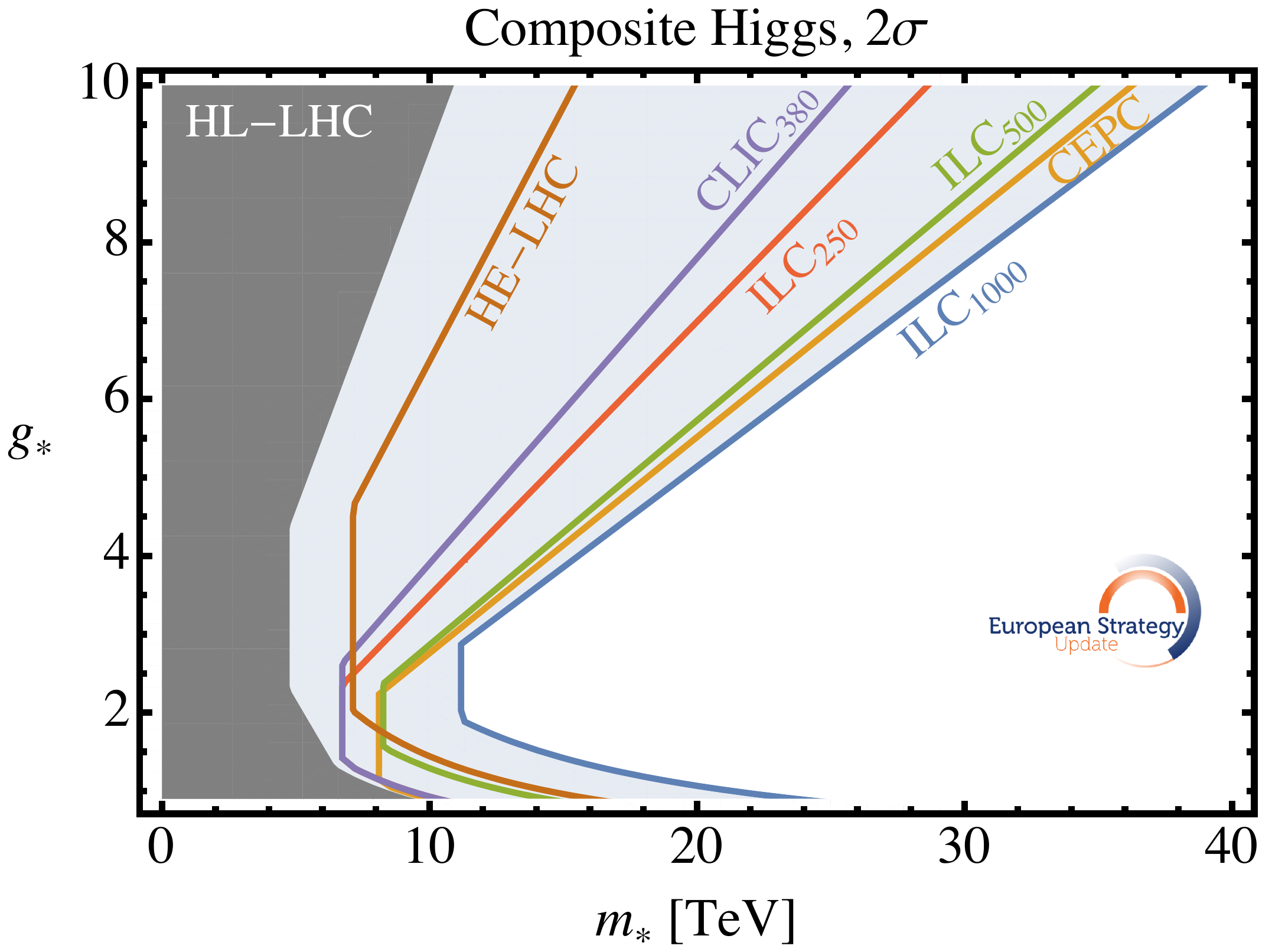}
    \caption{Left panel: exclusion reach on the Composite Higgs model parameters of FCC-hh, FCC-ee, and of the high-energy stages of CLIC. Right panel: the reach of HE-LHC, ILC, CEPC and CLIC$_{380}$. The reach of HL-LHC is the grey shaded region.}
    \label{fig:Composite_Higgs}
\end{figure}

Figure~\ref{fig:Composite_Higgs} shows the exclusion reach on $m_*$ and $g_*$ from the highly complementary probes on the operators $\mathcal{O}_{\phi}$, $\mathcal{O}_{W}$ and $\mathcal{O}_{2W}$ with different experimental strategies in different colliders. For the FCC project, $\mathcal{O}_{\phi}$ is most effective at large $g_*$, and it is well probed by Higgs couplings measurements at FCC-ee. However FCC-hh and FCC-eh further improve the reach on $c_\phi$ as shown in the figure. The reach on $c_\phi$ for all collider options is extracted from the summary Table~8 of Ref.~\cite{deBlas:2019rxi}, with the exception of HL-LHC for which a more conservative value of $c_\phi|_{1\sigma}=0.42/{\rm{TeV}}^2$ (also reported in Ref.~\cite{deBlas:2019rxi}) is employed. The operator $\mathcal{O}_{2W}$ is instead effective at low $g_*$, and it is probed by high-energy charged DY measurements at FCC-hh~\cite{Farina:2016rws}. The mass-reach from $\mathcal{O}_{W}$ is instead independent of $g_*$.  The reach of direct resonance searches is also shown in Fig.~\ref{fig:Composite_Higgs}, for the \FCChh and the \HLLHC. It represents the sensitivity to an EW triplet $\rho$ vector resonance, generically present in Composite Higgs models. The reach is extracted from ref.~\cite{Thamm:2015zwa,Golling:2016gvc,Mangano:2651294}, and it emerges from a combination of dilepton and diboson final state studies. Direct searches are more effective at \emph{low} $g_*$, which may seem surprising. The reason is that $g_*$ is the $\rho$ coupling to the Higgs boson, while the coupling of the $\rho$ to quarks, which drives the production, scales like $g_2^2/g_*$ and therefore increases for small $g_*$. Unfortunately, no direct reach projection is currently available for the \HELHC.

\begin{figure}[t]
    \centering
    \includegraphics[width=1.0\textwidth]{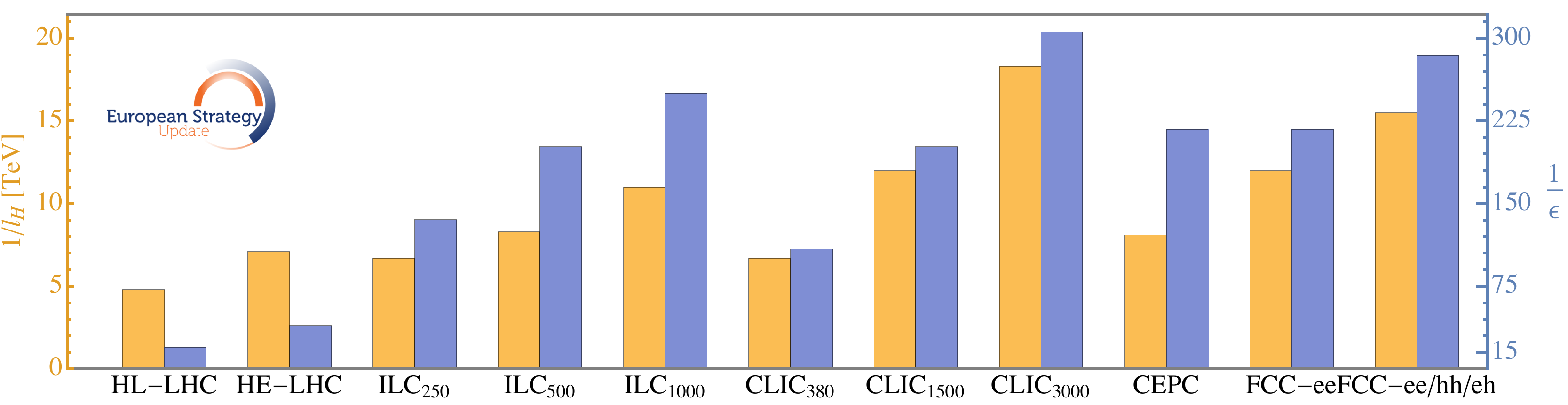}
    \caption{Exclusion reach of different colliders on the inverse Higgs length $1/\ell_H =m_*$ (orange bars, left axis) and the tuning parameter $1/\epsilon$ (blue bars, right axis), obtained by choosing the weakest bound valid for any value of the coupling constant $g_*$.}
    \label{fig:Composite_Higgs_Summary}
\end{figure}

The information in Fig.~\ref{fig:Composite_Higgs} can be projected into a single number, as displayed in Fig.~\ref{fig:Composite_Higgs_Summary}. The orange bars show the maximum $m_*$ (or, equivalently, the minimum Higgs size $\ell_H$) a given collider is sensitive to, independently of the value of $g_*$. The blue bars show the tuning parameter $1/\epsilon$ (which is equal to the conventional tuning parameter $\Delta$), obtained as follows. Higgs compositeness can address the naturalness problem, provided it emerges at a relatively low scale, but the parameter $m_*$ is not the most appropriate measure of the degree of fine-tuning required to engineer the correct Higgs mass and EWSB scale. A better measure is (see e.g., \cite{Matsedonskyi:2015dns}) $1/\epsilon>(m_T/500\,{\rm{GeV}})^2>m_*^2/g_*^2v^2$,
where $v=246$~GeV and $m_T$ is the top-partner mass. The second inequality provides the estimate of the reach on $\epsilon$ reported in Fig.~\ref{fig:Composite_Higgs_Summary}. The equation also displays the impact of fermionic top-partner searches on $\epsilon$. The discovery reach of these particles at HL-LHC, HE-LHC and FCC-hh are of $1.5$, $2$ and $4.7$~TeV, respectively. These correspond to a reach on $1/\epsilon$ of $10$, $16$ and $88$.

\newcommand{\gl}{\ensuremath{\tilde{g}}}
\newcommand{\tone}{\ensuremath{\tilde{t}_1}}
\newcommand{\neut}{\ensuremath{\tilde{\chi}}}
\newcommand{\stopq}{\ensuremath{\tilde{t}}}
\newcommand{\chiOneZero}{\ensuremath{\tilde{\chi}_1^0}}
\newcommand{\chiTwoZero}{\ensuremath{\tilde{\chi}_2^0}}
\newcommand{\chiOnePM}{\ensuremath{\tilde{\chi}_1^\pm}}
\newcommand{\chiOneMP}{\ensuremath{\tilde{\chi}_1^\mp}}
\newcommand{\Ptmiss}{\ensuremath{p_T^{\rm miss}}}
\newcommand{\Dmstop}{\ensuremath{\Delta m(\stopq,\chiOneZero)}}

\section{Supersymmetry}
\label{sec:BSM-SUSY}

Supersymmetry (SUSY) remains the only known dynamical solution to the Higgs naturalness problem that can be extrapolated up to very high energies, in a consistent and calculable way. It gives a successful framework for gauge coupling unification; it automatically predicts a potential candidate for DM; it turns the Higgs mass into a calculable parameter; it gives a conceptual justification for EWSB; it offers a hint for a final reconciliation between gravity and gauge forces. In spite of this impressive list of attractive features, there is no firm experimental evidence that speaks in favour of SUSY. The results from the LHC have already constrained significantly a variety of models that implement SUSY at the EW scale and have contributed to shift the focus of the continuing experimental scrutiny. The questions that need to be answered are whether SUSY can still be the key to understand Higgs naturalness but its implementation at the EW scale is less simple than originally assumed, or whether SUSY is the key to understanding EWSB but its implementation comes at the price of some amount of tuning. Both these questions can only be answered by further experimental investigation. The most important targets to gather information about these questions are the SUSY particles that directly affect the Higgs mass and thus contribute more sizeably to the degree of tuning. These are the top squarks (stops) and the Higgsinos. Moreover, because of an accidentally large two-loop contribution to the stop mass, the gluino also plays a central role in the degree of naturalness of SUSY models. Nonetheless, the interest in understanding the full structure of the theory strongly motivates also the search for the other SUSY particles (squarks, sleptons, and EW gauginos), especially since the limits from the LHC on sleptons and EW gauginos are still rather weak in comparison to those on strongly interacting particles.   

SUSY may be realised in nature in a variety of different models, leading to distinct collider signals coming from production of superpartners. Coloured SUSY particles such as squarks and gluinos are produced via the strong interaction and have the highest cross sections at hadron colliders. EW gauginos and Higgsinos mix into neutralino and chargino mass states, which will be collectively referred to as electroweakinos (EWkinos). Their production cross sections depend on mixing parameters and are typically much smaller than those of coloured superpartners at hadron colliders. 
For this reason, the EW sector remains more difficult to test at hadron machines and searches at $e^{+}e^{-}$ colliders would complement the SUSY parameter space coverage. The null results from searches at the LHC motivate scrutiny of less conventional as well as more experimentally difficult processes where SUSY may lurk. These include the cases of nearly-degenerate mass spectra, non-prompt decays and exotic signatures.

\subsection{Gluinos and squarks} 
\begin{figure}[htb]
    \centering
\includegraphics[width=1.0\textwidth]{\main/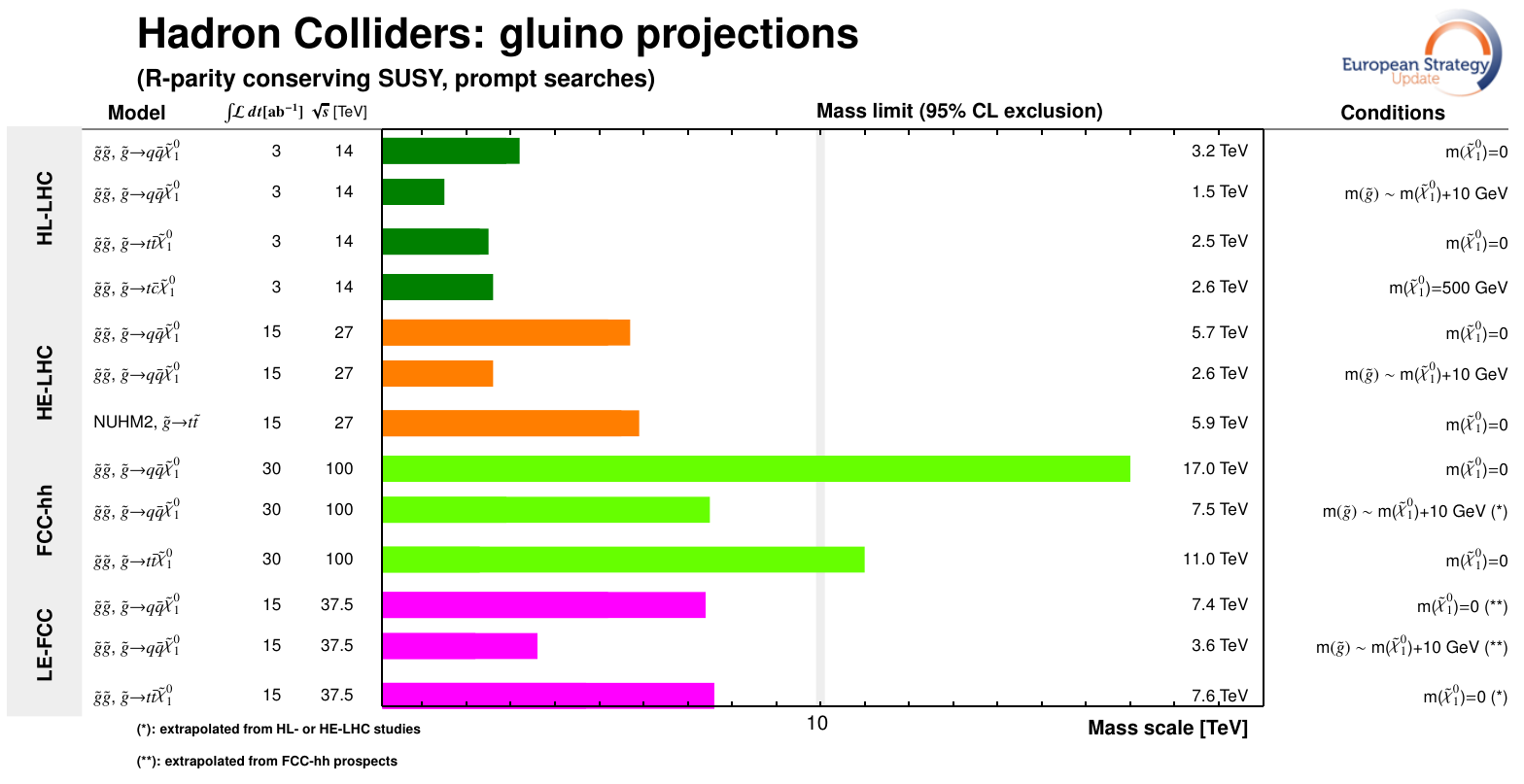}
    \caption{
    Gluino exclusion reach of different hadron colliders: HL- and HE-LHC~\cite{CidVidal:2018eel}, and FCC-hh~\cite{Golling:2016gvc, Abada:2019lih}. Results for low-energy \FCChh are obtained with a simple extrapolation.
    }
\label{fig:SUSY_gluino}
\end{figure}

The expected reach of gluino (\gl) searches at hadron colliders is shown in Fig.~\ref{fig:SUSY_gluino}, for $R$-parity conserving scenarios and under simplifying assumptions on the $\gl$ prompt decay mode.  In all cases, the lightest neutralino ($\tilde \chi_1^0$) is assumed to be the lightest supersymmetric particle (LSP). Exclusion limits at 95\% CL are presented; the corresponding definitive observation with a significance of $5\,\sigma$ is 5--10\% lower for each process. The most stringent gluino bounds are found for massless LSP. If the gluino is close in mass to the LSP, the topology is referred to as `compressed' and the amount of missing transverse momentum (\Ptmiss) in the event decreases. The typical multijet + \Ptmiss~SUSY searches are less effective and must be replaced with monojet-like analyses, where the gluinos recoil against an initial state radiation (ISR) jet. Lepton colliders are ineffective in the search for gluinos, which are neutral with respect to the EW interaction.

\begin{figure}[htb]
    \centering
\includegraphics[width=1.0\textwidth]{\main/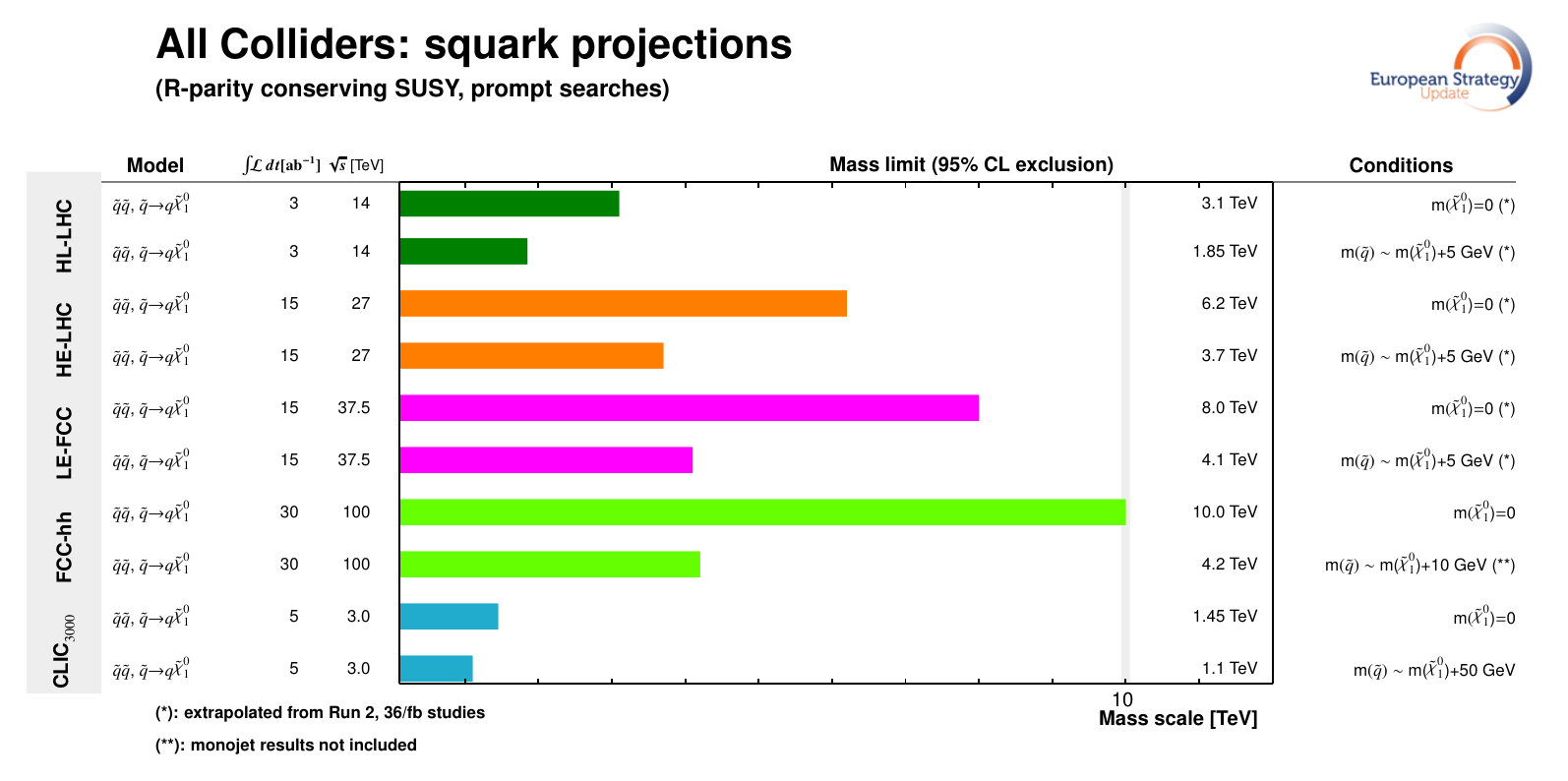}
    \caption{
    Exclusion reach of different hadron and lepton colliders for first- and second-generation squarks.}
\label{fig:SUSY_squarks}
\end{figure}

While there are no recent projections on searches for first- and second-generation squarks at \HLLHC~\cite{ATL-PHYS-PUB-2014-010} and \HELHC, the 95\% CL exclusion reach for models where squarks decay directly into a quark and a neutralino can be extrapolated from the recent ATLAS and CMS results using Run 2 data (36~fb$^{-1}$, see for example Refs.~\cite{PhysRevD.97.112001} and~\cite{Khachatryan2017}). Two scenarios and analysis approaches are considered: massless neutralino (from jets+\Ptmiss~searches) and mass splitting of 5~GeV between the squark and neutralino (inferred from monojet searches). The results are shown in Fig.~\ref{fig:SUSY_squarks}. Extrapolated prospects for the LE-FCC are also reported, as well as the reach for \CLICThreeThousand~\cite{CLIC:PrivateCommunication} and results of dedicated studies at the \FCChh~\cite{Golling:2016gvc}.

\begin{figure}[htb]
    \centering
\includegraphics[width=1.0\textwidth]{\main/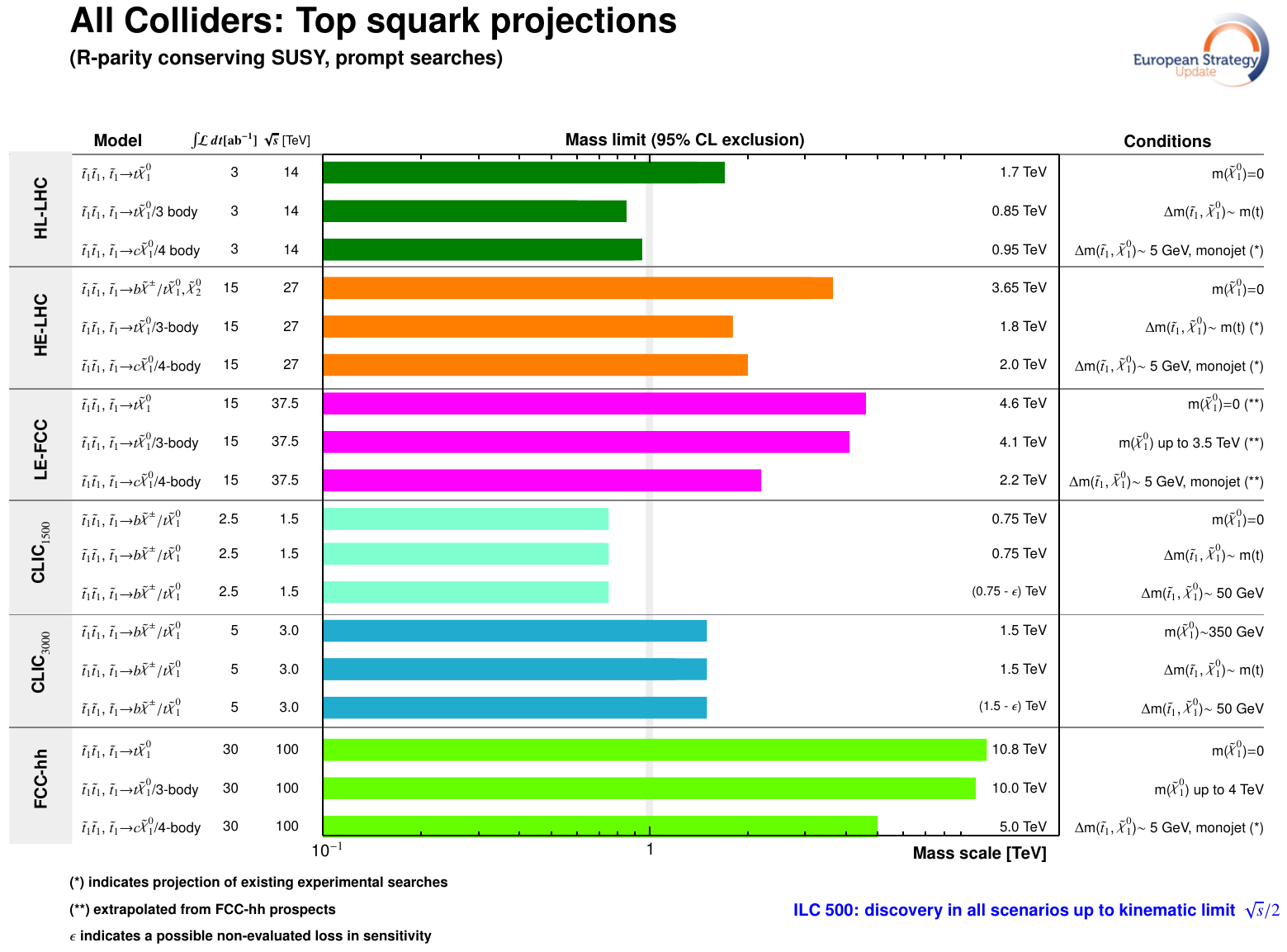}
    \caption{
    Top squark exclusion reach of different hadron and lepton colliders. All references are reported in the text. Results for CLIC have been communicated privately by the authors. Results for LE-FCC are extrapolated from HL- and HE-LHC studies. }
\label{fig:SUSY_stop}
\end{figure}

Most studies of top squark (\tone) pair-production at hadron colliders assume $\tone \to t \chiOneZero$ and fully hadronic or semi-leptonic final states with large \Ptmiss. The best experimental sensitivity is achieved for 
$m(\chiOneZero ) \approx 0$ (i.e.~$\Dmstop \gg m_{t}$), while the reach in $m_{\tilde{t}}$~degrades for larger \chiOneZero~masses. For this reason, high-energy lepton colliders, e.g.\ \CLICThreeThousand,  might become competitive with \HLLHC in these topologies, as their stop mass reach is close to $\sqrt{s}/2$ even for low $\Dmstop$. Lower centre-of-mass energy lepton facilities do not have sufficient kinematic reach. The exclusion limits are summarised in Fig.~\ref{fig:SUSY_stop}; 
the discovery potential in all channels is about 5\% lower. If the \stopq$-$\chiOneZero~mass splitting is such that final states include very off-shell $W$ and $b$-jets, \stopq~masses up to about 1 TeV can be excluded at the \HLLHC~\cite{CidVidal:2018eel}. A two-fold and five-fold increase in reach is expected for the \HELHC~\cite{CidVidal:2018eel} and \FCChh~\cite{Abada:2019lih} respectively, with potential of improvements, especially in very compressed scenarios, via optimisation of monojet searches~\cite{Aaboud2018}.  

\begin{table}[t]
    \caption{Estimates of the degree of fine tuning in SUSY theories that can be probed with measurements of stop and gluino masses. The fine-tuning parameter is defined as $1/\epsilon \equiv \Delta m_h^2 / m_h^2$~\cite{Barbieri:1987fn}, where $\Delta m_h^2$ is the contribution to the physical Higgs mass $m_h$, which for stops (at one-loop) and gluino (at two-loops) is given by $1/\epsilon_{\tilde t}=(3y_t^2m_{\tilde t}^2/2\pi^2 m_h^2)\, \ln ({\Lambda }/ m_{\tilde t})$ and $1/\epsilon_{\gl}=(4y_t^2\alpha_s m_{\gl}^2/\pi^3 m_h^2)\, \ln^2 ({\Lambda }/ m_{\gl})$ in leading-log approximation. For high-scale SUSY-breaking mediation $\ln ({\Lambda }/ m_{\tilde t, \gl})\approx 30$ is taken, while for low-scale mediation $\ln ({\Lambda }/ m_{\tilde t, \gl})\approx 1$ is used.}
    \centering
               {\setlength{\extrarowheight}{14pt}
    \begin{tabular}{c||c|c}
       $\epsilon$  & High-scale mediation & Low-scale mediation \\
         \hline
         \hline 
         stop & $5\times 10^{-5} \left( \frac{\rm 10~TeV}{m_{\tilde t}}\right)^2$
           & $2\times 10^{-3} \left( \frac{\rm 10~TeV}{m_{\tilde t}}\right)^2$ \\
        gluino & $7\times 10^{-6} \left( \frac{\rm 17~TeV}{m_{\gl}}\right)^2$
           & $6\times 10^{-3} \left( \frac{\rm 17~TeV}{m_{\gl}}\right)^2$
    \end{tabular}}
    \label{tab:susyft}
\end{table}

Future collider searches of gluinos and stops will be powerful probes on the role of naturalness in the Higgs sector, as shown in Table~\ref{tab:susyft}. For a SUSY-breaking mediation mechanism near the unification scale, gluino searches at \FCChh will probe naturalness at the level of $10^{-5}$ and, even in the case of low-scale mediation, naturalness can be tested at the level of $10^{-3}$ from the leading stop contribution. 
Independently of any naturalness consideration, the measured value of the Higgs mass can be used as an indicator of the scale of SUSY particle masses. Indeed, in the minimal SUSY model, the prediction of the Higgs mass agrees with the experimental value only for stops in the multi-TeV range or larger. The most relevant range of stop masses can therefore be probed only at future hadron colliders. 

\subsection{Charginos, neutralinos and sleptons}
\label{sec:BSM-SUSY-weak}

\begin{figure}[htb]
    \centering
\includegraphics[width=0.7\textwidth]{\main/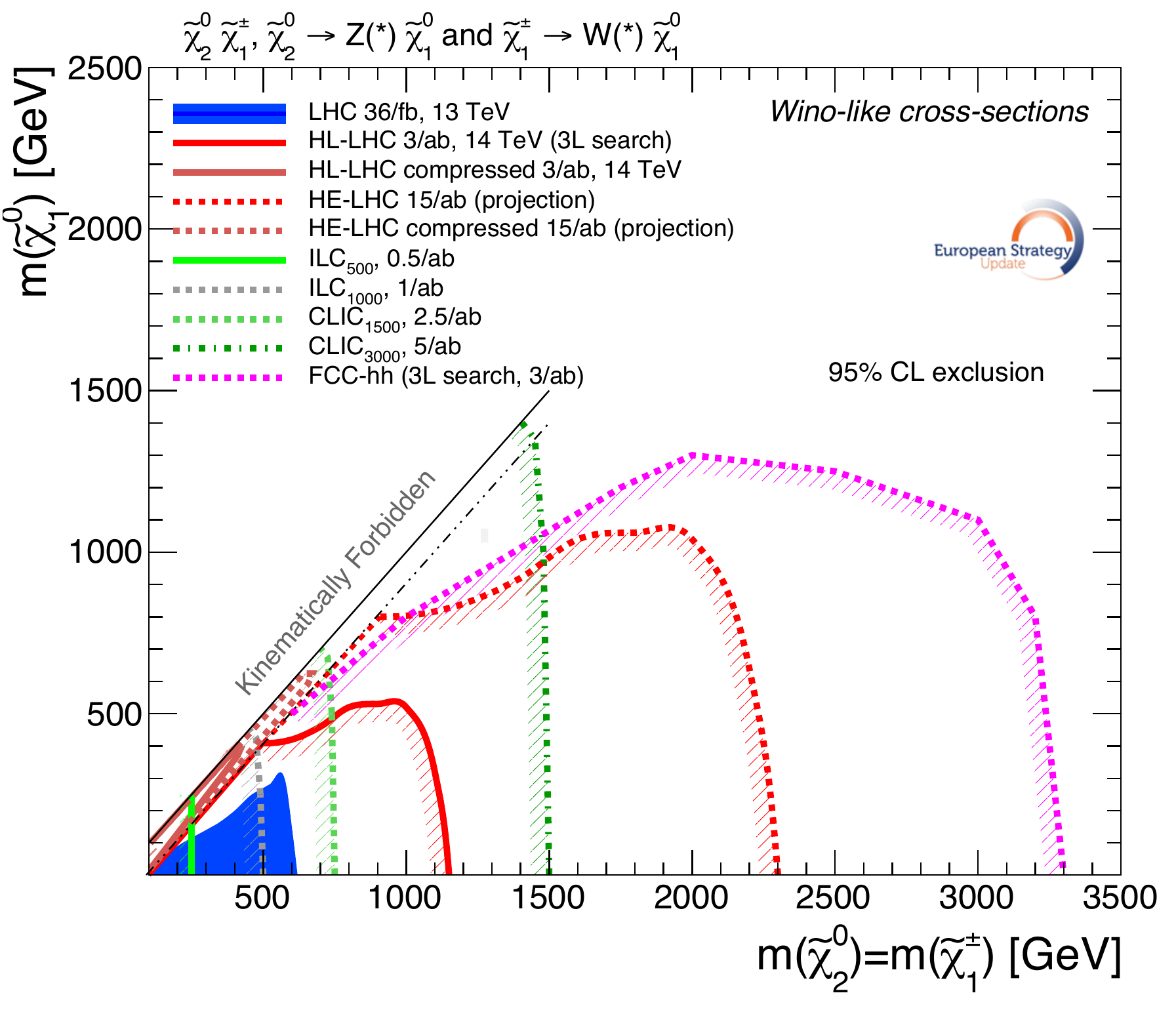}
    \caption{
    Exclusion reach for Wino-like lightest chargino (${\tilde \chi}_1^\pm$) and next-to-lightest neutralino (${\tilde \chi}_2^0$) from hadron and lepton colliders. 
    }
\label{fig:SUSY_Winos}
\end{figure}

In the context of $R$-parity conserving scenarios, the largest production rates for EWkinos in clean channels at hadron machines are obtained when the LSP is Bino-like and the lightest chargino (\chiOnePM) and next-to-lightest neutralino ($\tilde \chi_2^0$) are Wino-like, forming an approximately mass degenerate SU(2) triplet. 
Large $\tilde{\chi}_{2}^{0}\chiOnePM$ and $\chiOnePM\chiOneMP$ production rates and large mass splittings between the LSP and next-to-lightest SUSY states (NSLP) allow for the exploitation of multi-lepton final states at all facilities.
Figure~\ref{fig:SUSY_Winos} shows the exclusion reach for hadron and lepton colliders on the plane spanned by the relevant EWkino masses. For lepton colliders, the dominant processes are $e^+e^- \to$~\chiOnePM\chiOneMP, \chiTwoZero\chiOneZero; for hadron colliders only \chiTwoZero\chiOnePM ~production is considered here.
NLSP masses well above 1(2)~TeV could be reached by HL(HE)-LHC searches for a low-mass LSP~\cite{CidVidal:2018eel}; masses up to 3.3~TeV can be excluded at \FCChh~\cite{Golling:2016gvc}, with potential for improvements using optimised  selection criteria. Compressed SUSY spectra can be targeted at hadron colliders exploiting low-momentum leptons recoiling against an ISR jet. Lepton colliders analyses are competitive in this case: sensitivity up to EWkino masses equal to $\sqrt{s}/2$ are possible even for $\Delta m(\tilde{\chi}_{1}^{\pm}, \chi_1^0)$ as low as 1~\,GeV, with no loss in acceptance (ILC~\cite{Antusch:2017pkq}, CLIC~\cite{deBlas:2018mhx}).

\begin{figure}[htb]
    \centering
\includegraphics[width=0.7\textwidth]{\main/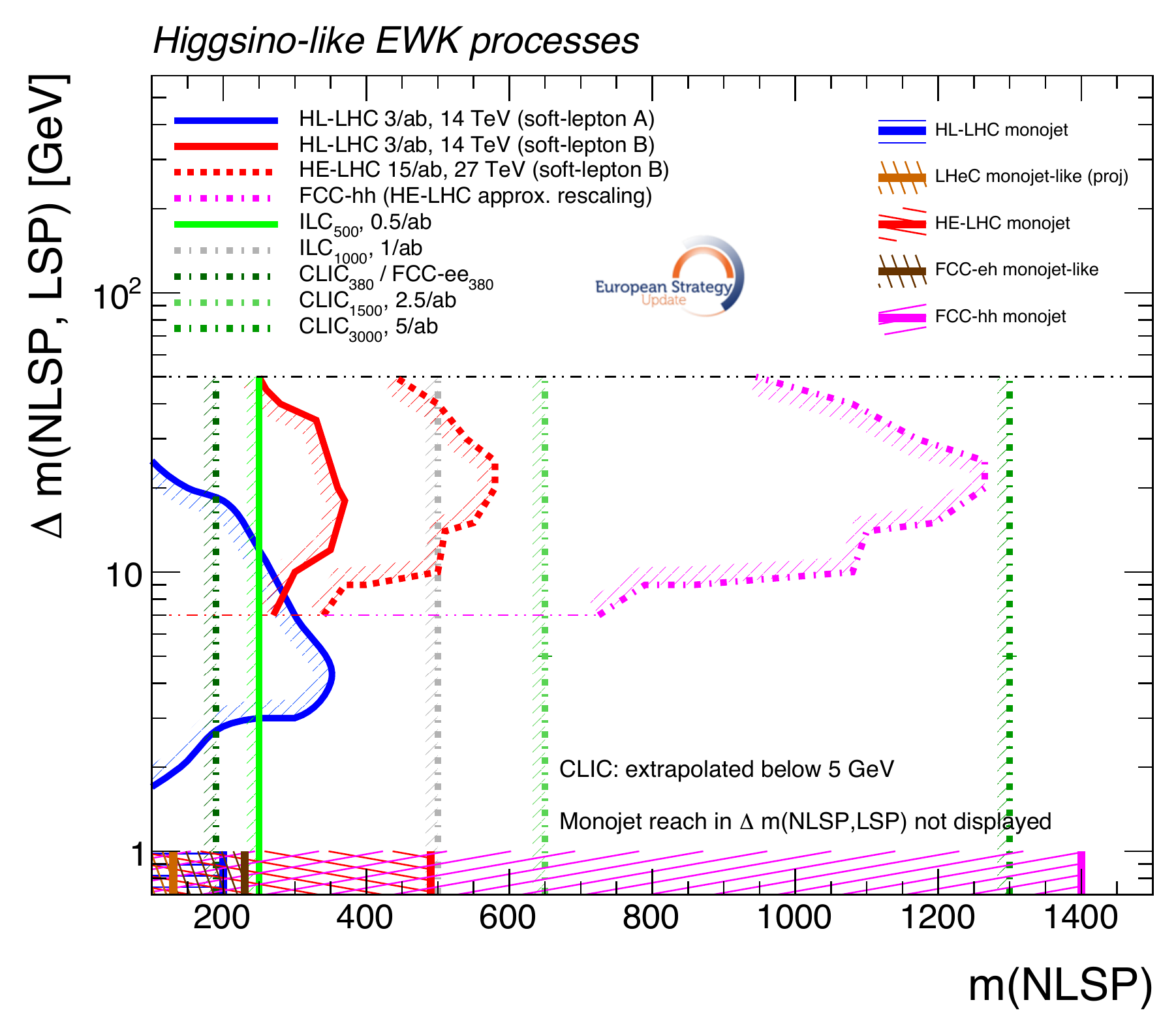}
    \caption{
    Exclusion reach for Higgsino-like charginos and next-to-lightest neutralinos with equal mass $m\,$(NLSP), as a function of the mass difference $\Delta m$ between NLSP and LSP. Exclusion reaches using monojet searches at $pp$ and $ep$ colliders are also superimposed (see text for details).}
\label{fig:SUSY_Higgsinos}
\end{figure}

If the Higgsino mass is much smaller than the gaugino masses, $\tilde \chi_{1,2}^0$ and \chiOnePM~form an approximately mass degenerate Dirac SU(2) doublet, and the EWkino spectrum is compressed. 
The EWkino production rates are smaller than in the previous case, making dedicated searches more challenging. Analyses exploiting ISR jets show good prospects at \HLLHC and hadron colliders in general (see Fig.~\ref{fig:SUSY_Higgsinos}):
\chiOnePM, \chiTwoZero~masses up to 350\,GeV can be probed at \HLLHC for mass splittings $\Delta m \equiv \Delta m (\tilde \chi_2^0,\tilde \chi_1^0 )\approx \Delta m (\tilde \chi_1^\pm,\tilde \chi_1^0 )$ between 1.6 and 50 GeV\footnote{The ``soft-lepton A'' analysis~\cite{ATL-PHYS-PUB-2018-031}, the expectation from which is also presented as a function of $m(\chiTwoZero)$, applies to the case $m(\chiOnePM)<m(\chiTwoZero)$ and $\chiOneZero$ is the NLSP. In this scenario, $m(\chiOnePM)$ down to 0.8 GeV can be probed.}, with a factor 1.5 increase expected at the \HELHC~\cite{CidVidal:2018eel}. 
\FCChh projections computed with simple extrapolations show that the 1 TeV boundary might be reached, with expected 95\% CL limits up to 1.3~TeV depending on $\Delta m$.
On the other hand, the sensitivity of lepton colliders depend only weakly on the nature of the LSP: \ILCFiveHundred and \ILCOneThousand~\cite{Antusch:2017pkq} can cover the full mass range up to $\sqrt{s}/2$ for $\Delta m$ as low as 0.5~GeV, while CLIC$^{}_{1500}$ and \CLICThreeThousand allow a reach up to 650\,GeV and 1.3\,TeV, respectively~\cite{CLIC:PrivateCommunication}.  
Monojet searches at hadron colliders can again complement the reach for scenarios with small $\Delta m$~\cite{CidVidal:2018eel}. The soft decay products of the NLSP are not reconstructed and the sensitivity solely depends on the production rate of EWkinos in association with an ISR jet. The reach of different colliders are illustrated by the hatched areas of Fig.~\ref{fig:SUSY_Higgsinos} for an indicative $\Delta m<1$~GeV. The sensitivity deteriorates at larger $\Delta m$, due to the requirements on additional leptons or jets. No attempt is made to evaluate this loss here, which is expected to become relevant for $\Delta m \approx 5$~GeV and above. Prospects for $ep$ colliders (LHeC and FCC-eh) performed using monojet-like signatures~\cite{Abada:2019lih} are also shown in Fig.~\ref{fig:SUSY_Higgsinos}.

A special case arises when the lightest neutralino is either pure Higgsino or Wino. The chargino-neutralino mass splitting is around 340 MeV and 160 MeV respectively, and the chargino has a correspondingly long lifetime, which can be as large as several picoseconds. The value of \Ptmiss~is small unless the pair-produced EWkinos recoil against an ISR jet. Taking advantage of the long lifetime of the charginos, which can result in decays in the active volume of the tracker detector, searches for disappearing charged tracks can be performed at hadron colliders~\cite{CidVidal:2018eel}. As an example, at the \HLLHC, studies using simplified models of \chiOnePM~production lead to exclusions of chargino masses up to $m_{{\tilde \chi}_1^\pm}$ = 750\,GeV (1100\,GeV) for lifetimes of 1\,ns for the Higgsino (Wino) hypothesis. When considering the lifetimes corresponding to the chargino-neutralino mass splittings given above (leading to thermal relic dark matter candidates and referred to as pure Higgsino and pure Wino, respectively), masses up to 300 (830)~GeV can be excluded. The reach for all facilities is illustrated in Sect.~\ref{sec:BSM-DM}. Analyses exploiting displaced decays of the charged SUSY state have been studied also for lepton colliders, e.g.\ \CLICThreeThousand (using charge stub tracks~\cite{deBlas:2018mhx}), and for $ep$ colliders (using disappearing tracks~\cite{Curtin2018}).  

 
Collider experiments have significant sensitivity also to sleptons. Searches for staus, superpartners of $\tau$ leptons, might be particularly challenging at $pp$ facilities due to the complexity of identifying hadronically-decaying taus and reject misidentified candidates. 
Analysis of events characterised by the presence of at least one hadronically-decaying $\tau$ and \Ptmiss~show that the \HLLHC will be sensitive to currently unconstrained pair-produced $\tilde{\tau}$ with discovery (exclusion) potential for $m_{\tilde{\tau}}$ up to around 550~(800)~GeV~\cite{CidVidal:2018eel}. 
The reach depends on whether one considers $\tilde{\tau}$ partners of the left-handed or the right-handed tau lepton ($\tilde{\tau}_R$ or $\tilde{\tau}_L$, respectively), with substantial reduction of the sensitivity in case of $\tilde{\tau}_R$.
The \HELHC would provide sensitivity up to 1.1~TeV~\cite{CidVidal:2018eel}, and an additional three-fold increase is expected for the \FCChh (extrapolation). 
Lepton colliders could again provide complementary sensitivity especially in compressed scenarios: \ILCFiveHundred~\cite{Antusch:2017pkq} would allow discovery of $\tilde{\tau}$ up to 230~GeV even with small datasets, whilst  \CLICThreeThousand would allow reach up to $m_{\tilde{\tau}}=1.25$~TeV and $\Delta m(\tilde{\tau},\chi_1^0)=50$~GeV~\cite{CLIC:PrivateCommunication}. 

\subsection{Non-prompt SUSY particles decays}
There are numerous examples of SUSY models where new particles can be long-lived and may travel macroscopic distances before decaying. Long lifetimes may be due to small mass splittings, as in the case of pure Higgsino/Wino scenarios, or due to small couplings, as in $R$-parity violating SUSY models, or due to heavy mediators, as in Split SUSY.
For \HLLHC~\cite{CidVidal:2018eel}, studies are available on long-lived gluinos and sleptons. Exclusion limits on gluinos with lifetimes $\tau>0.1$\,ns can reach about 3.5 TeV, using reconstructed massive displaced vertices. Muons displaced from the interaction point, such as found in SUSY models with $\tilde{\mu}$ lifetimes of $c\tau > 25$ cm, can be excluded at 95\% CL at the \HLLHC. New fast timing detectors will also be sensitive to displaced photon signatures arising from long-lived particles in the $0.1 < c\tau < 300$ cm range.

Complementarities in long-lived particle searches and enhancements in sensitivity might be achieved if new proposals for detectors and experiments such as MATHUSLA200, FASER, CODEX-b, MilliQan and LHeC are realised in parallel to the HL-LHC. As an example, with a zero-background hypothesis, MATHUSLA200~\cite{Curtin:2018mvb} would offer a coverage complementary to \HLLHC in terms of new particles with $c\tau$ in the range 100~m--20~km, targeting $R$-parity violating decays of gluinos, top squarks as well as sleptons and Higgsinos.

\section{Extended Higgs sectors and high-energy flavour dynamics}
\label{sec:BSM-ExtendedScalars}

This section presents the physics reach of future collider facilities in the areas of extended Higgs sectors, neutral naturalness, and high-energy flavour dynamics. Substantial improvements with respect to \HLLHC are possible for all included physics topics, and both direct and indirect searches typically provide complementary information. In several cases the combined interpretation of measurements in hadron and lepton collisions, at different stages of energy, enhance the physics potential even further.

\subsection{Extended Higgs sectors}
The scalar sector of the SM consists of one isospin doublet of scalar fields $H$ which contains the minimal set of required degrees of freedom: longitudinal polarisations of the $W$ and $Z$ bosons and the Higgs particle. However, many new-physics scenarios predict extended Higgs sectors, with one particle closely resembling the SM Higgs boson, and additional scalars.

A simple possibility is the extension of the SM scalar potential by a singlet massive scalar field $S$ with interactions $V= \lambda_{S}S^{4}/4+\lambda_{HS} |H|^{2}S^{2}$. 
This scenario is of particular interest, not just for its minimality, but also because  it can change the nature of the EW phase transition (see Chapter~\ref{chap:ew}). In the SM, the phase transition is a smooth crossover, whereas a strong first-order phase transition could open up new exciting possibilities, such as observable gravitational waves in future facilities or a framework for a mechanism generating the cosmic baryon asymmetry.

\begin{figure}[ht]
\centering
\includegraphics[width=0.49\textwidth]{\main/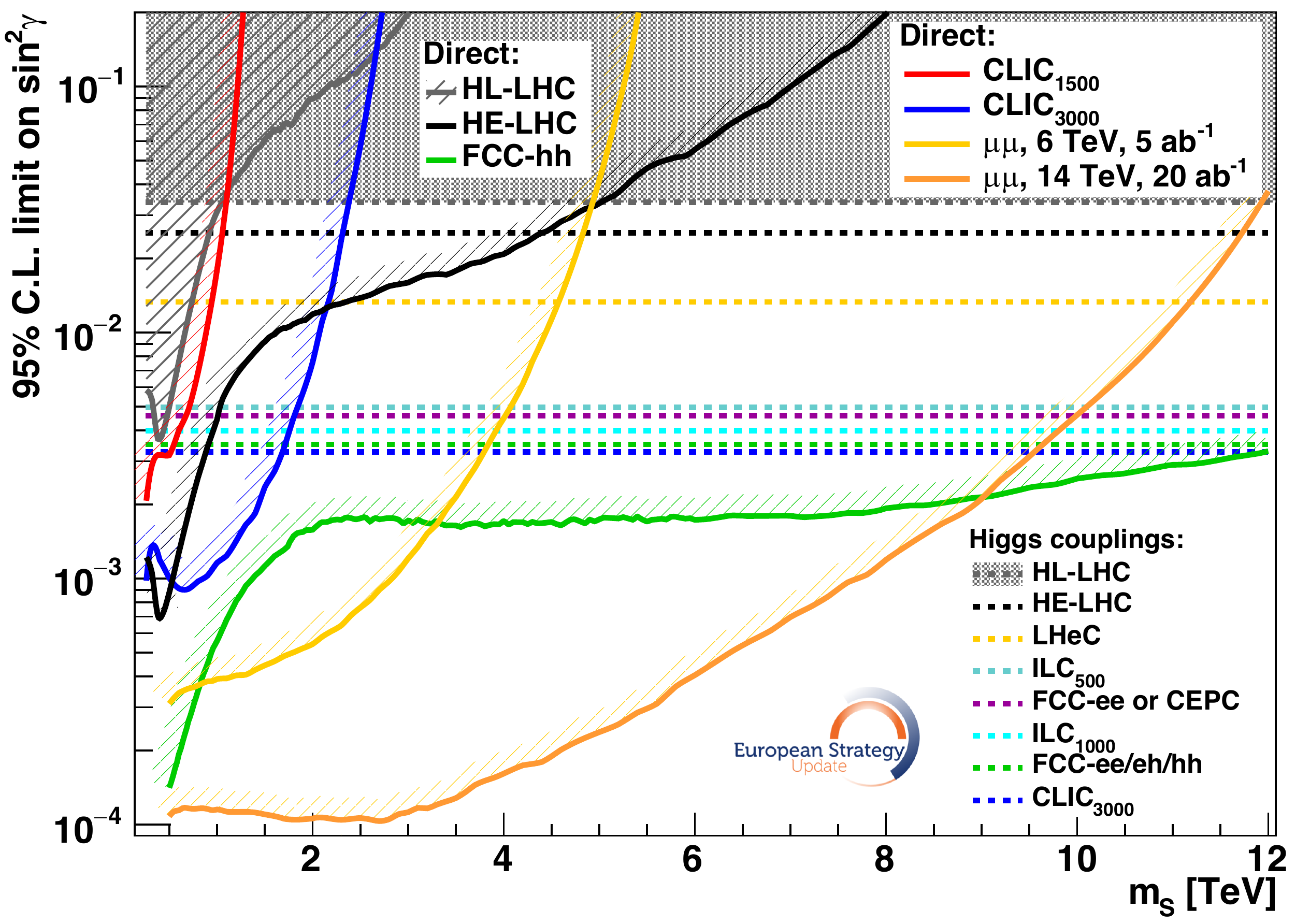}
\includegraphics[width=0.49\textwidth]{\main/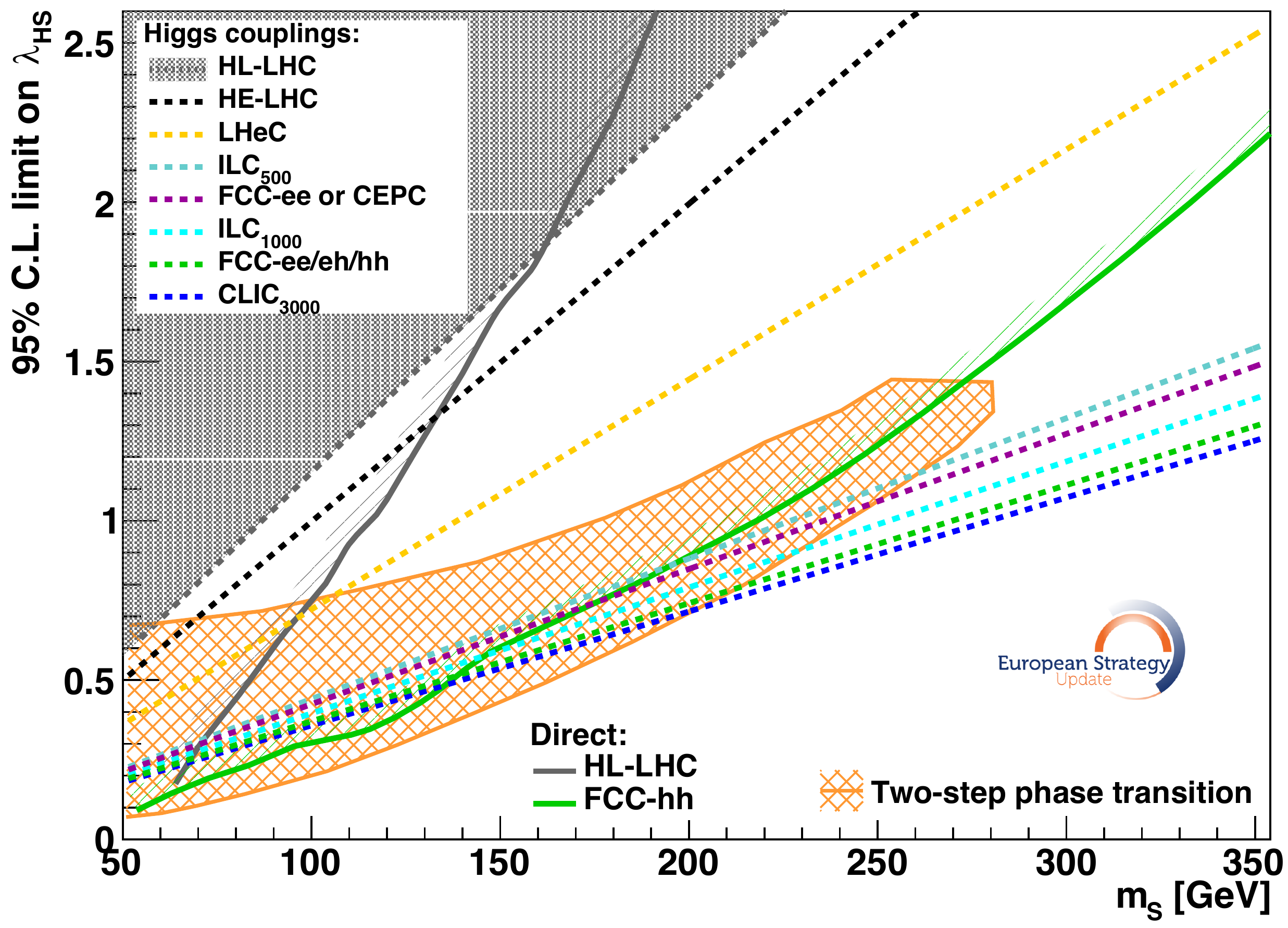}
\caption{Direct and indirect sensitivity at 95\% CL to a heavy scalar singlet mixing with the SM Higgs boson (left) and in the no-mixing limit (right). The hatched region shows the parameters compatible with a strong first-order EW phase transition.}
\label{fig:singlets}
\end{figure}

From the point of view of collider searches, there are two distinct singlet scenarios, depending on whether or not the singlet mixes with the Higgs. In the case where a heavy singlet mixes with the SM Higgs boson, direct limits on the mixing parameter $\sin \gamma$ for high-energy lepton~\cite{Buttazzo:2018qqp} and hadron colliders~\cite{Sirunyan:2018qlb, CidVidal:2018eel} are compared in Fig.~\ref{fig:singlets} (left). The overall scaling of the Higgs boson couplings~\cite{deBlas:2019rxi} provides an indirect probe even at colliders with $\sqrt{s}$ below the mass of the heavy singlet. At \HLLHC, \HELHC and CLIC direct and indirect searches provide complementary information, whereas the direct reach at \FCChh exceeds the sensitivity from Higgs-coupling measurements for masses up to 12\,TeV. Precision EW observables are not competitive with Higgs measurements and the corresponding reach is not shown in the figure.

In the no-mixing case, in which $S$ does not acquire a vacuum expectation value and the $Z_{2}$ symmetry remains unbroken, the new scalar is stable and escapes undetected. Figure~\ref{fig:singlets} (right) shows the indirect sensitivity on the coupling $\lambda_{HS}$ from the overall scaling of the Higgs couplings. The figure also shows the reach using the VBF process $pp \rightarrow SSjj$ at ~\cite{Craig:2014} and \FCChh~\cite{Curtin:2014jma}, which provides the best direct sensitivity in hadron collisions. 

It is interesting to note that a large fraction of the region compatible with a first-order phase transition could be probed by the full CLIC or FCC programmes. For illustration purposes, Fig.~\ref{fig:singlets} shows an example of the region compatible with a two-step phase transition, where the singlet supports the Higgs in delivering a strong first-order phase transition~\cite{Chala:2018opy}. 
Strongly first-order phase transitions are particularly interesting as they could also lead to sizeable gravitational wave signals at future experiments like LISA, linking discoveries at Earth-based colliders with space interferometry (see Chapter~\ref{chap:cosm}). The case of a light singlet scalar, with mass lower than 125 GeV, is discussed extensively in the section on feebly interacting particles~\ref{sec:BSM-FIPs}.

\begin{figure}[ht]
\centering
\includegraphics[width=0.7\textwidth]{\main/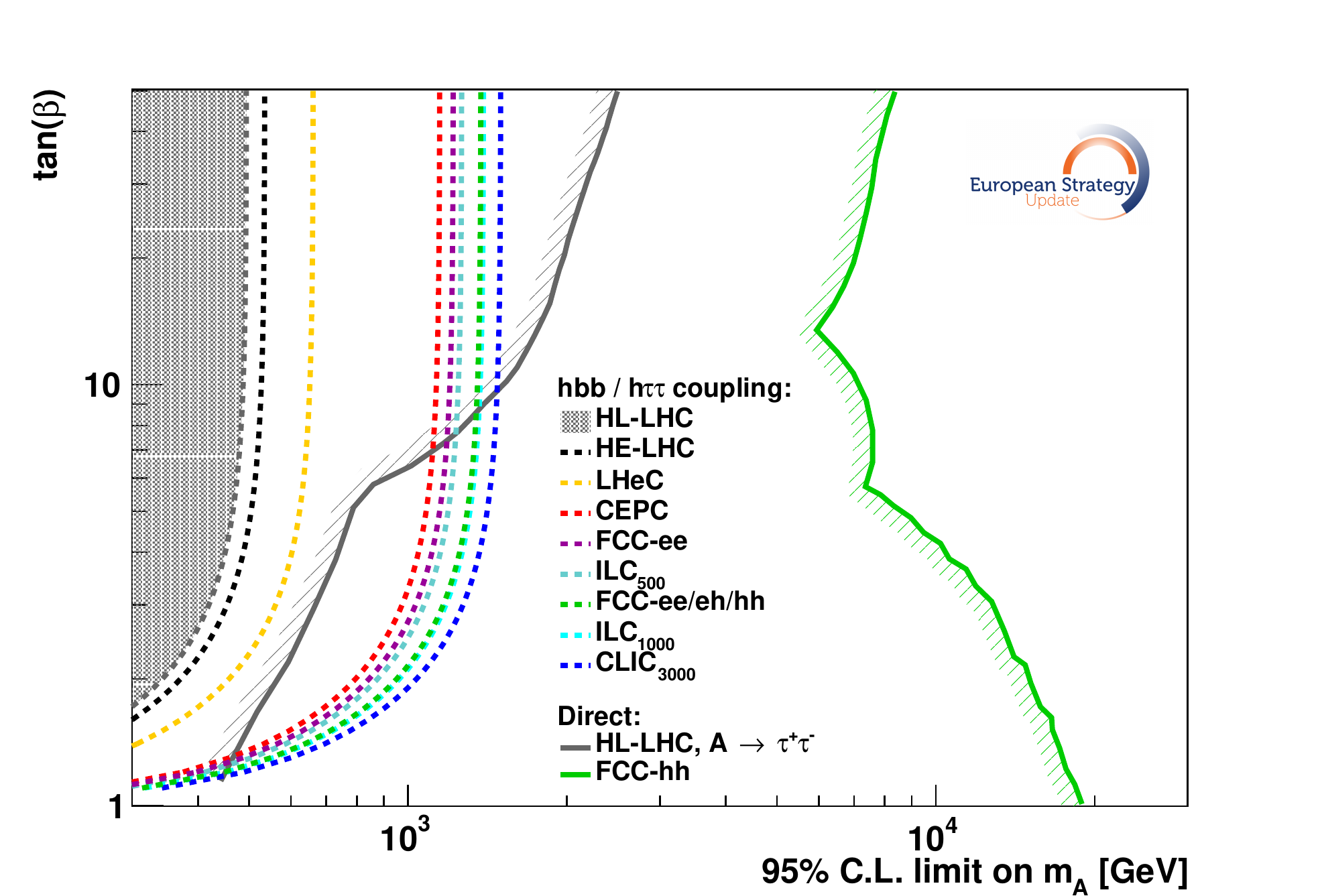}
\caption{Direct and indirect sensitivity at 95\% CL to heavy neutral scalars in minimal SUSY. 
}
\label{fig:2hdm}
\end{figure}

Another common extension of the SM Higgs sector is the addition of a second SU(2) doublet, which naturally appears in supersymmetric extensions of the Higgs sector or in  models with a non-minimal pattern of symmetry breaking. In this case, the scalar sector contains two CP-even scalars $h$ and $H$, one CP-odd scalar $A$ and a charged scalar $H^{\pm}$. The direct mass reach of lepton colliders for these scalars is generally close to $\sqrt{s}/2$ independent of $\tan{\beta}$, mainly using $e^{+}e^{-} \rightarrow H^{+}H^{-}$ and $e^{+}e^{-} \rightarrow AH$. The cross sections for these processes are almost independent of the heavy Higgs boson masses for a given $\sqrt{s}$. As expected, the highest mass reach is obtained at hadron colliders. 

A well-studied example of a Type-II Two-Higgs Doublet Model is the minimal SUSY extension of the SM. The \HLLHC is sensitive to heavy neutral scalars up to 2.5~TeV for $\tan{\beta} > 50$ using $A/H \rightarrow \tau^{+}\tau^{-}$ decays~\cite{Curtin:2014jma} while the region of low $\tan{\beta}$ can be probed with decays to top-quark pairs~\cite{Carena:2016npr}. At \FCChh, the expected 95\% CL exclusion limits for the $H/A$ states are generally better than 5\,TeV and extend up to 20\,TeV for low $\tan{\beta}$~\cite{Craig:2016ygr}. These projections are compared in Fig.~\ref{fig:2hdm} to the indirect sensitivity on $m_{A}$ from Higgs couplings to third-generation fermions~\cite{Gunion:2002zf, Gorbahn:2015gxa}. While for a given lepton collider $g_{hbb}$ is the most precise option, hadron colliders provide better accuracy on $g_{h\tau\tau}$~\cite{deBlas:2019rxi}. Hence the more precise of the two couplings was chosen to calculate the limit for each collider scenario. A global analysis of all Higgs observables would improve the indirect sensitivity further. The corresponding \FCChh sensitivity for the charged $H^{\pm}$ particles are in the range from 10 to 15\,TeV~\cite{Hajer:2015gka}.

Doubly-charged Higgs bosons exist in Type-II see-saw models, where a new scalar triplet could couple to the SM leptons to produce the light neutrino masses. For low values of the vacuum expectation value of the triplet ($v_{\Delta}$) the doubly-charged Higgs bosons would decay leptonically. Polarisation measurements in decays to $\tau$ leptons can help discriminate between different heavy scalar mediated neutrino mass models. An interesting case is the region where $v_{\Delta}$ is relatively large and the doubly-charged Higgs would decay into two same-sign $W$ bosons. Lepton colliders could reach masses almost up to the kinematic limit of $\sqrt{s}/2$ using $e^{+}e^{-} \rightarrow H^{++}H^{--}$ events~\cite{Agrawal:2018pci}. The \FCChh would be sensitive to $H^{++}H^{--} \rightarrow W^{+}W^{+}W^{-}W^{-}$ for doubly-charged Higgs masses up to 1.7\,TeV and $v_{\Delta} > 10^{-4}$\,GeV~\cite{Du:2018eaw}.

\subsection{Neutral naturalness}
Neutral naturalness (NN) describes the class of theories in which the top-quark partner, needed to regulate the leading SM quantum corrections to the Higgs mass, is colour neutral. This makes NN theories particularly elusive to LHC direct searches. The prototype model of NN is the Twin Higgs~\cite{Chacko:2005pe}, which introduces a mirror copy of the SM with twin particles and a twin gauge group. The SM and its mirror copy are related by a discrete symmetry that ensures equal coupling constants between the two sectors. As a result of the doubling of states, the scalar potential has an enlarged accidental global symmetry which, after spontaneous symmetry breaking, leads to a Goldstone boson with the right quantum numbers to be identified with the SM Higgs. Since the accidental symmetry is not exact in the full theory, the Higgs acquires a mass at the loop level, but its value is smaller than the cutoff thanks to an approximate cancellation between the contributions from SM and twin-sector particles. In the Twin Higgs, the top-quark partner is a new EW-neutral fermion, but NN variations can turn the top-partner into an EW-charged scalar (Folded SUSY~\cite{Burdman:2006tz}), an EW-charged fermion (Quirky Little Higgs~\cite{Cai:2008au}), or an EW-neutral scalar (Hyperbolic Higgs~\cite{Cohen:2018mgv,Cheng:2018gvu}). 

The NN top-partners are expected to have masses around $m_T \approx \sqrt{1/\epsilon}$~500~GeV, where $\epsilon$ is the degree of fine tuning. If they carry EW charge, the NN top-partners can be produced at colliders via DY-like processes. However, because of their unavoidable couplings with the Higgs, a more robust way of detecting their presence is through Higgs precision measurements. Since the parametric scaling of Higgs coupling modifications in NN is identical to the case of Composite Higgs, the reach of future colliders on the degree of tuning $\epsilon$ of NN theories can be read from Fig.~\ref{fig:Composite_Higgs_Summary} (blue bars, right axis).

NN theories can address the naturalness problem without coloured particles at the weak scale, but require a completion at a nearby mass scale $m_{\rm NN}$, where new coloured states are expected. This mass scale is model dependent, but it is generally in the range $m_{\rm NN}\approx \sqrt{0.1/\epsilon}$~3--5~TeV. This shows that, for moderate tuning, the NN coloured particles are beyond the reach of the LHC but are perfect targets for high-energy future colliders. 
The states in the twin sector that communicate with the SM only through the Higgs can lead to unusual signatures. An interesting case are the twin glueballs, which are expected to be light and typically long-lived. They can be produced in Higgs decays and then decay back into SM states, leading to distinguishing vertex displacements that can be hunted for in main detectors or, depending on the lifetime, with dedicated detectors far from the interaction point such as MATHUSLA200. Searches for invisible Higgs decays also contribute to the exploration of the parameter space.

\subsection{High-energy flavour dynamics}
Heavy new physics can induce, through the exchange of virtual particles, processes that are extremely rare in the SM, such as FCNC effects in the top-quark sector (see also Chapter~\ref{chap:flav}). Experimental projections on searches for rare FCNC top-quark decays are available for many of the proposed projects. Complementary sets of measurements using various top-quark production processes and decays are accessible in $e^+e^-$, $ep$ and $pp$ collisions. The projections are summarised in Table~\ref{tab:top_fcnc_decays}. While not all possibilities have been explored yet, generally improvements of 1--2 orders of magnitude are possible compared to \HLLHC.

\begin{table}
\caption{Limits on FCNC top-quark decays at 95\% CL for various future colliders~\cite{Cerri:2018ypt, Bambade:2019fyw, Abramowicz:2018rjq, Khanpour:2014xla, Abada:2019lih}. Results are also given for flavour-inclusive final states with $q = u, c$. Empty entries correspond to cases in which studies are not available at the time of writing.}
\centering
\begin{tabular}{ c | c | c | c | c | c | c | c | c }
$BR \times 10^{5}$ & \HLLHC & \HELHC & \ILCFiveHundred & \CLICThreeHundredEighty & LHeC & \FCCee & \FCChh & \FCCeh \\
\hline
$t \rightarrow Hc$ & & & $\approx 3$ & 15 & & & 1.6 & \\
$t \rightarrow Hu$ & & & & & 150 & & & 22 \\
$t \rightarrow Hq$ & 10 & & & & & & 2.8 & \\
\hline
$t \rightarrow Zq$ & 2.4 - 5.8 & & & & 4 & 2.4 & $\approx 0.1$ & 0.6 \\
\hline
$t \rightarrow \gamma c$ & 7.4 & & $\approx 1$ & 2.6 & & & 0.024 & \\
$t \rightarrow \gamma u$ & 0.86 & & & & & & 0.018 & \\
$t \rightarrow \gamma q$ & & & & & 1 & 1.7 & & 0.085 \\
\hline
$t \rightarrow gc$ & 3.2 & 0.19 & & & & & & \\
$t \rightarrow gu$ & 0.38 & 0.056 & & & & & & \\
\end{tabular}
\label{tab:top_fcnc_decays}
\end{table}

At lepton colliders, the process $e^{+}e^{-} \rightarrow tj$ is much more powerful compared to tests on top-quark decays, which are limited by statistics. In particular, operation at the highest possible energies improves the sensitivity to four-fermion operators. The full CLIC programme would be sensitive to operator scales in the region of 50--100~TeV~\cite{deBlas:2018mhx}.

\begin{figure}[ht]
\centering
\includegraphics[width=0.7\textwidth]{\main/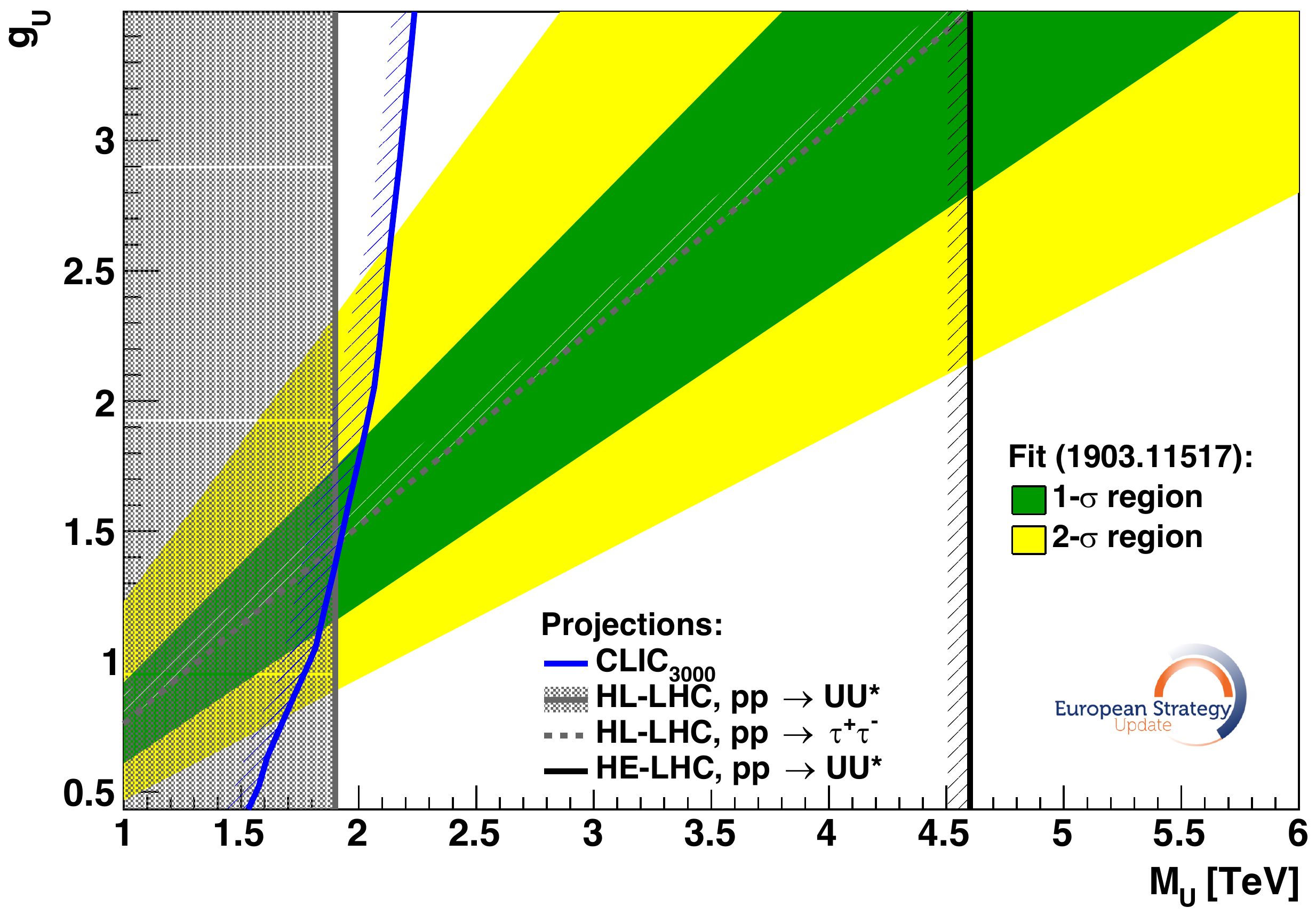}
\caption{
Direct and indirect sensitivity at 95\% CL for the vector leptoquark $U_{1}$ in the mass versus coupling plane.}
\label{fig:leptoquarks}
\end{figure}

Renewed interest in leptoquarks was triggered by recent results in rare $B$ decays, which show discrepancies with respect to the SM predictions. For example, the $b \rightarrow c\tau\nu$ results are presently compatible with a rather light leptoquark coupled predominantly to the third generation. The mass reach from QCD pair-production of the scalar leptoquark $S_{3}$ or vector leptoquark $U_{1}$ at \HLLHC is 1.5--2.0~TeV~\cite{CidVidal:2018eel}, independently of the coupling to the lepton-quark current. Complementary information would be provided in the large-coupling region by the process $pp \rightarrow \tau^{+}\tau^{-}$. A modest improvement in sensitivity for the $S_{3}$ would be provided by CLIC~\cite{deBlas:2018mhx}. The highest mass reach is available at hadron colliders. For example, \HELHC would improve the direct mass reach by more than a factor two compared to \HLLHC. Projections for the different colliders in the $U_{1}$ case are shown and compared to a recent global flavour fit~\cite{Cornella:2019hct} in Fig.~\ref{fig:leptoquarks}.

\documentclass[../report.tex]{subfiles}
\providecommand{\main}{..}
\IfEq{\jobname}{\currfilebase}{\AtEndDocument{\biblio}}{}

\newcommand{\sigv}{\langle \sigma v_{\rm{rel}} \rangle}
\newcommand{\MET}{\slashed{E}_T}
\newcommand{\mDM}{m_{\rm{DM}}}
\newcommand{\mmed}{M_{\rm{med}}}
\newcommand{\mMed}{M_{\rm{med}}}
\newcommand{\gDM}{\ensuremath{g_{\mathrm{DM}}}}
\newcommand{\gq}{\ensuremath{g_q}}
\newcommand{\gSM}{\ensuremath{g_q}}
\newcommand{\gf}{\ensuremath{g_f}}

\newcommand{\ifb}{\rm{fb}^{-1}}

\newcommand{\mA}{M_{A}}
\newcommand{\GamA}{\Gamma_{A}}

\begin{document}

\section{Dark Matter}
\label{sec:BSM-DM}



For an introduction to dark matter (DM) and a general study of its experimental consequences, see Chapter~\ref{chap:dm}. This section presents the ways in which the nature of DM and its interactions can be probed at future colliders, complementing experiments and observations from astroparticle physics. 

The thermal freeze-out mechanism provides a cosmological clue for the generation of the observed DM density, suggesting that DM particles have masses in the range from multi-keV to about 100~TeV, and couplings to SM particles of comparable or weaker strength than EW interactions. High-energy colliders could produce DM particles within this mass range in controlled conditions. The main signature at colliders is the missing transverse momentum carried by the DM particle, which remains invisible to detectors due to the presumed weak strength of its interaction with SM particles. If the DM particle is lighter than $m_h/2$ and it is coupled to the Higgs, an interesting exploration channel is the invisible Higgs decay width (see Chapters~\ref{chap:ew} and \ref{chap:dm}). An alternative signature is the detection of the mediator particles whose exchange may be responsible for the annihilation processes that determine the DM particle abundance. Mediators can lead to a variety of collider signatures in visible channels, although their discovery would not provide evidence for DM until invisible channels are identified as well.
 
There are many possible thermal freeze-out scenarios, each with their own unique experimental signals. Here we present only some interesting examples of heavy DM particles and discuss the prospects for their searches at future colliders. The case of light DM particles is addressed in Sect.~\ref{sec:BSM-FIPs} and non-collider searches for DM are discussed in Chapter~\ref{chap:dm}.

\subsection{Higgsinos and Winos}

The most straightforward example of a DM thermal relic is a massive particle with EW gauge interactions only. Of special interest are the cases of spin-1/2 particles transforming as doublets or triplets under SU(2) symmetry, which are usually referred to as Higgsino and Wino respectively. Although this terminology is borrowed from SUSY, these examples should be regarded as standalone DM models.

\begin{figure}[htb]
    \centering
    \includegraphics[width=0.48\textwidth]{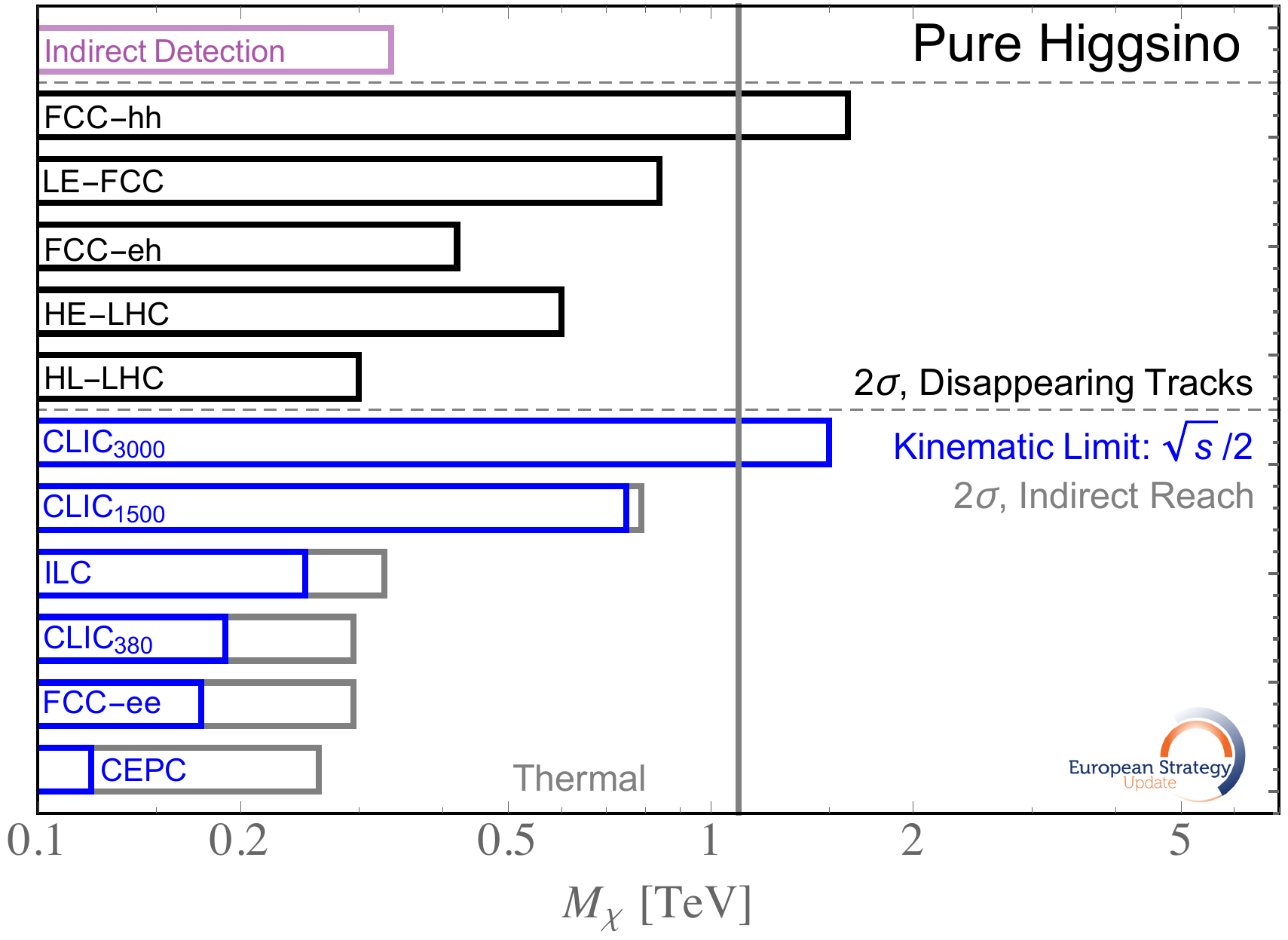}
   ~ \includegraphics[width=0.48\textwidth]{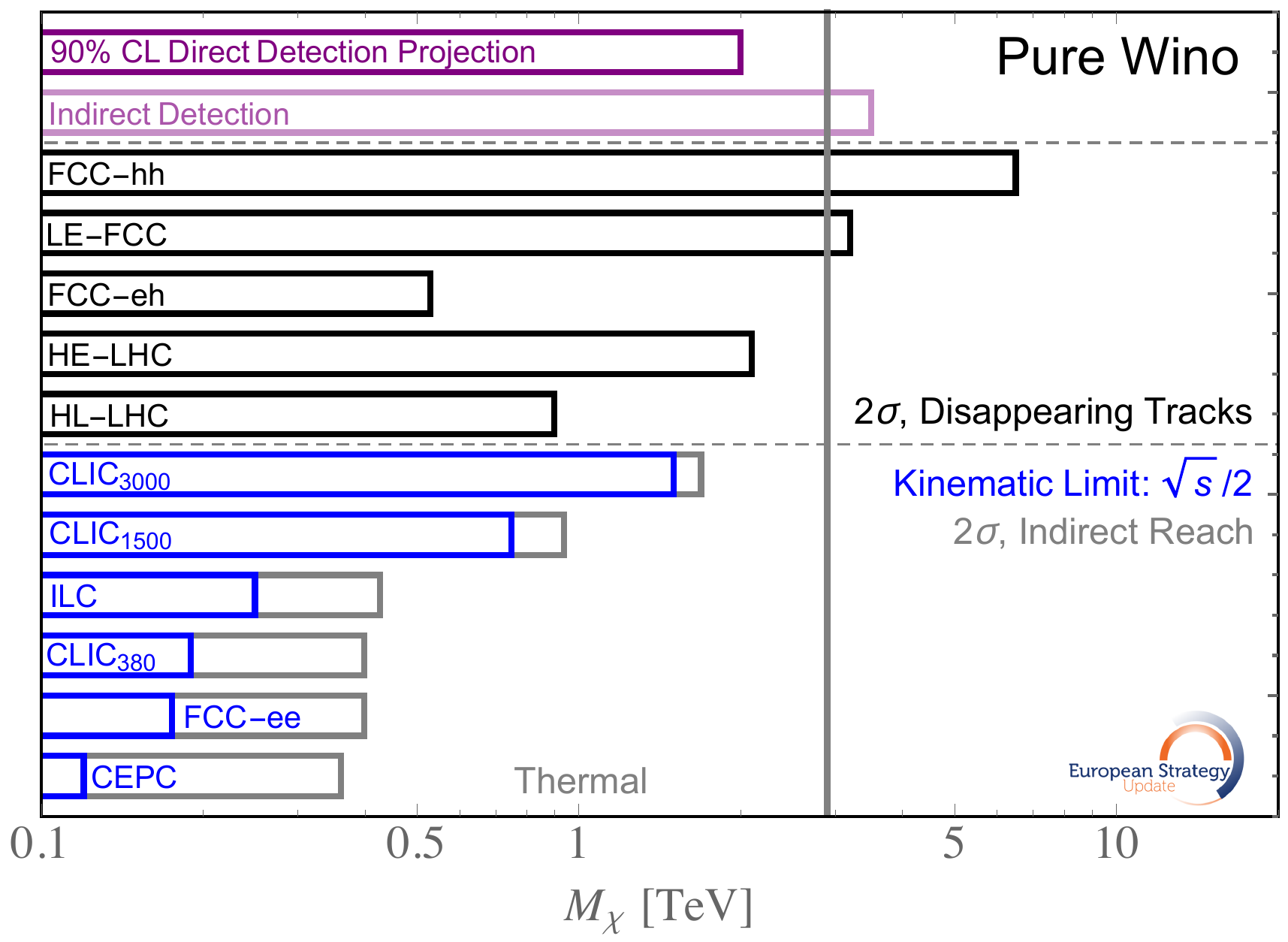}
    \caption{Summary of $2\sigma$ sensitivity reach to pure Higgsinos and Winos at future colliders. Current indirect DM detection constraints (which suffer from unknown halo-modelling uncertainties) and projections for future direct DM detection (which suffer from uncertainties on the Wino-nucleon cross section) are also indicated. The vertical line shows the mass corresponding to DM thermal relic.}
    \label{fig:HiggsinoWino}
\end{figure}

A summary of the reach of future colliders for Higgsinos and Winos is presented in Fig.~\ref{fig:HiggsinoWino} (for a more detailed discussion of the relevant collider signatures, see Sect.~\ref{sec:BSM-SUSY-weak}).  
A study of the preferred mass range from thermal freeze-out can be found in \cite{Krall:2017xij}.  Current direct detection limits are not shown as, for a pure Higgsino with a small inelastic splitting, the spin-independent cross section is well below the irreducible neutrino background.  For the Wino the cross section is above the neutrino background, but only by less than an order of magnitude \cite{Chen:2018uqz}.  DARWIN, a future direct detection experiment, is expected to probe the Wino scenario up to masses of $2\substack{+3 \\ -1}$ TeV, where the large error reflects the theoretical uncertainties on the Wino-proton cross section~\cite{Chen:2018uqz}.  This projection is shown in Fig.~\ref{fig:HiggsinoWino}. Current indirect detection limits were taken from \cite{Krall:2017xij}, which compiled both FERMI and H.E.S.S.\ telescope constraints.  
In summary, Higgsino DM with a thermal mass is likely within the coverage of CTA, while it is not constrained by H.E.S.S., and thermal Wino DM is strongly constrained by H.E.S.S.\ ~\cite{Hryczuk2019-mv}.
We note that the indirect detection constraints are subject to large systematic uncertainties from halo-shape modelling.  These uncertainties are not reflected in the bars shown in Fig.~\ref{fig:HiggsinoWino} and should be kept in mind when comparing with collider searches. 

The reach for disappearing track searches at \HLLHC, \HELHC and \FCChh is taken from \cite{CidVidal:2018eel}, while the result for low-energy \FCChh is obtained with a simple extrapolation. The reach for Higgsinos at \FCCeh is taken from Vol.~1 of the FCC CDR \cite{Abada:2019lih}, while the \FCCeh reach for Winos was provided in a private communication by the collaboration. Due to the cleanliness of the interactions, direct searches at lepton colliders typically come close to the kinematical limit (shown in blue in Fig.~\ref{fig:HiggsinoWino}), but do not surpass it.  On the other hand, the sensitivity to the EW parameters $W$ and $Y$ allows for a reach that can extend beyond the kinematic limit, albeit in an indirect manner.  These constraints, shown in grey, are calculated using the formulae of~\cite{DiLuzio:2018jwd} updated with the sensitivity reported in the ECFA Higgs report \cite{deBlas:2019rxi}. It is important to remark that \FCChh can conclusively test the hypothesis of thermal DM for both the Higgsino and Wino scenarios, while \CLIC can cover the Higgsino case. These results were obtained assuming the EW one-loop neutral-charged mass splitting (see Sect.~\ref{sec:BSM-SUSY-weak}). \FCChh searches could become ineffective under the special circumstances of heavy-particle contributions that accidentally cancel the loop-induced mass splitting.

\subsection{Simplified Models: axial vector and scalar mediators}

Simplified DM Models are a schematic way to parametrise classes of theories without committing to specific dynamical constructions or introducing too many unknown parameters. The Simplified Models considered here introduce one DM particle and one mediator, and their free parameters are the masses of these two particles and the coupling constants of the interactions mediator/DM and mediator/SM particles. The mediator can be either a SM particle itself (e.g. the Higgs or the $Z$ boson) or a new BSM particle. Depending on the nature of the DM particle and the mediator, one can construct a large variety of Simplified Models and here two representative examples~\cite{Abercrombie:2015wmb} are chosen.

In both cases, the DM particle is a massive Dirac fermion ($\chi$). In the first example, the mediator is a spin-1 particle ($Z^\prime$) coupled to an axial-vector current in the Lagrangian as $-Z^\prime_\mu (g_{\rm DM} \, \bar{\chi}\gamma^{\mu}\gamma_5\chi + g_f \sum_{f}  \bar{f}\gamma^{\mu}\gamma_5f )$, where $f$ are SM fermions. This model is particularly interesting for collider searches because the reach of direct DM searches is limited, as the  interaction in the non-relativistic limit is purely spin-dependent.  In the second example, the mediator is a spin-0 particle ($\phi$) with interactions $\phi (g_{\rm DM}\, \bar{\chi}\chi-  g_f   \sum_{f} y_f \bar f f /\sqrt{2})$. This model can serve as a prototype for various extensions of the SM involving enlarged Higgs sectors.

\begin{figure}[t]
    \centering
    \includegraphics[width=0.48\textwidth]{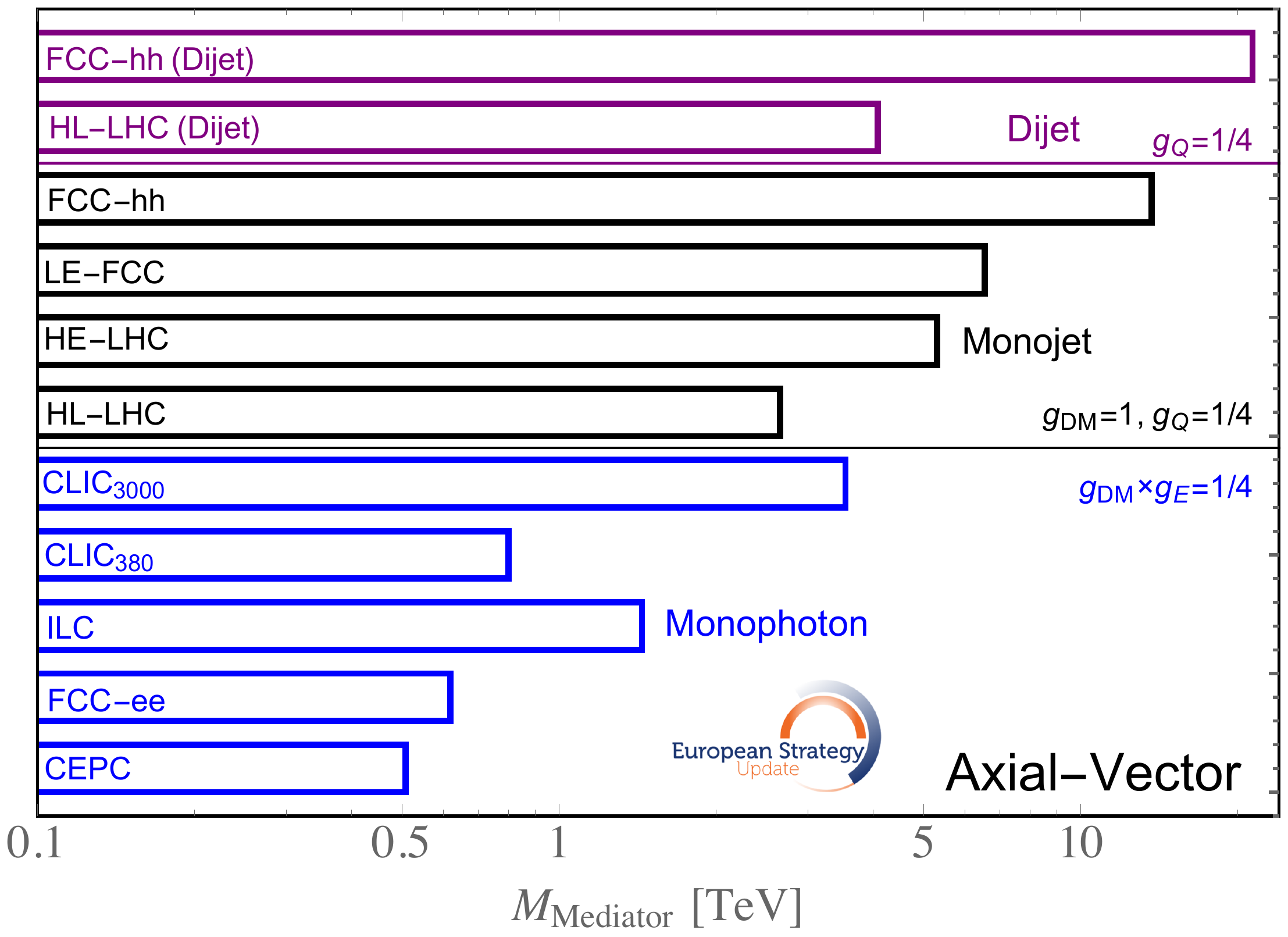}
    \includegraphics[width=0.48\textwidth]{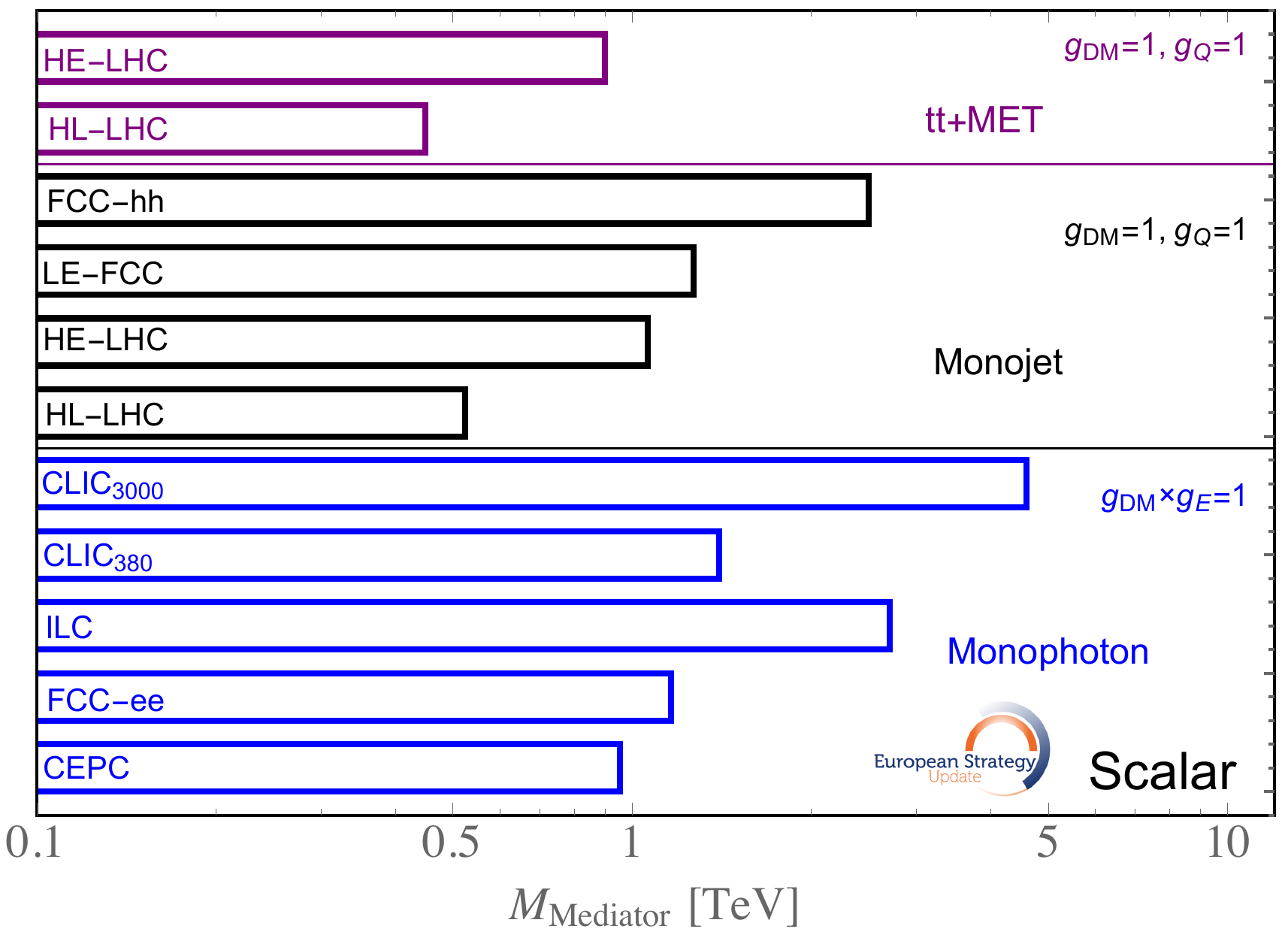}
    
\caption{Summary of $2\,\sigma$ sensitivity to axial-vector and scalar simplified models at future colliders for a DM mass of $M_{\rm{DM}}=1$ GeV and for the couplings shown in the figure. References and details on the estimates included in these plots can be found in the text. }
    \label{fig:AxialVectorScalar}
\end{figure}

In Fig.~\ref{fig:AxialVectorScalar} a compilation of future collider sensitivities to the two Simplified Models under consideration, with a choice of couplings of ($g_{\rm f}=0.25$, $g_{\rm DM}=1.0$) for the axial-vector model and ($g_{\rm f}=1.0$, $g_{\rm DM}=1.0$) for the scalar model, are shown. The reach of collider experiments to this kind of models is strongly dependent on the choice of couplings. As an example, the sensitivity of dijet and monojet searches decreases significantly with decreased quark couplings: with 36 fb$^{-1}$ of LHC data~\cite{Aaboud:2019yqu} and assuming a DM mass of 300 GeV and $g_{\rm DM}=1.0$, the limits from dijet searches on the axial-vector mediator mass decrease from 2.6 TeV for a quark coupling of $g_{\rm q}=0.25$ to 900 GeV for $g_{\rm q}=0.1$, while the monojet limits decrease from 1.6 TeV ($g_{\rm q}=0.25$) to 1 TeV ($g_{\rm q}=0.1$). 

The mono-photon constraints at lepton colliders result from the mediator coupling to leptons, whereas at hadron colliders only the quark couplings are relevant.  As a result, the two cases cannot be compared like-for-like, although the results illustrate the relevant strengths for exploring the dark sector in a broad sense.  Furthermore, mono-photon constraints apply in a general EFT context, hence additional complementary coupling-dependent constraints, such as on four-electron interactions, may be relevant. 

Constraints for \HLLHC and \HELHC are taken from \cite{CidVidal:2018eel, Haisch:2018bby}.  The \FCChh monojet constraints for the axial-vector model are estimated using the collider reach tool, with results consistent with the analysis performed in \cite{Abada:2019lih}. Estimates for \FCChh, in the case of the scalar model, are taken from~\cite{Harris:2015kda}.  Estimates for low-energy FCC-hh (LE-FCC) are generated from the collider reach tool alone. Complementary dijet-resonance constraints for the axial-vector model are taken from \cite{Golling:2016gvc,Harris:2015kda}.  For the lepton colliders, the \CLIC monophoton estimates were provided privately by the CLICdp collaboration and all other lepton collider estimates are taken from \cite{Habermehl:2018yul}. For CEPC estimates, without considering systematic uncertainties, see~\cite{Liu:2019ogn}.
It is clear from these estimates that future colliders can provide sensitive probes of DM, potentially revealing evidence for invisible particle production, even for very massive mediators. 

Searches at high-energy hadron colliders have the best reach for the visible decays of multi-TeV mediator particles. Going beyond the \HLLHC reach for those same resonances in the mass region between 10 GeV and 1 TeV is still possible with an increased dataset at hadron colliders (see Sect.~\ref{sec:BSM-FIPs} and e.g.\ Ref.~\cite{Curtin:2014cca}), but it is inherently more challenging than for lepton colliders. It is often the case that signatures of sub-TeV resonances at hadron colliders are indistinguishable from those of their high-rate backgrounds, especially considering the impact of simultaneous $pp$ interactions on searches for hadronically decaying resonances at high-luminosity hadron colliders. Since it is generally not possible to record all events in their entirety for further analysis, as doing so would saturate the experiment data-acquisition and trigger systems, maintaining the sensitivity for sub-TeV resonances at hadron colliders requires the employment of specific data-taking and analysis techniques~\cite{2018arXiv180208640A} (see also Chapter~\ref{chap:inst}).

The discovery of invisible particles at a collider experiment does not imply that those invisible particles constitute the cosmological dark matter; for that, it would be necessary to compare collider results to direct and indirect detection experiment, as well as to astrophysical observations (e.g.\ the dark matter relic density). The comparison of the sensitivity of experiments at future colliders and direct/indirect detection experiments searching for dark matter for the models in this section can be found in Chapter~\ref{chap:dm}. 

\end{document}


\section{Feebly-interacting particles}
\label{sec:BSM-FIPs}

Unknown particles or interactions are needed to explain a number of observed phenomena and outstanding questions in particle physics, astrophysics and cosmology. While there is a vast landscape of theoretical models that try to address these puzzles, on the experimental side most of the efforts have so far concentrated on the search for new particles with sizeable couplings to SM particles and masses above the EW scale.
An alternative possibility, largely unexplored, is that particles responsible for the still unexplained phenomena are below the EW scale and have not been detected because they interact too feebly with SM particles. 
These particles would belong to an entirely new sector, the so-called {\it hidden} or {\it dark sector}. While masses and interactions of particles in the dark sector are largely unknown, the mass range between the MeV and tens of GeV appears particularly interesting, both theoretically and experimentally, and is the subject of this section.

An important motivation for new physics in this mass range is DM (see Chapter~\ref{chap:dm}), which could be made of light particles, with either a thermal or non-thermal cosmological origin. Thermal DM in the MeV--GeV range with SM interactions is overproduced in the early Universe and therefore viable scenarios require additional SM neutral mediators to deplete the overabundance~\cite{Lee:1977ua,Boehm:2002yz,Boehm:2003hm,Pospelov:2007mp,Feng:2008ya,ArkaniHamed:2008qn}. 
These mediators, which must be singlets under the SM
gauge symmetry, can lead to couplings of {\it feebly-interacting particles} to the SM through {\it portal} operators.

\subsection{The formalism of portals }
\label{ssec:fips_portals}

Portals are the lowest canonical-dimension operators that mix new dark-sector states with gauge-invariant (but not necessarily Lorentz-invariant) combinations of SM fields. Following closely the scheme used in the Physics Beyond Colliders study~\cite{Beacham:2019nyx}, four types of portal are considered:
\begin{center}
\begin{tabular}{c|l}
Portal & Coupling \\
 \hline
Vector (Dark Photon, $A_{\mu}$) &  $-\frac{\epsilon}{2 \cos\theta_W} F^\prime_{\mu\nu} B^{\mu\nu}$ \\ 
Scalar (Dark Higgs, $S$) & $(\mu S + \lambda_{HS} S^2) H^{\dagger} H$  \\ 
Fermion (Sterile Neutrino, $N$) & $y_N L H N$ \\ 
Pseudo-scalar (Axion, $a$) &  $\frac{a}{f_a} F_{\mu\nu}  \tilde{F}^{\mu\nu}$, $\frac{a}{f_a} G_{i, \mu\nu}  \tilde{G}^{\mu\nu}_{i}$,
$\frac{\partial_{\mu} a}{f_a }\overline{\psi} \gamma^{\mu} \gamma^5 \psi$
\end{tabular}
\end{center}
Here $F^\prime_{\mu \nu}$ is the field strength for the {\it dark photon}, which mixes with the hypercharge field strength
$B^{\mu\nu}$; $S$ is the {\it dark Higgs}, a new scalar singlet that couples to the SM Higgs doublet
$H$;
and $N$ is a {\it heavy neutral lepton} (HNL)
that couples to the SM left-handed leptons. These three cases are the only possible renormalisable portal interactions. While many new operators can be written at the non-renormalisable level, a particularly important example is provided by the {\it axion} (or axion-like) particle $a$ that couples to gauge and fermion fields at dimension five.

\subsection{Experimental sensitivities}
\label{ssec:fips_sensitivities}
The portal framework is used to define some benchmark cases, for which sensitivities of different experimental proposals are evaluated and compared with each other. 
Unless otherwise stated, all limits presented in this section correspond to 90\% CL, since the majority of the literature has been using this standard.

\begin{figure}[t]
    \centering
    \includegraphics[width=\textwidth]{\main/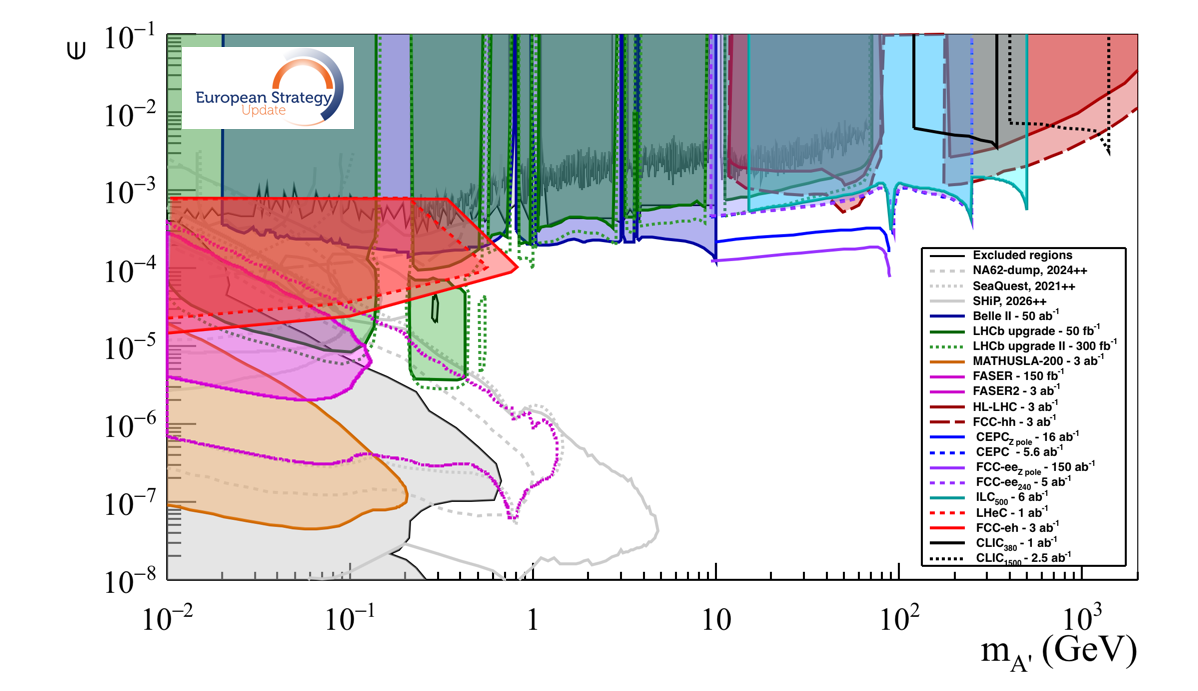}
    \caption{Sensitivity for Dark Photons in the plane mixing parameter $\epsilon$ versus Dark Photon mass. HL-LHC, CEPC, FCC-ee and FCC-hh curves correspond to 95\% CL exclusion limits, LHeC and FCC-eh curves correspond to the observation of 10 signal events, and all other curves are expressed as 90\% CL exclusion limits. The sensitivity of future colliders, mostly covers the large-mass, large-coupling range, and is fully complementary to the the low-mass, very low-coupling regime where beam-dump and fixed-target experiments are most sensitive.
    }
    \label{fig:FIPs-DarkPhoton}
\end{figure}

\noindent
{\bf Vector portal} \\
New light vector particles mixed with the photon are not uncommon in BSM models containing hidden sectors, possibly related to the DM problem.
The parameters describing this class of models are $\epsilon$, $\alpha_D$, $m_{A'}$ and $m_{\chi}$,
where $\epsilon$ is the mixing parameter between the dark and ordinary photon; $\alpha_D = g^2_D/ 4 \pi$ is the coupling strength of the dark photon with DM; and $m_{A^\prime}$ and $m_{\chi}$ are the dark photon and DM particle mass, respectively.
The study of experimental sensitivities at future colliders is performed in the plane of $\epsilon$ versus $m_{A^\prime}$,
assuming $\alpha_D$ to be negligible with respect to $\epsilon$.
It is important to note that only minimal Dark Photon models have been considered in this study. 
Non-minimal models used by, e.g.\ the \HLLHC experiments~\cite{CidVidal:2018eel} and other future facilities, are not addressed here.
The results are shown in Fig.~\ref{fig:FIPs-DarkPhoton}.

Visible decays of vector mediators are mostly constrained from searches for di-electron or di-muon resonances and from the re-interpretation of data from fixed target or neutrino experiments in the low (< 1 GeV) mass region.
NA48/2~\cite{Batley:2015lha}, A1~\cite{Merkel:2014avp} and BaBar~\cite{Lees:2014xha} experiments put the strongest bounds for 
$\epsilon > 10^{-3}$ in the $0.01 - 10$ GeV mass range.
These results are complemented by those from beam dump experiments, such as E141~\cite{Riordan:1987aw} and E137~\cite{Bjorken:1988as, Batell:2014mga} at SLAC, E774 at Fermilab~\cite{Bross:1989mp}, CHARM~\cite{Bergsma:1985qz} and NuCal~\cite{Blumlein:1990ay}.

The low-mass range (0.01--1 GeV, see Chapter~\ref{chap:dm}) is best covered by beam-dump experiments (SHiP~\cite{Anelli:2015pba}, NA62 in dump mode~\cite{Lanfranchi:2018xrz}), and by FASER at the ATLAS interaction point~\cite{Ariga:2018pin} in the very low-coupling regime ($\epsilon < 10^{-4}$). These are complemented by the LHCb Upgrade~\cite{Ilten:2015hya} and Belle-II~\cite{Kou:2018nap}. Future collider experiments (HL-LHC~\cite{Curtin:2014cca}, \CEPC~\cite{CEPCStudyGroup:2018ghi}, \FCCee~\cite{Karliner:2015tga}, \FCCeh~\cite{donofrio2019searching}, \FCChh~\cite{Curtin:2014cca}, {\ILCFiveHundred}) have unique coverage in the high-mass range ($>$ 10 GeV) down to $\epsilon \sim 10^{-4}$. \FCCeh
could fill the gap left by LHCb in the low-mass region. There is an interesting complementarity between future collider experiments, which cover the high-mass large-coupling regime, and beam-dump experiments, which cover the low-mass, very low-coupling regime.

\vskip 2mm
\noindent
    {\bf Scalar portal} \\
In the {\it scalar} or {\it Higgs portal}, the dark sector is coupled to the Higgs boson via the bilinear $H^{\dagger} H$ operator of the SM. The minimal scalar portal model operates with one extra
singlet field $S$ and two types of couplings, $\mu$ (or $\sin \theta$)  and $\lambda_{HS}$~\cite{OConnell:2006rsp}. 
The coupling constant $\lambda_{HS}$ leads to pair-production of $S$ but cannot induce its decay, which requires a non-vanishing $\sin \theta$.
This portal has several theoretical motivations. The new scalar can generate the baryon asymmetry of the Universe~\cite{Cohen:1987vi} and play the role of mediator
between SM particles and light DM in case of secluded annihilations ($\chi \chi \to \phi \phi$, where $\chi$ is the light DM particle and $\phi$ the light scalar mediator) ~\cite{Krnjaic:2015mbs}.
It can also address the Higgs fine-tuning problem (via the {\it relaxion} mechanism ~\cite{Graham:2015cka}), which generically leads to relaxion-Higgs mixing~\cite{Flacke:2016szy} 
and provides an alternative baryogenesis mechanism~\cite{Abel:2018fqg} and a DM candidate~\cite{Fonseca:2018xzp,Banerjee:2018xmn}. 

\begin{figure}[htb]
    \centering
    \includegraphics[width=\textwidth]{\main/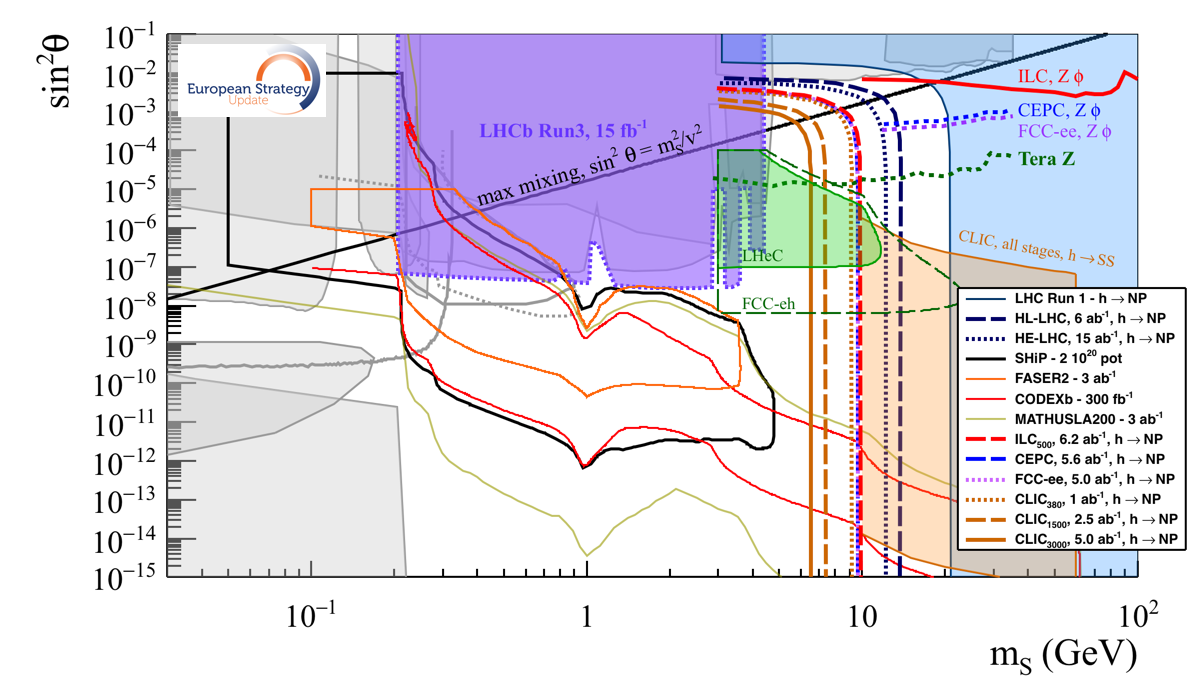}
    \caption{Exclusion limits for a Dark Scalar mixing with the Higgs boson. LHeC, FCC-eh, CLIC (all stages) curves and the vertical lines correspond to 95\% CL exclusion limits, while all others to 90\% CL exclusion limits. See text for details.
    }
    \label{fig:FIPs-DarkHiggs}
\end{figure}

The experimental sensitivities are shown in Fig.~\ref{fig:FIPs-DarkHiggs}.
Shaded grey areas are already excluded, as detailed in Ref.~\cite{Beacham:2019nyx}.
The low-mass ($< 10$ GeV, see Chapter~\ref{chap:dm}), low-coupling range is optimally covered by SHiP at the Beam Dump Facility and MATHUSLA200. 
FASER2, with 3~ab$^{-1}$ will explore the region above few GeV compatible with that of CODEX-b. MATHUSLA200 has a unique reach in the high-mass and very low-coupling regime.
Vertical lines correspond to the bounds on the Higgs/dark-Higgs quartic coupling $\lambda_{HS}$ and on $m^2_S/v^2$ from the projections for the untagged-Higgs at future colliders~\cite{deBlas:2019rxi} (see discussion in~\cite{Frugiuele:2018coc}). The mass range above a few GeV can be explored also by CLIC and LHeC/FCC-eh using the displaced-vertex technique. The large-coupling regime is covered by $e^+ e^-$ colliders using the recoil technique ($e^+ e^- \to Z S$) or running at the $Z$-pole, via the process $e^+ e^- \to Z \to S \ell^+ \ell^-$.

\vskip 2mm
\noindent
In the limit of small mixing angle, one can bound the Higgs/dark-Higgs quartic coupling $\lambda_{HS}$ via the Higgs invisible width, which is naturally expected to satisfy the relation $\lambda_{HS}\lesssim m_S^2/v^2$. 
In Table~\ref{table:scalarInv} projections for the constraints on $\lambda_{HS}$ and the scalar mass for various future collider options are provided. 

\begin{figure}[htb]
    \centering
    \includegraphics[width=\textwidth]{\main/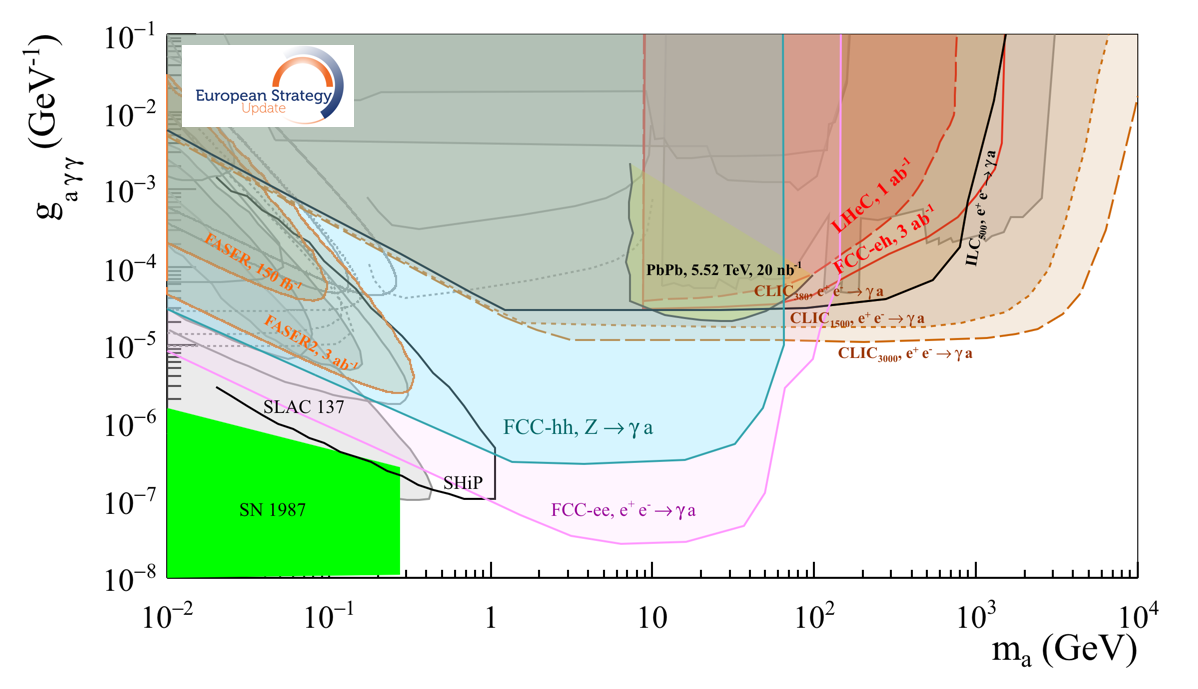}
    \caption{Exclusion limits for ALPs coupled to photons. All curves correspond to 90\% CL exclusion limits, except for LHeC/FCC-eh (95\% CL exclusion limits), FCC-ee (observation of four signal events) and FCC-hh (observation of 100 signal events). See text for details.}
    \label{fig:FIPs-ALP}
\end{figure}

\begin{table}[h!]
\caption{\small Bounds on the Higgs/dark-Higgs quartic coupling $\lambda_{HS}$, and on the scalar mass $m_S$. The projection for the Higgs invisible width are taken from~\cite{deBlas:2019rxi}.}
\begin{center}
 \begin{tabular}{|c| c c c c c c c c|} 
 \hline
  Collider &{\small HL-LHC} & {\small HL-LHC}& {\small HL-LHC} &  {\small ILC$_{500}$}& {\small  CLIC$_{3000}$}&{\small CEPC} & {\small FCC} & {\small FCC} \\
   & &  {\small + LHeC} &  {\small + HE-LHC} & & & & {\small ee} & {\small ee/eh/hh}
 \\ [0.5ex] 
 \hline\hline
 ${\lambda_{HS}/{10^{-3}}}$ & 2.0& 1.5& 1.8& 0.69& 1.2& 0.77& 0.64& 0.23 \\ 
 \hline
$m_S\,$[GeV] & 11& 9.7& 10& 6.5& 8.4& 6.8& 6.2& 3.7\\
 \hline

\end{tabular}
\end{center}
\label{table:scalarInv}
\end{table}
\vskip 2mm
\noindent
{\bf Pseudo-Scalar portal} \\
QCD axions are a central idea in particle physics 
to solve the strong CP problem (see Chapter~\ref{chap:dm}).
Current QCD axion models are restricted to the sub-eV mass range. Less constrained are generalisations to axion-like particles (ALPs)~\cite{Jaeckel:2010ni}, which can act as
mediators between light DM and SM particles.
Figure~\ref{fig:FIPs-ALP} shows the sensitivity of future collider experiments to ALPs interacting with photons. 

The typical production mechanisms are: the Primakoff effect and $\pi^0$/$\eta$ decays (see Chapter~\ref{chap:dm}) in beam-dump experiments; DY production followed by $Z \to a \gamma$ for hadron colliders; $e^+ e^-  \to (Z) \to a \gamma$ with $a \to \gamma \gamma$ for lepton colliders.
Three mass regions can be clearly identified. The region $m_a < 1$~GeV is dominated by beam-dump experiments in the very low-coupling regime and by FASER2~\cite{Ariga:2018pin}. In this mass regime, ALPs have a lifetime long enough to escape direct detection in experiments at future colliders, and can be identified only via missing energy.
The intermediate mass region (1~GeV $< m_a < $ 90~GeV) can be optimally explored by $e^+ e^-$ colliders (\CEPC~\cite{CEPCStudyGroup:2018ghi}, \CLIC~\cite{deBlas:2019rxi}, \ILC, 
\FCCee~\cite{Bauer:2018uxu}) running at the $Z$-pole and by hadron colliders (\FCChh~\cite{Bauer:2018uxu}) via $Z$ decays. In most of this mass range, the two photons from $a$ decays are not resolved and, hence, the ALP mass cannot be determined. Finally the high-mass region (from tens of GeV to a few TeV) can be optimally explored  by $e^+ e^-$ linear colliders (\ILC and \CLIC) and $ep$ colliders (LHeC and \FCCeh~\cite{Yue:2019gbh}).
Future collider experiments can also search for ALPs with fermion and gluon couplings but the corresponding sensitivity curves have been evaluated only in a few cases and are not considered in this study. 

\vskip 2mm
\noindent
{\bf Fermion portal} \\
The physics case for {\it Heavy Neutral Leptons} (HNL) is discussed in Chapter~\ref{chap:neut} and here only a summary of projections on the experimental reach is presented.

\begin{figure}[htb]
    \centering
    \includegraphics[width=\textwidth]{\main/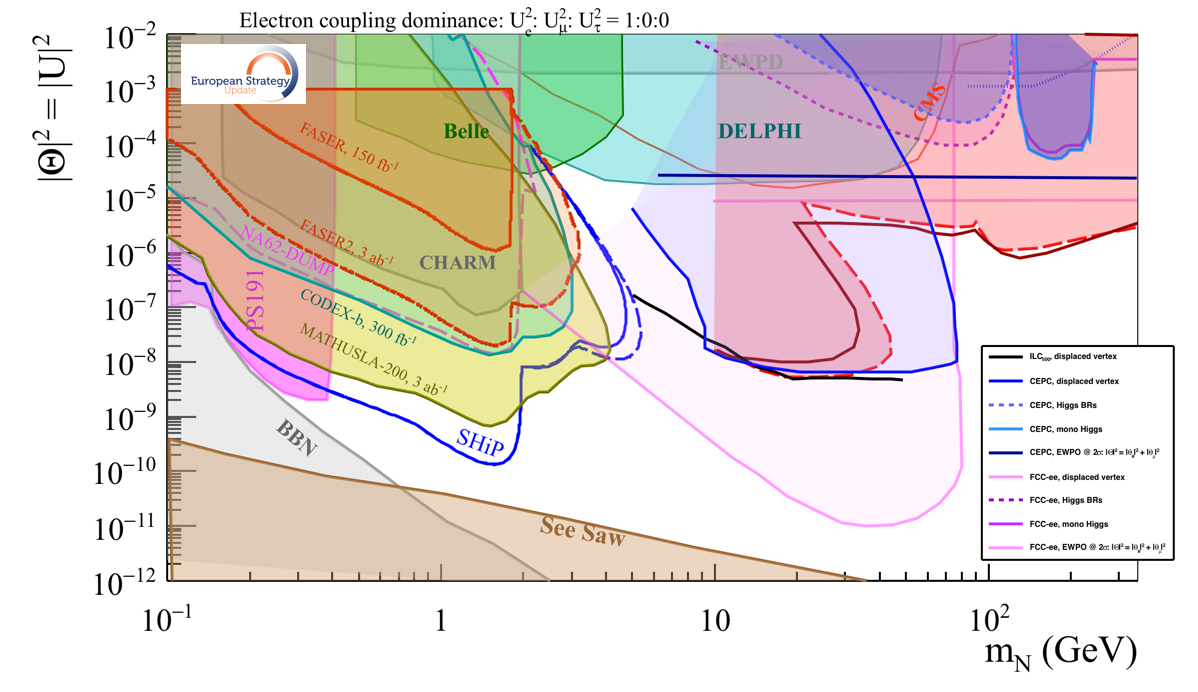}
    \caption{90\% CL exclusion limits for a Heavy Neutral Lepton mixed with the electron neutrino. See text for details.
    }
    \label{fig:FIPs-HNL}
\end{figure}

Figure~\ref{fig:FIPs-HNL} shows the sensitivity of experiments at current and future accelerators to the mixing parameter between the electron neutrino and HNL
in the mass range 0.1--100 GeV. The low-mass range ($< 5$ GeV) is dominated by SHiP at the Beam Dump Facility, followed by experiments at the LHC interaction points, MATHUSLA200, FASER, CODEX-b~\cite{Beacham:2019nyx}. The mass region between 5 and about 90 GeV can be explored by \ILC~\cite{Antusch:2017pkq}, \CEPC~\cite{CEPCStudyGroup:2018ghi} and \FCCee~\cite{Abada:2019zxq}, being dominated by the \FCCee running at the $Z$-pole~\cite{Abada:2019zxq}.
HNL with masses above 90 GeV can be directly searched for using 
displaced-vertex techniques at \FCChh~\cite{Antusch:2016ejd}. 
FCC-eh can also explore HNLs in the same mass range~\cite{Antusch:2016ejd} but sensitivity plots have been produced only for couplings to the second neutrino generation and are not included in Fig.~\ref{fig:FIPs-HNL}.
Among indirect techniques, EW precision measurements allow the sensitivity to HNL to be extended up to very high masses, well beyond what shown in Fig.~\ref{fig:FIPs-HNL}. Finally, SHiP and \FCCee running at the $Z$-pole have the potential to exclude the region of masses and couplings compatible with leptogenesis~\cite{Eijima:2018qke} almost down to the see-saw limit.  The sensitivity to Heavy Neutral Leptons coupled predominantly to the second and third generation is shown in Figs.~\ref{fig:sec3_LLP}  and~\ref{fig:dark:daulepton}, respectively.

%


\def\circa#1{\,\raise.3ex\hbox{$#1$\kern-.75em\lower1ex\hbox{$\sim$}}\,}

\section{Summary and conclusions}
\label{sec:BSM-Conclusions}

In recent years, the scene of BSM research has been evolving rapidly, thanks to a wealth of new experimental data in particle and astroparticle physics. On the theoretical front, less emphasis has been given to unified frameworks able to deal simultaneously with many key questions in particle physics, and more attention has been given to models that address individual shortcomings of the SM or simply single unexplained facts. This has created a more fragmented landscape of research activity, where there is no single dominating trend, but multiple approaches pursuing different directions. The need to look for new theoretical paradigms is making today's research in particle physics very exciting, rich with opportunities for alternative and revolutionary ideas. In this situation, more than ever, an intense and diversified programme of new experimental projects is needed to unravel the many mysteries left unresolved by the SM and to provide clues for progress in theoretical speculations.

The current report reflects broadly the present state of the field. Instead of giving a comprehensive account of all BSM model variations and their phenomenological signatures, the analysis has focused on a representative set of cases that allow for an informative comparison of the reach of future experimental projects. At the beginning of the ESPP physics activities, four fundamental questions that would serve as a leitmotif for the BSM studies were identified and presented to the physics community at the Open Symposium in Granada. This chapter is concluded with a presentation, in the form of a summary, of those questions and the answers that have emerged from the  study.

\medskip
\noindent{\bf 1.~~To what extent can we tell whether the Higgs boson is fundamental or composite?}

\noindent Undoubtedly the Higgs boson is the centrepiece of today's BSM physics. Its discovery has led to an unprecedented situation in physics, since no fundamental scalar particles and no fundamental forces different from gauge forces had ever been observed prior to the Higgs. These facts are not mere curiosities, but are at the core of the main puzzles confronting particle physics today. 
Progress with these issues requires an experimental programme targeted at precision measurements of Higgs interactions and EW observables. This programme is a {\it clear priority for the future of particle physics}. Higgs precision measurements are especially efficient in testing strongly-interacting EW breaking sectors (such as in composite Higgs models), theories for EW breaking in which there are no weak-scale coloured particles associated with the Higgs (such as Neutral Naturalness), and theories in which the Higgs is mixed with other scalar states.

A central question for the precision programme is the nature of the Higgs boson, i.e. whether it is a fundamental or composite particle. Theories like SUSY suggest that the Higgs boson is as fundamental as any other SM particle, while models based on approximate Goldstone symmetries suggest that the Higgs has a composite structure, much like the pion in QCD. As shown in Sect.~\ref{sec:BSM-EWKNR}, this question can be quantitatively addressed by future colliders, which can test the `size' of the Higgs up to inverse distances $1/\ell_H \sim 10-20$~TeV, more than four orders of magnitude below the size of a proton. To put this result in perspective, we define the degree of compositeness $\delta$ of a particle with mass $m$ as the ratio between its effective size and its Compton wavelength $\lambda_C = 2\pi \hbar /m c$ (which is a measure of the particle's quantum nature). For a proton, which is a fully composite object, one finds $\delta_p \approx m_p/(2\pi \Lambda_{\rm QCD})\approx 1$. For a pion, which is a composite particle but emerges as a Goldstone boson below the QCD scale, one finds $\delta_\pi \approx m_\pi /(2\pi m_\rho ) =0.03$. Future colliders will be able to probe the Higgs degree of compositeness at the level of $10^{-3}$. Knowledge about the fundamental nature of the Higgs will give us decisive indications on the directions to pursue in future BSM research.

\medskip
\noindent{\bf 2.~~Are there new interactions or new particles around or above the electroweak scale?}

\noindent All dynamical frameworks that address the open problems of EW symmetry breaking predict that the Higgs boson must be accompanied by new particles or new phenomena. This conclusion is quite generic. In SUSY, the Higgs is elevated to a supermultiplet and, furthermore, the scalar structure is doubled. Composite Higgs, Little Higgs, neutral naturalness predict accompanying particles as well. Any dynamical explanation of Higgs naturalness requires partners to the top quark and gauge bosons with appropriate properties. Therefore, the hunt for new high-energy phenomena is an essential route towards progress in our understanding of particle physics. Future colliders are superb explorers of this route, as discussed in Sects.~\ref{sec:BSM-EWKNR}, \ref{sec:BSM-SUSY} and \ref{sec:BSM-ExtendedScalars}. 
Searches at the highest energies are especially efficient in testing theories for EW breaking with new coloured particles (such as SUSY) and a great variety of new phenomena not necessarily associated with the Higgs (such as additional gauge bosons and scalars, heavy resonances, or exotic particles such as leptoquarks). The power of the high-energy frontier lies in its versatility for exploring the unknown, but also extends to the programme of precision measurements (especially for energy-growing effects) and to searches for rare processes that benefit from high luminosity. 

The collider exploration of short distances can proceed through {\it direct} or {\it indirect} searches. Proposed future colliders can explore new physics extensively, up to scales of tens of TeV through {\it direct} searches. The direct exploration with colliders operating at higher energies is the only way to have hands-on access to new phenomena and to inspect their microscopic nature. As shown in this chapter, a variety of new-physics scenarios can be effectively tested in this way. For example (see Sect.~\ref{sec:BSM-SUSY}), direct searches translate into a probe of the degree of naturalness of SUSY theories down to a level of $10^{-5}$, testing deeply one of the guiding principles of particle physics. 


An alternative experimental strategy is based on {\it indirect} searches. A particularly interesting class of indirect probes are tests of accidental symmetries or cancellation mechanisms in the SM, especially in the flavour sector (see Chapter~\ref{chap:flav}).
A second class of indirect probes are Higgs precision measurements (see Chapter~\ref{chap:ew}). Quite generically, modifications of Higgs couplings are proportional to the degree of fine-tuning of the theory, with a fully natural theory predicting ${\mathcal O}(1)$ effects in Higgs couplings. This link provides the basis for a comparison between the relative effectiveness of direct versus indirect searches. The third class of indirect probes are EW precision measurements (see Chapter~\ref{chap:ew}). The expected intensity and accuracy of future lepton colliders at the $Z$ and $WW$ thresholds will allow for improvements of indirect sensitivity of several orders of magnitude. The distinction between EW and Higgs precision tests is purely historical, as they are really two aspects of the same question. 

The fourth class of indirect probes is the study of deviations in SM scattering processes (examples are the contact interactions studied in Sect.~\ref{sec:BSM-EWKNR}). The important difference of these observables, with respect to the other three classes of indirect tests, is that their effects grow with powers of $s/\Lambda^2$ and therefore these measurements benefit not only from the statistics of high luminosity, but also from gains in the collider energy. For this reason, indirect tests are not only the domain of lepton colliders. Hadron colliders with very high luminosities are also effective and complementary in this respect, in particular when looking for processes that grow with energy or for rare processes, for which the large production rates at hadron colliders are essential.

The complementarity between direct and indirect searches can be illustrated with the example of a new resonance with mass $M$ and couplings $g_{Z^\prime}$ to SM particles (see Fig.~\ref{fig:Universal_Zp}). With {\it direct} searches, high-energy colliders can explore larger $M$ by increasing $\sqrt{s}$, and smaller $g_{Z^\prime}$ by increasing the luminosity. Virtual effects of the resonance can be detected {\it indirectly} by measuring deviations from SM predictions in the high-energy tails of distributions. These measurements can probe masses beyond the collider kinematic limit, but are sensitive only to the ratio $g_{Z^\prime}/M$. Higher energies allow for more effective probes of the ratio $g_{Z^\prime}/M$. This example shows the complementarity between the two experimental strategies, with {\it direct} searches being in general more effective in the weakly-coupled regime (i.e.\ small $g_{Z^\prime}$) and {\it indirect} searches in the strongly-coupled regime (i.e.\ large $g_{Z^\prime}$). Although this distinction provides a good general guideline, a precise comparison between the two strategies can only be performed on a model-by-model basis. 

{\it Indirect} searches have the advantage of probing particle masses well beyond the collider kinematical limit, but cannot identify the specific source of new physics.
Only discoveries in {\it direct} searches can give first-hand access to the microscopic structure of new phenomena. However, any new discovery, irrespective of whether it stems from direct or indirect searches, will certainly motivate a scientific programme of dedicated precision measurements. This was the case for the LEP programme which followed the discovery of the $W$ and $Z$ bosons, and it is today the case for a Higgs factory which is proposed to follow after the discovery of the Higgs boson.

In this report, strong emphasis has been placed on the interplay between direct and indirect searches at colliders in the exploration of the high-energy frontier. This interplay allowed us to make quantitative comparisons between the exploratory power of different experimental projects and to show the great complementarity of the two methods in the search for new physics. 
Nevertheless, addressing the mysteries of particle physics at the EW scale requires a bold step in the exploration of Nature at the smallest possible distances.  In general, hadron colliders offer a higher mass reach, but lepton colliders can guarantee an almost full coverage of the kinematic region accessible to them. And while Higgs and EW precision measurements at an $e^+e^-$ collider provide an imperative and well-motivated physics programme with guaranteed deliverables, only colliders that break new ground in the high-energy domain offer direct exploration of the unknown and firsthand observations of Nature's behaviour at distance scales that have never been probed before.
Of the two currently-proposed colliders at the energy frontier, \CLICThreeThousand offers a general exploration of any new particle with EW interactions up to the kinematic limit, which corresponds to masses of 1.5 TeV for pair-production processes, while FCC-hh can explore some cases up to much larger masses, e.g.\ gluinos up to 17 TeV, top squarks up to 10 TeV, and new Higgs bosons from a second EW doublet up to 5--20 TeV, depending on the value of $\tan\beta$.

Bold and pioneering exploration of the unknown has always characterised the history and successes of particle physics. Once again, the quest to understand the fundamental physical laws is driving particle physics toward ever smaller distance scales. The exploration of the high-energy frontier remains the {\it most promising option as well as the most pressing priority for the future of particle physics}. High-energy colliders are the unique and irreplaceable tool to explore directly phenomena at very short distances and to confront the open questions related to the EW scale.




\medskip
\noindent{\bf 3.~~What cases of thermal-relic particles are still unprobed and can be fully covered by future collider searches?}

\noindent Dark Matter provides a fascinating link between large-scale astronomy and short-distance particle physics (see Chapter~\ref{chap:dm}). While the realm of possibilities for DM is still enormous, there are well-defined windows where particle physics can contribute in a unique way. An interesting prototype for DM is a heavy weakly-interacting particle. In Sect.~\ref{sec:BSM-DM} it has been shown how future colliders can test this hypothesis, demonstrating that the early-Universe thermal origin of DM in the form of an EW doublet can be conclusively proven or ruled out at both \CLICThreeThousand and \FCChh, and the corresponding EW triplet can be probed conclusively at the \FCChh. Other benchmark models for DM can also be effectively tested at future colliders and examples have been provided.

Other forms of DM, constituted of much lighter particles, can also be explored by particle-physics experiments (see Chapter~\ref{chap:dm}). With a variety of collider-based, beam-dump and fixed-target experiments, it is possible to probe the existence of new light particles (such as axions, ALPs, sterile neutrinos, etc.) which, in some cases, may be related to the DM problem, besides being motivated by other particle-physics or astrophysics considerations (see Sect.~\ref{sec:BSM-FIPs}).

\medskip
\noindent{\bf 4.~~To what extent can current or future accelerators probe feebly-interacting sectors?}

\noindent The absence, so far, of unambiguous signals of new physics from direct searches at the LHC, indirect searches in flavour physics and direct DM detection experiments invigorates the need for broadening the experimental effort in the quest for new physics and in exploring ranges of interaction strengths and masses different from those already covered by existing or planned projects. {\it While exploration of the high-mass frontier remains an essential target, other research directions have valid theoretical motivations and deserve equal attention}. Feebly-interacting particles (see Sect.~\ref{sec:BSM-FIPs}) represent an alternative paradigm with respect to the traditional BSM physics explored at the LHC. The full investigation of this paradigm over a large range of couplings and masses requires a great variety of experimental facilities. In this context, the physics reach of experiments at future colliders is complemented by beam-dump facilities which typically cover the range of low masses and extremely feeble couplings.

\chapter{Dark Matter and Dark Sectors}
\label{chap:dm}
\noindent This chapter is based on material submitted by the particle physics community to the  ESPP update process that is relevant for dark matter and dark sectors exploration. Section 9.1 provides an introduction to the topic. Section 9.2 briefly highlights current results and potential of dark matter astrophysical probes.
Dark Matter (DM)  and Dark Sector (DS) searches at colliders are discussed in Chapter~\ref{chap:bsm}, and are briefly touched upon in Sect.~9.3 of this chapter through their complementarity with dark matter direct and indirect  detection experiments. Section 9.4 concentrates on accelerator based DM/DS searches at fixed target and beam dump experiments. Section 9.5 discusses axions and Axion-Like Particle (ALP) searches. Section 9.6 concludes on the main findings of the preparatory group exercise for this chapter. More detailed information can be found in a supporting note~\cite{DM-note}.

\section{Introduction}


There is compelling evidence from galactic and cosmological observations that DM exists, and detecting DM in the laboratory is one of the greatest challenges of particle physics \cite{bertone:2004pz}. Since DM  - if made of particles or compact objects - is the dominant form of matter in the universe, it is highly plausible that there is also a richer 
Hidden Sector (HS).
The constituents of such HS could include multiple species of massive particles, one or more which might mix with Standard Model (SM) particles such as the Higgs boson, the photon or neutrinos, via so called HS-SM portal operators (see \cite{Battaglieri:2017aum} for
a review).
A hidden sector that contains dark matter is more generically called a DS, and includes at least a mediator connecting  HS particles and SM particles. The SM operator interacting/mixing with the mediator is often referred to as a portal (e.g.\ Higgs portal).
In the most simplified  DS realization, with just  DM as its only component, a new, beyond the Standard Model (BSM) particle may act as the mediator.
Such a mediator can also be the force carrier of a new gauge group under which the
SM particles are charged, e.g.\ U(1)$_{B-L}$ \cite{Ilten:2018crw,Bauer:2018onh}.




While the observational evidence for dark matter is exceptionally convincing, our current level of ignorance of the basic properties of dark matter is remarkable 
\cite{Bertone:2016nfn}. The mass of dark matter particles could be anything from as light as $10^{-22}$ eV \cite{Hu:2000ke} to as heavy as primordial black holes of tens of solar masses \cite{Bird:2016dcv}. The lower mass limit comes from the requirement that dark matter particles can have a sufficiently short de Broglie wavelength to form dwarf galaxies and galactic sub-halos, while the upper limit is set by observational limits on massive compact halo objects (MACHOs) and CMB anisotropies \cite{Brandt:2016aco,Nakama:2017xvq,Ali-Haimoud:2016mbv}. Within this very large mass range it is useful to define some distinct families of possibilities. 

Considering the DM production mechanism in the early Universe provides a useful classification. We shall concentrate  here on  two prominent examples. 
Assuming the DM is produced thermally, through interactions with the SM in the early Universe, 
narrows the viable DM masses significantly, to the few keV to 100 TeV range (e.g.\ WIMPS and hidden sector particles). 
In the case of standard cosmology, theoretical models become highly predictive; however, in nonstandard cosmology a wide range of models can produce the observed dark 
matter abundance.
Alternatively, ultralight particles in the sub-eV range must be produced non-thermally (e.g.\ QCD axion and axion-like particles), also allowing for different cosmological histories. 

We are equally ignorant of dark matter interactions. As part of a dark sector, 
dark matter may interact only gravitationally with SM matter or it may interact via DS mediators (neutral under the SM gauge group) interacting/mixing through a SM portal, as well as through new particles charged under the SM gauge group that couple to the dark matter directly.
Alternatively, dark matter may carry SM charges itself as  part of an extended BSM sector with many new states carrying SM quantum numbers. In such case  a preserved  symmetry distinguishing between the SM and BSM particles  renders the DM particle stable (e.g.\ as in supersymmetry). 
In this latter approach, inspired by BSM theories built to solve additional mysteries of particle physics, we usually refer to the dark matter as Weakly Interacting Massive Particles (WIMPs).

The history of dark matter direct detection  experiments has been dominated by WIMP searches, motivated by the 
 so-called ``WIMP miracle'': the qualitative observation that particles with roughly weak scale $\cal{O}$(100) GeV masses, and weak scale interactions with SM particles, will end up with roughly the observed thermal relic density after freeze out in standard Big Bang cosmology (see \cite{Jungman:1995df} for a review). WIMP scenarios, of $\cal{O}$(100) GeV mass dark matter particles interacting with detectors via $Z$ boson exchange, have already been strongly constrained by the impressive current limits of multiple overlapping direct WIMP searches. Current and proposed WIMP direct detection experiments will push the sensitivity 
 down to the so-called ``neutrino floor'', where background from known astrophysical neutrino sources will swamp the expected signal of nuclear recoils from interactions with WIMPs from the galactic halo \cite{Billard:2013qya}. New strategies are being developed to conquer the neutrino floor and enable exploration beyond it.
 Collider searches for  WIMP candidates, such as those provided by models of supersymmetry,  are a main focus of interest at the HL-LHC and provide a strong case for a future high-energy hadron collider.

WIMP dark matter is not the only example of plausible dark matter particles that can be explained as thermal relics from early Universe cosmology. 
Hidden sector dark matter can provide such thermal relics in the mass range 
between about a few keV \cite{Safarzadeh:2018hhg}  and  about 100~TeV, with unitarity constraints yielding the upper bound \cite{griest:1989wd}. In principle, such relic dark matter could interact with laboratory detectors via new exotic, feeble interactions.
Laboratory detection of keV--GeV thermal dark matter, often referred to as light dark matter (LDM), requires deploying novel detector technologies and/or new detection strategies, since existing strategies typically fail below the GeV scale. 
Interestingly enough, the phenomenology of low-mass region (keV--GeV) thermal dark matter is quite different from the standard WIMP. In particular, in order to 
 deplete the DM density at freeze out to agree with the currently observed values, 
it typically demands the existence of light mediator(s) 
that also give rise to novel laboratory searches independently of their connection to DM. 
New accelerator-based searches, in which a high-intensity, relativistic beam of particles (electrons, protons, muons) impact on a dense target, offer a   compelling path to probe  LDM and DS portals. These accelerator-based fixed target experiments can exploit beam-dump and missing energy/momentum techniques and will have unprecedented sensitivity to DS realizations at the CERN SPS and/or the SLAC LCLS-II facilities.  
They are complementary to electron recoil signals at dedicated low-mass direct detection experiments. 

 In the case of ultralight, non thermal DM with mass less than about an eV, the measured galactic DM density implies that the occupation numbers of these ultralight dark states are very high. Therefore such ultralight DM (lighter that about 0.7 keV) must be bosonic \cite{Tremaine:1979we}, and 
 the effects of halo DM in terrestrial experiments are best described as wavelike disturbances. 
 A particularly compelling candidate for ultralight DM is the QCD axion, whose mass can be in the range 
$10^{-12}$ to ${\rm  10^{-3}eV}$ (see \cite{Marsh:2015xka} for a review). 
If QCD axions are a major component of DM, they could  be detected as coherent waves from the halo. Quite generally, axions can also be produced as higher energy particles by our Sun and detected by terrestrial experiments. Many BSM scenarios contain axion-like particles and other sub-eV dark sector particles that might eventually also be produced at the laboratory 
without relying on cosmological and astrophysical assumptions. Ultralight DM searches motivate a broad variety of non-accelerator experiments, which however profit crucially from technologies developed in accelerator-based research \cite{Battaglieri:2017aum}.


%
%








\section{Astrophysical Probes of Dark Matter}

The understanding of the dark matter properties will demand signals from multiple experiments, together with compatibility with indirect measurements from DM annihilation or decay products. This section  briefly reports on the status and prospects for DM Direct Detection (DD) and Indirect Detection (ID) searches, with emphasis on masses greater that about 1 GeV. For lower mass regimes see Sects. \ref{subsec:DM-BDandFT-Complementarity} and \ref{subsec:DM-Axions-Complementarity}.

\subsection{Direct detection}

\begin{figure}[t]
\centering
  \includegraphics[width=0.85\linewidth,angle=0,origin=c]{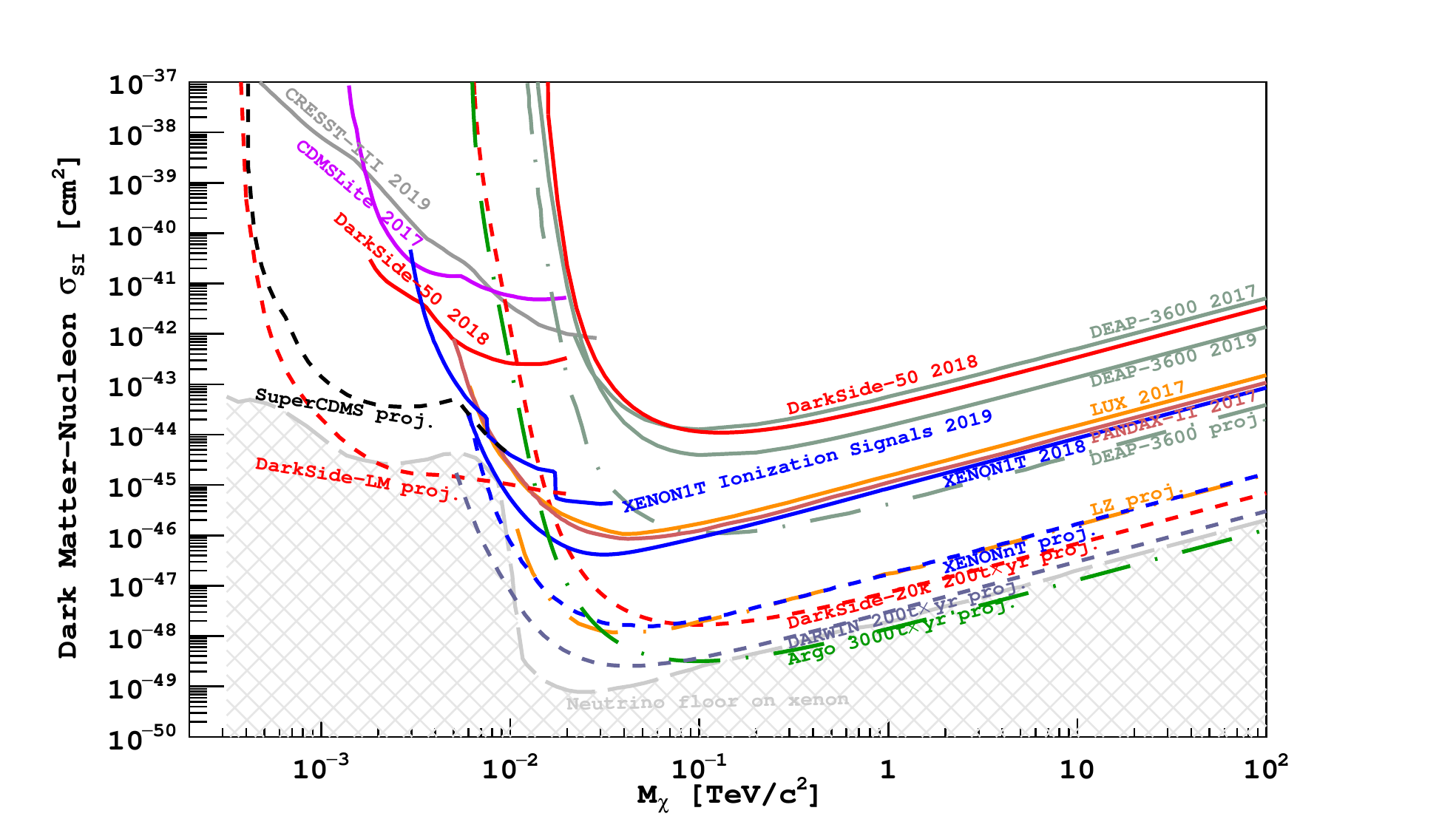}
\caption{90\% CL exclusion limits showing leading results from direct detection (continuous lines, Refs.~\cite{Angloher:2012kl,Akerib:2017kg,Cui:2017kg,Aprile:2018ct,PhysRevD.98.102006,Agnes:2018fg}). 
Sensitivities of future Ge-, Xe-, and Ar-based direct searches are also shown with dashed lines, Ref.~\cite{Nelson:2014wy,Kudryavtsev:2015hy,Aprile:2015wv,Boulay:2017tn,Agnese:2017fn}). 
The neutrino floor curve follows the definition of Refs.~\cite{Billard:2014cx}. 
}
\label{fig:sensitivity}
\end{figure}

%
At present, direct detection searches have excluded spin-independent dark matter-nucleon cross sections as low as ~10$^{-46}$ cm$^2$ \cite{Cui:2017kg,Aprile:2018ct}, 
shown as solid curves in Fig.~\ref{fig:sensitivity}, and spin-dependent cross sections as low as 10$^{-41}$ cm$^2$
\cite{Amole:2017iz}.  
For interactions with heavy mediators, collider searches may have  comparable sensitivity for the subset of models and parameters chosen in Chapter~\ref{chap:bsm};
specific examples are shown in Sect.~\ref{subsec:complementarity_collider}.  In Fig.~\ref{fig:sensitivity}, the leading results in the 5 GeV range and below come from the DarkSide-50 LAr TPC low-mass search and from cryogenic solid-state detectors, while at higher masses from cryogenic noble liquids, led for the past decade by the pioneering XENON programme at LNGS.  There have also been several experiments that are consistent with a dark matter signal, both from direct 
\cite{Bernabei:2010gs, Aalseth:2014wc, Agnese:2013hy, Angloher:2012kl} 
and indirect
\cite{Aguilar:2014cj,Ajello:2016ge} 
searches. These results demonstrate that confirmation by multiple experiments, with independent detection strategies and targets, is essential for a convincing discovery.
%
%
%

Projected sensitivities of near-future direct detection dark matter searches are shown in Fig.~\ref{fig:sensitivity} as dashed curves.  Three mid-term searches using Xe TPCs---LZ \cite{dobsonLZ}, 
PANDA, and XENON-nT 
\cite{grandiXENON}
---all aim to reach ~10$^{-48}$ cm$^2$ scale sensitivity at 30 GeV dark matter mass.  Searches at colliders for scalar dark matter expect to reach relevant  sensitivities for the low-mass range below 10~GeV,
discussed further in Sec.~\ref{subsec:complementarity_collider}. The DarkSide-20k experiment expects to reach the 10$^{-47}$ cm$^2$ scale at 1 TeV.  Long-term future searches using Xe (DARWIN) and Ar (ARGO) project reaching beyond 10$^{-48}$ cm$^2$ in the next decade. For spin-dependent interactions, near-term future experiments using Xe and CF$_3$ targets project to reach sensitivity to 10$^{-42}$ cm$^2$ WIMP-neutron
\cite{grandiXENON} 
and WIMP-proton cross sections, at $\sim$\,50 GeV.
\cite{sigmaSD-prospects}.
At low mass (around 1 to 10 GeV), solid state experiments, e.g.\ SuperCDMS, expect to achieve 10$^{-42}$ cm$^2$ cross section reach on a 5 year time scale.

Major challenges to future direct detection experiments come from: (a)  $\nu-e$ and coherent elastic scattering backgrounds from solar and atmospheric neutrinos, which is known as the ``neutrino floor'' and shown in the grey hatched region in Fig.~\ref{fig:sensitivity}; (b) neutrino flux uncertainties on these backgrounds; and, (c) technology scaling to increase in mass over current searches by factors of 10 or more whilst improving background rejection and lowering radioactivity.

In consideration of the strong synergy between direct dark matter detection and the programme for its production and discovery in high-energy collisions at accelerators as well as in accelerator-based fixed target experiments, discussions at the Open Symposium in Granada highlighted that CERN's support for selected direct dark matter search programmes that can take critical advantage of technology developed at CERN can deliver a decisive boost of their sensitivity.

\subsection{Indirect detection}

Indirect astrophysical searches for the annihilation products of dark matter (namely gamma-rays, neutrinos, antimatter) provide important and complementary constraints on DM models that are searched for at particle colliders and fixed-target/beam-dump experiments, as well as on models with axion-like particles.   Annihilation searches are sensitive to the thermal cross-section for DM masses close  to 100 GeV (depending on the channel), with prospects to reach 10 TeV within less than ten years.  No conclusive signals have been found so far.

\begin{figure}
    \centering
    \includegraphics[width=0.8\linewidth]{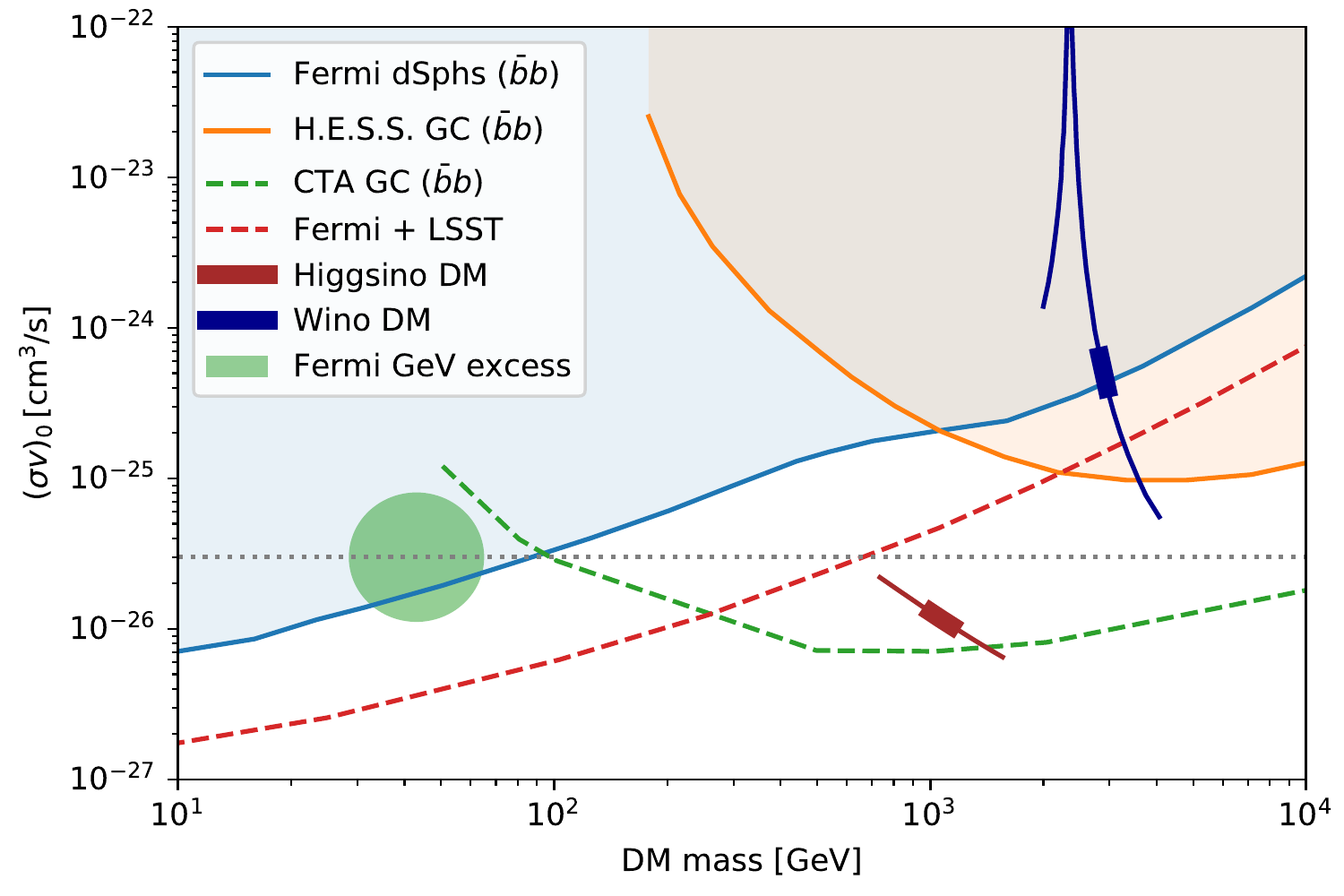}
    \caption{Current constraints on the DM self-annihilation cross section into $\bar bb$ from Fermi LAT (dSph)~\cite{Albert2017-hn} and H.E.S.S.~Galactic Centre (GC)~\cite{Abdallah:2016ygi}, and expected future reach with the CTA and additional dwarfs found by LSST~\cite{Drlica-Wagner:2019xan}. 
    Also shown as red (blue) lines the annihilation cross sections for pure Higgsino (Wino) DM in the vecinity of DM masses yielding the correct relic density (the thick regions indicating the correct relic density, from~\cite{Ibe:2015tma}), as well as the parameter range preferred by the Fermi GeV excess~\cite{Ackermann:2017gx, Albert2017-hn}. The H.E.S.S.~constraints weaken significantly if the DM profile at the Galactic centre is cored, leaving pure Wino DM consistent with H.E.S.S.~and Fermi observations (see text for details).}
    \label{fig:ID}
\end{figure}

The arguably cleanest constraints on DM annihilation in the GeV--TeV mass range come from gamma-ray observations of dwarf spheroidal (dSph) galaxies. Recent analyses (based on a few dozen dSphs) exclude $s$-wave annihilating WIMPs into $\bar b b$ with masses between about 5 and 80 GeV~\cite{Albert2017-hn} (Fig.~\ref{fig:ID}).  Arguably the best example for a DM annihilation signal \emph{candidate} is the so-called `Fermi GeV excess'~\cite{Ackermann:2017gx}, an excess emission of GeV photons in the inner Galaxy.  It is marginally consistent with dSph and antiproton constraints.  Astrophysical interpretations of the excess will be probed with upcoming radio observations (millisecond pulsar searches~\cite{Calore:2015bsx}), while collider experiments can test the dark matter origin.  For instance, if interpreted in terms of 60 GeV dark matter in supersymmetric models (consistent with dSphs, see Fig.~\ref{fig:ID}), the decays of heavy CP-odd and CP-even scalars into $\tau$-pair provides a possible target for the HL-LHC~\cite{Carena:2019pwq}.  In a simple model with a scalar mediator and fermionic Dirac DM with $\bar bb$ annihilation channel, effects on the Higgs signal strength and exotic Higgs decay can be probed with prospective future colliders such as  ILC and FCC-hh~\cite{Kim2018-yk}.  Furthermore, DM annihilation into $\bar bb$ or other hadronic final states contributes significantly to the cosmic-ray antinuclei flux observed at Earth~\cite{Giesen:2015ufa}. Future probes of DM annihilation into hadronic final states (sensitive to the Fermi GeV excess), will come from anti-deuteron measurements with the balloon experiment GAPS (around 2021~\cite{Aramaki2016-to}) and AMS-02.

The future LSST has the potential to discover hundreds of additional dSphs~\cite{Drlica-Wagner:2019xan}, which together with Fermi LAT data can improve limits from dSph galaxies by a factor of around five.  The upcoming Cherenkov Telescope Array (CTA) is expected to strengthen current H.E.S.S.\ constraints by a factor of about ten~\cite{Drlica-Wagner2019-nr, Carr2016-vl}, see Fig.~\ref{fig:ID}.
Furthermore, neutrino observations of the Sun with IceCube/ANTARES ~\cite{Adrian-Martinez2016-yb, IceCube_Collaboration2016-fp} 
provide in many cases the most stringent constraints on the spin-dependent WIMP-proton cross section.  The ORCA detector of KM3NeT and the IceCube Upgrade are expected to further probe spin-dependent cross sections down to $3\times10^{-5}\,\text{pb}$, for DM masses below $\sim 100$ GeV.
These ranges will also be probed by upcoming direct detection experiments~\cite{sigmaSD-prospects}.

\section{Dark matter and Dark sectors at Colliders}
\label{sec:DM-Colliders}

\subsection{Overview}

Present and prospective future colliders offer a unique opportunity to create DM in the laboratory. 
DM can be searched for at colliders when produced directly in beam-beam collisions in association with other SM particle(s), or in the decays of SM particles or yet undiscovered BSM states. 
In all cases the DM signal will consist of significant missing transverse energy (in addition to that accounted for by standard neutrinos) plus highly energetic SM objects (for example jet(s), a $Z$-boson, a Higgs boson or a photon), or more complex SM final states in case of cascade decays. A discussion of the discovery potential for a few  benchmark models used for DM  searches at high-energy colliders, as well as the visible decays of the DM mediators, can be found in Chapter~\ref{chap:bsm}.
Moreover, thermally produced light DM particles typically  call for different types of light DS mediators that couple to the SM through portals. 
Summary plots of the complementarity among many different accelarator-based experiments are shown in Sect.~\ref{sec:BSM-FIPs}.

\subsection{Complementarity of high-energy collider results with Direct and Indirect Detection}
\label{subsec:complementarity_collider}

The discovery of DM at direct and indirect detection experiments is necessary to ascertain the cosmological connection of a collider discovery, which in turn could provide information on the nature of the DM--SM interaction. 
Collider results make no assumptions on the thermal history of the DM candidate that is produced via the mediator particle considered as benchmark. In some regions of the parameter space covered by collider searches, a non-standard cosmology is needed to achieve the observed relic density.

The comparison of sensitivities across DD, ID and collider experiments is possible within a given theoretical framework, and strongly depends on the choices of model and parameters. 
These comparisons nevertheless give an idea of the potential parameter space for a DM discovery if DM is realised in one of those example models. 

Firstly, this section discusses the complementarity between the collider sensitivity to Wino and Higgsino, as shown in Chapter~\ref{chap:bsm}, and the ID projection for the same scenarios.
The comparison of DD and collider sensitivity for a Higgs portal model~\cite{Patt:2006fw,Djouadi:2011aa} is then shown, considering the results of the Higgs decays into invisible particles from Chapter~\ref{chap:ew}. Finally, the scalar benchmark model that has been discussed in Chapter~\ref{chap:bsm} is considered for comparison of colliders and direct detection experiments,  and a version of the same model with pseudoscalar couplings is used for comparison of colliders and ID  experiments.

\subsubsection*{Wino and Higgsino}

The complementarity of collider and indirect detection searches within the Wino and Higgsino dark matter models is most evident at relatively high mass (see Fig.~\ref{fig:ID}). Above a DM mass of $\sim$ 0.5--1 TeV, ID provides strong constraints on DM annihilation, almost reaching the typical $s$-channel thermal cross section~\cite{Abdallah:2016ygi}, via observations of the inner Galaxy with the H.E.S.S.~Cherenkov Telescope. Pure Wino DM with a mass around 2.9 TeV~\cite{Beneke:2016ync}, which would account for all of dark matter, is constrained by these searches, as well as searches for $\gamma\gamma$ final states, if the dark matter distribution at our Galactic centre has a cusp-like profile~\cite{Cohen2013-ah, Cabrera-Catalan2015-os, Hryczuk2019-mv, Rinchiuso:2018ajn}.   However, if the distribution is more cored, the constraints become weaker (for details see supplemental note~\cite{DM-note}). As shown in Chapter~\ref{chap:bsm}, Wino DM can also be probed with the FCC-hh~\cite{CidVidal:2018eel} and LE-FCC, while slightly lighter Wino-like dark matter would be already accessible for HE-LHC~\cite{CidVidal:2018eel} and CLIC$_{3000}$~\cite{deBlas:2018mhx}.
The future Cherenkov Telescope Array (CTA) will improve on current H.E.S.S.~constraints by another order of magnitude~\cite{Drlica-Wagner2019-nr, Carr2016-vl, Silverwood2015-ff}.  In the case of a discovery, the combination of CTA and collider results would allow to determine the dark matter distribution at the Galactic centre, with important implications for the formation history and evolution of galaxies.  CTA will be furthermore sensitive to Higgsino DM with a mass about $1.1$ TeV~\cite{Cabrera-Catalan2015-os, Krall:2017xij, Hryczuk2019-mv}, which is also a target of the FCC-hh~\cite{CidVidal:2018eel} and CLIC$_{3000}$~\cite{deBlas:2018mhx} as shown in Chapter~\ref{chap:bsm}. Interestingly, the cross section of many of these models would be lower than the neutrino background for direct searches, emphasizing the need for different experimental approaches.

\subsubsection*{Higgs portal, scalar and pseudoscalar mediator models}

\begin{figure}[!htbp]
    \centering
    \includegraphics[width=0.78\textwidth]{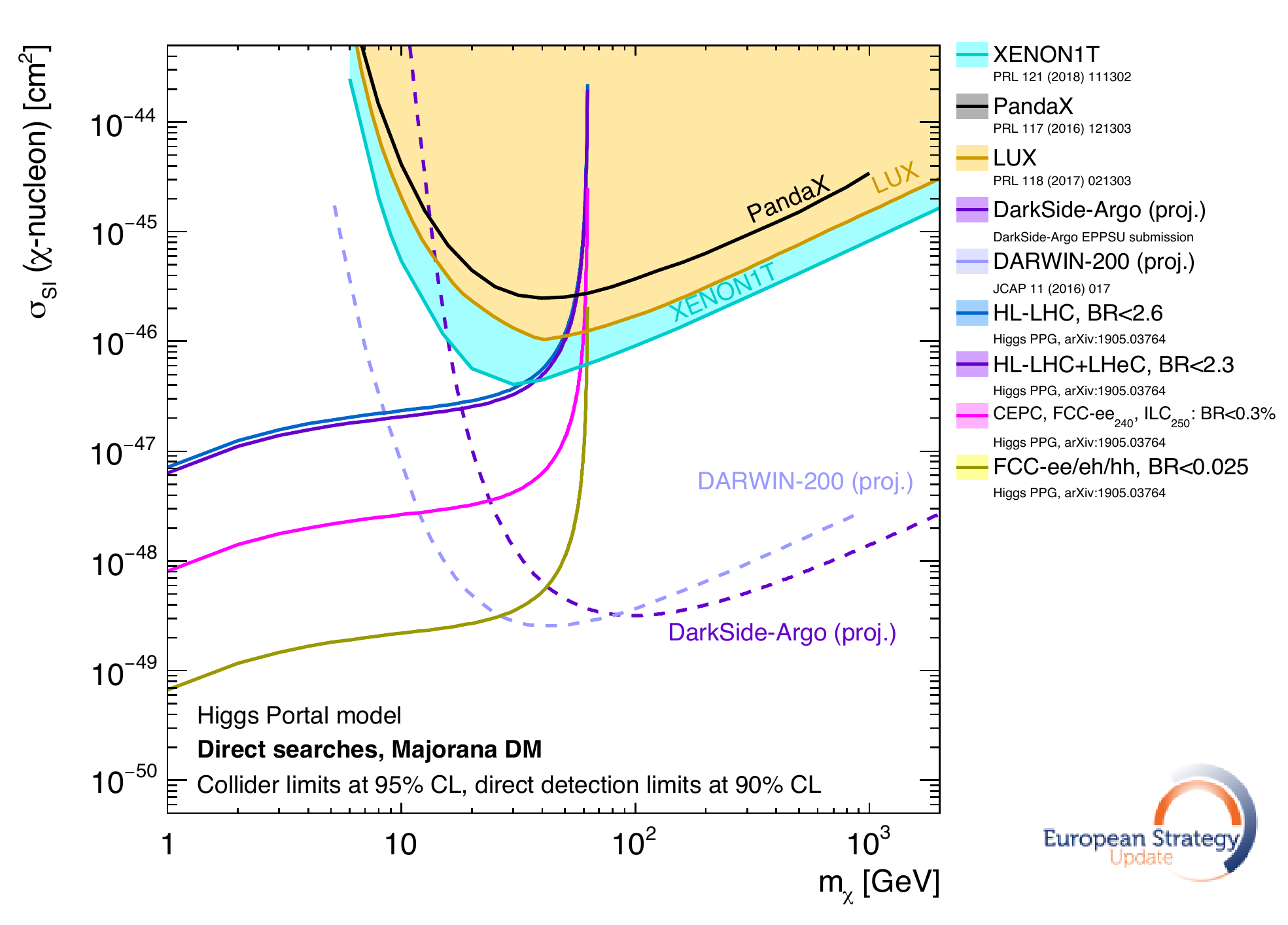}\\
    \includegraphics[width=0.78\textwidth]{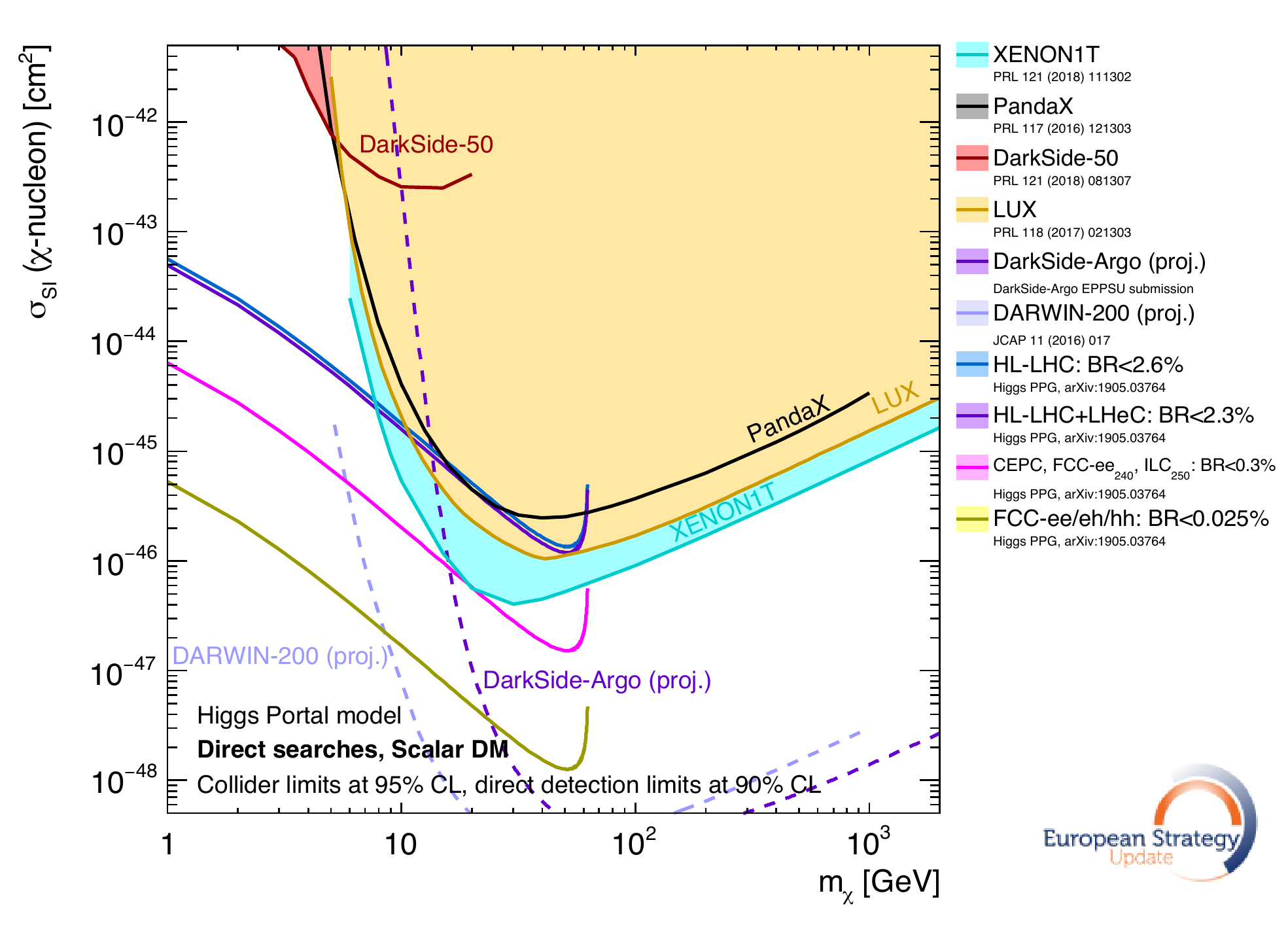}
    \caption{Comparison of projected limits from future colliders (direct searches for invisible decays of the Higgs boson) with constraints from current and future direct detection experiments on the spin-independent WIMP--nucleon scattering cross section for a simplified model with the Higgs boson decaying to invisible  (DM) particles, either Majorana (top) or scalar (bottom). Collider limits are shown at 95\% CL and direct detection limits at 90\% CL. 
 Collider searches and DD experiments exclude the areas above the curves. }
    \label{fig:Higgs_CollidersDD}
\end{figure}

Direct detection reach and future hadron collider reach for searches for invisible decays of the Higgs
are shown in Fig.~\ref{fig:Higgs_CollidersDD}, within the context of a Higgs portal model~\cite{Patt:2006fw,Djouadi:2011aa}. Collider results from Chapter~\ref{chap:ew} are translated to the nucleon-DM scattering plane using the procedure and factors in~\cite{Fox:2011pm, Hoferichter:2017olk}. 
The DM candidates considered as an example in these Higgs portal benchmarks are either a scalar, or a Majorana fermion.

\begin{figure}[!htbp]
    \centering
          \includegraphics[width=0.78\textwidth]{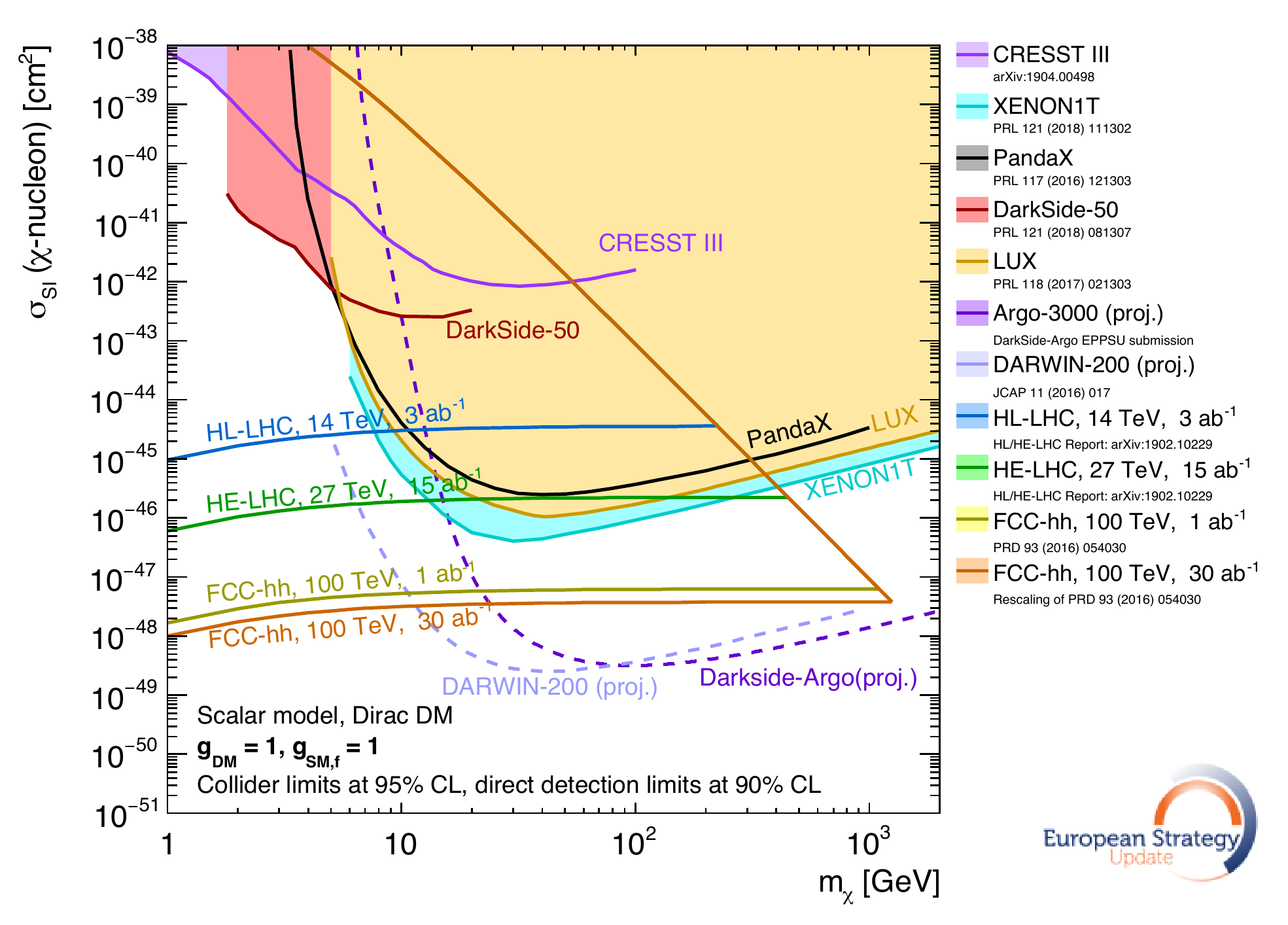}
        \includegraphics[width=0.78\textwidth]{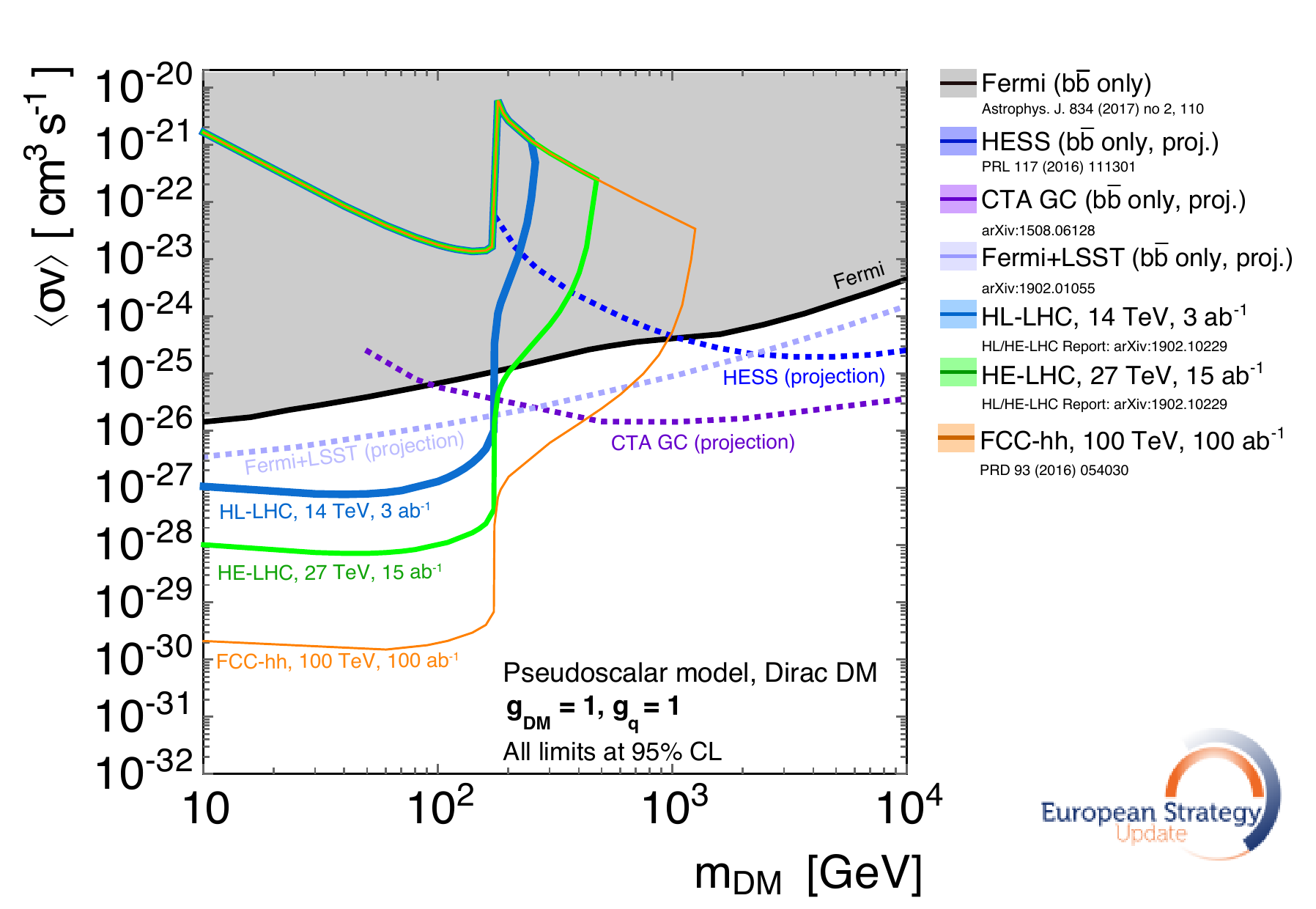}
    \caption{Top: Comparison of projected limits from future colliders with constraints from current and future DD experiments on the spin-independent WIMP--nucleon scattering cross section in the context of a simplified model where a scalar particle with unit couplings mediates the interaction between SM fermions and Dirac fermionic DM. 
    Collider limits are shown at 95\% CL and direct detection limits at 90\% CL. 
    Bottom: comparison of a selection of projected limits from future colliders with constraints from current and future indirect detection experiments in the context of a simplified model where a pseudoscalar particle with unit couplings mediates the interaction between SM fermions and Dirac fermionic DM. All limits are shown at 95\% CL.
    In both figures, collider searches and DD experiments exclude the areas above the curves~\cite{DM_IndirectDetectionPlots, DM_DirectDetectionPlots}.}
    \label{fig:Scalar_CollidersDD}
\end{figure}

The comparison of the reach of DD, ID and future hadron colliders for the benchmark models of a scalar or pseudoscalar mediator decaying into Dirac DM use the procedure in~\cite{Boveia:2016mrp}, with the coupling choices made in Chapter~\ref{chap:bsm}. The results are shown in Fig.~\ref{fig:Scalar_CollidersDD}. 
For the ID plot, the procedure in~\cite{Boveia:2016mrp} maps a velocity-averaged annihilation cross section to multiple values of LHC mediator mass--DM mass pairs.
Fig.~\ref{fig:Scalar_CollidersDD} considers that if a mapped velocity-averaged annihilation cross section value is obtained from a mediator mass -- DM mass pair reachable by the LHC, that cross section value is considered within LHC reach. 
Such a procedure highlights the potential of a 
 situation where a hint for invisible particles at colliders can guide ID searches and vice versa. 
The bounds from indirect detection experiment shown in Fig.~\ref{fig:Scalar_CollidersDD} only consider annihilation into $b$-quarks, while collider plots consider all channels. A full treatment would involve the calculation described in \cite{Carpenter:2016thc}. 

From Figs.~\ref{fig:Higgs_CollidersDD} and \ref{fig:Scalar_CollidersDD} (top), one can observe that above $\sim$10 GeV DM mass, next-decade DD experiments are more sensitive than future colliders to DM signals. Future collider experiments, instead, are well  suited to explore models with mediators decaying to lighter DM candidates as well as possibly reaching DM masses up to a TeV from the decays of multi-TeV-mass mediators.
From Fig.~\ref{fig:Scalar_CollidersDD} (bottom), it follows that collider searches have better sensitivity for DM masses below the top mass, while ID searches are more powerful for higher DM masses.  

Notably, it is in the intermediate DM mass range (roughly between 10 GeV and 1 TeV for the scalar/pseudoscalar cases, and between 10 GeV and half the Higgs mass for the Higgs portal models considered) that the combination of future high-energy colliders, direct detection and indirect detection programmes will complement each other and shed light on the nature of a DM candidate at reach in the next decades. 




\section{DM and DS at beam-dump and fixed-target experiments}
\label{sec:DM-BDandFT}

In this section we  summarize the physics case for $\sim$ few keV--GeV scale DM and other HS particles and identify key targets of opportunity for future experimental efforts. 

\subsection{Theoretical Motivation}
The lighter half of the thermal DM  mass range, few keV--GeV, is currently underexplored and typically inaccessible using traditional WIMP detection strategies \cite{Battaglieri:2017aum}; new techniques are necessary to comprehensively probe the predictive models in this window.
There are several key differences between traditional WIMPs and sub-GeV LDM candidates. Unlike WIMPs, which can carry electroweak charge, LDM must be neutral under all SM gauge interactions---otherwise it would have been discovered at LEP.\footnote{If $<$ GeV  DM were electrically millicharged, the size of the coupling necessary to avoid cosmological  overproduction is excluded by a variety of experiments \cite{Davidson:2000hf, Berlin:2018sjs}; additional forces are still necessary even if light DM is also millicharged.}
Furthermore, under the assumption of standard cosmology, in order to achieve the observed relic density for LDM particles, such models require couplings to the SM sector that are much larger than $G_{\rm F}$ \cite{Lee:1977ua}. This implies the existence of a light 
mediator between the SM and a new dark sector. If the mediators 
are produced on-shell in the laboratory, they can decay either to LDM or to SM particles through the  portal couplings (see Sect.~\ref{sec:BSM-FIPs} for a summary). Improving experimental sensitivity to both LDM and SM decay channels has important implications for different theoretical scenarios.

No single experiment or approach is sufficient to cover the vast parameter space that dark sectors 
can occupy \cite{Izaguirre:2015yja,Battaglieri:2017aum}. A comprehensive dark sector programme therefore relies on a complementary set of experiments and techniques that span the multitude of dark sector signatures and collectively cover the broad mass and coupling range:

\noindent{\bf Mediators decaying to LDM:} The most predictive class of thermal LDM models involves mediators that are heavier than  (twice) the LDM candidate and, therefore, decay to LDM pairs when produced on shell. Because the mediator is heavier than the LDM, the only kinematically accessible annihilation\footnote{When DM and mediator masses
are nearly degenerate, ``forbidden'' dark matter \cite{DAgnolo:2015ujb} 
serves as a notable counter-example to this claim.} channel for  early-Universe freeze-out is the $s$-channel {\it DM~DM~$\rightarrow$ SM SM} reaction which depends on the mediator's coupling to both dark and visible matter. 
Thus, the SM-mediator coupling occuring through a portal must have a minimum value to realise a thermal annihilation cross section, so mediators that decay to LDM feature predictive experimental-sensitivity targets; improving coverage to such mediators by 2--3 orders of magnitude in cross section can convincingly discover or falsify a broad class of thermal DM candidates whose relic density arise from annihilation directly into SM particles in the early Universe \cite{Battaglieri:2017aum,Izaguirre:2015yja}. 

\noindent{\bf Mediators decaying to SM particles:} If the mediator is {\it lighter} than the thermal LDM candidate, freeze-out occurs through {\it DM DM $\rightarrow$ mediator mediator} annihilation reactions, 
which are independent of the SM-mediator coupling. In this regime, the mediator decays  to SM particles and motivates searches for new forces. However, in the event of a discovery, there is no necessary connection to thermal LDM; new particles with SM decays can exist independently of any DM assumptions. Indeed, beyond the thermal DM motivation, such particles arise in various new physics scenarios related to  leptogenesis, electroweak baryogenesis, 
the electroweak hierarchy problem, neutrino mass generation, and non-thermal DM scenarios which do not feature predictive experimental sensitivity targets, but are nonetheless well-motivated theoretically.
In many of these other scenarios, the SM coupling is required to be very small, so if the mediator is a long-lived particle (LLP), it can decay at a macroscopic distance away from its production point, in an accelerator-based experiment (see Chapter~\ref{chap:bsm}).  
\subsection{Experimental techniques}
As the mediators of the dark sector couple weakly to the SM, facilities with high energy and intensity beams impinging on dense targets are essential in order to maximize the experimental sensitivity to such models. Facilities at CERN's SPS or SLAC's LCLS-II, will have unprecedented sensitivity to a wide range of dark sector models and offer unique opportunities to explore large portions of the relevant parameter space. The two main experimental techniques are beam-dump and fixed-target missing-energy/momentum searches. \\

\noindent {\bf Beam dump:} 
A high-intensity beam of relativistic particles (electrons, protons or muons) impinges on a thick, passive target. The detector, consisting of spectrometers and calorimeter systems are installed  10s--100s of metres downstream of the target. Shielding is placed between the target and detector to absorb SM particles. Signal candidates are reconstructed through electromagnetic energy depositions in the detector in time-coincidence with beam bunches.
If DS mediators are produced in the target (e.g.\ via meson decays,  Drell-Yan or bremsstrahlung), a beam-dump signal can arise in two main ways:
\begin{enumerate}
\item If the mediator decays to LDM particles, the LDM can pass through the shielding material, enter the detector, and scatter off detector atoms, whose relativistic recoils are detectable \cite{deNiverville:2011it,Izaguirre:2013uxa}. 
This search strategy is akin to direct detection with two important differences a) the DM is produced directly at the target (not from the halo) and  b) the DM is relativistic, so even light particle masses ($\ll$GeV) can induce highly energetic target-particle recoils well above detection thresholds. Since mediator production  and the LDM scattering both scale as the SM-mediator coupling squared, the signal is proportional to the fourth power of this coupling.
\item If, instead, the mediator decays to SM particles, beam-dump signals arise from the decay products of a Long-Lived mediator Particle (LLP) that decays in between the shielding and the downstream spectrometer. In this scenario, the signal rate depends on both mediator production (which scales as the SM-mediator coupling squared) and on its decay probability in the region of interest (which is also proportional to the SM-mediator coupling squared), so in total, this signal is proportional to the fourth power of the mediator-SM coupling. 
\end{enumerate}

The SHiP experiment~\cite{Anelli:2015pba,Alekhin:2015byh} is an example of a beam-dump experiment with a dual spectrometer to search for scenarios 1) and 2) simultaneously.

\noindent {\bf Missing-energy/momentum:} In this setup, a relativistic beam of particles (electrons or muons) also strikes a fixed-target, but the energy of each beam particle is measured both before and after it passes through the target. If the beam particle scatters inside the target and produces a new mediator which decays to LDM, the beam particle typically emerges with a much lower energy and there is no other SM particle production detected \cite{Izaguirre:2014bca,Banerjee:2016tad,Gninenko:2014pea,Kahn:2018cqs}. Both ECAL and HCAL systems are installed downstream of the target to veto on any other energy deposition aside from the recoiling beam particle. Since this process depends only on the production of the mediator and does not require the LDM to interact with a detector, the signal rate scales as the SM-mediator coupling squared. A mature, dedicated missing energy/momentum search with $\sim 10^{16}$ particles on target (e.g.\ LDMX \cite{Akesson:2018vlm,Berlin:2018bsc} or NA64$^{++}$~\cite{Beacham:2019nyx}, the upgraded version of NA64 \cite{Banerjee:2016tad}) offers a unique path towards achieving experimental sensitivity to vast spectra of 
predictive LDM thermal targets. Although this technique is optimized to search for mediators that decay to LDM, it can also offer some sensitivity to mediator decays into SM particles, but this reach is typically subdominant to that of beam-dump experiments. 

The ability to categorically determine that a potential signal is not an instrumental effect or an unaccounted background is paramount in searches for rare signatures. Beam-dump and fixed-target experiments are unique in that they enable the design of detectors with veto systems, spectrometer tracking, calorimetry and veto systems, while maximizing the geometrical acceptance of HS signatures. Therefore, experiments like LDMX and SHiP will be equipped with redundant systems for suppressing backgrounds and will  be able to define control regions in order to check the level of agreement with expected background rates. Furthermore, these experiments can provide independent confirmation of a potential signal originated from the target, by reconstructing the momenta of the decay products of the mediator (in the case of an LLP at a beam dump experiment) or of the beam particle itself (if the mediator decays to DM at a missing energy experiment), which can further validate the presence of an HS signature.  


\subsection{Recent/current experiments}
Beam-dump and missing-energy experiments are already underway at various facilities throughout the world. 
In 2018 the MiniBooNE experiment at Fermilab reported new limits on LDM produced in a dedicated beam dump mode with $10^{20}$ protons on target \cite{Aguilar-Arevalo:2018wea}, which improved existing limits on such models; future improvements exploiting the Fermilab Booster neutrino beam are possible but are not expected to cover thermal  sensitivity targets. The NA64 experiment at CERN recently set limits on mediators decaying to LDM with the missing-energy technique with $10^{10}$ electrons on target \cite{Banerjee:2017hhz}, improving coverage and superseding limits set by the NA62 collaboration that looked at LDM decays of dark photons from $\pi^0$ decays, but does not yet cover thermal sensitivity milestones \cite{CortinaGil:2019nuo}. The NA62 experiment using the SPS beam recently placed limits on neutrino-portal mediators produced in kaon decays improving on existing limits for HNL masses below the kaon mass. The NA62 experiment will further increase its sensitivity to DS by operating in beam-dump mode during parts of Run\,3 of the LHC and beyond (NA62$^{++}$). 

However, a dedicated beam-dump facility and fixed-target experiments are required in order to maximise the sensivity to couplings in the MeV--GeV mass range, and to broadly explore a vast range of HS portals.

\subsection{Future directions}
Extensive studies have been performed for a new general-purpose beam-dump facility (BDF) at the CERN SPS accelerator aimed at exploring the domain of DS models and performing measurements involving $\tau$ neutrinos and Lepton Flavour Violating decays of $\tau$ leptons. The high intensity of the SPS 400~GeV beam is essential to probe a wide variety of models containing long-lived exotic particles with masses below $\mathcal{O}(10)$\,GeV. A new beamline, target complex and experimental hall at the North Area of CERN is foreseen to deliver $4\times10^{13}$ protons on target during 1~second spills using slow extraction of the beam~\cite{Ahdida:2650896}. The Search for Hidden Particles (SHiP) experiment~[ID12] is a major use-case of the BDF. 
The SHiP collaboration's Technical Proposal~\cite{Anelli:2015pba} 
underwent successful scientific review by the SPSC in 2016.
Since then, the CERN BDF team and the SHiP collaboration recently completed a first Comprehensive Design Study demonstrating good control over the slow extraction and the design of the target complex~\cite{Ahdida:2650896}, and good control of the expected physics performance based on the active muon shield and SHiP sub-detectors~\cite{Ahdida:2654870}. A Conceptual Design Report is foreseen by the end of 2019 at which point the project will be mature enough for an implementation decision to be made. The first physics run could take place during Run\,4 of the LHC, starting in 2026.

The SLAC LCLS-II beamline is currently being built to run at 4~GeV with first light by 2020 and with a future upgrade for an 8~GeV running. In early 2019, the LDMX collaboration submitted an experimental proposal to the US Department of Energy Office of Science for detector R\&D, and SLAC management has also expressed support for a dedicated extraction of the LCLS-II beam for dark sector experiments. If LDMX is realised at SLAC, an initial phase of running will deliver $10^{14}$ electrons on target at 4~GeV and future improvements could yield $10^{16}$ electrons on target running at 8~GeV beam energy. 

The NA64 experiment~[ID9] has performed a search for dark photons decaying to LDM with sample of $3\times10^{11}$ electrons on target (EOT) and a beam energy of 100~GeV delivered by the H4 beamline at the SPS. The detector is planned to be upgraded (NA64$^{++}$) in order to cope with a high-intensity beam. This involves replacing the ECAL electronics, the addition of a zero-degree HCAL, and the upgrade of the data acquisition system. Using this setup, NA64$^{++}$ aims at collecting $3\times10^{12}$\,EOT within a 6 to 8 month period. There is also a proposal to make use of the muon beam that currently serves the COMPASS experiment. This run could start as early as 2022 aiming at collecting $5\times10^{13}$ muons on target. This would complement searches for DS models that predominantly couple to the second generation. 

It has been proposed to use the SPS to accelerate and deliver 16~GeV electrons (eSPS) for the LDMX~[ID36] experiment using slow extraction at a rate of $10^{16}$ electrons per year. This can be achieved by leveraging CLIC technology to  accelerate electrons to 3.5~GeV before injecting into the eSPS. Initial studies demonstrate that such an operation is feasible, and would require reserving 30\% of the SPS duty cycle towards this project \cite{Akesson:2018yrp}.

REDTOP~[ID28] is a fixed target experiment to study rare $\eta^{(}$$'^{)}$ meson decays. As part of its research programme, it will search for dark photons and ALPs covering a selective but unique parameter space in  couplings and masses. The REDTOP experiment was originally proposed to be hosted at FNAL. Preliminary studies show that it could also be hosted at CERN, though the impact on the rest of the CERN physics programme could be significant.

AWAKE$^{++}$ is an R\&D programme for electron acceleration using plasma cells excited by proton bunches. This is one of the most promising next generation accelerator technologies. Since higher gradient of the electric field can be obtained, electrons can be accelerated in a shorter distance. It has been proposed to use the electron beam in an electron beam-dump experiment to search for dark photons through their decays to electrons and muons.

\begin{figure}
    \centering
    \includegraphics[width=0.85\textwidth]{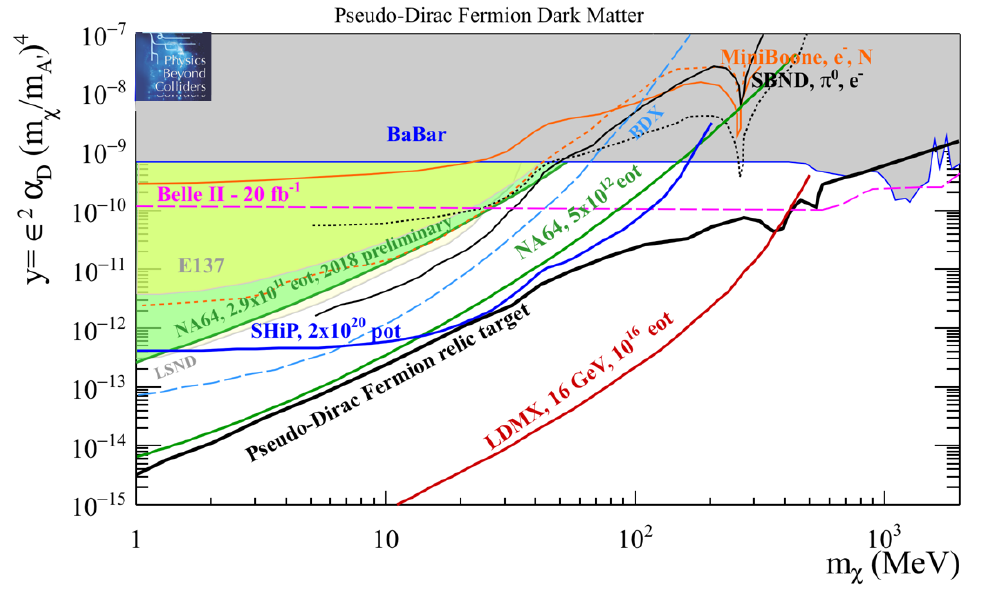}
    \includegraphics[width=0.85\textwidth]{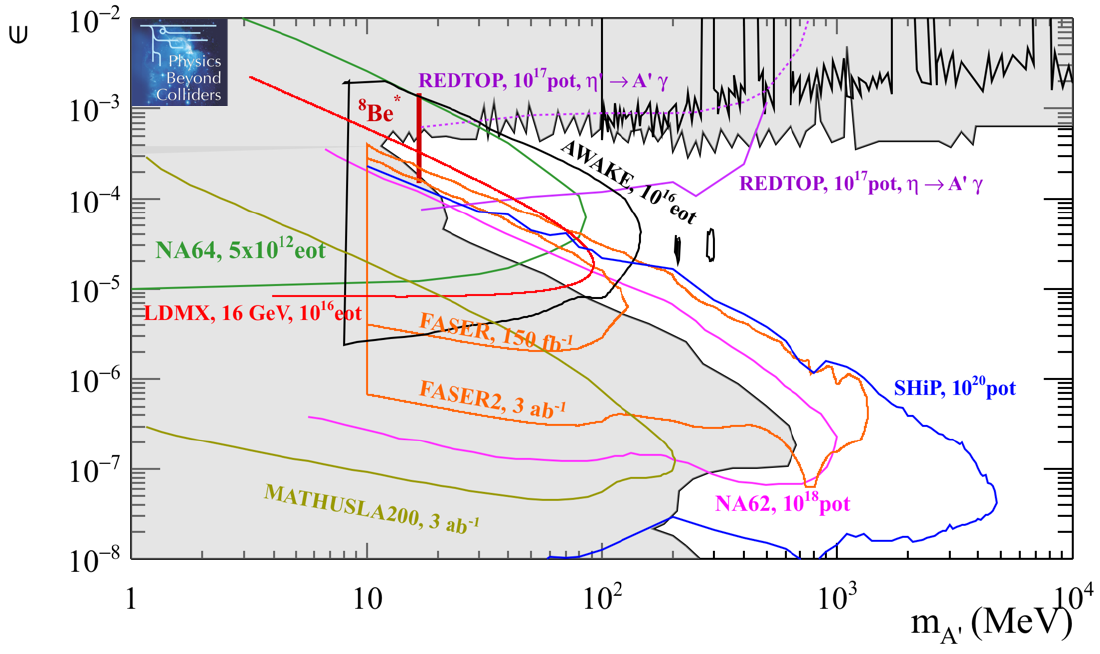}
    \caption{Current limits and expected sensitivities of proposed experiments relevant to this section for dark photon mediators decaying to LDM particles (top) and SM particles (bottom). Figures are from the PBC report~\cite{Beacham:2019nyx}. The top figure  assumes  $\alpha_D = 0.1$ and $m_{A'}/m_\chi = 3$.}
    \label{fig:sec3_sensitivity}
\end{figure}

Figure~\ref{fig:sec3_sensitivity} shows
the current limits and expected sensitivities of a comprehensive list of experiments for dark photon mediators decaying to LDM and SM particles as a function of $\epsilon$ (the mixing between the photon mediator and the SM photon that defines the SM-mediator coupling) and the dark matter or mediator mass, respectively. In the case of decay into LDM the figure shows results for a fixed value of the mediator-DM coupling, $\alpha_D$, and a fixed ratio of the mediator-DM masses, $m_{A'}/m_\chi$. 

\begin{figure}
   \centering
    \includegraphics[width=0.7\textwidth]{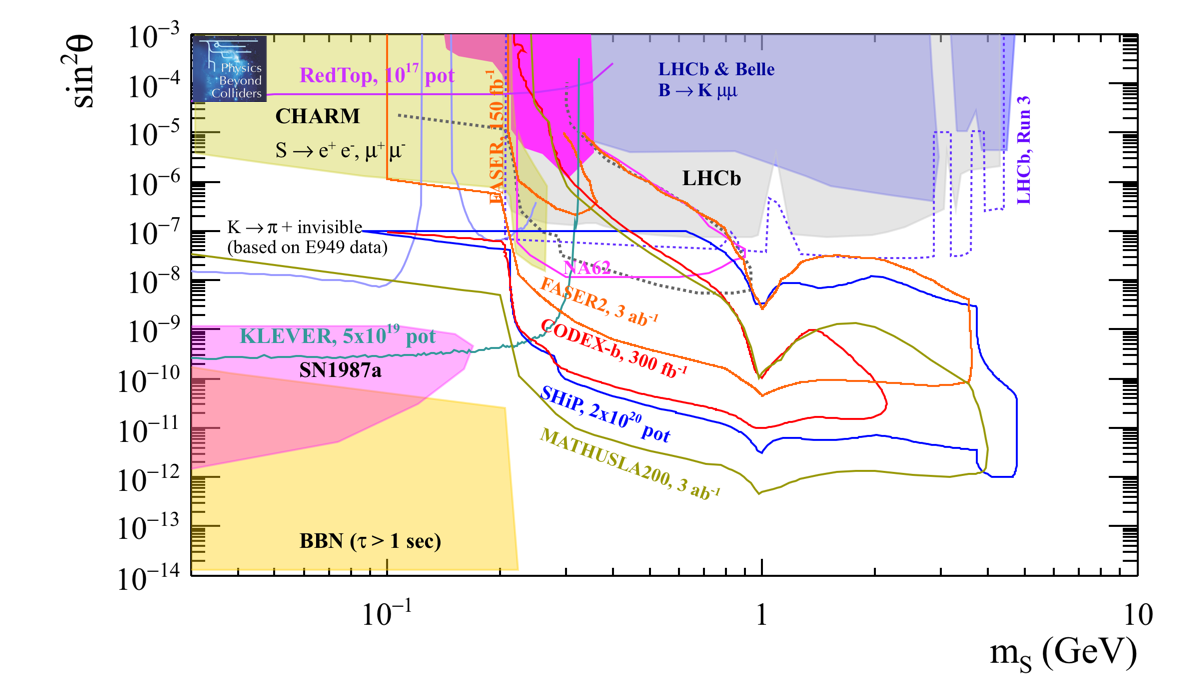}\\
    \includegraphics[width=0.7\textwidth]{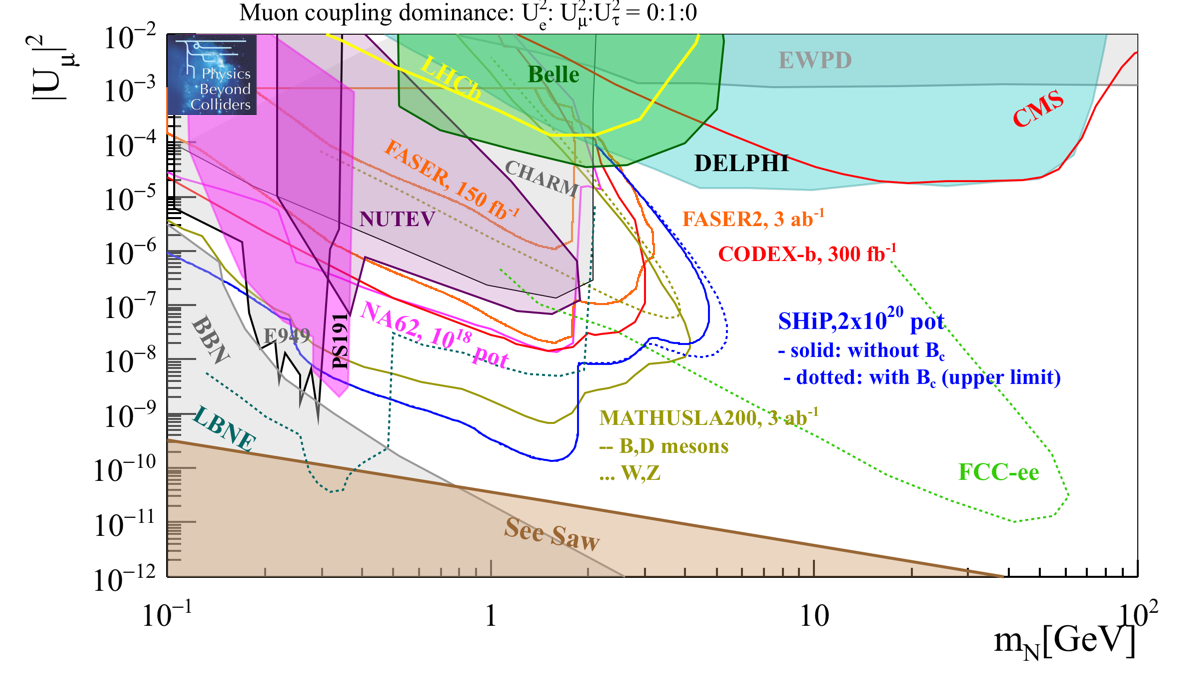} \\
    \includegraphics[width=0.7\textwidth]{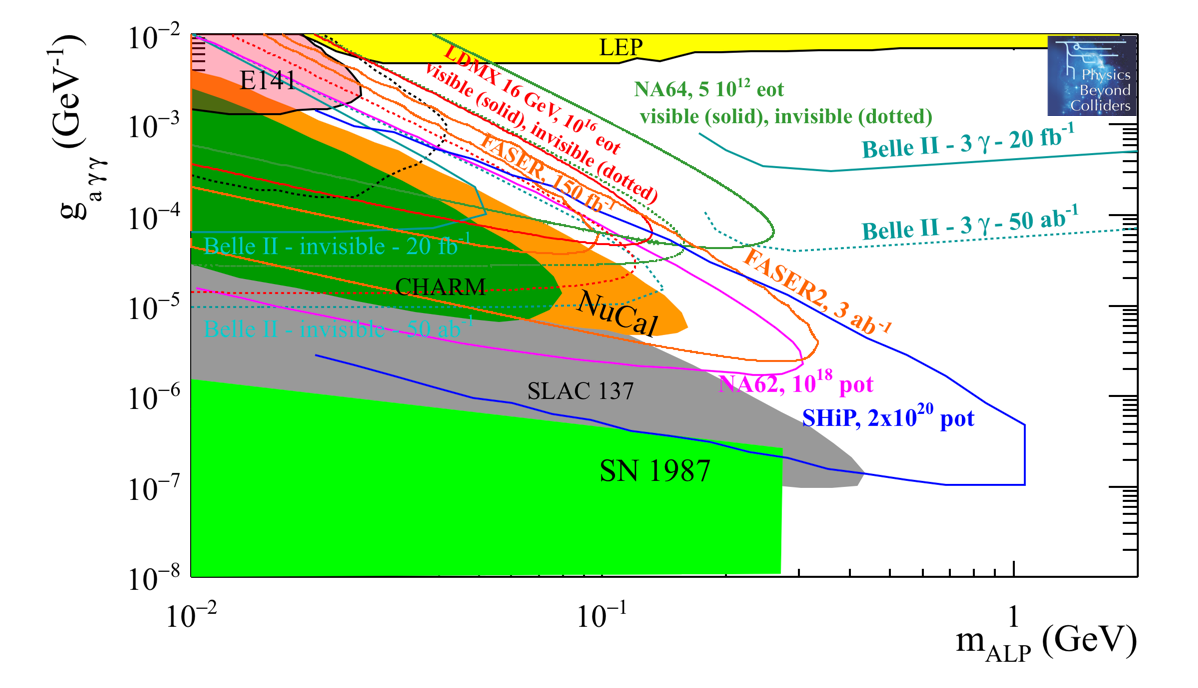}
   \caption{ Current limits and expected sensitivities of proposed accelerator-based  experiments 
   for a scalar particle with a Higgs portal (top figure), a heavy neutral lepton (HLN)  with a neutrino portal (middle figure) and an ALP pseudsoscalar that couples exclusively to photons (bottom figure). All figures are shown as a  function of the mediator's mass and the relevant parameter defining the mediator-SM portal mixing.
   Here  the scalar is assumed to decay to SM particles though its mixing with the Higgs. 
   The neutrino portal example   shows the mixing matrix element $|U_\mu|^2$ plotted against the HLN mass and assumes the latter HLN mixes only with muon flavoured neutrinos. All plots in this figure are taken from the PBC report~\cite{Beacham:2019nyx}.
  }
   \label{fig:sec3_LLP}
\end{figure}

Figure~\ref{fig:sec3_LLP} shows the expected sensitivities of a comprehensive list of experiments for different plausible DS mediators decaying into SM particles through a portal. This includes a scalar particle with a Higgs-mixing $\sin^2 \theta$ (Higgs portal) and zero quartic self coupling (top figure), a heavy neutral lepton (HNL) mixing with active neutrinos (lepton portal, middle figure) and an ALP pseudsoscalar that couples exclusively to photons (bottom figure). All figures are depicted  as a function of the mediator's mass and the relevant parameter defining the mediator-SM portal mixing.
In all portals, where the mediator mass lies in the MeV to GeV range, experiments exploiting the SPS and a future Beam Dump Facility have the largest reach.
A comprehensive comparison for all portals and a discussion on the various assumptions can be found in the supplemental note~\cite{DM-note}, as well as the PBC report~\cite{Beacham:2019nyx}. 
Figure~\ref{fig:sec3_LLP} highlights the reach for the mass region below about 10 GeV; in Chapter~\ref{chap:bsm} similar figures highlight the reach of complementary accelerator-based experiments for larger values of the DS mediator (FIP) masses.

\subsection{Complementarity with direct and indirect detection}
\label{subsec:DM-BDandFT-Complementarity}

In addition to the accelerator programme outlined here, there has been considerable progress developing new techniques for low-threshold direct detection searches, typically involving novel targets including electrons \cite{Essig:2011nj}, semiconductors~\cite{Essig:2015cda}, graphene~\cite{Hochberg:2016ntt}, superconductors \cite{Hochberg:2015fth}, and several others. These techniques are very promising if DM scatters in a velocity-independent, elastic manner. Currently running experiments, including SENSEI \cite{Tiffenberg:2017aac}, DAMIC \cite{Aguilar-Arevalo:2019wdi}, and CRESST \cite{Petricca:2017zdp} 
will reach the scalar LDM thermal sensitivity target for masses below $\sim$\,100\,MeV, as shown in the supporting document~\cite{DM-note}. However, if DM scatters inelastically or with velocity dependence in the non-relativistic limit, these experiments typically do not improve upon existing limits for these models and accelerator searches are essential for reaching important experimental milestones.

Beam-dump experiments will furthermore probe dark matter models with light $\sim$MeV mediators, which, e.g., can induce gamma-ray emission from the Sun, observable with Fermi LAT and water Cherenkov detectors like HAWC~\cite{Arina2018-gl}.  More generally, operators responsible for the annihilation/decay of sub-GeV dark matter particles, which can be explored with proposed future MeV missions like AMEGO~\cite{Moiseev2018-kg}, can potentially lead to observable signatures at beam-dump experiments~\cite{Kumar2018-be}.  Upcoming beam-dump experiments like SHiP or NA62++ can probe the two heavier right-handed neutrinos predicted by the $\nu$MSM models~\cite{Beacham:2019nyx}, which can explain the observed 3.5 keV feature in X-ray data in terms of sterile neutrino dark matter~\cite{Bulbul2014-on, Boyarsky2014-gy}.



\section{Axions and ALPs}
\label{sec:DM-Axions}


Very weakly coupled sub-eV mass particles have become an increasingly attractive option for new physics and a candidate for DM.
Experimentally there is a productive mix of new ideas combined with more mature proposals for medium scale experiments that have significant sensitivity to promising regions in parameter space and that could bring discovery in the next 10 to 20 years.
This section 
is based on community input documents [ID27, ID31, ID42, ID60, ID69, ID112, ID113, ID161] and on the  detailed reports~\cite{Alemany:2019vsk,Beacham:2019nyx,Siemko:2652165}.
It summarizes the physics case for axions, axion-like particles, dark photons and other very light (sub-eV) particles and discusses the current experimental developments, highlighting aspects where strategic support is needed.

\subsection{Theoretical discussion}
Theoretical model building as well as phenomenology (e.g.\ non-thermal dark matter candidates) provides motivation for a variety of different particle types in the sub-eV range. 
The main focus will be on the best motivated candidate, the axion as well as its closest relatives, axion-like particles (ALPs). However, most of the experiments discussed below are also sensitive to other light particles such as dark scalars and dark photons.

Pseudo-Goldstone bosons are a natural realization of physics that is at the same time very weakly coupled, but also exhibits very low-mass particles. The most compelling example is the axion
appearing as a consequence of the Peccei-Quinn solution
to the strong CP problem. It is the pseudo-Goldstone boson of the Peccei-Quinn U(1) symmetry, that is spontaneously broken by a vacuum expectation value at a scale $f_{a}$. As befits the underlying symmetry, all interactions are suppressed by $1/f_{a}$, where
 the decay constant $f_{a}$ is   a free parameter directly linked to an underlying scale of fundamental physics.
Determining $f_{a}$ could therefore give us a clear signal of the scale where a more fundamental completion of the Standard Model takes place.
The mass of the standard QCD axion is tied to its decay constant via $m_A=57.0(7)\,{\rm meV}\,(10^8\,{\rm GeV}/f_a)$,
and therefore, up to ${\mathcal{O}}(1)$ factors, also to its couplings to Standard Model particles. For the example of the two-photon coupling $g_{a\gamma\gamma}\sim \alpha/(4\pi f_{a})$ this is shown in Fig.~\ref{fig:alpsfig} by the light blue band, where its thickness is related to the details of the underlying models.
Terrestrial experiments and astrophysical observations currently restrict $g_{a\gamma\gamma}\lesssim 10^{-10}\,{\rm GeV}^{-1}$ for the standard QCD
axion.

Some models, e.g.~\cite{Agrawal:2017ksf,Gaillard:2018xgk} would allow for the existence of heavier QCD axions with smaller decay constants. 
More general axion-like particles with a more flexible mass/coupling ratio are also possible.
A very interesting region lies at the MeV--GeV mass scale and can be explored at fixed-target facilities as well as collider experiments, as discussed in more detail in the relevant sections.

In the sub-eV region the axion and ALPs are stable on cosmological time scales for $g_{a\gamma\gamma}\lesssim 10^{-10}\,{\rm GeV}^{-1}$,
and are natural dark matter candidates.
Sufficient production of QCD axions in the early Universe to account for the full dark matter content is naturally ensured by the misalignment mechanism for $10^{-16}\,{\rm GeV}^{-1}\lesssim g_{a\gamma\gamma}\lesssim 10^{-13}\,{\rm GeV}^{-1}$. But the formation and decay of cosmological defects, as well as other mechanisms could also give rise to a sufficient amount of cold dark matter for smaller decay constants, populating essentially the entire region of photon axion-couplings not already excluded by astrophysical observations. More general ALPs can populate even larger regions in parameter space.

Additional motivation for ALPs comes from the possible embedding into fundamental extensions of the Standard Model, e.g.\ based on string theory.
Generically, this allows for the existence of 
axions or axion-like particles, where the axion scale can be within reach of near future experiments. 
Moreover, axions and ALPs whose coupling to
photon is in the  $g_{a\gamma\gamma} \sim 10^{-11}{\rm GeV}^{-1}$ region, have been proposed to explain a number of astrophysical anomalies such as several stellar cooling anomalies and an anomalous transparency of the Universe for $\gamma$-rays.

Altogether, the theoretical and phenomenological considerations presented above select two particularly well motivated target regions. First, the QCD axion as the dark matter, in the mass range $0.1\,\mu{\rm eV}-0.1\,{\rm eV}$ corresponding to couplings $10^{-16}\,{\rm GeV}^{-1}\lesssim g_{a\gamma\gamma}\lesssim 10^{-13}\,{\rm GeV}^{-1}$. Second, axions and ALPs with couplings in the region $g_{a\gamma\gamma}\sim (10^{-12}-10^{-10})\,{\rm GeV}^{-1}$, suggested by astrophysical anomalies. As can be seen in Fig.~\ref{fig:alpsfig} the planned experimental searches are very well aligned to this.

\subsection{Experiments}

\begin{figure}[!t]
\centering
\includegraphics[width=0.7\textwidth]{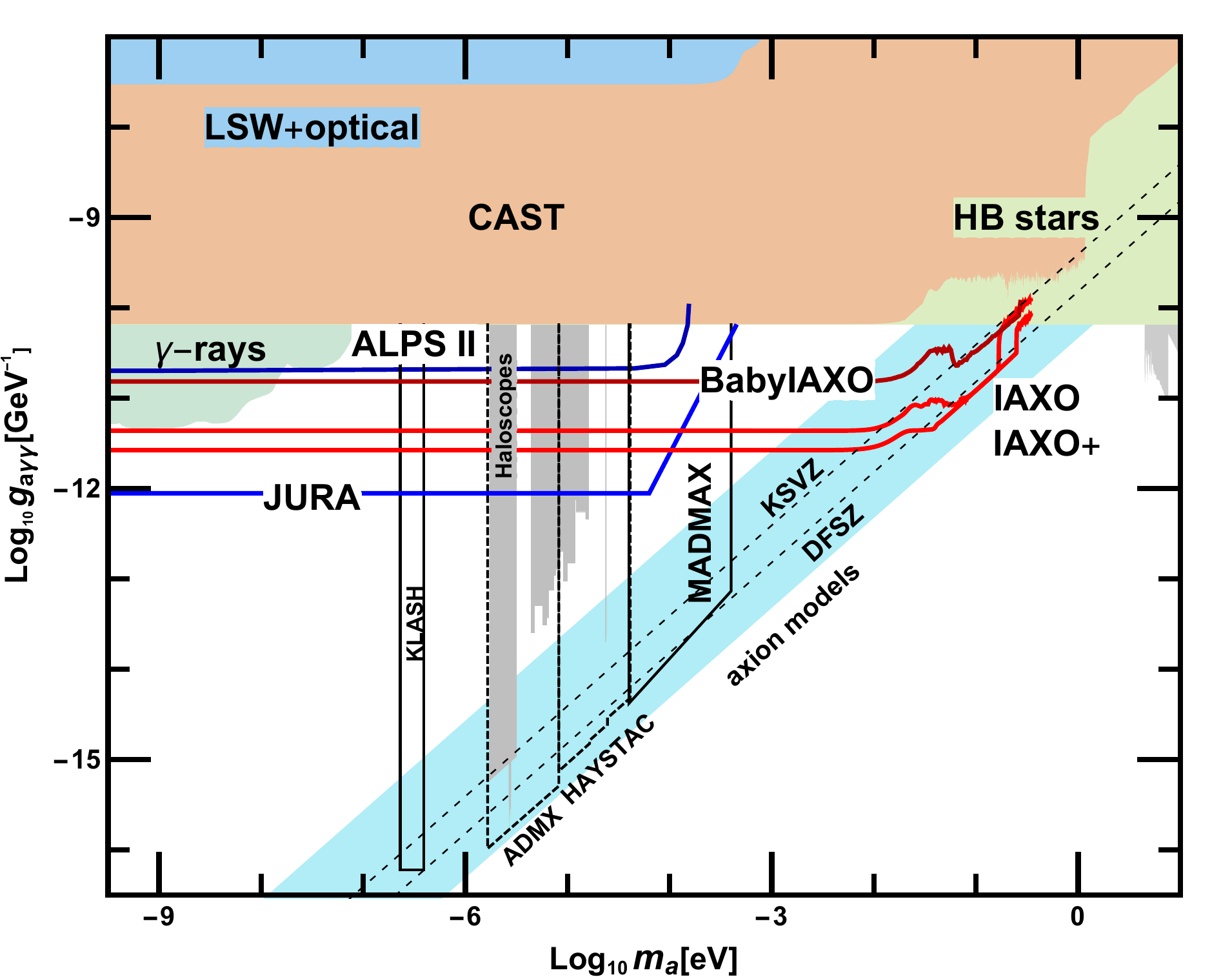}
\hspace*{1cm}
\caption{Current exclusion of ALPs and axions coupling to photons in the sub-eV
mass-scale (see, e.g.,~\cite{Jaeckel:2010ni,Irastorza:2018dyq} for details) with experimental prospects. Astrophysical limits are shown in green, pure laboratory experiments are indicated in blue, helioscopes in red and haloscopes in grey. The turquoise shaded region indicates the typical coupling range expected for QCD axion models. Couplings to other particles than photons are discussed in the supporting note~\cite{DM-note}.}
\label{fig:alpsfig}	
\end{figure}


Depending on the source of the ALPs there are currently three main search strategies:
\begin{enumerate}
\item[(i)] Direct detection with haloscopes~\cite{Sikivie:1983ip}, searching for ALPs being the dark matter;
\item[(ii)] Detection with helioscopes of ALPs being produced inside the Sun~\cite{Sikivie:1983ip};
\item[(iii)] Experiments that produce and detect ALPs in the laboratory and are therefore independent of astrophysical and cosmological assumptions.
\end{enumerate}
Current bounds on axions and ALPs and the reach of future experiments are  shown
in Fig.~\ref{fig:alpsfig}.
Many of these experiments have overlap in technological requirements. The overlap in technologies, details of the below mentioned experiments and their synergies
are reviewed in~\cite{Siemko:2652165}.

\noindent{\bf (i) Haloscopes}: Amongst the axion haloscopes the US-led ADMX
~\cite{Du:2018uak} 
and HAYSTAC
~\cite{Zhong:2018rsr} 
experiments are at the forefront.
ADMX is based on a resonant cavity approach, where axions are converted into photons via a magnetic field inside a
{radio-frequency} cavity. It {has started to} scan the lower mass part of the parameter space for the QCD axion. To go beyond the range explored by ADMX new geometries and technologies are being explored. 
In particular in Europe there are important developments with QUAX-a$\gamma$~\cite{Alesini:2019ajt}, RADES~\cite{Melcon:2018dba}, CAST-CAPP~\cite{Miceli:2015xas}, KLASH \cite{Gatti:2018ojx}, BRASS and CNRS/LNCMI-Grenoble~\cite{Siemko:2652165}.

At higher masses/frequencies, suggested in particular also by the post-inflationary axion scenario, a novel technique based on semi-resonant dielectric mirrors seems promising. The effort is named MadMax \cite{Majorovits:2017ppy} and is currently led by a collaboration between the MPI for Physics in Munich and DESY in Hamburg. 
The full scale version, that could reach axion sensitivity for a large fraction of the higher mass region, is in preparation. 


\noindent {\bf (ii) Helioscopes:} Axion helioscopes aim to detect axions produced in the Sun via photon-axion conversion. The signal is X-ray photons resulting from a 
re-conversion of axions into photons inside a strong magnet pointed at the Sun. Importantly this signal is independent of axions being the dark matter. 
Building on the experience of the CAST experiment at CERN, that currently provides the best limits on the axion-photon coupling for sub-meV masses, IAXO aims at improving the sensitivity towards smaller couplings by about two orders of magnitude. This will provide a significant sensitivity to meV QCD axions as well as to ALPs that are plausible explanations of several astrophysical anomalies. Moreover, IAXO can also yield additional information beyond the photon coupling. It could deliver crucial information on axion-electron couplings as well as their mass. The IAXO physics potential has recently been summarized in~\cite{Armengaud:2019uso}.
IAXO received crucial support from CERN in the area of magnet design. A possible siting at DESY has been discussed.
A smaller-sized version dubbed BabyIAXO is ready to be built and could deliver physics results within less than 5 
years. 

\noindent {\bf (iii) Pure Laboratory Experiments:} Currently ALPS-II \cite{Bahre:2013ywa} is exploiting the light-shining-through-walls (LSW) technique where laser photons are converted to axions and back to photons inside strong magnetic fields. For masses below about $0.1\,{\rm meV}$ it aims to exceed the CAST sensitivity by about one order of magnitude in the coming years. It could thereby provide a robust test of the suggested astrophysical anomalies.
Using magnets planned  for a future collider, a large scale LSW experiment (JURA) could exceed the sensitivity of IAXO  in the low-mass region by a factor of 3
or more in the future.
A ${\rm THz}$ photon source is proposed to be used in STAX (under R\&D now), instead of laser light,
profiting from a large photon number.

A different approach using Vacuum Magnetic Birefringence experiments has been proposed.
Their sensitivity for ALPs is, however, typically much smaller than LSW.
Running experiments based on this method are PVLAS and BMV, while in addition VMB@CERN has been proposed as new experiment. See e.g.~\cite{Siemko:2652165} for a detailed discussion.

\subsection{Complementarity with direct and indirect detection searches}
\label{subsec:DM-Axions-Complementarity}
Radio searches for the conversion of axion/ALP dark matter into photons inside the magnetosphere of neutron stars can have sensitivity
~\cite{Pshirkov:2007st,Huang:2018lxq,Hook:2018iia} 
for ALP masses in the range $\sim 0.2$--$40\,\mu$eV, and potentially above.  The signature is the emission of a narrow radio line from individual neutron stars, with a frequency that corresponds to the mass of the ALP.  Several of such searches are now underway, with expected sensitivities to the photon-ALP coupling down to $g_{a\gamma\gamma} \sim 10^{-12}\,\text{GeV}^{-1}$. The future SKA may have the ability to probe significant parts of the QCD axion parameter space~\cite{Safdi:2018oeu}.

Direct detection searches have sensitivity to ALP dark matter in the 1--100 keV mass region, via searches for an absorption triggered by the ALP-electron coupling. The currently best sensitivity is obtained by Xe detectors, which constrain the coupling of new pseudoscalars $g_{Ae} < 4 \times 10^{-13}$ in the 40--120 keV range
~\cite{Abe:2018owy}.
Mid-term future experiments project a factor of $\sim$\,20 improvement
~\cite{Aalbers:2016jon}.


\bigskip





\section{Conclusions}
Gravitational and cosmological observations provide overwhelming evidence of the existence of DM and, if made of particles or compact objects, of its predominance over other types of matter in the Universe. They also provide a compelling proof of physics beyond the SM. Given our high current level of ignorance about the basic properties of dark matter, a comprehensive suite of  experiments and techniques are required in order to cover the many possibilities.
In this chapter we have discussed experiments that can explore a broad range of dark matter masses, from ultralight DM (below eV) to thermal DM, either light (a few keV to GeV) or heavy/WIMP-like (GeV to 100 TeV).
Many of such experiments can also explore a DS with dark mediators and other feebly interacting particles than may be present within DM scenarios.

Accelerator-based, beam-dump and fixed-target experiments such as SHiP, LDMX, NA62$^{++}$ and NA64$^{++}$ can perform sensitive and comprehensive searches of sub-GeV DM and associated DS mediators. They will broadly test models of thermal LDM that are as yet underexplored. CERN has the opportunity to play a leading role in these searches by fully exploiting the opportunities offered by the SPS and the foreseen Beam Dump Facility.

Prospective future colliders (ILC/CLIC, FCC-ee/hh/eh and HL/HE-LHC) have excellent potential to explore models of thermal DM in the GeV--10 TeV mass range, notably including WIMPs as well as models with different types of DM mediators. 
Feebly interacting particles, plausibly produced at colliders, can also be searched for at prospective new detectors (e.g.\ FASER, CODEXb, MATHUSLA) further away from the beam interaction points. 
These new search strategies are complementary to accelerator-based, fixed target (beam dumps and missing energy/momentum) experiments as well as to standard collider searches depending on the mass region and the coupling strength between the SM and the DM/DS. 

The search for ultralight DM particles like the axion has gained significant momentum. IAXO  provides a compelling opportunity to extend the search for axions. In addition, haloscopes such as MadMax could directly detect axion dark matter, whereas ALPS-II and other prospective light-shining-through-wall experiments can provide competitive pure laboratory tests.

Complementary to the compelling experimental programme described above,
 astrophysical probes of DM through DD and ID searches cover a vast range of DM candidate masses through mature as well as rapidly developing, new technologies, and provide an essential handle for a convincing discovery.

\vspace*{3mm}
\noindent{\bf{Outlook on synergies}}: Focusing on the quest for DM in the coming decades, at the Granada Symposium
there was consensus in further developing synergies between the efforts of the high energy physics and  astrophysics communities. 
The discussion highlighted the need for enhanced communication between accelerator/collider-based, direct detection and indirect detection dark sector searches, as well as the potential benefits of common technology platforms (see Chapter~\ref{chap:inst}). 

Consensus on common search targets is important for a joint interpretation of results from different searches, and will be of fundamental importance to validate a putative DM discovery in different experiments and channels.
This can be facilitated by the existing LHC Dark Matter and Physics Beyond Collider working groups, and the newly established EuCAPT Astroparticle Theory Center as a joint venture of ECFA and APPEC, as well as by further discussions among the many experts in the field.

Vacuum over large volumes, cryogenics, photosensors, liquid argon detectors, design and operation of complex experiments---including software and data processing---are common themes within and beyond the communities engaged in DM and DS searches. Technological challenges related to these topics can benefit from new and existing platforms for joint discussion and collaboration.
The expertise present at CERN as the hub for the current largest collider programme worldwide, together with the expertise of other large European National labs and the complementary expertise of innovative small-scale 
experiments, can stimulate knowledge transfer and add guidance and coherence to the overall DM programme.

\chapter{Accelerator Science and Technology}
\label{chap:acc}


This chapter presents a summary of accelerator science and technology related submissions. Both state-of-the-art and challenges for the main technologies are highlighted. A summary of the expected performance of the future colliders considered in this document is given in Table~\ref{table:future-colliders}.
The parameters and comparisons of the projects are based on the inputs submitted to the European Strategy Update, unless stated otherwise. Common assumptions have been made \cite{Bordry:2018gri} for the annual operating schedule of colliders proposed at CERN, but note that different assumptions have been made for other colliders.
\begin{table}
\caption{Summary of the future colliders considered in this report. The number of detectors given is the number of detectors running concurrently, and only counting those relevant to the entire Higgs physics programme. The instantaneous luminosity per detector and the integrated luminosity provided are those used in the individual reports. For $e^+e^-$ colliders the integrated luminosity corresponds to the sum of those recorded by all the detectors. 
For HL-LHC this is also the case, while for HE-LHC and FCC-hh it corresponds to 75\% of that. The values for $\sqrt{s}$ are approximate, e.g.\ when a scan is proposed as part of the programme  this is included in the closest value (most relevant for the $Z$, $W$ and $t$ programme). For the polarisation, the values given correspond to the electron and positron beam, respectively. For HL-LHC, HE-LHC, FCC, CLIC and LHeC the instantaneous and integrated luminosity values are taken from Ref.~\cite{Bordry:2018gri}. For these colliders, the operation time per year, listed in the penultimate column, is assumed to be 
$1.2\times 10^{7}$\,s, based on CERN experience~\cite{Bordry:2018gri} (this is reduced by a margin of 10--18\% in the projections presented for physics results from FCC-ee). CEPC (ILC) assumes $1.3\times 10^{7}$ ($1.6\times 10^{7}$)\,s for the annual integrated luminosity calculation. When two values for the instantaneous luminosity are given these are before and after a luminosity upgrade planned.
 Abbreviations are used in this report for the various stages of the programmes, by adding the energy (in GeV) as a subscript, e.g.\ CLIC$_{380}$; when the entire programme is discussed, the highest energy value label is used, e.g.\ CLIC$_{3000}$; this is always inclusive, i.e.\ includes the results of the lower-energy versions of that collider. Also given are the shutdowns (SDs) needed between energy stages of the machine; SDs planned during a run at a given energy are included in the respective energy line.
\label{table:future-colliders}}
\begin{center}
\begin{tabular}{|l|ccccc|cc|l|} 
\hline\hline
Collider & Type & $\sqrt{s}$ & $\cal{P}$ [\%] & $N_{\rm Det}$& $\cal{L}_{\rm inst}$/Det. & $\cal{L}$ & Time  & Ref.\\
& &  & [$e^-$/$e^+$] & & \small [$10^{34}$cm$^{-2}$s$^{-1}]$ & [ab$^{-1}$] & [years] &\\\hline
\rule{0pt}{1.0em}%
HL-LHC & $pp$ & 14\,TeV & -- & 2 & 5 & 6.0 & 12 & \cite{Cepeda:2019klc} \\\hline
HE-LHC & $pp$ & 27\,TeV & -- & 2 & 16 & 15.0 & 20 & \cite{Cepeda:2019klc}  \\\hline
FCC-hh & $pp$ & 100\,TeV & -- & 2 & 30 & 30.0 & 25 & \cite{Mangano:FCC2018}  \\\hline
FCC-ee & $ee$ & $M_Z$ & 0/0 & 2 & 100/200 & 150 & 4 & \cite{Mangano:FCC2018}   \\
  & & $2M_W$ & 0/0 & 2 & 25 & 10 & 1-2 & \\
  & & $240$\,GeV & 0/0 & 2 & 7 & 5 & 3 & \\
  & & $2m_{top}$ & 0/0 & 2 & 0.8/1.4 & 1.5 & 5 & \\
    &  \multicolumn{4}{c}{(1y SD before $2m_{top}$ run)} &  & & (+1)  & \\\hline
  ILC & $ee$ & $250$~GeV & $\pm 80$/$\pm 30$ & 1 & 1.35/2.7 & 2.0 & 11.5 & \cite{Fujii:2017vwa} \\
  & & $350$~GeV & $\pm 80$/$\pm 30$ & 1 & 1.6 & 0.2 & 1 & \cite{Bambade:2019fyw} \\
  & & $500$~GeV & $\pm 80$/$\pm 30$ & 1 & 1.8/3.6 & 4.0 & 8.5 & \\
  &  \multicolumn{4}{c}{(1y SD after 250 GeV run)} & & & (+1) & \\\hline
   CEPC & $ee$ & $M_Z$ & 0/0 & 2 & 17/32 & 16 & 2 & \cite{CEPCStudyGroup:2018ghi} \\
  & & $2M_W$ & 0/0 & 2 & 10 & 2.6 & 1 & \\
  & & $240$~GeV & 0/0 & 2 & 3 & 5.6 & 7 &  \\\hline
    CLIC & $ee$ & $380$~GeV & $\pm 80$/$0$ & 1 & 1.5 & 1.0 & 8 & \cite{Roloff:2018dqu}  \\
  & & $1.5$~TeV & $\pm 80$/$0$ & 1 & 3.7 & 2.5 & 7 & \\
  & & $3.0$~TeV & $\pm 80$/$0$ & 1 & 6.0 & 5.0 & 8 & \\
  &  \multicolumn{4}{c}{(2y SDs between energy stages)} & & & (+4) & \\\hline
   LHeC & $ep$ & 1.3\,TeV & -- & 1 & 0.8 & 1.0 & 15  & \cite{Bordry:2018gri} \\\hline
   HE-LHeC & $ep$ & 1.8 \,TeV & -- & 1 & 1.5 & 2.0 & 20 &  \cite{Mangano:FCC2018} \\\hline
   FCC-eh & $ep$ & 3.5\,TeV & -- & 1 & 1.5 &  2.0 & 25 & \cite{Mangano:FCC2018} \\
\hline\hline
\end{tabular}
\end{center}
\end{table}

\section{Present state of accelerator technology for HEP}

Both circular electron-positron and hadron colliders have operated or are operating at peak luminosities above 1--2$\times 10^{34}$~cm$^{-2}$s$^{-1}$. 
PEP-II \cite{pep2}, KEKB \cite{kekb,kekb2} and LHC \cite{lhcdesign,lhcrun2} have all exceeded their design specifications in terms of peak performance. 

Today's colliders all operate with bunched beams.
Assuming the collision of beams with identical parameters,
and bunches colliding at an average frequency 
$f_{\rm coll}$, with $N_{b}$ particles per bunch, 
a basic expression for the luminosity is 
\begin{equation} \label{accel:eq:lum}
    L = f_{\rm coll} \frac{N_b^2 }{4\pi\sigma_x^{\ast} \sigma_y^{\ast}}
    =  f_{\rm coll} \frac{N_b^2}{4\pi \sqrt{\varepsilon_x\,\beta_x^*\,\varepsilon_y\,\beta_y^*}}
\end{equation}
where $\sigma_x^{\ast}$ and $\sigma_y^{\ast}$ designate the rms transverse beam 
sizes in the horizontal (bend) and vertical directions at the interaction point (IP), which can also be expressed in terms of the geometric emittances  and IP beta functions.     
In the above form, 
it is assumed that the bunches are identical in transverse profile,
that the profiles are Gaussian and independent of position along the bunch, and the particle
distributions are not altered during the 
bunch collision. 
Nonzero beam crossing angles $\theta_{c}$ in the horizontal plane and long bunches 
(rms bunch length $\sigma_{z}$) will reduce the luminosity from the above value, 
e.g.~by a factor $1/(1+\phi^{2})^{1/2}$, 
where the parameter
$ \phi \equiv \theta_{c} \sigma_{z}/(2 \sigma_{x}^{\ast})$ is known as the Piwinski angle,
but a large Piwinski angle angle may also allow for smaller beta function and higher bunch population. 
The disruption or pinch effects and the dynamic changes of beta functions and emittance (due to the collision) also modify the luminosity in linear and circular colliders, respectively. 
Various phenomena may limit the luminosity,
such as beamstrahlung, disruption, beam-beam tune shift, achievable beam power etc.
The above formulae can be recast in different forms according to the most relevant constraints.

For example, in the case of modern hadron colliders, like the HL-LHC, the luminosity is often limited by the maximum acceptable detector pile-up, and the luminosity then needs to be levelled at this value. 

In the case of circular $e^+e^-$ 
colliders the luminosity may be limited by the total synchrotron radiation power $P_{\rm}$ (determining the total beam current $I_{b}$) and the maximum beam-beam tune shift $\xi_y$, 
in which case the
luminosity formula can be rewritten as  
\begin{equation}
    L \propto \frac{\xi_y}{\beta_{y}^{\ast}} 
    \frac{P_{\rm SR}}{E_b^{3}} \; \, \, \ \text{(circular $e^+e^-$ colliders)}\, ,
\end{equation}
where $E_{b}$ denotes the beam energy. 
The maximum beam-beam tune shift $\xi_y$ increases
with beam energy, and also depends on the collision scheme.

In the case of linear $e^+e^-$ 
colliders it is convenient
to rewrite the luminosity expression as
\begin{equation}
L \propto H_{D} \frac{N_{b}}{\sigma_{x}^\ast}
\frac{1}{\sqrt{\beta_{y}^{\ast}\varepsilon_y}} I_{b}
\; \, \, \text{(linear $e^+e^-$ colliders)}\, ,
\end{equation}
where $H_{D}$ denotes the luminosity enhancement factor, which includes the geometry of the collision and the
beam-beam effects, and is of the order of one.
The factor $N_{b}/\sigma_{x}^{\ast}$ is proportional to the number of beamstrahlung photons emitted per beam particle; it determines the purity of the luminosity energy spectrum, which is constrained by
experimental requirements.
The last term $I_{b}$ 
is the average beam current $I_{b}=f_{\rm coll}  N_{b} e$ (with $e$ the electron charge),
and the collision rate $f_{\rm coll}$ equals the product of linac pulse 
rate and the number of bunches per pulse.

{\bf $pp$ colliders:} Between  2015 and 2018 (`Run 2') the LHC has accumulated 190 fb$^{-1}$ in proton-proton collisions per detector in ATLAS and CMS. 
The goal for HL-LHC is to deliver about 0.25 ab$^{-1}$ per year with the aim of integrating a total luminosity in the range of 3 to 4.5 ab$^{-1}$ by the late 2030s. 
The beam current is above 0.55\,A and the beta functions at the collision point are as low as 25\,cm. Dipole {\bf magnets} of 11 T and quadrupole magnets with a peak pole field of nearly 12 T, based on Nb$_3$Sn superconductor, are under development for HL-LHC. As part of the US DOE magnet development programme, a short model dipole magnet exceeded a field of 14 T at FNAL in late spring 2019. In parallel superconducting wires are being developed worldwide. In 2018-19, two independent US teams developed advanced Nb$_3$Sn cables with artificial pinning centres; these offer a 50\% higher critical current density than the HL-LHC cable and fulfil the target requirements for FCC. New suppliers in Japan, Korea and Russia have produced Nb$_3$Sn cables that meet the HL-LHC requirements, widening the base of potential manufacturers.

{\bf $e^+e^-$ colliders:} Between 1999 and 2008, the {\bf circular colliders (CC)} PEP-II and KEKB accumulated a total of almost 1.6 ab$^{-1}$, with beam currents at PEP-II as high as 2.1 A ($e^-$) and 3.2 A ($e^+$) and a vertical IP beta function at KEKB as low as 6 mm; their integrated luminosity  was greatly increased by the introduction of {\bf top-up} injection. SuperKEKB is presently being commissioned with a peak luminosity design goal of 
$8\times 10^{35}$~cm$^{-2}$s$^{-1}$. The {\bf crab-waist (CW)} collision scheme was demonstrated at DA$\Phi$NE around 2008, where it substantially increased the peak luminosity, and is incorporated into all 
future circular $e^+e^-$ proposals. SLC at SLAC (1988--1998) is the first and only {\bf linear collider (LC)} implemented to date.   Since then significant R\&D has been performed on  high-gradient/high-frequency NC RF for CLIC and on SC RF for the ILC (see below). Based on these technologies several high-energy linacs have been built to serve as X-ray FELs (SwissFEL, EU-XFEL, LCLS) and significant experience has been gained in beam tuning/focusing at test facilities (FACET, ATF/ATF2), demonstrating `nanobeam' feasibility for a future LC. ATF2 has achieved the scaled ILC vertical spot size of $\sim$~40 nm, 
albeit with a relaxed optics and at roughly 1/10 of the design bunch charge; the charge was reduced to mitigate wakefield effects. 

The world record for {\bf positron} production rates is still held by the SLC positron source. 
LCs require much higher positron production rates than SLC (CLIC about 20 times more, ILC baseline about 40 times, ILC upgrade about 160 times). The CLIC design incorporates a conventional positron source while the ILC baseline (for polarised positrons) passes the high-energy electron beam through a $\sim$200 m long undulator, generating photons that hit a rapidly rotating target to produce $e^+e^-$  pairs.


\section{Technologies for electroweak sector} 

\paragraph*{$e^+e^-$ Higgs factories}

The Higgs production process in $e^+e^-$ colliders peaks at different energies according to the different channels, as shown in the chapter on Electroweak Physics (see Fig.~\ref{fig:xsec_vs_cme}). We call here Higgs factories the $e^+e^-$ colliders with c.m.~energies optimized for the maximum corresponding physics reach.

One can be confident that the energy goal can be reached for all the considered configurations. Any remaining design issues can be mitigated during the project preparation phase. LEP operated at centre-of-mass collision energies above 200 GeV, and similar technologies, at larger scale, are the basis of FCC-ee [ID132] \cite{fccee}  or CEPC [ID51] \cite{cepc}. RF technology for both LC projects is considered mature, thanks to the intensive R\&D of last decades carried out by both HEP and photon source communities. The novel drive-beam scheme for generating the RF power for the CLIC main linacs has been demonstrated at CTF3, where the critical technical systems that are required have also been tested.

 \begin{figure}[ht]
 \centering
 \includegraphics[width=0.75\textwidth]{\main/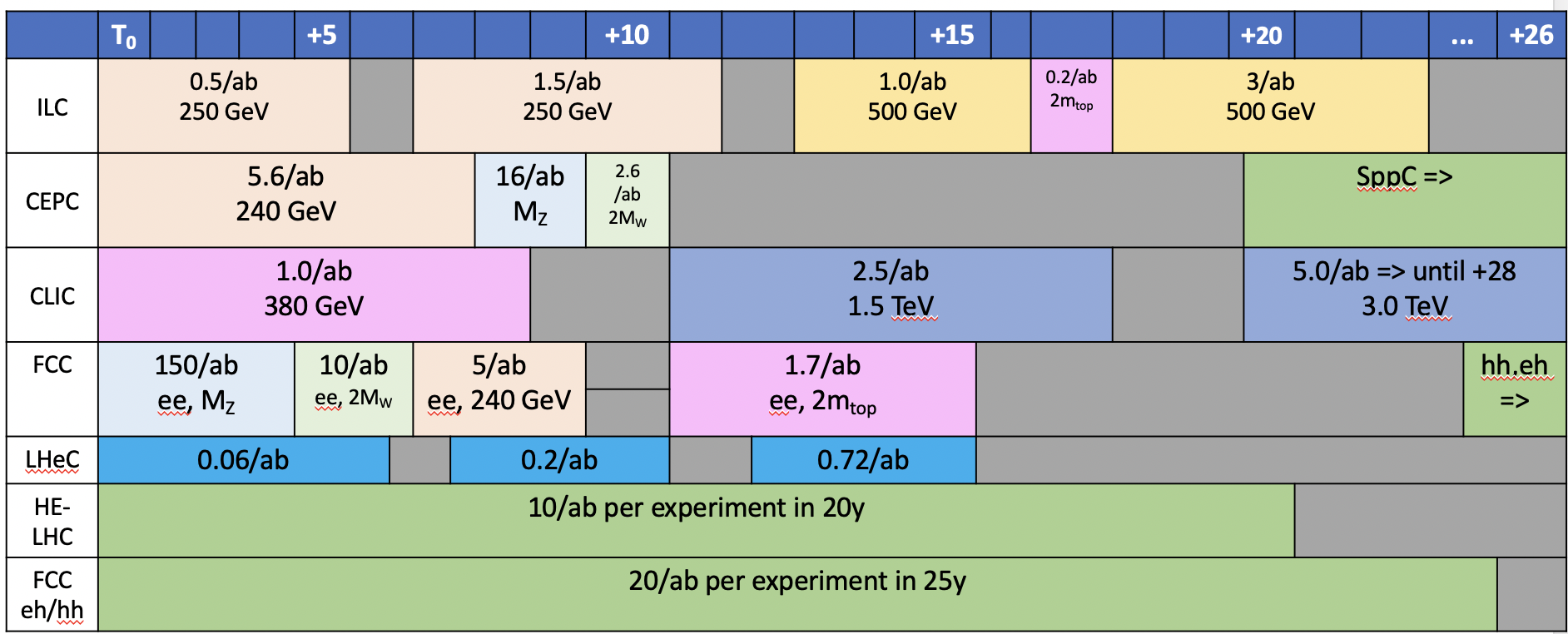}
 \caption{Time-lines of various collider projects, in years from start-time $T_0$ \protect\cite{deBlas:2019rxi}. 
 }
\label{fig:schedules}
\end{figure}

All proposals (Figure \ref{fig:schedules}) have ambitious luminosity targets, based on a combination of extrapolations from previous facilities (LEP, B-factories, DA$\Phi$NE, SLC, light sources and FELs), test-facility results, and theoretical predictions. The design luminosities naturally have larger uncertainties than the target energies since they rely on the integrated performance of each facility. 

\begin{figure}[ht]
\centering
\includegraphics[width=0.9\textwidth]{\main/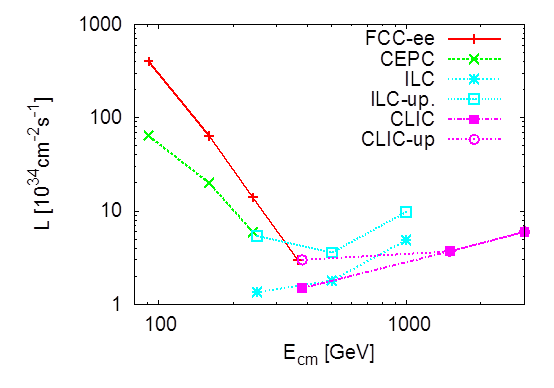}
\caption{Luminosity versus c.m.~energy for $e^+e^-$ Higgs Factories.
Two IPs are assumed for the circular colliders FCC-ee and CEPC.}
\label{luminosities}
\end{figure}

Figure \ref{luminosities} shows design luminosity as a function of energy for the $e^+e^-$ Higgs factories. The CC performances are influenced by the synchrotron radiation power which can be handled. Since this power is proportional to 
$I_b E_b^4$, the beam current $I_b$ must be reduced as the beam energy $E_b$ is increased; higher luminosities are hence obtained at lower energies, with the luminosity roughly proportional to $E_b^{-3.5}$.  The LCs provide higher luminosities at higher energies; the luminosity per unit beam current is roughly proportional to $E$.  The luminosity-performance crossover is in the region of 250 to 400 GeV. While one can be confident that the luminosity targets of the proposed colliders can be reached in principle, important feasibility work remains during the project preparation phase.

In order to achieve the design luminosity, all proposed $e^+e^-$ 
colliders rely on small beam sizes at collision  (FCC-ee 30--70 nm, ILC 3--8 nm, CLIC 1--3 nm), below those achieved at existing facilities. This requires very small beam emittances and ambitious focusing. Nanobeams are addressed via design and specifications, benchmarked simulations, low-emittance ring progress and studies, extensive prototyping and method developments (for alignment, stabilization, instrumentation and feedback systems, and algorithms), and in system and facility tests (FACET, light-sources, FEL linacs, ATF2).

In FCC-ee and CEPC the required emittance is achieved in the collider ring itself. FCC-ee and CEPC are based on a combination of concepts that have been proven and used in previous and present colliders. Some theoretical and experimental studies have been performed of critical effects, such as beam lifetime, beam-beam, impedances and electron cloud. Also, effects that have not been present in previous colliders have been studied, in particular the impact of beamstrahlung on the beam lifetime and instabilities. During the technical design phase, more complete studies, such as simulations of strong beam-beam effects with lattice imperfections, 
will be required to confirm this.

In ILC and CLIC the beam is produced in an injector and then cooled to small emittance in damping rings. The emittance targets for the damping rings have been achieved (and exceeded) with electron beams at modern light sources. A combination of technologies such as high-accuracy alignment, active magnet stabilization and beam-based alignment is designed to ensure that the emittance remains within budget during the beam transport to the collision point. Prototype tests at SLAC and KEK confirm the performance of the required beam-based alignment and tuning techniques, beam orbit and collision feedback systems, and beam focusing systems. 

In LCs 80\% polarisation of the electron beam is planned based on demonstrated performance in the SLC. ILC considers two alternative positron sources: a novel design based on undulators aimed at producing $\sim$30\% positron polarisation, and a conventional design for unpolarised positrons. In CCs no polarisation of the colliding bunches is foreseen; 
however, self-polarisation at the Z and W operation points is exploited for extremely precise energy calibration at the $10^{-6}$ level via resonant depolarisation.

Figure\ref{fig:schedules} shows possible schedules for the different facilities, including the option LHeC [ID159]. The LHeC proposal considers collisions of 7 TeV protons circulating in the LHC with 60 GeV electrons from a multi-turn high-current energy recovery linac (ERL). A corresponding ERL test facility (PERLE) [ID147] is planned at LAL/Orsay. A similar $ep$ collider, based on the FCC-hh, is part of the FCC design (FCC-eh). In parallel, R\&D is presently proceeding for an electron-ion collider in the US [ID74]. 
A possible variant or upgrade of FCC-ee using an ERL scheme has recently been proposed \cite{bnl-erl}, promising higher luminosity at higher energy, 
but this is still at a very preliminary stage.

The main challenges for both LCs and CCs are the RF (energy) and nanobeam (luminosity) performances, but here there are strong synergies with modern synchrotron and FEL light-source requirements. The importance of these connections among electron accelerators with their broader science applications, and impact of collaborations among key European (and overseas) partner national institutes cannot be overstated.

\paragraph*{Linear colliders}

\begin{figure}[ht]
\centering
\includegraphics[width=0.32\textwidth]{\main/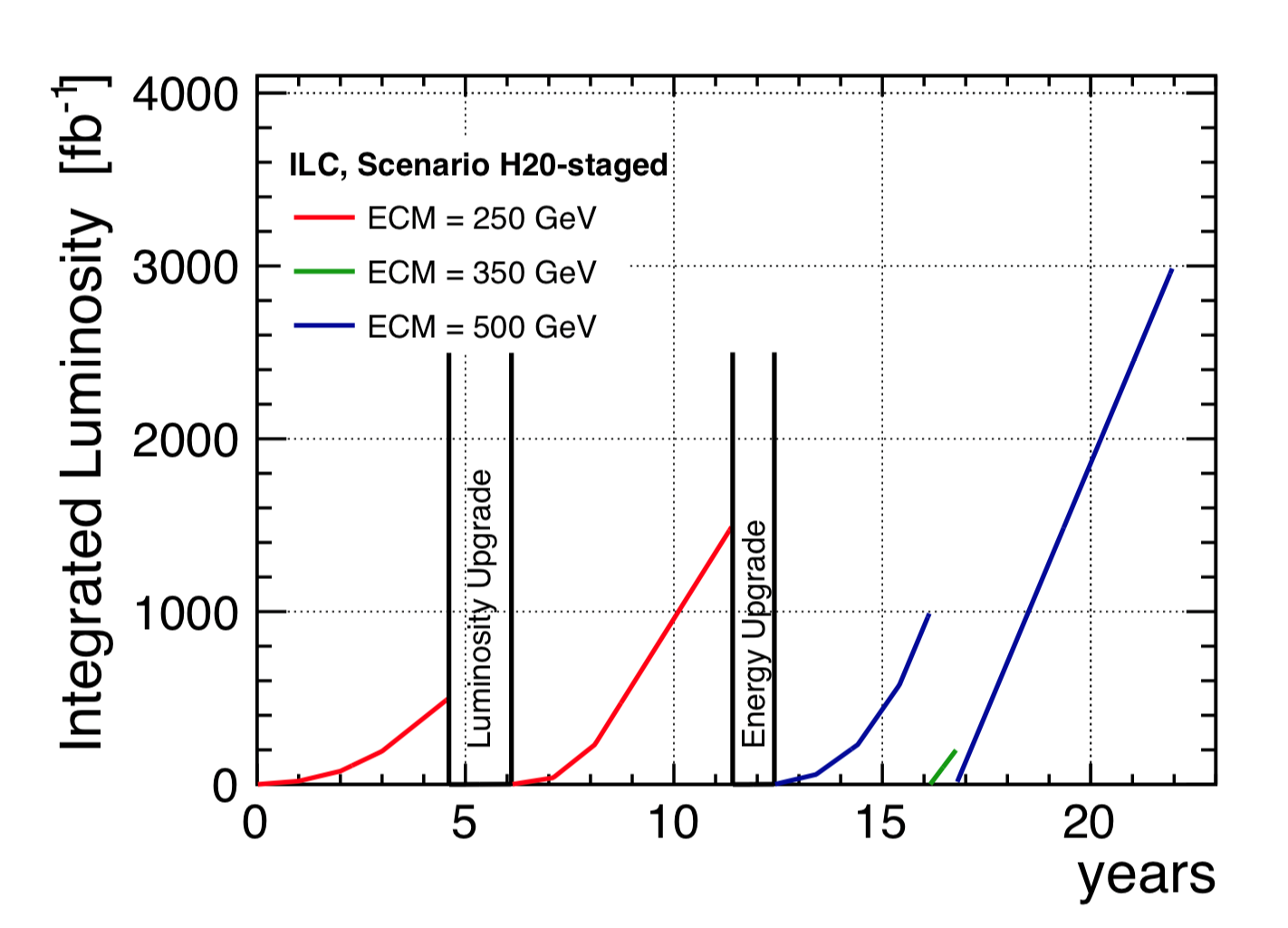}
\includegraphics[width=0.32\textwidth]{\main/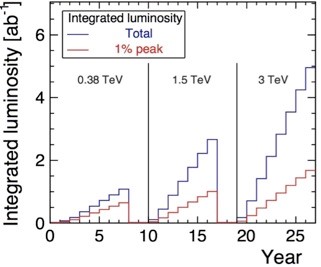}
\includegraphics[width=0.32\textwidth]{\main/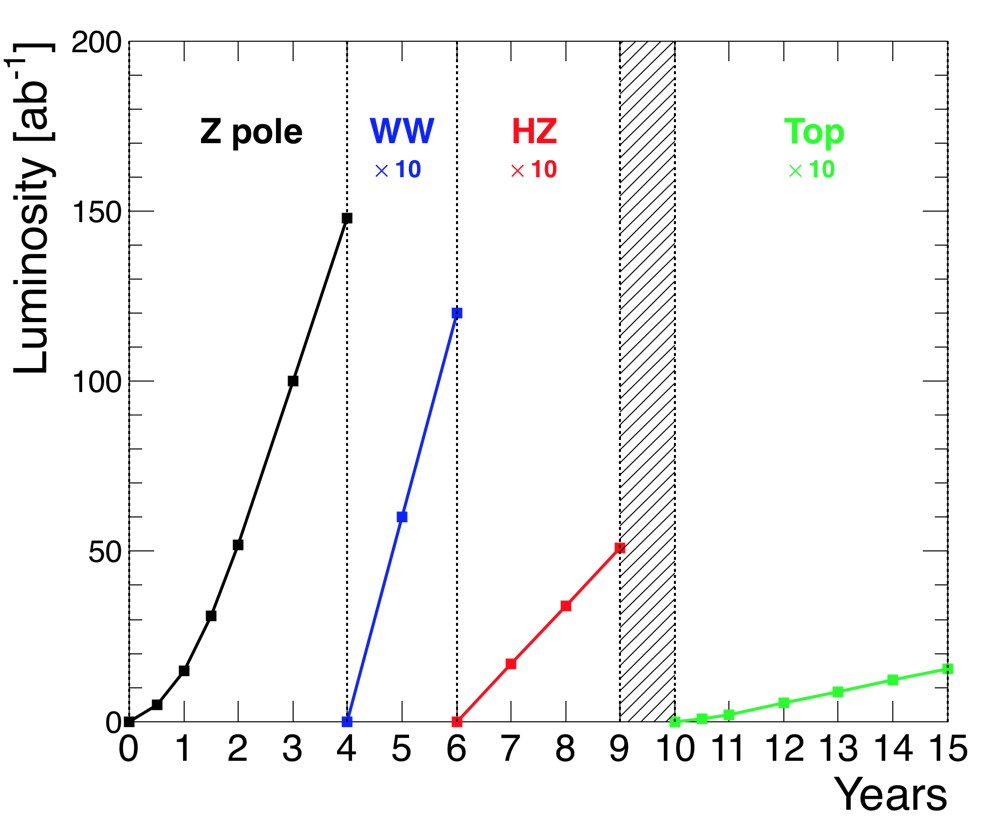}
\caption{Foreseen integrated luminosities of ILC [ID77], CLIC [ID146] and FCC-ee [ID132, ID135] versus year.}
\label{fig:int-lumi-ee}
\end{figure}

The LC initial stage provides a cost-effective and fast access to $e^+e^-$ collisions by 2035 for Higgs, top-quark and both SM and BSM studies. Such a machine leaves the door open for study of higher energy machines (LC extensions, and circular proton/muon colliders), for possible implementation on the 2040--50 timescale. The interplay between an evolving LC facility and a circular hadron, or possibly muon, collider---optimized in terms of technology, cost, size and first and foremost physics capability---would provide the global particle physics community with powerful tools for the foreseeable future. This approach also establishes a number of key accelerator R\&D goals for the next 1--2 decades. The main design parameters of the two LC options are given in Table~\ref{tab:ilc-clic}  and Fig.~\ref{fig:int-lumi-ee} shows the corresponding foreseen integrated luminosities.

\begin{table}[htbp]
\caption{Parameters for ILC and CLIC stages ($x$ = horizontal, $y$ = vertical).}
\label{tab:ilc-clic}
\begin{center}
\begin{tabular}{l|ccc|ccc}
\hline\hline
 &  \multicolumn{3}{c|}{ILC} &
 \multicolumn{3}{c}{CLIC} \\
 \hline
  & initial & $L$ upgr. & 500 GeV &  stage 1 & stage 2 & stage 3 \\
 \hline
 c.m.~energy [GeV] & 250 & 250 & 500 & 380 &
 1500 & 3000 \\
 rep.~rate [Hz] & 5 & 5/10 & 5 & 50 & 50 & 50 \\
 no.~bunches / pulse & 1312 & 2625 & 1312/ & 352 & 312 & 312 \\
  & & & 2625 &  &  &  \\
 bunch population [$10^{9}$] & 20 & 20 & 20 & 5.2 
 &  3.7 & 3.7
 \\
av.~beam current $I_b$ [$\mu$A] & 21  & 21/42 & 21/42 
& 15 & 9  & 9  \\ 
IP beta function $\beta_{x}^{\ast}$[mm]
& 13 & 13 &11 & 8 & 8 & 6 \\
IP beta function $\beta_{y}^{\ast}$ [mm]
& 0.41 & 0.41 & 0.48 &  0.1 & 0.1 & 0.07 \\
IP geometric emittance $\varepsilon_{x}$ [pm]
& 20 & 20 & 20 & 3 & 0.4 & 0.2  \\
IP geometric emittance $\varepsilon_{y}$ [fm]
& 140  & 140 &  70 & 80 & 14 &  7\\
rms IP beam size $x$ [nm] & 516 & 516 & 474 & 
149 & $\sim$60 &  $\sim$40\\
rms IP beam size $y$ [nm] & 7.7 & 7.7 & 5.9 & 
2,9 & $\sim$1.5 &  $\sim$1\\
luminosity enhancement $H_{D}$ & 2.55 & 2.55 & 2.26 & & &  \\
total luminosity $L$ [$10^{34}/$cm$^{2}$s$^{1}$] &
1.35 & 2.7/5.4 & 1.8/3.6 & 1.5 & 3.7 & 5.9\\
luminosity in top 1\% $L_{0.01}/L$ & 
73\% & 73\% & 58.3\% &  60\% & 38\% & 34\% \\
electrical site power [MW] &  115 & 135/185 &
163 & 168 & 364 & 589 \\
helium inventory [t] & 43 & 43/85 & 85 & -- & -- & -- \\
site length [km]  &  20.5 & 20.5/31 & 31 & 
11.4 & 29.0 & 50.1  \\
integrated luminosity [fb$^{-1}$/yr]	
& 100 & 300 & 600 & 180 & 444 & 708\\
\hline\hline
\end{tabular}
\end{center}
\end{table}

The proposals for CLIC [ID146] and ILC [ID77] are mature and complete, covering design of the accelerators, including their upgrades to higher energies, technical developments for critical systems, performance verification studies and detailed project implementation plans. There are strong communities supporting both project studies and proposals, as well as comprehensive detector and physics studies. Key features of these proposals include: the initial stages have costs and power budgets on a similar scale as LHC, making them well suited for rapid implementation; they can be readily expanded both in terms of energy---with existing, improved or novel RF technologies---and luminosity; polarised beams are foreseen; they can also be operated at lower energies (for example at the $Z$-pole, albeit with much lower luminosity than the CCs) and gamma-gamma collisions are possible. The physics performance is covered elsewhere in this document. Following discussions at the Granada symposium CLIC and ILC have submitted updated information about their $Z$-pole performances and luminosity upgrade options  \cite{Latina:2687090,Yokoya:2019rhx}. 

{\bf CLIC} is proposed to be implemented as a CERN-hosted international project (following the LHC and HL-LHC models) in three energy stages, 380, 1500 and 3000 GeV in the centre of mass (Table~\ref{tab:ilc-clic}) with design luminosities between 1.5 and 5.9$\times 10^{34}$~cm$^{-2}$s$^{-1}$. A recent study for the initial stage \cite{Latina:2687090} shows that increasing the bunch train frequency by a factor 2 could double the luminosity, with only modest increases in the power 
($\sim 50$ MW) and cost ($\sim 5$\%). The CLIC timeline includes a preparation phase 2020--2025, followed by a 7-year construction and commissioning period, in order to be ready for data-taking before 2035. The CLIC$_{380}$ cost is estimated at 5.9 BCHF, with upgrade costs of +5 and +7 BCHF for the two further stages. The AC power of the initial (final) stage is estimated at about 170 MW (580~MW). Only the CLIC$_{380}$ power estimate has so far been optimized.

{\bf Dual-beam acceleration} has been demonstrated at CTF3 
\cite{ctf3}. Gradients of up to 200~MV/m have been achieved with {\bf normal-conducting RF} Cu technology; numerous CLIC X-band cavities have been operated at the design gradient of 100 MV/m and X-band RF technology is now well established and industrially available.  There is also underpinning experience with C-band for large-scale systems, e.g.~SwissFEL, including the production processes for the accelerator structures with demonstrated design performance. 

{\bf ILC } is proposed to be built for initial operation at 250 GeV, with a direct upgrade path to 500 GeV, and a possible further upgrade to 1000 GeV (Table~\ref{tab:ilc-clic}). The luminosities foreseen are 1.35--5.4$\times 10^{34}$~cm$^{-2}$s$^{-1}$;  increasing the bunch train frequency by a factor 2 for the initial stage could double the luminosity while increasing the power by only 20--30 MW and the cost by $\sim$8\% \cite{michizono}. The timeline includes a preparation phase of 4 years, followed by a 9--10 year construction and commissioning period, in order to be ready before 2035. The ILC$_{250}$ cost is estimated to be 4.8--5.3 BILCU (ILCU = 2012 USD), while ILC$_{500}$ would be around 8 BILCU. The AC power is estimated at about 115 MW (300 MW) for 250 GeV (1000 GeV). The ILC is foreseen to be constructed as a Japan-hosted international project. Relevant European capabilities and the scope of possible participation have been presented [ID66]. 

Over the last decades excellent progress has been has made with {\bf superconducting RF} technology, driven primarily by and for TESLA, ILC and then successfully implemented at EU-XFEL \cite{XFEL} at DESY. The EU-XFEL represents a large-scale deployment of SC cavities  made by industry; most  cavities exceed the design gradient of $\sim$23 MV/m, and some reach over 40 MV/m, which exceeds the ILC design gradient of 35 MV/m. EU-XFEL, together with the on-going construction of LCLS-II at SLAC provide a development and testing ground for key elements (e.g.~magnets, instrumentation, controls, and vacuum systems), with parameters close to those needed for ILC. Further technology optimization is ongoing, linked to evolving SCRF R\&D for improved cavity gradient and Q values.

\paragraph*{Circular Colliders}

Circular $e^+e^-$ collider proposals build upon 50 years of experience. The designs of FCC-ee 
[ID132] and CEPC [ID51] exploit the historical knowledge from LEP (high energy), KEKB (high current, high luminosity, strong $e^+$ source), PEP-II (high current), DA$\Phi$NE (crab waist), and SuperKEKB (extremely low $\beta_{y}^{\ast}$, large Piwinski angle), and SLC (damping rings, powerful $e^+$  source). CCs can accommodate several interaction points (IPs); two IPs are assumed in the FCC-ee and CEPC baselines and an alternative design with 4 IPs is under development for FCC-ee.

Parameters for the stages of FCC-ee and CEPC are summarized in Table \ref{tab:fcc-cepc}. 
FCC-ee foresees starting on the $Z$ pole and then upgrading the RF systems in steps with optimized machine configuration for $Z$, $WW$, $ZH$, and $t\overline{t}$ working points. CEPC plans initially to install the full RF system for Higgs production, and later to operate on the $Z$ pole and then at the $WW$ threshold. CCs have the potential for an extremely high production rate of $Z$ bosons (Tera-$Z$ factory) and exquisite precision energy calibration at the $Z$ pole (100 keV) and at the $WW$ threshold using resonant depolarization.
The total FCC-ee construction cost (for $Z$, $W$ and $H$ working points) is estimated to be 10.5 BCHF [ID132]; operation at the $t\bar{t}$ working point will require later installation of additional RF cavities and associated cryogenic cooling infrastructure for an additional cost of 1.1 BCHF. The precision of the overall cost estimate is at the 30\% level. The annual energy consumption is similar to that of HL-LHC, and varies between 1 and 2 TWh. The total cost of CEPC was reported as 5 billion USD \cite{wang,cepc}. It should be noted that in each case both civil engineering and technical infrastructures can largely be reused for a subsequent hadron collider (FCC-hh [ID133, ID135] or SppC).

Each of the CC main concepts and parameters has been demonstrated in a previous collider (see earlier). Therefore, the CC designs are mature and R\&D is focussed on engineering optimization towards easing operability, machine efficiency, and maintainability aiming at efficient and cost-effective exploitation. R\&D includes highly efficient SC RF systems, high-power RF couplers and high-efficiency RF power production all of which profit from past investments in a range of RF user communities. For FCC-ee a hybrid-technology solution has been found using Nb-sputtered Cu cavities for lower energies (as those first developed for LEP) operated at 4.5 K, combined with bulk Nb cavities for higher energies (profiting from ILC R\&D), operated at 1.8 K.  Other CC R\&D includes novel ultra-thin NEG coatings for the vacuum system, radiation shielding of accelerator components (coils, bellows, flanges), a feasible design of the machine-detector interface (MDI) and an energy efficient, cost effective magnet system for the collider arcs.

A performance upgrade of FCC-ee (luminosity increase by about 80\%) could be implemented via doubling the number of IPs from 2 to 4, without any effect on the circulating beam current and minimal impact on power consumption 
\cite{shatilov-2019,oide-2019,zimmermann-pol}. 
CEPC [ID51] considers a luminosity upgrade through increasing the SR power per beam from 30 to 50 MW (i.e.\ to the FCC-ee design value). A potential upgrade option with much higher gains both in luminosity (up to a factor 100) and energy (beyond 500 GeV) through conversion to an ERL-based collider was proposed recently \cite{bnl-erl}; its feasibility and cost must be studied in greater detail.

\begin{table}[htbp]
\caption{Parameters for FCC-ee [ID132] and CEPC [ID51] stages.}
\label{tab:fcc-cepc}
\begin{center}
\begin{tabular}{l|cccc|ccc}
\hline\hline
 & \multicolumn{4}{c|}{FCC-ee} &
 \multicolumn{3}{c}{CEPC} \\
 \hline
 stage &  1 &  2 &  3 &  4 &  1 &  2 &  3 \\
 \hline
 c.m.~energy [GeV] & 91 & 160 & 240 & 350 / 365 &
 240 & 91 & 160 \\
 beam current [mA] & 1390 & 147 & 29 & 6.4 / 5.4
 &  17.4 & 461	& 87.9 \\ 
 emittance $\varepsilon_{x}$ [nm]& 0.27 & 0.84 & 0.63& 1.34 /  1.46 & 1.21 & 0.18 & 0.54 \\
 IP beta fn.\ $\beta_{y}^{\ast}$ [mm]
& 0.8 & 1.0 & 1.0 & 1.6 & 1.5& 1.0& 1.5\\
RF voltage [GV]	& 0.1 & 0.75 & 2.0 & 4.0+5.4/6.9 &  2.17 & 0.10& 0.47 \\
RF frequency [MHz]& 400	& 400 & 400 &  400 + 800 & 	\multicolumn{3}{c}{650} \\ 
RF cavities & 1-cell & 4-cell & 4-cell & 4-cell + 5-cell & 
\multicolumn{3}{c}{2-cell}\\
RF cavity material & Nb/Cu & Nb/Cu & Nb/Cu & Nb/Cu + bulk Nb & 
\multicolumn{3}{c}{bulk Nb}\\
RF cavity temp.\ [K]& 4.5& 	4.5	& 4.5 & 4.5 + 1.9 & 
\multicolumn{3}{c}{1.9}\\
peak lumi.\ (for 2 IPs) & 460 & 56& 17& 3.6 / 3.1
& 6 & 64 & 20 \\
\; \;  [$10^{34}$~ cm$^{-2}$s$^{-1}$] & & & & & & & \\
SR power/beam [MW] & 50 & 50 & 50 & 50 & 30 & 30 & 30 \\
electrical power [MW] &  259 & 277 &
282 &	$\sim$350 & 270 & 149 & 223 \\
 helium inventory [t] & 8 & 8 & 18 & 32 & 6 & 6 & 6 \\
run time [years] &  4 & 1--2 & 3 &1 / 4 &	
7 & 2 & 1 \\
total int.\ lumi.\ [ab$^{-1}$]	
& 150 & 10 & 5& 0.2 / 1.5 & 5.6 & 16 & 2.6\\
\hline\hline
\end{tabular}
\end{center}
\end{table}

\paragraph*{Complementary circular colliders}

{\bf SuperKEKB}~\cite{skekb,skekb2}, the KEKB upgrade, is presently under commissioning. Its design is based on the nanobeam collision scheme in a large Piwinski angle regime. It aims at $L = 80 \times 10^{34}$~cm$^{-2}$s$^{-1}$, increasing 
beam currents up to 3.6 and 2.6 A ($e^-$ and $e^+$ respectively), decreasing $\beta_{x,y}^{\ast}$  
to the level of 30 and 0.3 mm (H and V respectively) and decreasing the emittance to about 4 nm. 
SuperKEKB will test, and go beyond, many of the parameters of FCC-ee [ID132] and CEPC [ID51], such as the vertical design beta function, the Touschek positron beam lifetime (3 minutes, to be compared with an expected beam lifetime of 20--60 min in FCC-ee due to radiative Bhabha scattering), the higher beam currents and the higher production rate of positrons.

The design of the {\bf Super Charm-Tau Factory} (SCT) [ID49] 
\cite{stc} at BINP, Novosibirsk, is a collider with luminosity of 
$10^{35}$~cm$^{-2}$s$^{-1}$ at centre-of-mass energy between 2  and 6~GeV with the possibility to exploit longitudinally polarised electrons at the interaction point, for production of charmonium and tau leptons. It is based, as the FCC-ee, on the Crab Waist collision scheme.

\section{Path towards highest energies}
For the foreseeable future, proton-proton collisions appear to offer the greatest collision-energy reach, up to roughly 100 TeV. Table \ref{tab:ppcoll} summarizes the major parameters and technical challenges for possible future proton colliders. 
All the proposed facilities would also support a corresponding (heavy) ion collision programme \cite{fcchh,helhc,cepc}, 
at significantly higher energy than the LHC.  
The realization of future hadron colliders depends both on high-field superconducting magnets, for which major R\&D is required, and on the provision of a large circular tunnel. A preliminary implementation study of a 100 km tunnel is included in the FCC CDR. 
Such a tunnel could, in a first phase, house a high-luminosity circular $e^+e^-$ collider with a fully complementary physics programme, as proposed both in the FCC integrated programme and in 
the CEPC-SppC approach, 
or a lower-energy proton collider based on e.g.~6 T single-layer Nb-Ti magnets at 1.9 K. 
The cost complement for constructing the FCC-hh based on the FCC-ee infrastructure is estimated at 17 BCHF. 
A high-energy linear $e^+e^-$ collider, i.e.~ILC$_{1000}$ or CLIC$_{3000}$, described above, can also push the physics reach in lepton collisions towards and beyond that of the LHC (see Chapters~\ref{chap:ew} and \ref{chap:bsm}). Electron-hadron colliders, i.e.~LHeC or FCC-eh, could incorporate one ring of the proton collider and a new electron linac based on e.g.\ LC linac technology or an ERL. Muon colliders need substantial R\&D (see Section \ref{sec:muon}); if realizable, they have the potential to reach collision energies up to a few tens of TeV. Plasma-based wakefield acceleration also offers long-term possibilities for reaching high energies (see Sect.~\ref{sec:plasma}).

\begin{table}[htbp]
\caption{Parameters of proposed future high-energy hadron colliders HE-LHC [ID136] \cite{helhc}, FCC-hh [ID133] \cite{fcchh} and SppC [ID51] \cite{cepc}.
\label{tab:ppcoll}
}
\begin{center}
\begin{tabular}{lccc}
\hline\hline
&  HE-LHC & FCC-hh & SppC  \\
\hline
beam energy [TeV]&
        13.5 & {50} &
        {37.5}\\
circumference [km] &
        26.7 & {97.75} &\\
interaction regions  &
	 2 (+2) & 2+2 &	
        {2}   \\ \hline
int.\ lumi.\ per main experiment [ab$^{-1}$/yr]  &
0.5 & 
        {0.2--1.0} &
        {0.4}  
         \\ 
peak luminosity [10$^{34}$/cm$^{2}$/s] & 
16 & 
        {5--30}&  
        {10} 
 \\
 electrical site power [MW] &  
 162 (with scSPS) & 580 (550 opt.) & 700 \\
 helium inventory [t] & 163 & 880 & not available  \\
time between collisions [ns]&
25 & 25 & 25 \\
energy spread [rms, $10^{-3}$]&
0.1 &
        {0.1} &
        {0.2} \\
bunch length [rms, mm]&
80 & 
        {80}&
        {75.5} \\
rms IP beam size [$\mu$m]&
6.6 &
        {6.8 (initial)}& 
        {6.8 (initial)}
 \\
injection energy [TeV]&
1.3 & 
	{3.3}&
	{2.1}  \\ 
rms transverse geom.\ emittance [nm]&
0.17 (init.) & 
        {0.04 (init.)}&
        {0.06 (init.)}
        \\ 
$\beta^*$ at IP [cm] &
45 & 
        {110--30 }&        
        {75} \\ \hline
beam-beam parameter/IP [$10^{-3}]$ &
12 & 
         {5--15}&
         {7.5}\\
RF frequency [MHz]&
400 & 
	{400}&
	{400/200} \\
particles per bunch [$10^{10}]$ &
22 & 
        {10}&
        {15}0 \\
bunches per beam  &
2808 & 
        {10600}&
        {10080} \\
average beam current [mA]  &
1120 & 
        {500}&
        {730}\\
length of standard arc cell [m]  &
107 or 137 & 
        {213}&
        {148}\\ \hline
peak magnetic field [T]  &
16 & 
        {16}&
        {12}\\
SR power loss/beam  [MW] &
0.1 & 
	{2.4}&
	{1.1} \\ \hline\hline
	\end{tabular}
	\end{center}
\end{table}

The most critical requirement for a high-energy collider is energy reach and the proposed ILC$_{1000}$ and CLIC$_{3000}$, or HE-LHC [ID136], FCC-hh [ID133] and SppC [ID51], offer the highest energies in electron and proton collisions, respectively. Relevant criteria for feasibility of implementation are cost, AC power, and the required R\&D effort. The construction costs and AC power estimates are all within a factor 2--3 of each another. Nominally the lowest cost is for HE-LHC, followed by ILC$_{1000}$, CLIC$_{3000}$ and FCC-hh. The AC site power consumption is the lowest for HE-LHC, followed by ILC$_{1000}$, FCC-hh and CLIC$_{3000}$.  In terms of the required duration/scale of the R\&D effort to reach a TDR level of readiness, CLIC$_{3000}$ and a version of HE-LHC based on HL-LHC 12 T magnet technology are ahead of other proposals, even if still require R\&D/design optimization.

\begin{figure}[ht]
 \centering
 \includegraphics[width=0.75\textwidth]{\main/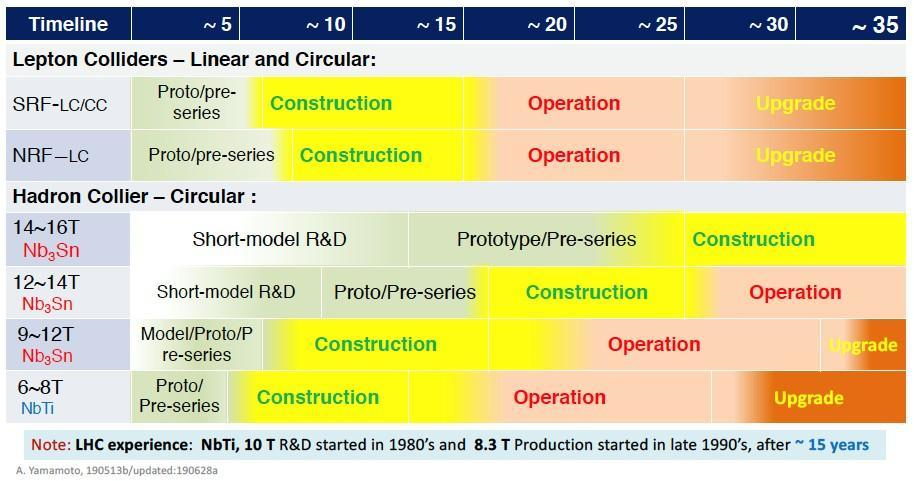}
\caption{ A relative time-line expected for realizing future lepton and hadron colliders (from A.~Yamamoto, presented at the Open Symposium in Granada, and updated based on the discussion that followed).}
\label{fig:hctl}
\end{figure}

The hardest challenge for the proton colliders is the development of a representative magnet currently aiming towards 16 T field. There are fundamental challenges in obtaining the required current density in superconductors and in dealing with the ultimate magnetic pressures and mechanical stresses in the superconductor and associated components. One can estimate the timescale needed to innovate new approaches/technology and overcome these limits through continual R\&D efforts (see also Fig.~\ref{fig:hctl}), as follows:

\begin{enumerate}

\item 
    Nb$_3$Sn, 14--16 T (25-28 TeV @ LHC, 90-100 TeV @ FCC-hh):  10--15 years for short-model R\&D (already on-going in the HL-LHC framework), and the following 10 to 15 years for prototype/pre-series with industry, resulting  in 20-30 years needed before a construction start.

\item Nb$_3$Sn, 12--14 T (21-25 TeV @ LHC, 75-90 TeV @ FCC-hh):  7--10 years for short-model R\&D, and an additional 7--10 years for prototype/pre-series with industry, resulting in 15 to 20 years before a construction start. The technical feasibility to reach 14 T has been demonstrated recently via the short-model FRESCA-2 project in Europe and the MDP programme in the US, yielding an accelerator-type cos-theta short dipole model with 4-layer coil geometry (as for the FCC 16 T design).

\item Nb$_3$Sn, 9--12 T (16-21 TeV @ LHC, 55-75 TeV @ FCC-hh): based on experience with the HL-LHC 11 T dipole and IR quadrupole magnets, a few years for short-model R\&D, and an additonal few years for prototype/pre-series with industry, resulting in construction feasibility in 5--10 years.

\item  NbTi, 6--8 T (35-50 TeV @ FCC-hh): NbTi 8 T dipole magnets are very similar to the current LHC main dipole with two-layer coils, and 6 T magnets may be the cheapest option by using a single-layer cos-theta coil winding. After reasonable prototyping work, including optimization based on the existing proven technology, both options could be available for construction within 5--7 years. 
\end{enumerate}

Because of its much higher current density capability, High Temperature Superconductor (HTS) technology will inevitably be required to reach fields beyond 16 T. A critical limitation of HTS today is its much higher cost, even compared with Nb$_3$Sn, however Nb$_3$Sn is already available as an industrial product, and HTS technology will presumably mature in the future.

The stored beam energy and the corresponding beam power at future hadron colliders is significantly higher than for the LHC, setting challenging requirements for machine protection, collimation, and beam abort system. 

Helium inventories for all colliders using superconducting technologies (magnets or RF) are listed in Tables \ref{tab:ilc-clic}, \ref{tab:fcc-cepc} and \ref{tab:ppcoll}.
The present LHC helium inventory of 135 t 
corresponds to less than a \% of the annual production worldwide. 
The FCC-hh inventory will represent a few \% of the annual production.  
If 10\% of the FCC-hh helium inventory is lost per year, about 
90 t would need to be purchased every year; the associated cost 
(on the order of 5~MCHF per year at the present supply price) 
should  be foreseen in the operation budget. 
Helium inventory numbers for SppC are not available yet; they 
will depend on the superconductor finally 
chosen for the magnets.


\section{Muon Colliders}
\label{sec:muon}
A muon collider [ID120] has the potential to reach attractive luminosities in very high-energy lepton collisions. It would be a powerful discovery machine, since, in contrast to a proton collider, the full collision energy is available in the centre of mass rather than shared among constituent partons. The higher muon mass (than that of the electron) reduces synchrotron radiation emission and allows for the acceleration and collision of the beams in a circular facility, permitting a much lower integrated RF voltage per turn and efficient use of the beams for luminosity production. A muon collider promises a linear increase of the luminosity per unit beam power with increasing collision energy, in contrast to an LC where it is independent of the collision energy. A circular collider also allows the possibility of multiple collision points, thereby mitigating the large AC power consumption at very high energies. In addition, the luminosity spectrum could be substantially better than for an $e^+e^-$ LC due to the strongly reduced beamstrahlung and initial-state radiation. 

Two main muon collider concepts have been proposed: in one the muons are generated using protons (MAP) \cite{Delahaye:2013jla, accel:muon3-sr1} 
in the other using positrons (LEMMA) \cite{Antonelli:2015nla, accel:lemma2019}. The proton-driven scheme was the object of a well-supported study, mainly in the US, but the effort was suspended about five years ago. The recently proposed positron-driven scheme is being studied with a limited effort mainly at INFN. Since no organised collaboration exists for muon colliders, a review group has been charged to assess their perspectives and status. This review is based on the material made available by the MAP and LEMMA studies and on some additional calculations. 
Figure \ref{fig9} shows the dependence of the luminosity per unit beam power for the proton-based muon collider in comparison with $e^+e^-$ colliders. 

The proton-driven scheme is based on classical muon production via pion decay. The study has addressed the global collider parameters and several key technical issues, such as cooling of the muon beams at different stages, fast-ramping magnets and RF cavities in a high magnetic field. Although it has not reached the level of a CDR, it is sufficiently complete to give confidence in the collider parameters. In the positron-driven scheme, 45 GeV positrons impinging on electrons at rest in a target produce muon pairs close to the reaction threshold, hence with a very low emittance. Two issues in the original LEMMA study have recently been identified that potentially reduce the luminosity by orders of magnitude. The LEMMA team is performing a redesign of the collider concept but it is too early to assess the results.

The decays of the accelerated muons drive critical issues:
\begin{itemize}
\item	
At the collision points, the decay electrons induce a large background of electrons and photons. A first simulation study with realistic conditions indicates that this background can be mitigated by suitable shielding, detector design, and analysis, such that it would not damage the physics capability.
\item	
The neutrinos from muon decays along the ring produce showers in the Earth. This leads to some radiation at the location where the plane of the collider ring intercepts the Earth's surface. At very high energies beyond 6 TeV, this could ultimately limit the achievable luminosity for the proton-based scheme. The positron-driven scheme would be particularly attractive in this respect since its smaller emittance requires much smaller beam current and thus reduces the neutrino dose, enhancing further the possible energy reach.
\end{itemize}


Other options of muon production can also be explored. For example, a possible direct source of low-emittance muon or intense positron beams could be the ``Gamma Factory'', where partially stripped heavy ions stored in the LHC (or in the FCC-hh) are collided with a laser pulse to generate intense bursts of X-rays [ID6].

\begin{figure}[ht]
 \centering
 \includegraphics[width=0.75\textwidth]{\main/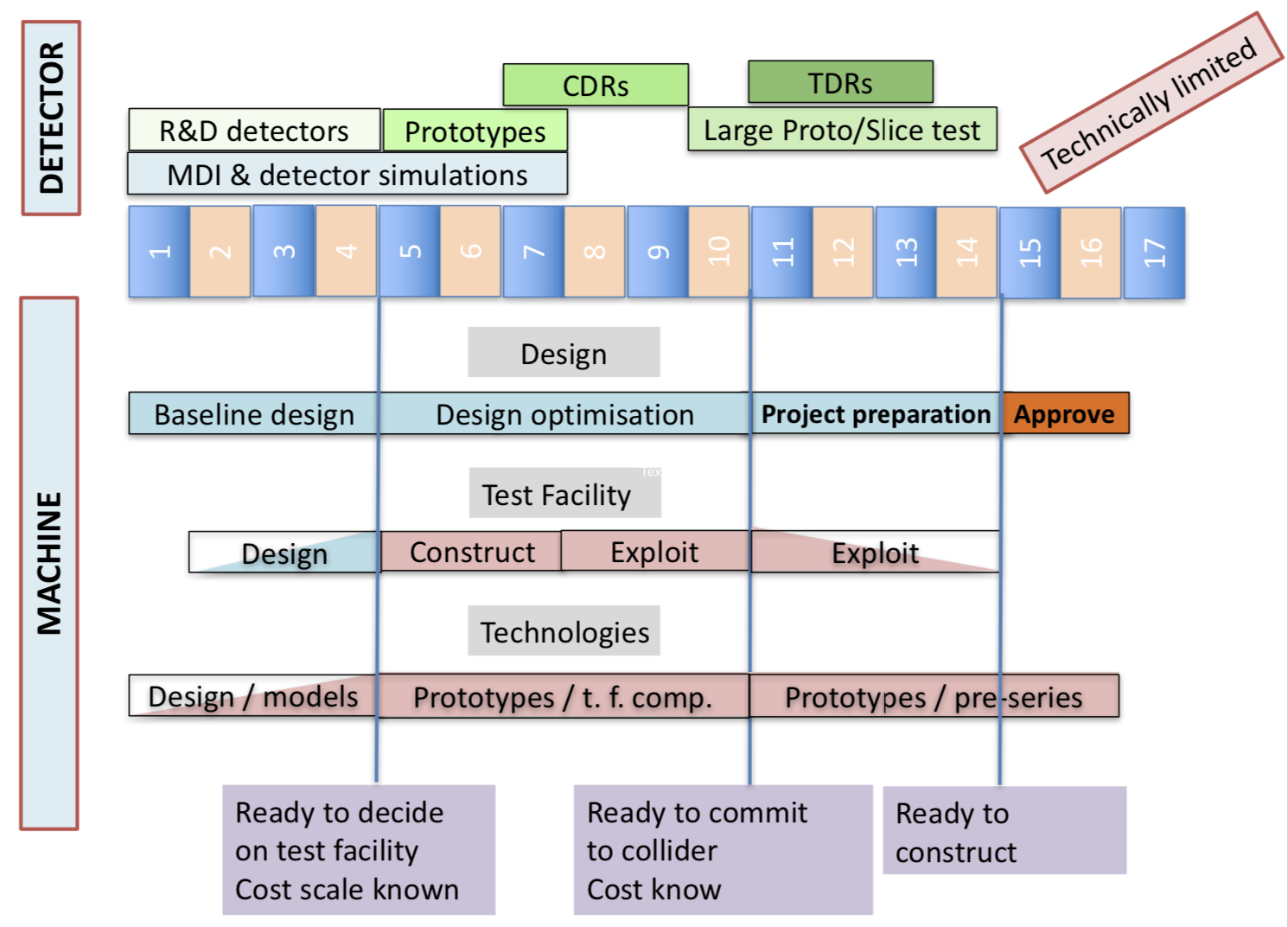}
 \caption{Potential technically-limited time-line for a muon collider.}
\label{fig:muonrd}
\end{figure}

A conceptual R\&D programme is illustrated in Figure \ref{fig:muonrd}. In the first stage, the baseline collider concept would be developed in parallel with the specification of a major R\&D project that could address the key technical issues, possibly including some physics goals using high-intensity muon or neutrino beams. This phase would require relatively modest resources. A consortium of interested institutes, including CERN, is starting to form. Due to the challenging design issues the project provides an excellent opportunity to nurture new ideas and  skills for the future. Depending on the results of this stage one could launch the second stage in which one or more test facilities could be built and operated. The collider design could be further optimised and a CDR developed, including costing, and used as the basis for a decision on whether to proceed to the next TDR stage, at which point a decision on construction could be taken. On this basis it is possible to imagine operation of a muon collider on the time scale of 30 years from now. The formation of a global collaboration will be essential to carry out the work coherently and efficiently. 

Although a muon collider offers the potential to push the energy frontier beyond the capabilities of any other conventional approach currently considered, the concept is not mature enough to be considered for construction today. A strong R\&D programme would be needed to develop it as a possible candidate for a high-energy physics project. This would be synergistic with R\&D on topics such as high-field superconducting magnets, fast-ramping magnets, efficient superconducting RF, and normal-conducting high-field RF; other topics, such as crystal collimation, might also be important.
Most critical issues like the cooling of the muon beams should be demonstrated at dedicated test facilities. 

\section{Plasma acceleration}
\label{sec:plasma}
Accelerating gradient is one of the key elements in determining the energy reach and the size of any accelerator. RF cavities accelerating gradients range from few MeV/m up to the value demonstrated in CLIC test facilities of over 100 MeV/m. Plasma-based particle accelerators [ID95, ID109], where accelerating fields are created by the collective motion of plasma electrons driven by lasers or particle beams, have shown capability of reaching an order of magnitude higher gradients. A myriad of applications of such technology would benefit from the compactness of the devices, HEP of course being  one of the interested fields.

\paragraph*{Present status of the field}

In the past decade significant progress has been made on both laser- and particle-beam-driven plasma accelerators. Although the achieved beam quality parameters have come closer to the design requirements, not all of them were demonstrated simultaneously. 
The European community of laboratories and universities active in plasma acceleration development has recently joined forces for the design study of EuPRAXIA (European Plasma Research Accelerator with Excellence in Applications)[ID109], aiming at designing a high-energy plasma based accelerator with high quality beam for multiple applications. Both laser-driven and particle-driven configurations are being considered, and the project is complemented by several test facilities. This study is of high interest also for possible future HEP applications.

\noindent{\it Laser-driven:} With the emergence of PW class lasers, electrons up to 8 GeV have been generated from 20 cm long plasma structures \cite{wim2} and staging of two independently powered modules has been demonstrated at the 100 MeV energy level \cite{wim3}. Methods for reduction of electron-beam energy spread from initially few percent to a few tenths of a percent have been developed \cite{wim4,wim5}, and emittances of 0.2\,$\mu$m have been measured. Strong plasma-based focusing elements have been developed and optimized which provide symmetric high gradients at the few kT/m level \cite{wim6,wim7,wim8} that are emittance preserving \cite{wim9}. Sophisticated diagnostics that are able to probe plasma profiles, accelerating and focusing fields as well as the properties of the emerging femtosecond electron beams, with high temporal resolution \cite{wim10} are being developed. Continuous operation of laser plasma accelerators operating at the 1 Hz repetition rate with energy stability at the few percent level has been demonstrated over extended times (>24 hrs), and an understanding of the origin of fluctuations has been obtained, permitting a path to feedback stabilization when increasing the repetition rate to kHz and beyond \cite{wim11}.

\noindent{\it Electron-driven:} Electron-driven plasma acceleration has shown energy doubling of a fraction of a beam from 42 GeV to 85 GeV in an 85 cm-long plasma column \cite{wim12}. Single-stage acceleration of a full electron beam by 9 GeV has been achieved \cite{wim13}, and sub-percent energy spread and 30\% energy transfer efficiency from the drive to the accelerated beam has been demonstrated \cite{wim14}. The acceleration of positron bunches was realised in a quasi-linear wake-field \cite{wim15} and energy gain of 5 GeV with an energy spread at the percent level was reached \cite{wim16}.

\noindent{\it Proton-driven:} The possibility of using relativistic proton beams as a driver was demonstrated at the CERN AWAKE experiment \cite{wim17}: although the proton beams have much longer bunch length, it was shown \cite{wim18,wim19} that they self-modulate in plasma to micro-bunches which can resonantly drive wakefields. The acceleration of electrons to multi-GeV energies in a 10\,m-long plasma was demonstrated \cite{wim17}.

For colliders, the ``hosing'' mechanism, equivalent to beam break-up, has been analyzed in depth and mitigation strategies have been developed to ensure suppression of this important instability \cite{wim20,wim21}. Tolerance studies on alignment of beams and structures are being conducted to understand the operational parameter regimes and challenges that need to be overcome \cite{wim22}. Advances have been made in speeding up the computational tools, with the aim of reducing the computation time from multiple days to minutes. This has been achieved via the development of advanced computational methods, reduced models that capture the essential physics, and through the emergence of higher-speed computers.

The latest generation of beam-driven plasma accelerators is utilizing superconducting accelerator technology and deploying feedback systems and advanced controls for the synthesis of finely tuneable and stable drive beams. They are now entering the era of precision measurements with sub-percent beam energy stability and fine beam-loading control for energy-spread minimization.  These activities are accompanied by the exploration of repetition rate limits for plasma wake-field processes facilitated by accelerator technology for multi-MHz and high average power operation.

\paragraph*{Challenges}

The next ten years of advanced accelerator development will focus on addressing challenges identified by the community:
\begin{enumerate}
\item	
Demonstration of reliable 24/7 operation of GeV-class plasma-based accelerators producing high quality electron beams with low energy spread (<\,0.5\%), low emittance (<\,1\,$\mu$m) and high charge per bunch (>\,30 pC) in femtosecond bunches.
\item	
Higher energy staging of electron acceleration with independent drive beams, equal energy, and charge-preserving beam capture; optimization of external injection methods.
\item	
Understanding mechanisms for emittance growth and developing methods for achieving emittances that are compatible with colliders.
\item Energy efficiency studies and optimization.
\item Demonstration of high average power operation for both beam- and laser-driven plasma accelerators;
\item Completion of a single electron acceleration stage at higher energy (10 GeV).
\item Demonstration and understanding of methods to accelerate a positron bunch with good quality.
\item Demonstration of scalable electron acceleration to 10s of GeV, with emittance control, via proton-driven plasma wake-field acceleration, leading to first high-energy physics applications.
\item Construction of dedicated advanced and novel accelerator facilities in order to deliver reliably high-quality, multi-GeV electron beams from a small number of stages.
\item Energize the advanced accelerator community, including the HEP community, towards an advanced collider.
\item Continual development of a comprehensive and realistic operational parameter set for a multi-TeV collider.
\item For the laser-driven scheme, the development of a MW-level (average power) laser system, which is 4--5 orders of magnitude higher than what is available today.
\item Further studies of beamstrahlung effects for c.m. energies up to 10 TeV. 
\end{enumerate} 

\paragraph*{Advanced Accelerator Concept roadmap}

 The primary long-term goal of a multi-TeV collider sets a timescale for the Advanced Accelerator Concepts (AAC) roadmap for completion of a TDR in the 2035--2040 interval \cite{wim1}.
 
\noindent{\it Near-term goals:} completion of a TDR for a potential early application in the 2025--2030 interval. During the innovation and discovery phase, the focus will be on generating and preserving high-quality electron bunches, methods for producing and efficient capture of positron beams, long-distance acceleration and scalability of plasmas for particle beam driven wake-fields, studies on energy efficiency, suppression of instabilities, and many other topics. Early applications include compact and possibly transportable Thomson scattering based gamma-ray sources, compact FELs, medical radiation delivering devices, and radiation-based inspection accelerators. In addition, laser plasma accelerators could be considered as injectors for next generation diffraction limited light sources.

\noindent{\it Mid-term goal:} the AWAKE technology could provide particle physics applications: AWAKE Phase 2 could be used for fixed-target experiments for dark photon searches as well as to deliver future electron-proton or electron-ion collisions with low luminosity.

\noindent{\it Long-term goal:} design of a high-energy electron/positron/gamma linear collider based on laser- and/or beam-driven plasma wake-field acceleration.

\section{Accelerators Beyond Colliders}

\paragraph*{Accelerator-based Neutrino Beams}

High-energy and high-beam-power accelerators are extensively used for neutrino physics research. The cost of leading facilities is second only to colliders. (For reference, total project cost (TPC) of J-PARC [ID76] is about 1.7 B\$, the cost estimate of the proposed ESS$\nu$SB [ID98] is 1.3 B\,Euro.)

At present, the leading operational facilities are the Fermilab Main Injector complex that delivers over 0.75 MW of 120 GeV protons on the neutrino target, and the J-PARC facility in Japan which recently approached 0.5 MW of the 30 GeV proton beam power.

Both facilities have multi-MW upgrade plans: Fermilab---through construction of a new 0.8 GeV PIP-II linac (to achieve over 1.2 MW by 2026) and then PIP-III (either an 8-12 GeV RCS or an SRF linac to achieve >2.4 MW in mid-2030's) [ID167, ID150]; 
J-PARC---through new faster magnet power supplies to reduce the cycle from 2.48s to 1.32s and RF upgrade to e.g. 1.3 MW by 2028
[ID76, ID158]. Far future plans of the J-PARC team include the construction of a new 8\,GeV Booster in addition to their existing 3 GeV RCS to attain 3.2 MW out of the Main Ring (MR), and even, still later, a new 9 MW 9\,GeV proton driver consisting of three SRF linacs (1.2 GeV, 3.3 GeV and 6.2 GeV) in the straight sections of the KEKB tunnel which will be available after the conclusion of the SuperKEKB operation. It is of note that the proton driver for the energy frontier muon colliders, like the proposed 14 TeV c.m. energy muon collider in the LHC tunnel, will need to operate at 2 to 4 MW average power level. 

Two issues are currently not resolved so that one cannot yet claim feasibility of these upgrades or any other multi-MW facilities: a) target; b) beam losses.
The long list of issues associated with high-power targets is further compounded by the fact that the required beam impacts are very short---1 to 10 $\mu$s. As a result, the countermeasures against radiation damage (DPAs) and thermal shock-waves at the existing neutrino targets and horns work only up to $\sim$0.8 MW of beam power. MW and multi-MW targets are under active development and prototyping. Ongoing R\&D programmes include studies of material properties, new forms (foams, fibers), new target designs (e.g., rotating or liquid targets). This R\&D activity has common elements with target R\&D for other HEP frontier projects (dark sector searches, linear collider, positron-based muon collider/LEMMA) and requires coordinated support.

The other most stringent limits on the beam power are set by the need to lower the fractional beam losses while increasing the beam power. The tolerable uncontrolled beam loss in accelerator enclosures is typically about 1 W/m, so the fractional beam loss must be kept under $\Delta N/N\sim C({\rm ircumference})\times 1 (W/m) /P({\rm ower})$. 
Such demands are in gross contradiction with the commonly observed increase of the $\Delta N/N$ with intensity, caused, e.g.\ by space-charge effects. 
These issues are very serious, are being addressed \cite{Shiltsev19}, and require long-term support of dedicated accelerators, machine studies, theory and modeling efforts.

There are four new proposals submitted to the EPPSU which show significant scientific promise and should be further studied---Protvino/ORCA [ID124], ENUBET [ID57], $\nu$STORM [ID154] and ESS$\nu$SB [ID98]. The first two call for moderate expansion of existing facilities and operation at sub-MW power levels. $\nu$STORM (cost estimate 160~M\,Euros if built at CERN) would produce beams of electron and muon antineutrinos from the decay of muons confined within a 580 m racetrack-shape storage ring. It requires only 156\,kW of 100 GeV protons and the major challenges of the proposal are the necessity of large aperture (0.5 m) magnets to accept most of the secondary muons and a sophisticated focusing lattice which should assure survival of about 60\% of muons with 10\% rms momentum spread over 100 turns. $\nu$STORM represents a very promising approach with great potential to boost R\&D toward energy frontier muon colliders.

ESS$\nu$SB is a technically very challenging proposal with a total cost of some 1.3 B Euros. It calls for an increase of the ESS beam power from the world record design value of 5 MW to 10 MW by increasing the accelerator duty cycle from 4\% to 8\%; the additional 5 MW are used to generate a uniquely intense neutrino Super Beam for the measurement of leptonic CP violation. Beside the upgrade of the SRF linac repetition rate from 14 Hz to 28 Hz, ESS should switch from operation with protons to operation with H$^-$ particles. An accumulator ring with 400 m circumference will need to be built to compress to the beam pulse to one microsecond. Due to very short beam pulse, the required 5\,MW neutrino target station will be much more challenging than the 5 MW ESS neutron spallation target. One should also expect---and address---very strong space charge effects both in the linac and in the accumulator ring. High-power H$^-$ stripping would also be a challenge.

\paragraph*{Gamma Factory}

The Gamma Factory [ID6] proposes to use the large relativistic boost of partially stripped ions stored in the LHC to convert a laser photon beam into a high-intensity gamma-ray beam. 
Domains of interest could be many, and the potential of the Gamma Factory via the different primary and secondary beams would be applicable to a wide range of research programmes. A concrete possibility being explored in the short term is the potential physics gains of transversely cooled iso-scalar ion beams in the HL-LHC. 
A first practical milestone was passed by the Gamma Factory team with the successful storage and acceleration of partially stripped ions in the LHC, paving the way to a proof-of-principle experiment at the SPS. The collaboration proposes a comprehensive experimental program for deployment and execution between 2021 and 2024.  A Letter of Intent has been submitted to the SPSC~\cite{GF-LOI}.

\paragraph*{BSM Searches with Accelerators}

\begin{figure}[t]
 \centering
 \includegraphics[width=0.75\textwidth]{\main/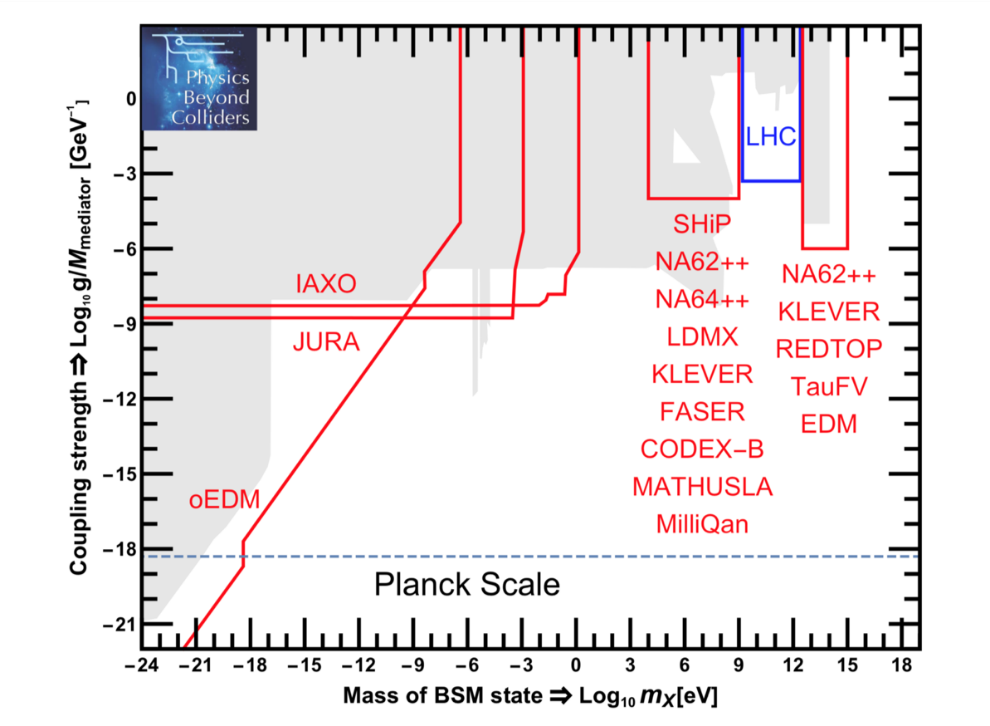}
\caption{BSM parameter space---PBC options shown, general breakdown maps onto worldwide options.}
\label{fig:bsm}
\end{figure}

Many of the BSM experiments could be based on existing accelerator facilities, like SPS, LHC, PSI, FNAL, etc. A few others would require new dedicated accelerators, such as an EDM ring. 
The use of ESS [ID164] for search for baryon number violation (NNBAR~[ID156]) is also proposed.
As per the conclusions of the Physics Beyond Colliders (PBC)/BSM reports [arXiv:1902.00260, ID20, ID42, ID60], and illustrated in Figure \ref{fig:bsm},  the BSM proposals broadly break down along the lines of:

\noindent{\bf \it  Sub-eV Axion/ALP searches} [ID112]: Here the main thrusts are well-established:
\begin{itemize}
\item	Helioscopes (BabyIAXO/IAXO); \vspace*{-2mm}
\item	Haloscopes (ADMX, MADMAX);\vspace*{-2mm}
\item Light-shining-through-wall experiments (JURA, STAX);\vspace*{-2mm}
\item Oscillating EDMs in protons or deuterons in an electrostatic ring (an idea rather than a proposal at the moment, cf.\ CPEDM measurement proposal below) [ID18].
\end{itemize}

\noindent{\bf \it MeV-GeV mass range:}
\begin{itemize}
\item	Direct detection WIMP searches;\vspace*{-2mm}
\item	Proton beam dump: new proposals (SHiP [ID12, ID129]), re-purposed existing experiments (NA62, MiniBooNE, SeaQuest);\vspace*{-2mm}
\item	Electron beam dump: NA64 [ID9], LDMX [ID36], BDX, etc.  \vspace*{-2mm}
\item	Long-lived particles at the LHC (FASER, MATHUSLA, CODEX-b, MilliQan).
\end{itemize}

\noindent{\bf \it > TeV mass range:}
\begin{itemize}
\item Ultra-rare or forbidden decays (KLEVER [ID153], TauFV [ID102], Mu3e, MEG, REDTOP 
[ID28], etc.);\vspace*{-2mm}
\item Search for a permanent EDM in protons/deuterons (CPEDM) or in strange/charmed baryons (LHC-FT).\vspace*{-2mm}
\end{itemize}

Proposed BSM/PBC experiments at CERN and elsewhere, e.g.~COMET at J-PARC [ID38], are compiled in Tables \ref{tab:BSMsummary} and \ref{tab:BSMOthers}, 
 respectively. The competitiveness of all CERN PBC options are explored in depth in the BSM paper \cite{Alemany:2019vsk}, which includes wide-ranging evaluation covering 11 benchmark cases. Also see the PBC summary report for the European Strategy Update [ID20].

Concerning the compatibility of experiments with beam from the CERN SPS, an SPS operation model has been fully developed for the North Area (NA). In general, the SPS could only support one major user in addition to standard NA operation. The operation of BDF/SHiP/TauFV [ID12, ID129, ID102] is compatible with standard NA operation (and LHC, AWAKE, HiRadMat, MD). The operation of BDF/SHiP/TauFV is compatible with standard NA/KLEVER operation with some compromise. Similar conclusions hold for NA/KLEVER with eSPS/LDMX. The eSPS/LDMX is not compatible with full BDF/SHiP operation. And either $\nu$STORM or ENUBET would not be compatible with BDF---a temporal separation will be necessary.

As for the possible time line of SPS-based experiments, the conventional SPS beam programme is foreseen for execution over LHC Run 3/Run 4. KLEVER and COMPASS$^{++}$ with RF separated beams would both require significant investment and development. Data taking would be in Run 4 at the earliest.
If approved, BDF/SHiP would target construction starting circa 2025. Ideally, BDF would make use of the injectors' LS3 (2025) to perform key civil engineering work during the associated NA stop. Realistically the earliest SHiP could start taking data is mid-Run 4.
If approved, eSPS would ideally target construction in the next 5 years. eSPS could potentially execute its programme before SHiP data taking, but this would require strong commitment by CERN within the next 2 years or so, followed by prompt technical design, approval, and execution.
$\nu$STORM and ENUBET, given their present limited implementation studies, and CERN's other commitments, cannot be envisaged to start construction until after 2030.

\begin{table}[h]
\begin{center}
\caption{Projects considered in the PBC-BSM working group categorised in terms of their sensitivity to a set of benchmark models in
a given mass range. The characteristics of the required beam lines, whenever applicable, are also displayed. Taken from the BSM report. }
\vspace*{5mm}
\label{tab:BSMsummary}
\begin{footnotesize}
\begin{tabular}{lllcc}
\hline
Proposal &    Main Physics Cases  & Beam Line  & Beam Type  & Beam Yield \\
\hline
\multicolumn{5}{c}{\textbf{sub-eV mass range:}} \\
\hline
IAXO  & Axions/ALPs (photon coupling)  & \multicolumn{2}{c}{axions from sun} &  -- \\
JURA  & Axions/ALPs (photon coupling)  & laboratory  & eV photons  & --  \\
CPEDM  & $p$, $d$ EDMs  & EDM ring  & $p$, $d$  & --  \\
 &  Axions/ALPs (gluon coupling) &  &  $p$, $d$  & -- \\
 
LHC-FT  & charmed hadrons EDMs  & LHCb IP  & 7 TeV $p$  &  -- \\
\hline
\multicolumn{5}{c}{\textbf{MeV-GeV mass range:}} \\
\hline
SHiP  & ALPs, Dark Photons, Dark Scalars &  BDF, SPS  & 400 GeV $p$ & $2
\times 10^{20}$/5 yr \\
 & LDM, HNLs, lepto-phobic DM, ... &  &  & \\
 
NA62++  & ALPs, Dark Photons,  & K12, SPS  & 400 GeV $p$  & up to $3
\times 10^{18}$/yr \\
 & Dark Scalars, HNLs  &  &  &  \\
 
NA64++  & ALPs, Dark Photons,  & H4, SPS  & 100 GeV   $e^\text{-}$  & 
$5\times 10^{12}$ eot/yr \\
& Dark Scalars, LDM  &  &  & \\
& + L$_\mu$ -- L$_\tau$  & M2, SPS  & 160 GeV $\mu$  & 
$10^{12}$--$10^{13}$ mot/yr \\
& + CP, CPT, leptophobic DM  & H2--H8, T9  &  ~40 GeV $\pi$, $K$, $p$  &  
$5\times 10^{12}$/yr \\

LDMX  & Dark Photon, LDM, ALPs,...  & SLAC/eSPS  &  8/16 GeV $e^\text{-}$  & $10^{16}$--$10^{18}$ eot/yr \\

AWAKE++  & Dark Photon  & AWAKE   & 30-50 GeV $e^\text{-}$  & $10^{16}$ eot/yr \\

RedTop  & Dark Photon, Dark scalar, ALPs  & CERN PS  & 1.8/3.5 GeV $p$ & $10^{17}$ pot \\

MATHUSLA  & weak-scale LLPs, Dark Scalar,  & ATLAS or  & 14 TeV $p$  &  3000 fb\textsuperscript{-1} \\
& Dark Photon, ALPs, HNLs  &  CMS IP &  & \\

FASER  & Dark Photon, Dark Scalar, ALPs,  & ATLAS IP &  14 TeV $p$  & 3000 fb\textsuperscript{-1} \\
 & HNLs, B--L gauge bosons &  &  & \\

MilliQan  & milli charge &  CMS IP &  14 TeV $p$  & 300-3000 fb\textsuperscript{-1} \\

CODEX-b  & Dark Scalar, HNLs, ALPs,  & LHCb IP  & 14 TeV $p$  & 300 fb\textsuperscript{-1} \\
 & LDM, Higgs decays  &  &  & \\
\hline
\multicolumn{5}{c}{\textbf{> TeV mass range:}} \\
\hline
KLEVER & $K_L \rightarrow \pi^0\nu\bar{\nu}$  & P42/K12  & 400 GeV $p$  & 
$5\times 10^{19}$ pot/5 yr \\
TauFV  & LFV $\tau$ decays  & BDF  & 400 GeV $p$  & $\cal{O}$(2\%) of BDF $p$ \\

CPEDM  & $p$, $d$ oEDMs &  EDM ring  & $p$, $d$ &  - \\
 & Axions/ALPs (gluon coupling)  & & $p$, $d$  & - \\

LHC-FT  & charmed hadrons MDMs, EDMs  & LHCb IP  & 7 TeV  $p$ & \\

\hline
\end{tabular}
\end{footnotesize}
\end{center}
\end{table}


\begin{table}[t]
\begin{center}
\caption{Selection of projects complementary to those considered by PBC-BSM working group. Note that the experiments are in different phases: proposals; construction; upgrades. (BD -- beam dump; SX -- slow extraction; DD -- direct detection) }
\vspace*{3mm}
\label{tab:BSMOthers}
\begin{footnotesize}
\begin{tabular}{lllcc}
\hline
Proposal &    Main Physics Cases  & Beam Line  & Beam Type  & Beam Yield \\
\hline
\multicolumn{5}{c}{\textbf{sub-eV mass range:}} \\
\hline

MADMAX & Axions & Lab: dielectric/ & cosmos & -- \\
 &  & B field &  &  \\
STAX & ALPs  & LSW sub-THz  &  cosmos  & --\\
 &   &  photons &   & \\
MAGIS100$\rightarrow$1K & Dark sector & Atom interferom. & cosmos &  -- \\
 &  & (FNAL) &  &  \\
\hline
\multicolumn{5}{c}{\textbf{MeV-GeV mass range:}} \\
\hline

DARKSIDE-20k  & WIMP DD LAr & LNGS & cosmos & 200 t.yr \\
\quad$\rightarrow$Argo  &  &  &  &  $\rightarrow$ 3000 t.yr\\
DARWIN & WIMP DD LXe & possibly LNGS & cosmos & 200 t.yr \\
LUX-ZEPLIN & WIMP DD LXe & SURF & cosmos & 15 t.yr \\
XENONnT & WIMP DD LXe & LNGS  & cosmos &  20 t.yr\\
CRESST-III Ph.\ 2  & WIMP DD, $A'$ & LNGS & cosmos &  -- \\
  & CaWO$_4$ &  &  &   \\
SuperCDMS & WIMP DD Ge & SNOLAB & cosmos & -- \\
SEAQUEST BD & LDM & FNAL MI & 120 GeV $p$ SX &  
$1.44\times 10^{18}$
 pot/2 yr \\ 
MiniBooNE-DM & LDM & FNAL Booster & 8 GeV $p$ & 
$1.9\times 10^{20}$ pot \\
BDX & LDM, $A'$ & JLAB  & 11 GeV $e$  & 
$10^{22}$ eot \\
DarkLight &	 $A'$ & JLAB  & 100 MeV $e$ on $p$ & 5 mA \\
SENSEI & LDM   & CCDs (FNAL/  & cosmos & -- \\
 &  & SNOLAB)  &  &  \\
MAGIX & $A'$ & MESA  & 150 MeV $e$  & $\sim 10^{35}$
cm\textsuperscript{-2}s\textsuperscript{-1} \\
MMAPS &  $e^\text{-}e^\text{+}\rightarrow \gamma A'$ & Cornell synch. & 5.3 GeV $e^\text{+}$ SX &  $L_{av} = 10^{34}$ 
cm\textsuperscript{-2}s\textsuperscript{-1}    \\	
BELLE-II & ALPS, $A'$ & \multicolumn{2}{c}{SuperKEKB  \hfill $e^\text{-}$$e^\text{+}$ $\sqrt{s}$ = 10.58 GeV}  &   50 ab\textsuperscript{-1}\\


\hline
\multicolumn{5}{c}{\textbf{> TeV mass range:}} \\
\hline

Mu3e I/II & CLFV $\mu^+\rightarrow e^+e^+e^-$  & PSI HiMB &  $\mu$ & 
$2\times 10^{9}$ stopped $\mu$/s \\

MEG II & CLFV $\mu^+\rightarrow e^+\gamma$  & PSI &  $\mu$ &  
$O$($10^{10}$
$\mu$/s) to exp. \\

KOTO(+)  & $K_L \rightarrow \pi^0\nu\bar{\nu}$ & J-PARC MR SX   & 30 GeV $p$    &  
$\sim 10^{8}$ K/spill   \\
  &  &   & 50$\rightarrow$100 kW   &  at 50 kW   \\

Mu2e/Mu2e-II &  CLFV $\mu^-N\rightarrow e^-N$ & FNAL & 8/<4 GeV  $p\rightarrow\mu$ & 7.7 kW pot (Ph.\ I)  \\
 &   &  &  & $6.7\times 10^{17}$~$\mu$  \\

COMET I/II & CLFV $\mu^-N\rightarrow e^-N$ & J-PARC MR  & 8 GeV $p\rightarrow\mu$ & 
$1.5\times 10^{16}$--$1.8\times 10^{18}$ \\
 &   &   &  &  stopped $\mu$ \\

\hline
\multicolumn{5}{c}{\textbf{Accelerator-driven $\nu$ experiments}} \\
\hline
DUNE & $\nu$ & \multicolumn{2}{c}{FNAL MR (PIP-II) \hfill 60-120 GeV $p$}  & 1.1$\rightarrow$1.9$\times 10^{21}$ pot/yr\\
T2HK & $\nu$ & J-PARC MR & 30 GeV $p$ & $\sim 10^{21}$ pot/yr \\
\hline
\end{tabular}
\end{footnotesize}
\end{center}\vspace*{-5mm}
\end{table}

\section{Energy management}

All proposed HEP projects will consume large amounts of energy, $\cal{O}$(TWh) per year. On the other hand, an increasing fraction of sustainable energy sources like wind and photovoltaics in the future energy mix will result in strong variations of the supply of electrical energy. It is the HEP community's responsibility to develop sustainable models and optimized technologies in terms of energy consumption, aiming also at exporting improved technologies for other applications in the society.

One way to mitigate the impact of HEP facilities is to actively manage their energy consumption. The aim should be to avoid high loads on the grid during low supply conditions, and instead using preferentially ``excess energy''. The possibilities of energy management using dynamic operation of facilities and energy storage systems should be studied in more detail.

It is necessary to invest R\&D effort into improving the energy efficiency of HEP facilities through critical technologies. In certain areas, such R\&D will have an immediate impact on research and other facilities operated today, and the savings in energy consumption may be used to co-finance the investments. Certain improved technologies may also serve society directly. These relevant fields of R\&D include (but are not limited to) optimized magnet design, high-efficiency RF power generation, improved cryogenics, lower loss or higher operating temperature SRF cavity technology, beam energy recovery, district heating using recovered heat, and energy storage.

\begin{figure}[ht]
 \centering
 \includegraphics[width=0.75\textwidth]{\main/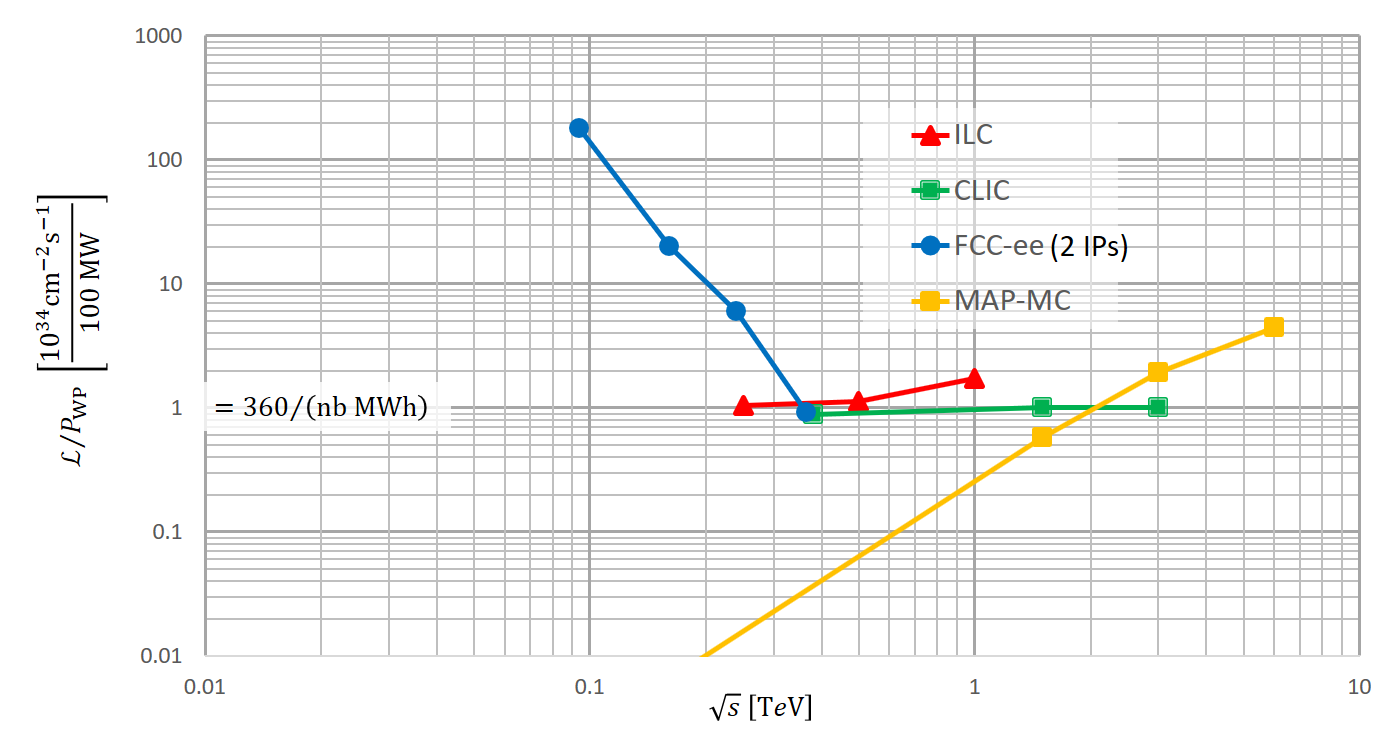}
\caption{One possible figure of merit for future colliders: luminosity per supplied primary energy (from E.~Jensen and M.~Seidel's presentation in Granada~\cite{jensen-granada}).}
\label{fig9}
\end{figure}

Proposals for future lepton colliders include linear colliders using normal-conducting or superconducting technology (CLIC and ILC), a large diameter synchrotron (FCC-ee) and muon colliders (MAP-MC). Defining the energy efficiency as integrated luminosity per supplied primary energy, these proposals have different dependencies on the targeted centre-of-mass energy. Figure \ref{fig9} shows the trends for these proposals, based on the numbers from the proposals submitted to the European Strategy Update. The absolute numbers presented should however be taken with caution, since the level of confidence and the degree of refinement for those proposals differ substantially. The overall trend however is based on fundamental principles and gives some indication. It shows the steep decrease of efficiency for the synchrotron radiation dominated FCC, and almost flat energy dependence for the linear colliders.

It should equally be taken into consideration that a highly optimized machine running   at its performance limit may have more significant downtime than a robust machine running very stably. In other words, energy efficiency of individual components is nothing without high system availability. The optimization should be on the overall performance, considering also the trade-offs between different subsystems and the effective Mean Time Between Failures.

\section{The role of National Laboratories in the European Strategy}
\label{sec:NationalLabs}

In the global vision of the network of European Research Infrastructures (RIs), national laboratories have achieved in the last decade a relevant role as centres providing facilities and specific competences which complement and support large scale projects for accelerators and experiments in particle, nuclear and astro-particle physics.

This collaboration scheme has been successfully exploited in several recent scientific European endeavors, such as at the LHC, EU-XFEL, ESS, and ITER. Present collaborations span over a large variety of activities: high intensity sources, new superconducting or HTS high-field magnets for future colliders, innovating technologies for accelerators, RF or plasma-based, only to mention few examples.

The grid of resources, in facilities and competences, available throughout Europe's RIs
represents a formidable asset. Not only, do they contribute to innovation with a rich variety of technological know-out, potentially transferable to high-tech European industry; they also provide a wide spectrum of activities in fundamental physics. This is of particular relevance when RIs in Europe are confronted with their capabilities to access Horizon Europe funds. Open Access laboratories collaborating with European industry boost the potential of applied physics research. Moreover, ingenuity of researchers fosters the development of local small-scale experiments, which span over a very large spectrum of questions in fundamental physics, providing the multi-faceted approach which makes vivid and competitive the research communities.

A strong collaboration between technological facilities at European laboratories and industry can be seminal for the realization of the several scientific projects on-going in Europe and elsewhere. As an example, the AMICI Horizon 2020 Design Study is charged to strengthen the capabilities of European companies to compete as qualified suppliers of components for accelerators, large superconductor magnets, and in the development of innovative applications in sectors such as healthcare, security, cultural heritage and space.

National laboratories play a vital role in particle physics research and innovation, and recent years have seen a strong growth in the number of RIs that are contributing to large European research projects. They must be recognized as long-term strategic investments at all levels, deeply rooted in the particle physics community, providing a model for their sustainability, and indispensable both for enabling and developing excellence in the European Strategy.
 
They also have an impact on skills and education agendas, increasing the competences of researchers, and students through their outreach activities. They steadily improve the perception and understanding of science and technology in society at large. 

National laboratories enrich the region where they are located and as such, they are important
as contributors to technology transfer and to competitiveness. The health of particle physics in Europe also depends on the vitality of these laboratories in the Member States: Systematic collaborations should be reinforced or established, making optimal use of the available infrastructures and human resources.

\section {Complementarities and synergies with other fields}
\label{sec:ACC-synergies}
Accelerator technology development has always been driven by HEP needs. A diversity of other applications benefit from and synergistically contribute to the advances in the field. Some of the most significant examples are mentioned here.

{\bf Fusion energy:} IFMIF and DONES accelerators, as complementary facilities to ITER project, are based on high-power and high-reliability proton linacs. They share with HEP project the challenges on high-intensity high-reliability proton and deuteron beam injectors; superconducting RF cavity technology in a high-power high-reliability context; high-intensity beam dynamics and beam halo understanding; high-power non-interceptive beam instrumentation; reliability modelling of particle accelerators. Their development has impact on industries building the systems, which reverts in capacity and provider availability.

{\bf HTS} \cite{HTSapps} technology has a wide variety of applications in medicine, science and power systems engineering on top of the high field magnets, these last being also used in fusion power plants. As an example, HTS can apply in the field of electric power systems in cables, motors, generators, and transformers where superconductors replace resistive conductors, plus superconducting magnetic energy storage (SMES) and fault-current limiters (FCLs).

{\bf Medical applications} of accelerators in isotope production, radio and hadron-therapy \cite{PTCOG} are profiting of developments being carried out in HEP laboratories, and examples are the latest designs of SC gantries \cite{gatoroid, Schippers}, or of medical detectors \cite{GEM, DECTRIS}. 

{\bf Photonics and neutronics} share with HEP technologies and challenges. Common to all these fields are the high reliability needs and the data analytics evolution which exponentially grow in parallel to the high-rep.\ and detector capabilities. In photonics the examples are: the diffraction-limited storage rings with very low emittance (nanobeam, stability, magnet and vacuum technology, etc.), the FELs (Linac technology, highly brilliant beam production and preservation, stability, etc.), the development of compact sources, where the CLIC technology has given birth to the  CompactFEL concept \cite{COMPACTLIGHT}. While in neutronics SNS, ESS or the China Spallation Neutron Source (CSNS) \cite{CSNS} share with HEP all challenges of high-power linacs and targets. 

{\bf Plasma acceleration}, as mentioned above, promises developments for compact facilities with a wide variety of applications, in medicine, photonics, etc., compatible with university capacities and small and medium sized laboratories.

An important aspect of this Strategy update is to recognize the potential impact of the development of accelerator and associated technology on the progress in other branches of science, such as astroparticle physics, cosmology and nuclear physics. Moreover, joint developments with applied fields in academia and industry have brought about benefits to fundamental research and may become indispensable for the progress in our field.



\chapter{Instrumentation and Computing}
\label{chap:inst}

Developments in the area of instrumentation and computing enable tool-driven revolutions that can open the door to future discoveries.  This is only made possible if appropriate support for innovation exists.  The type of support required includes not only financial support but also access to shared infrastructures, the existence of effective organizational networks, appropriate support and recognition of the workforce engaged in instrumentation and computing R\&D activities, and  structures through which the community can build a relationship with industry.  Furthermore, while university-based research teams do engage in R\&D activities~[ID68], strong support from national labs and larger institutions has been shown to be essential.
 
 The success of the LHC programme and the ongoing development activities in preparation for HL-LHC upgrades demonstrate the ability of the particle physics community to take on large and long term projects. The challenges of future experiments, however, often scale with the complexity of the physics goals.  Therefore, it is as important as ever that the particle physics community maintains and further strengthens a research and development ecosystem that stimulates and supports innovation in both instrumentation and computing.  

\section{Particle physics instrumentation}

\begin{figure}
\begin{center}
\includegraphics[width=0.99\textwidth]{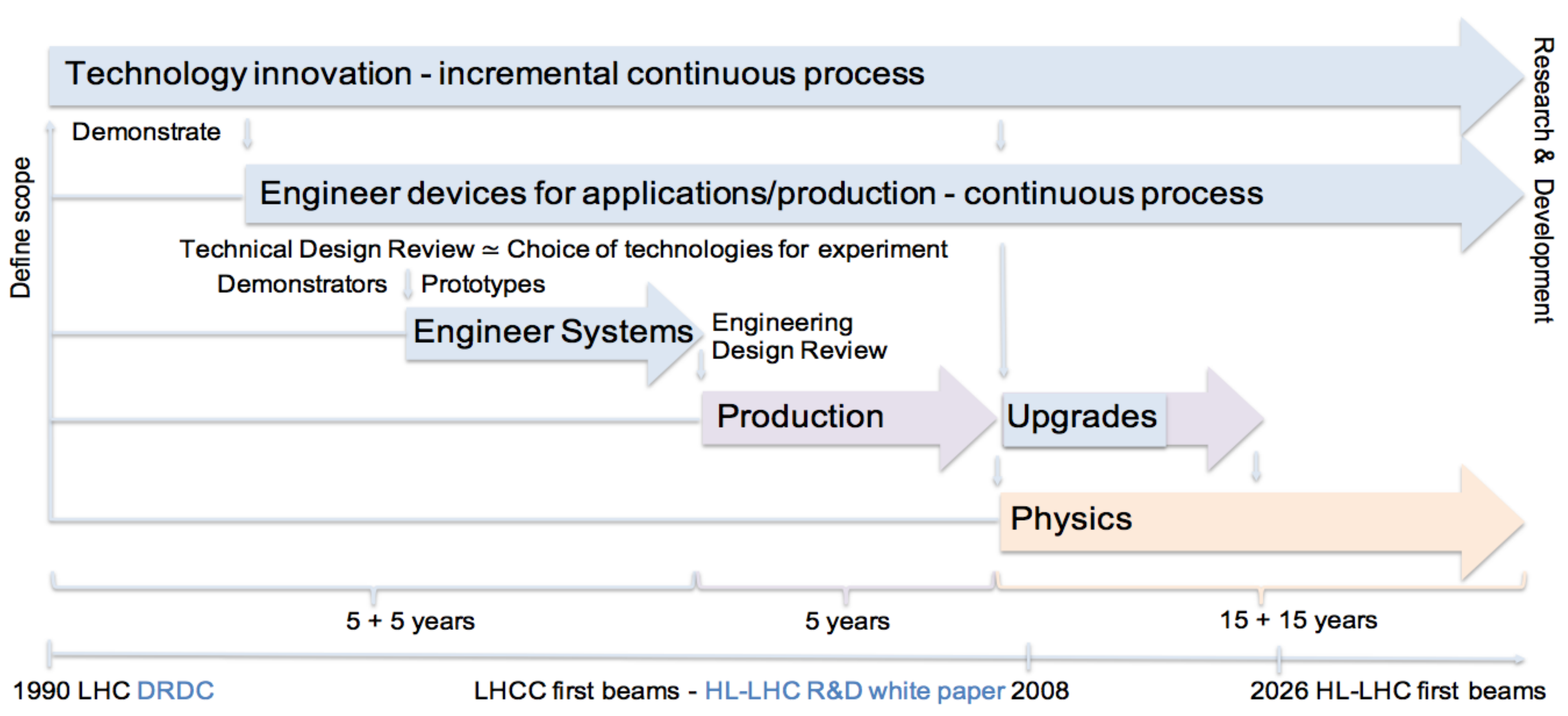}
\caption{\label{fig:RandD} Illustration of the typical timescale involved in the technology development cycle of particle physics experiments leading up to HL-LHC~\cite{bib:DC-talk}.}
\end{center}
\end{figure}

The use of a new detector technology in an experiment is the result of a development cycle that includes different types of activities, from generic R\&D activities, R\&D activities guided by the needs of future projects, 
focused R\&D activities for an approved experiment, production/industrialization and installation/commissioning of the technology.  As illustrated in Fig.~\ref{fig:RandD}, the work leading up to the HL-LHC detector upgrades indicates that the development cycle of a new technology is typically a decade or more.  It is therefore important for the community to plan ahead and ensure that appropriate support exists for all stages of a technology development cycle.  In 2018, an ECFA Detector Panel survey~[ID68] of the particle physics community found that 87\% of respondents reported engagement in R\&D activities, and that 75\% of these activities were carried out within the context of  present/future experiments (e.g.\ ATLAS, CMS, FAIR, etc.), 18\% within a consortium (e.g.\ AIDA, RDx, etc.), and 7\% of the activities were identified as generic R\&D. Looking ahead, the key to enabling new discoveries will be the community's ability to maintain an appropriately diversified portfolio of different types of R\&D activities that can evolve over time to adapt to the needs of the community.  For example, as HL-LHC upgrade activities move into construction and commissioning, it is imperative that the community maintains momentum in R\&D activities with a focus on solving the known technological challenges of the next generation of experiments.

\subsection{Challenges and technologies for next generation experiments}
\label{sec:detector:challenges_technologies}
Next generation experiments include those at colliders, fixed-target and beam-dump facilities, as well as dedicated projects for the search of elusive particles. Some of the proposed programmes exploit synergies between particle physics and other fields, in particular astroparticle physics and cosmology.  While the landscape of these proposed next-generation experiments is broad in terms of both physics goals and detector technologies, the technological challenges are well-defined~\cite{bib:LS-talk}. These include, depending on the physics goals and experimental conditions, micron scale spatial resolution and low mass, picosecond hit time resolution, high-performance photodetectors (also operating at cryogenic temperature and low dark current), radiation tolerance, large number of channels, high readout speed, and large sensitive area at low cost. 
Moreover, the need for combined features (4D tracking, 5D imaging) becomes more and more urgent.  The time scales spanned by these future programmes, ranging from few years to decades, constitutes a challenge in itself, in addition to the complexity and diversity of the required technologies.

Recent R\&D efforts have already highlighted promising technologies, some of which can apply to different detectors through specific engineering developments~\cite{bib:FS-talk}. Table~\ref{tab:tech} provides a summary of some of these promising technologies and their possible specific applications to address future experimental challenges.


\begin{table}
   \caption{Summary of promising technologies and their possible specific detector applications to address experimental challenges of approved and future projects.}
    \centering
    \begin{tabularx}{\textwidth}{>{\raggedright}XY|Y|Y|Y|Y|} 
     \multicolumn{1}{c}{} & \multicolumn{5}{c}{Technologies} \\ \hhline{~=====}
     \multicolumn{1}{c|}{} & Solid state & Gas & Scintillator & \makecell{Noble liquid } & Cherenkov  \\ \hhline{~=====}
     \noalign{\medskip}\noalign{\smallskip}
     Vertex /\newline Tracker & 
     \multicolumn{5}{l}{\makecell[l]{{\bf Challenges}: high spatial resolution, high rate/occupancy, fast/precise timing,\\ radiation hardness, low mass, 4D tracking.}} \\ \cline{2-6}
     \multicolumn{1}{c|}{} & \makecell{Planar, 3D,\\ (D)MAPS~\ref{f1}, \\LGAD~\ref{f2},\\ (HV-HR)\\ CMOS~\ref{f3}}  & 
     \makecell{TPC~\ref{f4}, \\DC~\ref{f5}} & \makecell{SciFi~\ref{f6} +\\ SiPM~\ref{f7}} & &  \\ \cline{2-6}
    \noalign{\medskip}\noalign{\smallskip}
     %
     Calorimeter &      \multicolumn{5}{l}{\makecell[l]{{\bf Challenges}: high granularity, radiation hardness, large volume, excellent\\hit timing, PFA/dual-readout capability, 5D imaging.}}\\ \cline{2-6}
     \multicolumn{1}{c|}{} &
    \makecell{Si sensors\\ sampling} & \makecell{RPC~\ref{f8} or\\ MPGD~\ref{f9} \\ sampling} & \makecell{Tile/fibers +\\ SiPM sampl., \\ homogeneous \\ crystals\\ (e.g. LYSO)} & \makecell{LAr\\ sampling} & \makecell{Quartz fibers\\ sampling in \\ dual-readout} \\  \cline{2-6}
    \noalign{\medskip}\noalign{\smallskip}
     Muon\newline detector & \multicolumn{5}{l}{\makecell[l]{{\bf Challenges}: large area, low cost, spatial resolution, high rate.}}\\ \cline{2-6}
    \multicolumn{1}{c|}{}  & &
    \makecell{MPGD, \\RPC, DT~\ref{f10} \\ MWPC~\ref{f11}} & \makecell{Scint+\\ WLS fibers \\+ SiPM} & &  \\  \cline{2-6}
    \noalign{\medskip}\noalign{\smallskip}
     PID /\newline TOF&  \multicolumn{5}{l}{\makecell[l]{{\bf Challenges}: high photon detection efficiency, large area photodetectors,\\ thinner radiator, timing resolution $\leq$~10~ps, radiation hardness.}}\\ \cline{2-6}
    \multicolumn{1}{c|}{}  & 
    \makecell{LGAD\\(timing)} & \makecell{TPC, DC,\vspace{0.2cm}\\MRPC~\ref{f12}\\(timing) } & \makecell{} & & \makecell{RICH~\ref{f13},\\ TOF~\ref{f14},\\TOP~\ref{f15},\\ DIRC~\ref{f16}}  \\  \cline{2-6}
    \noalign{\medskip}\noalign{\smallskip}
     Neutrino /\newline Dark Matter & \multicolumn{5}{l}{\makecell[l]{{\bf Challenges}: high photon detection efficiency, very large volume,radio \\ purity, cryogenic temperature, large area photodetectors.}} \\ \cline{2-6}
    \multicolumn{1}{c|}{}  & 
     \makecell{Si, Ge} & \makecell{TPC} & \makecell{liquid scint., \\ scint. tiles /\\ bars} & \makecell{single/dual-\\phase TPC} & \makecell{water/ice + \\ mPMT~\ref{f17} }\\  \cline{2-6}\noalign{\bigskip}\cline{1-2}
     \multicolumn{3}{@{}p{0.49\textwidth}@{}}{%
       \raggedright\footnotesize\vspace{-0.3cm}
        \begin{enumerate}[ref=\textsuperscript{\arabic*},parsep=-3pt]
        \item (Depleted) Monolithic Active Pixel Sensor\label{f1}
        \item Low Gain Avalanche Detector\label{f2}
        \item (High Voltage - High Resistivity) CMOS\label{f3}
        \item Time Projection Chamber\label{f4}
        \item Drift Chamber\label{f5}   
        \item Scintillating Fiber tracker\label{f6}
        \item Silicon Photomultiplier\label{f7}
        \item Resistive Plate Chamber\label{f8}
        \item MicroPattern Gaseous Detector\label{f9}
        \end{enumerate}} &
        \multicolumn{3}{@{}p{0.49\textwidth}@{}}{%
        \raggedright\footnotesize\vspace{-0.3cm}
        \begin{enumerate}[ref=\textsuperscript{\arabic*},parsep=-3pt]
        \setcounter{enumi}{8}
        \item Drift Tube\label{f10}
        \item Multi-Wire Proportional Chamber\label{f11}
        \item Multi-gap Resistive Plate Chamber\label{f12}
        \item Ring-Imaging Cherenkov detector\label{f13}
        \item Time-Of-Flight detector\label{f14}
        \item Time-Of-Propagation counter\label{f15}
        \item Detection of Internally Reflected Cherenkov\label{f16}
        \item multi-anode Photo-Multiplier Tube\label{f17}
        \end{enumerate}}
    \end{tabularx}
    \label{tab:tech}
\end{table}

Hybrid and monolithic pixel sensors for vertex and tracking detectors achieve high resolution through high granularity, low material, radiation resistance and high data rate~\cite{Garcia-Sciveres:2017ymt}. 
In particular, Monolithic Active Pixel Sensors (MAPS), developed starting from the CMOS imaging sensor technology, have been initially deployed in test-beam infrastructures and then adapted to the needs of high-energy physics experiments, finding applications in HL-LHC detector upgrades (ALICE [ID110, ID46]) and are subject to process optimization for most next generation experiments (see e.g.\ CLIC~[ID146]~\cite{Dannheim:2673779}). This technology is promising for its intrinsic possibility of reducing the thickness of sensors while maintaining high performance and is suited for large area assembly at low production cost.
A new avalanche silicon detector concept with a low gain, known as a Low Gain Avalanche Detector (LGAD), seems to be a favourable option to achieve, with a given pitch, the time resolution of 30~ps~\cite{Curras:2673324}.
However, further effort is necessary to be able to achieve the combination of required performance. 
The trend towards monolithic technologies also tends to blur the boundaries with ancillary components, such as read-out circuitry, and special care must be taken in optimizing them in combination with the sensitive components.

Present and future challenges in calorimetry include excellent photon detection, high granularity, radiation hardness, large volume, and ultimately the possibility of exploiting either the high granularity and particle flow technique~\cite{Sefkow:2015hna, THOMSON200925} or the dual read out calorimetry approach~\cite{RevModPhys.90.025002}.
Research and development in the field has resulted in the use of many different technologies, depending on the application~\cite{Sefkow:2015hna}: crystals, scintillator-based sampling, silicon-based sampling, and liquid noble gas calorimeters [ID16].
Homogeneous, shashlik-type~\cite{shashlik} and spaghetti-type~\cite{Acosta:1991ap} calorimeters are considered for the LHCb electromagnetic calorimeter upgrade, while a technology similar to the ATLAS tile calorimeter is under investigation for the hadronic barrel calorimeter of an \FCChh detector. The baseline choice for an electromagnetic calorimeter for an \FCChh detector is instead based on liquid-argon technology, which was also recently considered for a detector for \FCCee.
Another option being investigated for \FCCee is the dual readout fibre calorimeter.
A silicon-tungsten calorimeter is proposed for \CLIC, the CLD~\cite{CLD} detector for \FCCee, and is a promising option for a \FCChh calorimeter in areas with a radiation level compatible with the technology. Research and development for the silicon and tile calorimeter technology will profit from the experience of the CMS High Granularity Calorimeter (HGC)~\cite{CMSCollaboration:2015zni} for HL-LHC.

Particle identification is crucial for flavour physics, with the biggest challenge for the proposed future detectors being an extreme timing resolution.
The Belle\,II experiment at SuperKEKB [ID11] features a RICH with quartz radiator (Time-Of-Propagation or TOP) in the
barrel region, which targets a single photon timing resolution of 80~ps~\cite{Abe:2010gxa}.  The Cherenkov technology is also considered for the LHCb detector upgrade, which features a RICH detector with a time resolution down to 10~ps and a TORCH (Time Of interally Reflected
CHerenkov light) detector with a time resolution close to 70~ps per photon resulting in an effective time resolution of 15~ps per track~\cite{Aaij:2244311}.

In the realm of gaseous detectors, Micro Pattern Gas Detectors (MPGDs [ID87]) have proven successful in many LHC experiments (e.g.\ GEMs, Micromegas) and are foreseen for HL-LHC detector upgrades.   They remain a valid asset for addressing future experimental challenges such as high granularity and precise timing.  Several new MPGD designs are being developed to facilitate production and evolve the technology to pixel devices with integrated readout electronics.  Another example of gaseous detector is the glass RPC with semi-digital readout that has been designed to be used as part of a hadronic calorimeter for ILD~\cite{Grenier_2014}[ID107].

Experimental challenges of very large mass detectors, such as those used for neutrino physics and dark matter searches, include in many cases extremely high radiopurity, large volume cryogenic systems and high photon detection efficiency over very large area. Here, SiPMs will likely be one of the key enabling technologies for LAr (e.g.\ DarkSide-20k~[ID62], DarkSide-LM~[ID62], Argo~[ID62], DUNE~[ID126]) and LXe-based detectors (e.g.\ DARWIN~[ID97], nEXO), achieving higher photo-detection efficiency, much better single-photon resolution, and operating at lower bias voltage than traditional PMTs. Silicon photo-multipliers also feature better radiopurity and can be efficiently integrated into tiles that cover large areas.  
Neutrino physics experiments (e.g.\ Hyper-K~[ID158]) could significantly benefit from the development of novel multi-anode PMTs (mPMT) providing large sensitive area coverage per module with directional information.  The Global Argon Dark Matter Collaboration (GADMC) programme~[ID62] will require the use of large volumes of low-radioactivity argon depleted in $\rm {}^{39}Ar$, necessitating the development of an underground argon extraction and purification plant as well as a dedicated cryogenic distillation column.

For all types of particle detectors, the integration of advanced electronics and data transmission functionalities plays an increasingly important role and is a significant challenge in itself. Advances in enabling microelectronics and optoelectronics technologies are driven by industry.  Technologies used in particle physics often lag behind industry by many years, at the risk of becoming obsolete.  
The use of new microelectronic technologies for particle physics applications has a high development cost and long learning and qualification time. In this context it is important for the community to focus on the development of a limited number of technology nodes (e.g.\ 28 nm, and 17 nm or lower) expected to be available for a long time.  It is equally important to put in place a critical mass of expert personnel in support for the community. Finally, development activities must be coordinated to share prototyping and production costs.  This includes, for example, community support for design tools and foundry submissions through international coordination programmes.
Rapid development in FPGA processor power and I/O bandwidths are also enabling complex online feature extraction and increased readout flexibility; however, this comes at a cost of increased firmware development complexity which requires high-level expertise.

Future experimental projects also face a large number of diverse engineering challenges, for example, in the areas of system integration, power distribution, cooling, mechanical support structures and production techniques.  Concerning mechanics and cooling infrastructures, depending on hadronic or leptonic colliders, the general requirements span from high radiation hardness to low mass and large dimension. Gas and microchannel cooling is considered, when possible, for tracking systems.  Carbon fibre cryostats and ultrathin superconducting magnets are also being investigated.

With such a plethora of options for technological development, it is mandatory to keep in mind some of the lessons learned from past R\&D~\cite{bib:DC-talk}. First, it must be recognized that efforts in R\&D activities are never a loss, i.e.\ even if a technological innovation is not retained for a specific experimental programme, it is nevertheless available for other projects.  As such, it is important to note that a close interplay between future project studies, the R\&D collaborations, and European Commission projects is needed to ensure an efficient strategic alignment whilst preserving accessibility of results and know-how to the wider community.  In the context of R\&D activities guided by the needs of future projects, technological specifications must be well-defined and documented, and need to consider all aspects of the final applications: compatibility with other systems and their global performance, operating conditions including their uncertainties and required margins, production capabilities and costs. For common components developed to address the needs of several detectors and experiments, anticipating the implementation of different configurations is particularly important. 
Thorough definition and documentation of test protocols, including test conditions as close as possible to final operation, and earliest possible start of radiation tolerance tests are fundamental to minimize risks.
Finally, given that components may be discontinued by vendors or become prohibited due to new legislation, the market behaviour must be constantly monitored and some degree of diversity of solutions must be maintained.

\subsection{Generic detector R\&D}

Generic detector R\&D refers to research activities that seek to push back the limits of technology, and is not uniquely driven by the needs to fulfill specific experimental requirements.  Since generic detector R\&D has the potential to bring about tool-driven revolutions, it is essential that the community maintains the ability to carry out these types of research activities, even through the design, planning and execution of large experimental projects.  An example of this is the outcome of continued improvements made in solid-state sensors which has led to the possibility to add timing detectors to LHC experiments upgrades; an improvement that was not originally foreseen and that is now opening up the ability to achieve better pile-up rejection.  

Generic technology innovation often emerges from synergies within the field of particle physics, with other fields of science, or with industry, and in most cases, is an incremental, continuous and long-term process.  Furthermore, generic R\&D, by itself, can typically be costly and often requires access to various specialized infrastructures (see Sect.~\ref{sec:detector:infrastructure}).  As a result, programmes such as AIDA2020~\cite{bib:AIDA2020} and ATTRACT~\cite{bib:ATTRACT} play an important role in enabling and supporting generic R\&D activities. 
It is therefore important to ensure that these types of programme be appropriately supported and have also the ability to expand, as needed, in order to preserve and stimulate the community's potential for innovation.

\subsection{Test facilities, infrastructures and tools}
\label{sec:detector:infrastructure}

The development of novel particle physics
instruments requires specialized infrastructures, tools and access to test
facilities that all come at a substantial
cost.  National labs and large institutions play a central and 
important role in supporting the community
by providing access to, and user support for, these facilities, infrastructures and tools.  

\paragraph*{Test-beam and irradiation facilities}
Test-beam facilities are vital for the characterization, calibration and commissioning of new instruments.  A list of test-beam facilities\footnote{An AIDA2020-supported database of available test-beam facilities world wide is under development~\cite{bib:testbeamDB}.} in the world currently available to the particle physics community is presented in Table~\ref{tab:testbeam}.  These facilities offer different particle species, energies and beam structures that are complementary to each other.   In the past few years, the demand for access to the largest test-beam facilities such as those at CERN, DESY, Fermilab and SLAC has remained high.  The CERN test-beam facilities, for example, are at present used at full capacity.  It is expected that the development of instrumentation for approved and future projects will maintain, even possibly increase, the need of the community to have access to these types of facilities.  Yet, the future medium- to long-term availability of some of the test-beam facilities currently used by the community is at present uncertain.  Furthermore, parts of these facilities, for example at CERN, are also ageing and will require adequate maintenance and/or upgrade in the coming years to continue to support the community.

\begin{table}
    \caption{List of test-beam facilities in the world, as of May 2019. Only beam lines with beams of energies higher than 100~MeV are included [Courtesy of C. Rembser and H. Wilkens].}
    \centering
    \footnotesize
    \begin{tabularx}{\textwidth}{XYYYY} \hline\hline
    Laboratory & Number of\newline beam lines\phantom{X} & Particles & Energy  & Availability \\ \hline\hline
    CERN/PS\newline (CH) & 
    2 & 
    $e, h, \mu {\rm (sec.)}$ & 
    0.5 - 10~GeV &
    \multirow{2}{2.5cm}{9 months/year continuous except winter shutdown, typical duty cycle \\ PS$\sim$ 1-3\% \\ SPS 20-40\%}
    \\  \cline{1-4}
    CERN/SPS\newline (CH) & 
    4 & 
    $p$ (prim.) \newline $e,h$,$\mu$ (sec.)\newline $e,h$ (tert.) \newline Pb ions (prim.) \newline other ion species \newline (out of fragmented primary Pb ions)  &
    400~GeV\newline 10-400~GeV \newline 10-200~GeV \newline \newline 20-400~GeV\newline proton equiv.\newline (z=1) & 
    \\ \hline
    DAFNE BTF\newline Frascati\newline(IT) &
    2 &
    $\rm e^+/e^-$ both prim. and sec. &
    25-750~MeV\newline Rep. rate 50~Hz\newline 1-40~ns\newline $\rm 1-10^{10}$~part./pulse &
    typically 25-35 weeks/year \\ \hline
    DESY\newline (D) & 
    3 &
    $\rm e^+,e^-$(sec.) &
    1-6~GeV\newline max 100~kHz &
    11 months/year,\newline duty cycle $\sim$ 50\% \\ \hline
    ELPH (Sendai)\newline (JP) &
    2 &
    $\gamma$ (tagged)\newline $\rm e^+,e^-$(conv.) \newline \newline &
    0.7-1.2~GeV \newline0.1-1.0~GeV\newline rate < 500~kHz\newline (typ. 2~kHz) &
    2 months/year \\ \hline
    FERMILAB FTBF \newline (US) &
    2 &
    $p$ (prim.) \newline $e,h$,$\mu$ (sec.)\newline $h$ (tert.) &
    120~GeV\newline 1-66~GeV \newline 200-500~MeV &
    24~hrs/day, duty cycle 6\% \\ \hline
    IHEP (Beijing)\newline (CH) &
    2 & 
    $e$ (prim.)\newline $e$ (sec.) \newline $p$,$\pi$ (sec.) &
    1.1-2.5~GeV\newline 100-300~MeV \newline 0.4-1.2~GeV &
    3 months/yr, duty cycle depends on BEPCII operation mode \\ \hline
    IHEP (Protvino) \newline (RU) & 
    5 & 
    $p$ (prim.)\newline $p,K,\pi,\mu, e$ (sec.)\newline C-12 (prim) &
    70~GeV\newline1-45~GeV\newline 6-300~GeV &
    2 months/yr,\newline duty cycle(U-70 machine) 15-30\% \\ \hline
    PSI/SBL \newline (CH) &
    2-4 &
    $\pi^{\pm},\mu^\pm,e^\pm,p$ &
    50-450~MeV\newline rate < $10^9$~Hz\newline 20~ns structure &
    6-8 months/yr \\ \cline{2-5}
    
    PSI/PIF \newline (CH) &
    1 & $p$ & 5 - 230 MeV\newline $\rm I_{\rm max}$ = 2-5 nA & 11 months/year, typ. weekends \\ \hline
    
    SLAC\newline (US) &
    0 & 
    $e$ (prim.)\newline $e$ (sec.)\phantom{xxx} & 
    2.5-15~GeV\newline 1-14~GeV &
    9 months/yr, \newline duty cycle 50\% \newline [No beam in 2019] \\ \hline
    SPRING-8\newline Compton Facility\newline (JP) &
    1 & 
    $\gamma$ (tagged)\newline $e^\pm$(conv.)\phantom{xx} &
    1.5-3.0~GeV\newline 0.4-3.0~GeV &
    $>$60 days/yr \\ \hline
    University of Bonn\newline ELSA \newline (D) &
    1 &
    $e^-$ &
    1.2-3.2~GeV\newline rate$\sim$ 500~Hz-\newline 1~GHz &
    upon request, \newline typ. $\sim$ 30 days/yr \\ \hline
    Univ.\ of Mainz \newline MAMI \newline (D) &
    3 &
    \phantom{XX}{$e^-$}\newline $\gamma$ \phantom{xX}&
    $\rm E_{e^-,\gamma}$ < 1.6~GeV \newline $\rm I_{e^-}$ < 100~$\mu$A &
    upon request, \newline typ. $\sim$ 30 days/yr \\ \hline
    \end{tabularx}
    \label{tab:testbeam}
\end{table}

Irradiation facilities are used to characterize the radiation hardness, ageing effects and performance of detectors, electronics components, systems and materials, under high particle flux and fluence such as those expected at accelerators and in accelerator-based experiments.  Table~\ref{tab:irradiation} summarizes information about a representative sample of irradiation facilities in the world used by the particle physics community for the testing and qualification of detector and accelerator components.  In addition, many other facilities are available to the community and information about these facilities can be found in the AIDA2020-supported world-wide database of irradiation facilities~\cite{bib:db}.  Current irradiation facilities fulfill, and have been optimized in some cases for, the HL-LHC radiation environment.  The particle fluence expected at some of the future accelerator projects will however be more than a factor of 10 larger (e.g. fluence of $10^{17}$ particles/$\rm cm^2$ at \FCChh for 30~ab$^{-1}$).  In the future, the testing of materials, devices, systems and electronics under this type of extreme radiation environment will be critical and require access to irradiation facilities with increased particle fluxes, larger radiation areas for simultaneous tests of many components, and longer testing campaigns~\footnote{For example, the recently formed CERN Radiation Test Facilities Steering Group (RTF-SG) is charged with facilitating the information exchange at CERN concerning the operation and possible future extension of the radiation test facilities at CERN~\cite{bib:RTF-SG}.}.

\begin{table}
    \caption{Representative list of some of the different types of irradiation facilities in the world commonly used by the particle physics community for the testing and qualification of detector and accelerator components~\cite{bib:cern-radiation,bib:irradiation-talk}. This is not an exhaustive list since irradiation services can be sometimes supplied by some member institutions of an R\&D project, the community makes use of industrial services or of medical accelerators for reasons of convenience, availability, etc., and in other cases, irradiation must be performed with specific particles/energies.
    Several other facilities are available to the community and can be found in the AIDA2020-supported world-wide database of irradiation facilities~\cite{bib:db}.}
    \centering
    \footnotesize
\begin{tabularx}{\textwidth}{SSYM}\hline\hline
    Access provider\newline and infrastructure  & Particles & Energy, flux, etc.  & Availability \\ \hline\hline
    CERN/IRRAD \newline (CH) &
    $p$ &
    23~GeV\newline flux: $\rm 1-3\times 10^{10}~p/cm^2/s$ &
    May-November \newline [Closed during LS2] \\ 
     & & & \href{http://ps-irrad.web.cern.ch/}{Link to more info} \\ \hline
    KIT/KAZ\newline (D) &
    \makecell{$p$\\ \\ X-ray}&
    25.3~MeV \newline current 2~$\mu$A\newline flux$\rm\sim 2.5\times 10^{13}~p/cm^2/s$ \newline dose rate up to 18~kGy/h&  \href{https://www.etp.kit.edu/english/irradiation_center.php}{Link to more info}
     \\ \hline
    UoB/MC40\newline cyclotron\newline (UK) &
    $p$ &
    up to 40~MeV, typ. 27~MeV \newline current < 2~$\mu$A, typ. 0.1-0.5~$\mu$A \newline flux $\rm\sim 10^{13} p/cm^2/s$ & \href{http://aida2020.web.cern.ch/content/uob}{Link to more info}
     \\ \hline
    CERN/CC60 \newline (CH)&
    $\gamma$ &
    1.17~MeV, 1.33~MeV\newline (10~TBq $\rm {}^{60}Co$ source)\newline $\rm\sim~3~Gy/h$ at 1~m & 
    All year \newline \href{https://hse.cern/content/main-features-new-calibration-hall}{Link to more info}
     \\ \hline
    CERN/GIF++ \newline (CH) &
    \makecell{$\gamma$\\ \\ \\ (+$\mu$)} &
    0.662~MeV\newline (14~TBq $\rm {}^{137}Cs$ source)\newline dose-rate $\sim$0.5~Gy/h at 1~m\newline selectable flux\newline typ. 100~GeV, $\rm \sim 10^4$ part./spill & All year \newline (+ 6-8 weeks/yr \newline of SPS operation) \newline \href{https://gif-irrad.web.cern.ch/gif-irrad/}{Link to more info} 
    \\ \hline
     CERN/CHARM \newline (CH) &
     mixed-field \newline (23~GeV $p$ \newline on Cu/Al \newline target) & n: thermal - HE\newline HEH: > 100~MeV\newline Lateral: $\rm 10^7-10^{10}~HEH/cm^2/h$\newline Long.:$\rm 10^8 - 10^{12}~HEH/cm^2/h$\newline  homogeneous field over $\rm \sim 1~m^2$ &
     May-November \newline [Closed during LS2] \newline \href{http://charm.web.cern.ch/}{Link to more info} 
     \\ \hline
     UCL/CRC \newline (BE) &
     \makecell{heavy \\ ions} &
     high LET or high penetration cocktail\newline flux $\rm\sim$ few - $\rm 10^4~part/cm^2/s$ & \href{https://uclouvain.be/en/research-institutes/irmp/crc/parameters-and-available-particles.html}{Link to more info}
     \\ \hline
     JSI/TRIGA\newline reactor \newline (SI) &
     $n$ &
     fast \& thermal,\newline flux $\rm 1-4\times 10^{12}~n/cm^2/s$\newline thermal, regenerated fast,\newline flux $\rm < 5\times 10^8~n/cm^2/s$ &\href{http://www.rcp.ijs.si/ric/reactor-a.htm}{Link to more info}
     \\ \hline
     Los Alamos/\newline Sandia\newline (US) &
     \phantom{XXX}$p$ \newline \newline \phantom{XXX}$n$\newline \newline $\gamma$ &
     800~MeV, \newline $\rm 5\times 10^11$~part./pulse at 1~Hz \newline 0.1~MeV to > 600~MeV, \newline $\rm 10^6~n/cm^2/s$ above 1~MeV per pulse \newline dose rate $10^{-8}$ - 10~Gy/s & \href{https://www.sandia.gov/research/facilities/annular_core_research_reactor.html}{Link to more info}
     \\ \hline
    \end{tabularx}
    \label{tab:irradiation}
\end{table}

The recent and ongoing testing campaign of novel instruments in preparation for HL-LHC has shown that the time availability to testbeam and irradiation facilities can in some cases be a limiting factor.  While access to many facilities are available to the community through transnational access programmes such as that provided by AIDA2020, others are user-paying facilities.  The ability to make extensive use of these facilities also importantly depends on the quality of user and technical support, as well as the availability of high-quality beam instrumentation, test set-ups, and related facility management software infrastructure (e.g.\ see~\cite{bib:testbeam-software}). For example, the beam telescopes and associated DAQ infrastructure developed under the EUDET~\cite{bib:EUDET} and AIDA~\cite{bib:AIDA} frameworks have proved invaluable to the community, providing common tools for test-beam operation, which significantly lower the cost and complexity of performing these tests.  
Continued and coordinated support of a European network of test-beam and specialized irradiation facilities, including user support and the development of common tools, is of the utmost importance for the community.

Established in 2014, the CERN neutrino platform provides the neutrino community with support for detector R\&D, tests, construction, as well as access to a charged particle test beam infrastructure.  This initiative, along with the creation of a neutrino group within the CERN EP department, is an important complementary contribution to the US and Japanese neutrino programmes and is providing coherence to the European neutrino community.  As such, this type of initiative could be considered for other complementary research programmes.



\paragraph*{Infrastructures}
Typical infrastructures required for instrumentation development
include, but are not limited to, clean rooms, probe stations,
bonding/packaging tools, radioactive/laser sources, 
spectro-photometer and spectro-fluorometers, specialized electronics and data acquisition equipment, equipment for impurity/dopant, content and profiles in solid-state devices, etc. While some of this type of infrastructure can be available
at some institutions, the cost can be prohibitive to many and often can be
single-use for a particular project.  National labs and large institutions play a critical role in providing access to various specialized infrastructures required for detector R\&D.  The sharing of such infrastructures 
is important for the community.   Furthermore, efficient usage of existing
infrastructures requires the existence of platform(s) 
disseminating the information, and potentially coordinating the
infrastructure usage.   One crucial element in the use of such specialized infrastructures is the ability to support the technical personnel required to run them.  This is a particularly difficult problem to address given that a significant portion of research budget is moving into project-based funding that only lasts for the duration of the project and must be finalized at its end. To make the use of existing infrastructures effective, laboratories and universities need to invest in people who can run and maintain them. Also, the economic model for accessing the infrastructures needs to be adapted to the changing scenarios.

\paragraph*{Specialized tools} In the development of novel technologies, the use of specialized tools is
essential to minimize, for example, the number of prototype
iterations.  These specialized tools include modelling tools, advanced
detector simulation tools, tools to estimate radiation doses, and
design and validation of electronics system and mechanical 
components.  While some of these tools are free, others have costly
licenses.  National labs and larger organizations can play an important role in
providing the community with access to, and expertise in the use of, some of these specialized tools.  Here again, the challenge of supporting the technical personnel required to efficiently exploit complex tools exists and must be addressed. It is also important to note that the particle physics community has
also a role to play in the 
development of some of these tools (e.g.\ GEANT4~\cite{bib:geant1,bib:geant2,bib:geant3}), ensuring that the
tools meet the needs of the community.  A strong synergy exists between access to test-beam facilities providing well-defined conditions and the ability to improve simulation models of particle interactions with materials that are important for the design of novel technologies.





\subsection{R\&D coordination}
\label{sec:coordination}

Results of the 2018 ECFA Detector Panel survey~[ID68] of the particle physics community clearly demonstrate that expertise in instrumentation R\&D is distributed over many institutions (universities, national and international laboratories).  As such, international coordination of R\&D activities is critical to maximize the scientific outcomes of these activities and make the most efficient use of resources. More specifically, the coordination of activities provides
\begin{itemize}
\item the ability to efficiently share and streamline work thereby limiting duplication of effort;
\item opportunities for the development of, and access to, common tools, infrastructures and services;
\item access to a large network of information and world-wide expertise in different areas;
\item wide dissemination of knowledge and results;
\item unique environment for the training of the next generation of experts;
\item a visible framework for institutions.
\end{itemize}

The community has a strong track record of successful implementation of different collaborative structures used to coordinate R\&D activities.  These include, for example,
\begin{itemize}
\item CERN R\&D programmes such as the DRDC and White Paper Theme 3 R\&D programmes which laid the foundations for the construction and subsequent upgrades of the LHC detectors;
\item European Commission funded projects such as EUDET, AIDA, AIDA2020 and ATTRACT which provide precious support to the particle physics community to develop new devices, common ancillary electronics, beam test and irradiation facilities, and also play an important community-building role in establishing coherence at the European level;
\item other R\&D collaborations (e.g.\ CALICE, ProtoDUNE) that, in some cases, have the flexibility to evolve in scope.
\end{itemize}

In addition, while strictly speaking not a coordinating body, it is important to note the existence of the ECFA Detector Panel~\cite{bib:EFCA-detector} that has a mandate to take on a reviewing role and provide advice on detector development efforts for projects at accelerator and non-accelerator experiments in particle and astroparticle physics.  It is primarily concerned with large detector R\&D projects involving many laboratories and requiring significant resources.

Looking ahead, there is a clear need to strengthen existing R\&D collaborative structures, and create new ones, to address future experimental challenges of the field post HL-LHC. Consistent with this view, CERN has proposed an R\&D programme [ID16] that concentrates on advancing key technologies rather than on developing specialized detector applications. 
Many of the developments are proposed to be be carried out jointly with external groups, also exploiting existing collaborative structures like the RD50 and RD51 collaborations, and with industrial partners.
In addition, it is generally acknowledged that the community must continue to vigorously pursue the development of future EC-funded projects, which have played a central role in enabling past R\&D activities.  
These projects all have strong leverage on matching funds. 

Some of the challenges in maintaining the community's ability to carry out detector R\&D research through collaborative structures include:
\begin{itemize}
    \item Evaluation of, and ability to secure, the appropriate level of funding.  For example, there is a feeling among some stakeholders that more funding for innovation could have enabled devices with improved performance and/or lower cost for HL-LHC (e.g.\ more radiation tolerant Crystals, SiPMs, CMOS sensors with fast readout, etc.).  In general, R\&D has high added value compared to the overall cost of the scientific programmes.
    \item Resources management. While collaborative structures can build on several smaller-scale national and local initiatives with various funding sources, this implies that (at least some) resources are in the hands of contributing institutions.  As a result, estimating efforts needed/invested is not easy and unforeseen withdrawals of efforts can impact schedules. More formal agreements throughout different R\&D phases could help address this challenge.
    \item Maintaining effectiveness.  Experience has shown that an international independent R\&D review process of consortia and programmes is important for maintaining their effectiveness and securing funding.
    \item Integration of R\&D not targeted at CERN-hosted projects.
    \item Engagement with the world-wide (non-European) community.
\end{itemize}

To help address some of these challenges, it would be advantageous 
for the community to define a global R\&D roadmap that could be used to support proposals at the European and national levels.  This community roadmap could, for example, identify grand challenges to guide the R\&D process on the medium- and long-term timescales, as well as define technology nodes, broad enough to be used as a basis for creating R\&D platforms.  The preparation of this document is an important step in this direction, on the European level and even beyond. 
\subsection{Synergies and opportunities}
\label{sec:detector-synergies}

Instrumentation development for particle physics is both a driver for, and a beneficiary of, progress made in other areas, within the field of particle physics, in other fields of science, and industry. 

Within the field of particle physics, technologies developed under generic R\&D programmes or with the aim to address common experimental challenges often provide a boost in innovation that suits the needs of detector upgrades and/or novel designs. Examples of this include the development of liquid noble-gas Time Projection Chambers (TPCs) for dark matter and neutrino experiments (DarkSide [ID62], DARWIN [ID97], DUNE [ID126, ID131]), liquid scintillator with photomultipliers (JUNO [ID19]), and pure water Cherenkov with photomultipliers (SuperKamiokande). A highly-granular electromagnetic calorimeter, addressing similar needs as the CMS electromagnetic calorimeter upgrade (HGC), is foreseen for the Light Dark Matter eXperiment (LDMX) at eSPS [ID42, ID36]. A fast radiation-hard electromagnetic calorimeter using GAGG crystals, similar to the electromagnetic calorimeter for LHCb Upgrade II, is designed for the fixed-target experiment TauFV [ID42, ID102]. The two experiments also share the design of the TORCH particle-identification detector mentioned in  Sect.~\ref{sec:detector:challenges_technologies}.

Across other fields of science, in particular with the astrophysics and cosmology communities, strong synergies for detector R\&D as well as large civil, vacuum  and cryogenic infrastructures have been identified~\cite{bib:CDV-talk} (also see e.g.\ [ID60, ID64]). 
The ECFA Detector Panel, with its advisory mandate spanning both particle and astro-particle physics, is uniquely positioned to identify such synergies.  Particle physics has also been the driver of important developments in the fields of medicine (e.g.\ medical imaging, radiation treatments, etc.) and biology. For example, the Medipix and Timepix chips~\cite{bib:medipix} originally developed to meet requirements of solid-state tracking detectors and TPCs in particle physics have found applications in medical imaging, space dosimetry, material analysis (e.g.\ characterisation of pharmaceuticals, evaluation and synthesis of new materials, detection of counterfeit drugs). To facilitate the cross-fertilization of innovations across disciplines, the development of technology-centred R\&D programmes, as well as platforms for the exchange of information and ideas, should be encouraged.  


The relationship with external partners is as bidirectional as within scientific fields. 
Products of particle physics are highly specific with relatively small and cyclic markets. Their developments and production rely on partners working in other scientific and societal fields using similar devices, particularly for imaging technologies, for electronics and mechanical systems. In some cases, collaboration and knowledge transfer with industrial partners (e.g.\ access to the bases of fabrication process, understanding the mechanisms of industrialization) are absolutely necessary for ensuring mass production and for reducing the risks of market instability.
Successful initiatives of collaboration and cooperation with industry partners (e.g.\ ATTRACT~\cite{bib:ATTRACT}, ERDIT~\cite{bib:ERDIT}) should be encouraged.
Valuable candidates for partnership range from material science laboratories, foundations or start-ups in technology innovations for relatively small productions, to medium size companies for larger productions and scientific/technical knowledge transfer (e.g.\ Hamamatsu Photonics for providing large quantities of silicon sensors for HL-LHC, TowerJazz for producing CMOS MAPs).
It is subject of debate whether involving industrial partners at an earlier stage of the scientific R\&D projects would be beneficial to particle physics. 
For an effective synergy, forms of partnership with industry that go beyond the client-supplier model should be explored, carefully considering regulatory and patent issues.

\section{Computing and software for particle physics}

The scientific outcomes of an experiment are made possible by the development of an efficient computing and software infrastructure. Historically, the development of these infrastructures for particle physics has benefited from Moore's law, the use of commodity hardware, and highly  distributed systems.  This, however, no longer holds true. Nowadays, the landscape of computing infrastructure for particle physics is rapidly changing and becoming more diverse.  The particle physics community faces important challenges in this area and the current computing and software models must evolve to meet the needs of approved and future experiments. 

\subsection{Computing challenges of current and next generation experiments, and evolution}
Since early 2000s the particle physics community has adopted a large-scale distributed computing model whereby Grid technologies integrate computer centres world-wide to form a single computing and storage infrastructure supporting, for example, the LHC experiments~\cite{bib:IB-talk}. At the same time, the use of heterogeneous facilities consisting of Cloud computing, HPC and High Level Trigger (HLT) farms has offered extra opportunistic capacity, though often at the cost of significant development effort and suited to a subset of work-flows. As a result, presently, the particle physics community has access to an amount of computing resources meeting the needs of its physics programmes.
 
\begin{figure}
    \centering
    \includegraphics[width=0.49\textwidth]{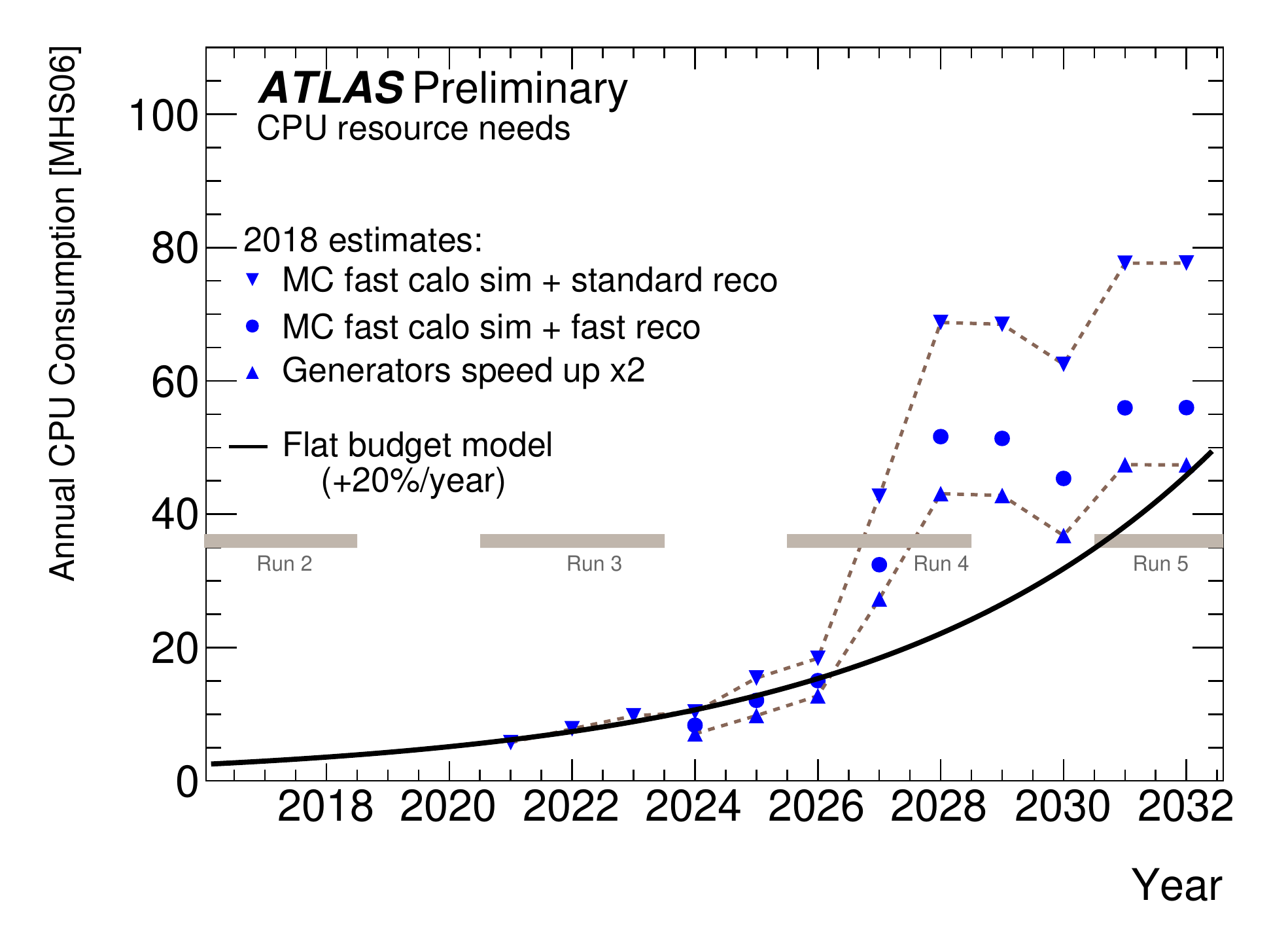}\hfill
    \includegraphics[width=0.49\textwidth]{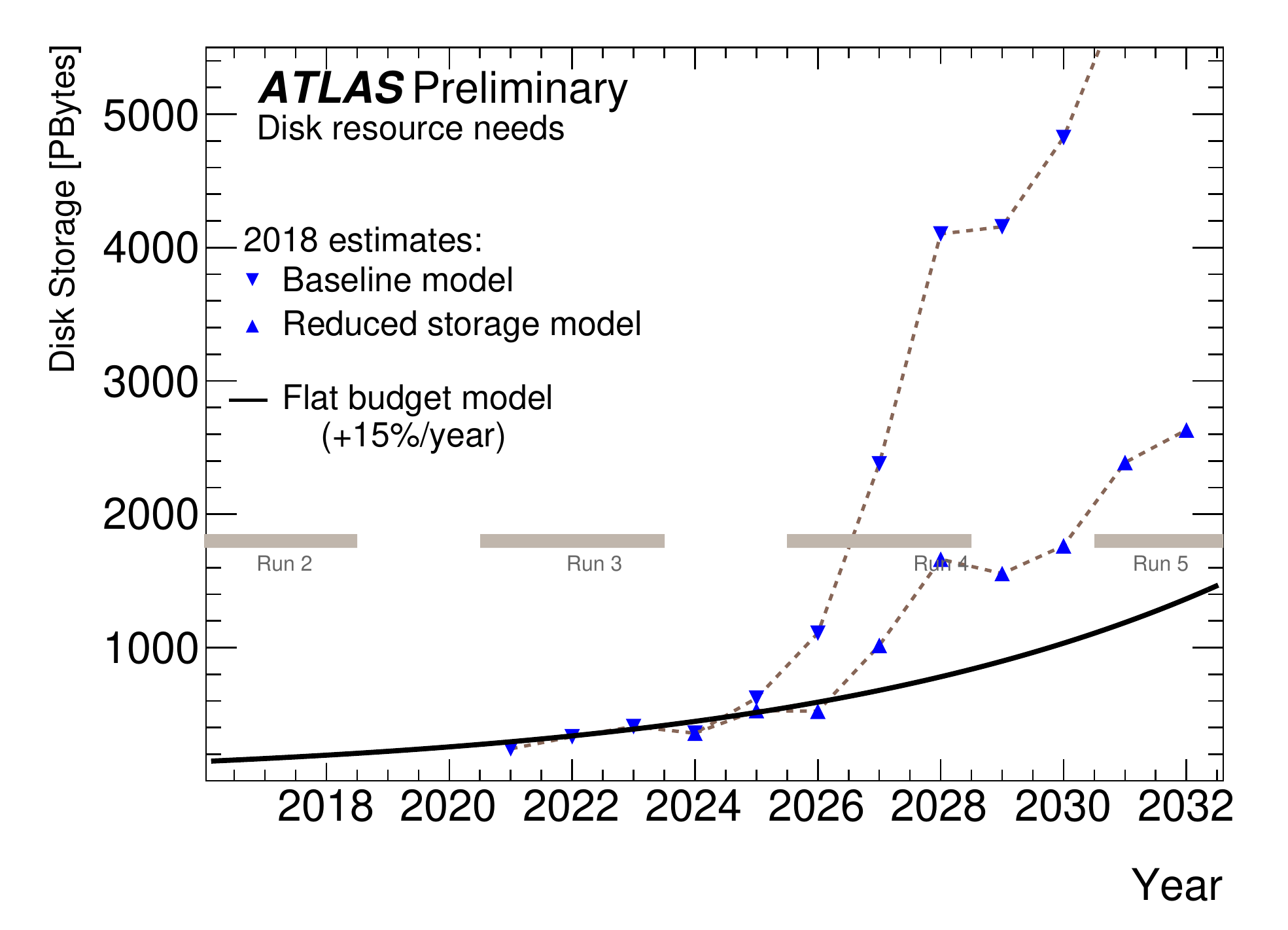}
    \caption{Estimated CPU (left) and total disk resources (right) needed by the ATLAS experiment as a function of time for both data and simulation processing.  Plots taken from~\cite{bib:atlas-computing}.
    }
    \label{fig:atlas-computing}
\end{figure}

In the coming years, however, the science programmes at the HL-LHC Run 4 and beyond, Belle-II at SuperKEKB, future circular and linear colliders (\FCC, \CEPC~[ID29], \ILC, \CLIC, etc.), and large neutrino experiments (DUNE, JUNO, etc.)~[ID126], will together require about an order of magnitude more 
computing resources presently available, while increase in funding for computing is not expected~\cite{bib:MK-talk}.  For example, Fig.~\ref{fig:atlas-computing} shows the estimated CPU and disk resource needs of ATLAS during LHC Run 4 and beyond~\cite{bib:atlas-computing}.  Within this context, the community faces a number of challenges that needs to be urgently addressed, now, in order to enable the physics outcomes of future experiments. 
\begin{itemize}
\item Cost : Predictability of hardware cost has become difficult as the price trends are driven by commercial markets and the increased performance that was seen for the same amount of money due to Moore's Law no longer holds.  
\item Hardware evolution: Computing resources are nowadays becoming increasingly heterogeneous, placing requirements on both the code base and on the experiment computing systems.
\item Data storage, management and preservation: Data storage needs of future projects exceed the predicted capacity and affordability of the current data storage and management model.  The current model also does not provide the adaptability required to efficiently exploit heterogeneous compute infrastructures. 
There is also an uncertainty in the future availability of tape storage which could have a significant impact on the data storage capacity available to experiments.  The tape market now depends on a single manufacturer, increasing the risk of market collapse.  Data preservation (reproducibility, accessibility) and Open access both lack consistent policies in the experiments and across the funding agencies.  This must be addressed urgently, and new proposals must include data management plans.  Open access is also typically not explicitly funded although expected by many funding agencies.

\item Software challenges: The large volume of legacy software used in particle physics requires important improvements in memory usage and throughput.  The software must also be upgraded to make the most efficient use of different computing platforms, e.g. to take advantage of increased processing power from accelerators and GPU use, tuning on CPUs, etc. The current legacy software also offers many opportunities for algorithmic enhancement, and should also be upgraded to take advantage of developments in machine learning and AI applications~[ID5].
\item The increasing skills gap between early career physicists and the profile needed for programming on new compute architectures~[ID114] necessitates the needs for professional training, development and career opportunities for computing-software professionals [ID5, ID34, ID53, ID59, ID68, ID69, ID114, ID127, ID150].

\end{itemize}

As approved projects enter into operation and future projects are being developed, it is clear that there will be unprecedented pressure on computing resources availability for particle physics in the 2020s. 
Past experience has shown that development time-scales are long, and therefore, these challenges must be addressed now to maximize the scientific outputs of tomorrow.

\subsection{R\&D for computing and software}

To meet the challenges laid out in this document, the particle physics community must carry out carefully planned and coordinated R\&D programmes~[ID53, ID64, ID79, ID84, ID117, ID126, ID127, ID128, ID134, ID150, ID162] that will adopt new hardware, take advantage of industrial trends and emerging technologies, improve the software, and position HEP computing and software for revolutionary and disruptive technologies. Past experience has shown that it is important to plan, from the start, the development of an agile computing and software infrastructure, as technologies will continue to evolve~\cite{bib:RJ-talk}.  It is also equally important to plan for an infrastructure that requires less hardware and less effort to maintain and operate as an experiment matures~\cite{bib:RJ-talk}. The following highlights areas of R\&D activities that must be pursued~\cite{bib:MG-talk,bib:GS-talk}.

\begin{itemize}
    \item Tools and applications development for effective use of capacity provided by heterogeneous hardware and specialized architectures.
    For example, these include tools and applications to take advantage of
    GPU improvements for extremely parallel applications that offload the demand on CPUs, to provide work-flow and task specific acceleration, to enable pattern recognition and data transformation with FPGAs, to provide a cost effective way to enhance the memory bandwidth and low-power consumption with TPUs\footnote{Tensor Processing Unit -- an ASIC accelerator for AI applications.}, and to provide common provisioning mechanisms transparent to users.
    \item Application and data access tools development at HPC facilities and using commercial Clouds, which can deliver extra capacity to particle physics.  
    
    \item Continued R\&D on data organization, infrastructure, management and access in the face of technological changes and cost increase due to large data volume, data preservation and the data open-access requirement~[ID77, ID79, ID150]. 
 
    \item Software R\&D, which is expected to provide one of the biggest opportunities to address the needs of the particle physics community.  While inefficiently designed software is very costly in terms of resource usage, good and clever software allows for greater physics opportunities within the same fixed resource budget.  For the HL-LHC experiments, the software and the data formats will be improved to allow for more efficient processing and storage. To best use the technologies, algorithms and HEP software need to be redesigned and written, respectively. Common software framework, turnkey stacks can be developed through the inter-experiment collaboration for the HL-LHC and future collider projects. There are great opportunities for HEP to improve and generate new software by organizing the community, reaching out to industry, software engineers and other sciences~[ID16, ID34, ID43, ID53, ID59, ID64, ID77, ID79, ID108, ID126, ID127, ID150, ID162].
    
    \item Hardware infrastructure development and support. The particle physics community should seize the opportunity to be involved in the planning stage of future multidisciplinary research infrastructures, such as large HPC systems and Clouds (e.g.\ European Open Science Cloud (EOSC)), in order to ensure that these systems will be best equipped to address the needs of the community.  Likewise, the maintenance of existing, and development of new, computing centres for HEP is expected to continue to be important.

    \item Preparation and follow-up for the innovative, new technologies on the horizon, such as quantum computing and neuromorphic computing~[ID59, ID128].  These may revolutionize HEP computing, despite the fact that they have a great deal of uncertainty. The particle physics community must position itself to be able to migrate seamlessly to the new technology when it is ready.

\end{itemize}

To effectively carry out the above activities, the field needs more skilled developers, and a significant investment is required here~[ID5, ID114]. In order to assess the correct balance between hardware and development expenditure, a holistic view of computing and software is required.  Furthermore, R\&D in the area of software and computing must also consist of a balanced portfolio that includes activities targeted at the immediate needs of running experiments, activities focused on addressing the needs of future experiments, and pure R\&D activities that anticipate the use of disruptive technologies.

\subsection{Synergies and opportunities}
In order to provide a sustainable future for software and computing in the field, synergies with the experiments, other disciplines and with industry are vital~[ID53, ID64, ID70, ID77, ID84, ID117, ID126, ID128, ID150]. Having led the field of data intensive science, driven by the needs of the LHC and others, the subject must transition to being an important player in a wider ecosystem.

Internal synergies are illustrated by the number of submissions that highlight the intention to leverage computing and software developments from the LHC. The diversity of development projects in the various LHC experiments was useful to prototype various ideas, but the transition to the use of common tools now the experiments are mature must continue to reduce the operational and development costs. New projects should investigate the available tools where they exist rather than developing in-house solutions, minimizing later development costs. The move to the WLCG supporting the wider particle physics communities has been underway for a long time in some regions, and has become more formal with DUNE taking an associate status for developments. 

Synergies exist with other science disciplines. Data management issues are similar in new big astronomy projects and Rucio, already widely used in particle physics, is being used by LIGO and seriously evaluated by SKA, LSST and CTA.  The CERNVM-FS software and configuration distribution tool has an even wider use in the scientific community and beyond. There is also a fruitful collaboration between SKA and the LHC community on GEANT.  Unique opportunities of exploiting synergies between astronomy and particle physics in the area of data stewardship also exist within the H2020 ESCAPE project~\cite{bib:ESCAPE}. 

Synergies also exist with the commercial world. There have been fruitful direct collaborations between commercial Cloud providers and individual experiments, and also through Helix-Nebula. These may be the best route to accessing non-standard architectures and burst capacity. Much can be gained from the use of their management and data mining tools. CERN OpenLab~\cite{bib:openlab} will remain an important element of the engagement with industry.

There are many vehicles for these synergies to be discovered and exploited. In distributed computing, the WLCG will continue to play a central role; however, the introduction of new user communities raises issues of governance, with a clear separation between the development aspects and the resource allocation aspects. This challenge must be addressed in order to benefit from the availability and use of common tools.  The HEP Software Foundation (HSF)~\cite{bib:HSF} provides an important focus for common software development tasks and long-term planning~\cite{bib:HSF-plan} and is reaching out beyond particle physics.  There are also regional projects that will be important in this space; the WLCG and HSF will have an important role promoting coherence in these activities.  Moving forward, a strong synergy with other fields and the commercial world is essential to have access to the resources need for the next-generation experiments.
\vfill
\section{Interplay between instrumentation and computing}
The requirement of efficiency to extract the maximum physics potential from the available computing resources requires an increasingly ``holistic'' approach to the design of experiments and their associated computing and software systems. 

For example, the detector design decisions can either ease or place a huge burden on the offline computing and simulation. Full simulation studies are required not only to assess the level of background and optimize the detector performance,  but also to understand the computational costs in simulation and reconstruction arising from decisions on  the geometry, segmentation and situation of detectors. Subsequent evaluation of the detector design must include the computing burden as a metric. 

Online processing has in important role to play in this holistic approach. Many submissions illustrate the trend towards moving more complex algorithmic processing into the online systems~[e.g.\ ID5, ID162]. This is in parallel with an increasing trend to monolithic devices that integrate TDAQ functions into the devices as it reduces the offline burdens. Experiment design should consider the use of detector electronics to do  processing and data reduction at an early stage. In addition, offline-like reconstruction is increasingly possible in online triggering systems, with further reductions in the offline data volume without loss of physics performance.

There is also a need for a more holistic approach within the computing and software systems. For example, the design of the overall computing system must take into account both the hardware costs and the costs of operation. Equally, there is a trade off between effort in the development of efficient software and event models and consequent resource requirements. While these may fall into different accounting categories (e.g.\ recurrent versus capital costs), they need to be tensioned. This in turn necessitates realistic estimates of required software development effort.

Organizationally, computing and detector R\&D have had little cross-talk. Indeed, computing and software has often been considered as a secondary activity after detector design. In future, it would be desirable that projects view computing and software responsibilities on an equal footing with sub-detector responsibilities.

\section{Developing and preserving knowledge and expertise}


The scientific questions tackled by the particle physics community are long-term inquiries, and as such, in order to reach the community's scientific goals, human factors need to be carefully considered.  
One of the challenges faced by the particle physics community is an adequate level of development and preservation of expertise in instrumentation and computing-related R\&D activities.  For the benefit of the entire research field, it is of utmost importance that both types of activities be recognized correctly as fundamental research activities bearing a large impact on the final physics results.  This requires, now more than ever, a major change in paradigm in the community.

\subsection{Recruitment}

For the future of our field, it is essential to attract brilliant young physicists to the interesting challenges of instrumentation and computing-related R\&D.  To effectively do so, these activities must be recognized not only as a means to allow scientists to do physics analyses, but as research areas requiring a high level of imagination and creativity.  Attractiveness of these research areas could also be increased by more effectively communicating to prospective students that innovative ideas in these areas will often have far reaching impact on industry and society.  Furthermore, in outreach and public relations, it would be highly beneficial and desirable that the community more systematically highlights the technological dimensions of particle physics.  Particle physics challenges have the potential to continue to attract the best and brightest students.

\subsection{Training}

The knowledge, specialization and expertise required in the present and future field of particle physics are extensive.  Students and postdocs often lack basic knowledge in detector technologies, electronics, mechanics, software and simulation.  One of the reasons is that these specializations are rapidly evolving and university courses become insufficient to prepare students in these technical aspects.  
In addition, excellence in instrumentation development is often not valued and sufficiently recognized at the universities so as to attract the interest of students in this branch of research.
As a consequence, it also becomes difficult to attract young people to work in R\&D for instrumentation and computing. Yet, paradoxically, detector (system) prototyping is an excellent fertile training ground for young particle physicists.

It would be profitable to enhance, already at the level of university training and/or degree requirements, the basic knowledge required for applied physics activities.
One way would be to set up common training sessions for physicists and engineers, for example through platforms that can bring universities and laboratories together~[ID65]. Such activities would increase the training classes and find root in the departments, by creating also a positive feedback in terms of increased opportunities for careers in academies. 
In this respect, the community could consider creating a document targeting universities and laboratories, advocating for the training specificity of instrumentation and detector development.
The means that universities and small laboratories have, however, are typically limited. It remains therefore extremely important to preserve initiatives of internships at CERN and other large laboratories~[ID65].

With the long time-scale often associated with particle physics experimental projects, opportunities for students to participate in all phases of an experiment are becoming more and more scarce.  Investment in the specialized education of young physicists in the form of schools in particle physics instrumentation and/or scientific computing (e.g.\ EDIT school) is highly beneficial~[ID65].

\subsection{Expertise preservation and career opportunities}

Results from the 2018 ECFA Detector Panel survey~[ID68] of the community show that career perspectives for detector experts were perceived to exist in research, industry and a tertiary sector requiring advanced software development skills by 39\%, 66\% and 80\% of respondents.  This perception is driven by the reality of a flawed academic hiring model in which data analysis contributions are given more weight than instrumentation and/or computing-related research activities.  Whether it is the cause, or a consequence, of this hiring model, the reality is that the particle physics community has a strong tendency to under-sell the intellectual challenges---and satisfaction---of detector and computing-related R\&D work. 
Furthermore, compounded to this perception is the reality that except for a very few geniuses, individuals cannot typically be expert and innovate simultaneously in all areas of physics analysis, detectors, computing, teaching, outreach, etc.  As a result, few attractive career opportunities currently exist for individuals with specific instrumentation or computing expertise.   

The success and future of particle physics will directly depend on the community's ability to greatly improve in a systematic way the recognition and career opportunities of detector and computing experts.  One avenue to address this challenge is for the community to strengthen its effort to advocate for the recognition of these research areas in universities' hiring plans, possibly exploiting synergies and potential interests with engineering and computing science departments.  National and international laboratories also have the opportunity to play a major role in the community, for example, by creating prestigious career paths for physicists with particular expertise in instrumentation or computing-related research.  In doing so, care should be taken to avoid the development of a two-tier career system composed of ``scientific'' and ``applied'' streams with little or no cross-breeding, thereby only reinforcing the current {\it status quo}.  Other avenues to address this challenge could be to setup specialized and attractive grants for instrumentation and computing R\&D, as well as prizes recognizing both young and experienced scientists.





\section{Summary of key points}

The following summarizes some of the key points discussed in this chapter.

\begin{itemize}
    \item It is critically important, more than ever, for the community to maintain a strong focus on instrumentation R\&D and to foster an environment that stimulates innovation, with the primary goal of addressing the well-defined technological challenges of future experimental programmes. 
    
    \item Computing challenges are immediate and need to be addressed now through a vigorous R\&D programme. A cross-computing view tensioning new hardware costs with the costs of software development is required.
    
    \item It is becoming increasingly vital to take a more holistic approach to detector design which includes impact on computing resources; however, the detector and the computing/software communities have been drifting apart, and individuals that can bridge the growing gap are rare. This is a challenge to the community. 
        
    \item Both detector and computing development efforts benefit from the existence of networks and  consortia, within particle physics as well as extending to other disciplines and with industry.
    There is a clear need to strengthen existing R\&D collaborative structures, and create new ones, to address future experimental challenges of the field post HL-LHC.
    

    \item While producing experts who then go on to shape the wider world is a significant benefit to society,  
    a limited amount of success in attracting, developing and retaining instrumentation and computing experts poses a growing risk to the field of particle physics.  It is of utmost importance that both instrumentation and computing development activities be recognized correctly as fundamental research areas bearing a large impact on the final physics results.



\end{itemize}
\vfill

\addcontentsline{toc}{chapter}{Appendices}

\pagestyle{myheadings}
\appendix
\section{Glossary}
\markboth{Appendix A}{Glossary}
\begin{flushleft}
\begin{small}
\begin{tabular}{ll}
AD      &   Antimatter Decelerator, facility at CERN \\
ALP 	&	Axion-like Particle \\
APPEC   &   Astro Particle Physics European Consortium \\
BDF     &   Beam Dump Facility, proposed at the CERN SPS \\
BSM     &   Beyond the SM, i.e.\ new physics \\
CC      &   Circular Collider \\
CEPC    &   Circular Electron Positron Collider, proposed $e^+e^-$ collider (sited in China) \\
CH      &   Composite Higgs \\
CKM 	& 	Cabibbo-Kobayashi-Maskawa, the quark mixing matrix\\
CL      &   Confidence Level \\
CLIC    &   Compact Linear Collider, proposed $e^+e^-$ collider (sited at CERN) \\
CMOS    &   Technique for fabricating integrated circuits in silicon \\
CP      &   Combination of discrete symmetries: Charge-conjugation (C) and Parity (P) \\
CPV     &   CP Violation \\
CR      &   Cosmic Ray \\
DIS     &   Deep Inelastic Scattering \\
DM		&	Dark Matter \\
DS      &   Dark Sector \\
ECFA    &   European Committee for Future Accelerators \\
EDM	    & 	Electric Dipole Moment \\
EIC     &   Electron-ion Collider \\
ERL     &   Energy Recovery Linac \\
ESPPU   &   European Strategy for Particle Physics Update, also sometimes EPPSU or ESU\\
EW		&	Electroweak \\
EWPO    &   Electroweak Precision Observables \\
EWSB    &   Electroweak Symmetry Breaking \\
FCC     &   Future Circular Collider, proposed 100-km scale collider (sited at CERN) \\
FCC-ee  &   Version of FCC with $e^+e^-$ collisions \\
FCC-eh  &   Version of FCC with electron-hadron collisions \\
FCC-hh  &   Version of FCC with hadron collisions (proton or heavy-ion) \\
FCNC	&	Flavour Changing Neutral Current \\
FEL     &   Free Electron Laser, light source \\
FIP     &   Feebly Interacting Particle \\
GIM		& 	Glashow-Iliopoulos-Maiani, mechanism suppressing some decays\\
GPD     &   Generalised Parton Distribution (or General Purpose Detector, depending on context) \\
GUT     &   Grand Unified Theory \\
HE-LHC  &   High Energy LHC, proposed collider with $\sim$\,double LHC energy in same tunnel \\
HEP     &   High Energy Physics \\
HL-LHC  &   High Luminosity LHC, approved upgrade of the LHC to provide higher luminosity \\
HLT     &   High Level Trigger \\
HNL     &   Heavy Neutral Lepton \\
HPC     &   High Performance Computing \\
HTS     &   High Temperature Superconductor \\
HV      &   High Voltage \\
ID      &   Indirect Detection (or Identification, depending on context) \\
\end{tabular}
\end{small}
\end{flushleft}
\begin{flushleft}
\begin{small}
\begin{tabular}{ll}
ILC     &   International Linear Collider, proposed $e^+e^-$ collider (sited in Japan) \\
IP      &   Interaction Point \\
IR      &   Infrared, i.e.\ low energy limit \\
KEKB    &   B factory $e^+e^-$ collider in Japan \\
LC      &   Linear Collider \\
LDM     &   Light Dark Matter \\
LEP     &   $Z$ factory $e^+e^-$ collider at CERN, used the same tunnel now occupied by LHC\\
LFUV	&	Lepton Flavour Universality Violation \\
LFV		& 	Lepton Flavour Violation \\
LHC     &   Large Hadron Collider, hadron collider at CERN \\
LHeC    &   Proposed electron-hadron collider using hadrons from the LHC plus an ERL \\
LLP     &   Long-lived Particle \\
LNV		& 	Lepton Number Violation \\
LQCD    &   Lattice QCD \\
LSP     &   Lightest Supersymmetric Particle \\
MFV	    &	Minimal Flavour Violation \\
MSSM    &   Minimal Supersymmetric Model \\
NLL     &   Next to Leadling Logarithm \\
NLO     &   Next to Leading Order \\
NP		&	New Physics, i.e.\ physics beyond the Standard Model \\
PBC     &   Physics Beyond Colliders, a study \\
PDF     &   Parton Distribution Function \\
POT     &   Protons On Target \\
pQCD    &   Perturbative QCD \\
QCD	    & 	Quantum Chromodynamics, theory of the strong interaction \\
QED     &   Quantum Electrodynamics, theory of the electromagnetic interaction \\
QFT     &   Quantum Field Theory \\
QGP     &   Quark Gluon Plasma \\
RCS     &   Rapid-cycling Synchrotron \\
RF      &   Radio Frequency \\
SC      &   Superconducting \\
SCT		& 	Super Charm Tau (also known as Tau-Charm factory, STC), proposed $e^+e^-$ collider \\
SD      &   Shutdown \\
SKA     &   Square Kilometer Array, radio telescope array \\
SLC     &   Linear Collider previously operating at SLAC \\
SPS     &   Super Proton Synchrotron, accelerator at CERN \\
SM		& 	Standard Model (of particle physics) \\
SM-EFT	&	Standard Model Effective Field Theory \\
SPPC    &   Proposed hadron collider to follow the CEPC using the same tunnel \\
SUSY    &   Supersymmetry, proposed model of NP \\
TDAQ    &   Trigger and Data Aquisition \\
TDR     &   Technical Design Report \\
TPC     &   Time Projection Chamber (or Total Project Cost, depending on context) \\
UHE     &   Ultra High Energy \\
UT		&	Unitarity Triangle, relationship between quark mixing matrix elements \\
WIMP    &   Weakly Interacting Massive Particle, a candidate for DM \\
WLCG    &   Worldwide LHC Computing Grid \\
\end{tabular}
\end{small}
\end{flushleft}

\section{Open Symposium scientific programme}
\markboth{Open Symposium scientific programme}{Appendix B}
The full agenda and presentations of the Open Symposium in Granada (13-16 May 2019) are available from the meeting  \href{https://cafpe.ugr.es/eppsu2019/}{website}. The presentations making up the scientific programme are listed here. There were also many discussion sessions. The symposium ended with summaries of the parallel sessions, presented by the Conveners.
\vspace*{3mm}

\begin{flushleft}
\begin{tabular}{lr}
\bf Title & \bf Presented by \\ \hline 
\vspace*{-3mm} & \\
\multicolumn{2}{c}{\bf Opening plenary} \\
\it Goals of the Symposium & Halina Abramowicz \\
\it Implementation of the 2013 European Strategy Update & Fabiola Gianotti \\
\it Outstanding questions in Particle Physics & Pilar Hernandez\\
\it State of the art and challenges in accelerator technology  & Akira Yamamoto \\
\it Future -- path to very high energies & Vladimir Shiltsev \\
\it Technological challenges of particle physics experiments & Francesco Forti \\
\it Computing challenges of the future & Simone Campana \\ \hline 
\vspace*{-3mm} & \\
\multicolumn{2}{c}{\bf Electroweak Physics} \\
\vspace*{-3mm} & \\
\it Prospects for Higgs and EW measurements at HL-LHC & Patrizia Azzi \\
\it QCD uncertainties on Higgs and EWK measurables & Fabrizio Caola \\
\it Theory perspective on direct \& indirect searches for new physics & Riccardo Rattazzi \\
\it Overview and technical challenges of proposed Higgs factories & Daniel Schulte \\
\it Capability of future machines for precision Higgs physics & Maria Cepeda \\
\vspace*{-3mm} & \\
\it Electroweak Precision Measurements at future experiments & Mark Lancaster \\
\it Precision EW calculations (Giga-$Z$, $WW$, Higgs BRs, etc.) & Stefan Dittmaier \\
\it The Higgs potential and its cosmological histories & Geraldine Servant \\
\it Path towards measuring the Higgs potential & Elisabeth Petit \\
\it Interpretation of Higgs and EWK data in EFT framework & Jorge de Blas \\ \hline 
\vspace*{-3mm} & \\
\multicolumn{2}{c}{\bf Strong Interactions} \\
\vspace*{-3mm} & \\
\it Scientific aspirations of the community in strong interactions & Thomas Gehrmann \\
\it Experimental QCD physics at future $pp$ and $e^+e^-$ colliders & David d'Enterria \\
\it Theoretical path for QCD physics & Gavin Salam \\
\it Strong int.\ physics with (HL-)LHC pre-accelerator complex & Gunar Schnell \\
\it Precision QCD physics at low energies & Klaus Kirch \\
\it Lattice QCD: challenges and opportunities & Hartmut Wittig \\ 
\vspace*{-3mm} & \\
\it Theory challenges for Heavy Ion physics & Urs Wiedemann \\
\it Heavy Ion collisions at (HL-)LHC & Johanna Stachel \\
\it Strong interaction physics at future $eA$ colliders & \quad Nestor Armesto Perez \\
\it Emerging facilities around the world for strong int.\ physics & Tetyana Galatyuk \\
\it Synergies with astroparticle, nuclear and neutrino physics & Tanguy Pierog \\
\it Strong interaction physics at future $ep$ colliders & Uta Klein \\
\it What strong int.\ physics can one do with LHC after HL-LHC? & Daniel Boer \\ \hline 
\end{tabular}
\vfill \newpage
\begin{tabular}{lr}
\bf Title & \bf Presented by \\ \hline 
\vspace*{-3mm} & \\
\multicolumn{2}{c}{\bf Flavour Physics} \\
\it Overview on Flavour Physics & Yosef Nir \\
\it Flavour and CP searches with kaons & Marco Sozzi \\
\it Flavour and CP searches with heavy flavours & Marie-Helene Schune \\
\it Precision prospects for the lattice input to CKM determinations & Carlos Pena \\
\it Theoretical perspective on EDMs and the strong CP problem & Michael Dine \\
\it The EDM hunt: neutron, charged hadrons, leptons  & Stephan Paul \\
\vspace*{-3mm} & \\
\it CP violation in Higgs and in gauge boson couplings & Stefania Gori \\
\it Flavour and CP violation in the dark sectors & Jure Zupan \\
\it Invisible channels \& long-lived particles: comparison of potential & Augusto Ceccucci \\
\it Physics prospects with muons & Yoshikata Kuno \\
\it Physics prospects with taus & Alberto Lusiani \\
\it Lepton universality: $B$ and $K$ anomalies & Svetlana Fajfer \\ 
\vspace*{-3mm} & \\
\multicolumn{2}{c}{{\it Discussion leaders} \hfill Gudrun Hiller, Gino Isidori, Ana Teixeira} \\
\hline
\vspace*{-3mm} & \\
\multicolumn{2}{c}{\bf Neutrino Physics \& Cosmic Messengers} \\
\vspace*{-3mm} & \\
\it Theories of neutrino masses and leptonic CP violation & Silvia Pascoli \\
\it Precision determination of neutrino mass-mixing parameters & Eligio Lisi \\
\it Prospects for measurement of neutrino mass ordering \& leptonic CPV & Mauro Mezzetto \\
\it Measurements of Neutrino-nucleus cross sections and neutrino flux & Federico Sanchez \\
\it Measurements of the neutrino mass & Susanne Mertens \\
\it Prospects for the search of sterile neutrinos & Bonnie Fleming \\
\it Prospects for the search of Heavy Neutral Leptons & Nicola Serra \\
\vspace*{-3mm} & \\
\it Cosmic ray physics & Andreas Haungs \\
\it Neutrino astroparticle physics & Francis Halzen \\
\multicolumn{2}{c}{{\it Gravitational waves} \hfill Bangalore Sathyaprakash} \\
\it Multimessenger physics & Marek Kowalski \\ \hline
\vspace*{-3mm} & \\
\multicolumn{2}{c}{\bf Dark Matter and Dark Sector} \\
\vspace*{-3mm} & \\
\it Dark Sectors and DM Models: from ultralight to ultra heavy & Hitoshi Murayama \\
\it Dark Matter Direct Detection Searches & Jocelyn Monroe \\
\it Indirect DM detection overview & Christoph Weniger \\
\it How can Direct \& Indirect DM searches guide accel.\ searches? & Mariangela Lisanti \\
\it Theory: DM at Colliders & Matthew Mccullough \\
\it Collider Search: DM at Colliders & Caterina Doglioni \\
\vspace*{-3mm} & \\
\it Ultra-light DM (ALPS) Theory and Overview & Prateek Agrawal \\
\it ALPS: Lab searches & Axel Lindner \\
\it ALPS: Helioscope searches & Igor Garcia Irastorza \\
\it Dark Sector searches with Beam Dumps: Theory \& Overview & Claudia Frugiuele \\
\it Lepton Beams: LDMX@eSPS (NA64++, AWAKE++) & Ruth Pottgen \\
\it Proton Beams: SHIP@BDF & Elena Graverini \\
\it General Perspective & Claude Vallee \\ \hline
\end{tabular}
\vfill \newpage
\begin{tabular}{lr}
\bf Title & \bf Presented by \\ \hline 
\vspace*{-3mm} & \\
\multicolumn{2}{c}{\bf Beyond the Standard Model} \\
\vspace*{-3mm} & \\
\it EWSB dynamics and resonances: what expect from experiments & Juan Alcaraz Maestre \\
\it EWSB dynamics and resonances: implications for theory & Andrea Wulzer \\
\it Supersymmetry: what we can expect from experiments & Monica D'Onofrio \\
\it Supersymmetry: implications for theory & Andreas Weiler \\
\it Extended Higgs sectors \& HE flavour dynamics: expected from exp. & Philipp Roloff \\
\vspace*{-3mm} & \\
\it Feebly interacting particles: theory landscape & Gilad Perez \\
\it Feebly interacting particles: what we can expect from experiments & Gaia Lanfranchi \\ \hline
\vspace*{-3mm} & \\
\multicolumn{2}{c}{\bf Accelerator Science and Technology} \\
\it LHC future & Lucio Rossi \\
\it Future Circular Colliders & Michael Benedikt \\
\it Future Linear Colliders & Steinar Stapnes \\
\it Overview \& technological challenges of proposed Higgs Factories & Daniel Schulte \\
\it Capability of future machines for precision Higgs physics & Maria Cepeda \\
\vspace*{-3mm} & \\
\it Muon collider & Daniel Schulte \\
\it Accelerator-based Neutrino beams & Vladimir Shiltsev \\
\it Energy efficiency of HEP infrastructures & Erk Jensen \\
\it Current plasma acceleration projects & Edda Gschwendtner \\
\it Challenges of plasma acceleration & Wim Leemans \\
\it Beyond colliders & Mike Lamont \\ \hline
\vspace*{-3mm} & \\
\multicolumn{2}{c}{\bf Instrumentation and Computing} \\
\vspace*{-3mm} & \\
\it Lessons learned from past instrumentation R\&D & Didier Contardo \\
\it Detector challenges of future HEP experiments & Lucie Linssen \\
\it Detector R\&D for future HEP experiments & Felix Sefkow \\
\it Technological synergies in inst.\ R\&D with non-HEP exp.\ \& industry & Cinzia Da Via \\
\vspace*{-3mm} & \\
\it Current HEP computing model & Ian Bird \\
\it Lessons learned from development of current HEP computing model & Roger Jones \\
\it Future challenges of HEP computing & Matthias Kasemann \\
\it HEP computing infrastructure R\&D & Maria Girone \\
\it HEP Computing Software R\&D & Graeme Stewart \\ 
\vspace*{-3mm} & \\
\multicolumn{2}{c}{{\it Panel discussion members} \hfill Amber Boehnlein, Simone Campana, Ariella Cattai, } \\
\multicolumn{2}{c}{\hfill Giuliana Fiorillo, Francesco Forti, Weidong Li \& the speakers} \\ \hline
\vspace*{-3mm} & \\
\multicolumn{2}{c}{\bf Closing plenary} \\
\vspace*{-3mm} & \\
\it Perspective on the European Strategy from the Americas & Young-Kee Kim \\
\it Perspective on the European Strategy from Asia & Geoffrey Taylor \\
\it APPEC Roadmap & Teresa Montaruli \\
\it NuPPEC long term plan & Marek Lewitowicz \\
\it Programmes of Large European and National Labs & Pierluigi Campana \\
\it Overview of National Inputs to the Strategy Update & Siegfried Bethke \\
\multicolumn{2}{c}{{\it Education, Communication and Outreach} \hfill Perrine Royole-Degieux} \\ \hline
\end{tabular}
\end{flushleft}
\newpage
\section{European Strategy Update contributions}
\markboth{European Strategy Update contributions}{Appendix C}
There were 167 documents submitted to the European Strategy Update by the deadline of 18 December 2018, of which 7 were withdrawn (mostly due to multiple submission). The remaining 160 submissions are also available via the meeting website, and are listed below (some titles abridged to fit).  The submissions are referenced in the text of this document using the form [ID$n$], and some have addenda which give further details of the community involved, schedule and cost (available via the website). The contributions can be accessed directly by clicking on the ID number in the list below.

\begin{flushleft}
\begin{tabular}{llr} 
\bf ID & \bf Title & \bf Submitted by \\ \hline 
\vspace*{-3mm} & & \\
\href{https://indico.cern.ch/event/765096/contributions/3295508/attachments/1785100/2905996/cern.pdf}{1} & \it Searches for Sterile Neutrinos at CERN & Robert Shrock\\
\href{https://indico.cern.ch/event/765096/contributions/3295509/attachments/1785101/2905999/Yock.pdf}{2} & \it Newtonian Test of the Standard Model & Philip Yock \\
\href{https://indico.cern.ch/event/765096/contributions/3295511/attachments/1785103/2906003/TIARA_ES2020-contribution_final.pdf}{4} & \it TIARA contribution to the ESPPU & Roy Aleksan\\
\href{https://indico.cern.ch/event/765096/contributions/3295512/attachments/1785106/2906008/A_European_Data_Science_Institute_for_Particle_Physics.pdf}{5} & \it A European Data Science Institute for Fundamental Physics\hspace*{1cm} & Maurizio Pierini\\
\href{https://indico.cern.ch/event/765096/contributions/3295513/attachments/1785107/2906010/Gamma_Factory.pdf}{6} & \it Gamma Factory for CERN & Mieczyslaw Krasny \\
\href{https://indico.cern.ch/event/765096/contributions/3295514/attachments/1785110/2906015/2018_ALEGRO_ESPP.pdf}{7} & \it Advanced LinEar collider study GROup (ALEGRO) Input & Brigitte Cros\\
\href{https://indico.cern.ch/event/765096/contributions/3295516/attachments/1785111/2906018/European_Strategy.pdf}{8} & \it DM follies and the `Krisis' of particle physics & Francois Richard\\
\href{https://indico.cern.ch/event/765096/contributions/3295517/attachments/1785115/2906025/na64-epps-2018.pdf}{9} & \it Prospects for exploring the Dark Sector with NA64 & Sergei Gninenko\\
\href{https://indico.cern.ch/event/765096/contributions/3295519/attachments/1785116/2906032/main1.pdf}{11} & \it The Belle II experiment at SuperKEKB & Bostjan Golob\\
\href{ https://indico.cern.ch/event/765096/contributions/3295520/attachments/1785117/2906034/2018_12_11_SHiP_ESPP_v23.pdf}{12} & \it The SHiP experiment at the SPS Beam Dump Facility & Andrei Golutvin \\
\href{ https://indico.cern.ch/event/765096/contributions/3295521/attachments/1785120/2906038/espp_summary_long.pdf}{13} & \it Proposal from the NA61/SHINE Collaboration for the ESPPU & Marek Gazdzicki \\
\href{ https://indico.cern.ch/event/765096/contributions/3295522/attachments/1785121/2906041/NEVOD.pdf}{14} & \it Complex NEVOD for multi-component investig.\ of cosmic rays & Anatoly Petrukhin \\
\href{ https://indico.cern.ch/event/765096/contributions/3295523/attachments/1785122/2906045/EuropeStrategy.pdf}{15} & \it Contribution of the French Physics Society & Yannis Karyotakis \\
\href{ https://indico.cern.ch/event/765096/contributions/3295524/attachments/1785124/2906048/RD-ESPP-2018-final.pdf}{16} & \it Strategic R\&D Prog.\ on Technologies for Future Experiments & Christian Joram \\
\href{ https://indico.cern.ch/event/765096/contributions/3295526/attachments/1785125/2906049/EUROPEAN_STRATEGY_vESFPP.pdf}{17} & \it Status \& perspectives of neutron time-of-flight facility n$_{TOF}$ ... & Enrico Chiaveri \\
\href{ https://indico.cern.ch/event/765096/contributions/3295545/attachments/1785126/2906052/Storage_Ring_EDM.pdf}{18} & \it Feasibility Study for an EDM Storage Ring & Hans Stroeher \\
\href{ https://indico.cern.ch/event/765096/contributions/3295546/attachments/1785127/2906053/JUNO_EPPSU.pdf}{19} & \it The JUNO Experiment & Marcos Dracos \\
\href{ https://indico.cern.ch/event/765096/contributions/3295547/attachments/1785128/2906056/PBC-CB_Executive_Summary.pdf}{20} & \it PBC Conventional Beams Executive Summary & Lau Gatignon \\
\href{ https://indico.cern.ch/event/765096/contributions/3295548/attachments/1785130/2906059/INFN_Hadr_Phys_Contribution.pdf}{21} & \it Initial contribution of the INFN Hadron Physics Community& Mauro Taiuti \\
\href{ https://indico.cern.ch/event/765096/contributions/3295549/attachments/1785131/2906061/Communicating_particle_physics_matters_-_EPPCN_contribution_to_EPPSU.pdf}{22} & 
\multicolumn{2}{c}{{\it Communicating particle physics matters} \hfill Perrine Royole Degieux} \\
\href{ https://indico.cern.ch/event/765096/contributions/3295550/attachments/1785132/2906063/ASP-ESG-Main-Document.pdf}{24} & 
\multicolumn{2}{c}{{\it The Biennial African School on Fund.\ Phys.\ and Appl.} \hfill Ketevi A.\  Assamagan} \\
\href{ https://indico.cern.ch/event/765096/contributions/3295551/attachments/1785134/2906066/CLFV_Muons.pdf}{25} & \it Charged LFV using Intense Muon Beams at Future Facilities & Andre Schoening \\
\href{ https://indico.cern.ch/event/765096/contributions/3295552/attachments/1785135/2906067/INFN_Initial_Input.pdf}{26} & \it Initial INFN input on the ESPPU & Fabio Zwirner \\
\href{ https://indico.cern.ch/event/765096/contributions/3295553/attachments/1785136/2906068/IAXO_ESPP.pdf}{27} & \it The Int.\ Axion Observatory (IAXO): case, status and plans & Igor G.\ Irastorza \\
\href{ https://indico.cern.ch/event/765096/contributions/3295554/attachments/1785137/2906070/REDTOP_Overview_v11.pdf}{28} & \it REDTOP: Rare Eta Decays with a TPC for Optical Photons & Corrado Gatto \\
\href{ https://indico.cern.ch/event/765096/contributions/3295557/attachments/1785139/2906073/MainText_-_CEPC_Input_to_the_ESPP.pdf}{29} & \it CEPC Input to the ESPPU -- Physics and Detector & Manqi Ruan \\
\href{ https://indico.cern.ch/event/765096/contributions/3295574/attachments/1785141/2906077/INR-ESPP.pdf}{30} & \it Large-scale neutrino detectors: from INR, Russian Acad.\ of Sci.\ & Sergey Troitskiy \\
\href{ https://indico.cern.ch/event/765096/contributions/3295595/attachments/1785142/2906079/ESPP2020-ES.pdf}{31} & \it Input from the Spanish Particle Physics Community & Teresa Rodrigo \\
\href{ https://indico.cern.ch/event/765096/contributions/3295596/attachments/1785144/2906081/KET_ESPP_Statement_2018.pdf}{33} & \it Statement by the German Particle Physics Community & Ulrich Uwer \\
\href{ https://indico.cern.ch/event/765096/contributions/3295598/attachments/1785145/2906082/Israeli_input_European_Strategy_for_Particle_Physics.pdf}{34} & \it Israeli Input to the European Strategy for Particle Physics & Gilad Perez \\
\href{ https://indico.cern.ch/event/765096/contributions/3295599/attachments/1785146/2906084/main.pdf}{35} & \it AWAKE: On the path to Particle Physics Applications & Allen Caldwell \\
\href{ https://indico.cern.ch/event/765096/contributions/3295600/attachments/1785148/2906086/eBeam-Light-Dark-Matter-acceleratorRD.pdf}{36} & \it Dark Sector Phys.\ with a Primary Electron Beam Facility ... & Torsten Akesson \\
\href{ https://indico.cern.ch/event/765096/contributions/3295601/attachments/1785150/2906089/Statement_strategy-heavy-ions.pdf}{37} & \it Future of Heavy Ion Physics at Colliders & Johanna Stachel \\
\href{ https://indico.cern.ch/event/765096/contributions/3295602/attachments/1785154/2906094/COMET-EuropeanStrategy.pdf}{38} & \it COMET & Yoshitaka Kuno \\
\href{ https://indico.cern.ch/event/765096/contributions/3295603/attachments/1785156/2906098/ISOLDE-EPICproject-ESPPupdate.pdf}{39} & \it EPIC: Exploiting the Potential of ISOLDE at CERN & Gerda Neyens \\
\href{ https://indico.cern.ch/event/765096/contributions/3295604/attachments/1785158/2906102/np-sec-dec16-18.pdf}{40} & \it Input of Nuclear Physics Section, Russian Acad.\ of Sci. & Valery Rubakov \\
\end{tabular}
\end{flushleft}
\newpage
~\\ \vspace*{-10mm}
\begin{flushleft}
\begin{tabular}{llr} 
\bf ID & \bf Title & \bf Submitted by \\ \hline 
\vspace*{-3mm} & & \\
\href{ https://indico.cern.ch/event/765096/contributions/3295605/attachments/1785161/2906107/Further_searches_of_the_Higgs_scalar_sector.pdf}{41} & \it Further searches of the Higgs scalar sector & Carlo Rubbia \\
\href{ https://indico.cern.ch/event/765096/contributions/3295606/attachments/1785162/2906109/CERN-PBC-study.pdf}{42} & \it The Physics Beyond Colliders Study at CERN & Claude Vallee \\
\href{ https://indico.cern.ch/event/765096/contributions/3295608/attachments/1785163/2906110/I43-research-plans-norway.pdf}{43} & \it Research Plans of the Norwegian Physics Communities till 2025 & Gerald Eigen \\
\href{ https://indico.cern.ch/event/765096/contributions/3295609/attachments/1785165/2906113/HFLAV_European_Strategy_Review_2018.pdf}{44} & \it HFLAV input to the ESPPU & Ulrik Egede \\
\href{ https://indico.cern.ch/event/765096/contributions/3295618/attachments/1785166/2906114/Neutrino_Town_Meeting__Summary_Document2018-12-18.pdf}{45} & \it Future Opportunities in Accelerator-based Neutrino Phys. & Joachim Kopp \\
\href{ https://indico.cern.ch/event/765096/contributions/3295619/attachments/1785168/2906116/SPbSU-contribution-to_ESPP-17122018.pdf}{46} & \it Heavy-flavour prod.\ in relativistic heavy-ions ... at CERN \& JINR  & Grigori Feofilov \\
\href{ https://indico.cern.ch/event/765096/contributions/3295620/attachments/1785169/2906119/ESPP-fixedtarget-ALICE.pdf}{47} & 
\multicolumn{2}{c}{{\it Physics opportunities for a fixed-target programme in ALICE} \hfill Laure M.\ Massacrier} \\
\href{ https://indico.cern.ch/event/765096/contributions/3295622/attachments/1785171/2906122/TownMeetingHI.pdf}{48} & \it Conclusions of the Town Meeting: Relativistic Heavy Ion Coll.\ & Urs Wiedemann \\
\href{ https://indico.cern.ch/event/765096/contributions/3295623/attachments/1785173/2906126/ES-SCT.pdf}{49} & \it Precision experiments at $e^+e^-$ collider Super Charm-Tau Factory & Vitaly Vorobyev \\
\href{ https://indico.cern.ch/event/765096/contributions/3295624/attachments/1785176/2906131/AWAKE_Applications_ReSubmit.pdf}{50} & \it Particle physics applications of the AWAKE acceleration scheme & Matthew Wing \\
\href{ https://indico.cern.ch/event/765096/contributions/3295627/attachments/1785177/2906133/CEPC_European_strategy_accelerator-v9Submit_version.pdf}{51} & \it CEPC Input to the ESPPU -- Accelerator & Jie Gao \\
\href{ https://indico.cern.ch/event/765096/contributions/3295629/attachments/1785178/2906134/alps2018-v5.pdf}{52} & \it Future colliders -- Linear and circular & Roman Poeschl \\
\href{ https://indico.cern.ch/event/765096/contributions/3295632/attachments/1785179/2906136/HEP-Computing-Evolution.pdf}{53} & \it HEP Computing Evolution & Ian Bird \\
\href{ https://indico.cern.ch/event/765096/contributions/3295658/attachments/1785181/2906138/2018_12_17_CERN_view_on_KT_ESPP.pdf}{54} & \it CERN's view on Knowledge Transfer as input for the ESPPU& Giovanni Anelli \\
\href{ https://indico.cern.ch/event/765096/contributions/3295660/attachments/1785182/2906140/HIP-Div-EPPSUcontribution2018.pdf}{55} & \it Particle Physics in Finland & Katri Huitu \\
\href{ https://indico.cern.ch/event/765096/contributions/3295662/attachments/1785183/2906145/ESPPU_Italy_HI_Final_Dec18.pdf}{56} & \it Ultra-relativistic Heavy-Ion Collisions: Italian community input & Andrea Dainese \\
\href{https://indico.cern.ch/event/765096/contributions/3295663/attachments/1785185/2906149/ENUBET_EU19_strategy.pdf}{57} & \it A high precision neutrino beam for a new gen.\ of SBL exp. & Andrea Longhin \\
\href{ https://indico.cern.ch/event/765096/contributions/3295663/attachments/1785185/2906149/ENUBET_EU19_strategy.pdf}{58} & 
\multicolumn{2}{c}{{\it AWAKE++ ... for New Particle Physics Experiments at CERN} \hfill Edda Gschwendtner} \\
\href{ https://indico.cern.ch/event/765096/contributions/3295667/attachments/1785187/2906153/INFN-Software-Computing.pdf}{59} & 
\multicolumn{2}{c}{{\it Initial INFN input on the ESPPU: software and computing} \hfill Donatella Lucchesi} \\
\href{ https://indico.cern.ch/event/765096/contributions/3295669/attachments/1785191/2906157/PBC-technonlogy-subgroup-report_10p_2018-12-17.pdf}{60} & \it PBC technology subgroup report & Andrzej Siemko \\
\href{ https://indico.cern.ch/event/765096/contributions/3295670/attachments/1785193/2906160/Denmark-ESPP-input181218.pdf}{61} & 
\multicolumn{2}{c}{{\it Input to the ESPPU from the Danish community} \hfill Jens-Jorgen Gaardhoeje} \\
\href{ https://indico.cern.ch/event/765096/contributions/3295671/attachments/1785196/2906164/DarkSide-Argo_ESPP_Dec_17_2017.pdf}{62} & \it Future Dark Matter Searches with Low-Radioactivity Argon & Cristiano Galbiati \\
\href{ https://indico.cern.ch/event/765096/contributions/3295672/attachments/1785198/2906168/Japan_strategy.pdf}{63} & \it Japan's Updated Strategy for Future Projects in PP & Toshinori Mori \\
\href{ https://indico.cern.ch/event/765096/contributions/3295673/attachments/1785200/2906171/GW3G-ET.pdf}{64} & \it Gravitational Waves in the ESPPU & Michele Punturo \\
\href{ https://indico.cern.ch/event/765096/contributions/3295675/attachments/1785201/2906174/European-Training.pdf}{65} & \it The JUAS and ESIPAP graduate schools & Johann Collot \\
\href{ https://indico.cern.ch/event/765096/contributions/3295677/attachments/1785203/2906178/ILC_EuropeanPerspective.pdf}{66} & \it The International Linear Collider: a European Perspective & Juan Fuster Verd\'{u} \\
\href{ https://indico.cern.ch/event/765096/contributions/3295678/attachments/1785204/2906180/FTP4LHC-ESPP2020-FINAL-ENDORSED.pdf}{67} & 
\multicolumn{2}{c}{{\it Community Support for a Fixed-Target Programme for LHC} \hfill Jean-Philippe Lansberg} \\
\href{ https://indico.cern.ch/event/765096/contributions/3295679/attachments/1785205/2906181/ECFA_detector_panel_ESPPU_input_Dec2018.pdf}{68} & \it ECFA Detector Panel Report & Doris Eckstein \\
\href{ https://indico.cern.ch/event/765096/contributions/3295680/attachments/1785207/2906183/kat_input_european_strategy2018_final_version.pdf}{69} & 
\multicolumn{2}{c}{{\it Statement by the German Astroparticle Physics Community} \hfill Christian Weinheimer} \\
\href{ https://indico.cern.ch/event/765096/contributions/3295688/attachments/1785209/2906185/gnn_for_cern_181218.pdf}{70} & \it A memorandum by the Global Neutrino Network & Uli Katz \\
\href{ https://indico.cern.ch/event/765096/contributions/3295689/attachments/1785210/2906186/RO_Input_ESPP_2020.pdf}{73} & 
\multicolumn{2}{c}{{\it Romanian input to the ESPPU} \hfill Alexandru M.\ Bragadireanu} \\
\href{ https://indico.cern.ch/event/765096/contributions/3295690/attachments/1785212/2906190/EIC-Acc-European-update-r20.pdf}{74} & \it Electron Ion Collider Accelerator Science and Technology & Andrei Seryi \\
\href{ https://indico.cern.ch/event/765096/contributions/3295695/attachments/1785214/2906193/MATHUSLA_detector.pdf}{75} & \it MATHUSLA & Henry Lubatti \\
\href{ https://indico.cern.ch/event/765096/contributions/3295701/attachments/1785216/2906195/Input_from_J-PARC_181218.pdf}{76} & 
\multicolumn{2}{c}{{\it Input from J-PARC to the ESPPU} \hfill Takeshi Komatsubara} \\
\href{ https://indico.cern.ch/event/765096/contributions/3295702/attachments/1785218/2906198/ILC_Global_Project.pdf}{77} & \it The International Linear Collider: a Global Project & Juan Fuster Verd\'{u} \\
\href{ https://indico.cern.ch/event/765096/contributions/3295704/attachments/1785219/2906199/ESPP-Si.pdf}{78} & \it Slovenian input to the ESPPU & Bostjan Golob \\
\href{ https://indico.cern.ch/event/765096/contributions/3295708/attachments/1785220/2906200/espp-hsf-submission.pdf}{79} & \it The Importance of Software and Computing to Particle Physics & Graeme Stewart \\
\href{ https://indico.cern.ch/event/765096/contributions/3295710/attachments/1785223/2906203/JINR.Input.K.pdf}{80} & \it Input of Joint Institute for Nuclear Research & Boris Sharkov \\
\href{ https://indico.cern.ch/event/765096/contributions/3295711/attachments/1785225/2906206/I81-View_ESPP.pdf}{81} & \it A View on the European Strategy for Particle Physics & Christian Fabjan \\
\href{ https://indico.cern.ch/event/765096/contributions/3295713/attachments/1785228/2906210/ESGv3.pdf}{82} & \it Submission from Univ.\ of Liverpool Exp.\ PP Group & Themis Bowcock \\
\href{ https://indico.cern.ch/event/765096/contributions/3295714/attachments/1785230/2906212/InputEuropeanStrategyUpdate_ClimateChange.pdf}{83} & \it Ensuring the future of PP in a more sustainable world & Veronique Boisvert \\
\href{ https://indico.cern.ch/event/765096/contributions/3295716/attachments/1785232/2906216/APPEC_input_PP_strategy_update_fin.pdf}{84} & \it APPEC Contribution to the ESPPU & Job de Kleuver \\
\href{ https://indico.cern.ch/event/765096/contributions/3295718/attachments/1785233/2906219/LPNHE_ESPP.pdf}{85} & 
\multicolumn{2}{c}{{\it LPNHE scientific perspectives for the ESPPU} \hfill Reina C.\ Camacho Toro} \\
\href{ https://indico.cern.ch/event/765096/contributions/3295719/attachments/1785236/2906230/PIK-reactor-complex.pdf}{86} & \it Particle Physics at PIK Reactor Complex & Vladimir Voronin \\
\end{tabular}
\end{flushleft}
\newpage
\begin{flushleft}
\begin{tabular}{llr} 
\bf ID & \bf Title & \bf Submitted by \\ \hline 
\vspace*{-3mm} & & \\
\href{ https://indico.cern.ch/event/765096/contributions/3295721/attachments/1785237/2906233/RD51-MPGD_Overview_ESU_18122018_FINAL.pdf}{87} & \it Development of Micro-Pattern Gaseous Detectors (RD51) & Maksym Titov \\
\href{ https://indico.cern.ch/event/765096/contributions/3295722/attachments/1785238/2906235/update_cz.pdf}{88} & \it Inputs to ESPPU by the Czech particle physics community & Tomas Davidek \\
\href{ https://indico.cern.ch/event/765096/contributions/3295723/attachments/1785239/2906236/HiggsSelfCouplingFutureStrategies.pdf}{89} & \it Future strategies for ... measurement of Higgs self coupling & Patrick Janot \\
\href{ https://indico.cern.ch/event/765096/contributions/3295724/attachments/1785240/2906238/NA60-ESPPU-2018-12-18.pdf}{90} & \it Study of hard and EM processes at CERN-SPS energies ... & Gianluca Usai \\
\href{ https://indico.cern.ch/event/765096/contributions/3295725/attachments/1785241/2906239/final_EPP_2018.pdf}{91} & \it Synthesis on the ELN Project by A.\ Zichichi & Horst Wenninger \\
\href{ https://indico.cern.ch/event/765096/contributions/3295726/attachments/1785242/2906242/ILC-IN2P3.pdf}{92} & \it Prospect of the IN2P3 Community involved in the ILC & Marc Winter \\
\href{ https://indico.cern.ch/event/765096/contributions/3295727/attachments/1785244/2906245/NICA_application-ESU.pdf}{93} & 
\multicolumn{2}{c}{{\it Nuclotron-based Ion Collider Facility at JINR (NICA Complex)} \hfill Dmitri Peshekhonov} \\
\href{ https://indico.cern.ch/event/765096/contributions/3295728/attachments/1785247/2906251/FASER_ESPP.pdf}{94} & \it FASER: ForwArd Search ExpeRiment at the LHC & Jonathan Lee Feng \\
\href{ https://indico.cern.ch/event/765096/contributions/3295729/attachments/1785250/2906256/2018-12-18_Statement-from-the-European-Network-for-Novel-Accelerators-out3.pdf}{95} & \it On the Prospect \& Vision of UH Gradient Plasma Acc.\ for HEP & Ralph Assmann \\
\href{ https://indico.cern.ch/event/765096/contributions/3295730/attachments/1785251/2906258/STFC_v3.pdf}{96} & \it STFC input on Innovation/Tech.\ Transfer, Knowledge Exchange & Elizabeth Bain \\
\href{ https://indico.cern.ch/event/765096/contributions/3295732/attachments/1785253/2906260/darwin_espp_dec2018.pdf}{97} & \it Input from the DARWIN collaboration to the ESPPU & Laura Baudis \\
\href{ https://indico.cern.ch/event/765096/contributions/3295734/attachments/1785255/2906263/The_European_Spallation_Source_neutrino_Super_Beam_ESSSB.pdf}{98} & \it The European Spallation Source neutrino Super Beam ESS$\nu$SB & Tord Ekelof \\
\href{ https://indico.cern.ch/event/765096/contributions/3295735/attachments/1785257/2906268/EICdocumentforESPPU.pdf}{99} & \it Synergies between a US-based Electron-Ion Coll.\ \& European PP & Daniel Boer \\
\href{ https://indico.cern.ch/event/765096/contributions/3295741/attachments/1785265/2906282/main.pdf}{100} & \it Precision calculations for high-energy collider processes & Thomas Gehrmann \\
\href{ https://indico.cern.ch/event/765096/contributions/3295742/attachments/1785267/2906286/theory-esu-2018-12-18-submitted.pdf}{101} & \it Theory Requirements and Possibilities for the FCC-ee ... & Alain Blondel \\
\href{ https://indico.cern.ch/event/765096/contributions/3295745/attachments/1785268/2906288/taufv_eppsu_final.pdf}{102} & \it TauFV: a fixed-target exp.\ to search for flavour viol.\ in tau decays & Guy Wilkinson \\
\href{ https://indico.cern.ch/event/765096/contributions/3295746/attachments/1785271/2906291/dis-related-subjects-submitted.pdf}{103} & \it The DIS and Related Subjects Strategy Document ... & Aharon Levy \\
\href{ https://indico.cern.ch/event/765096/contributions/3295747/attachments/1785273/2906293/EPPSU_IPPOG_final.pdf}{104} & \it Future Challenges in Particle Physics Education \& Outreach & Hans Peter Beck \\
\href{ https://indico.cern.ch/event/765096/contributions/3295748/attachments/1785274/2906295/Open_SC_lab.proposal.pdf}{105} & \it An Open Lab for Development of Technical Superconductors & Julia Double \\
\href{ https://indico.cern.ch/event/765096/contributions/3295749/attachments/1785275/2906296/Input_for_the_Update_of_the_European_Strategy_of_Particle_Physics.pdf}{106} & \it R\&D for Near Detector Systems for LBL Neutrino Experiments & Stefania Bordoni \\
\href{ https://indico.cern.ch/event/765096/contributions/3295752/attachments/1785276/2906298/ILD_European_Strategy_Document.pdf}{107} & \it The ILD Detector at the ILC & Ties Behnke \\
\href{ https://indico.cern.ch/event/765096/contributions/3295753/attachments/1785277/2906299/RoadmapA.pdf}{108} & \it Research Strategy of the Austrian Particle Physics Community & Manfred Jeitler \\
\href{ https://indico.cern.ch/event/765096/contributions/3295754/attachments/1785278/2906300/2018-12-18_EuPRAXIA_input_out3.pdf}{109} & \it The EuPRAXIA Research Infrastructure ... Plasma Accelerators & Ralph Assmann \\
\href{ https://indico.cern.ch/event/765096/contributions/3295755/attachments/1785279/2906302/Next_gen_hi.pdf}{110} & \it A next-generation LHC heavy-ion experiment & Federico Antinori \\
\href{ https://indico.cern.ch/event/765096/contributions/3295757/attachments/1785283/2906307/LHCSpin.pdf}{111} & \it The LHCSpin Project & Pasquale Di Nezza \\
\href{ https://indico.cern.ch/event/765096/contributions/3295758/attachments/1785286/2906312/Axions-for-ESPP.pdf}{112} & \it A European Strategy Towards Finding Axions and Other WISPs & Axel Lindner \\
\href{ https://indico.cern.ch/event/765096/contributions/3295759/attachments/1785289/2906317/Rizzo_BMV.pdf}{113} & \it Vacuum Mag.\ Birefringence \& Axion search using pulsed magnets & Carlo Rizzo \\
\href{ https://indico.cern.ch/event/765096/contributions/3295772/attachments/1785291/2906321/MonteCarlo.pdf}{114} & \it Monte Carlo event generators for HEP event simulation & Mike Seymour \\
\href{ https://indico.cern.ch/event/765096/contributions/3295775/attachments/1785292/2906324/Hadron_Physics_Opportunities_in_Europe_Maas.pdf}{115} & \it Hadron Physics Opportunities in Europe & Frank Maas \\
\href{ https://indico.cern.ch/event/765096/contributions/3295776/attachments/1785293/2906325/IN2P3_contribution_to_ESPP_2020.pdf}{116} & \it IN2P3 contribution for the ESPPU & Patrice Verdier \\
\href{ https://indico.cern.ch/event/765096/contributions/3295777/attachments/1785295/2906327/Statement_of_the_Pierre_Auger_Collaboration_as_input_for_the_European_Strategy_for_Particle_Physics_submitted.pdf}{117} & \it Statement of the Pierre Auger Collaboration as ESPPU input & Ralph Engel \\
\href{ https://indico.cern.ch/event/765096/contributions/3295779/attachments/1785296/2906331/ES_MUonE_document.pdf}{118} & \it The MUonE experiment & Clara Matteuzzi \\
\href{ https://indico.cern.ch/event/765096/contributions/3295782/attachments/1785297/2906333/GRAND_whitepaper.pdf}{119} & \it GRAND: Giant Radio Array for Neutrino Detect.\ ESPPU input & Sijbrand De Jong \\
\href{ https://indico.cern.ch/event/765096/contributions/3295784/attachments/1785298/2906335/MuonCollider_ESPP_18dec18.pdf}{120} & \it Muon Colliders & Nadia Pastrone \\
\href{ https://indico.cern.ch/event/765096/contributions/3295786/attachments/1785299/2906336/PSI_EuropeanStrategy_Input_Dec2018.pdf}{121} & \it Particle Physics \& Related Topics at the Paul Scherrer Institute & Klaus Kirch \\
\href{ https://indico.cern.ch/event/765096/contributions/3295787/attachments/1785300/2906338/Belgian_input_ESG_HEP.pdf}{122} & \it Belgian national input to the ESPPU & Nick Van Remortel \\
\href{ https://indico.cern.ch/event/765096/contributions/3295788/attachments/1785301/2906339/ESPP_Final.pdf}{123} & 
\multicolumn{2}{c}{{\it Electric dipole moment community input to ESPPU} \hfill Philipp Schmidt-Wellenburg} \\
\href{ https://indico.cern.ch/event/765096/contributions/3295791/attachments/1785302/2906341/P2O.pdf}{124} & \it Neutrino Beam from Protvino to KM3NeT/ORCA & Juergen Brunner \\
\href{ https://indico.cern.ch/event/765096/contributions/3295798/attachments/1785304/2906345/PL_DocumentUESPP_v.2.pdf}{125} & \it Polish input to the ESPPU & Jan Krolikowski \\
\href{ https://indico.cern.ch/event/765096/contributions/3295800/attachments/1785305/2906347/submission-2020_DUNE.pdf}{126} & 
\multicolumn{2}{c}{{\it Deep Underground Neutrino Experiment (DUNE)} \hfill Stefan Soldner-Rembold} \\
\href{ https://indico.cern.ch/event/765096/contributions/3295801/attachments/1785307/2906349/NationalSubmissionSwedenPPUpdate.pdf}{127} & \it Submission from the Swedish Particle Physics Community & David Milstead \\
\href{ https://indico.cern.ch/event/765096/contributions/3295802/attachments/1785308/2906350/QC-HEP-2020.pdf}{128} & \it Quantum Computing for High Energy Physics & Federico Carminati \\
\href{ https://indico.cern.ch/event/765096/contributions/3295803/attachments/1785310/2906353/SPS-Beam-Dump-Facility.pdf}{129} & \it SPS Beam Dump Facility & Mike Lamont \\
\href{ https://indico.cern.ch/event/765096/contributions/3295804/attachments/1785311/2906356/ESPP2020_LNF_Campana.pdf}{130} & \it Input to the Strategy Process from LNF-INFN & Pierluigi Campana \\
\end{tabular}
\end{flushleft}
\newpage
\begin{flushleft}
\begin{tabular}{llr} 
\bf ID & \bf Title & \bf Submitted by \\ \hline 
\vspace*{-3mm} & & \\
\href{ https://indico.cern.ch/event/765096/contributions/3295805/attachments/1785313/2906358/ESG-duneND.pdf}{131} & \it Enhancing the LBNF/DUNE Physics Programme & Roberto Petti \\
\href{ https://indico.cern.ch/event/765096/contributions/3298137/attachments/1786065/2907896/132_ESPP18_FCCee_180311-FCC_V0500_MainText.pdf}{132} & \it Future Circular Collider -- Lepton Collider (FCC-ee) & Michael Benedikt \\
\href{ https://indico.cern.ch/event/765096/contributions/3298184/attachments/1786069/2907901/133_ESPP18_FCChh_181115-FCC_V0600_MainText.pdf}{133} & \it Future Circular Collider -- Hadron Collider (FCC-hh) & Michael Benedikt \\
\href{ https://indico.cern.ch/event/765096/contributions/3295806/attachments/1785314/2906360/UK_input_ESPP_2018.pdf}{134} & 
\multicolumn{2}{c}{{\it UK input to the ESPPU} \hfill Claire Shepherd-Themistocleous} \\
\href{ https://indico.cern.ch/event/765096/contributions/3298216/attachments/1786071/2907909/135_EPPSU18_FCCintegral_1812111-FCC_V0500_MainText.pdf}{135} & \it Future Circular Collider -- Integrated Programme (FCC-int) & Michael Benedikt \\
\href{ https://indico.cern.ch/event/765096/contributions/3298229/attachments/1786073/2907913/136_ESPP18_HELHC_181119-FCC_V0500_MainText.pdf}{136} & \it Future Circular Collider -- High-Energy LHC (HE-LHC) & Michael Benedikt \\
\href{ https://indico.cern.ch/event/765096/contributions/3295807/attachments/1785315/2906362/SBN_Program_European_Strategy.pdf}{137} & \it The Short-Baseline Neutrino Program at Fermilab & Regina Abby Rameika \\
\href{ https://indico.cern.ch/event/765096/contributions/3295808/attachments/1785318/2906365/INFN-CSN1_EUStrategy2018_18dec18.pdf}{138} & \it INFN National Scientific Committee for HEP with Acc. & Nadia Pastrone \\
\href{ https://indico.cern.ch/event/765096/contributions/3295810/attachments/1785320/2906367/BrumStrategy-final.pdf}{139} & \it Birmingham Particle Physics Group Submission & Paul R.\ Newman \\
\href{ https://indico.cern.ch/event/765096/contributions/3295812/attachments/1785322/2906373/Saleh_ESPP2020.pdf}{140} & \it Energy frontier lepton-hadron coll., VLQ/leptons, preons ... & Saleh Sultansoy \\
\href{ https://indico.cern.ch/event/765096/contributions/3295813/attachments/1785324/2906376/European_Strategy_Swiss_input.pdf}{142} & \it Swiss input for the ESPPU & Tatsuya Nakada \\
\href{ https://indico.cern.ch/event/765096/contributions/3295862/attachments/1785327/2906380/COMASS_AMBER_ESPP2018.pdf}{143} & \it A New QCD Facility at the M2 beam line of the CERN SPS & Vincent Andrieux \\
\href{ https://indico.cern.ch/event/765096/contributions/3295975/attachments/1785328/2906381/UK_STFC_Input.pdf}{144} & \it ESPPU: Input from UK National Laboratories & Dave Newbold \\
\href{ https://indico.cern.ch/event/765096/contributions/3295976/attachments/1785329/2906383/CLIC-Physics.pdf}{145} & \it The Compact Linear $e^+e^-$ Coll.\ (CLIC): Physics Potential & Philipp Roloff \\
\href{ https://indico.cern.ch/event/765096/contributions/3295978/attachments/1785331/2906390/CLIC-AcceleratorDetector.pdf}{146} & \it The Compact Linear $e^+e^-$ Coll.\ (CLIC): Acc.\ and Detector & Aidan Robson \\
\href{ https://indico.cern.ch/event/765096/contributions/3295979/attachments/1785332/2906391/es18perle.pdf}{147} & \it PERLE: A High Power Energy Recovery Facility for Europe & Max Klein \\
\href{ https://indico.cern.ch/event/765096/contributions/3295981/attachments/1785333/2906392/Nupecc_ESPP2018.pdf}{148} & \it Nuclear physics and ESPPU & Eberhard Widmann \\
\href{ https://indico.cern.ch/event/765096/contributions/3295984/attachments/1785334/2906396/DPF-strategy.pdf}{149} & \it APS DPF: Community Planning and Science Drivers & Kate Scholberg \\
\href{ https://indico.cern.ch/event/765096/contributions/3295985/attachments/1785335/2906398/DPF-tools.pdf}{150} & \it APS DPF: Tools for Particle Physics & Kate Scholberg \\
\href{ https://indico.cern.ch/event/765096/contributions/3295988/attachments/1785336/2906400/New_Physics_and_Heavy_ions-EPPS.pdf}{151} & \it New physics searches with heavy-ion collisions at the LHC & David d'Enterria \\
\href{ https://indico.cern.ch/event/765096/contributions/3295995/attachments/1785339/2906404/HLLHC.pdf}{152} & \it The physics potential of HL-LHC & Michelangelo Mangano \\
\href{ https://indico.cern.ch/event/765096/contributions/3295999/attachments/1785342/2906408/klever_espp.pdf}{153} & \it KLEVER: An exp.\ to measure $K_L\rightarrow \pi^0 \nu \overline{\nu}$ & Matthew Moulson \\
\href{ https://indico.cern.ch/event/765096/contributions/3296001/attachments/1785344/2906412/nuSTORM_Executive_Summary.pdf}{154} & \it $\nu$STORM at CERN: Executive Summary & Kenneth R.\ Long \\
\href{ https://indico.cern.ch/event/765096/contributions/3296011/attachments/1785345/2906413/20181217_ESPP_CEA.pdf}{155} & \it CEA-Irfu contribution to the ESPPU & Anne-Isabelle Etienvre \\
\href{ https://indico.cern.ch/event/765096/contributions/3296012/attachments/1785346/2906416/HIBEAM-NNBAR-strategy-update-sub.pdf}{156} & \it The HIBEAM/NNBAR Experiment for the ESS & David A.\  Milstead \\
\href{ https://indico.cern.ch/event/765096/contributions/3296013/attachments/1785347/2906418/CanadianRoadMaps-Submission-to-European_Strategy-2018.pdf}{157} & \it Canadian Submission to the ESPPU & Michael Roney \\
\href{ https://indico.cern.ch/event/765096/contributions/3296014/attachments/1785348/2906420/JapanLBN.pdf}{158} & \it Opportunities in Accelerator-based Neutrino Phys.\ in Japan & Francesca Di Lodovico \\
\href{ https://indico.cern.ch/event/765096/contributions/3296015/attachments/1785349/2906421/es18lhec.pdf}{159} & \it Exploring the Energy Frontier with DIS at the LHC & Max Klein \\
\href{ https://indico.cern.ch/event/765096/contributions/3296016/attachments/1785350/2906423/HELHC.pdf}{160} & \it The physics potential of HE-LHC & Michelangelo Mangano \\
\href{ https://indico.cern.ch/event/765096/contributions/3296018/attachments/1785353/2906427/MAGIS_European_Strategy.pdf}{161} & \it MAGIS-1K: a 1000 m Atom Interferometer for DM \& GW & Jonathon Coleman \\
\href{ https://indico.cern.ch/event/765096/contributions/3296019/attachments/1785354/2906428/CERN_openlab_ESPP_Update.pdf}{162} & \it The Importance of Research-Ind.\ Collab.\ for Exascale Comp.\ & Alberto Di Meglio \\
\href{ https://indico.cern.ch/event/765096/contributions/3296021/attachments/1785355/2906430/EPPSU_QCD18dec18_8pm.pdf}{163} & \it QCD: Theory -- Input for the ESPPU & Francesco Hautmann \\
\href{ https://indico.cern.ch/event/765096/contributions/3296023/attachments/1785356/2906432/Late-ERIC-EUStrategy_final.pdf}{164} & \it The European Spallation Source ERIC -- ESPPU input & Valentina Santoro \\
\href{ https://indico.cern.ch/event/765096/contributions/3296025/attachments/1785357/2906434/Late-INFN-CSN2-Strategy.pdf}{165} & 
\multicolumn{2}{c}{{\it Initial INFN input on the ESPPU -- Astropart.\ Phys.\ Comm.\ 2 } \hfill Marco Pallavicini} \\
\href{ https://indico.cern.ch/event/765096/contributions/3296026/attachments/1785358/2906435/Late-Input-ESPPU-NL.pdf}{166} & \it Input from the Netherlands for the ESPPU & Stan Bentvelsen \\
\href{ https://indico.cern.ch/event/765096/contributions/3296027/attachments/1785359/2906436/Neutrino_ES_yellow_paper_Fermilab.pdf}{167} & \it Status of Fermilab's Neutrino Facilities & Louise Suter \\ \hline
\end{tabular}
\end{flushleft}

\newpage
\markboth{References}{References}
\addcontentsline{toc}{chapter}{References}
\begin{small}
\bibliographystyle{report}
\bibliography{\bibfiles}

\newcommand{\noopsort}[1]{} \newcommand{\printfirst}[2]{#1}
  \newcommand{\singleletter}[1]{#1} \newcommand{\switchargs}[2]{#2#1}
\providecommand{\href}[2]{#2}\begingroup\raggedright\begin{thebibliography}{100}

\bibitem{Dirac:1927dy}
P.~A.~M. Dirac, {\em {Quantum theory of emission and absorption of
  radiation}\/},
\href{http://dx.doi.org/10.1098/rspa.1927.0039}{Proc. Roy. Soc. Lond. {\bf
  A114} (1927)  243}.

\bibitem{Fermi:1934hr}
E.~Fermi, {\em {An attempt of a theory of beta radiation. 1.}\/},
\href{http://dx.doi.org/10.1007/BF01351864}{Z. Phys. {\bf 88} (1934)
  161--177}.

\bibitem{Sudarshan:1958vf}
E.~C.~G. Sudarshan and R.~E. Marshak, {\em {Chirality invariance and the
  universal Fermi interaction}\/},
\href{http://dx.doi.org/10.1103/PhysRev.109.1860.2}{Phys. Rev. {\bf 109} (1958)
   1860--1860}.

\bibitem{Feynman:1958ty}
R.~P. Feynman and M.~Gell-Mann, {\em {Theory of Fermi interaction}\/},
\href{http://dx.doi.org/10.1103/PhysRev.109.193}{Phys. Rev. {\bf 109} (1958)
  193--198}.

\bibitem{Weinberg:1967tq}
S.~Weinberg, {\em {A Model of Leptons}\/},
\href{http://dx.doi.org/10.1103/PhysRevLett.19.1264}{Phys. Rev. Lett. {\bf 19}
  (1967)  1264--1266}.

\bibitem{Salam:1968rm}
A.~Salam, {\em {Weak and Electromagnetic Interactions}\/},
Conf. Proc. {\bf C680519} (1968)  367--377.

\bibitem{Rubbia:1985pv}
C.~Rubbia, {\em {Experimental Observation of the Intermediate Vector Bosons
  $W^+$, $W^-$, and $Z^0$}\/},
\href{http://dx.doi.org/10.1103/RevModPhys.57.699}{Rev. Mod. Phys. {\bf 57}
  (1985)  699--722}.

\bibitem{ALEPH:2005ab}
{ALEPH, DELPHI, L3, OPAL, SLD, LEP Electroweak Working Group, SLD Electroweak
  Group, SLD Heavy Flavour Group Collaboration}, S.~Schael et al., {\em
  {Precision electroweak measurements on the $Z$ resonance}\/},
  \href{http://dx.doi.org/10.1016/j.physrep.2005.12.006}{Phys. Rept. {\bf 427}
  (2006)  257--454},
\href{http://arxiv.org/abs/hep-ex/0509008}{{\tt arXiv:hep-ex/0509008
  [hep-ex]}}.

\bibitem{Aad:2012tfa}
{ATLAS Collaboration}, {\em {Observation of a new particle in the search for
  the Standard Model Higgs boson with the ATLAS detector at the LHC}\/},
  \href{http://dx.doi.org/10.1016/j.physletb.2012.08.020}{Phys. Lett. {\bf
  B716} (2012)  1--29},
\href{http://arxiv.org/abs/1207.7214}{{\tt arXiv:1207.7214 [hep-ex]}}.

\bibitem{Chatrchyan:2012xdj}
{CMS Collaboration}, {\em {Observation of a New Boson at a Mass of 125 GeV with
  the CMS Experiment at the LHC}\/},
  \href{http://dx.doi.org/10.1016/j.physletb.2012.08.021}{Phys. Lett. {\bf
  B716} (2012)  30--61},
\href{http://arxiv.org/abs/1207.7235}{{\tt arXiv:1207.7235 [hep-ex]}}.

\bibitem{Contino:2017moj}
R.~Contino, D.~Greco, R.~Mahbubani, R.~Rattazzi, and R.~Torre, {\em {Precision
  Tests and Fine Tuning in Twin Higgs Models}\/},
  \href{http://dx.doi.org/10.1103/PhysRevD.96.095036}{Phys. Rev. {\bf D96}
  (2017) no.~9, 095036},
\href{http://arxiv.org/abs/1702.00797}{{\tt arXiv:1702.00797 [hep-ph]}}.

\bibitem{Peskin:1990zt}
M.~E. Peskin and T.~Takeuchi, {\em {A new constraint on a strongly interacting
  Higgs sector}\/},
\href{http://dx.doi.org/10.1103/PhysRevLett.65.964}{Phys. Rev. Lett. {\bf 65}
  (1990)  964--967}.

\bibitem{Peskin:1991sw}
M.~E. Peskin and T.~Takeuchi, {\em {Estimation of oblique electroweak
  corrections}\/},
\href{http://dx.doi.org/10.1103/PhysRevD.46.381}{Phys. Rev. {\bf D46} (1992)
  381--409}.

\bibitem{Altarelli:1991fk}
G.~Altarelli, R.~Barbieri, and S.~Jadach, {\em {Toward a model independent
  analysis of electroweak data}\/},
  \href{http://dx.doi.org/10.1016/0550-3213(92)90376-M}{Nucl. Phys. {\bf B369}
  (1992)  3--32}.
[Erratum: Nucl. Phys. B376, 444 (1992)].

\bibitem{Golden:1990ig}
M.~Golden and L.~Randall, {\em {Radiative Corrections to Electroweak Parameters
  in Technicolor Theories}\/},
\href{http://dx.doi.org/10.1016/0550-3213(91)90614-4}{Nucl. Phys. {\bf B361}
  (1991)  3--23}.

\bibitem{Barbieri:2004qk}
R.~Barbieri, A.~Pomarol, R.~Rattazzi, and A.~Strumia, {\em {Electroweak
  symmetry breaking after LEP-1 and LEP-2}\/},
  \href{http://dx.doi.org/10.1016/j.nuclphysb.2004.10.014}{Nucl. Phys. {\bf
  B703} (2004)  127--146},
\href{http://arxiv.org/abs/hep-ph/0405040}{{\tt arXiv:hep-ph/0405040
  [hep-ph]}}.

\bibitem{atlasstop}
{ATLAS Collaboration}, {\em {Search for top-squark pair production in final
  states with one lepton, jets, and missing transverse momentum using 36
  fb$^{-1}$ of $\sqrt{s}=13 $ TeV pp collision data with the ATLAS
  detector}\/},  \href{http://dx.doi.org/10.1007/JHEP06(2018)108}{JHEP {\bf 06}
  (2018)  108},
\href{http://arxiv.org/abs/1711.11520}{{\tt arXiv:1711.11520 [hep-ex]}}.

\bibitem{cmsstop}
{CMS Collaboration}, {\em {Search for top squark pair production in pp
  collisions at $ \sqrt{s}=13 $ TeV using single lepton events}\/},
  \href{http://dx.doi.org/10.1007/JHEP10(2017)019}{JHEP {\bf 10} (2017)  019},
\href{http://arxiv.org/abs/1706.04402}{{\tt arXiv:1706.04402 [hep-ex]}}.

\bibitem{atlasvtop}
{ATLAS Collaboration}, {\em {Combination of the searches for pair-produced
  vector-like partners of the third-generation quarks at $\sqrt{s} =$ 13 TeV
  with the ATLAS detector}\/},
  \href{http://dx.doi.org/10.1103/PhysRevLett.121.211801}{Phys. Rev. Lett. {\bf
  121} (2018) no.~21, 211801},
\href{http://arxiv.org/abs/1808.02343}{{\tt arXiv:1808.02343 [hep-ex]}}.

\bibitem{cmsvtop}
{CMS Collaboration}, {\em {Search for vector-like quarks in events with two
  oppositely charged leptons and jets in proton-proton collisions at $\sqrt{s}
  =$ 13 TeV}\/},  \href{http://dx.doi.org/10.1140/epjc/s10052-019-6855-8}{Eur.
  Phys. J. {\bf C79} (2019) no.~4, 364},
\href{http://arxiv.org/abs/1812.09768}{{\tt arXiv:1812.09768 [hep-ex]}}.

\bibitem{atlashcomb2}
{ATLAS Collaboration}, {\em Combined measurements of Higgs boson production and
  decay using up to 80 inverse fb of proton--proton collision data at
  $\sqrt{s}= 13$~TeV collected with the ATLAS experiment\/},
  ATLAS-CONF-2019-005, CERN, Geneva, Mar, 2019.
\newblock \url{http://cdsweb.cern.ch/record/2668375}.

\bibitem{cmshcomb2}
{CMS Collaboration}, {\em {Combined measurements of Higgs boson couplings in
  proton-proton collisions at $\sqrt{s}=13$~TeV}\/},
  \href{http://dx.doi.org/10.1140/epjc/s10052-019-6909-y}{Eur. Phys. J. {\bf
  C79} (2019) no.~5, 421},
\href{http://arxiv.org/abs/1809.10733}{{\tt arXiv:1809.10733 [hep-ex]}}.

\bibitem{Cepeda:2019klc}
{HL/HE WG2 group Collaboration}, M.~Cepeda et al., {\em {Higgs Physics at the
  HL-LHC and HE-LHC}\/},
\href{http://arxiv.org/abs/1902.00134}{{\tt arXiv:1902.00134 [hep-ph]}}.

\bibitem{Abramowicz:2016zbo}
H.~Abramowicz et al., {\em {Higgs physics at the CLIC electron-positron linear
  collider}\/},  \href{http://dx.doi.org/10.1140/epjc/s10052-017-4968-5}{Eur.
  Phys. J. {\bf C77} (2017) no.~7, 475},
\href{http://arxiv.org/abs/1608.07538}{{\tt arXiv:1608.07538 [hep-ex]}}.

\bibitem{ALEPH:2010aa}
{ALEPH, CDF, D0, DELPHI, L3, OPAL, SLD, LEP Electroweak Working Group, Tevatron
  Electroweak Working Group, SLD Electroweak and Heavy Flavour Groups
  Collaborations}, L.~E.~W. Group, {\em {Precision Electroweak Measurements and
  Constraints on the Standard Model}\/},
\href{http://arxiv.org/abs/1012.2367}{{\tt arXiv:1012.2367 [hep-ex]}}.

\bibitem{LEP-2}
{The ALEPH, DELPHI, L3, OPAL Collaborations, the LEP Electroweak Working
  Group}, {\em {Electroweak Measurements in Electron-Positron Collisions at
  W-Boson-Pair Energies at LEP}\/},  Phys. Rept. {\bf 532} (2013)  119,
\href{http://arxiv.org/abs/1302.3415}{{\tt arXiv:1302.3415 [hep-ex]}}.

\bibitem{tevmw}
{CDF, D0 Collaboration}, T.~A. Aaltonen et al., {\em {Combination of CDF and D0
  $W$-Boson Mass Measurements}\/},
  \href{http://dx.doi.org/10.1103/PhysRevD.88.052018}{Phys. Rev. {\bf D88}
  (2013) no.~5, 052018},
\href{http://arxiv.org/abs/1307.7627}{{\tt arXiv:1307.7627 [hep-ex]}}.

\bibitem{tevmtop}
{CDF, D0 Collaboration}, T.~Aaltonen et al., {\em {Combination of the top-quark
  mass measurements from the Tevatron collider}\/},
  \href{http://dx.doi.org/10.1103/PhysRevD.86.092003}{Phys. Rev. {\bf D86}
  (2012)  092003},
\href{http://arxiv.org/abs/1207.1069}{{\tt arXiv:1207.1069 [hep-ex]}}.

\bibitem{Aaboud:2017svj}
{ATLAS Collaboration}, M.~Aaboud et al., {\em {Measurement of the $W$-boson
  mass in pp collisions at $\sqrt{s}=7$ TeV with the ATLAS detector}\/},
  \href{http://dx.doi.org/10.1140/epjc/s10052-018-6354-3,
  10.1140/epjc/s10052-017-5475-4}{Eur. Phys. J. {\bf C78} (2018) no.~2, 110},
  \href{http://arxiv.org/abs/1701.07240}{{\tt arXiv:1701.07240 [hep-ex]}}.
[Erratum: Eur. Phys. J.C78,no.11,898(2018)].

\bibitem{Tokar:2019wny}
{ATLAS, CMS Collaboration}, S.~Tokar, {\em {Top-quark mass at ATLAS and
  CMS}\/},  in {\em {11th International Workshop on Top Quark Physics (TOP2018)
  Bad Neuenahr, Germany, September 16-21, 2018}}.
\newblock 2019.
\newblock
\href{http://arxiv.org/abs/1901.04740}{{\tt arXiv:1901.04740 [hep-ex]}}.
\newblock

\bibitem{Aaboud:2018wps}
{ATLAS Collaboration}, M.~Aaboud et al., {\em {Measurement of the Higgs boson
  mass in the $H\rightarrow ZZ^* \rightarrow 4\ell$ and $H \rightarrow
  \gamma\gamma$ channels with $\sqrt{s}=13$ TeV $pp$ collisions using the ATLAS
  detector}\/},  \href{http://dx.doi.org/10.1016/j.physletb.2018.07.050}{Phys.
  Lett. {\bf B784} (2018)  345--366},
\href{http://arxiv.org/abs/1806.00242}{{\tt arXiv:1806.00242 [hep-ex]}}.

\bibitem{Sirunyan:2017exp}
{CMS Collaboration}, {\em {Measurements of properties of the Higgs boson
  decaying into the four-lepton final state in pp collisions at $ \sqrt{s}=13 $
  TeV}\/},  \href{http://dx.doi.org/10.1007/JHEP11(2017)047}{JHEP {\bf 11}
  (2017)  047},
\href{http://arxiv.org/abs/1706.09936}{{\tt arXiv:1706.09936 [hep-ex]}}.

\bibitem{Tanabashi:2018oca}
{Particle Data Group Collaboration}, M.~Tanabashi et al., {\em {Review of
  Particle Physics}\/},
\href{http://dx.doi.org/10.1103/PhysRevD.98.030001}{Phys. Rev. {\bf D98} (2018)
  no.~3, 030001}.

\bibitem{Baak:2014ora}
{Gfitter Group Collaboration}, M.~Baak et al., {\em {The global electroweak fit
  at NNLO and prospects for the LHC and ILC}\/},
  \href{http://dx.doi.org/10.1140/epjc/s10052-014-3046-5}{Eur. Phys. J. {\bf
  C74} (2014)  3046},
\href{http://arxiv.org/abs/1407.3792}{{\tt arXiv:1407.3792 [hep-ph]}}.

\bibitem{deBlas:2016ojx}
J.~de~Blas, M.~Ciuchini, E.~Franco, S.~Mishima, M.~Pierini, L.~Reina, and
  L.~Silvestrini, {\em {Electroweak precision observables and Higgs-boson
  signal strengths in the Standard Model and beyond: present and future}\/},
  \href{http://dx.doi.org/10.1007/JHEP12(2016)135}{JHEP {\bf 12} (2016)  135},
\href{http://arxiv.org/abs/1608.01509}{{\tt arXiv:1608.01509 [hep-ph]}}.

\bibitem{Caprini:2015zlo}
C.~Caprini et al., {\em {Science with the space-based interferometer eLISA. II:
  Gravitational waves from cosmological phase transitions}\/},
  \href{http://dx.doi.org/10.1088/1475-7516/2016/04/001}{JCAP {\bf 1604} (2016)
  no.~04, 001},
\href{http://arxiv.org/abs/1512.06239}{{\tt arXiv:1512.06239 [astro-ph.CO]}}.

\bibitem{Fujii:2019zll}
{LCC Physics Working Group Collaboration}, K.~Fujii et al., {\em {Tests of the
  Standard Model at the International Linear Collider}\/},
\href{http://arxiv.org/abs/1908.11299}{{\tt arXiv:1908.11299 [hep-ex]}}.

\bibitem{gigazclic}
J.-J. Blaising and P.~Roloff, {\em Electroweak couplings of the $Z$ boson at
  CLIC\/},  Private communication.

\bibitem{deBlas:2019rxi}
J.~de~Blas et al., {\em {Higgs Boson Studies at Future Particle Colliders}\/},
\href{http://arxiv.org/abs/1905.03764}{{\tt arXiv:1905.03764 [hep-ph]}}.

\bibitem{Zeller:2001hh}
{NuTeV Collaboration}, G.~P. Zeller et al., {\em {A Precise determination of
  electroweak parameters in neutrino nucleon scattering}\/},
  \href{http://dx.doi.org/10.1103/PhysRevLett.88.091802,
  10.1103/PhysRevLett.90.239902}{Phys. Rev. Lett. {\bf 88} (2002)  091802},
  \href{http://arxiv.org/abs/hep-ex/0110059}{{\tt arXiv:hep-ex/0110059
  [hep-ex]}}.
[Erratum: Phys. Rev. Lett.90,239902(2003)].

\bibitem{Beneke:2007zg}
M.~Beneke, P.~Falgari, C.~Schwinn, A.~Signer, and G.~Zanderighi, {\em
  {Four-fermion production near the W pair production threshold}\/},
  \href{http://dx.doi.org/10.1016/j.nuclphysb.2007.09.030}{Nucl. Phys. {\bf
  B792} (2008)  89--135},
\href{http://arxiv.org/abs/0707.0773}{{\tt arXiv:0707.0773 [hep-ph]}}.

\bibitem{Abi:2018dnh}
{DUNE Collaboration}, B.~Abi et al., {\em {The DUNE Far Detector Interim Design
  Report Volume 1: Physics, Technology and Strategies}\/},
\href{http://arxiv.org/abs/1807.10334}{{\tt arXiv:1807.10334
  [physics.ins-det]}}.

\bibitem{Accardi:2012qut}
A.~Accardi et al., {\em {Electron Ion Collider: The Next QCD Frontier}\/},
  \href{http://dx.doi.org/10.1140/epja/i2016-16268-9}{Eur. Phys. J. {\bf A52}
  (2016) no.~9, 268},
\href{http://arxiv.org/abs/1212.1701}{{\tt arXiv:1212.1701 [nucl-ex]}}.

\bibitem{Odom:2006zz}
B.~C. Odom, D.~Hanneke, B.~D'Urso, and G.~Gabrielse, {\em {New Measurement of
  the Electron Magnetic Moment Using a One-Electron Quantum Cyclotron}\/},
  Phys. Rev. Lett. {\bf 97} (2006)  030801.
[Erratum: Phys. Rev. Lett.99,039902(2007)].

\bibitem{Gabrielse:2006gg}
G.~Gabrielse, D.~Hanneke, T.~Kinoshita, M.~Nio, and B.~C. Odom, {\em {New
  Determination of the Fine Structure Constant from the Electron g Value and
  QED}\/},  \href{http://dx.doi.org/10.1103/PhysRevLett.97.030802,
  10.1103/PhysRevLett.99.039902}{Phys. Rev. Lett. {\bf 97} (2006)  030802}.
[Erratum: Phys. Rev. Lett.99,039902(2007)].

\bibitem{Volkov:2017xaq}
S.~Volkov, {\em {New method of computing the contributions of graphs without
  lepton loops to the electron anomalous magnetic moment in QED}\/},
  \href{http://dx.doi.org/10.1103/PhysRevD.96.096018}{Phys. Rev. {\bf D96}
  (2017) no.~9, 096018},
\href{http://arxiv.org/abs/1705.05800}{{\tt arXiv:1705.05800 [hep-ph]}}.

\bibitem{Blum:2013xva}
T.~Blum, A.~Denig, I.~Logashenko, E.~de~Rafael, B.~L. Roberts, T.~Teubner, and
  G.~Venanzoni, {\em {The Muon (g-2) Theory Value: Present and Future}\/},
\href{http://arxiv.org/abs/1311.2198}{{\tt arXiv:1311.2198 [hep-ph]}}.

\bibitem{Jegerlehner:2018zrj}
F.~Jegerlehner, {\em {The Muon g-2 in Progress}\/},
  \href{http://dx.doi.org/10.5506/APhysPolB.49.1157}{Acta Phys. Polon. {\bf
  B49} (2018)  1157},
\href{http://arxiv.org/abs/1804.07409}{{\tt arXiv:1804.07409 [hep-ph]}}.

\bibitem{Grange:2015fou}
{Muon g-2 Collaboration}, J.~Grange et al., {\em {Muon (g-2) Technical Design
  Report}\/},
\href{http://arxiv.org/abs/1501.06858}{{\tt arXiv:1501.06858
  [physics.ins-det]}}.

\bibitem{Abbiendi:2016xup}
G.~Abbiendi et al., {\em {Measuring the leading hadronic contribution to the
  muon g-2 via $\mu e$ scattering}\/},
  \href{http://dx.doi.org/10.1140/epjc/s10052-017-4633-z}{Eur. Phys. J. {\bf
  C77} (2017) no.~3, 139},
\href{http://arxiv.org/abs/1609.08987}{{\tt arXiv:1609.08987 [hep-ex]}}.

\bibitem{Blondel:2019vdq}
A.~Blondel et al., eds., {\em {Theory report on the 11th FCC-ee workshop}}.
\newblock 2019.
\newblock
\href{http://arxiv.org/abs/1905.05078}{{\tt arXiv:1905.05078 [hep-ph]}}.
\newblock

\bibitem{Janot:2015gjr}
P.~Janot, {\em {Direct measurement of $\alpha_{QED}(m_{Z}^{2})$ at the
  FCC-ee}\/},  \href{http://dx.doi.org/10.1007/JHEP02(2016)053,
  10.1007/JHEP11(2017)164}{JHEP {\bf 02} (2016)  053},
  \href{http://arxiv.org/abs/1512.05544}{{\tt arXiv:1512.05544 [hep-ph]}}.
[Erratum: JHEP11,164(2017)].

\bibitem{Sauter:1931zz}
F.~Sauter, {\em {Uber das Verhalten eines Elektrons im homogenen elektrischen
  Feld nach der relativistischen Theorie Diracs}\/},
\href{http://dx.doi.org/10.1007/BF01339461}{Z. Phys. {\bf 69} (1931)
  742--764}.

\bibitem{Heisenberg:1935qt}
W.~Heisenberg and H.~Euler, {\em {Consequences of Dirac's theory of
  positrons}\/},  \href{http://dx.doi.org/10.1007/BF01343663,
  10.1007/978-3-642-70078-1_9}{Z. Phys. {\bf 98} (1936) no.~11-12, 714--732},
\href{http://arxiv.org/abs/physics/0605038}{{\tt arXiv:physics/0605038
  [physics]}}.

\bibitem{Schwinger:1951nm}
J.~S. Schwinger, {\em {On gauge invariance and vacuum polarization}\/},
\href{http://dx.doi.org/10.1103/PhysRev.82.664}{Phys. Rev. {\bf 82} (1951)
  664--679}.

\bibitem{Caldwell:2018atq}
A.~Caldwell et al., {\em {Particle physics applications of the AWAKE
  acceleration scheme}\/},
\href{http://arxiv.org/abs/1812.11164}{{\tt arXiv:1812.11164
  [physics.acc-ph]}}.

\bibitem{Altarelli:2006zza}
M.~Altarelli et al., eds., \href{http://dx.doi.org/10.3204/DESY_06-097}{{\em
  {XFEL: The European X-Ray Free-Electron Laser. Technical design report}}}.
\newblock 2006.
\newblock
\url{http://www-library.desy.de/cgi-bin/showprep.pl?desy06-097}.
\newblock

\bibitem{SLAC:2016van}
SLAC, {\em {Preliminary Conceptual Design Report for the FACET-II Project at
  SLAC National Accelerator Laboratory}\/}, .
SLAC-R-1067.

\bibitem{Burke:1997ew}
D.~L. Burke et al., {\em {Positron production in multi-photon light by light
  scattering}\/},
\href{http://dx.doi.org/10.1103/PhysRevLett.79.1626}{Phys. Rev. Lett. {\bf 79}
  (1997)  1626--1629}.

\bibitem{Andersen:2012ea}
{CERN NA63 Collaboration}, K.~Andersen et al., {\em {Experimental
  investigations of synchrotron radiation at the onset of the quantum
  regime}\/},  \href{http://dx.doi.org/10.1103/PhysRevD.86.072001}{Phys. Rev.
  {\bf D86} (2012)  072001},
\href{http://arxiv.org/abs/1206.6577}{{\tt arXiv:1206.6577 [physics.acc-ph]}}.

\bibitem{Kouveliotou:1998ze}
C.~Kouveliotou et al., {\em {An X-ray pulsar with a superstrong magnetic field
  in the soft gamma-ray repeater SGR 1806-20.}\/},
\href{http://dx.doi.org/10.1038/30410}{Nature {\bf 393} (1998)  235--237}.

\bibitem{pomeranchuk}
I.~Pomeranchuk and Y.~Smorodinsky, {\em {On energy levels in systems with
  $Z>137$}\/},  J. Phys. USSR {\bf 9} (1945)  97.

\bibitem{Bell:1987rw}
M.~Bell and J.~S. Bell, {\em {End Effects in Quantum Beamstrahlung}\/},
Part. Accel. {\bf 24} (1988)  1.

\bibitem{Blankenbecler:1988te}
R.~Blankenbecler and S.~D. Drell, {\em {Quantum Beamstrahlung: Prospects for a
  Photon-photon Collider}\/},
  \href{http://dx.doi.org/10.1103/PhysRevLett.62.116.3,
  10.1103/PhysRevLett.61.2324}{Phys. Rev. Lett. {\bf 61} (1988)  2324}.
[Erratum: Phys. Rev. Lett.62,116(1989)].

\bibitem{LHCHiggsCrossSectionWorkingGroup:2012nn}
{LHC Higgs Cross Section Working Group Collaboration}, A.~David et al., {\em
  {LHC HXSWG interim recommendations to explore the coupling structure of a
  Higgs-like particle}\/},
\href{http://arxiv.org/abs/1209.0040}{{\tt arXiv:1209.0040 [hep-ph]}}.

\bibitem{LHCHXSWG3}
{LHC Higgs Cross Section Working Group Collaboration}, J.~R. Andersen et al.,
  {\em {Handbook of LHC Higgs Cross Sections: 3. Higgs Properties}\/},
\href{http://arxiv.org/abs/1307.1347}{{\tt arXiv:1307.1347 [hep-ph]}}.

\bibitem{Khachatryan:2014aep}
{CMS Collaboration}, V.~Khachatryan et al., {\em {Search for a standard
  model-like Higgs boson in the $\mu^+\mu^-$ and $e^+e^-$ decay channels at the
  LHC}\/},  \href{http://dx.doi.org/10.1016/j.physletb.2015.03.048}{Phys. Lett.
  {\bf B744} (2015)  184--207},
\href{http://arxiv.org/abs/1410.6679}{{\tt arXiv:1410.6679 [hep-ex]}}.

\bibitem{Abada:2019zxq}
{FCC Collaboration}, A.~Abada et al., {\em {FCC-ee: The Lepton Collider}\/}, .
[Eur. Phys. J. ST228,no.2,261(2019)].

\bibitem{McCullough:2013rea}
M.~McCullough, {\em {An Indirect Model-Dependent Probe of the Higgs
  Self-Coupling}\/},  \href{http://dx.doi.org/10.1103/PhysRevD.90.015001,
  10.1103/PhysRevD.92.039903}{Phys. Rev. {\bf D90} (2014) no.~1, 015001},
  \href{http://arxiv.org/abs/1312.3322}{{\tt arXiv:1312.3322 [hep-ph]}}.
[Erratum: Phys. Rev.D92,no.3,039903(2015)].

\bibitem{Degrassi:2016wml}
G.~Degrassi, P.~P. Giardino, F.~Maltoni, and D.~Pagani, {\em {Probing the Higgs
  self coupling via single Higgs production at the LHC}\/},
  \href{http://dx.doi.org/10.1007/JHEP12(2016)080}{JHEP {\bf 12} (2016)  080},
\href{http://arxiv.org/abs/1607.04251}{{\tt arXiv:1607.04251 [hep-ph]}}.

\bibitem{Bizon:2016wgr}
W.~Bizon, M.~Gorbahn, U.~Haisch, and G.~Zanderighi, {\em {Constraints on the
  trilinear Higgs coupling from vector boson fusion and associated Higgs
  production at the LHC}\/},
  \href{http://dx.doi.org/10.1007/JHEP07(2017)083}{JHEP {\bf 07} (2017)  083},
\href{http://arxiv.org/abs/1610.05771}{{\tt arXiv:1610.05771 [hep-ph]}}.

\bibitem{vanderBij:1985ww}
J.~J. van~der Bij, {\em {Does Low-energy Physics Depend on the Potential of a
  Heavy Higgs Particle?}\/},
\href{http://dx.doi.org/10.1016/0550-3213(86)90131-8}{Nucl. Phys. {\bf B267}
  (1986)  557--565}.

\bibitem{Degrassi:2017ucl}
G.~Degrassi, M.~Fedele, and P.~P. Giardino, {\em {Constraints on the trilinear
  Higgs self coupling from precision observables}\/},
  \href{http://dx.doi.org/10.1007/JHEP04(2017)155}{JHEP {\bf 04} (2017)  155},
\href{http://arxiv.org/abs/1702.01737}{{\tt arXiv:1702.01737 [hep-ph]}}.

\bibitem{Kribs:2017znd}
G.~D. Kribs, A.~Maier, H.~Rzehak, M.~Spannowsky, and P.~Waite, {\em
  {Electroweak oblique parameters as a probe of the trilinear Higgs boson
  self-interaction}\/},
  \href{http://dx.doi.org/10.1103/PhysRevD.95.093004}{Phys. Rev. {\bf D95}
  (2017) no.~9, 093004},
\href{http://arxiv.org/abs/1702.07678}{{\tt arXiv:1702.07678 [hep-ph]}}.

\bibitem{Blondel:2019yqr}
N.~Alipour~Tehrani et al., {\em {FCC-ee: Your Questions Answered}\/},  in {\em
  {CERN Council Open Symposium on the Update of European Strategy for Particle
  Physics (EPPSU) Granada, Spain, May 13-16, 2019}}, A.~Blondel and P.~Janot,
  eds.
\newblock 2019.
\newblock
\href{http://arxiv.org/abs/1906.02693}{{\tt arXiv:1906.02693 [hep-ph]}}.
\newblock

\bibitem{babyfcc}
M.~Mangano, {\em Physics potential of a low-energy FCC-hh\/},  July, 2019.

\bibitem{Blondel:2019qlh}
A.~Blondel, A.~Freitas, J.~Gluza, T.~Riemann, S.~Heinemeyer, S.~Jadach, and
  P.~Janot, {\em {Theory Requirements and Possibilities for the FCC-ee and
  other Future High Energy and Precision Frontier Lepton Colliders}\/},
\href{http://arxiv.org/abs/1901.02648}{{\tt arXiv:1901.02648 [hep-ph]}}.

\bibitem{Blondel:2018mad}
A.~Blondel et al., {\em {Standard Model Theory for the FCC-ee: The Tera-Z}\/},
  in {\em {Mini Workshop on Precision EW and QCD Calculations for the FCC
  Studies : Methods and Techniques CERN, Geneva, Switzerland, January 12-13,
  2018}}.
\newblock 2018.
\newblock
\href{http://arxiv.org/abs/1809.01830}{{\tt arXiv:1809.01830 [hep-ph]}}.
\newblock

\bibitem{Lepage:2014fla}
G.~P. Lepage, P.~B. Mackenzie, and M.~E. Peskin, {\em {Expected Precision of
  Higgs Boson Partial Widths within the Standard Model}\/},
\href{http://arxiv.org/abs/1404.0319}{{\tt arXiv:1404.0319 [hep-ph]}}.

\bibitem{deFlorian:2016spz}
{LHC Higgs Cross Section Working Group Collaboration}, D.~de~Florian et al.,
  {\em {Handbook of LHC Higgs Cross Sections: 4. Deciphering the Nature of the
  Higgs Sector}\/},
\href{http://arxiv.org/abs/1610.07922}{{\tt arXiv:1610.07922 [hep-ph]}}.

\bibitem{Freitas:2019bre}
A.~Freitas et al., {\em {Theoretical uncertainties for electroweak and
  Higgs-boson precision measurements at FCC-ee}\/},
\href{http://arxiv.org/abs/1906.05379}{{\tt arXiv:1906.05379 [hep-ph]}}.

\bibitem{Corcella:1999qn}
G.~Corcella et al., {\em {HERWIG 6.1 release note}\/},
\href{http://arxiv.org/abs/hep-ph/9912396}{{\tt arXiv:hep-ph/9912396
  [hep-ph]}}.

\bibitem{Sjostrand:2000wi}
T.~Sj{\"o}strand et al., {\em {High-energy physics event generation with PYTHIA
  6.1}\/},  \href{http://dx.doi.org/10.1016/S0010-4655(00)00236-8}{Comput.
  Phys. Commun. {\bf 135} (2001)  238--259},
\href{http://arxiv.org/abs/hep-ph/0010017}{{\tt arXiv:hep-ph/0010017
  [hep-ph]}}.

\bibitem{Frederix:2018nkq}
R.~Frederix et al., {\em {The automation of next-to-leading order electroweak
  calculations}\/},  \href{http://dx.doi.org/10.1007/JHEP07(2018)185}{JHEP {\bf
  07} (2018)  185},
\href{http://arxiv.org/abs/1804.10017}{{\tt arXiv:1804.10017 [hep-ph]}}.

\bibitem{Alwall:2014hca}
J.~Alwall et al., {\em {The automated computation of tree-level and
  next-to-leading order differential cross sections, and their matching to
  parton shower simulations}\/},
  \href{http://dx.doi.org/10.1007/JHEP07(2014)079}{JHEP {\bf 07} (2014)  079},
\href{http://arxiv.org/abs/1405.0301}{{\tt arXiv:1405.0301 [hep-ph]}}.

\bibitem{Ablinger:2017tan}
J.~Ablinger et al., {\em {The three-loop splitting functions $P_{qg}^{(2)}$ and
  $P_{gg}^{(2, N_F)}$}\/},
  \href{http://dx.doi.org/10.1016/j.nuclphysb.2017.06.004}{Nucl. Phys. {\bf
  B922} (2017)  1--40},
\href{http://arxiv.org/abs/1705.01508}{{\tt arXiv:1705.01508 [hep-ph]}}.

\bibitem{Behring:2019tus}
A.~Behring et al., {\em {The Polarized Three-Loop Anomalous Dimensions from
  On-Shell Massive Operator Matrix Elements}\/},
\href{http://arxiv.org/abs/1908.03779}{{\tt arXiv:1908.03779 [hep-ph]}}.

\bibitem{Mistlberger:2018etf}
B.~Mistlberger, {\em {Higgs boson production at hadron colliders at N$^{3}$LO
  in QCD}\/},  \href{http://dx.doi.org/10.1007/JHEP05(2018)028}{JHEP {\bf 05}
  (2018)  028},
\href{http://arxiv.org/abs/1802.00833}{{\tt arXiv:1802.00833 [hep-ph]}}.

\bibitem{Vogt:2004mw}
A.~Vogt, S.~Moch, and J.~A.~M. Vermaseren, {\em {The Three-loop splitting
  functions in QCD: The Singlet case}\/},
  \href{http://dx.doi.org/10.1016/j.nuclphysb.2004.04.024}{Nucl. Phys. {\bf
  B691} (2004)  129--181},
\href{http://arxiv.org/abs/hep-ph/0404111}{{\tt arXiv:hep-ph/0404111
  [hep-ph]}}.

\bibitem{Moch:2004pa}
S.~Moch, J.~A.~M. Vermaseren, and A.~Vogt, {\em {The Three loop splitting
  functions in QCD: The Nonsinglet case}\/},
  \href{http://dx.doi.org/10.1016/j.nuclphysb.2004.03.030}{Nucl. Phys. {\bf
  B688} (2004)  101--134},
\href{http://arxiv.org/abs/hep-ph/0403192}{{\tt arXiv:hep-ph/0403192
  [hep-ph]}}.

\bibitem{Butterworth:2015oua}
J.~Butterworth et al., {\em {PDF4LHC recommendations for LHC Run II}\/},
  \href{http://dx.doi.org/10.1088/0954-3899/43/2/023001}{J. Phys. {\bf G43}
  (2016)  023001},
\href{http://arxiv.org/abs/1510.03865}{{\tt arXiv:1510.03865 [hep-ph]}}.

\bibitem{Dulat:2018rbf}
F.~Dulat, A.~Lazopoulos, and B.~Mistlberger, {\em {iHixs 2 - Inclusive Higgs
  cross sections}\/},
  \href{http://dx.doi.org/10.1016/j.cpc.2018.06.025}{Comput. Phys. Commun. {\bf
  233} (2018)  243--260},
\href{http://arxiv.org/abs/1802.00827}{{\tt arXiv:1802.00827 [hep-ph]}}.

\bibitem{Dulat:2018bfe}
F.~Dulat, B.~Mistlberger, and A.~Pelloni, {\em {Precision predictions at
  N$^3$LO for the Higgs boson rapidity distribution at the LHC}\/},
  \href{http://dx.doi.org/10.1103/PhysRevD.99.034004}{Phys. Rev. {\bf D99}
  (2019) no.~3, 034004},
\href{http://arxiv.org/abs/1810.09462}{{\tt arXiv:1810.09462 [hep-ph]}}.

\bibitem{Cieri:2018oms}
L.~Cieri, X.~Chen, T.~Gehrmann, E.~W.~N. Glover, and A.~Huss, {\em {Higgs boson
  production at the LHC using the $q_T$ subtraction formalism at N$^3$LO
  QCD}\/},  \href{http://dx.doi.org/10.1007/JHEP02(2019)096}{JHEP {\bf 02}
  (2019)  096},
\href{http://arxiv.org/abs/1807.11501}{{\tt arXiv:1807.11501 [hep-ph]}}.

\bibitem{Chen:2016zka}
X.~Chen, J.~Cruz-Martinez, T.~Gehrmann, E.~W.~N. Glover, and M.~Jaquier, {\em
  {NNLO QCD corrections to Higgs boson production at large transverse
  momentum}\/},  \href{http://dx.doi.org/10.1007/JHEP10(2016)066}{JHEP {\bf 10}
  (2016)  066},
\href{http://arxiv.org/abs/1607.08817}{{\tt arXiv:1607.08817 [hep-ph]}}.

\bibitem{Jones:2018hbb}
S.~P. Jones, M.~Kerner, and G.~Luisoni, {\em {Next-to-Leading-Order QCD
  Corrections to Higgs Boson Plus Jet Production with Full Top-Quark Mass
  Dependence}\/},
  \href{http://dx.doi.org/10.1103/PhysRevLett.120.162001}{Phys. Rev. Lett. {\bf
  120} (2018) no.~16, 162001},
\href{http://arxiv.org/abs/1802.00349}{{\tt arXiv:1802.00349 [hep-ph]}}.

\bibitem{Cruz-Martinez:2018dvl}
J.~Cruz-Martinez, E.~W.~N. Glover, T.~Gehrmann, and A.~Huss, {\em {NNLO
  corrections to VBF Higgs boson production}\/},
  \href{http://dx.doi.org/10.22323/1.303.0003}{PoS {\bf LL2018} (2018)  003},
\href{http://arxiv.org/abs/1807.07908}{{\tt arXiv:1807.07908 [hep-ph]}}.

\bibitem{Cacciari:2015jma}
M.~Cacciari, F.~A. Dreyer, A.~Karlberg, G.~P. Salam, and G.~Zanderighi, {\em
  {Fully Differential Vector-Boson-Fusion Higgs Production at
  Next-to-Next-to-Leading Order}\/},
  \href{http://dx.doi.org/10.1103/PhysRevLett.115.082002,
  10.1103/PhysRevLett.120.139901}{Phys. Rev. Lett. {\bf 115} (2015) no.~8,
  082002}, \href{http://arxiv.org/abs/1506.02660}{{\tt arXiv:1506.02660
  [hep-ph]}}.
[Erratum: Phys. Rev. Lett.120,no.13,139901(2018)].

\bibitem{Figy:2007kv}
T.~Figy, V.~Hankele, and D.~Zeppenfeld, {\em {Next-to-leading order QCD
  corrections to Higgs plus three jet production in vector-boson fusion}\/},
  \href{http://dx.doi.org/10.1088/1126-6708/2008/02/076}{JHEP {\bf 02} (2008)
  076},
\href{http://arxiv.org/abs/0710.5621}{{\tt arXiv:0710.5621 [hep-ph]}}.

\bibitem{Dreyer:2016oyx}
F.~A. Dreyer and A.~Karlberg, {\em {Vector-Boson Fusion Higgs Production at
  Three Loops in QCD}\/},
  \href{http://dx.doi.org/10.1103/PhysRevLett.117.072001}{Phys. Rev. Lett. {\bf
  117} (2016) no.~7, 072001},
\href{http://arxiv.org/abs/1606.00840}{{\tt arXiv:1606.00840 [hep-ph]}}.

\bibitem{Ferrera:2017zex}
G.~Ferrera, G.~Somogyi, and F.~Tramontano, {\em {Associated production of a
  Higgs boson decaying into bottom quarks at the LHC in full NNLO QCD}\/},
  \href{http://dx.doi.org/10.1016/j.physletb.2018.03.021}{Phys. Lett. {\bf
  B780} (2018)  346--351},
\href{http://arxiv.org/abs/1705.10304}{{\tt arXiv:1705.10304 [hep-ph]}}.

\bibitem{Caola:2017xuq}
F.~Caola, G.~Luisoni, K.~Melnikov, and R.~R{\"o}ntsch, {\em {NNLO QCD
  corrections to associated $WH$ production and $H \to b \bar b$ decay}\/},
  \href{http://dx.doi.org/10.1103/PhysRevD.97.074022}{Phys. Rev. {\bf D97}
  (2018) no.~7, 074022},
\href{http://arxiv.org/abs/1712.06954}{{\tt arXiv:1712.06954 [hep-ph]}}.

\bibitem{Reina:2001sf}
L.~Reina and S.~Dawson, {\em {Next-to-leading order results for $t\bar{t}h$
  production at the Tevatron}\/},
  \href{http://dx.doi.org/10.1103/PhysRevLett.87.201804}{Phys. Rev. Lett. {\bf
  87} (2001)  201804},
\href{http://arxiv.org/abs/hep-ph/0107101}{{\tt arXiv:hep-ph/0107101
  [hep-ph]}}.

\bibitem{Beenakker:2002nc}
W.~Beenakker, S.~Dittmaier, M.~Kramer, B.~Plumper, M.~Spira, and P.~M. Zerwas,
  {\em {NLO QCD corrections to $t\bar{t}H$ production in hadron collisions}\/},
   \href{http://dx.doi.org/10.1016/S0550-3213(03)00044-0}{Nucl. Phys. {\bf
  B653} (2003)  151--203},
\href{http://arxiv.org/abs/hep-ph/0211352}{{\tt arXiv:hep-ph/0211352
  [hep-ph]}}.

\bibitem{Yu:2014cka}
Y.~Zhang, W.-G. Ma, R.-Y. Zhang, C.~Chen, and L.~Guo, {\em {QCD NLO and EW NLO
  corrections to $t\bar{t}H$ production with top quark decays at hadron
  collider}\/},  \href{http://dx.doi.org/10.1016/j.physletb.2014.09.022}{Phys.
  Lett. {\bf B738} (2014)  1--5},
\href{http://arxiv.org/abs/1407.1110}{{\tt arXiv:1407.1110 [hep-ph]}}.

\bibitem{Farina:2012xp}
M.~Farina, C.~Grojean, F.~Maltoni, E.~Salvioni, and A.~Thamm, {\em {Lifting
  degeneracies in Higgs couplings using single top production in association
  with a Higgs boson}\/},
  \href{http://dx.doi.org/10.1007/JHEP05(2013)022}{JHEP {\bf 05} (2013)  022},
\href{http://arxiv.org/abs/1211.3736}{{\tt arXiv:1211.3736 [hep-ph]}}.

\bibitem{Demartin:2015uha}
F.~Demartin, F.~Maltoni, K.~Mawatari, and M.~Zaro, {\em {Higgs production in
  association with a single top quark at the LHC}\/},
  \href{http://dx.doi.org/10.1140/epjc/s10052-015-3475-9}{Eur. Phys. J. {\bf
  C75} (2015) no.~6, 267},
\href{http://arxiv.org/abs/1504.00611}{{\tt arXiv:1504.00611 [hep-ph]}}.

\bibitem{Khalek:2018mdn}
A.~Khalek et al., {\em {Towards Ultimate Parton Distributions at the
  High-Luminosity LHC}\/},
  \href{http://dx.doi.org/10.1140/epjc/s10052-018-6448-y}{Eur. Phys. J. {\bf
  C78} (2018) no.~11, 962},
\href{http://arxiv.org/abs/1810.03639}{{\tt arXiv:1810.03639 [hep-ph]}}.

\bibitem{ATL-Wmass}
{ATLAS Collaboration}, {\em {Prospects for the measurement of the W boson mass
  at the HL and HE LHC}\/},  {ATL-PHYS-PUB-2018-026}.
\newblock \url{http://cdsweb.cern.ch/record/2645431}.

\bibitem{ATL}
{ATLAS Collaboration}, {\em {Weak Mixing Angle in $pp \rightarrow
  Z/\gamma*\rightarrow e^+e^-$ events with the ATLAS detector at the High
  Luminosity Large Hadron Collider}\/},  {ATL-PHYS-PUB-2018-037}.
\newblock \url{http://cdsweb.cern.ch/record/2649330}.

\bibitem{Bruening:2013bga}
O.~Bruening and M.~Klein, {\em {The Large Hadron Electron Collider}\/},
  \href{http://dx.doi.org/10.1142/S0217732313300115}{Mod. Phys. Lett. {\bf A28}
  (2013) no.~16, 1330011},
\href{http://arxiv.org/abs/1305.2090}{{\tt arXiv:1305.2090 [physics.acc-ph]}}.

\bibitem{dEnterria:2019its}
D.~d'Enterria et al., {\em {$\alpha_s$(2019): Precision measurements of the QCD
  coupling}\/},  in {\em {Workshop on precision measurements of the QCD
  coupling constant (alphas-2019) Trento, Trentino, Italy, February 11-15,
  2019}}.
\newblock 2019.
\newblock
\href{http://arxiv.org/abs/1907.01435}{{\tt arXiv:1907.01435 [hep-ph]}}.
\newblock

\bibitem{N.Cartiglia:2015gve}
{LHC Forward Physics Working Group Collaboration}, K.~Akiba et al., {\em {LHC
  Forward Physics}\/},
  \href{http://dx.doi.org/10.1088/0954-3899/43/11/110201}{J. Phys. G {\bf 43}
  (2016)  110201},
\href{http://arxiv.org/abs/1611.05079}{{\tt arXiv:1611.05079 [hep-ph]}}.

\bibitem{Bazavov:2019lgz}
{USQCD Collaboration}, A.~Bazavov, F.~Karsch, M.~Swagato, and P.~Petreczky,
  {\em {Hot-dense Lattice QCD: USQCD whitepaper 2018}\/},
\href{http://arxiv.org/abs/1904.09951}{{\tt arXiv:1904.09951 [hep-lat]}}.

\bibitem{Citron:2018lsq}
Z.~Citron et al., {\em {Future physics opportunities for high-density QCD at
  the LHC with heavy-ion and proton beams}\/},  in {\em {HL/HE-LHC Workshop:
  Workshop on the Physics of HL-LHC, and Perspectives at HE-LHC Geneva,
  Switzerland, June 18-20, 2018}}.
\newblock 2018.
\newblock
\href{http://arxiv.org/abs/1812.06772}{{\tt arXiv:1812.06772 [hep-ph]}}.
\newblock

\bibitem{Andronic:2017pug}
A.~Andronic, P.~Braun-Munzinger, K.~Redlich, and J.~Stachel, {\em {Decoding the
  phase structure of QCD via particle production at high energy}\/},
  \href{http://dx.doi.org/10.1038/s41586-018-0491-6}{Nature {\bf 561} (2018)
  no.~7723, 321--330},
\href{http://arxiv.org/abs/1710.09425}{{\tt arXiv:1710.09425 [nucl-th]}}.

\bibitem{Anderle:2017qwx}
D.~d'Enterria and P.~Z. Skands, eds., {\em {Proceedings, Parton Radiation and
  Fragmentation from LHC to FCC-ee}}.
\newblock 2017.
\newblock
\href{http://arxiv.org/abs/1702.01329}{{\tt arXiv:1702.01329 [hep-ph]}}.
\newblock

\bibitem{Hallman:2019}
T.~J. Hallman, {\em Perspectives from DOE NP, 2019 EIC User Group Meeting\/}, .
\newblock \url{https://indico.in2p3.fr/event/18281/contributions/70135/}.

\bibitem{Britzger:2017}
D.~Britzger and M.~Klein, {\em Electroweak physics with inclusive DIS at the
  LHeC and FCC eh\/},
\newblock 2017.
\newblock \url{https://indico.cern.ch/event/568360/contributions/252 626/}.
\newblock 25th International Workshop on Deep Inelastic Scattering and Related
  Topics, Birmingham.

\bibitem{dEnterria:2007jwb}
D.~d'Enterria, {\em {Experimental tests of small-x QCD}\/},  eConf {\bf
  C0706044} (2007)  17, \href{http://arxiv.org/abs/0706.4182}{{\tt
  arXiv:0706.4182 [hep-ex]}}.
[,359(2007)].

\bibitem{Chen:2018wyz}
X.~Chen, {\em {A Plan for Electron Ion Collider in China}\/},
  \href{http://dx.doi.org/10.22323/1.316.0170}{PoS {\bf DIS2018} (2018)  170},
\href{http://arxiv.org/abs/1809.00448}{{\tt arXiv:1809.00448 [nucl-ex]}}.

\bibitem{Brodsky:2012vg}
S.~J. Brodsky, F.~Fleuret, C.~Hadjidakis, and J.~P. Lansberg, {\em {Physics
  Opportunities of a Fixed-Target Experiment using the LHC Beams}\/},
  \href{http://dx.doi.org/10.1016/j.physrep.2012.10.001}{Phys. Rept. {\bf 522}
  (2013)  239--255},
\href{http://arxiv.org/abs/1202.6585}{{\tt arXiv:1202.6585 [hep-ph]}}.

\bibitem{Hadjidakis:2018ifr}
C.~Hadjidakis et al., {\em {A Fixed-Target Programme at the LHC: Physics Case
  and Projected Performances for Heavy-Ion, Hadron, Spin and Astroparticle
  Studies}\/},
\href{http://arxiv.org/abs/1807.00603}{{\tt arXiv:1807.00603 [hep-ex]}}.

\bibitem{Andronic:2015wma}
A.~Andronic et al., {\em {Heavy-flavour and quarkonium production in the LHC
  era: from proton-proton to heavy-ion collisions}\/},
  \href{http://dx.doi.org/10.1140/epjc/s10052-015-3819-5}{Eur. Phys. J. C {\bf
  76} (2016) no.~3, 107},
\href{http://arxiv.org/abs/1506.03981}{{\tt arXiv:1506.03981 [nucl-ex]}}.

\bibitem{Niemi:2015qia}
H.~Niemi, K.~J. Eskola, and R.~Paatelainen, {\em {Event-by-event fluctuations
  in a perturbative QCD + saturation + hydrodynamics model: Determining QCD
  matter shear viscosity in ultrarelativistic heavy-ion collisions}\/},
  \href{http://dx.doi.org/10.1103/PhysRevC.93.024907}{Phys. Rev. C {\bf 93}
  (2016) no.~2, 024907},
\href{http://arxiv.org/abs/1505.02677}{{\tt arXiv:1505.02677 [hep-ph]}}.

\bibitem{Schenke:2012wb}
B.~Schenke, P.~Tribedy, and R.~Venugopalan, {\em {Fluctuating Glasma initial
  conditions and flow in heavy ion collisions}\/},
  \href{http://dx.doi.org/10.1103/PhysRevLett.108.252301}{Phys. Rev. Lett. {\bf
  108} (2012)  252301},
\href{http://arxiv.org/abs/1202.6646}{{\tt arXiv:1202.6646 [nucl-th]}}.

\bibitem{Accardi:2009qv}
A.~Accardi, F.~Arleo, W.~K. Brooks, D.~D'Enterria, and V.~Muccifora, {\em
  {Parton Propagation and Fragmentation in QCD Matter}\/},
  \href{http://dx.doi.org/10.1393/ncr/i2009-10048-0}{Riv. Nuovo Cim. {\bf 32}
  (2010)  439--553},
\href{http://arxiv.org/abs/0907.3534}{{\tt arXiv:0907.3534 [nucl-th]}}.

\bibitem{Bazavov:2018mes}
{HotQCD Collaboration}, A.~Bazavov et al., {\em {Chiral crossover in QCD at
  zero and non-zero chemical potentials}\/},
  \href{http://dx.doi.org/10.1016/j.physletb.2019.05.013}{Phys. Lett. {\bf
  B795} (2019)  15--21},
\href{http://arxiv.org/abs/1812.08235}{{\tt arXiv:1812.08235 [hep-lat]}}.

\bibitem{Aaij:2018ogq}
{LHCb Collaboration}, R.~Aaij et al., {\em {First Measurement of Charm
  Production in its Fixed-Target Configuration at the LHC}\/},
  \href{http://dx.doi.org/10.1103/PhysRevLett.122.132002}{Phys. Rev. Lett. {\bf
  122} (2019) no.~13, 132002},
\href{http://arxiv.org/abs/1810.07907}{{\tt arXiv:1810.07907 [hep-ex]}}.

\bibitem{Aaij:2018svt}
{LHCb Collaboration}, R.~Aaij et al., {\em {Measurement of Antiproton
  Production in ${\rm p He}$ Collisions at $\sqrt{s_{NN}}=110$ GeV}\/},
  \href{http://dx.doi.org/10.1103/PhysRevLett.121.222001}{Phys. Rev. Lett. {\bf
  121} (2018) no.~22, 222001},
\href{http://arxiv.org/abs/1808.06127}{{\tt arXiv:1808.06127 [hep-ex]}}.

\bibitem{Adam:2019wnb}
{STAR Collaboration}, J.~Adam et al., {\em {Beam energy dependence of
  (anti-)deuteron production in Au + Au collisions at the BNL Relativistic
  Heavy Ion Collider}\/},
  \href{http://dx.doi.org/10.1103/PhysRevC.99.064905}{Phys. Rev. {\bf C99}
  (2019) no.~6, 064905},
\href{http://arxiv.org/abs/1903.11778}{{\tt arXiv:1903.11778 [nucl-ex]}}.

\bibitem{Adam:2019xmk}
{STAR Collaboration}, J.~Adam et al., {\em {Collision-energy dependence of
  second-order off-diagonal and diagonal cumulants of net-charge, net-proton,
  and net-kaon multiplicity distributions in Au + Au collisions}\/},
  \href{http://dx.doi.org/10.1103/PhysRevC.100.014902}{Phys. Rev. {\bf C100}
  (2019) no.~1, 014902},
\href{http://arxiv.org/abs/1903.05370}{{\tt arXiv:1903.05370 [nucl-ex]}}.

\bibitem{Aduszkiewicz:2017sei}
{NA61/SHINE Collaboration}, A.~Aduszkiewicz et al., {\em {Measurements of $\pi
  ^\pm $ , K$^\pm $ , p and ${\bar{\text {p}}}$ spectra in proton-proton
  interactions at 20, 31, 40, 80 and 158 $\text{ GeV}/c$ with the NA61/SHINE
  spectrometer at the CERN SPS}\/},
  \href{http://dx.doi.org/10.1140/epjc/s10052-017-5260-4}{Eur. Phys. J. {\bf
  C77} (2017) no.~10, 671},
\href{http://arxiv.org/abs/1705.02467}{{\tt arXiv:1705.02467 [nucl-ex]}}.

\bibitem{Berns:2018tap}
{NA61/SHINE Collaboration}, N.~Abgrall et al., {\em {Measurements of $\pi ^\pm
  $ , $K^\pm $ and proton double differential yields from the surface of the
  T2K replica target for incoming 31 GeV/c protons with the NA61/SHINE
  spectrometer at the CERN SPS}\/},
  \href{http://dx.doi.org/10.1140/epjc/s10052-019-6583-0}{Eur. Phys. J. {\bf
  C79} (2019) no.~2, 100},
\href{http://arxiv.org/abs/1808.04927}{{\tt arXiv:1808.04927 [hep-ex]}}.

\bibitem{Galatyuk:2019lcf}
T.~Galatyuk, {\em {Future facilities for high $\mu_B$ physics}\/},
\href{http://dx.doi.org/10.1016/j.nuclphysa.2018.11.025}{Nucl. Phys. {\bf A982}
  (2019)  163--169}.

\bibitem{Adamova:2019vkf}
D.~Adamov{\'a} et al., {\em {A next-generation LHC heavy-ion experiment}\/},
\href{http://arxiv.org/abs/1902.01211}{{\tt arXiv:1902.01211
  [physics.ins-det]}}.

\bibitem{Bruce:2018yzs}
R.~Bruce et al., {\em {New physics searches with heavy-ion collisions at the
  LHC}\/},
\href{http://arxiv.org/abs/1812.07688}{{\tt arXiv:1812.07688 [hep-ph]}}.

\bibitem{Dainese:2016gch}
A.~Dainese et al., {\em {Heavy ions at the Future Circular Collider}\/},
  \href{http://dx.doi.org/10.23731/CYRM-2017-003.635}{CERN Yellow Rep. (2017)
  no.~3, 635--692},
\href{http://arxiv.org/abs/1605.01389}{{\tt arXiv:1605.01389 [hep-ph]}}.

\bibitem{Abada:2019lih}
{FCC Collaboration}, A.~Abada et al., {\em {FCC Physics Opportunities}\/},
\href{http://dx.doi.org/10.1140/epjc/s10052-019-6904-3}{Eur. Phys. J. {\bf C79}
  (2019) no.~6, 474}.

\bibitem{Golovatyuk:2016zps}
V.~Golovatyuk, V.~Kekelidze, V.~Kolesnikov, O.~Rogachevsky, and A.~Sorin, {\em
  {The Multi-Purpose Detector (MPD) of the collider experiment}\/},
\href{http://dx.doi.org/10.1140/epja/i2016-16212-1}{Eur. Phys. J. {\bf A52}
  (2016) no.~8, 212}.

\bibitem{Dainese:2019xrz}
{QCD Working Group Collaboration}, A.~Dainese et al., {\em {Physics Beyond
  Colliders: QCD Working Group Report}\/},
\href{http://arxiv.org/abs/1901.04482}{{\tt arXiv:1901.04482 [hep-ex]}}.

\bibitem{Friman:2011zz}
B.~Friman, C.~Hohne, J.~Knoll, S.~Leupold, J.~Randrup, R.~Rapp, and P.~Senger,
  {\em {The CBM physics book: Compressed baryonic matter in laboratory
  experiments}\/},
\href{http://dx.doi.org/10.1007/978-3-642-13293-3}{Lect. Notes Phys. {\bf 814}
  (2011)  pp.1--980}.

\bibitem{Ablyazimov:2017guv}
{CBM Collaboration}, T.~Ablyazimov et al., {\em {Challenges in QCD matter
  physics --The scientific programme of the Compressed Baryonic Matter
  experiment at FAIR}\/},
  \href{http://dx.doi.org/10.1140/epja/i2017-12248-y}{Eur. Phys. J. {\bf A53}
  (2017) no.~3, 60},
\href{http://arxiv.org/abs/1607.01487}{{\tt arXiv:1607.01487 [nucl-ex]}}.

\bibitem{Bzdak:2019pkr}
A.~Bzdak, S.~Esumi, V.~Koch, J.~Liao, M.~Stephanov, and N.~Xu, {\em {Mapping
  the Phases of Quantum Chromodynamics with Beam Energy Scan}\/},
\href{http://arxiv.org/abs/1906.00936}{{\tt arXiv:1906.00936 [nucl-th]}}.

\bibitem{Azzi:2019yne}
{HL-LHC, HE-LHC Working Group Collaboration}, P.~Azzi et al., {\em {Standard
  Model Physics at the HL-LHC and HE-LHC}\/},
\href{http://arxiv.org/abs/1902.04070}{{\tt arXiv:1902.04070 [hep-ph]}}.

\bibitem{Bizon:2017rah}
W.~Bizon, P.~F. Monni, E.~Re, L.~Rottoli, and P.~Torrielli, {\em {Momentum
  space resummation for transverse observables and the Higgs p$_{\perp}$ at
  N$^{3}$LL~NNLO}\/},  \href{http://dx.doi.org/10.1007/JHEP02(2018)108}{JHEP
  {\bf 02} (2018)  108},
\href{http://arxiv.org/abs/1705.09127}{{\tt arXiv:1705.09127 [hep-ph]}}.

\bibitem{Aoki}
{Flavour Lattice Averaging Group Collaboration}, S.~Aoki et al., {\em {FLAG
  Review 2019}\/},
\href{http://arxiv.org/abs/1902.08191}{{\tt arXiv:1902.08191 [hep-lat]}}.

\bibitem{Cirigliano:2019jig}
{USQCD Collaboration}, V.~Cirigliano, Z.~Davoudi, T.~Bhattacharya, T.~Izubuchi,
  P.~E. Shanahan, S.~Syritsyn, and M.~L. Wagman, {\em {The Role of Lattice QCD
  in Searches for Violations of Fundamental Symmetries and Signals for New
  Physics}\/},
\href{http://arxiv.org/abs/1904.09704}{{\tt arXiv:1904.09704 [hep-lat]}}.

\bibitem{Lehner:2019wvv}
{USQCD Collaboration}, C.~Lehner et al., {\em {Opportunities for lattice QCD in
  quark and lepton flavor physics}\/},
\href{http://arxiv.org/abs/1904.09479}{{\tt arXiv:1904.09479 [hep-lat]}}.

\bibitem{DeGrand:2015zxa}
T.~DeGrand, {\em Lattice tests of beyond Standard Model dynamics\/},
  \href{http://dx.doi.org/10.1103/RevModPhys.88.015001}{Rev. Mod. Phys. {\bf
  88} (2016)  015001},
\href{http://arxiv.org/abs/1510.05018}{{\tt arXiv:1510.05018 [hep-ph]}}.

\bibitem{Boughezal:2015dva}
R.~Boughezal, C.~Focke, X.~Liu, and F.~Petriello, {\em {$W$-boson production in
  association with a jet at next-to-next-to-leading order in perturbative
  QCD}\/},  \href{http://dx.doi.org/10.1103/PhysRevLett.115.062002}{Phys. Rev.
  Lett. {\bf 115} (2015) no.~6, 062002},
\href{http://arxiv.org/abs/1504.02131}{{\tt arXiv:1504.02131 [hep-ph]}}.

\bibitem{Ridder:2015dxa}
A.~Gehrmann-De~Ridder, T.~Gehrmann, E.~W.~N. Glover, A.~Huss, and T.~A. Morgan,
  {\em {Precise QCD predictions for the production of a Z boson in association
  with a hadronic jet}\/},
  \href{http://dx.doi.org/10.1103/PhysRevLett.117.022001}{Phys. Rev. Lett. {\bf
  117} (2016) no.~2, 022001},
\href{http://arxiv.org/abs/1507.02850}{{\tt arXiv:1507.02850 [hep-ph]}}.

\bibitem{Czakon:2015owf}
M.~Czakon, D.~Heymes, and A.~Mitov, {\em {High-precision differential
  predictions for top-quark pairs at the LHC}\/},
  \href{http://dx.doi.org/10.1103/PhysRevLett.116.082003}{Phys. Rev. Lett. {\bf
  116} (2016) no.~8, 082003},
\href{http://arxiv.org/abs/1511.00549}{{\tt arXiv:1511.00549 [hep-ph]}}.

\bibitem{Czakon:2016dgf}
M.~Czakon, D.~Heymes, and A.~Mitov, {\em {Dynamical scales for multi-TeV
  top-pair production at the LHC}\/},
  \href{http://dx.doi.org/10.1007/JHEP04(2017)071}{JHEP {\bf 04} (2017)  071},
\href{http://arxiv.org/abs/1606.03350}{{\tt arXiv:1606.03350 [hep-ph]}}.

\bibitem{Currie:2013dwa}
J.~Currie, A.~Gehrmann-De~Ridder, E.~W.~N. Glover, and J.~Pires, {\em {NNLO QCD
  corrections to jet production at hadron colliders from gluon scattering}\/},
  \href{http://dx.doi.org/10.1007/JHEP01(2014)110}{JHEP {\bf 01} (2014)  110},
\href{http://arxiv.org/abs/1310.3993}{{\tt arXiv:1310.3993 [hep-ph]}}.

\bibitem{Currie:2016bfm}
J.~Currie, E.~W.~N. Glover, and J.~Pires, {\em {Next-to-Next-to Leading Order
  QCD Predictions for Single Jet Inclusive Production at the LHC}\/},
  \href{http://dx.doi.org/10.1103/PhysRevLett.118.072002}{Phys. Rev. Lett. {\bf
  118} (2017) no.~7, 072002},
\href{http://arxiv.org/abs/1611.01460}{{\tt arXiv:1611.01460 [hep-ph]}}.

\bibitem{Anastasiou:2015ema}
C.~Anastasiou, C.~Duhr, F.~Dulat, F.~Herzog, and B.~Mistlberger, {\em {Higgs
  Boson Gluon-Fusion Production in QCD at Three Loops}\/},
  \href{http://dx.doi.org/10.1103/PhysRevLett.114.212001}{Phys. Rev. Lett. {\bf
  114} (2015)  212001},
\href{http://arxiv.org/abs/1503.06056}{{\tt arXiv:1503.06056 [hep-ph]}}.

\bibitem{Baikov:2006ai}
P.~A. Baikov and K.~G. Chetyrkin, {\em {New four loop results in QCD}\/},
  \href{http://dx.doi.org/10.1016/j.nuclphysbps.2006.09.031}{Nucl. Phys. Proc.
  Suppl. {\bf 160} (2006)  76--79}.
[,76(2006)].

\bibitem{Velizhanin:2011es}
V.~N. Velizhanin, {\em {Four loop anomalous dimension of the second moment of
  the non-singlet twist-2 operator in QCD}\/},
  \href{http://dx.doi.org/10.1016/j.nuclphysb.2012.03.006}{Nucl. Phys. {\bf
  B860} (2012)  288--294},
\href{http://arxiv.org/abs/1112.3954}{{\tt arXiv:1112.3954 [hep-ph]}}.

\bibitem{Baikov:2015tea}
P.~A. Baikov, K.~G. Chetyrkin, and J.~H. Kuehn, {\em {Massless Propagators,
  $R(s)$ and Multiloop QCD}\/},
  \href{http://dx.doi.org/10.1016/j.nuclphysbps.2015.03.002}{Nucl. Part. Phys.
  Proc. {\bf 261-262} (2015)  3--18},
\href{http://arxiv.org/abs/1501.06739}{{\tt arXiv:1501.06739 [hep-ph]}}.

\bibitem{Davies:2016jie}
J.~Davies, A.~Vogt, B.~Ruijl, T.~Ueda, and J.~A.~M. Vermaseren, {\em
  {Large-$n_f$ contributions to the four-loop splitting functions in QCD}\/},
  \href{http://dx.doi.org/10.1016/j.nuclphysb.2016.12.012}{Nucl. Phys. {\bf
  B915} (2017)  335--362},
\href{http://arxiv.org/abs/1610.07477}{{\tt arXiv:1610.07477 [hep-ph]}}.

\bibitem{Moch:2017uml}
S.~Moch, B.~Ruijl, T.~Ueda, J.~A.~M. Vermaseren, and A.~Vogt, {\em {Four-Loop
  Non-Singlet Splitting Functions in the Planar Limit and Beyond}\/},
  \href{http://dx.doi.org/10.1007/JHEP10(2017)041}{JHEP {\bf 10} (2017)  041},
\href{http://arxiv.org/abs/1707.08315}{{\tt arXiv:1707.08315 [hep-ph]}}.

\bibitem{Dulat:2017prg}
F.~Dulat, B.~Mistlberger, and A.~Pelloni, {\em {Differential Higgs production
  at N$^{3}$LO beyond threshold}\/},
  \href{http://dx.doi.org/10.1007/JHEP01(2018)145}{JHEP {\bf 01} (2018)  145},
\href{http://arxiv.org/abs/1710.03016}{{\tt arXiv:1710.03016 [hep-ph]}}.

\bibitem{Currie:2018fgr}
J.~Currie, T.~Gehrmann, E.~W.~N. Glover, A.~Huss, J.~Niehues, and A.~Vogt, {\em
  {N$^{3}$LO corrections to jet production in deep inelastic scattering using
  the Projection-to-Born method}\/},
  \href{http://dx.doi.org/10.1007/JHEP05(2018)209}{JHEP {\bf 05} (2018)  209},
\href{http://arxiv.org/abs/1803.09973}{{\tt arXiv:1803.09973 [hep-ph]}}.

\bibitem{Catani:2007vq}
S.~Catani and M.~Grazzini, {\em {An NNLO subtraction formalism in hadron
  collisions and its application to Higgs boson production at the LHC}\/},
  \href{http://dx.doi.org/10.1103/PhysRevLett.98.222002}{Phys. Rev. Lett. {\bf
  98} (2007)  222002},
\href{http://arxiv.org/abs/hep-ph/0703012}{{\tt arXiv:hep-ph/0703012
  [hep-ph]}}.

\bibitem{Bozzi:2005wk}
G.~Bozzi, S.~Catani, D.~de~Florian, and M.~Grazzini, {\em {Transverse-momentum
  resummation and the spectrum of the Higgs boson at the LHC}\/},
  \href{http://dx.doi.org/10.1016/j.nuclphysb.2005.12.022}{Nucl. Phys. {\bf
  B737} (2006)  73--120},
\href{http://arxiv.org/abs/hep-ph/0508068}{{\tt arXiv:hep-ph/0508068
  [hep-ph]}}.

\bibitem{Bonciani:2015sha}
R.~Bonciani, S.~Catani, M.~Grazzini, H.~Sargsyan, and A.~Torre, {\em {The $q_T$
  subtraction method for top quark production at hadron colliders}\/},
  \href{http://dx.doi.org/10.1140/epjc/s10052-015-3793-y}{Eur. Phys. J. {\bf
  C75} (2015) no.~12, 581},
\href{http://arxiv.org/abs/1508.03585}{{\tt arXiv:1508.03585 [hep-ph]}}.

\bibitem{Boughezal:2015eha}
R.~Boughezal, X.~Liu, and F.~Petriello, {\em {$N$-jettiness soft function at
  next-to-next-to-leading order}\/},
  \href{http://dx.doi.org/10.1103/PhysRevD.91.094035}{Phys. Rev. {\bf D91}
  (2015) no.~9, 094035},
\href{http://arxiv.org/abs/1504.02540}{{\tt arXiv:1504.02540 [hep-ph]}}.

\bibitem{Gaunt:2015pea}
J.~Gaunt, M.~Stahlhofen, F.~J. Tackmann, and J.~R. Walsh, {\em {N-jettiness
  Subtractions for NNLO QCD Calculations}\/},
  \href{http://dx.doi.org/10.1007/JHEP09(2015)058}{JHEP {\bf 09} (2015)  058},
\href{http://arxiv.org/abs/1505.04794}{{\tt arXiv:1505.04794 [hep-ph]}}.

\bibitem{Czakon:2011ve}
M.~Czakon, {\em {Double-real radiation in hadronic top quark pair production as
  a proof of a certain concept}\/},
  \href{http://dx.doi.org/10.1016/j.nuclphysb.2011.03.020}{Nucl. Phys. {\bf
  B849} (2011)  250--295},
\href{http://arxiv.org/abs/1101.0642}{{\tt arXiv:1101.0642 [hep-ph]}}.

\bibitem{Boughezal:2011jf}
R.~Boughezal, K.~Melnikov, and F.~Petriello, {\em {A subtraction scheme for
  NNLO computations}\/},
  \href{http://dx.doi.org/10.1103/PhysRevD.85.034025}{Phys. Rev. {\bf D85}
  (2012)  034025},
\href{http://arxiv.org/abs/1111.7041}{{\tt arXiv:1111.7041 [hep-ph]}}.

\bibitem{Ger}
A.~Gehrmann-De~Ridder, T.~Gehrmann, and E.~W.~N. Glover, {\em {Antenna
  subtraction at NNLO}\/},
  \href{http://dx.doi.org/10.1088/1126-6708/2005/09/056}{JHEP {\bf 09} (2005)
  056},
\href{http://arxiv.org/abs/hep-ph/0505111}{{\tt arXiv:hep-ph/0505111
  [hep-ph]}}.

\bibitem{Ball:2017nwa}
{NNPDF Collaboration}, R.~D. Ball et al., {\em {Parton distributions from
  high-precision collider data}\/},
  \href{http://dx.doi.org/10.1140/epjc/s10052-017-5199-5}{Eur. Phys. J. {\bf
  C77} (2017) no.~10, 663},
\href{http://arxiv.org/abs/1706.00428}{{\tt arXiv:1706.00428 [hep-ph]}}.

\bibitem{Bauer:2000ew}
C.~W. Bauer, S.~Fleming, and M.~E. Luke, {\em {Summing Sudakov logarithms in B
  ---> X(s gamma) in effective field theory}\/},
  \href{http://dx.doi.org/10.1103/PhysRevD.63.014006}{Phys. Rev. {\bf D63}
  (2000)  014006},
\href{http://arxiv.org/abs/hep-ph/0005275}{{\tt arXiv:hep-ph/0005275
  [hep-ph]}}.

\bibitem{Bauer:2008jx}
C.~W. Bauer, A.~Hornig, and F.~J. Tackmann, {\em {Factorization for generic jet
  production}\/},  \href{http://dx.doi.org/10.1103/PhysRevD.79.114013}{Phys.
  Rev. {\bf D79} (2009)  114013},
\href{http://arxiv.org/abs/0808.2191}{{\tt arXiv:0808.2191 [hep-ph]}}.

\bibitem{Collins:1984kg}
J.~C. Collins, D.~E. Soper, and G.~F. Sterman, {\em {Transverse Momentum
  Distribution in Drell-Yan Pair and W and Z Boson Production}\/},
\href{http://dx.doi.org/10.1016/0550-3213(85)90479-1}{Nucl. Phys. {\bf B250}
  (1985)  199--224}.

\bibitem{Catani:1990rr}
S.~Catani, B.~R. Webber, and G.~Marchesini, {\em {QCD coherent branching and
  semiinclusive processes at large x}\/},
\href{http://dx.doi.org/10.1016/0550-3213(91)90390-J}{Nucl. Phys. {\bf B349}
  (1991)  635--654}.

\bibitem{Catani:1991kz}
S.~Catani, G.~Turnock, B.~R. Webber, and L.~Trentadue, {\em {Thrust
  distribution in $e^+ e^-$ annihilation}\/},
\href{http://dx.doi.org/10.1016/0370-2693(91)90494-B}{Phys. Lett. {\bf B263}
  (1991)  491--497}.

\bibitem{Bauer:2002nz}
C.~W. Bauer, S.~Fleming, D.~Pirjol, I.~Z. Rothstein, and I.~W. Stewart, {\em
  {Hard scattering factorization from effective field theory}\/},
  \href{http://dx.doi.org/10.1103/PhysRevD.66.014017}{Phys. Rev. {\bf D66}
  (2002)  014017},
\href{http://arxiv.org/abs/hep-ph/0202088}{{\tt arXiv:hep-ph/0202088
  [hep-ph]}}.

\bibitem{Banfi:2004nk}
A.~Banfi, G.~P. Salam, and G.~Zanderighi, {\em {Resummed event shapes at hadron
  hadron colliders}\/},
  \href{http://dx.doi.org/10.1088/1126-6708/2004/08/062}{JHEP {\bf 08} (2004)
  062},
\href{http://arxiv.org/abs/hep-ph/0407287}{{\tt arXiv:hep-ph/0407287
  [hep-ph]}}.

\bibitem{Becher:2010tm}
T.~Becher and M.~Neubert, {\em {{Drell-Yan} Production at Small $q_T$,
  Transverse Parton Distributions and the Collinear Anomaly}\/},
  \href{http://dx.doi.org/10.1140/epjc/s10052-011-1665-7}{Eur. Phys. J. {\bf
  C71} (2011)  1665},
\href{http://arxiv.org/abs/1007.4005}{{\tt arXiv:1007.4005 [hep-ph]}}.

\bibitem{Stewart:2010pd}
I.~W. Stewart, F.~J. Tackmann, and W.~J. Waalewijn, {\em {The Beam Thrust Cross
  Section for Drell Yan at NNLL Order}\/},
  \href{http://dx.doi.org/10.1103/PhysRevLett.106.032001}{Phys. Rev. Lett. {\bf
  106} (2011)  032001},
\href{http://arxiv.org/abs/1005.4060}{{\tt arXiv:1005.4060 [hep-ph]}}.

\bibitem{Banfi:2011dx}
A.~Banfi, M.~Dasgupta, and S.~Marzani, {\em {QCD predictions for new variables
  to study dilepton transverse momenta at hadron colliders}\/},
  \href{http://dx.doi.org/10.1016/j.physletb.2011.05.028}{Phys. Lett. {\bf
  B701} (2011)  75--81},
\href{http://arxiv.org/abs/1102.3594}{{\tt arXiv:1102.3594 [hep-ph]}}.

\bibitem{Berger:2010xi}
C.~F. Berger, C.~Marcantonini, I.~W. Stewart, F.~J. Tackmann, and W.~J.
  Waalewijn, {\em {Higgs Production with a Central Jet Veto at NNLL NNLO}\/},
  \href{http://dx.doi.org/10.1007/JHEP04(2011)092}{JHEP {\bf 04} (2011)  092},
\href{http://arxiv.org/abs/1012.4480}{{\tt arXiv:1012.4480 [hep-ph]}}.

\bibitem{Jouttenus:2011wh}
T.~T. Jouttenus, I.~W. Stewart, F.~J. Tackmann, and W.~J. Waalewijn, {\em {The
  Soft Function for Exclusive N Jet Production at Hadron Colliders}\/},
  \href{http://dx.doi.org/10.1103/PhysRevD.83.114030}{Phys. Rev. {\bf D83}
  (2011)  114030},
\href{http://arxiv.org/abs/1102.4344}{{\tt arXiv:1102.4344 [hep-ph]}}.

\bibitem{Becher:2012qa}
T.~Becher and M.~Neubert, {\em {Factorization and NNLL Resummation for Higgs
  Production with a Jet Veto}\/},
  \href{http://dx.doi.org/10.1007/JHEP07(2012)108}{JHEP {\bf 07} (2012)  108},
\href{http://arxiv.org/abs/1205.3806}{{\tt arXiv:1205.3806 [hep-ph]}}.

\bibitem{Zhu:2012ts}
H.~X. Zhu, C.~S. Li, H.~T. Li, D.~Y. Shao, and L.~L. Yang, {\em
  {Transverse-momentum resummation for top quark pairs at hadron colliders}\/},
   \href{http://dx.doi.org/10.1103/PhysRevLett.110.082001}{Phys. Rev. Lett.
  {\bf 110} (2013) no.~8, 082001},
\href{http://arxiv.org/abs/1208.5774}{{\tt arXiv:1208.5774 [hep-ph]}}.

\bibitem{Banfi:2012jm}
A.~Banfi, P.~F. Monni, G.~P. Salam, and G.~Zanderighi, {\em {Higgs and Z boson
  production with a jet veto}\/},
  \href{http://dx.doi.org/10.1103/PhysRevLett.109.202001}{Phys. Rev. Lett. {\bf
  109} (2012)  202001},
\href{http://arxiv.org/abs/1206.4998}{{\tt arXiv:1206.4998 [hep-ph]}}.

\bibitem{Becher:2013xia}
T.~Becher, M.~Neubert, and L.~Rothen, {\em {Factorization and $N^{3}LL$ NNLO
  predictions for the Higgs cross section with a jet veto}\/},
  \href{http://dx.doi.org/10.1007/JHEP10(2013)125}{JHEP {\bf 10} (2013)  125},
\href{http://arxiv.org/abs/1307.0025}{{\tt arXiv:1307.0025 [hep-ph]}}.

\bibitem{Stewart:2013faa}
I.~W. Stewart, F.~J. Tackmann, J.~R. Walsh, and S.~Zuberi, {\em {Jet $pT$
  resummation in Higgs production at $NNLL$ $NNLO$}\/},
  \href{http://dx.doi.org/10.1103/PhysRevD.89.054001}{Phys. Rev. {\bf D89}
  (2014) no.~5, 054001},
\href{http://arxiv.org/abs/1307.1808}{{\tt arXiv:1307.1808 [hep-ph]}}.

\bibitem{Procura:2014cba}
M.~Procura, W.~J. Waalewijn, and L.~Zeune, {\em {Resummation of Double
  Differential Cross Sections and Fully Unintegrated Parton Distribution
  Functions}\/},  \href{http://dx.doi.org/10.1007/JHEP02(2015)117}{JHEP {\bf
  02} (2015)  117},
\href{http://arxiv.org/abs/1410.6483}{{\tt arXiv:1410.6483 [hep-ph]}}.

\bibitem{Li:2016ctv}
Y.~Li and H.~X. Zhu, {\em {Bootstrapping Rapidity Anomalous Dimensions for
  Transverse Momentum Resummation}\/},
  \href{http://dx.doi.org/10.1103/PhysRevLett.118.022004}{Phys. Rev. Lett. {\bf
  118} (2017) no.~2, 022004},
\href{http://arxiv.org/abs/1604.01404}{{\tt arXiv:1604.01404 [hep-ph]}}.

\bibitem{Monni:2016ktx}
P.~F. Monni, E.~Re, and P.~Torrielli, {\em {Higgs Transverse Momentum
  Resummation in Direct Space}\/},
  \href{http://dx.doi.org/10.1103/PhysRevLett.116.242001}{Phys. Rev. Lett. {\bf
  116} (2016) no.~24, 242001},
\href{http://arxiv.org/abs/1604.02191}{{\tt arXiv:1604.02191 [hep-ph]}}.

\bibitem{Lin:2017snn}
H.-W. Lin et al., {\em {Parton distributions and lattice QCD calculations: a
  community white paper}\/},
  \href{http://dx.doi.org/10.1016/j.ppnp.2018.01.007}{Prog. Part. Nucl. Phys.
  {\bf 100} (2018)  107--160},
\href{http://arxiv.org/abs/1711.07916}{{\tt arXiv:1711.07916 [hep-ph]}}.

\bibitem{Accardi:2016ndt}
A.~Accardi et al., {\em {A Critical Appraisal and Evaluation of Modern
  PDFs}\/},  \href{http://dx.doi.org/10.1140/epjc/s10052-016-4285-4}{Eur. Phys.
  J. {\bf C76} (2016) no.~8, 471},
\href{http://arxiv.org/abs/1603.08906}{{\tt arXiv:1603.08906 [hep-ph]}}.

\bibitem{Alekhin:2017kpj}
S.~Alekhin, J.~Bl{\"u}mlein, S.~Moch, and R.~Placakyte, {\em {Parton
  distribution functions, $\alpha_s$, and heavy-quark masses for LHC Run
  II}\/},  \href{http://dx.doi.org/10.1103/PhysRevD.96.014011}{Phys. Rev. {\bf
  D96} (2017) no.~1, 014011},
\href{http://arxiv.org/abs/1701.05838}{{\tt arXiv:1701.05838 [hep-ph]}}.

\bibitem{Ball:2014uwa}
{NNPDF Collaboration}, R.~D. Ball et al., {\em {Parton distributions for the
  LHC Run II}\/},  \href{http://dx.doi.org/10.1007/JHEP04(2015)040}{JHEP {\bf
  04} (2015)  040},
\href{http://arxiv.org/abs/1410.8849}{{\tt arXiv:1410.8849 [hep-ph]}}.

\bibitem{Dulat:2015mca}
S.~Dulat, T.-J. Hou, J.~Gao, M.~Guzzi, J.~Huston, P.~Nadolsky, J.~Pumplin,
  C.~Schmidt, D.~Stump, and C.~P. Yuan, {\em {New parton distribution functions
  from a global analysis of quantum chromodynamics}\/},
  \href{http://dx.doi.org/10.1103/PhysRevD.93.033006}{Phys. Rev. {\bf D93}
  (2016) no.~3, 033006},
\href{http://arxiv.org/abs/1506.07443}{{\tt arXiv:1506.07443 [hep-ph]}}.

\bibitem{Harland-Lang:2014zoa}
L.~A. Harland-Lang, A.~D. Martin, P.~Motylinski, and R.~S. Thorne, {\em {Parton
  distributions in the LHC era: MMHT 2014 PDFs}\/},
  \href{http://dx.doi.org/10.1140/epjc/s10052-015-3397-6}{Eur. Phys. J. {\bf
  C75} (2015) no.~5, 204},
\href{http://arxiv.org/abs/1412.3989}{{\tt arXiv:1412.3989 [hep-ph]}}.

\bibitem{AbdulKhalek:2019bux}
{NNPDF Collaboration}, R.~Abdul~Khalek et al., {\em {A First Determination of
  Parton Distributions with Theoretical Uncertainties}\/},
\href{http://arxiv.org/abs/1905.04311}{{\tt arXiv:1905.04311 [hep-ph]}}.

\bibitem{Ides159}
{LHeC and PERLE Collaborations}, O.~Br{\"u}ning and M.~Klein, {\em {Exploring
  the Energy Frontier with Deep Inelastic Scattering at the LHC}}.
\newblock CERN, Geneva, 2018.
\newblock \href{http://arxiv.org/abs/CERN LHeC-Note-001-2018}{{\tt CERN
  LHeC-Note-001-2018}}.
\newblock \url{https://indico.cern.ch/event/765096/contributions/3296015/}.

\bibitem{Klein:2018rhq}
M.~Klein, \href{http://dx.doi.org/10.1142/9789813238053_0015}{{\em {Future Deep
  Inelastic Scattering with the LHeC}\/}, } in {\em From My Vast Repertoire
  ...: Guido Altarelli's Legacy}, A.~Levy, S.~Forte, and G.~Ridolfi, eds.,
  pp.~303--347.
\newblock 2019.
\newblock
\href{http://arxiv.org/abs/1802.04317}{{\tt arXiv:1802.04317 [hep-ph]}}.
\newblock

\bibitem{Iancu:2002xk}
E.~Iancu, A.~Leonidov, and L.~McLerran, {\em {The Color glass condensate: An
  Introduction}\/},  in {\em {QCD perspectives on hot and dense matter.
  Proceedings, NATO Advanced Study Institute, Summer School, Cargese, France,
  August 6-18, 2001}}, pp.~73--145.
\newblock 2002.
\newblock
\href{http://arxiv.org/abs/hep-ph/0202270}{{\tt arXiv:hep-ph/0202270
  [hep-ph]}}.
\newblock

\bibitem{Petreska:2018cbf}
E.~Petreska, {\em {TMD gluon distributions at small x in the CGC theory}\/},
  \href{http://dx.doi.org/10.1142/S0218301318300035}{Int. J. Mod. Phys. {\bf
  E27} (2018) no.~05, 1830003},
\href{http://arxiv.org/abs/1804.04981}{{\tt arXiv:1804.04981 [hep-ph]}}.

\bibitem{Altinoluk:2019fui}
T.~Altinoluk, R.~Boussarie, and P.~Kotko, {\em {Interplay of the CGC and TMD
  frameworks to all orders in kinematic twist}\/},
  \href{http://dx.doi.org/10.1007/JHEP05(2019)156}{JHEP {\bf 05} (2019)  156},
\href{http://arxiv.org/abs/1901.01175}{{\tt arXiv:1901.01175 [hep-ph]}}.

\bibitem{Kusina:2017gkz}
A.~Kusina, J.~P. Lansberg, I.~Schienbein, and H.-S. Shao, {\em {Gluon Shadowing
  in Heavy Flavor Production at the LHC}\/},
  \href{http://dx.doi.org/10.1103/PhysRevLett.121.052004}{Phys. Rev. Lett. {\bf
  121} (2018) no.~5, 052004},
\href{http://arxiv.org/abs/1712.07024}{{\tt arXiv:1712.07024 [hep-ph]}}.

\bibitem{AbdulKhalek:2019mzd}
{NNPDF Collaboration}, R.~Abdul~Khalek, J.~J. Ethier, and J.~Rojo, {\em
  {Nuclear parton distributions from lepton-nucleus scattering and the impact
  of an electron-ion collider}\/},
  \href{http://dx.doi.org/10.1140/epjc/s10052-019-6983-1}{Eur. Phys. J. {\bf
  C79} (2019) no.~6, 471},
\href{http://arxiv.org/abs/1904.00018}{{\tt arXiv:1904.00018 [hep-ph]}}.

\bibitem{Patrignani:2016xqp}
{Particle Data Group Collaboration}, C.~Patrignani et al., {\em {Review of
  Particle Physics}\/},
\href{http://dx.doi.org/10.1088/1674-1137/40/10/100001}{Chin. Phys. {\bf C40}
  (2016) no.~10, 100001}.

\bibitem{Aoki:2016frl}
S.~Aoki et al., {\em {Review of lattice results concerning low-energy particle
  physics}\/},  \href{http://dx.doi.org/10.1140/epjc/s10052-016-4509-7}{Eur.
  Phys. J. {\bf C77} (2017) no.~2, 112},
\href{http://arxiv.org/abs/1607.00299}{{\tt arXiv:1607.00299 [hep-lat]}}.

\bibitem{Bazavov:2017lyh}
A.~Bazavov et al., {\em {$B$ and $D$ meson leptonic decay constants from
  four-flavor lattice QCD}\/},
  \href{http://dx.doi.org/10.1103/PhysRevD.98.074512}{Phys. Rev. {\bf D98}
  (2018) no.~7, 074512},
\href{http://arxiv.org/abs/1712.09262}{{\tt arXiv:1712.09262 [hep-lat]}}.

\bibitem{Bazavov:2018omf}
{Fermilab Lattice, MILC, TUMQCD Collaboration}, A.~Bazavov et al., {\em {Up-,
  down-, strange-, charm-, and bottom-quark masses from four-flavor lattice
  QCD}\/},  \href{http://dx.doi.org/10.1103/PhysRevD.98.054517}{Phys. Rev. {\bf
  D98} (2018) no.~5, 054517},
\href{http://arxiv.org/abs/1802.04248}{{\tt arXiv:1802.04248 [hep-lat]}}.

\bibitem{Dowdall:2019bea}
R.~J. Dowdall, C.~T.~H. Davies, R.~R. Horgan, G.~P. Lepage, C.~J. Monahan,
  J.~Shigemitsu, and M.~Wingate, {\em {Neutral B-meson mixing from full lattice
  QCD at the physical point}\/},
\href{http://arxiv.org/abs/1907.01025}{{\tt arXiv:1907.01025 [hep-lat]}}.

\bibitem{Alexandrou:2018sjm}
C.~Alexandrou et al., {\em {Proton and neutron electromagnetic form factors
  from lattice QCD}\/},
\href{http://arxiv.org/abs/1812.10311}{{\tt arXiv:1812.10311 [hep-lat]}}.

\bibitem{Alexandrou:2018lvq}
C.~Alexandrou et al., {\em {Connected and disconnected contributions to nucleon
  axial form factors using $N_f$ = 2 twisted mass fermions at the physical
  point}\/},  \href{http://dx.doi.org/10.1051/epjconf/201817506003}{EPJ Web
  Conf. {\bf 175} (2018)  06003},
\href{http://arxiv.org/abs/1807.11203}{{\tt arXiv:1807.11203 [hep-lat]}}.

\bibitem{Djukanovic:2019jtp}
D.~Djukanovic, K.~Ottnad, J.~Wilhelm, and H.~Wittig, {\em {Strange
  electromagnetic form factors of the nucleon with $N_f = 2 + 1$
  $\mathcal{O}(a)$-improved Wilson fermions}\/},
\href{http://arxiv.org/abs/1903.12566}{{\tt arXiv:1903.12566 [hep-lat]}}.

\bibitem{Aaij:2019evc}
{LHCb Collaboration}, R.~Aaij et al., {\em {Near-threshold $D\bar{D}$
  spectroscopy and observation of a new charmonium state}\/},
  \href{http://dx.doi.org/10.1007/JHEP07(2019)035}{JHEP {\bf 07} (2019)  035},
\href{http://arxiv.org/abs/1903.12240}{{\tt arXiv:1903.12240 [hep-ex]}}.

\bibitem{Piemonte:2019cbi}
S.~Piemonte, S.~Collins, D.~Mohler, M.~Padmanath, and S.~Prelovsek, {\em
  {Charmonium resonances with $J^{PC}=1^{--}$ and $3^{--}$ from $\bar DD$
  scattering on the lattice}\/},
\href{http://arxiv.org/abs/1905.03506}{{\tt arXiv:1905.03506 [hep-lat]}}.

\bibitem{Alexandrou:2017qyt}
C.~Alexandrou et al., {\em {Nucleon scalar and tensor charges using lattice QCD
  simulations at the physical value of the pion mass}\/},
  \href{http://dx.doi.org/10.1103/PhysRevD.96.099906,
  10.1103/PhysRevD.95.114514}{Phys. Rev. {\bf D95} (2017) no.~11, 114514},
  \href{http://arxiv.org/abs/1703.08788}{{\tt arXiv:1703.08788 [hep-lat]}}.
[erratum: Phys. Rev.D96,no.9,099906(2017)].

\bibitem{Bai:2018hqu}
Z.~Bai, N.~H. Christ, X.~Feng, A.~Lawson, A.~Portelli, and C.~T. Sachrajda,
  {\em {$K^+\to\pi^+\nu\bar{\nu}$ decay amplitude from lattice QCD}\/},
  \href{http://dx.doi.org/10.1103/PhysRevD.98.074509}{Phys. Rev. {\bf D98}
  (2018) no.~7, 074509},
\href{http://arxiv.org/abs/1806.11520}{{\tt arXiv:1806.11520 [hep-lat]}}.

\bibitem{Dragos:2019oxn}
J.~Dragos, T.~Luu, A.~Shindler, J.~de~Vries, and A.~Yousif, {\em {Confirming
  the Existence of the strong CP Problem in Lattice QCD with the Gradient
  Flow}\/},
\href{http://arxiv.org/abs/1902.03254}{{\tt arXiv:1902.03254 [hep-lat]}}.

\bibitem{Meyer:2018til}
H.~B. Meyer and H.~Wittig, {\em {Lattice QCD and the anomalous magnetic moment
  of the muon}\/},  \href{http://dx.doi.org/10.1016/j.ppnp.2018.09.001}{Prog.
  Part. Nucl. Phys. {\bf 104} (2019)  46--96},
\href{http://arxiv.org/abs/1807.09370}{{\tt arXiv:1807.09370 [hep-lat]}}.

\bibitem{Westin:2019tgc}
{CSSM/QCDSF/UKQCD Collaboration}, A.~Westin et al., {\em {Anomalous magnetic
  moment of the muon with dynamical QCD+QED}\/},
  \href{http://dx.doi.org/10.22323/1.334.0136}{PoS {\bf LATTICE2018} (2019)
  136},
\href{http://arxiv.org/abs/1902.01518}{{\tt arXiv:1902.01518 [hep-lat]}}.

\bibitem{Soltz:2015ula}
R.~A. Soltz, C.~DeTar, F.~Karsch, S.~Mukherjee, and P.~Vranas, {\em {Lattice
  QCD Thermodynamics with Physical Quark Masses}\/},
  \href{http://dx.doi.org/10.1146/annurev-nucl-102014-022157}{Ann. Rev. Nucl.
  Part. Sci. {\bf 65} (2015)  379--402},
\href{http://arxiv.org/abs/1502.02296}{{\tt arXiv:1502.02296 [hep-lat]}}.

\bibitem{Ding:2015ona}
H.-T. Ding, F.~Karsch, and S.~Mukherjee, {\em {Thermodynamics of strong
  interaction matter from Lattice QCD}\/},
  \href{http://dx.doi.org/10.1142/S0218301315300076}{Int. J. Mod. Phys. {\bf
  E24} (2015) no.~10, 1530007},
\href{http://arxiv.org/abs/1504.05274}{{\tt arXiv:1504.05274 [hep-lat]}}.

\bibitem{Ratti:2018ksb}
C.~Ratti, {\em {Lattice QCD and heavy ion collisions a review of recent
  progress}\/},  \href{http://dx.doi.org/10.1088/1361-6633/aabb97}{Rept. Prog.
  Phys. {\bf 81} (2018) no.~8, 084301},
\href{http://arxiv.org/abs/1804.07810}{{\tt arXiv:1804.07810 [hep-lat]}}.

\bibitem{Friman:2011pf}
B.~Friman, F.~Karsch, K.~Redlich, and V.~Skokov, {\em {Fluctuations as probe of
  the QCD phase transition and freeze-out in heavy ion collisions at LHC and
  RHIC}\/},  \href{http://dx.doi.org/10.1140/epjc/s10052-011-1694-2}{Eur. Phys.
  J. {\bf C71} (2011)  1694},
\href{http://arxiv.org/abs/1103.3511}{{\tt arXiv:1103.3511 [hep-ph]}}.

\bibitem{Karsch:2010ck}
F.~Karsch and K.~Redlich, {\em {Probing freeze-out conditions in heavy ion
  collisions with moments of charge fluctuations}\/},
  \href{http://dx.doi.org/10.1016/j.physletb.2010.10.046}{Phys. Lett. {\bf
  B695} (2011)  136--142},
\href{http://arxiv.org/abs/1007.2581}{{\tt arXiv:1007.2581 [hep-ph]}}.

\bibitem{Bethke:2017uli}
S.~Bethke, {\em {$\alpha_s$ 2016}\/},
\href{http://dx.doi.org/10.1016/j.nuclphysbps.2016.12.028}{Nucl. Part. Phys.
  Proc. {\bf 282-284} (2017)  149--152}.

\bibitem{Alemany:2019vsk}
R.~Alemany et al., {\em {Summary Report of Physics Beyond Colliders at
  CERN}\/},
\href{http://arxiv.org/abs/1902.00260}{{\tt arXiv:1902.00260 [hep-ex]}}.

\bibitem{Cheng:1987gp}
H.-Y. Cheng, {\em {The Strong CP Problem Revisited}\/},
\href{http://dx.doi.org/10.1016/0370-1573(88)90135-4}{Phys. Rept. {\bf 158}
  (1988)  1}.

\bibitem{Pospelov:2005pr}
M.~Pospelov and A.~Ritz, {\em {Electric dipole moments as probes of new
  physics}\/},  \href{http://dx.doi.org/10.1016/j.aop.2005.04.002}{Annals Phys.
  {\bf 318} (2005)  119--169},
\href{http://arxiv.org/abs/hep-ph/0504231}{{\tt arXiv:hep-ph/0504231
  [hep-ph]}}.

\bibitem{Engel:2013lsa}
J.~Engel, M.~J. Ramsey-Musolf, and U.~van Kolck, {\em {Electric Dipole Moments
  of Nucleons, Nuclei, and Atoms: The Standard Model and Beyond}\/},
  \href{http://dx.doi.org/10.1016/j.ppnp.2013.03.003}{Prog. Part. Nucl. Phys.
  {\bf 71} (2013)  21--74},
\href{http://arxiv.org/abs/1303.2371}{{\tt arXiv:1303.2371 [nucl-th]}}.

\bibitem{Semertzidis:2003iq}
{EDM Collaboration}, Y.~K. Semertzidis et al., {\em {A New method for a
  sensitive deuteron EDM experiment}\/},
  \href{http://dx.doi.org/10.1063/1.1664226}{AIP Conf. Proc. {\bf 698} (2004)
  no.~1, 200--204},
\href{http://arxiv.org/abs/hep-ex/0308063}{{\tt arXiv:hep-ex/0308063
  [hep-ex]}}.

\bibitem{Afach:2015sja}
J.~M. Pendlebury et al., {\em {Revised experimental upper limit on the electric
  dipole moment of the neutron}\/},
  \href{http://dx.doi.org/10.1103/PhysRevD.92.092003}{Phys. Rev. {\bf D92}
  (2015) no.~9, 092003},
\href{http://arxiv.org/abs/1509.04411}{{\tt arXiv:1509.04411 [hep-ex]}}.

\bibitem{Lenisa:2017okq}
P.~Lenisa and F.~Rathmann, {\em {COSY Prepares the First Measurement of the
  Deuteron Electric Dipole Moment}\/},
\href{http://dx.doi.org/10.1080/10619127.2017.1317175}{Nucl. Phys. News {\bf
  27} (2017) no.~3, 10--13}.

\bibitem{Chupp:2017rkp}
T.~Chupp, P.~Fierlinger, M.~Ramsey-Musolf, and J.~Singh, {\em {Electric dipole
  moments of atoms, molecules, nuclei, and particles}\/},
  \href{http://dx.doi.org/10.1103/RevModPhys.91.015001}{Rev. Mod. Phys. {\bf
  91} (2019) no.~1, 015001},
\href{http://arxiv.org/abs/1710.02504}{{\tt arXiv:1710.02504
  [physics.atom-ph]}}.

\bibitem{Mooser:2014vla}
{BASE Collaboration}, A.~Mooser et al., {\em {Direct high-precision measurement
  of the magnetic moment of the proton}\/},
  \href{http://dx.doi.org/10.1038/nature13388}{Nature {\bf 509} (2014)
  596--599},
\href{http://arxiv.org/abs/1406.4888}{{\tt arXiv:1406.4888 [physics.atom-ph]}}.

\bibitem{Smorra:2018syp}
C.~Smorra et al., {\em {350-fold improved measurement of the antiproton
  magnetic moment using a multi-trap method}\/},
\href{http://dx.doi.org/10.1007/s10751-018-1507-1}{Hyperfine Interact. {\bf
  239} (2018) no.~1, 47}.

\bibitem{Smorra:2016vxa}
{BASE Collaboration}, C.~Smorra et al., {\em {A parts-per-billion measurement
  of the antiproton magnetic moment}\/},
\href{http://dx.doi.org/10.1038/nature24048}{Nature {\bf 550} (2017) no.~7676,
  371--374}.

\bibitem{Afach:2014fha}
S.~Afach et al., {\em {A measurement of the neutron to $^{199}$Hg magnetic
  moment ratio}\/},
  \href{http://dx.doi.org/10.1016/j.physletb.2014.10.046}{Phys. Lett. {\bf
  B739} (2014)  128--132},
\href{http://arxiv.org/abs/1410.8259}{{\tt arXiv:1410.8259 [nucl-ex]}}.

\bibitem{Antognini:2015moa}
A.~Antognini et al., {\em {Experiments towards resolving the proton charge
  radius puzzle}\/},  \href{http://dx.doi.org/10.1051/epjconf/201611301006}{EPJ
  Web Conf. {\bf 113} (2016)  01006},
\href{http://arxiv.org/abs/1509.03235}{{\tt arXiv:1509.03235
  [physics.atom-ph]}}.

\bibitem{Pohl:2010zza}
R.~Pohl et al., {\em {The size of the proton}\/},
\href{http://dx.doi.org/10.1038/nature09250}{Nature {\bf 466} (2010)
  213--216}.

\bibitem{Antognini:2005fe}
A.~Antognini et al., {\em {The 2S Lamb shift in muonic hydrogen and the proton
  rms charge radius}\/},
\href{http://dx.doi.org/10.1063/1.2130175}{AIP Conf. Proc. {\bf 796} (2005)
  no.~1, 253--259}.

\bibitem{Bernauer:2010wm}
{A1 Collaboration}, J.~C. Bernauer et al., {\em {High-precision determination
  of the electric and magnetic form factors of the proton}\/},
  \href{http://dx.doi.org/10.1103/PhysRevLett.105.242001}{Phys. Rev. Lett. {\bf
  105} (2010)  242001},
\href{http://arxiv.org/abs/1007.5076}{{\tt arXiv:1007.5076 [nucl-ex]}}.

\bibitem{Antognini:1900ns}
A.~Antognini et al., {\em {Proton Structure from the Measurement of $2S-2P$
  Transition Frequencies of Muonic Hydrogen}\/},
\href{http://dx.doi.org/10.1126/science.1230016}{Science {\bf 339} (2013)
  417--420}.

\bibitem{Gilman:2013eiv}
{MUSE Collaboration}, R.~Gilman et al., {\em {Studying the Proton "Radius"
  Puzzle with $\mu$ p Elastic Scattering}\/},
\href{http://arxiv.org/abs/1303.2160}{{\tt arXiv:1303.2160 [nucl-ex]}}.

\bibitem{Denisov:2018unj}
B.~Adams et al., {\em {Letter of Intent: A New QCD facility at the M2 beam line
  of the CERN SPS (COMPASS++/AMBER)}\/},
\href{http://arxiv.org/abs/1808.00848}{{\tt arXiv:1808.00848 [hep-ex]}}.

\bibitem{Markisch:2018ndu}
B.~Markisch et al., {\em {Measurement of the Weak Axial-Vector Coupling
  Constant in the Decay of Free Neutrons Using a Pulsed Cold Neutron Beam}\/},
  \href{http://dx.doi.org/10.1103/PhysRevLett.122.242501}{Phys. Rev. Lett. {\bf
  122} (2019) no.~24, 242501},
\href{http://arxiv.org/abs/1812.04666}{{\tt arXiv:1812.04666 [nucl-ex]}}.

\bibitem{Hill:2017wgb}
R.~J. Hill, P.~Kammel, W.~J. Marciano, and A.~Sirlin, {\em {Nucleon Axial
  Radius and Muonic Hydrogen: A New Analysis and Review}\/},
  \href{http://dx.doi.org/10.1088/1361-6633/aac190}{Rept. Prog. Phys. {\bf 81}
  (2018) no.~9, 096301},
\href{http://arxiv.org/abs/1708.08462}{{\tt arXiv:1708.08462 [hep-ph]}}.

\bibitem{Adeva:2018fwr}
{DIRAC Collaboration}, B.~Adeva et al., {\em {First measurement of a long-lived
  $\pi^+\pi^-$ atom lifetime}\/},
  \href{http://dx.doi.org/10.1103/PhysRevLett.122.082003}{Phys. Rev. Lett. {\bf
  122} (2019) no.~8, 082003},
\href{http://arxiv.org/abs/1811.08659}{{\tt arXiv:1811.08659 [hep-ex]}}.

\bibitem{DIRAC:2016rpv}
{DIRAC Collaboration}, B.~Adeva et al., {\em {Observation of $\pi^- K^+$ and
  $\pi^+ K^-$ atoms}\/},
  \href{http://dx.doi.org/10.1103/PhysRevLett.117.112001}{Phys. Rev. Lett. {\bf
  117} (2016) no.~11, 112001},
\href{http://arxiv.org/abs/1605.06103}{{\tt arXiv:1605.06103 [hep-ex]}}.

\bibitem{Kampert:2012mx}
K.-H. Kampert and M.~Unger, {\em {Measurements of the Cosmic Ray Composition
  with Air Shower Experiments}\/},
  \href{http://dx.doi.org/10.1016/j.astropartphys.2012.02.004}{Astropart. Phys.
  {\bf 35} (2012)  660--678},
\href{http://arxiv.org/abs/1201.0018}{{\tt arXiv:1201.0018 [astro-ph.HE]}}.

\bibitem{dEnterria:2011twh}
D.~d'Enterria, R.~Engel, T.~Pierog, S.~Ostapchenko, and K.~Werner, {\em
  {Constraints from the first LHC data on hadronic event generators for
  ultra-high energy cosmic-ray physics}\/},
  \href{http://dx.doi.org/10.1016/j.astropartphys.2011.05.002}{Astropart. Phys.
  {\bf 35} (2011)  98--113},
\href{http://arxiv.org/abs/1101.5596}{{\tt arXiv:1101.5596 [astro-ph.HE]}}.

\bibitem{Mangano:2016jyj}
M.~L. Mangano et al., {\em {Physics at a 100 TeV pp Collider: Standard Model
  Processes}\/},  \href{http://dx.doi.org/10.23731/CYRM-2017-003.1}{CERN Yellow
  Rep. (2017) no.~3, 1--254},
\href{http://arxiv.org/abs/1607.01831}{{\tt arXiv:1607.01831 [hep-ph]}}.

\bibitem{dEnterria:2016oxo}
D.~d'Enterria and T.~Pierog, {\em {Global properties of proton-proton
  collisions at $\sqrt{s}$ = 100 TeV}\/},
  \href{http://dx.doi.org/10.1007/JHEP08(2016)170}{JHEP {\bf 08} (2016)  170},
\href{http://arxiv.org/abs/1604.08536}{{\tt arXiv:1604.08536 [hep-ph]}}.

\bibitem{Dembinski:2019uta}
{EAS-MSU, IceCube, KASCADE-Grande, NEVOD-DECOR, Pierre Auger, SUGAR, Telescope
  Array, Yakutsk EAS Array Collaboration}, H.~P. Dembinski et al., {\em {Report
  on Tests and Measurements of Hadronic Interaction Properties with Air
  Showers}\/},  \href{http://dx.doi.org/10.1051/epjconf/201921002004}{EPJ Web
  Conf. {\bf 210} (2019)  02004},
\href{http://arxiv.org/abs/1902.08124}{{\tt arXiv:1902.08124 [astro-ph.HE]}}.

\bibitem{Pierog:2019}
T.~Pierog et al., {\em {Collective hadronization and air showers: can LHC data
  solve the muon puzzle ?}\/},  PoS {\bf ICRC2019} (2019)  .

\bibitem{Nir_Private}
D.~Aloni et al., {\em Private communication from D.Aloni, A. Dery, M.B. Gavela
  and Y. Nir (see repository)\/},  2019.

\bibitem{Chivukula:1987py}
R.~S. Chivukula and H.~Georgi, {\em {Composite Technicolor Standard Model}\/},
\href{http://dx.doi.org/10.1016/0370-2693(87)90713-1}{Phys. Lett. {\bf B188}
  (1987)  99--104}.

\bibitem{Buras:2000dm}
A.~J. Buras et al., {\em {Universal unitarity triangle and physics beyond the
  standard model}\/},
  \href{http://dx.doi.org/10.1016/S0370-2693(01)00061-2}{Phys. Lett. {\bf B500}
  (2001)  161--167},
\href{http://arxiv.org/abs/hep-ph/0007085}{{\tt arXiv:hep-ph/0007085
  [hep-ph]}}.

\bibitem{DAmbrosio:2002vsn}
G.~D'Ambrosio et al., {\em {Minimal flavor violation: An Effective field theory
  approach}\/},  \href{http://dx.doi.org/10.1016/S0550-3213(02)00836-2}{Nucl.
  Phys. {\bf B645} (2002)  155--187},
\href{http://arxiv.org/abs/hep-ph/0207036}{{\tt arXiv:hep-ph/0207036
  [hep-ph]}}.

\bibitem{Cirigliano:2005ck}
V.~Cirigliano et al., {\em {Minimal flavor violation in the lepton sector}\/},
  \href{http://dx.doi.org/10.1016/j.nuclphysb.2005.08.037}{Nucl. Phys. {\bf
  B728} (2005)  121--134},
\href{http://arxiv.org/abs/hep-ph/0507001}{{\tt arXiv:hep-ph/0507001
  [hep-ph]}}.

\bibitem{Andreev:2018ayy}
{ACME Collaboration}, V.~Andreev et al., {\em {Improved limit on the electric
  dipole moment of the electron}\/},
\href{http://dx.doi.org/10.1038/s41586-018-0599-8}{Nature {\bf 562} (2018)
  no.~7727, 355--360}.

\bibitem{Graner:2016ses}
B.~Graner et al., {\em {Reduced Limit on the Permanent Electric Dipole Moment
  of Hg199}\/},  \href{http://dx.doi.org/10.1103/PhysRevLett.119.119901,
  10.1103/PhysRevLett.116.161601}{Phys. Rev. Lett. {\bf 116} (2016) no.~16,
  161601}, \href{http://arxiv.org/abs/1601.04339}{{\tt arXiv:1601.04339
  [physics.atom-ph]}}.
[Erratum: Phys. Rev. Lett.119, no.11, 119901(2017)].

\bibitem{Bennett:2008dy}
{Muon (g-2) Collaboration}, G.~W. Bennett et al., {\em {An Improved Limit on
  the Muon Electric Dipole Moment}\/},
  \href{http://dx.doi.org/10.1103/PhysRevD.80.052008}{Phys. Rev. {\bf D80}
  (2009)  052008},
\href{http://arxiv.org/abs/0811.1207}{{\tt arXiv:0811.1207 [hep-ex]}}.

\bibitem{PaulESPP19}
S.~Paul, {\em Presentation at the Flavour Session of the CERN Council Open
  Symposium on the ESPPU, Granada, 13-16 May 2019\/},  2019.

\bibitem{Anastassopoulos:2015ura}
V.~Anastassopoulos et al., {\em {A Storage Ring Experiment to Detect a Proton
  Electric Dipole Moment}\/},  \href{http://dx.doi.org/10.1063/1.4967465}{Rev.
  Sci. Instrum. {\bf 87} (2016) no.~11, 115116},
\href{http://arxiv.org/abs/1502.04317}{{\tt arXiv:1502.04317
  [physics.acc-ph]}}.

\bibitem{Jedi_exp}
{JEDI Collaboration}, Y.~Senichev,
  \href{http://dx.doi.org/10.18429/JACoW-RuPAC2016-MOZMH03}{{\em {Search for
  the Charged Particle Electric Dipole Moments in Storage Rings}\/}, }
in {\em {25th Russian Particle Accelerator Conference (RuPAC 2016) Saint
  Petersburg, Russia, November 21-25, 2016}}.
\newblock

\bibitem{Rathmann:2019vtb}
F.~Rathmann and N.~N. Nikolaev, {\em {Electric dipole moment searches using
  storage rings}\/},  in {\em {23rd International Symposium on Spin Physics
  (SPIN 2018) Ferrara, Italy, September 10-14, 2018}}.
\newblock 2019.
\newblock
\href{http://arxiv.org/abs/1904.13166}{{\tt arXiv:1904.13166 [nucl-ex]}}.
\newblock

\bibitem{Chislett:2016jau}
{Muon g-2 Collaboration}, R.~Chislett, {\em {The muon EDM in the g-2 experiment
  at Fermilab}\/},
\href{http://dx.doi.org/10.1051/epjconf/201611801005}{EPJ Web Conf. {\bf 118}
  (2016)  01005}.

\bibitem{Sato:2017sdn}
{E34 Collaboration}, Y.~Sato, {\em {Muon g-2/EDM experiment at J-PARC}\/},
\href{http://dx.doi.org/10.22323/1.294.0006}{PoS {\bf KMI2017} (2017)  006}.

\bibitem{Crivellin:2018qmi}
A.~Crivellin, M.~Hoferichter, and P.~Schmidt-Wellenburg, {\em {Combined
  explanations of $(g-2)_{\mu,e}$ and implications for a large muon EDM}\/},
  \href{http://dx.doi.org/10.1103/PhysRevD.98.113002}{Phys. Rev. {\bf D98}
  (2018) no.~11, 113002},
\href{http://arxiv.org/abs/1807.11484}{{\tt arXiv:1807.11484 [hep-ph]}}.

\bibitem{Bondar:2013cja}
{Charm-Tau Factory Collaboration}, A.~E. Bondar et al., {\em {Project of a
  Super Charm-Tau factory at the Budker Institute of Nuclear Physics in
  Novosibirsk}\/},  \href{http://dx.doi.org/10.1134/S1063778813090032}{Phys.
  Atom. Nucl. {\bf 76} (2013)  1072--1085}.
[Yad. Fiz.76,no.9,1132(2013)].

\bibitem{Luo:2018njj}
Q.~Luo and D.~Xu,
  \href{http://dx.doi.org/10.18429/JACoW-IPAC2018-MOPML013}{{\em {Progress on
  Preliminary Conceptual Study of HIEPA, a Super Tau-Charm Factory in
  China}\/}, } in {\em {Proceedings, 9th International Particle Accelerator
  Conference (IPAC 2018): Vancouver, BC Canada}}, p.~MOPML013.
\newblock
2018.
\newblock

\bibitem{Peng:2018}
H.-P. Peng, {\em {High Intensity Electron Positron Accelerator (HIEPA), Super
  Tau Charm Facility STCF in China}\/},  in {\em {Charm2018, Novosibirsk,
  Russia, May 21 - 25, 2018}}.
\newblock 2018.

\bibitem{Baryshevsky:2019vou}
V.~G. Baryshevsky, {\em {Electromagnetic dipole moments and time reversal
  violating interactions for high energy charged baryons in bent crystals at
  LHC}\/},
\href{http://dx.doi.org/10.1140/epjc/s10052-019-6857-6}{Eur. Phys. J. {\bf C79}
  (2019) no.~4, 350}.

\bibitem{Mazzolari:2018hsu}
A.~Mazzolari et al., {\em {Bent crystals for efficient beam steering of multi
  TeV-particle beams}\/},
\href{http://dx.doi.org/10.1140/epjc/s10052-018-6196-z}{Eur. Phys. J. {\bf C78}
  (2018) no.~9, 720}.

\bibitem{Baldini:2018nnn}
{MEG II Collaboration}, A.~M. Baldini et al., {\em {The design of the MEG II
  experiment}\/},  \href{http://dx.doi.org/10.1140/epjc/s10052-018-5845-6}{Eur.
  Phys. J. {\bf C78} (2018) no.~5, 380},
\href{http://arxiv.org/abs/1801.04688}{{\tt arXiv:1801.04688
  [physics.ins-det]}}.

\bibitem{Blondel:2013ia}
A.~Blondel et al., {\em {Research Proposal for an Experiment to Search for the
  Decay $\mu \to eee$}\/},
\href{http://arxiv.org/abs/1301.6113}{{\tt arXiv:1301.6113 [physics.ins-det]}}.

\bibitem{Adamov:2018vin}
{COMET Collaboration}, G.~Adamov et al., {\em {COMET Phase-I Technical Design
  Report}\/},
\href{http://arxiv.org/abs/1812.09018}{{\tt arXiv:1812.09018
  [physics.ins-det]}}.

\bibitem{Bartoszek:2014mya}
{Mu2e Collaboration}, L.~Bartoszek et al., {\em {Mu2e Technical Design
  Report}\/},
\href{http://arxiv.org/abs/1501.05241}{{\tt arXiv:1501.05241
  [physics.ins-det]}}.

\bibitem{KunoESPP19}
Y.~Kuno, {\em Presentation at the Flavour Session of the CERN Council Open
  Symposium on the ESPPU. Granada, 13-16 May\/},  {\it 2019}.

\bibitem{Knoepfel:2013ouy}
{Mu2e Collaboration}, K.~Knoepfel et al., {\em {Feasibility Study for a
  Next-Generation Mu2e Experiment}\/},  in {\em {Proceedings, 2013 Community
  Summer Study on the Future of U.S. Particle Physics: Snowmass on the
  Mississippi (CSS2013): Minneapolis, MN, USA, July 29-August 6, 2013}}.
\newblock 2013.
\newblock \href{http://arxiv.org/abs/1307.1168}{{\tt arXiv:1307.1168
  [physics.ins-det]}}.
\newblock
\url{http://lss.fnal.gov/archive/2013/conf/fermilab-conf-13-254.pdf}.
\newblock

\bibitem{Abusalma:2018xem}
{Mu2e Collaboration}, F.~Abusalma et al., {\em {Expression of Interest for
  Evolution of the Mu2e Experiment}\/},
\href{http://arxiv.org/abs/1802.02599}{{\tt arXiv:1802.02599
  [physics.ins-det]}}.

\bibitem{Buras:2015qea}
A.~J. Buras et al., {\em {$ {K}^{+}\to {\pi}^{+}\nu \overline{\nu} $ and $
  {K}_L\to {\pi}^0\nu \overline{\nu} $ in the Standard Model: status and
  perspectives}\/},  \href{http://dx.doi.org/10.1007/JHEP11(2015)033}{JHEP {\bf
  11} (2015)  033},
\href{http://arxiv.org/abs/1503.02693}{{\tt arXiv:1503.02693 [hep-ph]}}.

\bibitem{SozziESPP19}
M.~Sozzi, {\em Presentation at the Flavour Session of the CERN Council Open
  Symposium on the ESPPU. Granada, 13-16 May\/},  {\it 2019}.

\bibitem{Artamonov:2009sz}
{BNL-E949 Collaboration}, A.~V. Artamonov et al., {\em {Study of the decay
  $K^+\to\pi^+\nu \bar\nu$ in the momentum region $140 < P_\pi < 199$
  MeV/c}\/},  \href{http://dx.doi.org/10.1103/PhysRevD.79.092004}{Phys. Rev.
  {\bf D79} (2009)  092004},
\href{http://arxiv.org/abs/0903.0030}{{\tt arXiv:0903.0030 [hep-ex]}}.

\bibitem{NA62:2019}
{NA62 Collaboration}, {\em Latest measurement of $K^+\to \pi^+ \nu \bar\nu$
  with the NA62 experiment at CERN\/},  2019.
\newblock \url{https://indico.cern.ch/event/769729/sessions/318725/#20190910}.

\bibitem{NA62:Lazzeroni}
C.~Lazzeroni, {\em Contribution to the Flavour Session of the CERN Council Open
  Symposium on the ESPPU (NA62 private communication)\/},  2019.

\bibitem{Ahn:2018mvc}
{KOTO Collaboration}, J.~K. Ahn et al., {\em {Search for the $K_L \!\to\! \pi^0
  \nu \overline{\nu}$ and $K_L \!\to\! \pi^0 X^0$ decays at the J-PARC KOTO
  experiment}\/},
  \href{http://dx.doi.org/10.1103/PhysRevLett.122.021802}{Phys. Rev. Lett. {\bf
  122} (2019) no.~2, 021802}, \href{http://arxiv.org/abs/1810.09655}{{\tt
  arXiv:1810.09655 [hep-ex]}}.

\bibitem{KOTO_prospects}
T.~Yamanaka, {\em Presentation at the 26th J-PARC Program Advisory Committee,
  18 July 2018\/},  2018.
\newblock \url{https://kds.kek.
  jp/indico/event/28286/contribution/11/material/slides/1.pdf}.

\bibitem{Ambrosino:2019qvz}
{KLEVER Project Collaboration}, F.~Ambrosino et al., {\em {KLEVER: An
  experiment to measure BR($K_L\to\pi^0\nu\bar{\nu}$) at the CERN SPS}\/},
\href{http://arxiv.org/abs/1901.03099}{{\tt arXiv:1901.03099 [hep-ex]}}.

\bibitem{KOTO_II}
{KOTO Collaboration}, J.~Comfort et al., {\em J-PARC proposal E14, 2 May
  2006\/},  2006.
\newblock \url{http://koto.kek.jp/pub/p14.pd}.

\bibitem{Cirigliano:2011ny}
V.~Cirigliano et al., {\em {Kaon Decays in the Standard Model}\/},
  \href{http://dx.doi.org/10.1103/RevModPhys.84.399}{Rev. Mod. Phys. {\bf 84}
  (2012)  399},
\href{http://arxiv.org/abs/1107.6001}{{\tt arXiv:1107.6001 [hep-ph]}}.

\bibitem{CortinaGil:2019dnd}
{NA62 Collaboration}, E.~Cortina~Gil et al., {\em {Searches for lepton number
  violating $K^+$ decays}\/},
\href{http://arxiv.org/abs/1905.07770}{{\tt arXiv:1905.07770 [hep-ex]}}.

\bibitem{JPARC_TREK}
{\em The TREK Experiment (Time Reversal Experiment with Kaons): J-PARC
  Experiment E06\/},  2006.
\newblock \url{http://trek.kek.jp/trek.html}.

\bibitem{Kelly:2019yxg}
C.~Kelly and T.~Wang, {\em {Update on the improved lattice calculation of
  direct CP-violation in K decays}\/},
\href{http://dx.doi.org/10.22323/1.334.0277}{PoS {\bf LATTICE2018} (2019)
  277}.

\bibitem{Lehner:2015jga}
C.~Lehner, E.~Lunghi, and A.~Soni, {\em {Emerging lattice approach to the
  K-Unitarity Triangle}\/},
  \href{http://dx.doi.org/10.1016/j.physletb.2016.04.064}{Phys. Lett. {\bf
  B759} (2016)  82--90},
\href{http://arxiv.org/abs/1508.01801}{{\tt arXiv:1508.01801 [hep-ph]}}.

\bibitem{Buras:2015yca}
A.~J. Buras, D.~Buttazzo, and R.~Knegjens, {\em {$ K\to \pi \nu \overline{\nu}
  $ and $\varepsilon^\prime/ \varepsilon$ in simplified new physics models}\/},
   \href{http://dx.doi.org/10.1007/JHEP11(2015)166}{JHEP {\bf 11} (2015)  166},
\href{http://arxiv.org/abs/1507.08672}{{\tt arXiv:1507.08672 [hep-ph]}}.

\bibitem{Zupan:2019uoi}
J.~Zupan, {\em {Introduction to flavour physics}\/},  in {\em {2018 European
  School of High-Energy Physics (ESHEP2018) Maratea, Italy, June 20-July 3,
  2018}}.
\newblock 2019.
\newblock
\href{http://arxiv.org/abs/1903.05062}{{\tt arXiv:1903.05062 [hep-ph]}}.
\newblock

\bibitem{CMS:2014xfa}
{CMS, LHCb Collaboration}, V.~Khachatryan et al., {\em {Observation of the rare
  $B^0_s\to\mu^+\mu^-$ decay from the combined analysis of CMS and LHCb
  data}\/},  \href{http://dx.doi.org/10.1038/nature14474}{Nature {\bf 522}
  (2015)  68--72},
\href{http://arxiv.org/abs/1411.4413}{{\tt arXiv:1411.4413 [hep-ex]}}.

\bibitem{Aaij:2018uns}
{LHCb Collaboration}, R.~Aaij et al., {\em {Measurement of the CKM angle
  $\gamma$ using $B^\pm\to DK^\pm$ with $D\to K_\text{S}^0\pi^+\pi^-$,
  $K_\text{S}^0K^+K^-$ decays}\/},
  \href{http://dx.doi.org/10.1007/JHEP10(2018)107,
  10.1007/JHEP08(2018)176}{JHEP {\bf 08} (2018)  176},
  \href{http://arxiv.org/abs/1806.01202}{{\tt arXiv:1806.01202 [hep-ex]}}.
[erratum: JHEP10,107(2018)].

\bibitem{ATLAS:2019akj}
{ATLAS Collaboration}, {\em {Measurement of the CP violation phase $\phi_{s}$
  in $B_{s}\to J/\psi \phi$ decays in ATLAS at 13 TeV}\/},
ATLAS-CONF-2019-009 (2019)  .

\bibitem{Govorkova:2019jlz}
{LHCb Collaboration}, E.~Govorkova, {\em {Mixing and time-dependent CP
  violation in beauty at LHCb}\/},
\href{http://arxiv.org/abs/1905.08465}{{\tt arXiv:1905.08465 [hep-ex]}}.

\bibitem{Aaij:2013iua}
{LHCb Collaboration}, R.~Aaij et al., {\em {First observation of $CP$ violation
  in the decays of $B^0_s$ mesons}\/},
  \href{http://dx.doi.org/10.1103/PhysRevLett.110.221601}{Phys. Rev. Lett. {\bf
  110} (2013) no.~22, 221601},
\href{http://arxiv.org/abs/1304.6173}{{\tt arXiv:1304.6173 [hep-ex]}}.

\bibitem{Aaij:2019kcg}
{LHCb Collaboration}, R.~Aaij et al., {\em {Observation of $C\!P$ violation in
  charm decays}\/},
\href{http://arxiv.org/abs/1903.08726}{{\tt arXiv:1903.08726 [hep-ex]}}.

\bibitem{Aaij:2015tga}
{LHCb Collaboration}, R.~Aaij et al., {\em {Observation of $J/\psi p$
  Resonances Consistent with Pentaquark States in $\Lambda_b^0 \to J/\psi K^-
  p$ Decays}\/},  \href{http://dx.doi.org/10.1103/PhysRevLett.115.072001}{Phys.
  Rev. Lett. {\bf 115} (2015)  072001},
\href{http://arxiv.org/abs/1507.03414}{{\tt arXiv:1507.03414 [hep-ex]}}.

\bibitem{Aaij:2019vzc}
{LHCb Collaboration}, R.~Aaij et al., {\em {Observation of a narrow pentaquark
  state, $P_c(4312)^+$, and of two-peak structure of the $P_c(4450)^+$}\/},
  Submitted to: Phys. Rev. Lett. (2019)  ,
\href{http://arxiv.org/abs/1904.03947}{{\tt arXiv:1904.03947 [hep-ex]}}.

\bibitem{Cerri:2018ypt}
A.~Cerri et al., {\em {Opportunities in Flavour Physics at the HL-LHC and
  HE-LHC}\/},
\href{http://arxiv.org/abs/1812.07638}{{\tt arXiv:1812.07638 [hep-ph]}}.

\bibitem{PhysRevLett.103.171801}
{Belle Collaboration}, J.-T. Wei et al.,
  \href{http://dx.doi.org/10.1103/PhysRevLett.103.171801}{{\em Measurement of
  the Differential Branching Fraction and Forward-Backward Asymmetry for
  $B\ensuremath{\rightarrow}{K}^{(*)}{l}^{+}{l}^{\ensuremath{-}}$\/}, Phys.
  Rev. Lett. {\bf 103} (Oct, 2009)  171801}.
  \url{https://link.aps.org/doi/10.1103/PhysRevLett.103.171801}.

\bibitem{PhysRevD.86.032012}
{BABAR Collaboration}, J.~P. Lees et al.,
  \href{http://dx.doi.org/10.1103/PhysRevD.86.032012}{{\em Measurement of
  branching fractions and rate asymmetries in the rare decays
  $B\ensuremath{\rightarrow}{K}^{(*)}{\ensuremath{\ell}}^{+}{\ensuremath{\ell}}^{\ensuremath{-}}$\/},
  Phys. Rev. D {\bf 86} (Aug, 2012)  032012}.
  \url{https://link.aps.org/doi/10.1103/PhysRevD.86.032012}.

\bibitem{PhysRevLett.113.151601}
{LHCb Collaboration}, R.~Aaij et al.,
  \href{http://dx.doi.org/10.1103/PhysRevLett.113.151601}{{\em Test of Lepton
  Universality Using
  ${B}^{+}\ensuremath{\rightarrow}{K}^{+}{\ensuremath{\ell}}^{+}{\ensuremath{\ell}}^{\ensuremath{-}}$
  Decays\/}, Phys. Rev. Lett. {\bf 113} (Oct, 2014)  151601}.
  \url{https://link.aps.org/doi/10.1103/PhysRevLett.113.151601}.

\bibitem{Aaij:2019wad}
{LHCb Collaboration}, R.~Aaij et al., {\em {Search for lepton-universality
  violation in $B^+\to K^+\ell^+\ell^-$ decays}\/},
  \href{http://dx.doi.org/10.1103/PhysRevLett.122.191801}{Phys. Rev. Lett. {\bf
  122} (2019) no.~19, 191801},
\href{http://arxiv.org/abs/1903.09252}{{\tt arXiv:1903.09252 [hep-ex]}}.

\bibitem{Abdesselam:2019wac}
{Belle Collaboration}, A.~Abdesselam et al., {\em {Test of lepton flavor
  universality in ${B\to K^\ast\ell^+\ell^-}$ decays at Belle}\/},
\href{http://arxiv.org/abs/1904.02440}{{\tt arXiv:1904.02440 [hep-ex]}}.

\bibitem{Abdesselam:2019dgh}
{Belle Collaboration}, A.~Abdesselam et al., {\em {Measurement of
  $\mathcal{R}(D)$ and $\mathcal{R}(D^{\ast})$ with a semileptonic tagging
  method}\/},
\href{http://arxiv.org/abs/1904.08794}{{\tt arXiv:1904.08794 [hep-ex]}}.

\bibitem{PhysRevD.97.072013}
{LHCb Collaboration}, R.~Aaij et al.,
  \href{http://dx.doi.org/10.1103/PhysRevD.97.072013}{{\em Test of lepton
  flavor universality by the measurement of the
  ${B}^{0}\ensuremath{\rightarrow}{{D}^{*}}^{\ensuremath{-}}{\ensuremath{\tau}}^{+}{\ensuremath{\nu}}_{\ensuremath{\tau}}$
  branching fraction using three-prong $\ensuremath{\tau}$ decays\/}, Phys.
  Rev. D {\bf 97} (Apr, 2018)  072013}.
  \url{https://link.aps.org/doi/10.1103/PhysRevD.97.072013}.

\bibitem{PhysRevLett.118.211801}
{Belle Collaboration}, S.~Hirose et al.,
  \href{http://dx.doi.org/10.1103/PhysRevLett.118.211801}{{\em Measurement of
  the $\ensuremath{\tau}$ Lepton Polarization and $R({D}^{*})$ in the Decay
  $\overline{B}\ensuremath{\rightarrow}{D}^{*}{\ensuremath{\tau}}^{\ensuremath{-}}{\overline{\ensuremath{\nu}}}_{\ensuremath{\tau}}$\/},
  Phys. Rev. Lett. {\bf 118} (May, 2017)  211801}.
  \url{https://link.aps.org/doi/10.1103/PhysRevLett.118.211801}.

\bibitem{PhysRevLett.115.111803}
{LHCb Collaboration}, R.~Aaij et al.,
  \href{http://dx.doi.org/10.1103/PhysRevLett.115.111803}{{\em Measurement of
  the Ratio of Branching Fractions
  $\mathcal{B}({\overline{B}}^{0}\ensuremath{\rightarrow}{D}^{*+}{\ensuremath{\tau}}^{\ensuremath{-}}{\overline{\ensuremath{\nu}}}_{\ensuremath{\tau}})/\mathcal{B}({\overline{B}}^{0}\ensuremath{\rightarrow}{D}^{*+}{\ensuremath{\mu}}^{\ensuremath{-}}{\overline{\ensuremath{\nu}}}_{\ensuremath{\mu}})$\/},
  Phys. Rev. Lett. {\bf 115} (Sep, 2015)  111803}.
  \url{https://link.aps.org/doi/10.1103/PhysRevLett.115.111803}.

\bibitem{PhysRevD.94.072007}
{Belle Collaboration}, Y.~Sato et al.,
  \href{http://dx.doi.org/10.1103/PhysRevD.94.072007}{{\em Measurement of the
  branching ratio of
  ${\overline{B}}^{0}\ensuremath{\rightarrow}{D}^{*+}{\ensuremath{\tau}}^{\ensuremath{-}}{\overline{\ensuremath{\nu}}}_{\ensuremath{\tau}}$
  relative to
  ${\overline{B}}^{0}\ensuremath{\rightarrow}{D}^{*+}{\ensuremath{\ell}}^{\ensuremath{-}}{\overline{\ensuremath{\nu}}}_{\ensuremath{\ell}}$
  decays with a semileptonic tagging method\/}, Phys. Rev. D {\bf 94} (Oct,
  2016)  072007}. \url{https://link.aps.org/doi/10.1103/PhysRevD.94.072007}.

\bibitem{PhysRevD.92.072014}
{Belle Collaboration}, M.~Huschle et al.,
  \href{http://dx.doi.org/10.1103/PhysRevD.92.072014}{{\em Measurement of the
  branching ratio of
  $\overline{B}\ensuremath{\rightarrow}{D}^{(*)}{\ensuremath{\tau}}^{\ensuremath{-}}{\overline{\ensuremath{\nu}}}_{\ensuremath{\tau}}$
  relative to
  $\overline{B}\ensuremath{\rightarrow}{D}^{(*)}{\ensuremath{\ell}}^{\ensuremath{-}}{\overline{\ensuremath{\nu}}}_{\ensuremath{\ell}}$
  decays with hadronic tagging at Belle\/}, Phys. Rev. D {\bf 92} (Oct, 2015)
  072014}. \url{https://link.aps.org/doi/10.1103/PhysRevD.92.072014}.

\bibitem{PhysRevD.88.072012}
{BABAR Collaboration}, J.~P. Lees et al.,
  \href{http://dx.doi.org/10.1103/PhysRevD.88.072012}{{\em Measurement of an
  excess of
  $\overline{B}\ensuremath{\rightarrow}{D}^{\mathbf{(}*\mathbf{)}}{\ensuremath{\tau}}^{\mathbf{\ensuremath{-}}}{\overline{\ensuremath{\nu}}}_{\ensuremath{\tau}}$
  decays and implications for charged Higgs bosons\/}, Phys. Rev. D {\bf 88}
  (Oct, 2013)  072012}.
  \url{https://link.aps.org/doi/10.1103/PhysRevD.88.072012}.

\bibitem{PhysRevLett.109.101802}
{BABAR Collaboration}, J.~P. Lees et al.,
  \href{http://dx.doi.org/10.1103/PhysRevLett.109.101802}{{\em Evidence for an
  Excess of
  $\overline{B}\ensuremath{\rightarrow}{D}^{(*)}{\ensuremath{\tau}}^{\ensuremath{-}}{\overline{\ensuremath{\nu}}}_{\ensuremath{\tau}}$
  Decays\/}, Phys. Rev. Lett. {\bf 109} (Sep, 2012)  101802}.
  \url{https://link.aps.org/doi/10.1103/PhysRevLett.109.101802}.

\bibitem{Aaij:2017uff}
{LHCb Collaboration}, R.~Aaij et al., {\em {Measurement of the ratio of the
  $B^0 \to D^{*-} \tau^+ \nu_{\tau}$ and $B^0 \to D^{*-} \mu^+ \nu_{\mu}$
  branching fractions using three-prong $\tau$-lepton decays}\/},
  \href{http://dx.doi.org/10.1103/PhysRevLett.120.171802}{Phys. Rev. Lett. {\bf
  120} (2018) no.~17, 171802},
\href{http://arxiv.org/abs/1708.08856}{{\tt arXiv:1708.08856 [hep-ex]}}.

\bibitem{Aaij:2015oid}
{LHCb Collaboration}, R.~Aaij et al., {\em {Angular analysis of the $B^{0} \to
  K^{*0} \mu^{+} \mu^{-}$ decay using 3 fb$^{-1}$ of integrated luminosity}\/},
   \href{http://dx.doi.org/10.1007/JHEP02(2016)104}{JHEP {\bf 02} (2016)  104},
\href{http://arxiv.org/abs/1512.04442}{{\tt arXiv:1512.04442 [hep-ex]}}.

\bibitem{Wehle:2016yoi}
{Belle Collaboration}, S.~Wehle et al., {\em {Lepton-Flavor-Dependent Angular
  Analysis of $B\to K^\ast \ell^+\ell^-$}\/},
  \href{http://dx.doi.org/10.1103/PhysRevLett.118.111801}{Phys. Rev. Lett. {\bf
  118} (2017) no.~11, 111801},
\href{http://arxiv.org/abs/1612.05014}{{\tt arXiv:1612.05014 [hep-ex]}}.

\bibitem{Sirunyan:2017dhj}
{CMS Collaboration}, A.~M. Sirunyan et al., {\em {Measurement of angular
  parameters from the decay $\mathrm{B}^0 \to \mathrm{K}^{*0} \mu^+ \mu^-$ in
  proton-proton collisions at $\sqrt{s} = $ 8 TeV}\/},
  \href{http://dx.doi.org/10.1016/j.physletb.2018.04.030}{Phys. Lett. {\bf
  B781} (2018)  517--541},
\href{http://arxiv.org/abs/1710.02846}{{\tt arXiv:1710.02846 [hep-ex]}}.

\bibitem{Aaboud:2018krd}
{ATLAS Collaboration}, M.~Aaboud et al., {\em {Angular analysis of $B^0_d
  \rightarrow K^{*}\mu^+\mu^-$ decays in $pp$ collisions at $\sqrt{s}= 8$ TeV
  with the ATLAS detector}\/},
  \href{http://dx.doi.org/10.1007/JHEP10(2018)047}{JHEP {\bf 10} (2018)  047},
\href{http://arxiv.org/abs/1805.04000}{{\tt arXiv:1805.04000 [hep-ex]}}.

\bibitem{Jager:2012uw}
S.~Jaeger and J.~Martin~Camalich, {\em {On $B \to V \ell \ell$ at small
  dilepton invariant mass, power corrections, and new physics}\/},
  \href{http://dx.doi.org/10.1007/JHEP05(2013)043}{JHEP {\bf 05} (2013)  043},
\href{http://arxiv.org/abs/1212.2263}{{\tt arXiv:1212.2263 [hep-ph]}}.

\bibitem{Ciuchini:2017mik}
M.~Ciuchini et al., {\em {On Flavourful Easter eggs for New Physics hunger and
  Lepton Flavour Universality violation}\/},
  \href{http://dx.doi.org/10.1140/epjc/s10052-017-5270-2}{Eur. Phys. J. {\bf
  C77} (2017) no.~10, 688},
\href{http://arxiv.org/abs/1704.05447}{{\tt arXiv:1704.05447 [hep-ph]}}.

\bibitem{Matias:2012xw}
J.~Matias et al., {\em {Complete Anatomy of $\bar{B}_d \to \bar{K}^{* 0} (\to K
  \pi)l^+l^-$ and its angular distribution}\/},
  \href{http://dx.doi.org/10.1007/JHEP04(2012)104}{JHEP {\bf 04} (2012)  104},
\href{http://arxiv.org/abs/1202.4266}{{\tt arXiv:1202.4266 [hep-ph]}}.

\bibitem{DescotesGenon:2012zf}
S.~Descotes-Genon et al., {\em {Implications from clean observables for the
  binned analysis of $B \to K^*\mu^+\mu^-$ at large recoil}\/},
  \href{http://dx.doi.org/10.1007/JHEP01(2013)048}{JHEP {\bf 01} (2013)  048},
\href{http://arxiv.org/abs/1207.2753}{{\tt arXiv:1207.2753 [hep-ph]}}.

\bibitem{Bobeth:2017vxj}
C.~Bobeth et al., {\em {Long-distance effects in $B\rightarrow K^*\ell \ell $
  from analyticity}\/},
  \href{http://dx.doi.org/10.1140/epjc/s10052-018-5918-6}{Eur. Phys. J. {\bf
  C78} (2018) no.~6, 451},
\href{http://arxiv.org/abs/1707.07305}{{\tt arXiv:1707.07305 [hep-ph]}}.

\bibitem{HFLAVSPRING19}
HFLAV, {\em Average of R(D) and R(D*+) for Spring 2019\/},  2019.
\newblock
  \url{https://hflav-eos.web.cern.ch/hflav-eos/semi/spring19/html/RDsDsstar/RDRDs.html}.

\bibitem{Bediaga:2018lhg}
{LHCb Collaboration}, R.~Aaij et al., {\em {Physics case for an LHCb Upgrade II
  - Opportunities in flavour physics, and beyond, in the HL-LHC era}\/},
\href{http://arxiv.org/abs/1808.08865}{{\tt arXiv:1808.08865}}.

\bibitem{Aebischer:2019mlg}
J.~Aebischer et al., {\em {$B$-decay discrepancies after Moriond 2019}\/},
\href{http://arxiv.org/abs/1903.10434}{{\tt arXiv:1903.10434 [hep-ph]}}.

\bibitem{Kamenik:2017ghi}
J.~F. Kamenik et al., {\em {Lepton polarization asymmetries in rare
  semi-tauonic $ b \rightarrow s $ exclusive decays at FCC-$ee$}\/},
  \href{http://dx.doi.org/10.1140/epjc/s10052-017-5272-0}{Eur. Phys. J. {\bf
  C77} (2017) no.~10, 701},
\href{http://arxiv.org/abs/1705.11106}{{\tt arXiv:1705.11106 [hep-ph]}}.

\bibitem{Irles:2019xny}
A.~Irles et al., {\em {Complementarity between ILC250 and ILC-GigaZ}\/},  in
  {\em {Linear Collider Community Meeting Lausanne, Switzerland, April 8-9,
  2019}}.
\newblock 2019.
\newblock
\href{http://arxiv.org/abs/1905.00220}{{\tt arXiv:1905.00220 [hep-ex]}}.
\newblock

\bibitem{LusianiESPP19}
A.~Lusiani, {\em Presentation at the Flavour Session of the CERN Council Open
  Symposium on the ESPPU. Granada, 13-16 May 2019\/},  2019.

\bibitem{Abe:2010gxa}
{Belle-II Collaboration}, T.~Abe et al., {\em {Belle II Technical Design
  Report}\/},
\href{http://arxiv.org/abs/1011.0352}{{\tt arXiv:1011.0352 [physics.ins-det]}}.

\bibitem{Kou:2018nap}
{Belle-II Collaboration}, W.~Altmannshofer et al., {\em {The Belle II Physics
  Book}\/},
\href{http://arxiv.org/abs/1808.10567}{{\tt arXiv:1808.10567 [hep-ex]}}.

\bibitem{Baer:2013cma}
H.~Baer et al., {\em {The International Linear Collider Technical Design Report
  - Volume 2: Physics}\/},
\href{http://arxiv.org/abs/1306.6352}{{\tt arXiv:1306.6352 [hep-ph]}}.

\bibitem{Fujii:2017vwa}
K.~Fujii et al., {\em {Physics Case for the 250 GeV Stage of the International
  Linear Collider}\/},
\href{http://arxiv.org/abs/1710.07621}{{\tt arXiv:1710.07621 [hep-ex]}}.

\bibitem{Nir:2016zkd}
Y.~Nir, \href{http://dx.doi.org/10.5170/CERN-2015-001.123}{{\em {Flavour
  Physics and CP Violation}\/}, } in {\em {Proceedings, 7th
  CERN--Latin-American School of High-Energy Physics (CLASHEP2013): Arequipa,
  Peru, March 6-19, 2013}}, pp.~123--156.
\newblock 2015.
\newblock
\href{http://arxiv.org/abs/1605.00433}{{\tt arXiv:1605.00433 [hep-ph]}}.
\newblock

\bibitem{Heinemann:2019trx}
B.~Heinemann and Y.~Nir, {\em {The Higgs program and open questions in particle
  physics and cosmology}\/},
\href{http://arxiv.org/abs/1905.00382}{{\tt arXiv:1905.00382 [hep-ph]}}.

\bibitem{deBlas:2018mhx}
J.~de~Blas et al., {\em {The CLIC Potential for New Physics}\/},
\href{http://arxiv.org/abs/1812.02093}{{\tt arXiv:1812.02093 [hep-ph]}}.

\bibitem{Bambade:2019fyw}
P.~Bambade et al., {\em {The International Linear Collider: A Global
  Project}\/},
\href{http://arxiv.org/abs/1903.01629}{{\tt arXiv:1903.01629 [hep-ex]}}.

\bibitem{Bird:2004ts}
C.~Bird et al., {\em {Search for dark matter in $b \to s$ transitions with
  missing energy}\/},
  \href{http://dx.doi.org/10.1103/PhysRevLett.93.201803}{Phys. Rev. Lett. {\bf
  93} (2004)  201803},
\href{http://arxiv.org/abs/hep-ph/0401195}{{\tt arXiv:hep-ph/0401195
  [hep-ph]}}.

\bibitem{Kamenik:2011vy}
J.~F. Kamenik and C.~Smith, {\em {FCNC portals to the dark sector}\/},
  \href{http://dx.doi.org/10.1007/JHEP03(2012)090}{JHEP {\bf 03} (2012)  090},
\href{http://arxiv.org/abs/1111.6402}{{\tt arXiv:1111.6402 [hep-ph]}}.

\bibitem{Gorbunov:2007ak}
D.~Gorbunov and M.~Shaposhnikov, {\em {How to find neutral leptons of the
  $\nu$MSM?}\/},  \href{http://dx.doi.org/10.1007/JHEP11(2013)101,
  10.1088/1126-6708/2007/10/015}{JHEP {\bf 10} (2007)  015},
  \href{http://arxiv.org/abs/0705.1729}{{\tt arXiv:0705.1729 [hep-ph]}}.
[Erratum: JHEP11,101(2013)].

\bibitem{Calibbi:2016hwq}
L.~Calibbi et al., {\em {Minimal axion model from flavor}\/},
  \href{http://dx.doi.org/10.1103/PhysRevD.95.095009}{Phys. Rev. {\bf D95}
  (2017) no.~9, 095009},
\href{http://arxiv.org/abs/1612.08040}{{\tt arXiv:1612.08040 [hep-ph]}}.

\bibitem{Ema:2016ops}
Y.~Ema et al., {\em {Flaxion: a minimal extension to solve puzzles in the
  standard model}\/},  \href{http://dx.doi.org/10.1007/JHEP01(2017)096}{JHEP
  {\bf 01} (2017)  096},
\href{http://arxiv.org/abs/1612.05492}{{\tt arXiv:1612.05492 [hep-ph]}}.

\bibitem{Wilczek:1982rv}
F.~Wilczek, {\em {Axions and Family Symmetry Breaking}\/},
\href{http://dx.doi.org/10.1103/PhysRevLett.49.1549}{Phys. Rev. Lett. {\bf 49}
  (1982)  1549--1552}.

\bibitem{OConnell:2006rsp}
D.~O'Connell, M.~J. Ramsey-Musolf, and M.~B. Wise, {\em {Minimal Extension of
  the Standard Model Scalar Sector}\/},
  \href{http://dx.doi.org/10.1103/PhysRevD.75.037701}{Phys. Rev. {\bf D75}
  (2007)  037701},
\href{http://arxiv.org/abs/hep-ph/0611014}{{\tt arXiv:hep-ph/0611014
  [hep-ph]}}.

\bibitem{Batell:2009jf}
B.~Batell, M.~Pospelov, and A.~Ritz, {\em {Multi-lepton Signatures of a Hidden
  Sector in Rare B Decays}\/},
  \href{http://dx.doi.org/10.1103/PhysRevD.83.054005}{Phys. Rev. {\bf D83}
  (2011)  054005},
\href{http://arxiv.org/abs/0911.4938}{{\tt arXiv:0911.4938 [hep-ph]}}.

\bibitem{Winkler:2018qyg}
M.~W. Winkler, {\em {Decay and detection of a light scalar boson mixing with
  the Higgs boson}\/},
  \href{http://dx.doi.org/10.1103/PhysRevD.99.015018}{Phys. Rev. {\bf D99}
  (2019) no.~1, 015018},
\href{http://arxiv.org/abs/1809.01876}{{\tt arXiv:1809.01876 [hep-ph]}}.

\bibitem{Jaeckel:2010ni}
J.~Jaeckel and A.~Ringwald, {\em {The Low-Energy Frontier of Particle
  Physics}\/},
  \href{http://dx.doi.org/10.1146/annurev.nucl.012809.104433}{Ann. Rev. Nucl.
  Part. Sci. {\bf 60} (2010)  405--437},
\href{http://arxiv.org/abs/1002.0329}{{\tt arXiv:1002.0329 [hep-ph]}}.

\bibitem{Feng:2017uoz}
J.~L. Feng et al., {\em {ForwArd Search ExpeRiment at the LHC}\/},
  \href{http://dx.doi.org/10.1103/PhysRevD.97.035001}{Phys. Rev. {\bf D97}
  (2018) no.~3, 035001},
\href{http://arxiv.org/abs/1708.09389}{{\tt arXiv:1708.09389 [hep-ph]}}.

\bibitem{Gonnella:2017hsz}
{NA62 Collaboration}, F.~Gonnella, {\em {The NA62 experiment at CERN}\/},
\href{http://dx.doi.org/10.1088/1742-6596/873/1/012015}{J. Phys. Conf. Ser.
  {\bf 873} (2017) no.~1, 012015}.

\bibitem{Banerjee:2016tad}
{NA64 Collaboration}, D.~Banerjee et al., {\em {Search for invisible decays of
  sub-GeV dark photons in missing-energy events at the CERN SPS}\/},
  \href{http://dx.doi.org/10.1103/PhysRevLett.118.011802}{Phys. Rev. Lett. {\bf
  118} (2017) no.~1, 011802},
\href{http://arxiv.org/abs/1610.02988}{{\tt arXiv:1610.02988 [hep-ex]}}.

\bibitem{Berlin:2018pwi}
A.~Berlin et al., {\em {Dark Sectors at the Fermilab SeaQuest Experiment}\/},
  \href{http://dx.doi.org/10.1103/PhysRevD.98.035011}{Phys. Rev. {\bf D98}
  (2018) no.~3, 035011},
\href{http://arxiv.org/abs/1804.00661}{{\tt arXiv:1804.00661 [hep-ph]}}.

\bibitem{Beacham:2019nyx}
J.~Beacham et al., {\em {Physics Beyond Colliders at CERN: Beyond the Standard
  Model Working Group Report}\/},
\href{http://arxiv.org/abs/1901.09966}{{\tt arXiv:1901.09966 [hep-ex]}}.

\bibitem{Gligorov:2017nwh}
V.~V. Gligorov et al., {\em {Searching for Long-lived Particles: A Compact
  Detector for Exotics at LHCb}\/},
  \href{http://dx.doi.org/10.1103/PhysRevD.97.015023}{Phys. Rev. {\bf D97}
  (2018) no.~1, 015023},
\href{http://arxiv.org/abs/1708.09395}{{\tt arXiv:1708.09395 [hep-ph]}}.

\bibitem{Akesson:2018vlm}
{LDMX Collaboration}, T.~Akesson et al., {\em {Light Dark Matter eXperiment
  (LDMX)}\/},
\href{http://arxiv.org/abs/1808.05219}{{\tt arXiv:1808.05219 [hep-ex]}}.

\bibitem{Chou:2016lxi}
J.~P. Chou, D.~Curtin, and H.~J. Lubatti, {\em {New Detectors to Explore the
  Lifetime Frontier}\/},
  \href{http://dx.doi.org/10.1016/j.physletb.2017.01.043}{Phys. Lett. {\bf
  B767} (2017)  29--36},
\href{http://arxiv.org/abs/1606.06298}{{\tt arXiv:1606.06298 [hep-ph]}}.

\bibitem{Alpigiani:2018fgd}
{MATHUSLA Collaboration}, C.~Alpigiani et al., {\em {A Letter of Intent for
  MATHUSLA: A Dedicated Displaced Vertex Detector above ATLAS or CMS.}\/},
\href{http://arxiv.org/abs/1811.00927}{{\tt arXiv:1811.00927
  [physics.ins-det]}}.

\bibitem{Alekhin:2015byh}
S.~Alekhin et al., {\em {A facility to Search for Hidden Particles at the CERN
  SPS: the SHiP physics case}\/},
  \href{http://dx.doi.org/10.1088/0034-4885/79/12/124201}{Rept. Prog. Phys.
  {\bf 79} (2016) no.~12, 124201},
\href{http://arxiv.org/abs/1504.04855}{{\tt arXiv:1504.04855 [hep-ph]}}.

\bibitem{Boiarska:2019vid}
I.~Boiarska et al., {\em {Light scalar production from Higgs bosons and FASER
  2}\/},
\href{http://arxiv.org/abs/1908.04635}{{\tt arXiv:1908.04635 [hep-ph]}}.

\bibitem{Akhmedov:1998qx}
E.~K. Akhmedov, V.~A. Rubakov, and A.~{\relax Yu}. Smirnov, {\em {Baryogenesis
  via neutrino oscillations}\/},
  \href{http://dx.doi.org/10.1103/PhysRevLett.81.1359}{Phys. Rev. Lett. {\bf
  81} (1998)  1359--1362},
\href{http://arxiv.org/abs/hep-ph/9803255}{{\tt arXiv:hep-ph/9803255
  [hep-ph]}}.

\bibitem{Asaka:2005pn}
T.~Asaka and M.~Shaposhnikov, {\em {The nuMSM, dark matter and baryon asymmetry
  of the universe}\/},
  \href{http://dx.doi.org/10.1016/j.physletb.2005.06.020}{Phys. Lett. {\bf
  B620} (2005)  17--26},
\href{http://arxiv.org/abs/hep-ph/0505013}{{\tt arXiv:hep-ph/0505013
  [hep-ph]}}.

\bibitem{Nelson:2019fln}
A.~E. Nelson and H.~Xiao, {\em {Baryogenesis from B Meson Oscillations}\/},
\href{http://arxiv.org/abs/1901.08141}{{\tt arXiv:1901.08141 [hep-ph]}}.

\bibitem{Elor:2018twp}
G.~Elor, M.~Escudero, and A.~Nelson, {\em {Baryogenesis and Dark Matter from
  $B$ Mesons}\/},  \href{http://dx.doi.org/10.1103/PhysRevD.99.035031}{Phys.
  Rev. {\bf D99} (2019) no.~3, 035031},
\href{http://arxiv.org/abs/1810.00880}{{\tt arXiv:1810.00880 [hep-ph]}}.

\bibitem{Cabibbo:1963yz}
N.~Cabibbo, {\em {Unitary Symmetry and Leptonic Decays}\/},
  \href{http://dx.doi.org/10.1103/PhysRevLett.10.531}{Phys. Rev. Lett. {\bf 10}
  (1963)  531--533}.
[,648(1963)].

\bibitem{Kobayashi:1973fv}
M.~Kobayashi and T.~Maskawa, {\em {CP Violation in the Renormalizable Theory of
  Weak Interaction}\/},
\href{http://dx.doi.org/10.1143/PTP.49.652}{Prog. Theor. Phys. {\bf 49} (1973)
  652--657}.

\bibitem{Gambino:2019sif}
P.~Gambino, M.~Jung, and S.~Schacht, {\em {The $V_{cb}$ puzzle: An update}\/},
  \href{http://dx.doi.org/10.1016/j.physletb.2019.06.039}{Phys. Lett. {\bf
  B795} (2019)  386--390},
\href{http://arxiv.org/abs/1905.08209}{{\tt arXiv:1905.08209 [hep-ph]}}.

\bibitem{Esteban:2018azc}
I.~Esteban, M.~C. Gonzalez-Garcia, A.~Hernandez-Cabezudo, M.~Maltoni, and
  T.~Schwetz, {\em {Global analysis of three-flavour neutrino oscillations:
  synergies and tensions in the determination of $\theta_{23}, \delta_{\rm
  CP}$, and the mass ordering}\/},
  \href{http://dx.doi.org/10.1007/JHEP01(2019)106}{JHEP {\bf 01} (2019)  106},
\href{http://arxiv.org/abs/1811.05487}{{\tt arXiv:1811.05487 [hep-ph]}}.

\bibitem{Capozzi:2018ubv}
F.~Capozzi, E.~Lisi, A.~Marrone, and A.~Palazzo, {\em {Current unknowns in the
  three neutrino framework}\/},
  \href{http://dx.doi.org/10.1016/j.ppnp.2018.05.005}{Prog. Part. Nucl. Phys.
  {\bf 102} (2018)  48--72},
\href{http://arxiv.org/abs/1804.09678}{{\tt arXiv:1804.09678 [hep-ph]}}.

\bibitem{lisi:2019EPSSU_Granada}
E.~Lisi, {\em {talk at EPSSU open symposium, Granada, 2019}\/}, .

\bibitem{deSalas:2017kay}
P.~F. de~Salas, D.~V. Forero, C.~A. Ternes, M.~Tortola, and J.~W.~F. Valle,
  {\em {Status of neutrino oscillations 2018: 3$\sigma$ hint for normal mass
  ordering and improved CP sensitivity}\/},
  \href{http://dx.doi.org/10.1016/j.physletb.2018.06.019}{Phys. Lett. {\bf
  B782} (2018)  633--640},
\href{http://arxiv.org/abs/1708.01186}{{\tt arXiv:1708.01186 [hep-ph]}}.

\bibitem{Giganti:2017fhf}
C.~Giganti, S.~Lavignac, and M.~Zito, {\em {Neutrino oscillations: the rise of
  the PMNS paradigm}\/},
  \href{http://dx.doi.org/10.1016/j.ppnp.2017.10.001}{Prog. Part. Nucl. Phys.
  {\bf 98} (2018)  1--54},
\href{http://arxiv.org/abs/1710.00715}{{\tt arXiv:1710.00715 [hep-ex]}}.

\bibitem{Abe:2016tez}
{T2K Collaboration}, K.~Abe et al., {\em {Sensitivity of the T2K
  accelerator-based neutrino experiment with an Extended run to
  $20\times10^{21}$ POT}\/},
\href{http://arxiv.org/abs/1607.08004}{{\tt arXiv:1607.08004 [hep-ex]}}.

\bibitem{Abe:2019whr}
{T2K Collaboration}, K.~Abe et al., {\em {T2K ND280 Upgrade - Technical Design
  Report}\/},
\href{http://arxiv.org/abs/1901.03750}{{\tt arXiv:1901.03750
  [physics.ins-det]}}.

\bibitem{Abe:2018uyc}
{Hyper-Kamiokande Collaboration}, K.~Abe et al., {\em {Hyper-Kamiokande Design
  Report}\/},
\href{http://arxiv.org/abs/1805.04163}{{\tt arXiv:1805.04163
  [physics.ins-det]}}.

\bibitem{Cao:2015ita}
{ICFA Neutrino Panel Collaboration}, J.~Cao et al., {\em {On the
  complementarity of Hyper-K and LBNF}\/},
\href{http://arxiv.org/abs/1501.03918}{{\tt arXiv:1501.03918
  [physics.acc-ph]}}.

\bibitem{Liao:2016orc}
J.~Liao, D.~Marfatia, and K.~Whisnant, {\em {Nonstandard neutrino interactions
  at DUNE, T2HK and T2HKK}\/},
  \href{http://dx.doi.org/10.1007/JHEP01(2017)071}{JHEP {\bf 01} (2017)  071},
\href{http://arxiv.org/abs/1612.01443}{{\tt arXiv:1612.01443 [hep-ph]}}.

\bibitem{Miura:2016krn}
{Super-Kamiokande Collaboration}, K.~Abe et al., {\em {Search for proton decay
  via $p \to e^+\pi^0$ and $p \to \mu^+\pi^0$ in 0.31 megaton years exposure of
  the Super-Kamiokande water Cherenkov detector}\/},
  \href{http://dx.doi.org/10.1103/PhysRevD.95.012004}{Phys. Rev. {\bf D95}
  (2017) no.~1, 012004},
\href{http://arxiv.org/abs/1610.03597}{{\tt arXiv:1610.03597 [hep-ex]}}.

\bibitem{Bogomilov:2014koa}
M.~Bogomilov et al., {\em {Neutrino Factory}\/},
\href{http://dx.doi.org/10.1103/PhysRevSTAB.17.121002}{Phys. Rev. ST Accel.
  Beams {\bf 17} (2014) no.~12, 121002}.

\bibitem{Alvarez-Ruso:2017oui}
L.~Alvarez-Ruso et al., {\em {NuSTEC White Paper: Status and challenges of
  neutrino-nucleus scattering}\/},
  \href{http://dx.doi.org/10.1016/j.ppnp.2018.01.006}{Prog. Part. Nucl. Phys.
  {\bf 100} (2018)  1--68},
\href{http://arxiv.org/abs/1706.03621}{{\tt arXiv:1706.03621 [hep-ph]}}.

\bibitem{pascoli}
A.~Giuliani et al., {\em {Double Beta Decay APPEC Committee Report}\/},
\href{http://arxiv.org/abs/1910.04688}{{\tt arXiv:1910.04688 [hep-ex]}}.

\bibitem{Aghanim:2018eyx}
{Planck Collaboration}, N.~Aghanim et al., {\em {Planck 2018 results. VI.
  Cosmological parameters}\/},
\href{http://arxiv.org/abs/1807.06209}{{\tt arXiv:1807.06209 [astro-ph.CO]}}.

\bibitem{Laureijs:2011gra}
{EUCLID Collaboration}, R.~Laureijs et al., {\em {Euclid Definition Study
  Report}\/},
\href{http://arxiv.org/abs/1110.3193}{{\tt arXiv:1110.3193 [astro-ph.CO]}}.

\bibitem{Boyle:2017lzt}
A.~Boyle and E.~Komatsu, {\em {Deconstructing the neutrino mass constraint from
  galaxy redshift surveys}\/},
  \href{http://dx.doi.org/10.1088/1475-7516/2018/03/035}{JCAP {\bf 1803} (2018)
   035},
\href{http://arxiv.org/abs/1712.01857}{{\tt arXiv:1712.01857 [astro-ph.CO]}}.

\bibitem{Boyle:2018rva}
A.~Boyle, {\em {Understanding the neutrino mass constraints achievable by
  combining CMB lensing and spectroscopic galaxy surveys}\/},
  \href{http://dx.doi.org/10.1088/1475-7516/2019/04/038}{JCAP {\bf 1904} (2019)
  no.~04, 038},
\href{http://arxiv.org/abs/1811.07636}{{\tt arXiv:1811.07636 [astro-ph.CO]}}.

\bibitem{KamLAND-Zen:2016pfg}
{KamLAND-Zen Collaboration}, A.~Gando et al., {\em {Search for Majorana
  Neutrinos near the Inverted Mass Hierarchy Region with KamLAND-Zen}\/},
  \href{http://dx.doi.org/10.1103/PhysRevLett.117.109903,
  10.1103/PhysRevLett.117.082503}{Phys. Rev. Lett. {\bf 117} (2016) no.~8,
  082503}, \href{http://arxiv.org/abs/1605.02889}{{\tt arXiv:1605.02889
  [hep-ex]}}.
[Addendum: Phys. Rev. Lett.117,no.10,109903(2016)].

\bibitem{zsigmond}
A.~J. Zsigmond, {\em New results from GERDA Phase II\/},  Jun, 2018.
\newblock Talk at XXVIII International Conference on Neutrino Physics and
  Astrophysics, 4-9 June 2018, URL: https://doi.org/10.5281/zenodo.1287604.

\bibitem{DAndrea:2019umy}
{LEGEND Collaboration}, V.~D'Andrea, {\em {Neutrinoless Double Beta Decay
  Search with $^{76}$Ge: Status and Prospect with LEGEND}\/},  in {\em
  {(Moriond EW 2019) La Thuile, Italy, March 16-23, 2019}}.
\newblock
\href{http://arxiv.org/abs/1905.06572}{{\tt arXiv:1905.06572 [hep-ex]}}.
\newblock

\bibitem{Agostini:2018tnm}
{GERDA Collaboration}, M.~Agostini et al., {\em {Improved Limit on Neutrinoless
  Double-$\beta$ Decay of $^{76}$Ge from GERDA Phase II}\/},
  \href{http://dx.doi.org/10.1103/PhysRevLett.120.132503}{Phys. Rev. Lett. {\bf
  120} (2018) no.~13, 132503},
\href{http://arxiv.org/abs/1803.11100}{{\tt arXiv:1803.11100 [nucl-ex]}}.

\bibitem{Aalseth:2017btx}
{Majorana Collaboration}, C.~E. Aalseth et al., {\em {Search for Neutrinoless
  Double-$\beta$ Decay in $^{76}$Ge with the Majorana Demonstrator}\/},
  \href{http://dx.doi.org/10.1103/PhysRevLett.120.132502}{Phys. Rev. Lett. {\bf
  120} (2018) no.~13, 132502},
\href{http://arxiv.org/abs/1710.11608}{{\tt arXiv:1710.11608 [nucl-ex]}}.

\bibitem{Alvis:2019sil}
{Majorana Collaboration}, S.~I. Alvis et al., {\em {A Search for Neutrinoless
  Double-Beta Decay in $^{76}$Ge with 26 kg-yr of Exposure from the MAJORANA
  DEMONSTRATOR}\/},
\href{http://arxiv.org/abs/1902.02299}{{\tt arXiv:1902.02299 [nucl-ex]}}.

\bibitem{Albert:2017hjq}
{nEXO Collaboration}, J.~B. Albert et al., {\em {Sensitivity and Discovery
  Potential of nEXO to Neutrinoless Double Beta Decay}\/},
  \href{http://dx.doi.org/10.1103/PhysRevC.97.065503}{Phys. Rev. {\bf C97}
  (2018) no.~6, 065503},
\href{http://arxiv.org/abs/1710.05075}{{\tt arXiv:1710.05075 [nucl-ex]}}.

\bibitem{Albert:2017owj}
{EXO Collaboration}, J.~B. Albert et al., {\em {Search for Neutrinoless
  Double-Beta Decay with the Upgraded EXO-200 Detector}\/},
  \href{http://dx.doi.org/10.1103/PhysRevLett.120.072701}{Phys. Rev. Lett. {\bf
  120} (2018) no.~7, 072701},
\href{http://arxiv.org/abs/1707.08707}{{\tt arXiv:1707.08707 [hep-ex]}}.

\bibitem{Alvarez:2011my}
{NEXT Collaboration}, V.~Alvarez et al., {\em {The NEXT-100 experiment for
  neutrinoless double beta decay searches (Conceptual Design Report)}\/},
\href{http://arxiv.org/abs/1106.3630}{{\tt arXiv:1106.3630 [physics.ins-det]}}.

\bibitem{Aalbers:2016jon}
{DARWIN Collaboration}, J.~Aalbers et al., {\em {DARWIN: towards the ultimate
  dark matter detector}\/},
  \href{http://dx.doi.org/10.1088/1475-7516/2016/11/017}{JCAP {\bf 1611} (2016)
   017},
\href{http://arxiv.org/abs/1606.07001}{{\tt arXiv:1606.07001 [astro-ph.IM]}}.

\bibitem{Azzolini:2018tum}
{CUPID Collaboration}, O.~Azzolini et al., {\em {CUPID-0: the first array of
  enriched scintillating bolometers for $0\nu\beta\beta$ decay
  investigations}\/},
  \href{http://dx.doi.org/10.1140/epjc/s10052-018-5896-8}{Eur. Phys. J. {\bf
  C78} (2018) no.~5, 428},
\href{http://arxiv.org/abs/1802.06562}{{\tt arXiv:1802.06562
  [physics.ins-det]}}.

\bibitem{Alduino:2017ehq}
{CUORE Collaboration}, C.~Alduino et al., {\em {First Results from CUORE: A
  Search for Lepton Number Violation via $0\nu\beta\beta$ Decay of
  $^{130}$Te}\/},
  \href{http://dx.doi.org/10.1103/PhysRevLett.120.132501}{Phys. Rev. Lett. {\bf
  120} (2018) no.~13, 132501},
\href{http://arxiv.org/abs/1710.07988}{{\tt arXiv:1710.07988 [nucl-ex]}}.

\bibitem{Artusa:2016mat}
D.~R. Artusa et al., {\em {Enriched TeO$_2$ bolometers with active particle
  discrimination: towards the CUPID experiment}\/},
  \href{http://dx.doi.org/10.1016/j.physletb.2017.02.011}{Phys. Lett. {\bf
  B767} (2017)  321--329},
\href{http://arxiv.org/abs/1610.03513}{{\tt arXiv:1610.03513
  [physics.ins-det]}}.

\bibitem{Azzolini:2018dyb}
{CUPID-0 Collaboration}, O.~Azzolini et al., {\em {First Result on the
  Neutrinoless Double-$\beta$ Decay of $^{82}Se$ with CUPID-0}\/},
  \href{http://dx.doi.org/10.1103/PhysRevLett.120.232502}{Phys. Rev. Lett. {\bf
  120} (2018) no.~23, 232502},
\href{http://arxiv.org/abs/1802.07791}{{\tt arXiv:1802.07791 [nucl-ex]}}.

\bibitem{Bekker:2014tfa}
T.~B. Bekker et al., {\em {Above-ground test of an advanced Li$_2$MoO$_4$
  scintillating bolometer to search for neutrinoless double beta decay of
  $^{100}$Mo}\/},
  \href{http://dx.doi.org/10.1016/j.astropartphys.2015.06.002}{Astropart. Phys.
  {\bf 72} (2016)  38--45},
\href{http://arxiv.org/abs/1410.6933}{{\tt arXiv:1410.6933 [physics.ins-det]}}.

\bibitem{Drex13}
G.~Drexlin, V.~Hannen, S.~Mertens, and C.~Weinheimer, {\em Current direct
  neutrino mass experiments\/},  Adv. High Energy Phys. {\bf 2013} (2013)
  no.~293986, 36pp, \href{http://arxiv.org/abs/1307.0101}{{\tt
  arXiv:1307.0101}}.

\bibitem{Boser:2019rta}
S.~Boeser, C.~Buck, C.~Giunti, J.~Lesgourgues, L.~Ludhova, S.~Mertens,
  A.~Schukraft, and M.~Wurm, {\em {Status of Light Sterile Neutrino
  Searches}\/},
\href{http://arxiv.org/abs/1906.01739}{{\tt arXiv:1906.01739 [hep-ex]}}.

\bibitem{Mer:2015a}
S.~Mertens, T.~Lasserre, S.~Groh, et al., {\em Sensitivity of next-generation
  tritium beta-decay experiments for keV-scale sterile neutrinos\/},
  \href{http://dx.doi.org/10.1088/1475-7516/2015/02/020}{J. Cosmol. Astropart.
  Phys. {\bf 2015} (2015) no.~02, 020}.

\bibitem{Angrik:2005ep}
{KATRIN Collaboration}, J.~Angrik et al., {\em KATRIN design report 2004\/}, .
\url{https://publikationen.bibliothek.kit.edu/270060419/3814644}.

\bibitem{Otten:2008zz}
E.~W. Otten and C.~Weinheimer, {\em {Neutrino mass limit from tritium beta
  decay}\/},  \href{http://dx.doi.org/10.1088/0034-4885/71/8/086201}{Rept.
  Prog. Phys. {\bf 71} (2008)  086201},
\href{http://arxiv.org/abs/0909.2104}{{\tt arXiv:0909.2104 [hep-ex]}}.

\bibitem{Lobashev:1985mu}
V.~M. Lobashev and P.~E. Spivak, {\em {A method for measuring the
  anti-electron-neutrino rest mass}\/},
\href{http://dx.doi.org/10.1016/0168-9002(85)90640-0}{Nucl. Instrum. Meth. {\bf
  A240} (1985)  305--310}.

\bibitem{PICARD1992345}
A.~Picard et al., {\em A solenoid retarding spectrometer with high resolution
  and transmission for keV electrons\/},
  \href{http://dx.doi.org/https://doi.org/10.1016/0168-583X(92)95119-C}{Nuclear
  Instruments and Methods in Physics Research Section B: Beam Interactions with
  Materials and Atoms {\bf 63} (1992) no.~3, 345 -- 358}.
  \url{http://www.sciencedirect.com/science/article/pii/0168583X9295119C}.

\bibitem{Gastaldo:2017edk}
L.~Gastaldo et al., {\em {The electron capture in $^{163}$Ho experiment -
  ECHo}\/},
\href{http://dx.doi.org/10.1140/epjst/e2017-70071-y}{Eur. Phys. J. ST {\bf 226}
  (2017) no.~8, 1623--1694}.

\bibitem{Giachero:2016xnn}
{HOLMES Collaboration}, A.~Giachero et al., {\em {Measuring the electron
  neutrino mass with improved sensitivity: the HOLMES experiment}\/},
  \href{http://dx.doi.org/10.1088/1748-0221/12/02/C02046}{JINST {\bf 12} (2017)
  no.~02, C02046},
\href{http://arxiv.org/abs/1612.03947}{{\tt arXiv:1612.03947
  [physics.ins-det]}}.

\bibitem{Esfahani:2017dmu}
{Project 8 Collaboration}, A.~Ashtari~Esfahani et al., {\em {Determining the
  neutrino mass with cyclotron radiation emission spectroscopy-Project 8}\/},
  \href{http://dx.doi.org/10.1088/1361-6471/aa5b4f}{J. Phys. {\bf G44} (2017)
  no.~5, 054004},
\href{http://arxiv.org/abs/1703.02037}{{\tt arXiv:1703.02037
  [physics.ins-det]}}.

\bibitem{Pati:1974yy}
J.~C. Pati and A.~Salam, {\em {Lepton Number as the Fourth Color}\/},
  \href{http://dx.doi.org/10.1103/PhysRevD.10.275,
  10.1103/PhysRevD.11.703.2}{Phys. Rev. {\bf D10} (1974)  275--289}.
[Erratum: Phys. Rev.D11,703(1975)].

\bibitem{Mohapatra:1974gc}
R.~N. Mohapatra and J.~C. Pati, {\em {A Natural Left-Right Symmetry}\/},
\href{http://dx.doi.org/10.1103/PhysRevD.11.2558}{Phys. Rev. {\bf D11} (1975)
  2558}.

\bibitem{Senjanovic:1975rk}
G.~Senjanovic and R.~N. Mohapatra, {\em {Exact Left-Right Symmetry and
  Spontaneous Violation of Parity}\/},
\href{http://dx.doi.org/10.1103/PhysRevD.12.1502}{Phys. Rev. {\bf D12} (1975)
  1502}.

\bibitem{LopezFogliani:2005yw}
D.~E. Lopez-Fogliani and C.~Munoz, {\em {Proposal for a Supersymmetric Standard
  Model}\/},  \href{http://dx.doi.org/10.1103/PhysRevLett.97.041801}{Phys. Rev.
  Lett. {\bf 97} (2006)  041801},
\href{http://arxiv.org/abs/hep-ph/0508297}{{\tt arXiv:hep-ph/0508297
  [hep-ph]}}.

\bibitem{Kersten:2007vk}
J.~Kersten and A.~{\relax Yu}. Smirnov, {\em {Right-Handed Neutrinos at CERN
  LHC and the Mechanism of Neutrino Mass Generation}\/},
  \href{http://dx.doi.org/10.1103/PhysRevD.76.073005}{Phys. Rev. {\bf D76}
  (2007)  073005},
\href{http://arxiv.org/abs/0705.3221}{{\tt arXiv:0705.3221 [hep-ph]}}.

\bibitem{Sirunyan:2018mtv}
{CMS Collaboration}, A.~M. Sirunyan et al., {\em {Search for heavy neutral
  leptons in events with three charged leptons in proton-proton collisions at
  $\sqrt{s} =$ 13 TeV}\/},
  \href{http://dx.doi.org/10.1103/PhysRevLett.120.221801}{Phys. Rev. Lett. {\bf
  120} (2018) no.~22, 221801},
\href{http://arxiv.org/abs/1802.02965}{{\tt arXiv:1802.02965 [hep-ex]}}.

\bibitem{ATLAS:2012ak}
{ATLAS Collaboration}, G.~Aad et al., {\em {Search for heavy neutrinos and
  right-handed $W$ bosons in events with two leptons and jets in $pp$
  collisions at $\sqrt{s}=7$ TeV with the ATLAS detector}\/},
  \href{http://dx.doi.org/10.1140/epjc/s10052-012-2056-4}{Eur. Phys. J. {\bf
  C72} (2012)  2056},
\href{http://arxiv.org/abs/1203.5420}{{\tt arXiv:1203.5420 [hep-ex]}}.

\bibitem{Aaij:2014aba}
{LHCb Collaboration}, R.~Aaij et al., {\em {Search for Majorana neutrinos in
  $B^- \to \pi^+\mu^-\mu^-$ decays}\/},
  \href{http://dx.doi.org/10.1103/PhysRevLett.112.131802}{Phys. Rev. Lett. {\bf
  112} (2014) no.~13, 131802},
\href{http://arxiv.org/abs/1401.5361}{{\tt arXiv:1401.5361 [hep-ex]}}.

\bibitem{Blondel:2014bra}
{FCC-ee study Team Collaboration}, A.~Blondel, E.~Graverini, N.~Serra, and
  M.~Shaposhnikov, {\em {Search for Heavy Right Handed Neutrinos at the
  FCC-ee}\/},  \href{http://dx.doi.org/10.1016/j.nuclphysbps.2015.09.304}{Nucl.
  Part. Phys. Proc. {\bf 273-275} (2016)  1883--1890},
\href{http://arxiv.org/abs/1411.5230}{{\tt arXiv:1411.5230 [hep-ex]}}.

\bibitem{Drewes:2016jae}
M.~Drewes, B.~Garbrecht, D.~Gueter, and J.~Klaric, {\em {Testing the low scale
  seesaw and leptogenesis}\/},
  \href{http://dx.doi.org/10.1007/JHEP08(2017)018}{JHEP {\bf 08} (2017)  018},
\href{http://arxiv.org/abs/1609.09069}{{\tt arXiv:1609.09069 [hep-ph]}}.

\bibitem{Antusch:2017pkq}
S.~Antusch, E.~Cazzato, M.~Drewes, O.~Fischer, B.~Garbrecht, D.~Gueter, and
  J.~Klaric, {\em {Probing Leptogenesis at Future Colliders}\/},
  \href{http://dx.doi.org/10.1007/JHEP09(2018)124}{JHEP {\bf 09} (2018)  124},
\href{http://arxiv.org/abs/1710.03744}{{\tt arXiv:1710.03744 [hep-ph]}}.

\bibitem{Curtin:2018mvb}
D.~Curtin et al., {\em {Long-Lived Particles at the Energy Frontier: The
  MATHUSLA Physics Case}\/},
\href{http://arxiv.org/abs/1806.07396}{{\tt arXiv:1806.07396 [hep-ph]}}.

\bibitem{Anelli:2005ju}
G.~Anelli et al., {\em {Proposal to measure the rare decay $K^+ \to \pi^+ \nu
  \overline{\nu}$ at the CERN SPS}\/}, .
CERN-SPSC-2005-013.

\bibitem{Anelli:2015pba}
{SHiP Collaboration}, M.~Anelli et al., {\em {A facility to Search for Hidden
  Particles (SHiP) at the CERN SPS}\/},
\href{http://arxiv.org/abs/1504.04956}{{\tt arXiv:1504.04956
  [physics.ins-det]}}.

\bibitem{SHiP:2018yqc}
{SHiP Collaboration}, C.~Ahdida et al., {\em {The experimental facility for the
  Search for Hidden Particles at the CERN SPS}\/},
  \href{http://dx.doi.org/10.1088/1748-0221/14/03/P03025}{JINST {\bf 14} (2019)
  no.~03, P03025},
\href{http://arxiv.org/abs/1810.06880}{{\tt arXiv:1810.06880
  [physics.ins-det]}}.

\bibitem{Abe:2019kgx}
{T2K Collaboration}, K.~Abe et al., {\em {Search for heavy neutrinos with the
  T2K near detector ND280}\/},
\href{http://arxiv.org/abs/1902.07598}{{\tt arXiv:1902.07598 [hep-ex]}}.

\bibitem{Ballett:2019bgd}
P.~Ballett, T.~Boschi, and S.~Pascoli, {\em {Heavy Neutral Leptons from
  low-scale seesaws at the DUNE Near Detector}\/},
\href{http://arxiv.org/abs/1905.00284}{{\tt arXiv:1905.00284 [hep-ph]}}.

\bibitem{Arguelles:2019xgp}
C.~A. Arguelles et al., {\em {White Paper on New Opportunities at the
  Next-Generation Neutrino Experiments (Part 1: BSM Neutrino Physics and Dark
  Matter)}\/},
\href{http://arxiv.org/abs/1907.08311}{{\tt arXiv:1907.08311 [hep-ph]}}.

\bibitem{Acharya:2017ttl}
{CTA Consortium Collaboration}, B.~S. Acharya et al.,
  \href{http://dx.doi.org/10.1142/10986}{{\em {Science with the Cherenkov
  Telescope Array}}}.
\newblock WSP, 2018.
\newblock
\href{http://arxiv.org/abs/1709.07997}{{\tt arXiv:1709.07997 [astro-ph.IM]}}.
\newblock

\bibitem{Ahlers:2018fkn}
M.~Ahlers and F.~Halzen, {\em {Opening a New Window onto the Universe with
  IceCube}\/},  \href{http://dx.doi.org/10.1016/j.ppnp.2018.05.001}{Prog. Part.
  Nucl. Phys. {\bf 102} (2018)  73--88},
\href{http://arxiv.org/abs/1805.11112}{{\tt arXiv:1805.11112 [astro-ph.HE]}}.

\bibitem{IceCube:2018dnn}
{IceCube, Fermi-LAT, MAGIC, AGILE, ASAS-SN, HAWC, H.E.S.S., INTEGRAL, Kanata,
  Kiso, Kapteyn, Liverpool Telescope, Subaru, Swift NuSTAR, VERITAS,
  VLA/17B-403 Collaboration}, M.~G. Aartsen et al., {\em {Multimessenger
  observations of a flaring blazar coincident with high-energy neutrino
  IceCube-170922A}\/},
  \href{http://dx.doi.org/10.1126/science.aat1378}{Science {\bf 361} (2018)
  no.~6398, eaat1378},
\href{http://arxiv.org/abs/1807.08816}{{\tt arXiv:1807.08816 [astro-ph.HE]}}.

\bibitem{Giudice:2007fh}
G.~F. Giudice, C.~Grojean, A.~Pomarol, and R.~Rattazzi, {\em {The
  Strongly-Interacting Light Higgs}\/},
  \href{http://dx.doi.org/10.1088/1126-6708/2007/06/045}{JHEP {\bf 06} (2007)
  045},
\href{http://arxiv.org/abs/hep-ph/0703164}{{\tt arXiv:hep-ph/0703164
  [hep-ph]}}.

\bibitem{Beneke:2014sba}
M.~Beneke, D.~Boito, and Y.-M. Wang, {\em {Anomalous Higgs couplings in angular
  asymmetries of $H \to Z\ell^{+} \ell^{-}$ and e$^{+}$ e$^{-} \to HZ$}\/},
  \href{http://dx.doi.org/10.1007/JHEP11(2014)028}{JHEP {\bf 11} (2014)  028},
\href{http://arxiv.org/abs/1406.1361}{{\tt arXiv:1406.1361 [hep-ph]}}.

\bibitem{Franceschini:2017xkh}
R.~Franceschini et al., {\em { Electroweak Precision Tests in High-Energy
  Diboson Processes}\/},  \href{http://dx.doi.org/10.1007/JHEP02(2018)111}{JHEP
  {\bf 02} (2018)  111},
\href{http://arxiv.org/abs/1712.01310}{{\tt arXiv:1712.01310 [hep-ph]}}.

\bibitem{Appelquist:2002mw}
T.~Appelquist, B.~Dobrescu, and A.~Hopper, {\em {Nonexotic Neutral Gauge
  Bosons}\/},  \href{http://dx.doi.org/10.1103/PhysRevD.68.035012}{Phys. Rev.
  {\bf D68} (2003)  035012},
\href{http://arxiv.org/abs/hep-ph/0212073}{{\tt arXiv:hep-ph/0212073
  [hep-ph]}}.

\bibitem{CidVidal:2018eel}
X.~Cid~Vidal et al., {\em {Beyond the Standard Model Physics at the HL-LHC and
  HE-LHC}\/},  \href{http://arxiv.org/abs/1812.07831}{{\tt arXiv:1812.07831
  [hep-ph]}}.
Working Group 3.

\bibitem{Jamin:2019mqx}
C.~Helsens et al., {\em {Heavy resonances at energy-frontier hadron
  colliders}\/},
\href{http://arxiv.org/abs/1902.11217}{{\tt arXiv:1902.11217 [hep-ph]}}.

\bibitem{Panico:2015jxa}
G.~Panico and A.~Wulzer, {\em {The Composite Nambu-Goldstone Higgs}\/},
  \href{http://dx.doi.org/10.1007/978-3-319-22617-0}{Lect. Notes Phys. {\bf
  913} (2016)  pp.1--316},
\href{http://arxiv.org/abs/1506.01961}{{\tt arXiv:1506.01961 [hep-ph]}}.

\bibitem{Farina:2016rws}
M.~Farina et al., {\em {Energy helps accuracy: electroweak precision tests at
  hadron colliders}\/},
  \href{http://dx.doi.org/10.1016/j.physletb.2017.06.043}{Phys. Lett. {\bf
  B772} (2017)  210--215},
\href{http://arxiv.org/abs/1609.08157}{{\tt arXiv:1609.08157 [hep-ph]}}.

\bibitem{Thamm:2015zwa}
A.~Thamm, T.~R., and A.~Wulzer, {\em {Future tests of Higgs compositeness:
  direct vs indirect}\/},
  \href{http://dx.doi.org/10.1007/JHEP07(2015)100}{JHEP {\bf 07} (2015)  100},
\href{http://arxiv.org/abs/1502.01701}{{\tt arXiv:1502.01701 [hep-ph]}}.

\bibitem{Golling:2016gvc}
T.~Golling et al., {\em {Physics at a 100 TeV pp collider: beyond the Standard
  Model phenomena}\/},
  \href{http://dx.doi.org/10.23731/CYRM-2017-003.441}{CERN Yellow Rep. (2017)
  no.~3, 441--634},
\href{http://arxiv.org/abs/1606.00947}{{\tt arXiv:1606.00947 [hep-ph]}}.

\bibitem{Mangano:2651294}
M.~Mangano et al., {\em {Future Circular Collider}\/},   CERN-ACC-2018-0056,
  CERN, Geneva, Dec, 2018.
\newblock \url{https://cds.cern.ch/record/2651294}.
\newblock Published in Eur. Phys. J. C.

\bibitem{Matsedonskyi:2015dns}
O.~Matsedonskyi, G.~Panico, and A.~Wulzer, {\em {Top Partners Searches and
  Composite Higgs Models}\/},
  \href{http://dx.doi.org/10.1007/JHEP04(2016)003}{JHEP {\bf 04} (2016)  003},
\href{http://arxiv.org/abs/1512.04356}{{\tt arXiv:1512.04356 [hep-ph]}}.

\bibitem{ATL-PHYS-PUB-2014-010}
{ATLAS Collaboration}, {\em {Search for Supersymmetry at the high luminosity
  LHC with the ATLAS experiment}\/},   ATL-PHYS-PUB-2014-010, CERN, Geneva,
  Jul, 2014.
\newblock \url{https://cds.cern.ch/record/1735031}.

\bibitem{PhysRevD.97.112001}
{ATLAS Collaboration}, M.~Aaboud et al.,
  \href{http://dx.doi.org/{10.1103/PhysRevD.97.112001}}{{\em {Search for
  squarks and gluinos in final states with jets and missing transverse momentum
  using $36\text{ }\text{ }{\mathrm{fb}}^{\ensuremath{-}1}$ of
  $\sqrt{s}=13\text{ }\text{ }\mathrm{TeV}$ $pp$ collision data with the ATLAS
  detector}\/}, {Phys. Rev. D} {\bf {97}} ({Jun}, {2018})  {112001}}.
  \url{{https://link.aps.org/doi/10.1103/PhysRevD.97.112001}}.

\bibitem{Khachatryan2017}
{CMS Collaboration}, V.~Khachatryan et al.,
  \href{http://dx.doi.org/10.1140/epjc/s10052-017-4787-8}{{\em {A search for
  new phenomena in pp collisions at $\sqrt{s} = 13$ TeV in final states with
  missing transverse momentum and at least one jet using the
  $\alpha_\mathrm{T}$ variable}\/}, The European Physical Journal C {\bf 77}
  (May, 2017)  294}. \url{https://doi.org/10.1140/epjc/s10052-017-4787-8}.

\bibitem{CLIC:PrivateCommunication}
{Private communication by the CLIC Collaboration}.

\bibitem{Aaboud2018}
{ATLAS collaboration}, \href{http://dx.doi.org/10.1007/JHEP09(2018)050}{{\em
  Search for supersymmetry in final states with charm jets and missing
  transverse momentum in 13 TeV pp collisions with the ATLAS detector\/},
  Journal of High Energy Physics {\bf 2018} (Sep, 2018)  50}.
  \url{https://doi.org/10.1007/JHEP09(2018)050}.

\bibitem{Barbieri:1987fn}
R.~Barbieri and G.~Giudice, {\em {Upper Bounds on Supersymmetric Particle
  Masses}\/},
\href{http://dx.doi.org/10.1016/0550-3213(88)90171-X}{Nucl. Phys. {\bf B306}
  (1988)  63--76}.

\bibitem{ATL-PHYS-PUB-2018-031}
{ATLAS Collaboration}, {\em {ATLAS sensitivity to winos and higgsinos with a
  highly compressed mass spectrum at the HL-LHC}\/},   ATL-PHYS-PUB-2018-031,
  CERN, Geneva, Nov, 2018.
\newblock \url{https://cds.cern.ch/record/2647294}.

\bibitem{Curtin2018}
D.~Curtin et al., \href{http://dx.doi.org/10.1007/JHEP07(2018)024}{{\em New
  physics opportunities for long-lived particles at electron-proton
  colliders\/}, Journal of High Energy Physics {\bf 2018} (Jul, 2018)  24}.
  \url{https://doi.org/10.1007/JHEP07(2018)024}.

\bibitem{Buttazzo:2018qqp}
D.~Buttazzo et al., {\em {Fusing Vectors into Scalars at High Energy Lepton
  Colliders}\/},  \href{http://dx.doi.org/10.1007/JHEP11(2018)144}{JHEP {\bf
  11} (2018)  144},
\href{http://arxiv.org/abs/1807.04743}{{\tt arXiv:1807.04743 [hep-ph]}}.

\bibitem{Sirunyan:2018qlb}
{CMS Collaboration}, A.~Sirunyan et al., {\em {Search for a new scalar
  resonance decaying to a pair of Z bosons in proton-proton collisions at
  $\sqrt{s}=13 $ TeV}\/},  \href{http://dx.doi.org/10.1007/JHEP06(2018)127,
  10.1007/JHEP03(2019)128}{JHEP {\bf 06} (2018)  127},
  \href{http://arxiv.org/abs/1804.01939}{{\tt arXiv:1804.01939 [hep-ex]}}.
[Erratum: JHEP03,128(2019)].

\bibitem{Craig:2014}
N.~Craig et al., {\em {The Higgs Portal Above Threshold}\/},
  \href{http://dx.doi.org/10.1007/JHEP02(2016)127}{JHEP {\bf 1602} (2016)
  127},
\href{http://arxiv.org/abs/1412.0258}{{\tt arXiv:1412.0258 [hep-ph]}}.

\bibitem{Curtin:2014jma}
D.~Curtin, P.~Meade, and C.-T. Yu, {\em {Testing Electroweak Baryogenesis with
  Future Colliders}\/},  \href{http://dx.doi.org/10.1007/JHEP11(2014)127}{JHEP
  {\bf 11} (2014)  127},
\href{http://arxiv.org/abs/1409.0005}{{\tt arXiv:1409.0005 [hep-ph]}}.

\bibitem{Chala:2018opy}
M.~Chala, M.~Ramos, and M.~Spannowsky, {\em {Gravitational wave and collider
  probes of a triplet Higgs sector with a low cutoff}\/},
  \href{http://dx.doi.org/10.1140/epjc/s10052-019-6655-1}{Eur. Phys. J. {\bf
  C79} (2019) no.~2, 156},
\href{http://arxiv.org/abs/1812.01901}{{\tt arXiv:1812.01901 [hep-ph]}}.

\bibitem{Carena:2016npr}
M.~Carena and Z.~Liu, {\em {Challenges and opportunities for heavy scalar
  searches in the $ t\overline{t} $ channel at the LHC}\/},
  \href{http://dx.doi.org/10.1007/JHEP11(2016)159}{JHEP {\bf 11} (2016)  159},
\href{http://arxiv.org/abs/1608.07282}{{\tt arXiv:1608.07282 [hep-ph]}}.

\bibitem{Craig:2016ygr}
N.~Craig et al., {\em {Heavy Higgs bosons at low $\tan \beta$: from the LHC to
  100 TeV}\/},  \href{http://dx.doi.org/10.1007/JHEP01(2017)018}{JHEP {\bf 01}
  (2017)  018},
\href{http://arxiv.org/abs/1605.08744}{{\tt arXiv:1605.08744 [hep-ph]}}.

\bibitem{Gunion:2002zf}
J.~Gunion and H.~Haber, {\em {The CP conserving two Higgs doublet model: The
  approach to the decoupling limit}\/},
  \href{http://dx.doi.org/10.1103/PhysRevD.67.075019}{Phys. Rev. {\bf D67}
  (2003)  075019},
\href{http://arxiv.org/abs/hep-ph/0207010}{{\tt arXiv:hep-ph/0207010
  [hep-ph]}}.

\bibitem{Gorbahn:2015gxa}
M.~Gorbahn, J.~No, and V.~Sanz, {\em {Benchmarks for Higgs Effective Theory:
  Extended Higgs Sectors}\/},
  \href{http://dx.doi.org/10.1007/JHEP10(2015)036}{JHEP {\bf 10} (2015)  036},
\href{http://arxiv.org/abs/1502.07352}{{\tt arXiv:1502.07352 [hep-ph]}}.

\bibitem{Hajer:2015gka}
J.~Hajer et al., {\em {Heavy Higgs Bosons at 14 TeV and 100 TeV}\/},
  \href{http://dx.doi.org/10.1007/JHEP11(2015)124}{JHEP {\bf 11} (2015)  124},
\href{http://arxiv.org/abs/1504.07617}{{\tt arXiv:1504.07617 [hep-ph]}}.

\bibitem{Agrawal:2018pci}
P.~Agrawal et al., {\em {Probing the Type-II Seesaw Mechanism through the
  Production of Higgs Bosons at a Lepton Collider}\/},
  \href{http://dx.doi.org/10.1103/PhysRevD.98.015024}{Phys. Rev. {\bf D98}
  (2018) no.~1, 015024},
\href{http://arxiv.org/abs/1803.00677}{{\tt arXiv:1803.00677 [hep-ph]}}.

\bibitem{Du:2018eaw}
Y.~Du et al., {\em {Type-II Seesaw Scalar Triplet Model at a 100 TeV $pp$
  Collider: Discovery and Higgs Portal Coupling Determination}\/},
  \href{http://dx.doi.org/10.1007/JHEP01(2019)101}{JHEP {\bf 01} (2019)  101},
\href{http://arxiv.org/abs/1810.09450}{{\tt arXiv:1810.09450 [hep-ph]}}.

\bibitem{Chacko:2005pe}
Z.~Chacko, H.-S. Goh, and R.~Harnik, {\em {The Twin Higgs: Natural electroweak
  breaking from mirror symmetry}\/},
  \href{http://dx.doi.org/10.1103/PhysRevLett.96.231802}{Phys. Rev. Lett. {\bf
  96} (2006)  231802},
\href{http://arxiv.org/abs/hep-ph/0506256}{{\tt arXiv:hep-ph/0506256
  [hep-ph]}}.

\bibitem{Burdman:2006tz}
G.~Burdman et al., {\em {Folded supersymmetry and the LEP paradox}\/},
  \href{http://dx.doi.org/10.1088/1126-6708/2007/02/009}{JHEP {\bf 02} (2007)
  009},
\href{http://arxiv.org/abs/hep-ph/0609152}{{\tt arXiv:hep-ph/0609152
  [hep-ph]}}.

\bibitem{Cai:2008au}
H.~Cai, H.-C. Cheng, and J.~Terning, {\em {A Quirky Little Higgs Model}\/},
  \href{http://dx.doi.org/10.1088/1126-6708/2009/05/045}{JHEP {\bf 05} (2009)
  045},
\href{http://arxiv.org/abs/0812.0843}{{\tt arXiv:0812.0843 [hep-ph]}}.

\bibitem{Cohen:2018mgv}
T.~Cohen et al., {\em {The Hyperbolic Higgs}\/},
  \href{http://dx.doi.org/10.1007/JHEP05(2018)091}{JHEP {\bf 05} (2018)  091},
\href{http://arxiv.org/abs/1803.03647}{{\tt arXiv:1803.03647 [hep-ph]}}.

\bibitem{Cheng:2018gvu}
H.-C. Cheng et al., {\em {Singlet Scalar Top Partners from Accidental
  Supersymmetry}\/},  \href{http://dx.doi.org/10.1007/JHEP05(2018)057}{JHEP
  {\bf 05} (2018)  057},
\href{http://arxiv.org/abs/1803.03651}{{\tt arXiv:1803.03651 [hep-ph]}}.

\bibitem{Abramowicz:2018rjq}
{CLICdp Collaboration}, H.~Abramowicz et al., {\em {Top-Quark Physics at the
  CLIC Electron-Positron Linear Collider}\/},
\href{http://arxiv.org/abs/1807.02441}{{\tt arXiv:1807.02441 [hep-ex]}}.

\bibitem{Khanpour:2014xla}
K.~Khanpour et al., {\em {Single top quark production as a probe of anomalous
  $tq\gamma$ and $tqZ$ couplings at the FCC-ee}\/},
  \href{http://dx.doi.org/10.1016/j.physletb.2017.10.047}{Phys. Lett. {\bf
  B775} (2017)  25--31},
\href{http://arxiv.org/abs/1408.2090}{{\tt arXiv:1408.2090 [hep-ph]}}.

\bibitem{Cornella:2019hct}
C.~Cornella, J.~Fuentes-Martin, and G.~Isidori, {\em {Revisiting the vector
  leptoquark explanation of the B-physics anomalies}\/},
\href{http://arxiv.org/abs/1903.11517}{{\tt arXiv:1903.11517 [hep-ph]}}.

\bibitem{Krall:2017xij}
R.~Krall and M.~Reece, {\em {Last Electroweak WIMP Standing: Pseudo-Dirac
  Higgsino Status and Compact Stars as Future Probes}\/},
  \href{http://dx.doi.org/10.1088/1674-1137/42/4/043105}{Chin. Phys. {\bf C42}
  (2018) no.~4, 043105},
\href{http://arxiv.org/abs/1705.04843}{{\tt arXiv:1705.04843 [hep-ph]}}.

\bibitem{Chen:2018uqz}
C.-Y. Chen, R.~J. Hill, M.~P. Solon, and A.~M. Wijangco, {\em {Power
  Corrections to the Universal Heavy WIMP-Nucleon Cross Section}\/},
  \href{http://dx.doi.org/10.1016/j.physletb.2018.04.021}{Phys. Lett. {\bf
  B781} (2018)  473--479},
\href{http://arxiv.org/abs/1801.08551}{{\tt arXiv:1801.08551 [hep-ph]}}.

\bibitem{Hryczuk2019-mv}
A.~Hryczuk et al., {\em Testing dark matter with Cherenkov light - prospects of
  {H.E.S.S}. and {CTA} for exploring minimal supersymmetry\/},
  \href{http://arxiv.org/abs/1905.00315}{{\tt arXiv:1905.00315 [hep-ph]}}.
  \url{http://arxiv.org/abs/1905.00315}.

\bibitem{DiLuzio:2018jwd}
L.~Di~Luzio, R.~Grober, and G.~Panico, {\em {Probing new electroweak states via
  precision measurements at the LHC and future colliders}\/},
  \href{http://dx.doi.org/10.1007/JHEP01(2019)011}{JHEP {\bf 01} (2019)  011},
\href{http://arxiv.org/abs/1810.10993}{{\tt arXiv:1810.10993 [hep-ph]}}.

\bibitem{Abercrombie:2015wmb}
D.~Abercrombie et al., {\em {Dark Matter Benchmark Models for Early LHC Run-2
  Searches: Report of the ATLAS/CMS Dark Matter Forum}\/},
\href{http://arxiv.org/abs/1507.00966}{{\tt arXiv:1507.00966 [hep-ex]}}.

\bibitem{Aaboud:2019yqu}
{ATLAS Collaboration}, {\em {Constraints on mediator-based dark matter and
  scalar dark energy models using $\sqrt s = 13$ TeV $pp$ collision data
  collected by the ATLAS detector}\/},
  \href{http://dx.doi.org/10.1007/JHEP05(2019)142}{JHEP {\bf 05} (2019)  142},
\href{http://arxiv.org/abs/1903.01400}{{\tt arXiv:1903.01400 [hep-ex]}}.

\bibitem{Haisch:2018bby}
U.~Haisch and G.~Polesello, {\em {Searching for production of dark matter in
  association with top quarks at the LHC}\/},
  \href{http://dx.doi.org/10.1007/JHEP02(2019)029}{JHEP {\bf 02} (2019)  029},
\href{http://arxiv.org/abs/1812.00694}{{\tt arXiv:1812.00694 [hep-ph]}}.

\bibitem{Harris:2015kda}
P.~Harris et al., {\em {Closing up on Dark Sectors at Colliders: from 14 to 100
  TeV}\/},  \href{http://dx.doi.org/10.1103/PhysRevD.93.054030}{Phys. Rev. {\bf
  D93} (2016) no.~5, 054030},
\href{http://arxiv.org/abs/1509.02904}{{\tt arXiv:1509.02904 [hep-ph]}}.

\bibitem{Habermehl:2018yul}
M.~Habermehl, \href{http://dx.doi.org/10.3204/PUBDB-2018-05723}{{\em {Dark
  Matter at the International Linear Collider}}}.
\newblock PhD thesis, DESY, Hamburg,
2018.
\newblock

\bibitem{Liu:2019ogn}
Z.~Liu, Y.-H. Xu, and Y.~Zhang, {\em {Probing dark matter particles at
  CEPC}\/},  \href{http://dx.doi.org/10.1007/JHEP06(2019)009}{JHEP {\bf 06}
  (2019)  009},
\href{http://arxiv.org/abs/1903.12114}{{\tt arXiv:1903.12114 [hep-ph]}}.

\bibitem{Curtin:2014cca}
D.~Curtin et al., {\em {Illuminating Dark Photons with High-Energy
  Colliders}\/},  \href{http://dx.doi.org/10.1007/JHEP02(2015)157}{JHEP {\bf
  02} (2015)  157},
\href{http://arxiv.org/abs/1412.0018}{{\tt arXiv:1412.0018 [hep-ph]}}.

\bibitem{2018arXiv180208640A}
J.~Albrecht et al., {\em {HEP Community White Paper on Software trigger and
  event reconstruction: Executive Summary}\/}, arXiv e-prints (Feb, 2018)
  arXiv:1802.08640, \href{http://arxiv.org/abs/1802.08640}{{\tt
  arXiv:1802.08640 [physics.comp-ph]}}.

\bibitem{Lee:1977ua}
B.~Lee and S.~Weinberg, {\em {Cosmological Lower Bound on Heavy Neutrino
  Masses}\/},
\href{http://dx.doi.org/10.1103/PhysRevLett.39.165}{Phys. Rev. Lett. {\bf 39}
  (1977)  165--168}.

\bibitem{Boehm:2002yz}
C.~Boehm, T.~Ensslin, and J.~Silk, {\em {Can Annihilating dark matter be
  lighter than a few GeVs?}\/},
  \href{http://dx.doi.org/10.1088/0954-3899/30/3/004}{J. Phys. {\bf G30} (2004)
   279--286},
\href{http://arxiv.org/abs/astro-ph/0208458}{{\tt arXiv:astro-ph/0208458
  [astro-ph]}}.

\bibitem{Boehm:2003hm}
C.~Boehm and P.~Fayet, {\em {Scalar dark matter candidates}\/},
  \href{http://dx.doi.org/10.1016/j.nuclphysb.2004.01.015}{Nucl. Phys. {\bf
  B683} (2004)  219--263},
\href{http://arxiv.org/abs/hep-ph/0305261}{{\tt arXiv:hep-ph/0305261
  [hep-ph]}}.

\bibitem{Pospelov:2007mp}
M.~Pospelov, A.~Ritz, and M.~Voloshin, {\em {Secluded WIMP Dark Matter}\/},
  \href{http://dx.doi.org/10.1016/j.physletb.2008.02.052}{Phys. Lett. {\bf
  B662} (2008)  53--61},
\href{http://arxiv.org/abs/0711.4866}{{\tt arXiv:0711.4866 [hep-ph]}}.

\bibitem{Feng:2008ya}
J.~Feng and J.~Kumar, {\em {The WIMPless Miracle: Dark-Matter Particles without
  Weak-Scale Masses or Weak Interactions}\/},
  \href{http://dx.doi.org/10.1103/PhysRevLett.101.231301}{Phys. Rev. Lett. {\bf
  101} (2008)  231301},
\href{http://arxiv.org/abs/0803.4196}{{\tt arXiv:0803.4196 [hep-ph]}}.

\bibitem{ArkaniHamed:2008qn}
N.~Arkani-Hamed et al., {\em {A Theory of Dark Matter}\/},
  \href{http://dx.doi.org/10.1103/PhysRevD.79.015014}{Phys. Rev. {\bf D79}
  (2009)  015014},
\href{http://arxiv.org/abs/0810.0713}{{\tt arXiv:0810.0713 [hep-ph]}}.

\bibitem{Batley:2015lha}
{NA48/2 Collaboration}, J.~R. Batley et al., {\em {Search for the dark photon
  in $\pi^0$ decays}\/},
  \href{http://dx.doi.org/10.1016/j.physletb.2015.04.068}{Phys. Lett. {\bf
  B746} (2015)  178--185},
\href{http://arxiv.org/abs/1504.00607}{{\tt arXiv:1504.00607 [hep-ex]}}.

\bibitem{Merkel:2014avp}
H.~Merkel et al., {\em {Search at the Mainz Microtron for Light Massive Gauge
  Bosons Relevant for the Muon g-2 Anomaly}\/},
  \href{http://dx.doi.org/10.1103/PhysRevLett.112.221802}{Phys. Rev. Lett. {\bf
  112} (2014) no.~22, 221802},
\href{http://arxiv.org/abs/1404.5502}{{\tt arXiv:1404.5502 [hep-ex]}}.

\bibitem{Lees:2014xha}
{BaBar Collaboration}, J.~P. Lees et al., {\em {Search for a Dark Photon in
  $e^+e^-$ Collisions at BaBar}\/},
  \href{http://dx.doi.org/10.1103/PhysRevLett.113.201801}{Phys. Rev. Lett. {\bf
  113} (2014) no.~20, 201801},
\href{http://arxiv.org/abs/1406.2980}{{\tt arXiv:1406.2980 [hep-ex]}}.

\bibitem{Riordan:1987aw}
E.~M. Riordan et al., {\em {A Search for Short Lived Axions in an Electron Beam
  Dump Experiment}\/},
\href{http://dx.doi.org/10.1103/PhysRevLett.59.755}{Phys. Rev. Lett. {\bf 59}
  (1987)  755}.

\bibitem{Bjorken:1988as}
J.~D. Bjorken, S.~Ecklund, W.~R. Nelson, A.~Abashian, C.~Church, B.~Lu, L.~W.
  Mo, T.~A. Nunamaker, and P.~Rassmann, {\em {Search for Neutral Metastable
  Penetrating Particles Produced in the SLAC Beam Dump}\/},
\href{http://dx.doi.org/10.1103/PhysRevD.38.3375}{Phys. Rev. {\bf D38} (1988)
  3375}.

\bibitem{Batell:2014mga}
B.~Batell, R.~Essig, and Z.~Surujon, {\em {Strong Constraints on Sub-GeV Dark
  Sectors from SLAC Beam Dump E137}\/},
  \href{http://dx.doi.org/10.1103/PhysRevLett.113.171802}{Phys. Rev. Lett. {\bf
  113} (2014) no.~17, 171802},
\href{http://arxiv.org/abs/1406.2698}{{\tt arXiv:1406.2698 [hep-ph]}}.

\bibitem{Bross:1989mp}
A.~Bross, M.~Crisler, S.~H. Pordes, J.~Volk, S.~Errede, and J.~Wrbanek, {\em {A
  Search for Shortlived Particles Produced in an Electron Beam Dump}\/},
\href{http://dx.doi.org/10.1103/PhysRevLett.67.2942}{Phys. Rev. Lett. {\bf 67}
  (1991)  2942--2945}.

\bibitem{Bergsma:1985qz}
{CHARM Collaboration}, F.~Bergsma et al., {\em {Search for Axion Like Particle
  Production in 400-{GeV} Proton - Copper Interactions}\/},
\href{http://dx.doi.org/10.1016/0370-2693(85)90400-9}{Phys. Lett. {\bf 157B}
  (1985)  458--462}.

\bibitem{Blumlein:1990ay}
J.~Bl{\"u}mlein et al., {\em {Limits on neutral light scalar and pseudoscalar
  particles in a proton beam dump experiment}\/},
\href{http://dx.doi.org/10.1007/BF01548556}{Z. Phys. {\bf C51} (1991)
  341--350}.

\bibitem{Lanfranchi:2018xrz}
{NA62 Collaboration}, G.~Lanfranchi, {\em {Search for exotic particles at
  NA62}\/},
\href{http://dx.doi.org/10.22323/1.330.0010}{PoS {\bf ALPS2018} (2018)  010}.

\bibitem{Ariga:2018pin}
{FASER Collaboration}, A.~Ariga et al., {\em {Technical Proposal for FASER:
  ForwArd Search ExpeRiment at the LHC}\/},
\href{http://arxiv.org/abs/1812.09139}{{\tt arXiv:1812.09139
  [physics.ins-det]}}.

\bibitem{Ilten:2015hya}
P.~Ilten et al., {\em {Dark photons from charm mesons at LHCb}\/},
  \href{http://dx.doi.org/10.1103/PhysRevD.92.115017}{Phys. Rev. {\bf D92}
  (2015) no.~11, 115017},
\href{http://arxiv.org/abs/1509.06765}{{\tt arXiv:1509.06765 [hep-ph]}}.

\bibitem{CEPCStudyGroup:2018ghi}
{CEPC Study Group Collaboration}, M.~Dong et al., {\em {CEPC Conceptual Design
  Report: Volume 2 - Physics \& Detector}\/},
\href{http://arxiv.org/abs/1811.10545}{{\tt arXiv:1811.10545 [hep-ex]}}.

\bibitem{Karliner:2015tga}
M.~Karliner et al., {\em {Radiative return capabilities of a high-energy,
  high-luminosity $e^+e^-$ collider}\/},
  \href{http://dx.doi.org/10.1103/PhysRevD.92.035010}{Phys. Rev. {\bf D92}
  (2015) no.~3, 035010},
\href{http://arxiv.org/abs/1503.07209}{{\tt arXiv:1503.07209 [hep-ph]}}.

\bibitem{donofrio2019searching}
M.~D'Onofrio, O.~Fischer, and Z.~S. Wang, {\em Searching for Dark Photons at
  the LHeC and FCC-he\/},  2019.
\newblock \href{http://arxiv.org/abs/1909.02312}{{\tt arXiv:1909.02312
  [hep-ph]}}.

\bibitem{Cohen:1987vi}
A.~Cohen and D.~Kaplan, {\em {Thermodynamic Generation of the Baryon
  Asymmetry}\/},
\href{http://dx.doi.org/10.1016/0370-2693(87)91369-4}{Phys. Lett. {\bf B199}
  (1987)  251--258}.

\bibitem{Krnjaic:2015mbs}
G.~Krnjaic, {\em {Probing Light Thermal Dark-Matter With a Higgs Portal
  Mediator}\/},  \href{http://dx.doi.org/10.1103/PhysRevD.94.073009}{Phys. Rev.
  {\bf D94} (2016) no.~7, 073009},
\href{http://arxiv.org/abs/1512.04119}{{\tt arXiv:1512.04119 [hep-ph]}}.

\bibitem{Graham:2015cka}
P.~Graham, D.~Kaplan, and S.~Rajendran, {\em {Cosmological Relaxation of the
  Electroweak Scale}\/},
  \href{http://dx.doi.org/10.1103/PhysRevLett.115.221801}{Phys. Rev. Lett. {\bf
  115} (2015) no.~22, 221801},
\href{http://arxiv.org/abs/1504.07551}{{\tt arXiv:1504.07551 [hep-ph]}}.

\bibitem{Flacke:2016szy}
T.~Flacke et al., {\em {Phenomenology of relaxion-Higgs mixing}\/},
  \href{http://dx.doi.org/10.1007/JHEP06(2017)050}{JHEP {\bf 06} (2017)  050},
\href{http://arxiv.org/abs/1610.02025}{{\tt arXiv:1610.02025 [hep-ph]}}.

\bibitem{Abel:2018fqg}
S.~Abel, R.~Gupta, and J.~Scholtz, {\em {Out-of-the-box Baryogenesis During
  Relaxation}\/},
\href{http://arxiv.org/abs/1810.05153}{{\tt arXiv:1810.05153 [hep-ph]}}.

\bibitem{Fonseca:2018xzp}
N.~Fonseca, E.~Morgante, and G.~Servant, {\em {Higgs relaxation after
  inflation}\/},  \href{http://dx.doi.org/10.1007/JHEP10(2018)020}{JHEP {\bf
  10} (2018)  020},
\href{http://arxiv.org/abs/1805.04543}{{\tt arXiv:1805.04543 [hep-ph]}}.

\bibitem{Banerjee:2018xmn}
A.~Banerjee, H.~Kim, and G.~Perez, {\em {Coherent relaxion dark matter}\/},
\href{http://arxiv.org/abs/1810.01889}{{\tt arXiv:1810.01889 [hep-ph]}}.

\bibitem{Frugiuele:2018coc}
C.~Frugiuele et al., {\em {Relaxion and light (pseudo)scalars at the HL-LHC and
  lepton colliders}\/},  \href{http://dx.doi.org/10.1007/JHEP10(2018)151}{JHEP
  {\bf 10} (2018)  151},
\href{http://arxiv.org/abs/1807.10842}{{\tt arXiv:1807.10842 [hep-ph]}}.

\bibitem{Bauer:2018uxu}
M.~Bauer et al., {\em {Axion-Like Particles at Future Colliders}\/},
  \href{http://dx.doi.org/10.1140/epjc/s10052-019-6587-9}{Eur. Phys. J. {\bf
  C79} (2019) no.~1, 74},
\href{http://arxiv.org/abs/1808.10323}{{\tt arXiv:1808.10323 [hep-ph]}}.

\bibitem{Yue:2019gbh}
C.-X. Yue, M.-Z. Liu, and Y.-C. Guo, {\em {Searching for axion-like particle at
  future $ep$ colliders}\/},
\href{http://arxiv.org/abs/1904.10657}{{\tt arXiv:1904.10657 [hep-ph]}}.

\bibitem{Antusch:2016ejd}
S.~Antusch, E.~Cazzato, and O.~Fischer, {\em {Sterile neutrino searches at
  future $e^-e^+$, $pp$, and $e^-p$ colliders}\/},
  \href{http://dx.doi.org/10.1142/S0217751X17500786}{Int. J. Mod. Phys. {\bf
  A32} (2017) no.~14, 1750078},
\href{http://arxiv.org/abs/1612.02728}{{\tt arXiv:1612.02728 [hep-ph]}}.

\bibitem{Eijima:2018qke}
S.~Eijima, M.~Shaposhnikov, and I.~Timiryasov, {\em {Parameter space of
  baryogenesis in the $\nu$MSM}\/},
\href{http://arxiv.org/abs/1808.10833}{{\tt arXiv:1808.10833 [hep-ph]}}.

\bibitem{DM-note}
{Supporting note for the Dark Matter and Dark Sectors chapter}. {CERN-ESU-012,
  available on CDS (the CERN Document System)}, January 2020.

\bibitem{bertone:2004pz}
G.~Bertone, D.~Hooper, and J.~Silk, {\em {Particle dark matter: Evidence,
  candidates and constraints}\/},
  \href{http://dx.doi.org/10.1016/j.physrep.2004.08.031}{Phys. Rept. {\bf 405}
  (2005)  279--390},
\href{http://arxiv.org/abs/hep-ph/0404175}{{\tt arXiv:hep-ph/0404175
  [hep-ph]}}.

\bibitem{Battaglieri:2017aum}
M.~Battaglieri et al., {\em {US Cosmic Visions: New Ideas in Dark Matter 2017:
  Community Report}\/},
\href{http://arxiv.org/abs/1707.04591}{{\tt arXiv:1707.04591 [hep-ph]}}.

\bibitem{Ilten:2018crw}
P.~Ilten, Y.~Soreq, M.~Williams, and W.~Xue, {\em {Serendipity in dark photon
  searches}\/},  \href{http://dx.doi.org/10.1007/JHEP06(2018)004}{JHEP {\bf 06}
  (2018)  004},
\href{http://arxiv.org/abs/1801.04847}{{\tt arXiv:1801.04847 [hep-ph]}}.

\bibitem{Bauer:2018onh}
M.~Bauer, P.~Foldenauer, and J.~Jaeckel, {\em {Hunting All the Hidden
  Photons}\/},  \href{http://dx.doi.org/10.1007/JHEP07(2018)094}{JHEP {\bf 07}
  (2018)  094},
\href{http://arxiv.org/abs/1803.05466}{{\tt arXiv:1803.05466 [hep-ph]}}.

\bibitem{Bertone:2016nfn}
G.~Bertone and D.~Hooper, {\em {History of dark matter}\/},
  \href{http://dx.doi.org/10.1103/RevModPhys.90.045002}{Rev. Mod. Phys. {\bf
  90} (2018) no.~4, 045002},
\href{http://arxiv.org/abs/1605.04909}{{\tt arXiv:1605.04909 [astro-ph.CO]}}.

\bibitem{Hu:2000ke}
W.~Hu, R.~Barkana, and A.~Gruzinov, {\em {Cold and fuzzy dark matter}\/},
  \href{http://dx.doi.org/10.1103/PhysRevLett.85.1158}{Phys. Rev. Lett. {\bf
  85} (2000)  1158--1161},
\href{http://arxiv.org/abs/astro-ph/0003365}{{\tt arXiv:astro-ph/0003365
  [astro-ph]}}.

\bibitem{Bird:2016dcv}
S.~Bird et al., {\em {Did LIGO detect dark matter?}\/},
  \href{http://dx.doi.org/10.1103/PhysRevLett.116.201301}{Phys. Rev. Lett. {\bf
  116} (2016) no.~20, 201301},
\href{http://arxiv.org/abs/1603.00464}{{\tt arXiv:1603.00464 [astro-ph.CO]}}.

\bibitem{Brandt:2016aco}
T.~D. Brandt, {\em {Constraints on MACHO Dark Matter from Compact Stellar
  Systems in Ultra-Faint Dwarf Galaxies}\/},
  \href{http://dx.doi.org/10.3847/2041-8205/824/2/L31}{Astrophys. J. {\bf 824}
  (2016) no.~2, L31},
\href{http://arxiv.org/abs/1605.03665}{{\tt arXiv:1605.03665 [astro-ph.GA]}}.

\bibitem{Nakama:2017xvq}
T.~Nakama, B.~Carr, and J.~Silk, {\em {Limits on primordial black holes from
  $\mu$ distortions in cosmic microwave background}\/},
  \href{http://dx.doi.org/10.1103/PhysRevD.97.043525}{Phys. Rev. {\bf D97}
  (2018) no.~4, 043525},
\href{http://arxiv.org/abs/1710.06945}{{\tt arXiv:1710.06945 [astro-ph.CO]}}.

\bibitem{Ali-Haimoud:2016mbv}
Y.~Ali-Ha{\"i}moud and M.~Kamionkowski, {\em {Cosmic microwave background
  limits on accreting primordial black holes}\/},
  \href{http://dx.doi.org/10.1103/PhysRevD.95.043534}{Phys. Rev. {\bf D95}
  (2017) no.~4, 043534},
\href{http://arxiv.org/abs/1612.05644}{{\tt arXiv:1612.05644 [astro-ph.CO]}}.

\bibitem{Jungman:1995df}
G.~Jungman, M.~Kamionkowski, and K.~Griest, {\em {Supersymmetric dark
  matter}\/},  \href{http://dx.doi.org/10.1016/0370-1573(95)00058-5}{Phys.
  Rept. {\bf 267} (1996)  195--373},
\href{http://arxiv.org/abs/hep-ph/9506380}{{\tt arXiv:hep-ph/9506380
  [hep-ph]}}.

\bibitem{Billard:2013qya}
J.~Billard, L.~Strigari, and E.~Figueroa-Feliciano, {\em {Implication of
  neutrino backgrounds on the reach of next generation dark matter direct
  detection experiments}\/},
  \href{http://dx.doi.org/10.1103/PhysRevD.89.023524}{Phys. Rev. {\bf D89}
  (2014) no.~2, 023524},
\href{http://arxiv.org/abs/1307.5458}{{\tt arXiv:1307.5458 [hep-ph]}}.

\bibitem{Safarzadeh:2018hhg}
M.~Safarzadeh, E.~Scannapieco, and A.~Babul, {\em {A limit on the warm dark
  matter particle mass from the redshifted 21 cm absorption line}\/},
  \href{http://dx.doi.org/10.3847/2041-8213/aac5e0}{Astrophys. J. {\bf 859}
  (2018) no.~2, L18},
\href{http://arxiv.org/abs/1803.08039}{{\tt arXiv:1803.08039 [astro-ph.CO]}}.

\bibitem{griest:1989wd}
K.~Griest and M.~Kamionkowski, {\em {Unitarity Limits on the Mass and Radius of
  Dark Matter Particles}\/},
\href{http://dx.doi.org/10.1103/PhysRevLett.64.615}{Phys. Rev. Lett. {\bf 64}
  (1990)  615}.

\bibitem{Tremaine:1979we}
S.~Tremaine and J.~E. Gunn, {\em {Dynamical Role of Light Neutral Leptons in
  Cosmology}\/},
\href{http://dx.doi.org/10.1103/PhysRevLett.42.407}{Phys. Rev. Lett. {\bf 42}
  (1979)  407--410}.

\bibitem{Marsh:2015xka}
D.~J.~E. Marsh, {\em {Axion Cosmology}\/},
  \href{http://dx.doi.org/10.1016/j.physrep.2016.06.005}{Phys. Rept. {\bf 643}
  (2016)  1--79},
\href{http://arxiv.org/abs/1510.07633}{{\tt arXiv:1510.07633 [astro-ph.CO]}}.

\bibitem{Cui:2017kg}
X.~Cui et al., \href{http://dx.doi.org/10.1103/PhysRevLett.119.181302}{{\em
  {Dark Matter Results from 54-Ton-Day Exposure of PandaX-II Experiment}\/},
  Phys. Rev. Lett. {\bf 119} (Oct., 2017)  181302}.
  \url{https://link.aps.org/doi/10.1103/PhysRevLett.119.181302}.

\bibitem{Aprile:2018ct}
E.~Aprile et al., \href{http://dx.doi.org/10.1103/PhysRevLett.121.111302}{{\em
  {Dark Matter Search Results from a One Ton-Year Exposure of XENON1T}\/},
  Phys. Rev. Lett. {\bf 121} (Sept., 2018)  111302}.
  \url{https://link.aps.org/doi/10.1103/PhysRevLett.121.111302}.

\bibitem{Amole:2017iz}
C.~Amole et al., \href{http://dx.doi.org/10.1103/PhysRevLett.118.251301}{{\em
  {Dark Matter Search Results from the PICO-60 C$_{3}$F$_{8}$ Bubble
  Chamber}\/}, Phys. Rev. Lett. {\bf 118} (June, 2017)  251301}.
  \url{http://link.aps.org/doi/10.1103/PhysRevLett.118.251301}.

\bibitem{Bernabei:2010gs}
R.~Bernabei et al.,
  \href{http://dx.doi.org/10.1140/epjc/s10052-010-1303-9}{{\em {New results
  from DAMA/LIBRA}\/}, Eur. Phys. J. C {\bf 67} (Mar., 2010)  39--49}.
  \url{http://www.springerlink.com/index/10.1140/epjc/s10052-010-1303-9}.

\bibitem{Aalseth:2014wc}
C.~E. Aalseth et al., {\em {Search for An Annual Modulation in Three Years of
  CoGeNT Dark Matter Detector Data}\/}, arXiv (Jan., 2014)  ,
  \href{http://arxiv.org/abs/1401.3295v1}{{\tt 1401.3295v1}}.
  \url{http://arxiv.org/abs/1401.3295v1}.

\bibitem{Agnese:2013hy}
R.~Agnese et al., \href{http://dx.doi.org/10.1103/PhysRevLett.111.251301}{{\em
  {Silicon Detector Dark Matter Results from the Final Exposure of CDMS II}\/},
  Phys. Rev. Lett. {\bf 111} (Dec., 2013)  251301}.
  \url{http://link.aps.org/doi/10.1103/PhysRevLett.111.251301}.

\bibitem{Angloher:2012kl}
G.~Angloher et al.,
  \href{http://dx.doi.org/10.1140/epjc/s10052-012-1971-8}{{\em {Results from
  730 kg days of the CRESST-II Dark Matter search}\/}, Eur. Phys. J. C {\bf 72}
  (Apr., 2012)  1971--22}.
  \url{http://link.springer.com/10.1140/epjc/s10052-012-1971-8}.

\bibitem{Aguilar:2014cj}
M.~Aguilar et al., \href{http://dx.doi.org/10.1103/PhysRevLett.113.121102}{{\em
  {Electron and Positron Fluxes in Primary Cosmic Rays Measured with the Alpha
  Magnetic Spectrometer on the International Space Station}\/}, Phys. Rev.
  Lett. {\bf 113} (Sept., 2014)  121102}.
  \url{http://link.aps.org/doi/10.1103/PhysRevLett.113.121102}.

\bibitem{Ajello:2016ge}
M.~Ajello et al., \href{http://dx.doi.org/10.3847/0004-637X/819/1/44}{{\em
  {Fermi-LAT Observations of High-Energy $\gamma$-ray Emission Toward the
  Galactic Center}\/}, Ap. J. {\bf 819} (Mar., 2016)  44}.
  \url{http://stacks.iop.org/0004-637X/819/i=1/a=44?key=crossref.1ad1bf72f7bfe54efdba92adda27eb57}.

\bibitem{Akerib:2017kg}
D.~S. Akerib et al.,
  \href{http://dx.doi.org/10.1103/PhysRevLett.118.021303}{{\em {Results from a
  Search for Dark Matter in the Complete LUX Exposure}\/}, Phys. Rev. Lett.
  {\bf 118} (Jan., 2017)  021303}.
  \url{http://link.aps.org/doi/10.1103/PhysRevLett.118.021303}.

\bibitem{PhysRevD.98.102006}
{DarkSide Collaboration}, P.~Agnes et al.,
  \href{http://dx.doi.org/10.1103/PhysRevD.98.102006}{{\em DarkSide-50 532-day
  dark matter search with low-radioactivity argon\/}, Phys. Rev. D {\bf 98}
  (Nov, 2018)  102006}.
  \url{https://link.aps.org/doi/10.1103/PhysRevD.98.102006}.

\bibitem{Agnes:2018fg}
P.~Agnes et al., \href{http://dx.doi.org/10.1103/PhysRevLett.121.081307}{{\em
  {Low-Mass Dark Matter Search with the DarkSide-50 Experiment}\/}, Phys. Rev.
  Lett. {\bf 121} (Aug., 2018)  081307}.
  \url{https://link.aps.org/doi/10.1103/PhysRevLett.121.081307}.

\bibitem{Nelson:2014wy}
H.~Nelson, {\em {LUX - Zeplin (LZ)}\/},  in {\em DM 2014}, For The LZ
  Collaboration.
\newblock Feb., 2014.
\newblock \url{https://hepconf.physics.ucla.edu/dm14/}.

\bibitem{Kudryavtsev:2015hy}
V.~A. Kudryavtsev, \href{http://dx.doi.org/10.1063/1.4927991}{{\em {Expected
  background in the LZ experiment}\/}, AIP Conf. Proc. {\bf 1672} (Aug., 2015)
  060003}.
  \url{http://scitation.aip.org/content/aip/proceeding/aipcp/10.1063/1.4927991}.

\bibitem{Aprile:2015wv}
E.~Aprile, {\em {XENONnT: an upgrade of XENON1T to reach 10$^{-48}$ cm$^{2}$
  sensitivity by 2022}\/},  in {\em LNGS Sci. Comm. Apr. 2015}, For The XENON
  Collaboration.
\newblock Apr., 2015.
\newblock \url{https://agenda.infn.it/conferenceDisplay.py?confId=9608}.

\bibitem{Boulay:2017tn}
M.~G. Boulay, {\em {Argon Dark Matter Searches: DarkSide-20K and Beyond}\/},
  in {\em New Ideas DM 2017}, For the DarkSide Collaboration.
\newblock Mar., 2017.
\newblock \url{https://indico.fnal.gov/conferenceDisplay.py?confId=13702}.

\bibitem{Agnese:2017fn}
R.~Agnese et al., \href{http://dx.doi.org/10.1103/PhysRevD.95.082002}{{\em
  {Projected sensitivity of the SuperCDMS SNOLAB experiment}\/}, Phys. Rev. D
  {\bf 95} (Apr., 2017)  215}.
  \url{http://link.aps.org/doi/10.1103/PhysRevD.95.082002}.

\bibitem{Billard:2014cx}
J.~Billard, E.~Figueroa-Feliciano, and L.~Strigari,
  \href{http://dx.doi.org/10.1103/PhysRevD.89.023524}{{\em {Implication of
  neutrino backgrounds on the reach of next generation dark matter direct
  detection experiments}\/}, Phys. Rev. D {\bf 89} (Jan., 2014)  023524}.
  \url{http://link.aps.org/doi/10.1103/PhysRevD.89.023524}.

\bibitem{dobsonLZ}
{J. Dobson (For the LZ Collaboration)} {Presentation at DM 2018} (2018)  .

\bibitem{grandiXENON}
{L. Grandi (For the XENON Collaboration)} {Presentation at DM 2018} (2018)  .

\bibitem{sigmaSD-prospects}
{PICO Collaboration}, K.~Kraus et al., {\em {Topics in Astroparticle and
  Underground Physics 2017 Presentation}\/}, .
  \url{https://indico.cern.ch/event/606690/contributions/2625263/attachments/1498237/2332347/PICO_TAUP_Sudbury_C_B_Krauss_July_2017.pdf}.

\bibitem{Albert2017-hn}
A.~Albert et al., {\em Searching for Dark Matter Annihilation in Recently
  Discovered Milky Way Satellites with {Fermi-LAT}\/},  Astrophys. J. {\bf 834}
  (2017) no.~2, 110. \url{http://dx.doi.org/10.3847/1538-4357/834/2/110}.

\bibitem{Abdallah:2016ygi}
{H.E.S.S. Collaboration}, H.~Abdallah et al., {\em {Search for dark matter
  annihilations towards the inner Galactic halo from 10 years of observations
  with H.E.S.S}\/},
  \href{http://dx.doi.org/10.1103/PhysRevLett.117.111301}{Phys. Rev. Lett. {\bf
  117} (2016) no.~11, 111301},
\href{http://arxiv.org/abs/1607.08142}{{\tt arXiv:1607.08142 [astro-ph.HE]}}.

\bibitem{Drlica-Wagner:2019xan}
{LSST Dark Matter Group Collaboration}, A.~Drlica-Wagner et al., {\em {Probing
  the Fundamental Nature of Dark Matter with the Large Synoptic Survey
  Telescope}\/},
\href{http://arxiv.org/abs/1902.01055}{{\tt arXiv:1902.01055 [astro-ph.CO]}}.

\bibitem{Ibe:2015tma}
M.~Ibe, S.~Matsumoto, S.~Shirai, and T.~T. Yanagida, {\em {Wino Dark Matter in
  light of the AMS-02 2015 Data}\/},
  \href{http://dx.doi.org/10.1103/PhysRevD.91.111701}{Phys. Rev. {\bf D91}
  (2015) no.~11, 111701},
\href{http://arxiv.org/abs/1504.05554}{{\tt arXiv:1504.05554 [hep-ph]}}.

\bibitem{Ackermann:2017gx}
M.~Ackermann et al., \href{http://dx.doi.org/10.3847/1538-4357/aa6cab}{{\em
  {The \emph{Fermi} Galactic Center GeV Excess and Implications for Dark
  Matter}\/}, Ap. J. {\bf 840} (May, 2017)  43}.

\bibitem{Calore:2015bsx}
F.~Calore, M.~Di~Mauro, F.~Donato, J.~W.~T. Hessels, and C.~Weniger, {\em
  {Radio detection prospects for a bulge population of millisecond pulsars as
  suggested by Fermi LAT observations of the inner Galaxy}\/},
  \href{http://dx.doi.org/10.3847/0004-637X/827/2/143}{Astrophys. J. {\bf 827}
  (2016) no.~2, 143},
\href{http://arxiv.org/abs/1512.06825}{{\tt arXiv:1512.06825 [astro-ph.HE]}}.

\bibitem{Carena:2019pwq}
M.~Carena, J.~Osborne, N.~R. Shah, and C.~E.~M. Wagner, {\em {The Return of the
  WIMP: Missing Energy Signals and the Galactic Center Excess}\/},
\href{http://arxiv.org/abs/1905.03768}{{\tt arXiv:1905.03768 [hep-ph]}}.

\bibitem{Kim2018-yk}
Y.~G. Kim, C.~B. Park, and S.~Shin, {\em Collider probes of singlet fermionic
  dark matter scenarios for the Fermi gamma-ray excess\/},
  \href{http://arxiv.org/abs/1809.01143}{{\tt arXiv:1809.01143 [hep-ph]}}.
  \url{http://arxiv.org/abs/1809.01143}.

\bibitem{Giesen:2015ufa}
G.~Giesen et al., {\em {AMS-02 antiprotons, at last! Secondary astrophysical
  component and immediate implications for Dark Matter}\/},
  \href{http://dx.doi.org/10.1088/1475-7516/2015/09/023,
  10.1088/1475-7516/2015/9/023}{JCAP {\bf 1509} (2015) no.~09, 023},
\href{http://arxiv.org/abs/1504.04276}{{\tt arXiv:1504.04276 [astro-ph.HE]}}.

\bibitem{Aramaki2016-to}
T.~Aramaki et al., {\em Antideuteron Sensitivity for the {GAPS} Experiment\/},
  Astropart. Phys. {\bf 74} (2016)  6--13.
  \url{http://dx.doi.org/10.1016/j.astropartphys.2015.09.001}.

\bibitem{Drlica-Wagner2019-nr}
A.~Drlica-Wagner and {Others}, {\em Probing the Fundamental Nature of Dark
  Matter with the Large Synoptic Survey Telescope\/},
  \href{http://arxiv.org/abs/1902.01055}{{\tt arXiv:1902.01055 [astro-ph.CO]}}.
  \url{http://arxiv.org/abs/1902.01055}.

\bibitem{Carr2016-vl}
J.~Carr et al., {\em Prospects for Indirect Dark Matter Searches with the
  Cherenkov Telescope Array ({CTA})\/},  PoS {\bf ICRC2015} (2016)  1203.
  \url{http://arxiv.org/abs/1508.06128}.

\bibitem{Adrian-Martinez2016-yb}
S.~Adrian-Martinez et al., {\em Limits on Dark Matter Annihilation in the Sun
  using the {ANTARES} Neutrino Telescope\/},  Phys. Lett. {\bf B759} (2016)
  69--74. \url{http://dx.doi.org/10.1016/j.physletb.2016.05.019}.

\bibitem{IceCube_Collaboration2016-fp}
{IceCube Collaboration}, {\em Search for annihilating dark matter in the Sun
  with 3 years of {IceCube} data\/},
  \href{http://arxiv.org/abs/1612.05949}{{\tt arXiv:1612.05949 [astro-ph.HE]}}.
  \url{http://arxiv.org/abs/1612.05949}.

\bibitem{Patt:2006fw}
B.~Patt and F.~Wilczek, {\em {Higgs-field portal into hidden sectors}\/},
\href{http://arxiv.org/abs/hep-ph/0605188}{{\tt arXiv:hep-ph/0605188
  [hep-ph]}}.

\bibitem{Djouadi:2011aa}
A.~Djouadi et al., {\em {Implications of LHC searches for Higgs--portal dark
  matter}\/},  \href{http://dx.doi.org/10.1016/j.physletb.2012.01.062}{Phys.
  Lett. {\bf B709} (2012)  65--69},
\href{http://arxiv.org/abs/1112.3299}{{\tt arXiv:1112.3299 [hep-ph]}}.

\bibitem{Beneke:2016ync}
M.~Beneke et al., {\em {Relic density of wino-like dark matter in the MSSM}\/},
   \href{http://dx.doi.org/10.1007/JHEP03(2016)119}{JHEP {\bf 03} (2016)  119},
\href{http://arxiv.org/abs/1601.04718}{{\tt arXiv:1601.04718 [hep-ph]}}.

\bibitem{Cohen2013-ah}
T.~Cohen, M.~Lisanti, A.~Pierce, and T.~R. Slatyer, {\em Wino Dark Matter Under
  Siege\/},  \href{http://arxiv.org/abs/1307.4082}{{\tt arXiv:1307.4082
  [hep-ph]}}. \url{http://arxiv.org/abs/1307.4082}.

\bibitem{Cabrera-Catalan2015-os}
M.~E. Cabrera-Catalan, S.~Ando, C.~Weniger, and F.~Zandanel, {\em Indirect and
  direct detection prospect for {TeV} dark matter in the nine parameter
  {MSSM}\/},  Phys. Rev. {\bf D92} (2015) no.~3, 035018.
  \url{http://dx.doi.org/10.1103/PhysRevD.92.035018}.

\bibitem{Rinchiuso:2018ajn}
L.~Rinchiuso, N.~L. Rodd, I.~Moult, E.~Moulin, M.~Baumgart, T.~Cohen, T.~R.
  Slatyer, I.~W. Stewart, and V.~Vaidya, {\em {Hunting for Heavy Winos in the
  Galactic Center}\/},
  \href{http://dx.doi.org/10.1103/PhysRevD.98.123014}{Phys. Rev. {\bf D98}
  (2018) no.~12, 123014},
\href{http://arxiv.org/abs/1808.04388}{{\tt arXiv:1808.04388 [astro-ph.HE]}}.

\bibitem{Silverwood2015-ff}
H.~Silverwood, C.~Weniger, P.~Scott, and G.~Bertone, {\em A realistic
  assessment of the {CTA} sensitivity to dark matter annihilation\/},  JCAP
  {\bf 1503} (2015) no.~03, 055.
  \url{http://dx.doi.org/10.1088/1475-7516/2015/03/055}.

\bibitem{Fox:2011pm}
P.~J. Fox et al., {\em {Missing Energy Signatures of Dark Matter at the
  LHC}\/},  \href{http://dx.doi.org/10.1103/PhysRevD.85.056011}{Phys. Rev. {\bf
  D85} (2012)  056011},
\href{http://arxiv.org/abs/1109.4398}{{\tt arXiv:1109.4398 [hep-ph]}}.

\bibitem{Hoferichter:2017olk}
M.~Hoferichter, P.~Klos, J.~Menendez, and A.~Schwenk, {\em {Improved limits for
  Higgs-portal dark matter from LHC searches}\/},
  \href{http://dx.doi.org/10.1103/PhysRevLett.119.181803}{Phys. Rev. Lett. {\bf
  119} (2017) no.~18, 181803},
\href{http://arxiv.org/abs/1708.02245}{{\tt arXiv:1708.02245 [hep-ph]}}.

\bibitem{DM_IndirectDetectionPlots}
B.~Gao, E.~Tolley, A.~Boveia, and L.~Carpenter. {Private Communication}, 2019.

\bibitem{DM_DirectDetectionPlots}
I.~John, M.~Rinoldi, F.~Ungaro, and U.~Schnoor. {Private Communication}, 2019.

\bibitem{Boveia:2016mrp}
G.~Busoni et al., {\em {Recommendations on presenting LHC searches for missing
  transverse energy signals using simplified $s$-channel models of dark
  matter}\/},
\href{http://arxiv.org/abs/1603.04156}{{\tt arXiv:1603.04156 [hep-ex]}}.

\bibitem{Carpenter:2016thc}
L.~Carpenter et al., {\em {Indirect Detection Constraints on s and t Channel
  Simplified Models of Dark Matter}\/},
  \href{http://dx.doi.org/10.1103/PhysRevD.94.055027}{Phys. Rev. {\bf D94}
  (2016) no.~5, 055027},
\href{http://arxiv.org/abs/1606.04138}{{\tt arXiv:1606.04138 [hep-ph]}}.

\bibitem{Davidson:2000hf}
S.~Davidson, S.~Hannestad, and G.~Raffelt, {\em {Updated bounds on millicharged
  particles}\/},  \href{http://dx.doi.org/10.1088/1126-6708/2000/05/003}{JHEP
  {\bf 05} (2000)  003},
\href{http://arxiv.org/abs/hep-ph/0001179}{{\tt arXiv:hep-ph/0001179
  [hep-ph]}}.

\bibitem{Berlin:2018sjs}
A.~Berlin, D.~Hooper, G.~Krnjaic, and S.~D. McDermott, {\em {Severely
  Constraining Dark Matter Interpretations of the 21-cm Anomaly}\/},
  \href{http://dx.doi.org/10.1103/PhysRevLett.121.011102}{Phys. Rev. Lett. {\bf
  121} (2018) no.~1, 011102},
\href{http://arxiv.org/abs/1803.02804}{{\tt arXiv:1803.02804 [hep-ph]}}.

\bibitem{Izaguirre:2015yja}
E.~Izaguirre, G.~Krnjaic, P.~Schuster, and N.~Toro, {\em {Analyzing the
  Discovery Potential for Light Dark Matter}\/},
  \href{http://dx.doi.org/10.1103/PhysRevLett.115.251301}{Phys. Rev. Lett. {\bf
  115} (2015) no.~25, 251301},
\href{http://arxiv.org/abs/1505.00011}{{\tt arXiv:1505.00011 [hep-ph]}}.

\bibitem{DAgnolo:2015ujb}
R.~T. D'Agnolo and J.~T. Ruderman, {\em {Light Dark Matter from Forbidden
  Channels}\/},  \href{http://dx.doi.org/10.1103/PhysRevLett.115.061301}{Phys.
  Rev. Lett. {\bf 115} (2015) no.~6, 061301},
\href{http://arxiv.org/abs/1505.07107}{{\tt arXiv:1505.07107 [hep-ph]}}.

\bibitem{deNiverville:2011it}
P.~deNiverville, M.~Pospelov, and A.~Ritz, {\em {Observing a light dark matter
  beam with neutrino experiments}\/},
  \href{http://dx.doi.org/10.1103/PhysRevD.84.075020}{Phys.Rev. {\bf D84}
  (2011)  075020}, \href{http://arxiv.org/abs/1107.4580}{{\tt arXiv:1107.4580
  [hep-ph]}}.

\bibitem{Izaguirre:2013uxa}
E.~Izaguirre, G.~Krnjaic, P.~Schuster, and N.~Toro, {\em {New Electron
  Beam-Dump Experiments to Search for MeV to few-GeV Dark Matter}\/},
  \href{http://dx.doi.org/10.1103/PhysRevD.88.114015}{Phys. Rev. {\bf D88}
  (2013)  114015},
\href{http://arxiv.org/abs/1307.6554}{{\tt arXiv:1307.6554 [hep-ph]}}.

\bibitem{Izaguirre:2014bca}
E.~Izaguirre, G.~Krnjaic, P.~Schuster, and N.~Toro, {\em {Testing GeV-Scale
  Dark Matter with Fixed-Target Missing Momentum Experiments}\/},
  \href{http://dx.doi.org/10.1103/PhysRevD.91.094026}{Phys. Rev. {\bf D91}
  (2015) no.~9, 094026},
\href{http://arxiv.org/abs/1411.1404}{{\tt arXiv:1411.1404 [hep-ph]}}.

\bibitem{Gninenko:2014pea}
S.~N. Gninenko, N.~V. Krasnikov, and V.~A. Matveev, {\em {Muon g-2 and searches
  for a new leptophobic sub-GeV dark boson in a missing-energy experiment at
  CERN}\/},  \href{http://dx.doi.org/10.1103/PhysRevD.91.095015}{Phys. Rev.
  {\bf D91} (2015)  095015},
\href{http://arxiv.org/abs/1412.1400}{{\tt arXiv:1412.1400 [hep-ph]}}.

\bibitem{Kahn:2018cqs}
Y.~Kahn, G.~Krnjaic, N.~Tran, and A.~Whitbeck, {\em {M$^{3}$: a new muon
  missing momentum experiment to probe $(g - 2)_\mu$ and dark matter at
  Fermilab}\/},  \href{http://dx.doi.org/10.1007/JHEP09(2018)153}{JHEP {\bf 09}
  (2018)  153},
\href{http://arxiv.org/abs/1804.03144}{{\tt arXiv:1804.03144 [hep-ph]}}.

\bibitem{Berlin:2018bsc}
A.~Berlin, N.~Blinov, G.~Krnjaic, P.~Schuster, and N.~Toro, {\em {Dark Matter,
  Millicharges, Axion and Scalar Particles, Gauge Bosons, and Other New Physics
  with LDMX}\/},  \href{http://dx.doi.org/10.1103/PhysRevD.99.075001}{Phys.
  Rev. {\bf D99} (2019) no.~7, 075001},
\href{http://arxiv.org/abs/1807.01730}{{\tt arXiv:1807.01730 [hep-ph]}}.

\bibitem{Aguilar-Arevalo:2018wea}
{MiniBooNE DM Collaboration}, A.~A. Aguilar-Arevalo et al., {\em {Dark Matter
  Search in Nucleon, Pion, and Electron Channels from a Proton Beam Dump with
  MiniBooNE}\/},  \href{http://dx.doi.org/10.1103/PhysRevD.98.112004}{Phys.
  Rev. {\bf D98} (2018) no.~11, 112004},
\href{http://arxiv.org/abs/1807.06137}{{\tt arXiv:1807.06137 [hep-ex]}}.

\bibitem{Banerjee:2017hhz}
{NA64 Collaboration}, D.~Banerjee et al., {\em {Search for vector mediator of
  Dark Matter production in invisible decay mode}\/},
  \href{http://dx.doi.org/10.1103/PhysRevD.97.072002}{Phys. Rev. {\bf D97}
  (2018) no.~7, 072002},
\href{http://arxiv.org/abs/1710.00971}{{\tt arXiv:1710.00971 [hep-ex]}}.

\bibitem{CortinaGil:2019nuo}
{NA62 Collaboration}, E.~Cortina~Gil et al., {\em {Search for production of an
  invisible dark photon in $\pi^0$ decays}\/},
  \href{http://dx.doi.org/10.1007/JHEP05(2019)182}{JHEP {\bf 05} (2019)  182},
\href{http://arxiv.org/abs/1903.08767}{{\tt arXiv:1903.08767 [hep-ex]}}.

\bibitem{Ahdida:2650896}
C.~C. Ahdida, M.~Calviani, B.~Goddard, R.~Jacobsson, and M.~Lamont, {\em {SPS
  Beam Dump Facility Comprehensive Design Study}\/},
  CERN-PBC-REPORT-2018-001, CERN, Geneva, Dec, 2018.
\newblock \url{https://cds.cern.ch/record/2650896}.

\bibitem{Ahdida:2654870}
{SHiP Collaboration}, C.~Ahdida et al., {\em {SHiP Experiment - Progress
  Report}\/},   CERN-SPSC-2019-010. SPSC-SR-248, CERN, Geneva, Jan, 2019.
\newblock \url{https://cds.cern.ch/record/2654870}.

\bibitem{Akesson:2018yrp}
T.~{\AA}kesson, Y.~Dutheil, L.~Evans, A.~Grudiev, Y.~Papaphilippou, and
  S.~Stapnes, {\em {A primary electron beam facility at CERN}\/},
\href{http://arxiv.org/abs/1805.12379}{{\tt arXiv:1805.12379
  [physics.acc-ph]}}.

\bibitem{Essig:2011nj}
R.~Essig, J.~Mardon, and T.~Volansky, {\em {Direct Detection of Sub-GeV Dark
  Matter}\/},  \href{http://dx.doi.org/10.1103/PhysRevD.85.076007}{Phys. Rev.
  {\bf D85} (2012)  076007},
\href{http://arxiv.org/abs/1108.5383}{{\tt arXiv:1108.5383 [hep-ph]}}.

\bibitem{Essig:2015cda}
R.~Essig et al., {\em {Direct Detection of sub-GeV Dark Matter with
  Semiconductor Targets}\/},
  \href{http://dx.doi.org/10.1007/JHEP05(2016)046}{JHEP {\bf 05} (2016)  046},
\href{http://arxiv.org/abs/1509.01598}{{\tt arXiv:1509.01598 [hep-ph]}}.

\bibitem{Hochberg:2016ntt}
Y.~Hochberg et al., {\em {Directional detection of dark matter with
  two-dimensional targets}\/},
  \href{http://dx.doi.org/10.1016/j.physletb.2017.06.051}{Phys. Lett. {\bf
  B772} (2017)  239--246},
\href{http://arxiv.org/abs/1606.08849}{{\tt arXiv:1606.08849 [hep-ph]}}.

\bibitem{Hochberg:2015fth}
Y.~Hochberg, M.~Pyle, Y.~Zhao, and K.~M. Zurek, {\em {Detecting Superlight Dark
  Matter with Fermi-Degenerate Materials}\/},
  \href{http://dx.doi.org/10.1007/JHEP08(2016)057}{JHEP {\bf 08} (2016)  057},
\href{http://arxiv.org/abs/1512.04533}{{\tt arXiv:1512.04533 [hep-ph]}}.

\bibitem{Tiffenberg:2017aac}
{SENSEI Collaboration}, J.~Tiffenberg et al., {\em {Single-electron and
  single-photon sensitivity with a silicon Skipper CCD}\/},
  \href{http://dx.doi.org/10.1103/PhysRevLett.119.131802}{Phys. Rev. Lett. {\bf
  119} (2017) no.~13, 131802},
\href{http://arxiv.org/abs/1706.00028}{{\tt arXiv:1706.00028
  [physics.ins-det]}}.

\bibitem{Aguilar-Arevalo:2019wdi}
A.~Aguilar-Arevalo et al., {\em {Constraints on Light Dark Matter Particles
  Interacting with Electrons from DAMIC at SNOLAB}\/},
\href{http://arxiv.org/abs/1907.12628}{{\tt arXiv:1907.12628 [astro-ph.CO]}}.

\bibitem{Petricca:2017zdp}
{CRESST Collaboration}, F.~Petricca et al., {\em {First results on low-mass
  dark matter from the CRESST-III experiment}\/},  in {\em {15th International
  Conference on Topics in Astroparticle and Underground Physics (TAUP 2017)
  Sudbury, Ontario, Canada, July 24-28, 2017}}.
\newblock 2017.
\newblock
\href{http://arxiv.org/abs/1711.07692}{{\tt arXiv:1711.07692 [astro-ph.CO]}}.
\newblock

\bibitem{Arina2018-gl}
C.~Arina, {\em Impact of cosmological and astrophysical constraints on dark
  matter simplified models\/},  Front. Astron. Space Sci. {\bf 5} (2018)  30.
  \url{http://dx.doi.org/10.3389/fspas.2018.00030}.

\bibitem{Moiseev2018-kg}
A.~Moiseev et al., {\em {All-Sky} Medium Energy Gamma-ray Observatory
  ({AMEGO})\/},  PoS {\bf ICRC2017} (2018)  798.
  \url{http://dx.doi.org/10.22323/1.301.0798}.

\bibitem{Kumar2018-be}
J.~Kumar, {\em Indirect detection of {sub-GeV} dark matter coupling to
  quarks\/}, Phys. Rev. D {\bf 98} (Dec., 2018)  116009.
  \url{https://link.aps.org/doi/10.1103/PhysRevD.98.116009}.

\bibitem{Bulbul2014-on}
E.~Bulbul et al., {\em Detection of An Unidentified Emission Line in the
  Stacked X-ray spectrum of Galaxy Clusters\/},  Astrophys. J. {\bf 789} (2014)
   13. \url{http://dx.doi.org/10.1088/0004-637X/789/1/13}.

\bibitem{Boyarsky2014-gy}
A.~Boyarsky, O.~Ruchayskiy, D.~Iakubovskyi, and J.~Franse, {\em Unidentified
  line in X-ray spectra of the Andromeda galaxy and Perseus galaxy cluster\/},
  Phys. Rev. Lett. {\bf 113} (Dec., 2014)  251301.
  \url{http://dx.doi.org/10.1103/PhysRevLett.113.251301}.

\bibitem{Siemko:2652165}
A.~Siemko, B.~Dobrich, G.~Cantatore, D.~Delikaris, L.~Mapelli, G.~Cavoto,
  P.~Pugnat, J.~Schaffran, P.~Spagnolo, H.~Ten~Kate, and G.~Zavattini, {\em
  {PBC technology subgroup report}\/},   CERN-PBC-REPORT-2018-006, CERN,
  Geneva, Dec, 2018.
\newblock \url{https://cds.cern.ch/record/2652165}.

\bibitem{Agrawal:2017ksf}
P.~Agrawal and K.~Howe, {\em {Factoring the Strong CP Problem}\/},
  \href{http://dx.doi.org/10.1007/JHEP12(2018)029}{JHEP {\bf 12} (2018)  029},
\href{http://arxiv.org/abs/1710.04213}{{\tt arXiv:1710.04213 [hep-ph]}}.

\bibitem{Gaillard:2018xgk}
M.~K. Gaillard, M.~B. Gavela, R.~Houtz, P.~Quilez, and R.~Del~Rey, {\em {Color
  unified dynamical axion}\/},
  \href{http://dx.doi.org/10.1140/epjc/s10052-018-6396-6}{Eur. Phys. J. {\bf
  C78} (2018) no.~11, 972},
\href{http://arxiv.org/abs/1805.06465}{{\tt arXiv:1805.06465 [hep-ph]}}.

\bibitem{Irastorza:2018dyq}
I.~G. Irastorza and J.~Redondo, {\em {New experimental approaches in the search
  for axion-like particles}\/},
  \href{http://dx.doi.org/10.1016/j.ppnp.2018.05.003}{Prog. Part. Nucl. Phys.
  {\bf 102} (2018)  89--159},
\href{http://arxiv.org/abs/1801.08127}{{\tt arXiv:1801.08127 [hep-ph]}}.

\bibitem{Sikivie:1983ip}
P.~Sikivie, {\em {Experimental Tests of the Invisible Axion}\/},
\href{http://dx.doi.org/10.1103/PhysRevLett.51.1415,
  10.1103/PhysRevLett.52.695.2}{Phys. Rev. Lett. {\bf 51} (1983)  1415--1417}.

\bibitem{Du:2018uak}
{ADMX Collaboration}, N.~Du et al., {\em {A Search for Invisible Axion Dark
  Matter with the Axion Dark Matter Experiment}\/},
  \href{http://dx.doi.org/10.1103/PhysRevLett.120.151301}{Phys. Rev. Lett. {\bf
  120} (2018) no.~15, 151301},
\href{http://arxiv.org/abs/1804.05750}{{\tt arXiv:1804.05750 [hep-ex]}}.

\bibitem{Zhong:2018rsr}
{HAYSTAC Collaboration}, L.~Zhong et al., {\em {Results from phase 1 of the
  HAYSTAC microwave cavity axion experiment}\/},
  \href{http://dx.doi.org/10.1103/PhysRevD.97.092001}{Phys. Rev. {\bf D97}
  (2018) no.~9, 092001},
\href{http://arxiv.org/abs/1803.03690}{{\tt arXiv:1803.03690 [hep-ex]}}.

\bibitem{Alesini:2019ajt}
D.~Alesini et al., {\em {Galactic axions search with a superconducting resonant
  cavity}\/},  \href{http://dx.doi.org/10.1103/PhysRevD.99.101101}{Phys. Rev.
  {\bf D99} (2019) no.~10, 101101},
\href{http://arxiv.org/abs/1903.06547}{{\tt arXiv:1903.06547
  [physics.ins-det]}}.

\bibitem{Melcon:2018dba}
A.~A. Melcon et al., {\em {Axion Searches with Microwave Filters: the RADES
  project}\/},  \href{http://dx.doi.org/10.1088/1475-7516/2018/05/040}{JCAP
  {\bf 1805} (2018) no.~05, 040},
\href{http://arxiv.org/abs/1803.01243}{{\tt arXiv:1803.01243 [hep-ex]}}.

\bibitem{Miceli:2015xas}
L.~Miceli, \href{http://dx.doi.org/10.3204/DESY-PROC-2015-02/miceli_lino}{{\em
  {Haloscope axion searches with the cast dipole magnet: the CAST-CAPP/IBS
  detector}\/}, } in {\em {Proceedings, 11th Patras Workshop on Axions, WIMPs
  and WISPs (Axion-WIMP 2015): Zaragoza, Spain, June 22-26, 2015}},
  pp.~164--168.
\newblock
2015.
\newblock

\bibitem{Gatti:2018ojx}
C.~Gatti et al., {\em {The Klash Proposal: Status and Perspectives}\/},  in
  {\em {14th Patras Workshop on Axions, WIMPs and WISPs (AXION-WIMP 2018)
  (PATRAS 2018) Hamburg, Germany, June 18-22, 2018}}.
\newblock 2018.
\newblock
\href{http://arxiv.org/abs/1811.06754}{{\tt arXiv:1811.06754
  [physics.ins-det]}}.
\newblock

\bibitem{Majorovits:2017ppy}
{MADMAX interest Group Collaboration}, B.~Majorovits et al., {\em {MADMAX: A
  new road to axion dark matter detection}\/},  in {\em {15th International
  Conference on Topics in Astroparticle and Underground Physics (TAUP 2017)
  Sudbury, Ontario, Canada, July 24-28, 2017}}.
\newblock 2017.
\newblock
\href{http://arxiv.org/abs/1712.01062}{{\tt arXiv:1712.01062
  [physics.ins-det]}}.
\newblock

\bibitem{Armengaud:2019uso}
{IAXO Collaboration}, E.~Armengaud et al., {\em {Physics potential of the
  International Axion Observatory (IAXO)}\/},
  \href{http://dx.doi.org/10.1088/1475-7516/2019/06/047}{JCAP {\bf 1906} (2019)
  no.~06, 047},
\href{http://arxiv.org/abs/1904.09155}{{\tt arXiv:1904.09155 [hep-ph]}}.

\bibitem{Bahre:2013ywa}
R.~Bahre et al., {\em {Any light particle search II Technical Design
  Report}\/},  \href{http://dx.doi.org/10.1088/1748-0221/8/09/T09001}{JINST
  {\bf 8} (2013)  T09001},
\href{http://arxiv.org/abs/1302.5647}{{\tt arXiv:1302.5647 [physics.ins-det]}}.

\bibitem{Pshirkov:2007st}
M.~S. Pshirkov and S.~B. Popov, {\em {Conversion of Dark matter axions to
  photons in magnetospheres of neutron stars}\/},
  \href{http://dx.doi.org/10.1134/S1063776109030030}{J. Exp. Theor. Phys. {\bf
  108} (2009)  384--388},
\href{http://arxiv.org/abs/0711.1264}{{\tt arXiv:0711.1264 [astro-ph]}}.

\bibitem{Huang:2018lxq}
F.~P. Huang, K.~Kadota, T.~Sekiguchi, and H.~Tashiro, {\em {Radio telescope
  search for the resonant conversion of cold dark matter axions from the
  magnetized astrophysical sources}\/},
  \href{http://dx.doi.org/10.1103/PhysRevD.97.123001}{Phys. Rev. {\bf D97}
  (2018) no.~12, 123001},
\href{http://arxiv.org/abs/1803.08230}{{\tt arXiv:1803.08230 [hep-ph]}}.

\bibitem{Hook:2018iia}
A.~Hook, Y.~Kahn, B.~R. Safdi, and Z.~Sun, {\em {Radio Signals from Axion Dark
  Matter Conversion in Neutron Star Magnetospheres}\/},
  \href{http://dx.doi.org/10.1103/PhysRevLett.121.241102}{Phys. Rev. Lett. {\bf
  121} (2018) no.~24, 241102},
\href{http://arxiv.org/abs/1804.03145}{{\tt arXiv:1804.03145 [hep-ph]}}.

\bibitem{Safdi:2018oeu}
B.~R. Safdi, Z.~Sun, and A.~Y. Chen, {\em {Detecting Axion Dark Matter with
  Radio Lines from Neutron Star Populations}\/},
  \href{http://dx.doi.org/10.1103/PhysRevD.99.123021}{Phys. Rev. {\bf D99}
  (2019) no.~12, 123021},
\href{http://arxiv.org/abs/1811.01020}{{\tt arXiv:1811.01020 [astro-ph.CO]}}.

\bibitem{Abe:2018owy}
{XMASS Collaboration}, K.~Abe et al., {\em {Search for dark matter in the form
  of hidden photons and axion-like particles in the XMASS detector}\/},
  \href{http://dx.doi.org/10.1016/j.physletb.2018.10.050}{Phys. Lett. {\bf
  B787} (2018)  153--158},
\href{http://arxiv.org/abs/1807.08516}{{\tt arXiv:1807.08516 [astro-ph.CO]}}.

\bibitem{Bordry:2018gri}
F.~Bordry, M.~Benedikt, O.~Bruning, J.~Jowett, L.~Rossi, D.~Schulte,
  S.~Stapnes, and F.~Zimmermann, {\em {Machine Parameters and Projected
  Luminosity Performance of Proposed Future Colliders at CERN}\/},
  \href{http://arxiv.org/abs/1810.13022}{{\tt arXiv:1810.13022
  [physics.acc-ph]}}.

\bibitem{Mangano:FCC2018}
{FCC Collaboration}, M.~Mangano et al., {\em {FCC Physics Opportunities}\/},
  \href{http://dx.doi.org/10.1140/epjc/s10052-019-6904-3}{Eur. Phys. J. {\bf
  C79} (2019) no.~6, 474}.

\bibitem{Roloff:2018dqu}
{CLIC, CLICdp Collaboration}, P.~Roloff, R.~Franceschini, U.~Schnoor, and
  A.~Wulzer, {\em {The Compact Linear e$^+$e$^-$ Collider (CLIC): Physics
  Potential}\/},  \href{http://arxiv.org/abs/1812.07986}{{\tt arXiv:1812.07986
  [hep-ex]}}.

\bibitem{pep2}
J.~Seeman, {\em {Last Year of PEP-II B-Factory Operation}\/},  Conf. Proc. {\bf
  C0806233} (2008)  TUXG01.

\bibitem{kekb}
K.~Oide, {\em {KEKB B-factory, the luminosity frontier}\/},
  \href{http://dx.doi.org/10.1143/PTP.122.69}{Prog. Theor. Phys. {\bf 122}
  (2009)  69--80}.

\bibitem{kekb2}
T.~Abe et al., {\em {Achievements of KEKB}\/},
  \href{http://dx.doi.org/10.1093/ptep/pts102}{PTEP {\bf 2013} (2013)  03A001}.

\bibitem{lhcdesign}
O.~S. Bruning, P.~Collier, P.~Lebrun, S.~Myers, R.~Ostojic, J.~Poole, and
  P.~Proudlock, \href{http://dx.doi.org/10.5170/CERN-2004-003-V-1}{{\em {LHC
  Design Report}}}.
\newblock CERN Yellow Reports: Monographs. CERN, Geneva, 2004.
\newblock \url{https://cds.cern.ch/record/782076}.

\bibitem{lhcrun2}
J.~Wenninger, {\em {Operation and Configuration of the LHC in Run 2}\/}, .
  \url{https://cds.cern.ch/record/2668326}.

\bibitem{fccee}
M.~Benedikt et al., {\em {FCC-ee: The Lepton Collider}\/}, The European
  Physical Journal Special Topics {\bf 228} (Jun, 2019)  261--623.
  \url{https://doi.org/10.1140/epjst/e2019-900045-4}.

\bibitem{cepc}
{\relax CEPC Study Group}, {\em CEPC Conceptual Design Report: Volume 1 -
  Accelerator\/},  \href{http://arxiv.org/abs/1809.00285}{{\tt arXiv:1809.00285
  [physics.acc-ph]}}.

\bibitem{bnl-erl}
V.~N. Litvinenko, T.~Roser, and M.~Chamizo~Llatas, {\em {Future High Energy
  Circular e+e- Collider using Energy-Recovery Linacs}\/},  {Input to the
  ESPPU, received after the Open Symposium in Granada} (2019)  ,
  \href{http://arxiv.org/abs/1909.04437}{{\tt arXiv:1909.04437
  [physics.acc-ph]}}.

\bibitem{Latina:2687090}
A.~Latina, D.~Schulte, and S.~Stapnes, {\em {CLIC study update August 2019}\/},
    CERN-ACC-2019-0051. CLIC-Note-1143, CERN, Geneva, Aug, 2019.
\newblock \url{http://cds.cern.ch/record/2687090}.

\bibitem{Yokoya:2019rhx}
K.~Yokoya, K.~Kubo, and T.~Okugi, {\em {Operation of ILC250 at the Z-pole}\/},
\href{http://arxiv.org/abs/1908.08212}{{\tt arXiv:1908.08212
  [physics.acc-ph]}}.

\bibitem{ctf3}
P.~K. Skowronski and R.~Corsini, {\em {Lessons from CTF3}\/},
  \href{http://dx.doi.org/10.22323/1.314.0537}{PoS {\bf EPS-HEP2017} (2018)
  537. 12 p}. \url{https://cds.cern.ch/record/2670656}.

\bibitem{michizono}
S.~Michizono. private communication.

\bibitem{XFEL}
W.~Decking et al.,
  \href{http://dx.doi.org/10.18429/JACoW-IPAC2019-TUPRB020}{{\em {Status of the
  European XFEL}\/}, } in {\em {Proceedings, 10th International Particle
  Accelerator Conference (IPAC2019): Melbourne, Australia, May 19-24, 2019}},
p.~TUPRB020.
\newblock

\bibitem{wang}
Y.~Wang. private communication.

\bibitem{shatilov-2019}
D.~Shatilov, {\em {Beam-Beam Effects with 4 IPs, FCC Week 2019, Brussels}\/}, .
  \url{https://indico.cern.ch/event/727555/contributions/3452787/}.

\bibitem{oide-2019}
K.~Oide, {\em {Issues for the next step, FCC Week 2019, Brussels}\/}, .
  \url{https://indico.cern.ch/event/727555/contributions/3452786/}.

\bibitem{zimmermann-pol}
F.~Zimmermann, {\em {FCC-ee Design Overview}\/},  {FCC Week 2019, Brussels}
  (2019)  . \url{https://indico.cern.ch/event/727555/contributions/3447588/}.

\bibitem{skekb}
{SuperKEKB Collaboration}, K.~Akai, K.~Furukawa, and H.~Koiso, {\em {SuperKEKB
  Collider}\/},  \href{http://dx.doi.org/10.1016/j.nima.2018.08.017}{Nucl.
  Instrum. Meth. {\bf A907} (2018)  188--199},
  \href{http://arxiv.org/abs/1809.01958}{{\tt arXiv:1809.01958
  [physics.acc-ph]}}.

\bibitem{skekb2}
K.~Akai et al., {\em {SuperKEKB status -- Belle II physics run started}\/},
  \href{http://dx.doi.org/10.22323/1.340.0701}{PoS {\bf ICHEP2018} (2019)
  701}.

\bibitem{stc}
P.~Piminov, {\em {Project for a Super Charm-Tau Factory at BINP}\/},
  \href{http://dx.doi.org/{10.1134/S1547477118070579}}{{Phys.~Part.~Nuclei~Lett.}
  {\bf 15} (2018)  732}.

\bibitem{fcchh}
M.~Benedikt et al., \href{http://dx.doi.org/10.1140/epjst/e2019-900087-0}{{\em
  FCC-hh: The Hadron Collider\/}, The European Physical Journal Special Topics
  {\bf 228} (Jul, 2019)  755--1107}.
  \url{https://doi.org/10.1140/epjst/e2019-900087-0}.

\bibitem{helhc}
F.~Zimmermann et al., {\em {Future Circular Collider}\/},   CERN-ACC-2018-0059,
  CERN, Geneva, Dec, 2018.
\newblock \url{https://cds.cern.ch/record/2651305}.
\newblock Published in Eur. Phys. J. ST.

\bibitem{Delahaye:2013jla}
J.-P. Delahaye et al., {\em {Enabling Intensity and Energy Frontier Science
  with a Muon Accelerator Facility in the US: A White Paper}\/},  in {\em
  {Proceedings, CSS2013 Minneapolis, US, July 29-August 6, 2013}}.
\newblock 2013.
\newblock \href{http://arxiv.org/abs/1308.0494}{{\tt arXiv:1308.0494
  [physics.acc-ph]}}.

\bibitem{accel:muon3-sr1}
{M.~Palmer, Discussion of the Scientific Potential of Muon Beams, FNAL, 18
  November 2015.}

\bibitem{Antonelli:2015nla}
M.~Antonelli, M.~Boscolo, R.~Di~Nardo, and P.~Raimondi, {\em {Novel proposal
  for a low emittance muon beam using positron beam on target}\/},
  \href{http://dx.doi.org/10.1016/j.nima.2015.10.097}{Nucl. Instrum. Meth. {\bf
  A807} (2016)  101--107}, \href{http://arxiv.org/abs/1509.04454}{{\tt
  arXiv:1509.04454 [physics.acc-ph]}}.

\bibitem{accel:lemma2019}
{M.~Biagini et al., Positron Driven Muon Source for a Muon Collider: Recent
  Developments, IPAC'19, Melbourne, MOZZPLS2 (2019).}

\bibitem{wim2}
A.~Gonsalves et al. PRL {\bf 122} (2019)  084801.

\bibitem{wim3}
S.~Steinke et al. Nature {\bf 530} (2016)  190.

\bibitem{wim4}
{R. D'Arcy et al.} Phys. Rev. Lett. {\bf 122} (2019)  034801.

\bibitem{wim5}
{V.~Shpakov et al.} Phys. Rev. Lett. {\bf 122} (2019)  114801.

\bibitem{wim6}
{J. van Tilborg et al.} Phys. Rev. Lett. {\bf 115}  184802.

\bibitem{wim7}
{R. Pompili et al.} Phys. Rev. Lett. {\bf 121} (2018)  174801.

\bibitem{wim8}
{J.-H. R{\"o}ckemann et al.} {PRAB} {\bf 21} (2018)  122801.

\bibitem{wim9}
{C.A.~Lindstrom et al.} Phys. Rev. Lett. {\bf 121} (2018)  194801.

\bibitem{wim10}
{D.~Marx et al.}, {\em {Simulation studies for characterizing ultrashort
  bunches using novel polarizable X-band transverse deflection structures}\/},
  {submitted} (2019)  .

\bibitem{wim11}
{A.R.~Maier et al.}, {\em {Continuous 24-hour operation of a laser-plasma
  accelerator}\/},  {in preparation}  .

\bibitem{wim12}
{I. Blumenfeld et al.} {Nature} {\bf 455} (2007)  741.

\bibitem{wim13}
{M.~Litos et al.} {Plasma Phys.~Control Fusion} {\bf 58} (2016)  034017.

\bibitem{wim14}
{M. Litos et al.} Nature (2014)  13882. \url{{10.1038/nature}}.

\bibitem{wim15}
{S.~Doche et al.} {Nat.~Sci.~Rep.} {\bf 7} (2017)  14180.

\bibitem{wim16}
{S.~Corde et al.} Nature {\bf 524} (2015)  552.

\bibitem{wim17}
{E.~Adli et al. (AWAKE Collaboration)} Nature {\bf 561} (2018)  {363--367}.

\bibitem{wim18}
{E.~Adli et al. (AWAKE Collaboration)} Phys. Rev. Lett. {\bf 122} (2019)
  054802.

\bibitem{wim19}
{M.~Turner et al. (AWAKE Collaboration)} Phys. Rev. Lett. {\bf 122} (2019)
  05801.

\bibitem{wim20}
{T.~Mehrling et al.} {Phys.~Plasmas} {\bf 25} (2018)  056703.

\bibitem{wim21}
{A. Martinez de la Ossa et al.} Phys. Rev. Lett. {\bf 121} (2018)  064803.

\bibitem{wim22}
{M.~Th{\'e}venet et al.} PRAB {\bf 22}  05130.

\bibitem{wim1}
{ALEGRO Collaboration} {arXiv:} {\bf {1901.10370v2}} (2019)  .

\bibitem{Shiltsev19}
V.~Shiltsev, {\em {Accelerator-based Neutrino beams}\/},  {Open Symposium
  ESPPU, Granada} (2019)  .
  \url{https://indico.cern.ch/event/808335/contributions/3365147/}.

\bibitem{GF-LOI}
{Gamma Factory study group}, {\em {Gamma Factory Proof-of-Principle
  Experiment}\/},
  \href{http://cds.cern.ch/record/2690736/files/SPSC-I-253.pdf}{{\color{blue}{CDS~link}}}
  (2019)  .

\bibitem{jensen-granada}
E.~Jensen, {\em {Energy efficiency of HEP infrastructures}\/},  {Open Symposium
  ESPPU, Granada} (2019)  .
  \url{https://indico.cern.ch/event/808335/contributions/3365146/}.

\bibitem{HTSapps}
M.~Noe et al., {\em {HTS Applications}\/},  {WAMSDO Proceedings} (2009)  .
  \url{http://cds.cern.ch/record/1163936/files/p94.pdf}.
  https://indico.cern.ch/event/775529/.

\bibitem{PTCOG}
{\em {Particle Therapy Co-Operative Group}\/}, .
  \url{https://www.ptcog.ch/index.php/facilities-in-operation}.

\bibitem{gatoroid}
L.~Bottura et al., {\em {GaToroid, A compact non-rotating gantry for charged
  particle therapy}\/},  {CERN Knowledge Transfer} (2019)  .
  \url{https://kt.cern/technologies/gatoroid}.

\bibitem{Schippers}
J.~M. Schippers, {\em {Miniaturizing proton therapy: a technical challenge with
  unclear clinical impact}\/},  {Int. J. Rad. Onc. Biol. Phys.} {\bf {no.~95}}
  (2016)  149--153. \url{http://dx.doi.org/10.1016/j.ijrobp.2016.02.030}.

\bibitem{GEM}
{\em GEMPix\/}, . \url{https://kt.cern/success-stories/gempix}.

\bibitem{DECTRIS}
DECTRIS, {\em {Pre-Clinical and Medical Research}\/}, .
  \url{https://www.dectris.com/applications/medical/preclinical-and-medical-research}.

\bibitem{COMPACTLIGHT}
{\em CompactLight\/}, . \url{http://www.compactlight.eu/Main/HomePage}.

\bibitem{CSNS}
S.~N. Fu and S.~Wang,
  \href{http://dx.doi.org/10.18429/JACoW-IPAC2019-MOZPLM1}{{\em {Operation
  status and upgrade of CSNS}\/}, } in {\em {Proceedings, 10th International
  Particle Accelerator Conference (IPAC2019): Melbourne, Australia, May 19-24,
  2019}}, p.~MOZPLM1.
\newblock 2019.
\newblock \url{https://doi.org/10.18429/JACoW-IPAC2019-MOZPLM1}.

\bibitem{bib:DC-talk}
D.~Contardo, {\em
  \href{https://indico.cern.ch/event/808335/contributions/3365103/attachments/1843583/3023789/DC_EPPSUGranada_14052019.pdf}{Lessons
  learned from past detector R\&D}\/},  Presentation, Open Symposium, Granada,
  Spain (13-16 May 2019)  .

\bibitem{bib:LS-talk}
L.~Linssen, {\em
  \href{https://indico.cern.ch/event/808335/contributions/3365102/attachments/1843581/3024457/Linssen_Granada_detectors_May2019.pdf}{Technological
  challenges of future HEP experiments}\/},  Presentation, Open Symposium,
  Granada, Spain (13-16 May 2019)  .

\bibitem{bib:FS-talk}
F.~Sefkow, {\em
  \href{https://indico.cern.ch/event/808335/contributions/3365101/attachments/1843661/3023942/DetRD-Granada19-FSefkow-v3.pdf}{Detector
  R\&D for future HEP experiments}\/},  Presentation, Open Symposium, Granada,
  Spain (13-16 May 2019)  .

\bibitem{Garcia-Sciveres:2017ymt}
M.~Garcia-Sciveres and N.~Wermes, {\em {A review of advances in pixel detectors
  for experiments with high rate and radiation}\/},
\href{http://dx.doi.org/10.1088/1361-6633/aab064}{Rept. Prog. Phys. {\bf 81}
  (2018) no.~6, 066101}.

\bibitem{Dannheim:2673779}
{CLICdp Collaboration}, \href{http://dx.doi.org/10.23731/CYRM-2019-001}{{\em
  {Detector Technologies for CLIC}}}.
\newblock CERN Yellow Reports: Monographs. May, 2019.
\newblock \url{http://cds.cern.ch/record/2673779}.

\bibitem{Curras:2673324}
E.~Curr\'{a}s et al., {\em {Inverse Low Gain Avalanche Detectors (iLGADs) for
  precise tracking and timing applications}\/},   arXiv:1904.02061, Apr, 2019.

\bibitem{Sefkow:2015hna}
F.~Sefkow et al., {\em {Experimental Tests of Particle Flow Calorimetry}\/},
  \href{http://dx.doi.org/10.1103/RevModPhys.88.015003}{Rev. Mod. Phys. {\bf
  88} (2016)  015003},
\href{http://arxiv.org/abs/1507.05893}{{\tt arXiv:1507.05893
  [physics.ins-det]}}.

\bibitem{THOMSON200925}
M.~Thomson, {\em Particle flow calorimetry and the PandoraPFA algorithm\/},
  \href{http://dx.doi.org/https://doi.org/10.1016/j.nima.2009.09.009}{Nuclear
  Instruments and Methods in Physics Research Section A: Accelerators,
  Spectrometers, Detectors and Associated Equipment {\bf 611} (2009) no.~1, 25
  -- 40}.

\bibitem{RevModPhys.90.025002}
S.~Lee, M.~Livan, and R.~Wigmans,
  \href{http://dx.doi.org/10.1103/RevModPhys.90.025002}{{\em Dual-readout
  calorimetry\/}, Rev. Mod. Phys. {\bf 90} (Apr, 2018)  025002}.
  \url{https://link.aps.org/doi/10.1103/RevModPhys.90.025002}.

\bibitem{shashlik}
G.~Atoyan et al., {\em {Lead-scintillator electromagnetic calorimeter with
  wavelength shifting fiber readout}\/},
  \href{http://dx.doi.org/0.1016/0168-9002(92)90773-W}{Nucl. Instrm. Meth. A
  {\bf 320} (1992)  144}.

\bibitem{Acosta:1991ap}
D.~Acosta et al., {\em {Electron, pion and multiparticle detection with a lead
  / scintillating - fiber calorimeter}\/},
\href{http://dx.doi.org/10.1016/0168-9002(91)90062-U}{Nucl. Instrum. Meth. {\bf
  A308} (1991)  481--508}.

\bibitem{CLD}
N.~Alipour~Tehrani et al., {\em CLD -- a detector concept for FCC-ee\/},
  LCD-Note-2018-004, December, 2018.

\bibitem{CMSCollaboration:2015zni}
{CMS Collaboration}, {\em {Technical Proposal for the Phase-II Upgrade of the
  CMS Detector}\/}, . CERN-LHCC-2015-010, LHCC-P-008, CMS-TDR-15-02.

\bibitem{Aaij:2244311}
{LHCb Collaboration}, {\em {Expression of Interest for a Phase-II LHCb Upgrade:
  Opportunities in flavour physics, and beyond, in the HL-LHC era}\/},
  CERN-LHCC-2017-003, CERN, Geneva, Feb, 2017.
\newblock \url{https://cds.cern.ch/record/2244311}.

\bibitem{Grenier_2014}
G.~Grenier, \href{http://dx.doi.org/10.1088/1748-0221/9/09/c09006}{{\em A
  hadronic calorimeter with Glass {RPC} as sensitive medium\/}, Journal of
  Instrumentation {\bf 9} (sep, 2014)  C09006}.

\bibitem{bib:AIDA2020}
\url{http://aida2020.web.cern.ch/}.

\bibitem{bib:ATTRACT}
\url{https://attract-eu.com/}.

\bibitem{bib:testbeamDB}
\url{http://www.cern.ch/tbdb}.

\bibitem{bib:db}
\url{http://cern.ch/irradiation-facilities}.

\bibitem{bib:RTF-SG}
\url{https://rtf-sg-info.web.cern.ch/}.

\bibitem{bib:cern-radiation}
B.~Gkotse et al., \href{http://dx.doi.org/10.1109/RADECS.2017.8696163}{{\em
  Irradiation Facilities at CERN\/}, } in {\em 2017 17th European Conference on
  Radiation and Its Effects on Components and Systems (RADECS)}, pp.~1--7.
\newblock Oct, 2017.

\bibitem{bib:irradiation-talk}
\url{https://indico.desy.de/indico/event/16161/session/12/contribution/19/material/slides/0.pptx}.

\bibitem{bib:testbeam-software}
B.~Gkotse et al., \href{http://dx.doi.org/10.1109/RADECS.2017.8696209}{{\em
  {Towards a Unified Environmental Monitoring, Control and Data Management
  System for Irradiation Facilities: the CERN IRRAD Use Case}\/}, } in {\em
  {Proceedings, 17th European Conference on Radiation and Its Effects on
  Components and Systems (RADECS 2017)}}, p.~279.

\bibitem{bib:EUDET}
\url{https://www.eudet.org/}.

\bibitem{bib:AIDA}
\url{http://aida2020.web.cern.ch/content/aida}.

\bibitem{bib:geant1}
S.~Agostinelli et al., {\em Geant4 -- A simulation toolkit\/},
  \href{http://dx.doi.org/10.1016/S0168-9002(03)01368-8}{Nucl.\ Instrum.\
  Meth.\ A {\bf 506} (2003) no.~3, 250}.

\bibitem{bib:geant2}
J.~Allison et al., {\em Geant4 developments and applications\/},
  \href{http://dx.doi.org/10.1109/TNS.2006.869826}{IEEE Transactions on Nuclear
  Science {\bf 53} (2006) no.~1, 270}.

\bibitem{bib:geant3}
J.~Allison et al., {\em Recent developments in Geant4\/},
  \href{http://dx.doi.org/10.1016/j.nima.2016.06.125}{Nuclear Instruments and
  Methods in Physics Research A {\bf 835} (2016)  186}.

\bibitem{bib:EFCA-detector}
\url{https://ecfa-dp.desy.de/}.

\bibitem{bib:CDV-talk}
C.~Da~Via, {\em
  \href{https://indico.cern.ch/event/808335/contributions/3365100/attachments/1843614/3023846/Technological_Synergies_in_Instrumentation_RD_and_Industry.pdf}{Technological
  synergies with non-HEP experimental programs and industry}\/},  Presentation,
  Open Symposium, Granada, Spain (13-16 May 2019)  .

\bibitem{bib:medipix}
\url{https://medipix.web.cern.ch/}.

\bibitem{bib:ERDIT}
\url{http://erdit.eu/}.

\bibitem{bib:IB-talk}
I.~Bird, {\em
  \href{https://indico.cern.ch/event/808335/contributions/3365098/attachments/1842604/3024260/ESPP-CurrentComputingModels-150519.pdf}{Current
  HEP computing model}\/},  Presentation, Open Symposium, Granada, Spain (13-16
  May 2019)  .

\bibitem{bib:MK-talk}
M.~Kasemann, {\em
  \href{https://indico.cern.ch/event/808335/contributions/3365096/attachments/1844453/3025607/Challenges.pdf}{Future
  challenges of HEP computing}\/},  Presentation, Open Symposium, Granada,
  Spain (13-16 May 2019)  .

\bibitem{bib:atlas-computing}
\url{https://twiki.cern.ch/twiki/bin/view/AtlasPublic/ComputingandSoftwarePublicResults}.

\bibitem{bib:RJ-talk}
R.~Jones, {\em
  \href{https://indico.cern.ch/event/808335/contributions/3365097/attachments/1843340/3023265/ComputingModelLessonsLearned.pdf}{Computing
  models: Lessons learned}\/},  Presentation, Open Symposium, Granada, Spain
  (13-16 May 2019)  .

\bibitem{bib:MG-talk}
M.~Girone, {\em
  \href{https://indico.cern.ch/event/808335/contributions/3367987/attachments/1844313/3025463/ESPP-MG.pdf}{R\&D:
  HEP computing infrastructure}\/},  Presentation, Open Symposium, Granada,
  Spain (13-16 May 2019)  .

\bibitem{bib:GS-talk}
G.~Stewart, {\em
  \href{https://indico.cern.ch/event/808335/contributions/3367988/attachments/1843865/3025660/eppsu-software-rd.pdf}{R\&D:
  HEP software}\/},  Presentation, Open Symposium, Granada, Spain (13-16 May
  2019)  .

\bibitem{bib:ESCAPE}
\url{https://cordis.europa.eu/project/rcn/219246/factsheet/en}.

\bibitem{bib:openlab}
\url{https://openlab.cern/}.

\bibitem{bib:HSF}
\url{https://hepsoftwarefoundation.org/}.

\bibitem{bib:HSF-plan}
{The HEP Software Foundation Collaboration}, J.~Albrecht et al., {\em {A
  Roadmap for HEP Software and Computing R\&D for the 2020s}\/},
  \href{http://dx.doi.org/10.1007/s41781-018-0018-8}{Comput. Softw. Big. Sci.
  {\bf 3} (2019)  7}.

\end{thebibliography}\endgroup
\end{small}
\end{document}